\documentclass[11pt,a4paper, english,usenames]{article}
\usepackage [latin1]{inputenc}
\usepackage{multicol}
\usepackage{xcolor}
\usepackage[margin=1in]{geometry}  
\usepackage[parfill]{parskip}    		
\usepackage{graphicx}			

\usepackage{amsmath,amssymb,amsthm,bm} 
\usepackage{enumerate}
\usepackage{mathtools}
\usepackage{empheq}
\usepackage{mhsetup}
\usepackage[color=yellow]{todonotes}
\usepackage{minitoc}
\usepackage{url}	
\usepackage{esint}
\usepackage[colorlinks]{hyperref}
\usepackage{outlines}[enumerate]
\usepackage{framed,comment}
\usepackage{shadow}
\usepackage{makeidx}
\usepackage{upgreek}
\usepackage[mathscr]{euscript}
\usepackage{pgfplots}
\usepackage{float}
\usepackage{tikz,tikz-cd}
\usepackage{rotating}
\usepackage{graphicx}
\usepackage{caption}
\usepackage{multicol}
\usepackage{cancel}
\usepackage{subcaption}
\usepackage[title]{appendix}
\usepackage{scalerel}
\usepackage{tikz-cd}
\usepackage{pgf, pgffor}
\extrafloats{100}
\numberwithin{equation}{section}

\definecolor{shadecolor}{cmyk}{0,0,0,0.1}
\definecolor{Violet}{cmyk}{0.21,0.22,0,0}
\definecolor{Plum}{cmyk}{0.50,1,0,0}
\definecolor{Periwinkle}{cmyk}{0.57,0.55,0,0}
\definecolor{ForestGreen}{cmyk}{0.91,0,0.88,0.12}
\definecolor{DarkGreen}{cmyk}{0.91,0,0.88,0.60}
\definecolor{OliveGreen}{cmyk}{0.64,0,0.95,0.40}
\definecolor{BrickRed}{cmyk}{0,0.89,0.94,0.28}
\definecolor{DarkOrchid}{cmyk}{0.40,0.80,0.20,0}
\definecolor{Fuchsia}{cmyk}{0.47,0.91,0,0.08}
\definecolor{Mulberry}{cmyk}{0.34,0.90,0,0.02}
\definecolor{Maroon}{cmyk}{0,0.87,0.68,0.32}
\definecolor{Mahogany}{cmyk}{0,0.85,0.87,0.35}
\definecolor{RawSienna}{cmyk}{0,0.72,1,0.45}
\definecolor{brown}{cmyk}{0,0.81,1,0.60}
\definecolor{MyLightMagenta}{cmyk}{0.1,0.8,0,0.1} 
\definecolor{MyDarkBlue}{rgb}{0,0.08,0.45} 

\makeatletter
\@addtoreset{equation}{section}

\makeatother

\makeindex

\renewcommand{\emph}[1]{\textit{#1}}

\newenvironment{answer}[1][Answer]
{
\centerline{\textcolor{shadecolor}{\rule[-2mm]{6.2in}{2mm}}}
\begin{quote}\noindent \textbf{#1.} }
{\mbox{ } \hfill{$\blacktriangle$}
\end{quote} 
\centerline{\textcolor{shadecolor}{\rule[0mm]{6.2in}{2mm}}}
}

\newenvironment{exercise}[1][Exercise]
{\begin{shaded} 
\begin{quote}\noindent 
\advance\leftskip-6pt 
\advance\rightskip6pt
\advance\linewidth-12pt
\textbf{#1.} }
{\mbox{ } \hfill{$\bigstar$}\vskip1sp
\end{quote} 
\end{shaded} }

\newenvironment{exercise-white}[1][Exercise]
{\begin{quote}\noindent 
\advance\leftskip-6pt 
\advance\rightskip6pt
\advance\linewidth-12pt
\textbf{#1.} }
{\mbox{ } \hfill{$\bigstar$}\vskip1sp
\end{quote} }

\newsavebox{\overlongequation}
\newenvironment{dontbotheriftheequationisoverlong}
 {\begin{displaymath}\begin{lrbox}{\overlongequation}$\displaystyle}
 {$\end{lrbox}\makebox[0pt]{\usebox{\overlongequation}}\end{displaymath}}

\newtheorem{theorem}{Theorem}[section]
\newtheorem{lemma}{Lemma}[section]
\newtheorem{corollary}{Corollary}[section]
\newtheorem{proposition}{Proposition}[section]
\newtheorem{remark}{Remark}[section]
\newtheorem{example}{Example}[section]
\theoremstyle{definition}
\newtheorem{definition}{Definition}[section]

\newcommand{\rem}[1]{}  
\newcommand{\bfi}{{\bfseries\itshape}}

\newcommand{\wh}[1]{\widehat{#1}}

\newcommand{\bs}[1]{\boldsymbol{#1}}

\newcommand{\mb}[1]{\mathbf{#1}}
\newcommand{\mbb}[1]{\mathbb{#1}}
\newcommand{\mc}[1]{\mathcal{#1}}
\newcommand{\mcal}[1]{\mc{#1}}
\newcommand{\mrm}[1]{\mathrm{#1}}
\newcommand{\op}[1]{\operatorname{#1}}
\newcommand{\pb}[2]{\left\{#1\,,\,#2\right\}}
\newcommand{\scp}[2]{\left<#1\,,\,#2\right>}

\newcommand{\ad}{\operatorname{ad}}
\newcommand{\Ad}{\operatorname{Ad}}
\newcommand{\dd}{\mathrm{~d}}

\def\AdMR{{\mathrm{Ad}}}
\def\adMR{{\mathrm{ad}}}

\def\p{{\partial}}
\def\pa{{\partial}}
\def\ep{{\epsilon}}
\def\de{{\delta}}
\def\zh{{\bs{\widehat{z}}}}
\def\nh{{\bs{\widehat{n}}}}

\def\bu{{\mathbf{u}}}

\def\bv{{\mathbf{v}}}
\def\bx{{\mathbf{x}}}

\def\Id{{\mathrm{Id}}}
\def\R{{\mathbb{R}}}
\def\curl{\mathrm{curl}}
\def\div{\mathrm{div}}
\def\tr{\mathrm{tr}}
\def\trace{\mathrm{tr}}
\def\Sym{\mathrm{Sym}}

\def\contract{\makebox[1.2em][c]{\mbox{\rule{.6em}
{.01truein}\rule{.01truein}{.6em}}}}

\DeclareFontFamily{U}{mathx}{}
\DeclareFontShape{U}{mathx}{m}{n}{<-> mathx10}{}
\DeclareSymbolFont{mathx}{U}{mathx}{m}{n}
\DeclareMathAccent{\widehat}{0}{mathx}{"70}
\DeclareMathAccent{\widecheck}{0}{mathx}{"71}
\renewcommand{\check}{\widecheck}
\renewcommand{\tilde}{\widetilde}

\def\p{\partial}

\newcommand{\sss}[1]{\scriptscriptstyle{#1}}

\begin{document}

\title{\textbf{
31 Lectures in Geometric Mechanics
}\\
}
\author{Darryl D. Holm\\
d.holm@imperial.ac.uk\\
Department of Mathematics, Imperial College London \\ SW7 2AZ, London, UK}
\date{}
\maketitle



\newpage

\section*{\centerline{Guiding principles of these lectures}}

  \begin{tabular*}{1.0\textwidth}%
     {@{\extracolsep{1cm}}l l}
\begin{minipage}[ ]{1.25in}
\begin{picture}(1,1.4)(0,60)
\includegraphics[width=3.5cm]{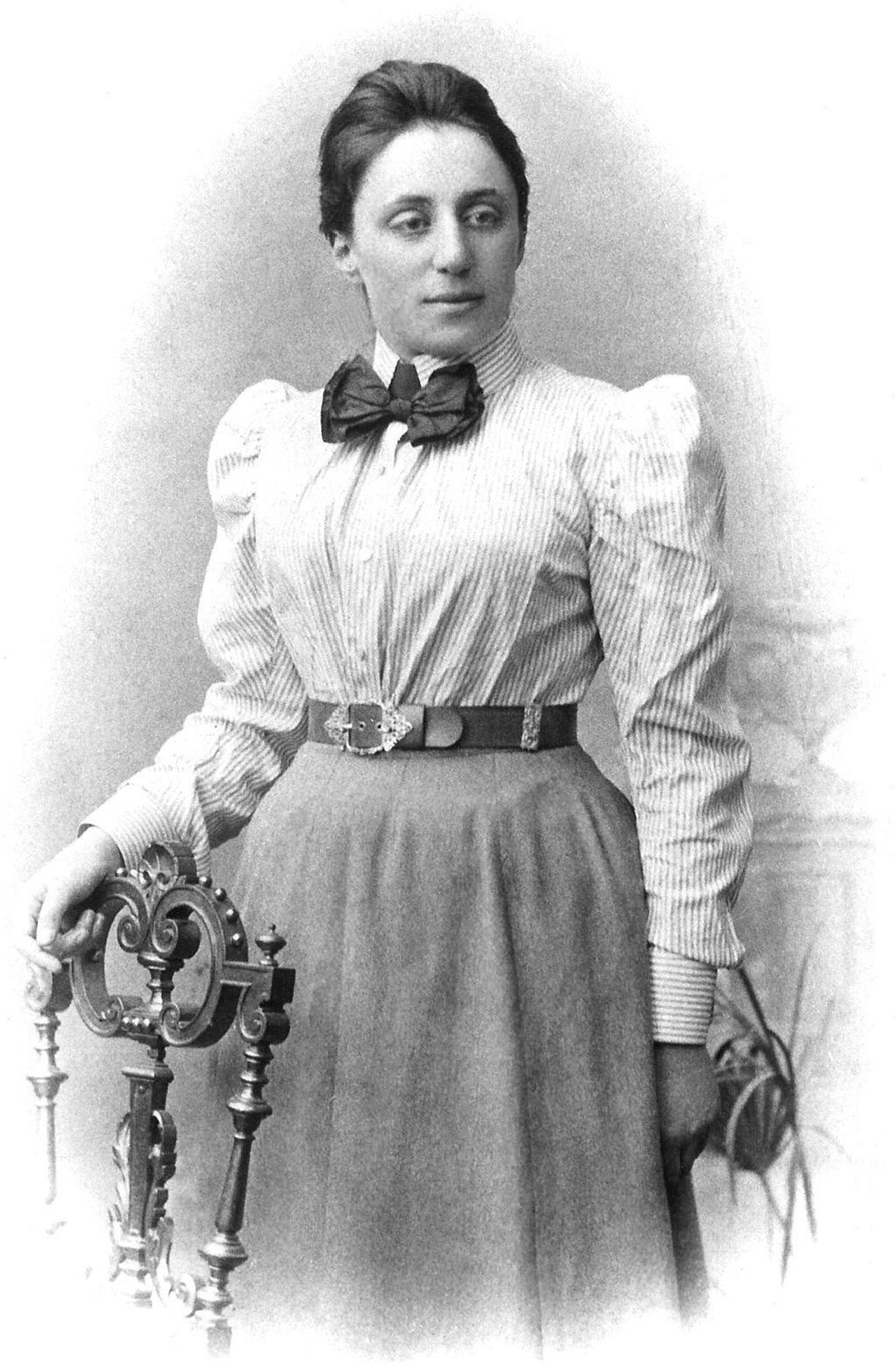} 
\put(-92,-13){Emmy Noether\\}
\end{picture}
\end{minipage}
& 
  \begin{minipage}[  ]{2.7in}\normalsize
  \vspace{2cm}

\color{black}
We are dealing with Lie group invariant variational principles.
\cite{noether1918invariante}
\vspace{2cm}
\color{black}

\vspace{3mm}
\end{minipage}
\end{tabular*}

  \begin{tabular*}{1.0\textwidth}%
     {@{\extracolsep{1cm}}l l}
\begin{minipage}[ ]{1.25in}
\begin{picture}(1,1.4)(0,70)
\includegraphics[width=3.5cm]{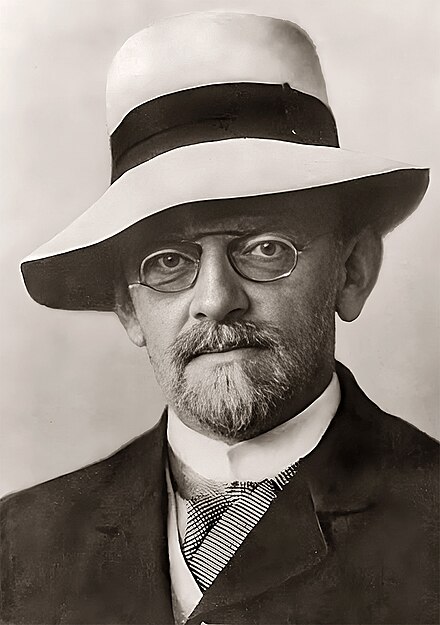} 
\put(-92,-13){David Hilbert\\}
\end{picture}
\end{minipage}
& 
  \begin{minipage}[  ]{2.75in}\normalsize

\color{black}
\vspace{1cm}Geometrical truths are not essentially different from physical ones
in any aspect and they are established in the same way. 
\cite{hilbert1935naturerkennen}
\vspace{15mm}
\color{black}

\vspace{1cm}
\end{minipage}
\end{tabular*}

  \begin{tabular*}{1.0\textwidth}%
     {@{\extracolsep{1cm}}l l}
\begin{minipage}[ ]{1.25in}
\begin{picture}(1,1.4)(0,60)
\includegraphics[width=3.5cm]{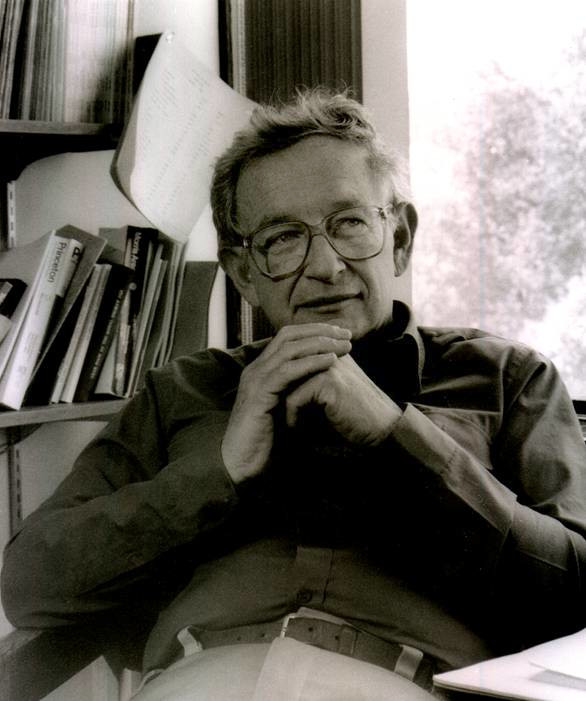} 
\put(-92,-13){Philip Anderson\\}
\end{picture}
\end{minipage}
\vspace{2cm}
& 
  \begin{minipage}[  ]{2.73in}\normalsize

\color{black}
\vspace{1cm}It is only slightly overstating the case to say that Physics is the study of Symmetry. 
\cite{anderson1972more}
\vspace{1cm}
\color{black}

\vspace{2cm}
\end{minipage}
\end{tabular*}

  \begin{tabular*}{1.0\textwidth}%
     {@{\extracolsep{1cm}}l l}
\begin{minipage}[ ]{1.25in}
\begin{picture}(1,1.4)(0,70)
\includegraphics[width=3.5cm]{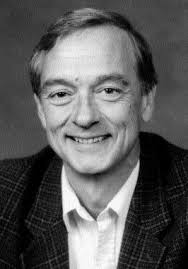} 
\put(-92,-13){Hermann Flaschka\\}
\end{picture}
\end{minipage}
& 
  \begin{minipage}[  ]{2.625in}\normalsize

\color{black}
The most difficult task is to think of workable examples
that will reveal something new. 
\cite{flaschka2015henry}
\color{black}

\vspace{3mm}
\end{minipage}
\end{tabular*}
\vspace{-2cm}

\newpage

\section*{Preface}

Geometric mechanics deals with dynamical systems
defined by Lie group invariant variational principles,
such as geodesic motion on a Lie group $G$ whose metric 
is invariant under the action of the Lie group $G$. An example is 
the variational formulation of Euler's rigid body equations 
in three dimensions whose solutions are then seen to be 
geodesics on the rotation group  $SO(3)$. 
Another example is the variational formulation of Euler's fluid equations 
in three dimensions whose solutions are then seen to be geodesics on 
the manifold of smooth invertible maps (diffeomorphisms) with respect to
the metric given by the fluid's kinetic energy. 

Writing about Geometric Mechanics (GM) is hard, though, because GM is something you do,
not something you simply say, or read about. 
Perhaps, GM should be regarded as a sport, which then would best be taught by practice. 
This is hard, too, though, as Hermann Flaschka told us: 
The most difficult task is to think of workable examples
that will reveal something new, \cite{flaschka2015henry}. 

The exercises and their explicit solutions are not optional reading in this textbook. 
They are essential to understanding the lectures.
The exercises are intended to reveal something new in the steps toward learning geometric mechanics.
The exercises are indented and marked with $\bigstar$
and $\blacktriangle$, respectively.  Moreover, the careful reader will find
that many of the exercises are answered in passing somewhere later in the
text in a more developed context.  

As reviewed in \cite{marle2013henri}, one may say that geometric mechanics 
arose in 1901 when Poincar\'e  in \cite{poincare1901forme} introduced 
a Lie group invariant variational principle which extended Euler's approach 
to the rigid body (resp. heavy top) from $SO(3)$ (resp. $SE(3)$) to arbitrary Lie groups. 
VI Arnold \cite{Ar1966} extended these results to include Euler's ideal fluid equations. 
Subsequently, many symmetry-breaking variants of Euler's ideal fluid equations were included in the 
realm of geometric mechanics, such as the equations of ocean and atmosphere dynamics, 
as well as the equations of ideal plasma physics which form the basis of astrophysics 
and guide the design of magnetic confinement devices such as tokamaks.

One answer to the question, ``Why study geometric mechanics?" besides its mathematical 
depth and beauty, is that geometric mechanics is a framework for investigating the geometric structure 
of many dynamical systems of considerable interest in physics. 

This textbook for learning geometric mechanics is written in the format of lecture notes.
These lecture notes in geometric mechanics are meant to convey insight through
clear definitions and workable examples.
The lecture notes have evolved over decades of teaching and research, as well as writing several other textbooks 
and surveys of geometric mechanics.
The lecture format adopted here is intended to convey the immediacy of the taught course and to 
be useful as a basis for other courses. The lecture notes comprise: 

AP = Applications of Pure maths, e.g., Noether's theorem: Lie group symmetry of Hamilton's variational principle implies conservation laws for its equations of motion.\smallskip

PA = Purifications of Applied maths, e.g., Euler fluid dynamics describes geodesic flow on the manifold of smooth invertible maps acting on the domain of flow. \smallskip

Both AP and PA appear here, though the difference is not mentioned.  It is left to the reader to decide 
whether it was AP or PA in each of the lectures containing almost two hundred solved exercises. 

Many excellent and more encyclopedic texts have been published on the
foundations of the subject of geometric mechanics and its links to symplectic and Poisson
geometry. See, for example, Abraham and Marsden~\cite{AbMa1978},
Arnold~\cite{Ar1979}, Guillemin and Sternberg~\cite{GuSt1984}, Jos\'e
and Saletan~\cite{JoSa98}, Libermann and Marle~\cite{LiMa1987}, Marsden
and Ratiu~\cite{MaRa1994}, McDuff and Salamon~\cite{McSa1995} and many
more. In fact, the scope encompassed by the modern literature on this
subject is a bit overwhelming. In following the theme of reduction by symmetry
from the Euler--Poincar\'e--Noether viewpoint of Lie group invariant variational principles in 
geometric mechanics, I have tried to select only the material that readers will find
absolutely necessary for solving the problems and exercises, at the
level of a beginning postgraduate student.  

The classic references for courses at this level are
are Marsden~\cite{Ma1992}, Marsden and Ratiu~\cite{MaRa1994},
Lee~\cite{Le2003}, Bloch~\cite{Bl2004}, and Ratiu, Tudoran, Sbano, Sousa
Dias and Terra~\cite{RaTuSbSoTe2005}. Other very useful references are
Arnold and Khesin~\cite{ArKh1998} and Olver~\cite{Ol2000}. The reader
may see the strong influences of all these references in these lecture
notes, but expressed at a considerably lower level of mathematical
sophistication than most of the originals. 
See also other geometric mechanics textbooks, such as
\cite{holm2011geometricI,holm2011geometricII,HoScSt2009,marsden1997lectures}.

The scope of the these lectures here is quite limited: a list of the topics in geometric mechanics 
not included in these lectures would fill volumes! For example, 
the necessary elements of calculus on smooth manifolds and the basics of
Lie group theory are only briefly described here. These topics
can be studied in more depth after learning the basics in the context of 
geometric mechanical examples explained here.

\begin{figure}[h!]
\begin{center}
\includegraphics*[width=\textwidth]{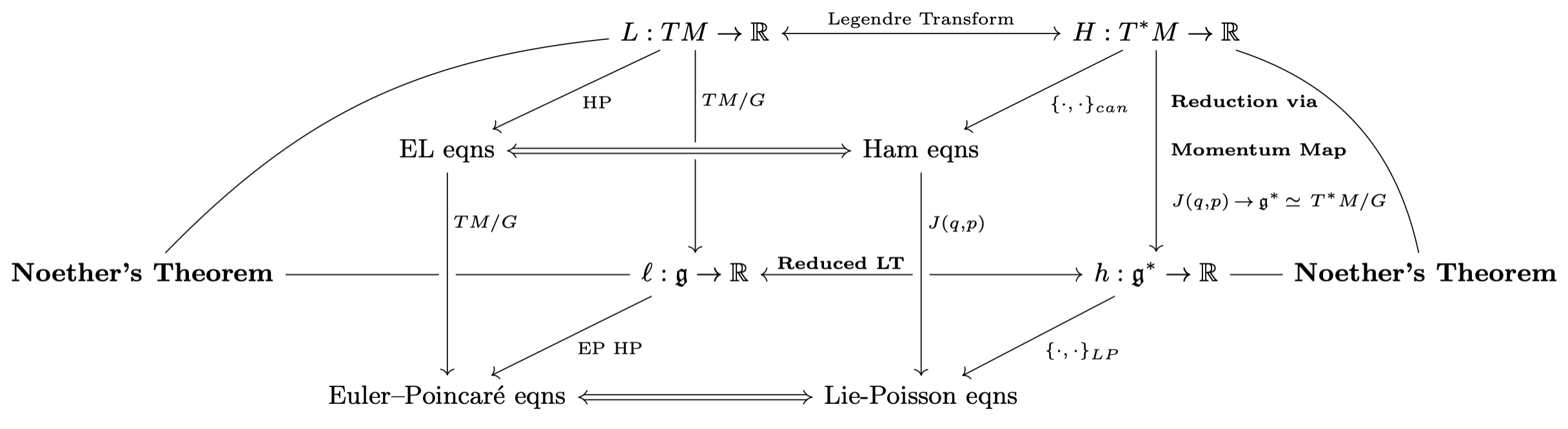}
\end{center}\vspace{-2mm}

\caption{Envision Geometric Mechanics as a cube of  six commuting diagrams.}
\label{GMcube-0}

\end{figure}

To envision geometric mechanics, imagine a 
cube whose six faces are each commutative diagrams of equivariant maps representing
relationships between Lagrangian and Hamiltonian reduction by Lie symmetry. 
See Figure \ref{GMcube-0}. This representation also facilitates teaching the 
fundamentals of geometric mechanics as a reference frame for a sequential course 
of lectures. 

In more detail, the notes consider Lie group invariant Lagrangians in Hamilton's principle 
defined on the tangent space $TG$ of a Lie group $G$. Invariance of such a Lagrangian
under the action of $G$ leads to the symmetry-reduced Euler--Lagrange
equations called the Euler--Poincar\'e equations. In this case, the
$G$-invariant Lagrangian is defined on the Lie algebra $\mathfrak{g}$ of the group and the
variables in its Euler--Poincar\'e equations are defined on the dual Lie algebra, $\mathfrak{g}^*$, where
dual is defined by the pairing obtained in the operation of taking variational derivative. 
On the Hamiltonian side, the Euler--Poincar\'e equations are expressed in terms of  
Lie--Poisson brackets among the accompanying momentum maps, which 
encode both the conservation laws and the geometry of their solution space. 

Thus, these lecture notes use Lie-Noether symmetries of Hamilton's variational principle 
to derive momentum maps leading to symmetry-reduced Euler--Poincar\'e equations 
whose dynamical variables evolve by coadjoint motion. This coadjoint motion endows all these 
symmetry-reduced equations with the same type of solution properties. 
The symmetry-reduced Legendre transformation provides the Hamiltonian formulation 
of these equations in terms of Lie-Poisson brackets. The Lie-Poisson brackets are 
defined on the dual of the Lie algebra of the Lie-Noether symmetry group that is used in 
reduction by symmetry of the Lagrangian in the starting Hamilton principle.%
\footnote{For an interesting discussion of Noether's theorems from the Hamiltonian side, 
see \cite{baez2020getting}.}

The standard Euler--Poincar\'e examples are treated in the lecture notes. These examples include
for example, particle dynamics, Foucault pendulum, rigid body, heavy top and geodesic motion on 
Lie groups. Additional topics deal with Lie symmetry reduction of Fermat's principle in geometric optics. 
The topic of nonlinear water waves deals with the completely integrable system known 
as the Camassa--Holm equation, whose soliton solutions are called \textit{peakons} because their 
velocity profile develops sharp $C^1$ peaks. 
The lectures also include the semidirect-product
Euler--Poincar\'e reduction theorem for ideal fluid dynamics. 
The Euler--Poincar\'e theorem leads to the Kelvin-Noether circulation theorem and 
the semidirect-product Lie--Poisson Hamiltonian formulation for incompressible 
and compressible motions of ideal fluid dynamics and plasma physics \cite{HoMaRa1998a}. 

The lecture series for this textbook is arranged in three Parts, as follows.
\begin{description}
\item
Part 1: Basic Elements of the geometric mechanics approach to 
finite-dimensional dynamical systems.
\item
Part 2: Geometric Mechanics on Manifolds. 
\item
Part 3: Euler--Poincar\'e framework of Continuum Partial Differential Equations. 
\end{description}

The First Part of these lecture notes is designed to introduce undergraduate mathematics and physics students
to the applications of geometric mechanics in finite dimensional dynamical systems of ordinary differential equations.

The Second Part discusses the minimal amount of the theory of manifolds and Lie groups needed to prepare senior undergraduates
and graduate students for the modern applications of geometric mechanics introduced in the Third Part. 

The Third Part 
discusses the geometric mechanics of the partial differential equations governing the dynamics of ideal continuum mechanics 
for fluids and plasmas at the modern research level. 

An aspect of modern applications emphasised here 
is the use of the composition of evolutionary maps for multi-physics, multi-timescale interactions including waves interacting 
with flows in the Euler--Poincar\'e framework in geophysical fluid dynamics (GFD) for ocean and atmosphere dynamics, 
and in magnetohydrodynamics (MHD) for applications in plasma physics such as magnetic confinement fusion (MFC)
and astrophysical processes such as Alfv\'en waves and gravity waves propagating on the Solar tachocline \cite{holm2023lagrangian}. 
The topics covered in each lecture can also be gleaned from its table of contents listed at the onset of each lecture. 

See Appendix \ref{course-outline-app} for a detailed topical outline of the course of lectures 
and some suggestions for postgraduate research projects in geometric mechanics.

\newpage

\section*{Acknowledgements}

I am enormously grateful to many friends and colleagues
whose encouragement, advice and support have helped sustain my interest
in this field. I am particularly grateful to my students and postdocs, for their
participation in my journey of learning a sequential path for teaching them geometric mechanics. 
On this journey, I have learned a lot and I have had a lot of fun. Thanks to them!

I also happily acknowledge the love, encouragement and support that my life partner, Justine Jones, has so 
generously given me to help me finish this effort.


\newpage 

\dosecttoc
\tableofcontents

\newpage

\part{Basic Elements}

\newpage
\vspace{4mm}\centerline{\textcolor{shadecolor}{\rule[0mm]{6.75in}{-2mm}}\vspace{-4mm}}
\section{Introduction}\vspace{-2mm}

\secttoc

\subsection{Geometric Mechanics Framework (GMF)} \label{sec:GMF}


\bfi{Classical mechanics may be visualised as a commuting diagram on the top face of the GMF cube.}

\begin{figure}[h!]
\begin{dontbotheriftheequationisoverlong}
\begin{tikzcd}[row sep=3em, column sep=small, scale=0.5]
& & \hyperref[sec:Ref1]{L:TM\to\mathbb{R}} \arrow[dl,"\text{Lagrange variational principle}"] \arrow[rr,  
"\text{Legendre Transform}" , leftrightarrow] 
& & \hyperref[sec:Ref2]{H:T^\ast M\to\mathbb{R}} \arrow[dl, "\text{Hamilton variational principle}"] \\
& \text{Euler-Lagrange eqns} \arrow[rr, crossing over, Leftrightarrow] & & \text{Hamilton's eqns}\\
\end{tikzcd}
\end{dontbotheriftheequationisoverlong}
\vspace{-15mm}
\caption{ \textbf{\textsf{Envision Classical Mechanics as a commuting diagram} } }
\end{figure}

\begin{figure}[h!]
\begin{center}
\includegraphics*[width=\textwidth]{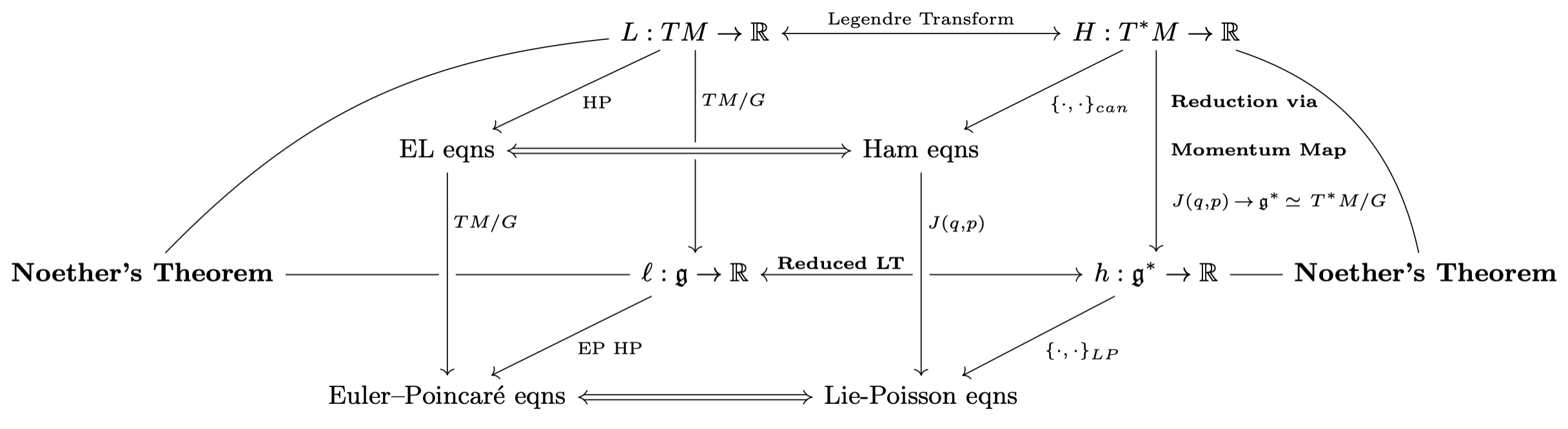}
\end{center}\vspace{-2mm}

\caption{ \textbf{\textsf{Envision Geometric Mechanics as a cube of  six commuting diagrams.}}}
\label{GMcube1}
\end{figure}

\medskip

\begin{definition}
As shown in Figure \ref{GMcube1}, Geometric Mechanics is a \emph{framework of relationships} in dynamical systems following from Lie group invariant variational principles.%
\footnote{A Lie group is a group of transformations that depends smoothly on a set of parameters. A variational principle extremalises an integral. Well known examples of Lie group invariant variational principles include Fermat's principle for geometric optics, Euler's equations for rigid body motion, Euler's equations for ideal fluid dynamics, and Einstein's equations for general relativity.}
\end{definition}
\begin{remark}$\,$\rm\\
$\bullet$ Geometric Mechanics can be used to formulate and analyse systems with constraints, controls, chaos, or stochasticity.\\ 
$\bullet$ Geometric Mechanics has applications in many fields, e.g., robotics, fluid dynamics, plasma physics, celestial mechanics, complex fluids -- including liquid crystals and superfluids -- as well as quantum molecular chemistry.
\end{remark}

\medskip

\begin{remark}\rm 
The reader will do well to remember this page and revisit the geometric mechanics framework 
sketched here occasionally to 
identify their present location in the framework as they progress in their study of these notes.
\end{remark}

\newpage


\section{Counterpoints  between Mathematics and Physics}

\secttoc

\begin{definition}$\,$

Mathematics is the study of emergent logical relationships. \smallskip

Physics is the study of experimentally observable phenomena.
\end{definition}

Because these definitions of these two endeavours appear to be fundamentally different, one might conclude that the two fields might not be related to each other at all! Hence, it is all the more {remarkable} that mathematical models should be so effective in \textit{describing} experimental observations such as the rotational dynamics of a rigid body, or the flow of water in the ocean, or the movement of air in the atmosphere
\cite{wigner1990unreasonable}.

In these notes, we will follow the guiding principles enunciated by Noether, Hilbert, Anderson, and Flaschka,  quoted before the Preface to take advantage of the \emph{counterpoints} between Geometric Mechanics and Physics by using each field of study to interpret the vocabulary of the other, as though Geometric Mechanics and Physics were complementary languages \cite{noether1918invariante,hilbert1935naturerkennen,anderson1972more}.  We will be dealing with Lie group invariant variational principles. We will blend mathematics and physics viewpoints, and in the process we will find workable examples that will reveal new things. 

\textbf{What is this lecture about?} This lecture comprises a series of different problem formulations and workable solutions which together are meant to welcome the brave new reader into the richly varied world of geometric mechanics. This lecture makes a fittingly eclectic introduction to the wide range of applications of geometric mechanics going on inside the rest of the book. It is meant to be \textit{explored} to provide context for \textit{future} understanding. The remaining lectures are laid out progressively as usual for textbooks. As they progress,  readers might consider returning to this lecture occasionally to measure for themselves how far their understanding has extended into the realm of geometric mechanics.

\vspace{4mm}\centerline{\textcolor{shadecolor}{\rule[0mm]{6.75in}{-2mm}}\vspace{-4mm}}
\subsection{Poisson Manifold - Phase Space of Rigid Body }
\index{manifold! Poisson manifold! Poisson bracket}

-- A \emph{Poisson manifold} \index{Poisson manifold} is a manifold equipped with a Poisson bracket structure, which maps two functions on the manifold into a third one. Poisson manifolds provide an arena for formulating classical mechanics, especially Hamiltonian mechanics.\\
-- Poisson brackets describe symmetries, conservation laws, and dynamics.\\
-- In the example of rigid body rotation, the state space of configurations of the rigid body can be recognised as being isomorphic to the Lie group $SO(3)$.\\
-- A Poisson bracket structure governs the dynamics of rigid bodies, whose Poisson manifold is isomorphic 
to the the dual Lie algebra $\mathfrak{so}(3)^*$ of the Lie group $SO(3)$.%
\footnote{As we shall see, the dual Lie algebra of a Lie group is a Poisson manifold.}

1. \textbf{Geometric Mechanics Concept -- Poisson manifold and Poisson bracket}\\ \index{Poisson manifold!Poisson bracket}
-- \textit{A \textbf{manifold} is a space which is locally indistinguishable from $\mathbb{R}^n$ and on which the rules of calculus apply.}\\
-- \textit{A \textbf{Poisson manifold} denoted as $(P,\{\cdot,\cdot\})$ is a manifold endowed with an operation \index{manifold}
\[\{\cdot,\cdot\}\colon C^\infty(P)\times C^\infty(P)\rightarrow C^\infty(P)\]
called the \textbf{Poisson bracket}, which satisfies
\begin{enumerate}
\item $\{F+G,H\}=\{F,H\}+\{G,H\}$ (bilinear)
\item $\{F,H\}=-\,\{H,F\}$ (skew symmetric)
\item $\{F,\{G,H\}\}+\{H,\{F,G\}\}+\{G,\{H,F\}\}=0$ (Jacobi identity)
\item $\{FG,H\}=\{F,H\}G+F\{G,H\}$ (Leibniz rule)
\end{enumerate}
}
\begin{definition}
\emph{Hamiltonian dynamics} \index{Hamiltonian dynamics} is defined on a Poisson manifold $(P,\{\cdot,\cdot\})$ as the relation 
\[\frac{dF}{dt}=\{F,H\}\]
for a certain Hamiltonian $H$ and an arbitrary function $F\in C^\infty(P)$ which may vary with time.
\end{definition}

2. \textbf{Corresponding Physical Concept}\\ 
-- \textbf{Poisson manifold:} The Poisson manifold in classical mechanics comprises the position variables $q$ in the configuration manifold $q\in Q$ along with their canonically conjugate  momentum variables $p$ in the cotangent bundle $(q,p)\in T^*Q$. The variables $(q,p)\in T^*Q$ satisfy the Poisson bracket relation $\{q,p\}=Id$ whose dynamics preserves the symplectic form $dq\wedge dp$. The Poisson manifold for angular momentum dynamics need not be symplectic, though. Instead, the Poisson manifold for angular momentum dynamics for angular momentum dynamics can be identified with the dual $\mathfrak{so}(3)^*$ of the Lie algebra $\mathfrak{so}(3)$, which is the tangent space at the identity of the rotation group $SO(3)$. This observation in Physics goes back to Pauli \cite{pauli1953hamiltonian} in the 1950's for quantum mechanics and to Sudarshan \cite{sudarshan1974classical} in the 1970's for classical mechanics.\\
-- \textbf{Poisson brackets:}
In rigid body rotation, Poisson brackets arise as relations among different physical quantities (such as angular momentum and energy) within the system. Poisson brackets provide a method for determining the dynamics of functions of physical quantities; in particular, for calculating and analysing conserved quantities in rigid body dynamics.

\begin{exercise}
Suppose the Poisson bracket for rigid body dynamics of body angular momentum $\bs{\Pi}\in\mathbb{R}^3$ may be written in $\mathbb{R}^3$ vector form for real functions defined on $\mathbb{R}^3$, denoted $C$, $F$, $H$ and volume element $d^3\Pi$ as
\begin{align*}
\frac{dF}{dt}\,d^3\Pi &= \{F,H\}\,d^3\Pi 
:= \frac{\p C\,}{\p \bs{\Pi}} \cdot \frac{\p F}{\p \bs{\Pi}} \times  \frac{\p H}{\p \bs{\Pi}}\,d^3\Pi
\\ &
= J\left( \frac{dC\,,\, dF\, ,\,dH\,}{d\Pi_1, d\Pi_2,d\Pi_3} \right)\,d^3\Pi 
= dC\wedge dF \wedge dH 
\,,
\end{align*}
where $d^3\Pi:= d\Pi_1\wedge d\Pi_2\wedge d\Pi_3$ for $\bs{\Pi}\in\mathbb{R}^3$ is the volume element.
The wedge $(\wedge)$ notation for differential forms is explained later in lectures \ref{sec-DiffForms} and \ref{diff-form-review}.
For now, one thinks in terms of this Poisson bracket as being associated with the Jacobian determinant 
for a change of variables $({d\Pi_1, d\Pi_2,d\Pi_3})\to({dC,\, dF,\,dH})$.
\\
$\bullet$\quad What is the value of the Poisson bracket $\{C,H\}$?
\\
$\bullet$\quad How would this Poisson bracket look if it were restricted to a level set of the function $C(\Pi_1,\Pi_2,\Pi_3)$?
\\
$\bullet$\quad Would this dynamics preserve the $\mathbb{R}^3$  volume element $d^3\Pi = d\Pi_1\wedge d\Pi_2\wedge d\Pi_3$?
\end{exercise}

\begin{answer}$\,$

$\bullet$\quad The Poisson bracket $\{C,H\}$ defined above vanishes for every function $H$.
\\
$\bullet$\quad When restricted to a level set of the function $C$ parameterised by level set coordinates $(x,y)$ this Poisson bracket would become 
$\{F,H\} = J(F,H) = F_xH_y - F_yH_x$.
 \\
$\bullet$\quad The dynamics of this Poisson bracket is given by
\[
\bs{V}:= \frac{d\bs{\Pi}}{d\;t} = - \,\frac{\p C}{\p \bs{\Pi}} \times  \frac{\p H}{\p \bs{\Pi}}
\,.\]
Since ${\rm div}_\Pi \bs{V} = 0$, this flow preserves volume in $\mathbb{R}^3$ with coordinates $(\Pi_1,\Pi_2,\Pi_3)$.
\end{answer}


\subsection{Hamilton's principle}\label{subsec-HPHP}

\index{Hamilton's principle!geodesics} 
 
1. \textbf{Geometric Mechanics Concept}\\
-- The Euler--Lagrange equations for the Lagrangian $L(q, v_q ): TQ \to \mathbb{R}$ for an arbitrary configuration manifold $Q$ with solution curve $q(t)\in Q$ follow from \textbf{Hamilton's principle}, as 
\index{Hamilton's principle!Euler--Lagrange equation} \index{Euler--Lagrange equations} 
\begin{align}
\begin{split}
0 = \delta S &= \delta \int_0^T L(q, v_q) + \scp{p}{\frac{dq}{dt} - v_q } \, dt
\\&=
\int_0^T \scp{\frac{\p L}{\p q} - \frac{dp}{dt}}{\delta q} + \scp{\frac{\p L}{\p v_q } - p}{\delta v_q } 
+ \scp{\delta p}{\frac{dq}{dt} - v_q }\,dt +\hspace{-4mm} \underbrace{\scp{p}{\delta q}\Big|_0^T}_{\hbox{\emph{Noether term}}}
\hspace{-5mm}
\,.\end{split} 
\label{HP-geodesic}
\end{align}
Here, the variation of the solution path $\delta q(t)$ for example is defined by the tangent to a local deformation $q(t,\ep)=\phi_\ep q(t)$ of the path 
parameterised by $\ep\in \mathbb{R}$ with $q(t,\ep)|_{\ep=0}=\phi_0 q(t)=q(t)$ at fixed time, $t$. That is, one defines the variation of the solution path $q(t)$ as,
\begin{align}
\delta q(t) := \frac{d q(t,\epsilon)}{d \epsilon}\Big|_{\ep=0} = \frac{d \phi_\ep }{d \epsilon}\Big|_{\ep=0}q(t)
\,,
\label{def-var}
\end{align}
and similarly for the other variations in Hamilton's principle \eqref{HP-geodesic}.

Also, the quantity
\begin{align}
\frac{\p L}{\p v_q }
\label{fibre-deriv}
\end{align}
arising in the functional variation of the action integral $S$ with respect to $v_q$ in equation \eqref{HP-geodesic} is called the \emph{fibre derivative}
of the Lagrangian $L$.
\index{Hamilton's principle!fibre derivative} 

Upon assuming that $\delta q$ vanishes at the endpoints in time, 
collecting terms in \eqref{HP-geodesic} yields the \emph{Euler--Lagrange equations} \index{Euler--Lagrange equations} 
\begin{equation}
\frac{d}{dt}\frac{\partial L}{\partial  v_q}\bigg|_{v_q = \frac{dq}{dt}}
-
\frac{\partial L}{\partial q}
=
0
\,.
\label{ELeqns}
\end{equation}
\begin{theorem}\label{Re: Nthm} 
\emph{Noether's theorem} \index{Noether's theorem} \cite{noether1918invariante} states the following. 

Suppose the Lagrangian $L(q, v_q)$
in Hamilton's principle \eqref{HP-geodesic} is invariant under the tangent lifted action $G: TQ\to TQ$  of a 1-parameter Lie group $G$, and $\delta q$ is given by the infinitesimal transformation $\delta q \in T_eG(q)$ of the Lie symmetry group $G$ acting on $q\in Q$ linearised around its identity transformation, $e$. 

That is, suppose the Lagrangian $L(q, v_q)$ is invariant under a particular $\delta q$ given by
\begin{equation}
\delta q :=\frac{d \phi_\ep }{d \epsilon}\Big|_{\ep=0}q(t) =: -\pounds_\xi q
\,,
\label{def: TangentLift}
\end{equation}
where $-\pounds_\xi q$ defines (minus) the Lie derivative of $q\in Q$ with respect to the Lie algebra element 
$\xi\in\mathfrak{g}\simeq T _eG$.%
\footnote{The reason for the minus sign in this definition of  Lie derivative will become clearer, later. Basically, the sign distinguishes between left and right actions of Lie groups on manifolds, but we are not there yet. For an example, see the paragraph about signs in the actions of body and space angular velocities on rotating rigid bodies discussed below in section \ref{SO(3)group-actions}. For those who already know manifold theory, the sign also distinguishes between tangents to pull-backs of smooth mappings of manifolds and tangents to their corresponding push-forwards.}
Then Hamilton's principle $\delta S = 0$ with $S:=\int_0^T L(q,\dot{q})dt$ implies conservation of the endpoint term $\scp{p}{\delta q}$ in \eqref{HP-geodesic}, provided the Euler--Lagrange equations in \eqref{ELeqns} hold.%
\footnote{One of the goals of this text is to explain what the mathematical terms in the statement of Noether's theorem mean in the context of geometric mechanics. For an interesting discussion of Noether's theorems from the Hamiltonian side, 
see \cite{baez2020getting}.}
\end{theorem}
\begin{proof}
Noether's theorem holds by inspection, since the other terms in  \eqref{HP-geodesic} vanish when the Euler--Lagrange equations in \eqref{ELeqns} hold.
\end{proof}

\index{Euler--Lagrange equations! Noether theorem! endpoint term}

\begin{remark}\rm
The Noether endpoint term in Hamilton's principle \eqref{HP-geodesic} may be rewritten equivalently as 
\begin{equation}
\scp{p}{\delta q}_{T^*Q\times TQ} := \scp{p}{\frac{dq(t,\epsilon)}{d\epsilon}\Big|_{\epsilon=0} }_{T^*Q\times TQ} 
=: \scp{p}{-\,\pounds_\xi q}_{T^*Q\times TQ}
\,.\label{NoetherTerm}
\end{equation}
The tangent lifted action of $G$ on $TQ$ denoted by $TQ\to \mathfrak{g}$ is defined by (minus) the  
Lie derivative $-\,\pounds_\xi q$ of $q\in Q$ with respect to the Lie algebra element $\xi\in\mathfrak{g}$ in \eqref{def: TangentLift}.  
In combination with the dual pairing $\scp{\,\cdot\,}{\,\cdot\,}_{T^*Q\times TQ} $ one induces the cotangent lifted action of 
$G$ on $T^*Q$ denoted by $T^*Q\to \mathfrak{g}^*$, as follows, 
\begin{equation}
\scp{\delta p}{ q}_{T^*Q\times TQ} := \scp{\frac{dp(t,\epsilon)}{d\epsilon}\Big|_{\epsilon=0}}{q }_{T^*Q\times TQ} 
= \scp{-\pounds^T_\xi p}{ q}_{T^*Q\times TQ}
\,.\label{DualNoetherTerm}
\end{equation}
\end{remark}

Upon introducing a natural real non-degenerate symmetric pairing $\scp{\,\cdot\,}{\,\cdot\,}_{\mathfrak{g}^*\times \mathfrak{g}}$, 
the Noether endpoint term in \eqref{NoetherTerm} induces a \emph{cotangent lift momentum map} denoted 
\begin{equation}
J(q,p) : T^*Q \to \mathfrak{g}^*
\,.\label{Cotangent-momap1}
\end{equation}
This is accomplished by defining the diamond $(\diamond)$ operation as
\begin{equation}
\scp{p}{-\,\pounds_\xi q}_{T^*Q\times TQ} =: \scp{p\diamond q}{\xi}_{\mathfrak{g}^*\times \mathfrak{g}} 
 =: \scp{J(q,p)}{\xi}_{\mathfrak{g}^*\times \mathfrak{g}} =: J^\xi(q,p)  
\,.\label{Cotangent-momap2}
\end{equation}

\begin{example}\rm 
The \emph{Noether term} \index{Noether endpoint term} for infinitesimal rotations of the Lie group $SO(3)$ acting on vectors $\mathbf{q}\in \mathbb{R}^3$
induces the following tangent and cotangents  lifts under the Euclidean pairing (dot product) on $\mathbb{R}^3$.
\begin{align}
\begin{split}
\scp{p}{\delta q}_{T^*\mathbb{R}^3\times T\mathbb{R}^3} 
&:=
\scp{p}{-\pounds_\xi q}_{T^*\mathbb{R}^3\times T\mathbb{R}^3} 
=
\mathbf{p}\cdot (-\boldsymbol{\xi} \times\mathbf{q})
=
-\boldsymbol{\xi}\cdot ( \mathbf{q} \times \mathbf{p} )
\,,\\
\scp{\delta p}{ q}_{T^*\mathbb{R}^3\times T\mathbb{R}^3} 
&=
\scp{-\pounds^T_\xi p}{ q}_{T^*\mathbb{R}^3\times T\mathbb{R}^3} 
=
\boldsymbol{\xi} \times \mathbf{p} \cdot \mathbf{q} 
=
(- \,\boldsymbol{\xi} \times \mathbf{p})\cdot \mathbf{q}
\,.\end{split} 
\label{cotang-R3}
\end{align}
\end{example}

\begin{exercise}
Show that the transformations in \eqref{cotang-R3} also follow by
computing the canonical equations for Hamiltonian 
$J^\xi(\mathbf{q},\mathbf{p}) = -\,\boldsymbol{\xi} \cdot (\mathbf{q}\times\mathbf{p})$,
\[
\delta \mathbf{q} = \frac{\p J^\xi}{\p \mathbf{p} } = -\,\boldsymbol{\xi} \times\mathbf{q}
\quad\hbox{and}\quad 
\delta \mathbf{p} = -\,\frac{\p J^\xi } { \p \mathbf{q} } = -\,\boldsymbol{\xi} \times\mathbf{p}
\,.\]
Explain why $\delta \mathbf{q}$ and $\delta \mathbf{p}$ should have the same form in this case. 
\end{exercise}

\subsubsection*{Example: Geodesics} \label{Re: Geodesic}

As mentioned earlier, a manifold is a space where the rules of calculus apply. A Riemannian manifold is a space on which geometric notions such as distance, metric, angles, length, volume, and curvature are also defined.\footnote{See, e.g., Wikipedia \url{https://en.wikipedia.org/wiki/Riemannian_manifold}} Optimal paths in a Riemannian configuration manifold are called geodesics.

\index{geodesic! Riemannian manifold ! metric ! kinetic energy}
When the configuration space $(Q, g)$ is a \emph{Riemannian manifold} with \emph{metric} \index{metric}  $g$, the natural Lagrangian on $(TQ, g)$ is given by the \emph{kinetic energy} \index{ kinetic energy} $K(v) : = \tfrac12 g(q)(v_q,v_q),$ for $q \in Q $ and $v_q \in T_q Q $. In
finite dimensions, in a local chart, this is written as $K(q, v_q)= \tfrac12 g_{ij}(q) v_q^i v_q^j$.

\begin{exercise}
Show that for  the kinetic energy $\frac12 g(v_q, v_q)$ with $v_q=\dot{q}$ and Riemannian metric $g$
the Euler--Lagrange equations \eqref{ELeqns} become (for finite dimensional $Q $ in
a local chart) the  \emph{geodesic equations},
\begin{align}
\ddot{q}^{\,i}+\Gamma_{jk}^{\,i} (q)\dot q^j\dot q^k=0\,,\quad  i
= 1\, ,\ldots, n,
\label{eqn: geodesic}
\end{align}
where one sums repeated indices over their range and the three-index quantities 
$ \Gamma_{jk}^{\,i} (q) =  \Gamma_{kj}^{\,i} (q)$ are symmetric in their lower indices and defined 
in terms of the metric by
\begin{align}
\Gamma_{jk}^{\,i} (q) = \tfrac12g^{il}\bigg(\frac{\partial g_{jl}}{\partial
q^k}+ \frac{\partial g_{kl}}{\partial q^j}-\frac{\partial
g_{jk}}{\partial q^l}\bigg)
,\quad\hbox{with}\quad g_{hi}g^{il}=\delta_h^l\,.
\label{def: ChristSym}
\end{align}
These are the \emph{Christoffel symbols} \index{Christoffel symbols} of the Levi-Civita connection on $(Q,g)$.
\end{exercise}

2. \textbf{Corresponding Physical Concept -- Motion on a curved surface}\\ -- 
Geodesics can be thought of as trajectories of free particles in a manifold. Indeed, the geodesic equation \eqref{eqn: geodesic} implies that the acceleration vector of the curve has no components in the direction of the surface (and therefore it is perpendicular to the tangent plane of the surface at each point of the curve). Consequently, the motion is determined by the bending of the surface. Geodesic motion in general is the natural path of motion in the absence of external forces, following Hamilton's principle of stationary action.  This is also the idea of general relativity, that particles move on geodesics and the bending of space-time is associated with gravity.

\begin{exercise}
Consider a free particle of mass $m$ moving on the Lobachevsky half-plane, $\mathbb{H}^2$, with coordinates $q=(x,y)$ and $y\ge0$. Its Lagrangian is the kinetic energy $K(q, v_q)= \tfrac12 g_{ij}(q) v_q^i v_q^j$ corresponding to the Lobachevsky metric, $g_{ij}=\delta_{ij}/y^2$.
Namely,
\begin{equation}
L=\frac{m}{2}\left (\frac{\dot x^2 +\dot y^2}{y^2}\right ).
\label{Lag-H2-intro}
\end{equation}

\begin{enumerate}
\item
Write the fibre derivative defined in \eqref{fibre-deriv} of the Lagrangian (\ref{Lag-H2-intro}) and 
\item
Compute its Euler-Lagrange equations. 

\item
Evaluate the Christoffel symbols. 
\end{enumerate}
\end{exercise}

\begin{answer}$\,$\smallskip

Fibre derivatives: 
\[
\frac{\partial L}{\partial \dot x} = \frac{m \dot x}{y^2} =: p_x
\quad\hbox{and}\quad
\frac{\partial L}{\partial \dot y} = \frac{m \dot y}{y^2} =: p_y
\,.\]
Euler-Lagrange equations $\frac{d}{dt}\frac{\partial L}{\partial \dot x}=\frac{\partial L}{\partial x}$ and $\frac{d}{dt}\frac{\partial L}{\partial \dot y}=\frac{\partial L}{\partial y}$ yield, respectively:
\begin{equation}
\frac{d}{dt}\left(\frac{ \dot x}{y^2}\right) = 0
\quad\hbox{and}\quad
\frac{d}{dt}\left(\frac{ \dot y}{y^2}\right) = 
-\,\frac{\dot x^2 +\dot y^2}{y^3}
\label{Lag-H2-eqns}
\end{equation}
Expanding these equations yield the Christoffel symbols for the geodesic motion,
\[
\ddot x - \frac{2}{y} \dot x \dot y =0
\,,\quad
\ddot y + \frac{1}{y} \dot x^2 -\,\frac{1}{y} \dot y^2 =0
\quad\Longleftrightarrow\quad
\Gamma^1_{12} = -\,\frac{2}{y} = \Gamma^1_{21} 
\,,\quad 
\Gamma^2_{11} = \frac{1}{y}
\,,\quad 
\Gamma^2_{22} = -\,\frac{1}{y}
\,,
\]
for $(p_1,p_2)=(p_x,p_y)$ and $(q_1,q_2)=(q_x,q_y)$ in $(x_1,x_2)=(x,y)$ planar components.
\end{answer}

As we will see later, it is easy to compute that the geodesics of the two-dimensional Riemannian manifold $\mathbb{H}^2$ are circles and straight lines perpendicular to the $x$-axis. See Exercise \ref{H2-halfplane}. \smallskip

The isometric transformations of the manifold $\mathbb{H}^2$ are linear fractional (Moebius) transformations of the 
complex plane with coordinates $z=x+iy$, 
\begin{equation}
z \to \frac{az+b}{cz+d} 
\quad\hbox{with}\quad
ad-bc = 1
\,,\label{Mobius-xform}
\end{equation}
with real coefficients, $a,b,c,d$,
This is the subgroup of the Moebius transformations that maps the upper half-plane $\mathbb{H}^2$ (given by $z = x + iy\,, y > 0$) into itself. 
\smallskip

These isometric transformations of $\mathbb{H}^2$ are significant in physics. 
For example, they correspond to Lorentz transformations of space-time 
associated with Huygens waves  \cite{sachkov2023lorentzian}.

\vspace{4mm}\centerline{\textcolor{shadecolor}{\rule[0mm]{6.75in}{-2mm}}\vspace{-4mm}}
\subsection{Lie Group $S^1$ -- Elroy \& his beanie -- Two planar coupled rigid bodies}

Continuing with the overview in this lecture of counterpoints between mathematics and physics, 
`Elroy and his beanie' is the title of a beautiful example in \cite{marsden1990reduction} 
which reveals that a mathematical property of geometric mechanics called \emph{holonomy} also conveys
the true nature of the geometric phase measured in the  Aharonov-Bohm effect in quantum physics.
This result recalls Hermann Flaschka's remark quoted before the Preface earlier.
\begin{quote}
The most difficult task is to think of workable examples that will reveal something new 
\cite{flaschka2015henry}. 
\end{quote}
 
1. \textbf{Geometric Mechanics Concept} -- \textbf{Holonomy of a connection 1-form}
\index{Elroy and his beanie,  holonomy}
Consider two \textit{planar} rigid bodies joined together by an axis linking their centres of masses. 

Let $I_1$ and $I_2$ be their principal moments of inertia, and let $\theta_1$ and $\theta_2$ be the angles they
make with a fixed direction in an inertial frame.\footnote{The principal moments of inertia $I_1$ and $I_2$ 
of an elliptical body are the second spatial moments of its area weighted by its density. For constant density, 
$I_1$ and $I_2$ are essentially the semi-major and semi-minor axes.}

Conservation of the total angular momentum for the planar rotations of this system states that 
\begin{align}
I_1\dot{\theta}_1 + I_2\dot{\theta}_2 = \mu
\,,\label{EB-ang}
\end{align}
where $\mu$ is a constant of the motion. The \emph{shape space} \index{shape space} of a system is the space which 
describes the configuration of the system. In this case, shape space is the circle $S^1$ 
specified by the relative angle $\psi=\theta_2 - \theta_1$.  
The formula \eqref{EB-ang} for conservation of angular momentum then reads
\begin{align}
I_1\dot{\theta}_1 + I_2(\dot{\psi}+\dot{\theta}_1) = \mu
\quad\hbox{or}\quad
A_{mech} := d\theta_1 + \frac{I_2}{I_1 + I_2}d\psi = \frac{\mu}{I_1 + I_2} dt
\label{A-mech}
\end{align}
Suppose that the total angular momentum $\mu$ is zero and body \#2 goes through one full revolution so that the relative angle $\psi$ increases from $0$ to $2\pi$. Then Body \#1 rotates by
\begin{align}
\theta_1 = - \,\frac{I_2}{I_1 + I_2} \int_0^{2\pi} d\psi = -\, \left(\frac{I_2}{I_1 + I_2}\right)2\pi 
\,.\label{net-rot}
\end{align}
Since $\theta_1$ is measured relative to a fixed frame, this is the amount that the entire system rotates relative to the fixed frame, each time Body \#2 completes one revolution. 

Here is a geometric interpretation of this calculation. The 1-form $A_{mech}$
in equation \eqref{A-mech} can be regarded as a flat connection for the trivial principal $S^1$ bundle 
$\mathcal{B}: S^1 \times S^1 \to S^1$ given by the projection $\mathcal{B}(\theta_1,\psi)=\theta_1$.
Formula \eqref{net-rot} is the \emph{holonomy} \index{holonomy} of the connection $A_{mech}$, when the shape variable 
(relative angle) makes a periodic traversal of the base circle $0\le\psi\le 2\pi$. 

2. \textbf{Corresponding Physical Concept -- Geometric phase / Berry-Hannay angle}\\ -- 
The connection $A_{mech}$ in equation \eqref{A-mech} is the same connection that appears in identifying \emph{holonomy} in the  Aharonov-Bohm effect in physics. 

-- Holonomy also can explain why a falling cat can always land on its feet (paws). It can reorient itself, even though it is falling with zero angular momentum. However, the holonomy of a falling cat is defined in the more complicated group bundle  $SO(3)\times SO(3)\to SO(3)$. The \emph{shape variable} for this case is again the relative angle, but this time in $SO(3)$ \cite{montgomery1993gauge}. In the next section we will find that a planar model of the Foucault spherical pendulum possesses a similar holonomy to that for Elroy \& his beanie. 
\begin{exercise}
\begin{itemize}
\item
Discuss holonomy of bundle $\mathcal{B}_2: S^1 \times S^1  \times S^1 \to S^1 \times S^1$, in which physically three  planar massive ellipses are pinned together at their centres of mass and there exist two shape parameters corresponding to differences in sequential  azimuthal angles. 
\item
Write the dynamical equations and determine the solution behaviour when the azimuthal shape variables of this system each has a linear restoring force whose natural frequency can resonate with the others.
\end{itemize}
\end{exercise}

\vspace{4mm}\centerline{\textcolor{shadecolor}{\rule[0mm]{6.75in}{-2mm}}\vspace{-4mm}}\vspace{4mm}\centerline{\textcolor{shadecolor}{\rule[0mm]{6.75in}{-2mm}}\vspace{-4mm}}
\subsection{$S^1$ Foucault pendulum - Planar oscillation/precession in a rotating frame} \index{Foucault pendulum!holonomy}

1. \textbf{Geometric Mechanics Concept -- Composition of maps $S_{oscillation}^1\circ S_{rotation}^1$ }\\ --
Hamilton's action principle for the Foucault pendulum is given in planar polar coordinates \index{composition of maps}
\index{Foucault pendulum!composition of maps}
$(r,\theta, \dot{r}, \dot{\theta})\in T\mathbb{R}^2$ by
\begin{align}
\begin{split}
0 = \delta S = \delta \int_0^T L \,dt &= \delta \int_0^T \frac{m}{2}\big(\dot{r}^2 + r^2(\dot{\theta}+\Omega)^2 - \omega^2r^2\big) dt
\,,\end{split}
\label{FP-action}
\end{align}
with constant parameters: rotation rate  $\Omega$ about the vertical axis (e.g., Earth's rotation) and oscillation frequency $\omega=\sqrt{g/l_0}$ with gravitational 
acceleration $g$ and pendulum length $l_0$. 

Taking variations $\delta r := \p_\ep r(t,\ep)|_{\ep=0}$ and 
$\delta \theta := \p_\ep \theta(t,\ep)|_{\ep=0}$ of the action integral $S$ yields
\begin{align}
\begin{split}
0 = \delta S &= m\int_0^T \scp{\dot{r}}{\delta \dot{r}} + \scp{r\Big((\dot{\theta}+\Omega)^2-\omega^2\Big)}{ \delta r} 
+ \scp{r^2(\dot{\theta}+\Omega)}{\delta \dot{\theta}}\,dt
\\& = m\int_0^T \scp{-\ddot{r} + r\Big((\dot{\theta}+\Omega)^2-\omega^2\Big)}{ \delta r} 
- \scp{\frac{d}{dt}\big( r^2(\dot{\theta}+\Omega)\big)}{\delta\theta}\,dt
\\&\hspace{1.5cm} 
\underbrace{
+ \scp{m\dot{r}}{\delta r}\Big|_0^T + \scp{mr^2(\dot{\theta}+\Omega)}{\delta \theta}\Big|_0^T
}_{\ \hbox{Noether endpoint terms}}
\,.\end{split}
\label{FP-var1}
\end{align}
The Noether endpoint terms define linear momentum and angular momentum given by
\begin{align}
\begin{split}
p_r &:= \frac{\p L}{\p \dot{r}} = m \dot{r}
\quad\hbox{and}\quad
p_\theta := \frac{\p L}{\p \dot{\theta}} = mr^2(\dot{\theta}+\Omega)
\,,
\\ \hbox{which satisfy}& \\
\frac{dp_r}{dt} &= mr\big((\dot{\theta}+\Omega)^2-\omega^2\big)
= \frac{p_\theta^2}{mr^3} -m\omega^2r
\quad\hbox{and}\quad
\frac{dp_\theta}{dt}  = 0
\,.
\end{split}
\label{FP-momaps}
\end{align}
The Legendre transform to the Hamiltonian $H(r,\theta,p_r,p_\theta)$ is given by
\begin{align}
\begin{split}
H(r,\theta,p_r,p_\theta) &:= \scp{p_r}{\dot{r}} + \scp{p_\theta}{\dot{\theta}} - L(r,\dot{r},\theta.\dot{\theta})
\\& = \frac{p_r^2}{2m} + \frac{p_\theta^2}{2mr^2} - \Omega p_\theta + \frac{m}{2}\omega^2r^2
\,.\end{split}
\label{FP-Ham}
\end{align}
Hamilton's canonical equations for the Foucault pendulum are given by
\begin{align}
\begin{split}
\dot{r} &= \frac{\p H}{\p p_r} = p_r/m \,, \quad \dot{\theta} = \frac{\p H}{\p p_\theta} = \frac{p_\theta}{mr^2} - \Omega\,,
\\
\dot{p}_r &= -\, \frac{\p H}{\p r} = \frac{p_\theta^2}{mr^3} -m\omega^2r
\,, \quad \dot{p}_\theta = -\,\frac{\p H}{\p \theta} = 0
\,.\end{split}
\label{FP-Ham-eqns}
\end{align}
\begin{exercise}
Show that the solutions of \eqref{FP-Ham-eqns} represent divergence-free flows in $\mathbb{R}^3$ preserving $d^3x= dp_\theta\wedge dr\wedge dp_r$ along the intersections of the level sets of $H$ and $p_\theta$. 
\end{exercise}
\begin{answer}
If we define $(x_1,x_2,x_3):=(r,p_r,p_\theta)$ and $C(\bs{x}) := x_3=p_\theta$ one may write equations \eqref{FP-Ham-eqns} as 
\begin{align}
\begin{split}
\frac{dF}{dt} &= \nabla C\cdot \nabla F \times \nabla H  = \bs{\hat e}_3 \cdot  \nabla F \times \nabla H 
\\& = F_{,1}H_{,2}-F_{,2}H_{,1} =: \{F,H\}_{x_3}
\,,\end{split}
\label{FH-Ham-eqns}
\end{align}
which represents canonical Hamiltonian dynamics of the radial degree of freedom on a level set of $x_3=p_\theta$.
After solving for the radial solution, $r(t)$, then one integrates the $\theta$-equation to reconstruct the azimuthal angle $\theta$ as a function of time. 
In the classical dynamics literature, the integral solution for $\theta(t)$ is called a \textit{quadrature}.
\end{answer}
In using conservation of $p_\theta$ to reduce the $\theta(t)$-solution to a quadrature, one finds the following periodic solution for $r(t)$ with period $T=\omega^{-1}$,
\begin{align}
r(t) = r(0)\sqrt{\cos^2(\omega t) + (\Omega^2/\omega^2) \sin^2(\omega t) }
\,.
\label{FP-r}
\end{align}
However, the quadrature solution for $\theta(t)$ given by integration as
\begin{align}
\int_0^T\dot{\theta}(t)dt  = \theta(T) - \theta(0) = \int_0^T \frac{p_\theta}{mr^2(t)} \,dt - \Omega T
\,,
\label{FP-r}
\end{align}
is \textit{not} periodic. Instead, at the end of each closed orbit $r(t)$ the angle $\theta(t)$ 
has precessed by the cumulative angle $\Delta \theta = \Omega T$.

The rotation of the moving frame with angular frequency $\Omega\ne0$ causes a perturbation in the evolution of the angular orbit $\theta(t)$ and as a result, when the radial orbit $r(t)$ closes at time $T=\omega^{-1}$, the angular orbit does not close. The discrepancy $\Delta \theta = \Omega T$ in the closure of the angular orbit after one period $T$ of the radial orbit arises because adding the term  $p_\theta \Omega$ to the Hamiltonian $H(r,\theta,p_r,p_\theta)$ places the Foucault pendulum into a rotating frame with frequency $\Omega$. Thus, one says the motion is a \emph{composition of maps} $S_{osc}^1\circ S_{rot}^1$ for oscillation and rotation. \index{composition of maps}

2. \textbf{Corresponding Physical Concept -- Geometric phase / Berry-Hannay angle}\\ -- 
The angular discrepancy $\Delta \theta = \Omega T$ in the angular orbit after one period $T$ of the radial orbit is called the Berry-Hannay angle corresponding to the angular displacement of the oscillation plane over one radial period. The physics concept underlying the Barry-Hannay deviation angle is known mathematically as \emph{holonomy}. Besides Foucault's measurement of the latitude on Earth, the physical phenomenon his pendulum demonstrates is that \textit{the vibrational degree of freedom in a dynamical system always excites the system's rotational degree of freedom}. 

\begin{exercise}
Use Hamilton's principle to derive the equations of motion and discuss the solution behaviour of a Foucault pendulum in planar polar coordinates when coupled to a planar azimuthal degree of freedom $\Theta(t)$ comprising a disc of fixed radius $R$ and mass $M$ with a linear angular restoring force given by
\[
M\ddot{\Theta}(t) + \kappa \Theta(t) = 0
\,,\]
for constant parameter $\varpi^2=\kappa/M$. In particular, what conditions are required for the planar Foucault pendulum in an oscillatory rotating frame to precess by regular steps in its azimuthal angle? 
\end{exercise}

\begin{answer}

The Lagrangian for this situation is independent of the azimuthal angle of the planar Foucault pendulum motion:
\[
L(\Theta,\dot{\Theta},r,\dot{r},\dot{\theta}) 
=
\frac12MR^2 \big( \dot{\Theta}^2 - \varpi^2\Theta^2 \big)  
+  \frac{m}{2}\big(
\dot{r}^2 + r^2(\dot{\theta} + \dot{\Theta})^2 - \frac12 \omega^2 r^2 \big)
\,.
\]
This Lagrangian implies two canonical momentum variables for the radial planar pendulum,
\[
p_\theta := \frac{\p L}{\p \dot{\theta}} = mr^2 \big( \dot{\theta} + \dot{\Theta} \big)
\quad\hbox{and}\quad
p_r := \frac{\p L}{\p \dot{r}} = m\dot{r}
\,.
\]
A Legendre transformation in these canonical pendulum variables produces 
the following Hamilton principle, cf. equation \eqref{FP-Ham},
\[
0 = \delta S = \delta \int_0^T 
\frac12MR^2 \big( \dot{\Theta}^2 - \varpi^2\Theta^2 \big) 
- p_\theta \dot{\Theta}
+ \frac{p_\theta^2}{2mr^2} + \frac{p_r^2}{2m}  + \frac{m}{2}\omega^2r^2
\,dt\,.
\]
The rest of the solution follows as above, but with the constant rotation rate $\Omega$ replaced by $\dot{\Theta}(t)$, 
so that finally the solution for $\theta(t)$ given by integration as
\begin{align}
\int_0^T\frac{d}{dt}\big(\theta(t) + \Theta(t) \big)dt = \int_0^T \frac{p_\theta}{mr^2(t)} \,dt 
\,.
\label{FP-r-Theta}
\end{align}
Thus, at the end of each closed periodic orbit of $r(t)$ the angle $\theta(t)$ 
has precessed by the cumulative angle $\Delta \theta = \Theta(T) - \Theta(0) $,
where $\Theta(t)$ is periodic with period $1/\varpi$.

Thus, the Foucault pendulum in an oscillatory rotating frame does not precess regularly,
except in the resonant case $n\omega + m\varpi = 0$ in which $m$ and $n$ are integers.

\end{answer}

\vspace{4mm}\centerline{\textcolor{shadecolor}{\rule[0mm]{6.75in}{-2mm}}\vspace{-4mm}}
\subsection{Lie Group $SO(3)$}\label{SO(3)group-actions} | \textbf{Rotational States of Rigid Body} | The Lie group $SO(3)$ represents all possible rotations of a rigid body, providing a global description of its rotational states.

1. \textbf{Geometric Mechanics Concept}\\
-- \textbf{Definition (group)}
\noindent
A group $G$ is a set of elements possessing
\begin{itemize}
\item a binary product $G\times G\rightarrow G$ such that
\begin{itemize} 
\item the product of $g$ and $h$ is written $gh$ for left group action ($hg$ for right group action);
\item the product is associative:  $(gh)k=g(hk)$ for left action; $k(hg)= (kh)g$ for right;
\end{itemize}
\item a unique identity element, denoted by $e$, such that $ge=eg=g\quad\quad\forall g\in G$;
\item an inverse operation $G\rightarrow G$ such that $g^{-1}g=gg^{-1}=e$.
\end{itemize}

\noindent
-- \textbf{Definition (Lie group)}
 \noindent
A \textbf{Lie group} is a smooth manifold which is also a group, so that the binary product and inversion are smooth functions. 

\medskip
\noindent
\textbf{Examples}
\begin{itemize}
\item The manifold of invertible square $n\times n$ real matrices is a Lie group denoted by GL$(n,\mathbb{R})$.
\item The manifold of invertible square $n\times n$ matrices with unit determinant is a Lie group denoted by SL$(n,\mathbb{R})$ and called the \textbf{special linear group}.
\item The manifold of rotation matrices in $n$ dimensions is a Lie group denoted by SO$(n,\mathbb{R})$ and called the \textbf{special orthogonal group}.
\end{itemize}
2. \textbf{Corresponding Physical Concept -- body \& space angular velocities}
\index{body angular velocity}\index{spatial angular velocity}

Recall that a  \textit{Lie group} is a group of transformations that depend smoothly on a set of parameters.

The Lie Group $SO(3)$ faithfully represents the domain of orientation states of a rigid body. Therefore, the Lie Group $SO(3)$ is the natural configuration manifold for rigid body dynamics. Consider a body whose centre of mass is fixed at a point $P\in\mathbb{R}^3$. The most general allowed motion is a rotation about $P$. To describe this motion, we specify positions in a fixed space frame with orthonormal basis $\mathbf{\tilde{e}}_a$, with $a=1,2,3$. One then transforms to a moving frame $\mathbf{e}_a(t)=O_{ab}(t)^{-1}\mathbf{\tilde{e}}_b$ in the body so that $\mathbf{e}_a(t)$ moves with the body as it rotates.

Both sets of axes are orthogonal, so we have
\[
\mathbf{\tilde{e}}_a \cdot \mathbf{\tilde{e}}_b =\delta_{ab}
\,, \quad \hbox{and}\quad
\mathbf{e}_a(t) \cdot  \mathbf{e}_b(t) =\delta_{ab}
\,.
\]
Any point $\mathbf{r}(t)$ in the moving body can be expanded in either the space frame or the body frame:
\[
\mathbf{r}(t) = \widetilde{r}_b(t)\mathbf{\tilde{e}}_b
\ \hbox{in the space frame and}\ \mathbf{r}(t) = r_a \mathbf{e}_a(t) \ \hbox{in the body frame}
\]
where  $\widetilde{r}_b(t) = r_aO_{ab} (t)$ with constant $r_a$.

In the body frame the moving and fixed bases are related by left action $\mathbf{e}_a(t) = O^{-1}_{ab}(t)\mathbf{\tilde{e}}_b$. Consequently, viewed from the body we have
\begin{align}
\frac{d\mathbf{e}_a(t)}{dt} = \frac{d}{dt}O^{-1}_{ab}(t)\mathbf{\tilde{e}}_b = - \big(O^{-1}\dot{O}\big)_{ab}O_{bc}^{-1}\mathbf{\tilde{e}}_c
= - \big(O^{-1}\dot{O}\big)_{ab}\mathbf{e}_b(t)   
\,.
\label{Omega-def}
\end{align}
In vector form, this equation is written as 
\begin{align}
\frac{d\mathbf{e}(t)}{dt} = -\,\bs{\Omega}\times \mathbf{e}(t)
\,,
\label{Omega-vector-def}
\end{align}
where $\bs{\Omega}$ is the \emph{left-invariant} body angular velocity vector \index{angular velocity! left-invariant} of a point in space as observed from the rotating body.

One transforms from motion in the body frame to motion in the spatial frame via the \emph{adjoint action} \index{adjoint action} $\Ad$ which defines how the Lie group $SO(3)$ acts on its Lie algebra $\mathfrak{so}(3)$, as follows.
Namely, the adjoint action $\Ad_{O(t)}$ of $SO(3)$ on the apparent rotation of the fixed spatial basis $\mathbf{\tilde{e}}$ as viewed from  the body basis  $\mathbf{e}_a(t) = O^{-1}_{ab}(t)\mathbf{\tilde{e}}_b\in \mathbb{R}^3$ yields the following expression for the rotation of the body, as viewed from the spatial basis, 
\[
\Big(\Ad_{O(t)}\mathbf{e}(t)\Big)_b := \Big(O(t)\mathbf{e}(t)O^{-1}(t)\Big)_b = \mathbf{\tilde{e}}_aO^{-1}_{ab}(t) =: \mathbf{E}_b(t)
\,.\]
Taking the time derivative yields the \emph{right-invariant} angular velocity \index{angular velocity! right-invariant} in the spatial frame given by
\begin{align}
\frac{d}{dt} \mathbf{E}_b(t) = \frac{d}{dt}  \Big(\mathbf{\tilde{e}}_aO^{-1}_{ab}(t) \Big)
= - \Big(\mathbf{E}_a(t) (\dot{O}O^{-1})_{ab}\Big)
\label{omega-def}
\end{align}
In vecotor form, this equation is written as, cf. equation \eqref{Omega-vector-def},
\begin{align}
\frac{d\mathbf{E}(t)}{dt} = \bs{\omega}\times \mathbf{E}(t)
\,,
\label{omega-vector-def}
\end{align}
where $ \bs{\omega}$ is the \emph{right-invariant} spatial angular velocity \index{angular velocity! right-invariant} of the rotating body as observed from fixed space.

\begin{remark}\rm  \textbf{Hat map.}\label{Remark-HatMap}
Notice that orthogonality $O^TO=Id$ implies that both angular velocities $O^{-1}\dot{O}$ and $\dot{O}O^{-1}$  are skew-symmetric $3\times3$ matrices, and the $\mathbb{R}^3$ vector cross-product is equivalent to skew matrix multiplication.
In particular, we write the \emph{hat map} $(\widehat{\,\cdot\,}): {\mathbb{R}^3}\mapsto\mathfrak{so}(3)$ as
\index{hat map! $\mathfrak{so}(3)$}\index{$\mathfrak{so}(3)$! hat map} \index{rigid body!angular velocity!left-invariant}
\[
\widehat{\Omega}_{ij} = - \,\ep_{ijk}\Omega^k
\quad\hbox{and}\quad
\bs{\widehat{\Omega}}  = \bs{\Omega}\times \,,
\quad\hbox{or}\quad
(\bs{\Omega}\times)_{ij} = {\widehat{\Omega}}_{ij}
\,.
\]
Here $\epsilon_{123}=1$ and $\epsilon_{213}=-1$, with cyclic permutations, and otherwise vanishes.
\end{remark}

\begin{exercise}\textbf{Skew-symmetry of the hat map.}
Show that the angular velocities $\widehat{\Omega}=O^{-1}\dot{O}$ and $\widehat{\omega}=\dot{O}O^{-1}$ are both skew-symmetric.
\end{exercise}

\begin{answer}
Since $O^TO=O^{-1}O=Id$ we have $\dot{O}^T= - O^T\dot{O}O^T$. Consequently,
\begin{align*}
(\widehat{\Omega})^T &= (O^T\dot{O})^T = \dot{O}^TO = - \,O^T\dot{O} = - \,\widehat{\Omega}\,,
\\
(\widehat{\omega})^T &= (\dot{O}O^T)^T = O \dot{O}^T = -\,\dot{O}\,O^T = - \,\widehat{\omega}\,,
\end{align*}
so both $\widehat{\Omega}=O^{-1}\dot{O}$ and $\widehat{\omega}=\dot{O}O^{-1}$ skew-symmetric,
although $\widehat{\Omega}$ is left-invariant and $\widehat{\omega}$ is right-invariant under the respective
actions of $SO(3)$. 
\end{answer}

\begin{exercise}\textbf{Properties of the hat map.} \index{hat map! properties}
The hat map arises in the infinitesimal rotations given by
\[\widehat{\Omega}_{jk}
= (O^{-1}dO/ds)_{jk}|_{s=0}
=-\,\Omega_i\epsilon_{ijk}\,.\] 
The hat map is an isomorphism between $\mathbb{R}^3$ and $\mathfrak{so}(3)$: 
\begin{eqnarray*}
(\mathbb{R}^3,\times)\mapsto(\mathfrak{so}(3),\,[\,\cdot\,,\,\cdot\,]\,)
\,.
\end{eqnarray*}
That is, the hat map identifies the composition of two vectors in $\mathbb{R}^3$ using 
the cross product with the commutator 
$[\widehat{P},\widehat{Q}]=\widehat{P}\widehat{Q}-\widehat{Q}\widehat{P}$ 
of two skew-symmetric $3\times3$ matrices, $\widehat{P}$ and $\widehat{Q}$.
Specifically, we write for any two vectors $\mathbf{Q},{\boldsymbol{\Omega}}\in\mathbb{R}^3$,
\[
-\,(\mathbf{Q}\times{\boldsymbol{\Omega}})_k
=
\epsilon_{klm}\Omega\,^lQ^m
=
{\widehat{\Omega}}_{km}\,Q^m
\,.
\]
Here upper (resp. lower) indices denote vector (resp. matrix) components. That is,
\[
{\boldsymbol{\Omega}}\times\mathbf{Q}
=
\widehat{\Omega}\,\mathbf{Q}
\quad\text{for all} \quad 
{\boldsymbol{\Omega}},\,\mathbf{Q}\in \mathbb{R}^3
\,.
\]
Verify the following formulas for $\mathbf{P}, \mathbf{Q}, {\boldsymbol{\Omega}}
\in
\mathbb{R}^3$:
\begin{align}
\begin{split}
\scp{\widehat{P}}{\widehat{Q}}
&=
- \,\frac{1}{2}
\operatorname{trace}\Big(\widehat{P}\widehat{Q}\Big)
= 
\mathbf{P}\cdot \mathbf{Q} 
\,,\\
\Big[\widehat{P},\widehat{Q}\Big]_{ik}
&=
- \,\ep_{ikj}(\mathbf{P} \times \mathbf{Q})^j
\,,\quad
\Big[\widehat{P},\widehat{Q}\Big]
= 
(\mathbf{P} \times \mathbf{Q}){\,\widehat{\, }}
\,,\\
\Big[\widehat{P},\widehat{Q}\Big]{\boldsymbol{\Omega}}
&=
(\mathbf{P} \times \mathbf{Q}) \times{\boldsymbol{\Omega}}
\,,\\
\Big[\widehat{Q},\widehat{\Omega}\Big]\mathbf{P}
&+
\Big[\widehat{\Omega},\widehat{P}\Big]\mathbf{Q}
\,+\,
\Big[\widehat{P},\widehat{Q}\Big]{\boldsymbol{\Omega}}
= 0\,.
\end{split}
\label{hat-exercises}
\end{align}
Hint: A useful identity here is $\ep_{ijk}\ep_{klm} = \delta_{il} \delta_{jm}  - \delta_{im} \delta_{jl} $.
\end{exercise}

\vspace{4mm}\centerline{\textcolor{shadecolor}{\rule[0mm]{6.75in}{-2mm}}\vspace{-4mm}}
\subsection{Lie Algebra $\mathfrak{so}(3)\simeq\mathbb{R}^3$} | \textbf{Infinitesimal Generators of Rotation} | Lie algebra $\mathfrak{so}(3)$ contains angular velocity vectors, representing infinitesimal rotations in the body's motion.

1. \textbf{Geometric Mechanics Concept}\\
--  If $G$ is a Lie group, then $T_eG$ (the tangent space of $G$ at the identity) is a vector space which possesses a remarkable structure called a \textbf{Lie algebra}.\\
\textbf{Definition (Lie algebra)}
A \textbf{Lie algebra} denoted $\mathfrak{g} = T_eG$ is a vector space endowed with a product known as the \textbf{commutator} (or \emph{Lie bracket}), 
\index{Lie bracket} which is a skew bilinear map
\begin{align*}
[\;\cdot\;,\;\cdot\;]\;:\;\mathfrak{g}\times \mathfrak{g}\rightarrow \mathfrak{g}
\end{align*}
that satisfies the Jacobi identity: $ [\xi\,,[ \eta\,,\,\zeta]\,] +  [\eta\,,[ \zeta\,,\,\xi]\,] +  [\zeta\,,[ \xi\,,\,\eta]\,] =0$ for $(\xi,\eta,\zeta)\in \mathfrak{g}$.

Examples of matrix Lie algebras under matrix commutator are:
\begin{enumerate}[(a)]
\item The Lie algebra $\mathfrak{gl}(n,\mathbb{R}):=T_e\textrm{GL}(n,\mathbb{R})$ can be identified with the vector space of real \textit{square} $n\times n$ matrices.
\item The Lie algebra $\mathfrak{sl}(n,\mathbb{R}):=T_e\textrm{SL}(n,\mathbb{R})$ can be identified with the vector space of real \textit{traceless} square matrices.
\item The Lie algebra $\mathfrak{so}(3)=T_e\textrm{SO}(3)$ can be identified with the vector space of real \textit{skew-symmetric} matrices.
\end{enumerate}

2. \textbf{Corresponding Physical Concept}\\ -- 
Lie algebra elements generate group trajectories by exponentiation. For $\xi \in\mathfrak{g}$, one has
\[
g_t=\exp{t[\xi, \cdot]} = Id + t[\xi, \cdot] + \frac{1}{2!} [\xi, [\xi, \cdot]\,] + \dots
\]
Thus, the Lie algebra is the linearisation of the Lie group near the identity and the Lie group transformation 
generated by a given Lie algebra element can be represented in exponential form.

\subsection{Left Action -- Composition of Maps}$\,$ \index{composition of maps}

1. \textbf{Geometric Mechanics Concept}
Let $M$ be a manifold and let $G$ be a Lie group.  A \textbf{(left) action} \index{action! left} of the Lie group $G$ 
on $M$ is a smooth mapping $\phi:G\times M\rightarrow M$ such that
\begin{enumerate}
\item $\phi(e,x)=x\quad\forall x\in M$.
\item $\phi(g,\phi(h,x))=\phi(gh,x)\quad\forall g,h\in G, \forall x\in M.$
\item For every $g\in G$ the map
\begin{align*}
\phi_g:M\rightarrow M
\,,\qquad
\phi_g(x)=\phi(g,x)
\end{align*}
is a diffeomorphism (i.e. smooth and invertible).
\end{enumerate}
\noindent
\textbf{Concatenation notation}:  We write $gx$ for $\phi(g,x)$ and think of the group element $g$ acting on the point $x$.  Then, (2) becomes $g(hx)=(gh)x$.

2. \textbf{Corresponding Physical Concept}
   - In the context of rigid body rotation, \emph{left action} \index{left action} is observed looking outward from the rigid body's own local reference frame. 
   For example, left action can represent the rotation of stars in the night sky relative to fixed coordinates on Earth. 

\subsection{Right Action -- Composition of Maps}$\,$ \index{right action!composition of maps}

1. \textbf{Geometric Mechanics Concept}
   - Right Lie group action is similar to left action but with multiplication from the opposite direction.
\noindent
A \textbf{right Lie group action} of $G$ on $M$ satisfies properties (1) and (3), while (2) is replaced by
\begin{align*}
\phi(g,\phi(h,x))=\phi(hg,x)\quad\forall g,h\in G, \forall x\in M.
\end{align*}

\medskip
\noindent
\textbf{Concatenation notation}:  $\phi(g,x)$ is denoted by $xg$, and (2) becomes $(xh)g=x(hg).$

\medskip
\noindent
\textbf{Example (Left vs right)}  Left action $(g,x)\mapsto gx$ becomes right action via $(g,x)\mapsto g^{-1}x$. 
That is, $g^{-1}x$ is a right action and $xg^{-1}$ is a left action.

2. \textbf{Corresponding Physical Concept}
-- Right action corresponds to observing the rotation of a rigid body from a fixed external reference frame.
For example, consider the rotation of a gyroscope, where each rotation is observed relative to the gyroscope's previous state in the fixed coordinate system. It can also be regarded as a change of frame in which the 2nd dynamics takes place in the frame of reference of the 1st dynamics. (An example is waves carried on currents in the ocean \cite{holm2023lagrangian}.)

\vspace{4mm}\centerline{\textcolor{shadecolor}{\rule[0mm]{6.75in}{-2mm}}\vspace{-4mm}}
\subsection{Lie Group Action on itself and on its Lie Algebra} | \textbf{Group Action on Rigid Body} | Lie group action, e..g., of SO(3) on a rigid body, governs the body's orientation in space.

1. \textbf{Geometric Mechanics Concept for matrix Lie groups}\\ --
\textbf{Definition (conjugation action)}
Let $g\in G$, with $G$ a matrix Lie group.  Then the \textbf{conjugation action} of $G$ on itself is given by the operation
\begin{align*}
\text{AD}_gh: h\mapsto ghg^{-1}\quad\quad\forall h\in G
\,.\end{align*}
Note that $\text{AD}_g$ changes a right action into a left action. 

Take an arbitrary curve $h(t)\in G$ such that $h(0)=e$.  Then, upon denoting
\begin{align*}
\xi=\dot{h}(0)\in T_eG
\quad\hbox{the derivative of $\text{AD}_gh$ defines}\quad
\frac{\text{d}}{\text{d}t} \Big|_{t=0} \text{AD}_g h(t)=:\text{Ad}_g\;\xi=g\xi g^{-1}\in T_eG.
\end{align*}

\noindent
\textbf{Definition (adjoint and coadjoint actions}
 of $G$ on $\mathfrak{g}$ and $\mathfrak{g}^*$) {\it The \textbf{adjoint action} of the matrix Lie group $G$ on its matrix Lie algebra $\mathfrak{g}$ is a map
\begin{align*}
{\AdMR}:\;G\times\mathfrak{g}\rightarrow\mathfrak{g}
\,,\quad
{\AdMR}_g\;\xi=g\xi g^{-1}
\,.
\end{align*}

\noindent
The dual map with symmetric, non-degenerate, real-valued pairing $\scp{\,\cdot\,}{\,\cdot\,}: \mathfrak{g}^*\times \mathfrak{g}\to \mathbb{R}$ 
enables one to introduce $\Ad^{\,*}_g$
\begin{align*}
\langle
{\AdMR}^{\,*}_g\;\mu,\;\xi
\rangle
=
\langle
\mu,\;{\AdMR}_g\,\xi
\rangle
\end{align*}
which is called the \textbf{coadjoint action} of $G$ on $\mathfrak{g}^*$, the dual to Lie algebra $\mathfrak{g}$ with respect to the pairing $\scp{\,\cdot\,}{\,\cdot\,}$.
}
Take $g(t)\in G$ such that $g(0)=e$ and denote 
\begin{align*}
\eta=\dot{g}(0)\in T_eG.
\end{align*}

\noindent
Then one defines
\begin{align*}
\textrm{ad}_\eta\;\xi:=
\left.
\frac{\textrm{d}}{\textrm{d}t}
\right\vert_{t=0\!}
{\AdMR}_{g(t)}\;\xi\quad\forall\xi\in\mathfrak{g}.
\end{align*}

\noindent
\textbf{Definition (adjoint and coadjoint action} of $\mathfrak{g}$ on $\mathfrak{g}$ and $\mathfrak{g}^*$) {\it 
The \textbf{adjoint action} of the matrix Lie algebra on itself is given as a map
\begin{align*}
{\adMR}:\mathfrak{g}\times\mathfrak{g}\rightarrow\mathfrak{g}
\\
{\adMR}_\eta\;\xi=[\eta,\xi] := \eta\xi - \xi\eta.
\end{align*}

\medskip
\noindent
The dual map
\begin{align*}
\langle
{\adMR}^*_\eta\,\mu,\;\xi
\rangle
=
\langle
\mu,\;{\adMR}_\eta\,\xi
\rangle
\end{align*}
is the \textbf{coadjoint action} of $\mathfrak{g}$ on $\mathfrak{g}^*$.
}
\begin{exercise}
Write these adjoint and coadjoint operations for the Lie group of upper triangular $3\times 3$ matrices.
Upper triangular $3\times 3$ matrices represent the action of the Heisenberg group. See, e.g., \cite{holm2011geometricII}
for more discussion of the adjoint and coadjoint actions of upper triangular $3\times 3$ matrices.
\end{exercise}

2. \textbf{Corresponding Physical Concept}\\ -- 
In rigid body rotation, coadjoint action describes the dynamics of angular momentum under rotation. It reflects transformations between two different reference frames, called the body and space frames. The angular velocities 
in the body and space frames are denoted 
\begin{align}
\hbox{Body:}\quad\widehat{\Omega} = O^{-1}\dot{O}\in \mathfrak{so}(3) 
\quad\hbox{and}\quad
\hbox{Space:}\quad\widehat{\omega} =\dot{O}O^{-1}\in \mathfrak{so}(3)
\,, 
\label{BodySpaceAngVel-Ad}
\end{align}
for $O(t)\in SO(3)$. 
\begin{exercise}
Show that the skew symmetric space and body angular velocities $\widehat{\omega}=\dot{O}O^{-1}$ and $\widehat{\Omega} = O^{-1}\dot{O}$, respectively, are related by $\widehat{\omega} = \text{Ad}_{O(t)}\widehat{\Omega}$.
\end{exercise}

\vspace{4mm}\centerline{\textcolor{shadecolor}{\rule[0mm]{6.75in}{-2mm}}\vspace{-4mm}}
\subsection{Coadjoint Motion - Angular Momentum State Space} \label{coad-left-inv-Lag}
1. \textbf{Geometric Mechanics Concept}\\ -- 
In geometric mechanics, \textbf{coadjoint motion} satisfies
$ \textrm{Ad}^*_{g^{-1}(t)\,}{\partial \ell}/{\partial\xi}(t)=const.$ for a left-invariant Lagrangian function defined on the Lie algebra $\ell: \mathfrak{g}\to \mathbb{R}$, $\xi(t)=g^{-1}\dot{g}(t)\in \mathfrak{g}$.

For a fixed $\zeta \in \mathfrak{g}$, one may calculate the time derivative of the following pairing
\begin{align}
\begin{split}
\frac{d}{dt}\left<\textrm{Ad}^*_{g^{-1}}\frac{\partial \ell}{\partial \xi},\zeta\right> &= \frac{d}{dt}\left<\frac{\partial \ell}{\partial \xi}(t),\textrm{Ad}_{g^{-1}(t)}\zeta\right>\\
&= \left<\frac{d}{dt} \frac{\partial \ell}{\partial \xi},\textrm{Ad}_{g^{-1}(t)}\zeta\right>+\left<\frac{\partial \ell}{\partial \xi},\frac{d}{dt}(g^{-1}(t)\zeta g(t))\right> \\
&= \left<\frac{d}{dt} \frac{\partial \ell}{\partial \xi},\textrm{Ad}_{g^{-1}(t)}\zeta\right> + \left<\frac{\partial \ell}{\partial \xi},-g^{-1}\dot{g}g^{-1}\zeta g + g^{-1}\zeta\dot{g}\right>\\
&=\left<\frac{d}{dt}\frac{\partial \ell}{\partial \xi},\textrm{Ad}_{g^{-1}(t)}\zeta\right> + \left<\frac{\partial \ell}{\partial \xi},-\xi\textrm{Ad}_{g^{-1}}\zeta+(g^{-1}\zeta g)(g^{-1}\dot{g})\right>\\
&= \left< \frac{d}{dt}\frac{\partial \ell}{\partial \xi},\textrm{Ad}_{g^{-1}(t)}\zeta\right> - \left<\frac{\partial \ell}{\partial \xi},
\textrm{ad}_{\xi}(\textrm{Ad}_{g^{-1}}\zeta)\right>\\
&=\left< \frac{d}{dt}\frac{\partial \ell}{\partial \xi}-\textrm{ad}^*_\xi\frac{\partial \ell}{\partial \xi},\textrm{Ad}_{g^{-1}}\zeta\right>\\
&=\left< \textrm{Ad}^*_{g^{-1}\!}\left(\frac{d}{dt}\frac{\partial \ell}{\partial \xi}-\textrm{ad}^*_\xi\frac{\partial \ell}{\partial \xi}\right),\zeta\right>
=0
\end{split}
\label{coadjoint-calc}
\end{align}
Since the fixed Lie-algebra element $\zeta$ is arbitrary, we have
$$\frac{d}{dt}\left(\textrm{Ad}^*_{g^{-1}(t)}\frac{\partial \ell}{\partial \xi}\right) 
=\textrm{Ad}^*_{g^{-1}(t)}\left(\frac{d}{dt}\frac{\partial \ell}{\partial \xi}-\textrm{ad}^*_\xi\frac{\partial \ell}{\partial \xi}\right) = 0\,.$$
Consequently, the solutions of the \textbf{Euler--Poincar\'e equation}, expressed as
\begin{align}
\left(\frac{d}{dt}\frac{\partial \ell}{\partial \xi}-\textrm{ad}^*_\xi\frac{\partial \ell}{\partial \xi}\right) = 0
\label{EP-eqn-intro}
\,,\end{align}
satisfy $ \textrm{Ad}^*_{g^{-1}(t)\,}{\partial \ell}/{\partial\xi}(t)=const$ and thus describe coadjoint motion. 
The solutions of the Euler--Poincar\'e equation in \eqref{EP-eqn-intro} comprise a set of \textbf{coadjoint orbits}.
\index{Euler--Poincar\'e equations!coadjoint orbits}
\begin{exercise}
Verify that 
\[
\frac{d}{dt}\Big|_{t=0} \Ad_{g_t^{-1}}\zeta = -\, \ad_{\xi} \zeta\,.
\]
for any \textit{fixed} $\zeta\in\mathfrak{g}$ and right action $\xi= \dot{g}_tg_t^{-1} \big|_{t=0}$.

Verify that 
\[
\frac{d}{dt} \Ad_{g_t^{-1}}\zeta = -\, \ad_{\xi_t} \Ad_{g_t^{-1}} \zeta\,.
\]
for any fixed $\zeta\in\mathfrak{g}$ and $\xi_t= \dot{g}_tg_t^{-1} $.
Hint: follow the calculations in \eqref{coadjoint-calc}.
\end{exercise}

\begin{answer}
One begins with the second part, because the first part follows by evaluating the result of the second part at $t=0$.
\begin{align*}
\frac{d}{dt} {\textrm{Ad}_{g_t^{-1}}\zeta} 
&= {\frac{d}{dt}(g_t^{-1}\zeta g_t)} = -(g_t^{-1}\dot{g_t})g_t^{-1}\zeta g_t + g_t^{-1}\zeta\dot{g}_t
\\&=
-\,\xi_t(\Ad_{g_t^{-1}}\zeta)+(g_t^{-1}\zeta g_t)(g_t^{-1}\dot{g}_t)
\\&=
-\,\xi_t(\Ad_{g_t^{-1}}\zeta)+(\Ad_{g_t^{-1}}\zeta)\xi_t
\\
\hbox{By \eqref{coadjoint-calc})}
&= -\,{\textrm{ad}_{\xi_t}(\textrm{Ad}_{g_t^{-1}}\zeta)}
\end{align*}
Then evaluating $\textrm{Ad}_{g_t^{-1}}\big|_{t=0} = Id$ yields the result for the first part. 
\end{answer}

As we shall discuss in detail later, the Euler--Poincar\'e equation \eqref{EP-eqn-intro} follows from a constrained Hamilton's variational principle. Namely, the \emph{Hamilton--Pontryagin} variational principle
which yields in this case \index{variational principle!Hamilton--Pontryagin}
\begin{align}
\begin{split}
0 = \delta S &= \delta \int_0^T \ell(\xi) + \scp{\mu}{g^{-1}\dot{g} - \xi} dt
\\&=  \int_0^T \scp{ \frac{\p \ell}{\p \xi} - \mu}{\delta \xi} 
+ \scp{\mu}{\frac{d \eta}{dt} + \text{ad}_{g^{-1}\dot{g}}\eta} 
+ \scp{\delta \mu}{g^{-1}\dot{g} - \xi} dt
\\&=  \int_0^T \scp{ \frac{\p \ell}{\p \xi} - \mu}{\delta \xi} 
- \scp{\frac{d \mu}{dt} - \text{ad}^*_{g^{-1}\dot{g}} \mu}{\eta} 
+ \scp{\delta \mu}{g^{-1}\dot{g} - \xi} dt
+ \scp{\mu}{\eta}\big|_0^T
\,.
\end{split}
\label{HPP}
\end{align}
where $\xi= g^{-1}\dot{g} $ and the quantity $\eta := g^{-1}\delta{g} $ vanishes 
at the endpoints in time and we have used the relation 
\begin{align}
\delta \xi = \frac{d \eta}{dt} + \text{ad}_{\xi}\eta
\,.
\label{EP-Var-Id}
\end{align}
\begin{exercise}
Derive the Euler--Poincar\'e relation in \eqref{EP-Var-Id}
\end{exercise}
\begin{answer}
Define `prime'  notation  $\eta := g^{-1}\delta{g} = g^{-1}g'$ and compute
\begin{align*}
\xi' &= (g^{-1}\dot{g})' = - (g^{-1} g' )( g^{-1}\dot{g}) + g^{-1} \dot{g}'= - \eta\xi + g^{-1} \dot{g}'
\,,\\
\dot{\eta} &= (g^{-1} g' )\dot{\,} = - (g^{-1} \dot{g})( g^{-1}g' ) + g^{-1} g'\dot{\,\,} = - \xi\eta + g^{-1} g'\dot{\,\,}
\,.\end{align*}
Subtracting the second equation from the first one, then noticing the cancellation of equal second-order cross 
derivatives $\dot{g}'=g'\dot{\,\,}$ and collecting terms yields the Euler--Poincar\'e equation for coadjoint motion in \eqref{EP-eqn-intro}.
\end{answer}

\begin{exercise}
Prove the following identities with $\xi_t= g_t^{-1}\dot{g}_t $ for \textit{fixed} $\eta\in \mathfrak{g}$ and $\mu\in \mathfrak{g}^*$: 
\begin{align}
\begin{split}
\scp{\frac{d}{dt}\big(\Ad_{g_t}\eta\big) }{\mu} 
&= \scp{\Ad_{g_t} \big(\ad_{g_t^{-1}\dot{g}_t}\eta\big) } {\mu }
= \scp{\Ad_{g_t} \big(\ad_{\xi_t}\eta\big) } {\mu }
\\
\scp{\frac{d}{dt}\big(\Ad_{g^{-1}_t}\eta\big) }{\mu} 
&= \scp{ -\, \ad_{g_t^{-1}\dot{g}_t}\big(\Ad_{g^{-1}_t}\eta\big) } {\mu }
= \scp{ -\, \ad_{\xi_t}\big(\Ad_{g^{-1}_t}\eta\big) } {\mu }
\,.\end{split}
\label{Ex-Var-Id}
\end{align}
\end{exercise}

\begin{answer}$\,$\\
By direct calculation,
\begin{align}
\begin{split}
\scp{\frac{d}{dt}\big(\Ad_{g_t}\eta\big) }{\mu} 
&= \scp{\frac{d}{dt}\big(g_t\eta g_t^{-1}\big) }{\mu} 
= \scp{\dot{g}_t\eta g_t^{-1} - g_t\eta g_t^{-1} \dot{g}_t g_t^{-1} }{ \mu  }
\\&= \scp{\Ad_{g_t} \big((g_t^{-1}\dot{g}_t) \eta \big) -  \Ad_{g_t} \big( \eta  (g_t^{-1}\dot{g}_t) \big)}{\mu }
\\&= \scp{\Ad_{g_t} \big( \ad_{g_t^{-1}\dot{g}_t} \eta \big) }{\mu }
= \scp{\Ad_{g_t} \big( \ad_{\xi_t} \eta \big) }{\mu }
\\ \hbox{Dual:}&= \scp{ \eta }{ \ad^*_{\xi_t} \big( \Ad^*_{g_t}  \mu \big) }
.\end{split}
\label{Ex-Var-Id-Ans}
\end{align}
A similar calculation yields $\scp{\frac{d}{dt}(\Ad_{g^{-1}_t}\eta) }{\mu} =\scp{ -\, \ad_{\xi_t}(\Ad_{g^{-1}_t}\eta) } {\mu }$.
\end{answer}

2. \textbf{Corresponding Physical Concept} \\ -
\emph{Euler--Poincar\'e Equations and Rigid Body Dynamics} -- Euler--Poincaré equations \index{Euler--Poincaré equations!rigid body} 
for rigid body rotation describe the dynamics considering the body's symmetry and frame of motion, leading in particular to the 
classical Euler rigid body equations in the body frame. The Euler--Poincaré equations will appear prominently in all the rest that follows in this text.
\index{variational principle!Euler--Poincar\'e}

\vspace{4mm}\centerline{\textcolor{shadecolor}{\rule[0mm]{6.75in}{-2mm}}\vspace{-4mm}}
\subsection{Reviewing the Diamond Operator and Cotangent Lift Momentum Map}$\,$
The diamond operator and cotangent lift momentum map were introduced in section \ref{subsec-HPHP} in the context of Noether's theorem 
for Lie symmetry of the Lagrangian in Hamilton's principle. Here, we say a bit more about their properties on the Hamiltonian side. 

1. \textbf{Geometric Mechanics Concept}
   - The diamond operator, denoted as $(\diamond)$, appears in Noether's endpoint term $\scp{p}{\delta q}\big|_0^T$ in Hamilton's principle \eqref{HP-geodesic} when the variation $\delta q$ is an infinitesimal left action of a Lie symmetry of the configuration space $(Q,q)$,
 \begin{align}
 \delta q := \tfrac{d}{d\epsilon}\big|_{\epsilon=0}q(t,\epsilon) =: -\,\pounds_\xi q 
\label{Left-Lie-action}
\end{align}
 where the notation $\pounds_\xi $ for the Lie derivative is defined as the tangent at the identity $\epsilon=0$ of the $\epsilon$-variational group action. One may compare relation \eqref{Left-Lie-action} for left Lie action to equations \eqref{Omega-def} and \eqref{omega-def} which contrast the left-invariant body representation and  right-invariant spatial representation of angular velocity.
 
 \begin{theorem}   
 \emph{Noether's theorem}:  \index{Noether's theorem} Let the Lagrangian $L: TQ\to \mathbb{R}$ in Hamilton's principle \eqref{HP-geodesic} be invariant under a Lie-group left-action $g_\epsilon q(t) = q(t,\epsilon): G\times Q \to Q$ where the curve $g_\epsilon\in G$ is the flow of $G$ parameterised by $\epsilon$ which becomes the identity transformation at $\epsilon=0$. The corresponding infinitesimal Lie $G$-symmetry $\delta q = -\,\pounds_\xi q$, in which the left Lie algebra action of $\xi\in\mathfrak{g}\simeq T _eG$ leaves the Lagrangian $L$ invariant and  implies conservation of  the endpoint term,
 \begin{align}\begin{split}
\scp{ \frac{\partial L}{\partial \dot{q}} }{\delta q }_{TQ} 
&=: \scp {p}{\delta q }_{TQ} =  \scp{ p }{ -\pounds_\xi q }_{TQ} 
\\&=: \scp {p\diamond q }{ \xi}_\mathfrak{g}
=: \scp{ J(q,p) }{ \xi }_\mathfrak{g} =: J^\xi(q,p)
\,, 
\end{split}
\label{Noether-Lie-term}
 \end{align}
where $\scp{\cdot}{\cdot}_{TQ}$ is a symmetric, non-degerate, real-valued pairing $T^*Q\times TQ \to \mathbb{R}$.  
The middle equation defines the \emph{diamond operator} \index{diamond operator} $(\diamond)$ which is of central importance. 
The quantity 
 \begin{align}
J(q,p)= p\diamond q = -\, q\diamond p
\label{diamond-def}
 \end{align}
is the \emph{cotangent-lift momentum map} $J: T^*(Q)\to \mathfrak{g}^*$ by left Lie algebra action.
 \end{theorem}   \index{momentum map! cotangent-lift momentum map}
 \index{momentum map! diamond operator! cotangent lift} 
 
\begin{remark}\rm   
From its definition, the properties of the diamond operator $(\diamond)$ are inherited from the properties of the Lie derivative.
In particular, the quantity 
\[
J^\xi(q,p) = \scp {p\diamond q }{ \xi}_\mathfrak{g} 
\]
is the Hamiltonian defined by the $\xi$-component of the momentum map $J(q,p)= p\diamond q\in \mathfrak{g}^*$. 

Under the \emph{canonical Poisson bracket}, one has \index{Poisson bracket! canonical} \index{canonical Poisson bracket}
\[
\delta q = \big\{ q,  J^\xi(q,p)\big\}_{can} = -\,\pounds_\xi q
\quad\hbox{and}\quad 
\delta p = \big\{ p,  J^\xi(q,p)\big\}_{can} = -\,\pounds^T_\xi  p
\,.\]
Here 
\[
\scp{p}{\delta q} := \scp{ p }{ -\pounds_\xi q }_{TG} 
= \scp{ -\pounds^T_\xi p }{ q }_{TG} = \scp{ \delta p }{ q }_{TG}
\]
and $\delta p$ is said to be the \emph{cotangent lift} of $\delta q$. Hence the name, \emph{cotangent-lift momentum map} for $J(q,p)$.
\end{remark}  

\begin{exercise}
Calculate the Poisson bracket $\{J^\xi(p,q),H(p,q)\}$ for $J^\xi(p,q)=\scp{p\diamond q}{\xi}$
and Hamiltonian $H(p,q)$. Explain how this result is related to Noether's theorem.%
\footnote{For an interesting discussion of Noether's theorems from the Hamiltonian side, 
see \cite{baez2020getting}.} 
\end{exercise}

\begin{answer}
\begin{align*}
-\,\frac{d}{dt}J^\xi(p,q) &=
\{H(p,q),J^\xi(p,q)\}_{can} = \frac{\p H}{\p q} \delta q + \frac{\p H}{\p p} \delta p
\\&=
- \,\frac{\p H}{\p q}\pounds_\xi q - \frac{\p H}{\p p} \pounds^T_\xi  p
=
-\,\pounds_\xi H(p,q)
\,.\end{align*}
\textbf{Hamiltonian Noether theorem.} Left invariance of the Hamiltonian $H(p,q)$ 
under the Lie transformation of phase space generated by the infinitesimal 
canonical transformation $-\,\pounds_\xi$ so that $\pounds_\xi H(p,q) = 0$ implies that the phase space function $J^\xi(p,q)$
which generates that Lie transformation by 
\[
\{(\,\cdot\,),\,J^\xi(p,q)\}_{can} = -\,\pounds_\xi (\,\cdot\,)
\] 
will be conserved
by the dynamics generated by the Hamiltonian $H(p,q)$ via the Poisson bracket $\{J^\xi(p,q),H(p,q)\}$.
Namely, the phase-space function $J^\xi(p,q)$ will be invariant under the Poisson-bracket operation for
Hamiltonian dynamics,
\[
\frac{d}{dt} = \{\,\cdot\,\,,H(p,q)\}
\,.
\]
\end{answer}


\begin{example}\rm [Momentum map for $SO(3)$ acting on $\mathbb{R}^3$] 
For $Q=\mathbb{R}^3$ and $\mathfrak{g}=\mathfrak{so}(3)$ one finds 
$\xi_Q(q) 
= -\, \pounds_\xi{q}
= -\, \xi\times q$
by the hat map and
\[
\Big\langle p\diamond q,{\xi}\Big\rangle
=
\Big\langle p,-\,\pounds_\xi{q}\Big\rangle_{TQ}
=
 -\, p \cdot (\xi\times q)
=
 -\,  (q\times p) \cdot \xi 
=
\Big\langle J(p,q),\,\xi\Big\rangle
= 
J^\xi(p,q)
\,,
\]
which is the Hamiltonian for an infinitesimal rotation around $\xi$ in $\mathbb{R}^3$.
In the case that $\mathfrak{g}=\mathfrak{so}(3)$, the pairing $\langle
\,\cdot\,,\,\cdot\,\rangle$ may be taken as dot products of vectors in $\mathbb{R}^3$, the momentum map 
$J(p,q)=p \diamond q = p\times q \in\mathbb{R}^3$ 
is the phase-space expression for angular momentum and the $\diamond$ operation is $\times$, which is  
the cross product of vectors in $\mathbb{R}^3$. 
The map $J(p,q)=p \diamond q= -\, q\times p\in\mathbb{R}^3$ is an example of a \emph{cotangent lift} momentum map. \index{cotangent lift!momentum map} 
\index{momentum map!cotangent lift}
\end{example}

2. \textbf{Corresponding Physical Concept}\\
   - In the case of particle motion for a rotationally left-invariant Lagrangian, one finds
 \[
 \delta \mathbf{q} = \frac{\p J^\xi}{\p \mathbf{p}} =  -\, \bs{\xi}\times \mathbf{q} 
 \,,\quad  
 \delta \mathbf{p} = -\,  \frac{\p J^\xi}{\p \mathbf{q}} = -\, \bs{\xi}\times \mathbf{p}
 \,,\quad\hbox{and}\quad 
 p\diamond q = \mathbf{p}\times \mathbf{q}
 \,.
 \]
Thus, the diamond operator for left Lie algebra action reduces to the vector cross product in $\mathbb{R}^3$ for the case of rotation and (minus) the particle angular momentum $\mathbf{J}=-\,\mathbf{q}\times \mathbf{p}$ emerges as an element in the dual space of the Lie algebra $\mathfrak{so}(3)^*\simeq \mathfrak{so}(3)\simeq \mathbb{R}^3$ for left Lie algebra action of $\mathfrak{so}(3)$ on $\mathbb{R}^3$. 

The cross product is very important in understanding the role of the angular momentum in particle dynamics for a rotationally invariant Lagrangian as in Newtonian gravitation. Likewise, we will see that the diamond operation and the cotangent-lift momentum map $J(q,p)=p\diamond q$ are very important in understanding Euler--Poincar\'e motion for other Lie group invariant Lagrangians. 

\begin{exercise}
Calculate the Poisson bracket $\{J^\xi(p,q),J^\zeta(p,q)\}$ for $J^\xi(p,q)=\scp{p\diamond q}{\xi}$
and $J^\zeta(p,q)=\scp{p\diamond q}{\zeta} = \scp{p}{-\,\pounds_\zeta q}$ for infinitesimal left action. 
\end{exercise}

\begin{answer}
\begin{align*}
\{J^\xi (q,p),J^\zeta (q,p)\}_{can} &= \{J^\xi (q,p),\langle  p\diamond q \,,\, \zeta \rangle_{\mathfrak{g} }\}_{can}
\\
\hbox{By product rule}\hspace{10mm}
&\hspace{-10mm}=\langle  \{J^\xi , p\}_{can} \diamond q +  p \diamond \{J^\xi , q \}_{can}
\,,\, \zeta\, \rangle_{\mathfrak{g} }
\\
&\hspace{-10mm}=\langle  (-\mc{L}_\xi^Tp)\diamond q +  p \diamond (\mc{L}_\xi q) \,,\, \zeta \,\rangle_{\mathfrak{g} }
\\
&\hspace{-10mm}=\langle  \mc{L}_\xi ( p \diamond q )\,,\, \zeta \rangle_{\mathfrak{g} }
= \langle  \mathrm{ad}^*_\xi  ( p \diamond q )\,,\, \zeta \,\rangle_{\mathfrak{g} }
\\
&\hspace{-10mm} = \langle    p \diamond q \,,\, \mathrm{ad}_\xi \zeta \rangle_{\mathfrak{g} }
=  \langle    p  \,,\, -\,\mc{L}_{[\xi , \zeta]}q \rangle_{TQ } 
\\
&\hspace{-10mm}= -\,J^{[\xi,\zeta]}(q,p)
\quad\hbox{(Anti-homomorphism)}
\,.\end{align*}
\end{answer}

\begin{exercise}
\begin{itemize}
\item
Derive for a \emph{right-invariant} Lagrangian function the corresponding results of Hamilton's principle in section \ref{coad-left-inv-Lag} 
which are derived there for a left-invariant Lagrangian function. 
\item
Transform the rigid body Lagrangian for left-invariant angular velocity in the body frame
into the \emph{same} Lagrangian written in terms of the right-invariant angular velocity in the spatial frame. 
\item
What differences emerge in Noether's theorem for the \emph{same} rigid body Lagrangian written in 
either the left-invariant or right-invariant angular velocities?
\end{itemize}
\end{exercise}

\begin{exercise}
Calculate the Poisson bracket $\{J^\xi(p,q),J^\zeta(p,q)\}$ for $J^\xi(p,q)=\scp{p\diamond q}{\xi}$
and $J^\zeta(p,q)=\scp{p\diamond q}{\zeta} = \scp{p}{-\,\pounds_\zeta q}$ for infinitesimal \emph{right} action. 
\end{exercise}

\vspace{4mm}\centerline{\textcolor{shadecolor}{\rule[0mm]{6.75in}{-2mm}}\vspace{-4mm}}




\vspace{4mm}\centerline{\textcolor{shadecolor}{\rule[0mm]{6.75in}{-2mm}}\vspace{-4mm}}
\section{Particle mechanics of Newton, Lagrange \& Hamilton}

\secttoc

\textbf{What is this lecture about?} This lecture is about the equivalences among the formulations of dynamics 
in the approaches of  Newton, Lagrange and Hamilton.

\noindent
\label{sec:13}
\subsection{Newton's equations for particle motion in Euclidean space} 
\emph{Newton's equations} in a fixed inertial frame \index{Newton's equations! force! acceleration}
\begin{equation}
m_i\mathbf{\ddot{q}}_i = \mathbf{F}_i,\quad
i=1,\ldots, N ,\quad\hbox{(no sum on }i)
\label{Newton1}
\end{equation}
describe the \emph{accelerations} $\mathbf{\ddot{q}}_i$ of $N$ particles
with
\begin{align*}
\text{Masses}\quad & m_i, \qquad i=1,\ldots, N ,\\
\text{Euclidean positions}\quad &
\mathbf{q}: =(\mathbf{q}_1, \dots , \mathbf{q}_N) \in
\mathbb{R}^{3N},
\end{align*}
in response to \emph{prescribed forces},
\begin{eqnarray*}
\mathbf{F} =(\mathbf{F}_1, \dots ,\mathbf{F}_N)
,
\end{eqnarray*}
acting on these particles.
Suppose the forces arise from a \emph{potential}. That is, let
\begin{equation}
\mathbf{F}_i(\mathbf{q})=
-\,\frac{\partial V(\{\mathbf{q}\})}{\partial \mathbf{q}_i}
,\qquad V{:\ } \mathbb{R}^{3N} \rightarrow \mathbb{R}
,
\label{force}
\end{equation}
where $\partial V/ \partial \mathbf{q}_i $ denotes the gradient of
the potential with respect to the variable $\mathbf{q}_i $.  Then Newton's
equations
\eqref{Newton1} become
\begin{equation}
m_i \mathbf{\ddot{q}}_i =
-\,\frac{\partial V}{\partial \mathbf{q}_i},\qquad
i=1,\ldots, N .
\label{Newton2}
\end{equation}
Such a Newtonian system in potential form is called a \emph{simple mechanical system}.
 \index{Newton's equations! simple mechanical system}
\begin{remark}\rm 
Newton (1620) introduced the gravitational potential for celestial
mechanics, now called the \emph{Newtonian potential},
\begin{equation}
 V(\{\mathbf{q}\})
=
\sum_{i,j=1}^N \frac{-\,Gm_im_j}{|\mathbf{q}_i-\mathbf{q}_j|}
.
\label{Newton-pot}
\end{equation}
\end{remark}

\subsection{Equivalence of Newton, Lagrange \& Hamilton dynamics}

\begin{theorem}[Lagrangian  and Hamiltonian
formulations]\label{thm:lhequiv}
Newton's equations in potential form,
\begin{equation}
m_i \mathbf{\ddot{q}}_i =
-\,\frac{\partial V}{\partial \mathbf{q}_i},\qquad
i=1,\ldots, N ,
\label{Newton-thm}
\end{equation}
for particle motion in Euclidean space $\mathbb{R}^{3N}$ are
\emph{equivalent} to the following four statements:
\begin{enumerate}
\item [(i)] 
The Euler--Lagrange equations
\begin{equation}
\frac{d}{dt}\left(\frac{\partial L}{\partial\mathbf{\dot{q}}_i}\right)
-\frac{\partial L}{\partial \mathbf{q}_i}=0, 
\qquad i = 1, \dots, N,
\label{EulerLag-eqs}
\end{equation}
hold for the Lagrangian
$
L {:\ }  \mathbb{R}^{6N} =
\{(\mathbf{q},\mathbf{\dot{q}}) \mid \mathbf{q}, \mathbf{\dot{q}} \in
\mathbb{R}^{3N} \} \rightarrow \mathbb{R}
,
$
defined by
\begin{equation}
L(\mathbf{q},\mathbf{\dot{q}}):=\sum_{i=1}^N\frac{m_i}{2}\,
\|\mathbf{\dot{q}}_i\|^2-V(\mathbf{q})
,\label{Lagrangian-thm}
\end{equation}
with $\|\mathbf{\dot{q}}_i\|^2=\mathbf{\dot{q}}_i\cdot\mathbf{\dot{q}}_i
=\dot{q}^j_i\dot{q}^k_i\delta_{jk}$ (no sum on $i$). Lagrangians 
of the separated form in \eqref{Lagrangian-thm} are said to govern 
\emph{simple mechanical systems}. \index{Newton's equations!simple mechanical systems}
\item [(ii)] 
\emph{Hamilton's principle of stationary action}, $\delta\mathcal{S}=0$,
holds  for the \emph{action functional} (dropping $i$ indices) \index{Hamilton's principle!momentum map}
\begin{equation}
\mathcal{S}[\mathbf{q}(\cdot)]: =\int_a^b
L(\mathbf{q}(t),\mathbf{\dot{q}}(t))
+ \scp{\mathbf{p}}{\mathbf{\dot{q}} - \frac{d\mathbf{q}}{dt}}
\, dt.
\label{Action-thm}
\end{equation}
\item [(iii)] 
\emph{Hamilton's equations of motion}, \index{Hamilton's equations}
\begin{equation}
\label{hameq-thm}
\mathbf{\dot{q}} = \frac{\partial H}{\partial \mathbf{p}}, \qquad
\mathbf{\dot{p}} =-\,\frac{\partial H}{\partial \mathbf{q}}
,
\end{equation}
hold for the \emph{Hamiltonian} \index{Hamiltonian} resulting from the 
\emph{Legendre transform}, \index{Legendre transform}
\begin{equation}
H(\mathbf{q},\mathbf{p}) :=
\mathbf{p}\cdot\mathbf{\dot{q}}(\mathbf{q},\mathbf{p}) -
L(\mathbf{q},\mathbf{\dot{q}}(\mathbf{q},\mathbf{p}))
,
\label{LegendreXform}
\end{equation}
where $\mathbf{\dot{q}}(\mathbf{q},\mathbf{p})$ solves for
$\mathbf{\dot{q}}$ from the definition $\mathbf{p}:=\partial
L(\mathbf{q},\mathbf{\dot{q}})/\partial\mathbf{\dot{q}}$.

In the case of Newton's equations in potential form \eqref{Newton-thm},
the Lagrangian in equation \eqref{Lagrangian-thm} yields
$\mathbf{p}_i=m_i\mathbf{\dot{q}}_i$ and the resulting Hamiltonian is
(restoring subscript $i$s)
\[
H = \underbrace{\sum_{i=1}^N\frac{1}{2m_i}\|\mathbf{p}_i\|^2
}_{\hbox{Kinetic energy}} + \underbrace{V(\mathbf{q})}_{\hbox{Potential}}
\,.\]
\item [(iv)] 
Hamilton's equations in their Poisson bracket formulation,
\begin{equation}
\dot{F}=\{F,H\}\quad\mbox{for all $F\in{\mathcal F}(P)$},
\label{PoissonEqs-thm}
\end{equation}
hold with \emph{Poisson bracket} \index{Poisson brackets}  defined by
\begin{equation}
\{F,G\} := \sum_{i =1}^N\left(\frac{\partial F}{\partial \mathbf{q}_i}\cdot
\frac{\partial G}{\partial \mathbf{p}_i} -\frac{\partial F}{\partial
\mathbf{p}_i} \cdot
\frac{\partial G}{\partial \mathbf{q}_i}\right)\quad \text{for all} \quad
F,G\in{\mathcal F}(P).
\label{Poisson-thm}
\end{equation}
\end{enumerate}
\end{theorem}

We will prove this theorem by proving a chain of linked
equivalence relations: 
\eqref{Newton-thm} $\Leftrightarrow$ (i)
$\Leftrightarrow$ (ii) $\Leftrightarrow$ (iii)
$\Leftrightarrow$ (iv) as propositions. (The symbol $\Leftrightarrow$
means ``equivalent to''.)

\begin{proof}[Step I: Proof that Newton's equations \eqref{Newton-thm}
$\Leftrightarrow$ (i)]
Check by direct verification.
\end{proof}

\begin{proof}[Step II: Proof that (i) $\Leftrightarrow$ (ii)]
The \emph{Euler--Lagrange equations \index{Euler--Lagrange equations!Hamilton's principle} 
\eqref{EulerLag-eqs} are equivalent to Hamilton's principle of stationary action}.

To simplify notation, we momentarily suppress the particle index $i$.

We need to prove the solutions of \eqref{EulerLag-eqs} are critical
points $\delta\mathcal{S}=0$ of the \emph{action functional} \index{action functional}
\begin{equation}
\mathcal{S}[\mathbf{q}(\cdot)]: =\int_a^b
L(\mathbf{q}(t),\mathbf{\dot{q}}(t))\, dt,
\label{Action}
\end{equation}
(where $\mathbf{\dot{q}} = d\mathbf{q}(t)/dt$) with respect to
variations on $C^\infty([a,b],\mathbb{R}^{3N})$, the space of
smooth trajectories $\mathbf{q} {:\ }  [a, b] \rightarrow
\mathbb{R}^{3N}$ with fixed endpoints $\mathbf{q}_a$,
$\mathbf{q}_b$.

In $C^\infty([a,b],\mathbb{R}^{3N})$ consider a \emph{deformation}
$\mathbf{q}(t, s)$, $s \in (-\epsilon, \epsilon)$, $\epsilon>0 $, with
fixed endpoints $\mathbf{q}_a $, $\mathbf{q}_b$, of a curve
$\mathbf{q}_0(t)$, that is, $\mathbf{q}(t, 0)=\mathbf{q}_0(t)$ for all
$t \in [a,b]$ and $\mathbf{q}(a, s) = \mathbf{q}_0(a) = \mathbf{q}_a $,
$\mathbf{q}(b, s) = \mathbf{q}_0(b) = \mathbf{q}_b $ for all $s \in
(-\epsilon, \epsilon)$.

Define a \emph{variation} \index{variation} of the curve $\mathbf{q}_0(\cdot)$ in
$C^\infty([a,b],\mathbb{R}^{3N})$ by
$$
\delta\mathbf{q}(\cdot):=\frac{d}{ds}\Big|_{s=0}\mathbf{q}(\cdot, s) \in
T_{\mathbf{q}_0(\cdot)}C^\infty([a,b],\mathbb{R}^{3N}),
$$
and define the \emph{first variation\/} \index{variation!first} of $\mathcal{S}$ at
$\mathbf{q}_0(t)$ to be the derivative
\begin{equation}
\delta\mathcal{S}
:=
\mathbf{D}\mathcal{S}[\mathbf{q}_0(\cdot)](\delta\mathbf{q}(\cdot))
:=\frac{d}{ds}\Big|_{s = 0} \mathcal{S}[\mathbf{q}(\cdot, s)].
\label{varA}
\end{equation}
Note that $\delta\mathbf{q}(a) = \delta\mathbf{q}(b) = \mathbf{0}$.
With these notations, \emph{Hamilton's principle of stationary action\/}
\index{Hamilton's principle! stationary action}
states that the curve
$\mathbf{q}_0(t)$ satisfies the Euler--Lagrange equations
\eqref{EulerLag-eqs} if and only if
$\mathbf{q}_0(\cdot)$ is a  critical point of the action functional,
that is,
$\mathbf{D}\mathcal{S}[\mathbf{q}_0(\cdot)] = 0 $. Indeed, using  the
equality of mixed partials, integrating by parts, and taking into
account that $\delta\mathbf{q}(a) = \delta\mathbf{q}(b) = 0 $, one finds
\begin{align}
\begin{split}
\delta\mathcal{S}
:=
\mathbf{D}\mathcal{S}[\mathbf{q}_0(\cdot)](\delta\mathbf{q}(\cdot))&=
\left.\frac{d}{ds}\right|_{s = 0} \mathcal{S}[\mathbf{q}(\cdot, s)]
=\left.\frac{d}{ds}\right|_{s = 0} \int_a^b
L(\mathbf{q}(t, s),\mathbf{\dot{q}}(t,s))\, dt \\
&= \sum_{i=1}^N\int_a^b\left[\frac{\partial L}{\partial \mathbf{q}_i} 
\cdot\delta \mathbf{q}_i(t,s) 
+ \Big(\frac{\partial L}{\partial \mathbf{\dot{q}}_i} 
\cdot\delta \mathbf{\dot{q}}_i \Big)\Big|_{\mathbf{\dot{q}}_i=\tfrac{d\mathbf{q}_i}{dt}} \right] dt\\
&= -\sum_{i=1}^N\int_a^b\left[
\frac{d}{dt}
\left(\frac{\partial L}{\partial
\mathbf{\dot{q}}_i}\Big|_{\mathbf{\dot{q}}_i=\tfrac{d\mathbf{q}_i}{dt}}  \right) 
-\frac{\partial L}{\partial \mathbf{q}_i}
\right]\cdot\delta\mathbf{q}_i \ dt
\\&\hspace{25mm} 
+ \underbrace{\
\Big(\frac{\partial L}{\partial \mathbf{\dot{q}}_i} \Big|_{\mathbf{\dot{q}}_i=\tfrac{d\mathbf{q}_i}{dt}}
\cdot\delta \mathbf{q}_i \Big) \bigg|_{t=a}^{t=b}
}_{\hbox{\emph{Noether quantity}}}
=0
\end{split}
\label{HamPrincNoether}
\end{align}
for all smooth $\delta \mathbf{q}_i(t)$ satisfying 
$\delta\mathbf{q}_i(a)  = \delta\mathbf{q}_i(b) = 0 $.
This proves the equivalence of (i) and
(ii), upon restoring particle index $i$ in the last two lines. 
Note that the Noether quantity here is written relative to 
fixed coordinates in the inertial frame of Newton's force law.
\end{proof}

\begin{definition}
The conjugate momenta for the Lagrangian in \eqref{Lagrangian-thm} are
defined as
\begin{equation}
\mathbf{p}_i:=\frac{\partial L}{\partial \mathbf{\dot{q}}_i}\Big|_{\mathbf{\dot{q}}_i=\tfrac{d\mathbf{q}_i}{dt}}  
= m_i
\frac{d\mathbf{q}_i}{dt}\in
\mathbb{R}^3, \quad i=1,\ldots, N 
,\quad\hbox{(no sum on }i).
\label{momenta}
\end{equation}
\end{definition}

\begin{definition}
The \emph{Hamiltonian} \index{Hamiltonian} is defined via the change of variables
$(\mathbf{q},\mathbf{\dot{q}})\mapsto (\mathbf{q},\mathbf{p})$, called
the \emph{Legendre transform}, \index{Legendre transform} where one assumes one may recover 
$\mathbf{\dot{q}}$ as a function of $(\mathbf{q},\mathbf{p})$,
\begin{align}
H(\mathbf{q},\mathbf{p}) :&=
\mathbf{p}\cdot\mathbf{\dot{q}}(\mathbf{q},\mathbf{p}) -
L(\mathbf{q},\mathbf{\dot{q}}(\mathbf{q},\mathbf{p}))
\nonumber\\
&=\sum_{i=1}^N\frac{m_i}{2}\|\mathbf{\dot{q}}_i\|^2+V(\mathbf{q})
\nonumber \\ 
&= 
\underbrace{\
\sum_{i=1}^N\frac{1}{2m_i}\|\mathbf{p}_i\|^2\
}_{\hbox{Kinetic energy}}
\
+\
\underbrace{\
V(\mathbf{q})\
}_{\hbox{Potential}}.
\label{hamiltonian}
\end{align}
\end{definition}
\begin{remark}\rm 
The value of the Hamiltonian coincides with the  \emph{total energy} of the
system. This value will be shown to remain constant under the
evolution of Euler--Lagrange equations \eqref{EulerLag-eqs}.
\end{remark}

\begin{remark}\rm 
The Hamiltonian $H$ may be obtained from the Legendre transformation as a
function of the variables $(\mathbf{q},\mathbf{p})$, provided one may solve
for $\mathbf{\dot{q}}(\mathbf{q},\mathbf{p})$. Solving for $\mathbf{\dot{q}}(\mathbf{q},\mathbf{p})$ requires the
Lagrangian to be \emph{regular} \index{Lagrangian!regular}. In particular, it requires
\[
\det\frac{\partial^2 L}
{\partial \mathbf{\dot{q}}_i\partial \mathbf{\dot{q}}_i}\ne0
\quad\hbox{(no sum on }i)
.
\] 
\end{remark}

\begin{proof}[Step III: Proof that (ii) $\Leftrightarrow$ (iii)]
(Hamilton's principle of stationary action is equivalent to Hamilton's
canonical equations.)
Lagrangian \eqref{Lagrangian-thm} is regular and the
derivatives of the Hamiltonian may be shown to satisfy,
\[
\frac{\partial H}{\partial \mathbf{p}_i}
= \frac{1}{m_i}\mathbf{p}_i
=\mathbf{\dot{q}}_i=\frac{d \mathbf{q}_i}{dt}
\quad\hbox{and}\quad
\frac{\partial H}{\partial \mathbf{q}_i}
= \frac{\partial V}{\partial \mathbf{q}_i}
= -\frac{\partial L}{\partial \mathbf{q}_i}
.
\]
Consequently, the Euler--Lagrange equations \eqref{EulerLag-eqs} imply
\[
\mathbf{\dot{p}}_i = \frac{d \mathbf{p}_i}{dt} = \frac{d}{dt}\left(
\frac{\partial L}{\partial\mathbf{\dot{q}}_i} \right)
= \frac{\partial L}{\partial \mathbf{q}_i} 
= -\,\frac{\partial H}{\partial \mathbf{q}_i}\,.
\]
These calculations show that the Euler--Lagrange equations
\eqref{EulerLag-eqs} are equivalent to Hamilton's canonical
equations:
\begin{equation}
\label{hameq}
\mathbf{\dot{q}}_i = \frac{\partial H}{\partial \mathbf{p}_i}, \qquad
\mathbf{\dot{p}}_i =-\,\frac{\partial H}{\partial \mathbf{q}_i}\,,
\end{equation}
where $\partial H/ \partial \mathbf{q}_i, \partial H / \partial
\mathbf{p}_i
\in \mathbb{R}^3 $ are the gradients of $H $ with respect to $\mathbf{q}_i,
\mathbf{p}_i \in \mathbb{R}^3$, respectively.
This proves the equivalence of (ii) and (iii).
\end{proof}

\begin{remark}\rm 
The \emph{Noether quantity} \index{Noether quantity} in \eqref{HamPrincNoether} may be expressed in terms of the canonically conjugate momentum $\mathbf{p}$ as (dropping the subscript $i$).
\[
\Big(\frac{\partial L}{\partial \mathbf{\dot{q}}} \Big|_{\mathbf{\dot{q}}=\tfrac{d\mathbf{q}}{dt}}
\cdot\delta \mathbf{q} \Big) \Big|_{t=a}^{t=b}
= 
\Big(\mathbf{p} \cdot \delta \mathbf{q} \Big) \Big|_{t=a}^{t=b}
\]
\end{remark}

\begin{theorem}
\emph{Noether's theorem and tangent lift momentum maps.} \index{Noether's theorem!tangent lift! momentum map}
In coordinate notation $(\dot{q},q)\in TQ$ consider a Lagrangian $L:TQ\to \mathbb{R}$ which is right-invariant under the tangent lift of the Lie group $G$. The tangent lift of the right action of $G$ on the fixed configuration manifold $Q$ to its tangent space $TQ$ induces an equivariant momentum map $J: T^*Q\to \mathfrak{g}^*$ given by the endpoint term in 
Noether's theorem as,
\[
\scp{J(v_q)}{\xi}_{\mathfrak{g}^*\times \mathfrak{g}} = \scp{\mathbb{F}L(v_q)}{\xi_Q(q)}_{T^*Q\times TQ} 
\,,
\]
where $\xi\in \mathfrak{g}$, $v_q\in T_qQ$, $\mathbb{F}L(v_q) := \p L/\p v_q\in T^*Q$ is the fibre derivative of the Lagrangian $L$ and the variation $\de q =\xi_Q(q)\in T_qQ$ is taken to be the right $G$-action on the configuration manifold $Q$. This infinitesimal action is explicitly given by the positive Lie derivative,
\[
\xi_Q(q) = \frac{d}{ds}\Big|_{s=0} g_s q = \pounds_\xi q 
\quad\hbox{for}\quad g_s\in G \quad\hbox{and}\quad  \xi \in T_eG \quad\hbox{for subscript}\quad e \quad\hbox{at}\quad s=0
\,.\]
Consequently, one may explicitly write
\[
\scp{J(v_q)}{\xi}_{\mathfrak{g}^*\times \mathfrak{g}} = \scp{\frac{\p L}{\p v_q}}{\pounds_\xi q}
=: \scp{q \diamond \frac{\p L}{\p v_q}}{\xi }_{\mathfrak{g}^*\times \mathfrak{g}}
\,,
\]
in which the last equality defines the diamond operation $\diamond: T^*_qQ\times T_qQ\to \mathfrak{g}^*$.
We may now write the tangent lift momentum map $J: T^*Q\to \mathfrak{g}^*$ for right action of $G$ on the fixed configuration manifold $Q$ in terms of $(\diamond)$ as
\[
J(v_q) = q \diamond \frac{\p L}{\p v_q}
\,.
\]
Hence, the infinitesimal action leads directly to the corresponding momentum map.

This theorem may be demonstrated in the familiar example of the angular momentum map which arises when the particle Lagrangian $L(q,\dot{q})$ for $(q,\dot{q})\in T\mathbb{R}^3$ is invariant under rotations by the Lie group $SO(3)$.

\end{theorem}

\begin{example}\rm 
When the Lagrangian $L(q,\dot{q})$ for $q\in \mathbb{R}^3$ is right-invariant under infinitesimal rotations about a certain direction $\bs{\xi} \in \mathbb{R}^3$, say, then the variation 
may be written as $\delta \mathbf{q}=\bs{\xi}\times \mathbf{q}$. This variation does not vanish at the endpoints in time. 
Instead, as Noether described, it is constant in time and yields the following conservation law in this case, upon substituting $\mathbb{F}L(v_q) := p_q$,
\[
\mathbf{p} \cdot \delta \mathbf{q} = \mathbf{p} \cdot \bs{\xi}\times \mathbf{q} 
= \mathbf{q}  \times \mathbf{p} \cdot \bs{\xi} =: \mathbf{J}(\mathbf{q},\mathbf{p})\cdot \bs{\xi} =: J^\xi(\mathbf{q},\mathbf{p})
\,.\] 
This formula implies conservation of the $\bs{\xi}$--component of the particle angular momentum
\[
\mathbf{J}(\mathbf{q},\mathbf{p}):=\mathbf{q}  \times \mathbf{p}
 \,.\]
Thus, according to Noether's theorem, invariance of the Lagrangian under infinitesimal rotations around the fixed vector 
$\bs{\xi} \in \mathbb{R}^3$ implies conservation of angular momentum in the direction of the rotation axis of symmetry, $\bs{\xi}$.
Quantities arising from the endpoint term for Noether symmetry this way are called \emph{cotangent-lift momentum maps}. The reason for that name can be explained by treating $ \mathbf{J}(\mathbf{q},\mathbf{p})\cdot \bs{\xi}$ as a Hamiltonian and calculating its canonical Poisson brackets with $\mathbf{q}$ and $\mathbf{p}$. Namely,
\begin{align}
\begin{split}
\delta \mathbf{q} &= \{\mathbf{q}, J^\xi\} = \frac{\partial J^\xi}{\partial \mathbf{p}} = \bs{\xi}\times \mathbf{q} 
\,,\\
\delta \mathbf{p} &= \{\mathbf{p}, J^\xi\} = -\,\frac{\partial J^\xi}{\partial \mathbf{q}} = \bs{\xi}\times \mathbf{p} 
\,.\end{split}
\label{cotanLiftCalc}
\end{align}
These equations imply that the infinitesimal right action $\delta \mathbf{q}$ to rotate the position variable $\mathbf{q}(t)$ in the fixed inertial frame of the configuration manifold $Q$ has been `lifted' to the corresponding action $\delta \mathbf{p}$ on the canonically conjugate momentum $\mathbf{p}(t)$ in phase space $T^*\mathbb{R}^3$ via the Poisson bracket operations with the momentum-map Hamiltonian. Moreover, this Hamiltonian is precisely Noether's end point term, $\mathbf{p} \cdot \delta \mathbf{q} $. Thus, the Noether endpoint quantity has produced the cotangent-lift momentum map for right action of the Lie group $SO(3)$ on $Q=\mathbb{R}^3$.
\end{example}

\begin{remark}\rm 
The Euler--Lagrange equations for a simple mechanical system are second order and they determine
\emph{curves in configuration space} $\mathbf{q}_i\in
C^\infty([a,b],\mathbb{R}^{3N})$. In contrast, Hamilton's equations are
first order and they determine \emph{curves in phase
space} $(\mathbf{q}_i,\mathbf{p}_i)\in
C^\infty([a,b],\mathbb{R}^{6N})$, a space whose dimension is twice the
dimension of the configuration space.
\end{remark}

\begin{proof}[Step IV: Proof that (iii) $\Leftrightarrow$ (iv)]
(Hamilton's canonical equations may be written using a Poisson
bracket.)

By the chain rule and Hamilton's canonical equations in \eqref{hameq} any $F\in{\mathcal
F}(P)$ satisfies
\begin{align*}
\frac{dF}{dt}
&=
\sum_{i =1}^N \left(\frac{\partial F}{\partial \mathbf{q}_i}\cdot
\mathbf{\dot{q}}_i +\frac{\partial F}{\partial \mathbf{p}_i} \cdot 
\mathbf{\dot{p}}_i\right)\\
&
= \sum_{i =1}^N \left(\frac{\partial F}{\partial
\mathbf{q}_i} \cdot \frac{\partial H}{\partial \mathbf{p}_i}
-\frac{\partial F}{\partial \mathbf{p}_i}
\cdot \frac{\partial H}{\partial \mathbf{q}_i}\right)
=\{F,H\}
.
\end{align*}
This finishes the proof of the theorem, by proving the
equivalence of (iii) and (iv). 
\end{proof}

\begin{remark}\rm [Energy conservation]
Since the Poisson bracket is skew symmetric, $\{H,F\}=-\{F,H\}$, one
finds that $\dot{H}=\{H,H\}=0$. Consequently, the value of the Hamiltonian
is preserved by the evolution. Thus, the Hamiltonian is said to be a
\emph{constant of the motion}.
\end{remark}

\begin{exercise}
Show that the canonical Poisson bracket is bilinear, skew symmetric, satisfies the
Jacobi identity and acts as derivation on products of functions in phase
space.
\end{exercise}

\begin{exercise}
Given two constants of motion, what does the Jacobi identity imply about
additional constants of motion?
\end{exercise}

\begin{exercise}
Compute the Poisson brackets among
\[
J_i= \epsilon_{ijk}q_jp_k = (\bs{\widehat{q}}\, \bs{p})_i = (\mathbf{q}\times\mathbf{p})_i
\]
in Euclidean space. What Lie algebra do these Poisson brackets recall to
you?
\end{exercise}

\begin{exercise}
Verify that Hamilton's equations determined by the function
$$\langle J(\mathbf{p},\mathbf{q}), \bs{\xi}\rangle 
= (\mathbf{q}\times\mathbf{p}) \cdot \bs{\xi}
$$
define infinitesimal rotations about the $\bs{\xi}$--axis. Thus, in the case of rotation,
the momentum map $J(\mathbf{p},\mathbf{q})$ is the \emph{angular momentum}. 
\end{exercise}

\begin{exercise}
{\bf Rotations and angular momentum}\\ 
Find the Hamiltonian vector field for \index{Hamiltonian vector field}
\begin{eqnarray*}
J^\xi=
{\boldsymbol{\xi}}\cdot(\mathbf{q}\times\mathbf{p})
\end{eqnarray*}
for ${\boldsymbol{\xi}}\in\mathbb{R}^3$ and 
$(\mathbf{q},\mathbf{p})\in T^*\mathbb{R}^3\simeq\mathbb{R}^3\times\mathbb{R}^3$.\smallskip

Interpret this vector field geometrically.
\end{exercise}

\begin{answer}
\emph{Notations and angular momentum}\\
The Hamiltonian vector field $X_{J^\xi}$ for
$J^\xi={\boldsymbol{\xi}}\cdot\mathbf{J}
={\boldsymbol{\xi}}\cdot(\mathbf{q}\times\mathbf{p})$
is obtained from the canonical Poisson bracket by setting
\begin{eqnarray*}
X_{J^\xi}
=\{\cdot\,,J^\xi\,\}
=
\{q_k\,,J^\xi\,\}\frac{\partial}{\partial q_k}
+
\{p_k\,,J^\xi\,\}\frac{\partial}{\partial p_k}
\,.\end{eqnarray*}

The Poisson brackets among the components of $\mathbf{J}$ may be computed in two stages. \\
First, use canonical
Poisson brackets $\{q_j,p_k\}=\delta_{jk}$ with the definition 
$J_l=\epsilon_{lmn}q_mp_n$ for $l,m,n=1,2,3$ to find 
\begin{eqnarray*}
\widehat{q}_{kl}
:=
\{q_k,J_l\}
=
\epsilon_{lmn}q_m\{q_k,p_n\}
=
\epsilon_{lmn}q_m\delta_{kn}
=
\epsilon_{lmk}q_m
= \epsilon_{klm}q_m
\,.\end{eqnarray*}
The skew symmetry of $\widehat{q}_{kl}=-\widehat{q}_{lk}$ arises from the skew symmetry of the
Poisson bracket. 
\\

A similar calculation of the Poisson bracket for the components of the
canonically conjugate momentum $\{\mathbf{p}\,,\,\mathbf{J}\}$ gives, 
\begin{eqnarray*}
\widehat{p}_{kl}
:=
\{p_k,J_l\}
= \epsilon_{klm}p_m
\end{eqnarray*}
Next, calculate
\begin{eqnarray*}
\{q_k,{\boldsymbol{\xi}}\cdot\mathbf{J}\}
=
\{q_k,\xi_l J_l\}
=
\xi_l\epsilon_{lmn}q_m\{q_k, p_n\}
=
\epsilon_{klm}\xi_l q_m
=
(\boldsymbol{\xi}\times\mathbf{q})_k
\end{eqnarray*}
Likewise for the conjugate momentum vector $\mathbf{p}$, so that
\begin{eqnarray*}
\{\mathbf{q},{\boldsymbol{\xi}}\cdot\mathbf{J}\}
=
\boldsymbol{\xi}\times\mathbf{q}
\quad\hbox{and}\quad
\{\mathbf{p},{\boldsymbol{\xi}}\cdot\mathbf{J}\}
=
\boldsymbol{\xi}\times\mathbf{p}
\end{eqnarray*}
The Hamiltonian vector field associated with \index{Hamiltonian vector field}
$J^\xi={\boldsymbol{\xi}}\cdot\mathbf{J}$ may now be expressed as
\begin{eqnarray*}
X_{J^\xi}
=\{\cdot\,,J^\xi\,\}
=
\{\mathbf{q}\,,J^\xi\,\}\cdot\frac{\partial}{\partial \mathbf{q}}
+
\{\mathbf{p}\,,J^\xi\,\}\cdot\frac{\partial}{\partial \mathbf{p}}
=
(\boldsymbol{\xi}\times\mathbf{q})\cdot\frac{\partial}{\partial \mathbf{q}}
+
(\boldsymbol{\xi}\times\mathbf{p})\cdot\frac{\partial}{\partial \mathbf{p}}
\end{eqnarray*}
\subsection{Geometrical interpretation}
The flow of the vector field $X_{J^\xi}$ is a rotation with angular
velocity
${\boldsymbol{\xi}}$. As might be expected, 
\begin{eqnarray*}
\{|\mathbf{q}|^2\,,J^\xi\,\}
=
\{|\mathbf{p}|^2\,,J^\xi\,\}
=
0
=
\{\mathbf{q}\cdot\mathbf{p}\,,J^\xi\,\}
\end{eqnarray*}
because rotations leave invariant the lengths and relative orientations of vectors
in $\mathbb{R}^3$. If ${\boldsymbol{\xi}}$ is a unit vector, the quantity
$J^\xi={\boldsymbol{\xi}}\cdot\mathbf{J}$ is the angular momentum in the direction
of ${\boldsymbol{\xi}}$. Rotationally invariant Hamiltonians Poisson-commute
with $J^\xi$. Therefore, rotationally invariant Hamiltonians conserve all three components of the angular momentum, $\mathbf{J}:=\mathbf{q}\times\mathbf{p}$.
\\ 
\end{answer}

\begin{remark}\rm 
\emph{Hat map}\index{hat map}
The table of Poisson brackets $\widehat{q}_{kl}=\{q_k,J_l\}$ produces the
linear invertible map,
\begin{eqnarray*}
\label{hat-map-coords}
\mathbf{q}: =(q_1, q_2, q _3) \in \mathbb{R}^3 
\quad\mapsto \quad
\{\mathbf{q}\,,\,\mathbf{J}\}
=
\widehat{\mathbf{q}}: 
=
\left[
\begin{array}{ccc}
0&-q_3&q_2\\
q_3&0&-q_1\\
-q_2&q_1&0
\end{array}
\right] 
\end{eqnarray*}
This \emph{hat map} $\widehat{q}_{ij} = -\,\epsilon_{ijk}q_k$ is a remarkable isomorphism because the skew-symmetric $3\times3$ matrices comprise the 
matrix Lie algebra $\mathfrak{so}(3)$ for the matrix Lie group $SO(3)$ of rotations by right action in three dimensions. \\
\end{remark}

\begin{exercise} \emph{Angular momentum Poisson brackets}\\ \index{Poisson brackets!angular momentum}
Compute the Poisson brackets among 
\begin{eqnarray*}
J_l=\epsilon_{lmn}q_mp_n
\quad\hbox{for}\quad
l,m,n=1,2,3,
\end{eqnarray*}
using the canonical Poisson brackets $\{q_k,p_m\}=\delta_{km}$. \\

Do the Poisson
brackets $\{J_l,J_m\}$ close among themselves? If so, what does this imply for the
restriction of the dynamics to functions of $(J_1,J_2,J_3)$? 
\\

Compute the
Hamiltonian vector field for $J^\xi={\boldsymbol{\xi}}\cdot\mathbf{J}$ in terms of $\mathbf{J}$. 
\end{exercise}
\bigskip

\begin{answer}
{\bf Angular momentum Poisson brackets}\\ 
The components of the angular momentum $J_l=\epsilon_{lmn}q_mp_n$ do Poisson-commute
among themselves. In particular, 
\begin{eqnarray*}
\widehat{J}_{kl}
:=
\{J_k,J_l\}
=\epsilon_{klm}J_m
\end{eqnarray*}
\textbf{Proof}
The Poisson brackets are computed as
\begin{eqnarray*}
\{\mathbf{q}\times\mathbf{p},{\boldsymbol{\xi}}\cdot\mathbf{J}\}
&=&
\{\mathbf{q},{\boldsymbol{\xi}}\cdot\mathbf{J}\}\times\mathbf{p}
+
\mathbf{q}\times\{\mathbf{p},{\boldsymbol{\xi}}\cdot\mathbf{J}\}
\\&=&
(\boldsymbol{\xi} \times \mathbf{q})\times\mathbf{p}
+
\mathbf{q}\times(\boldsymbol{\xi} \times \mathbf{p})
=
\boldsymbol{\xi}\times(\mathbf{q}\times\mathbf{p})
\,.\end{eqnarray*}
Consequently,
\begin{eqnarray*}
\{\mathbf{J},{\boldsymbol{\xi}}\cdot\mathbf{J}\}
=
\boldsymbol{\xi} \times \mathbf{J}
\end{eqnarray*}
and the result follows for $\widehat{J}_{kl}$. In addition, the Hamiltonian vector
field $X_{J^\xi}$ becomes,
\begin{eqnarray*}
X_{J^\xi}
=\{\cdot\,,J^\xi\,\}
=
\{\mathbf{J}\,,J^\xi\,\}\cdot\frac{\partial}{\partial \mathbf{J}}
=
-\,(\boldsymbol{\xi}\times\mathbf{J})\cdot\frac{\partial}{\partial \mathbf{J}}
\,.\end{eqnarray*}

Likewise, for any Hamiltonian $H(\mathbf{J})$ depending only on the angular momentum,
one finds by a similar calculation that,
\begin{eqnarray*}
\mathbf{\dot{J}} = \{\mathbf{J},H\}
=
- \,\mathbf{J}\times\frac{\partial H}{\partial\mathbf{J}}
\,.\end{eqnarray*}
Thus, by the \emph{product rule} \index{product rule} any function of the angular momentum
$F(\mathbf{J})$ satisfies the Poisson bracket relation,
\begin{eqnarray*}
\frac{dF}{dt} = \{F,H\}(\mathbf{J})
=
\mathbf{J}\cdot\frac{\partial F}{\partial\mathbf{J}}
\times\frac{\partial H}{\partial\mathbf{J}}
\hspace{2cm}(LPB)
\end{eqnarray*}
\end{answer}

\begin{remark}\rm 
Any function $C(|\mathbf{J}|)$ of the magnitude $|\mathbf{J}|$ satisfies
\begin{eqnarray*}
\{C,H\}(\mathbf{J})
=
C\,'(|\mathbf{J}|)\,\mathbf{J}\cdot
\frac{\partial |\mathbf{J}|}{\partial\mathbf{J}}
\times\frac{\partial H}{\partial\mathbf{J}}
=
C\,'(|\mathbf{J}|)\,\mathbf{J}\cdot
\frac{\mathbf{J}}{|\mathbf{J}|}
\times\frac{\partial H}{\partial\mathbf{J}}
=
0
\,,\quad\forall H
\,.
\end{eqnarray*}
Functions $C(|\mathbf{J}|)$ that are distinguished by Poisson-commuting with all
others under the Poisson bracket are called \emph{Casimirs}. \index{Casimirs}
\end{remark}

\begin{remark}\rm 
The map 
$J^\xi={\boldsymbol{\xi}}\cdot(\mathbf{q}\times\mathbf{p})$ is an example of a \emph{Poisson map} \index{Poisson map}.
This means the Lie-Poisson bracket in (LPB) satisfies the defining properties of a Poisson bracket, including the Jacobi identity. This may be recognized
by computing the Jacobi identity for the corresponding Hamiltonian vector fields 
$X_{\mathbf{J}}=\{\,\cdot\,,\,\mathbf{J}\}=-\,\mathbf{J}\times\frac{\partial}{\partial
\mathbf{J}}$. Another proof is to notice that the Lie-Poisson bracket $\{J_a\,,\,J_b\}=\epsilon_{abc}J_c$ may be expressed as a linear functional on a Lie algebra with basis elements $e_a$ and dual basis elements $e^d$ as
\begin{eqnarray*}
\{F\,,\,H\}(\mathbf{J})
&=&
\mathbf{J}\cdot\frac{\partial F}{\partial\mathbf{J}}
\times\frac{\partial H}{\partial\mathbf{J}}
\\
&=&
\Big\langle
J_d\,e^d\,,\,\frac{\partial F}{\partial J_a}\Big[e_a\,,\,
e_b\Big]\frac{\partial H}{\partial J_b}
\Big\rangle
\quad\hbox{with pairing}\quad
\langle\,e^d\,,\,e_c\,\rangle = \delta^d_c
\,.
\end{eqnarray*}
In this case the Lie algebra bracket $[\cdot\,,\,\cdot]$ with structure constants $C_{ab}^c=\epsilon_{abc}$ in \[[e_a\,,\,e_b]=C_{ab}^ce_c\] corresponds to $\mathfrak{so}(3)$, the Lie algebra for the rotation group in three dimensions.
\end{remark}

\begin{exercise}
Write hat maps for all three-dimensional matrix Lie algebras. 
\end{exercise}

\subsection{Phase-space action principle}
Hamilton's principle for stationary variations $\delta S=0$ of the action integral $S=\int_{t_a}^{t_b}L(q,\dot{q})dt$ on the tangent space $TM$ of a manifold $M$ may be augmented for clarity by imposing the relation $\dot{q}=dq/dt$ as an \emph{additional constraint} in terms of generalised coordinates $(q,\dot{q})\in T_qM$. The \emph{Hamilton--Pontryagin constrained action} is given by \index{Hamilton--Pontryagin principle!constrained action}
\begin{eqnarray}
S = \int_{t_a}^{t_b} L(q,\dot{q})
+ \scp{p}{\Big(\frac{dq}{dt}-\dot{q} \Big)}
\,dt
\,,
\label{constrained-action}
\end{eqnarray}
where $p$ is a \emph{Lagrange multiplier} for the \emph{constraint}, $\dot{q}=dq/dt$. 
\index{Lagrange multiplier! constraint}
The variations of this action result in 
\begin{align}
\begin{split}
\delta{S} 
&= \int_{t_a}^{t_b} 
\scp{\left(\frac{\partial L}{\partial q} - \frac{dp}{dt}\right)}{ \delta{q}}
+ \scp{\left(\frac{\partial L}{\partial \dot{q}}- p\right )}{ \delta\dot{q}}
\\&\hspace{15mm} +
\scp{ \Big(\frac{dq}{dt}-\dot{q} \Big)}{\delta{p}}\,dt
+\
 \scp{ p }{\delta{q}} \Big|_{t_a}^{t_b}
\,.
\end{split}
\label{PhaseSpaceLag}
\end{align}
The contributions at the endpoints $t_a$ and $t_b$ in time vanish, provided the variations $\delta{q}$ are assumed to vanish then.

Thus, stationarity of this action under these variations imposes the three relations
\begin{eqnarray}
\begin{split}
\delta{q}:&& \frac{\partial L}{\partial q} = \frac{dp}{dt}\,,
\\
\delta\dot{q}:&& \frac{\partial L}{\partial \dot{q}} = p\,,
\\
\delta{p}:&& \dot{q}=\frac{dq}{dt}
\,.
\end{split}
\label{LT}
\end{eqnarray}
\begin{itemize}
\item
Combining the first and second of these relations recovers the Euler--Lagrange equations, $[\,L\,]_{q}=0$. 
\item
The third relation constrains the variable $\dot{q}$ to be the time derivative of the trajectory $q(t)$ at any time $t$
after the variations have been taken.
\end{itemize}

Substituting the Legendre-transform relation in \eqref{LT} into the constrained action \eqref{constrained-action} yields the \emph{phase-space action} \index{phase-space action}
\begin{eqnarray}
S = \int_{t_a}^{t_b} 
\Big(p\frac{dq}{dt}-H(q,p) \Big)
\,dt
\,.
\label{phase-space-action}
\end{eqnarray}
Varying the phase-space action in (\ref{phase-space-action}) yields
\begin{eqnarray*}
\delta{S} 
&=& \int_{t_a}^{t_b} 
\left(\frac{dq}{dt}-\frac{\partial H}{\partial p}\right) \delta{p}
-
\left(\frac{dp}{dt}+\frac{\partial H}{\partial q}\right) \delta{q}
\,dt
+ 
\Big[ p\,\delta q  \Big]_{t_a}^{t_b}
\,.
\end{eqnarray*}
Provided the variations $\delta{q}$ vanish at the endpoints $t_a$ and $t_b$ in time, then the last term vanishes. Thus, stationary variations of the phase-space action in (\ref{phase-space-action}) recover Hamilton's canonical equations \eqref{hameq} directly.


Hamiltonian evolution along a curve $(q(t),p(t))\in T^*M$ satisfying
Equations \eqref{hameq} induces the evolution of a given function $F(q,p): T^*M\to \mathbb{R}$ on the phase space $T^*M$ of a manifold $M$, as
\begin{align}
\frac{dF}{dt}
&= 
\frac{\partial F}{\partial q}\frac{dq}{dt}
+
\frac{\partial H}{\partial q}\frac{dp}{dt} 
\nonumber\\&= 
\frac{\partial F}{\partial q}\frac{\partial H}{\partial p}
-
\frac{\partial H}{\partial q}\frac{\partial F}{\partial p} =:  \{F\,,\,H\}
\label{PB-def}\\&= 
\left(\frac{\partial H}{\partial p}\frac{\partial }{\partial q}
-
\frac{\partial H}{\partial q}\frac{\partial }{\partial p}\right)F =:  X_H F
\,.
\label{HamVF-def}
\end{align}
The second and third lines of this calculation introduce notation for two natural operations that will be investigated further in the next few sections. These are the \emph{canonical Poisson brackets} $ \{\,\cdot\,,\,\cdot\,\}$ in \eqref{PB-def} and the \emph{Hamiltonian vector field} $X_H=\{\,\cdot\,,\,H\,\}$ in \eqref{HamVF-def}. \index{Poisson brackets!canonical}
\index{vector field! Hamiltonian}

\newpage
\vspace{4mm}\centerline{\textcolor{shadecolor}{\rule[0mm]{6.75in}{-2mm}}\vspace{-4mm}}
\section{The rigid body in $\mathbb{R}^3$} 

\secttoc

\textbf{What is this lecture about?} This lecture reformulates Euler's equations for  
rigid body dynamics by using the complementary variational approaches of Lagrange and Hamilton.

\subsection{Euler's equations for the rigid body in $\mathbb{R}^3$}
In the absence of external torques, Euler's equations for rigid body
motion are:
\begin{equation}\label{rbe1}
\begin{aligned}
I_1\dot{\Omega}{_1}  &=  (I_2 -I_3)\Omega_2\Omega_3, \\
I_2\dot{\Omega}{_2}  &=  (I_3 -I_1)\Omega_3\Omega_1, \\
I_3\dot{\Omega}{_3}  &=  (I_1 -I_2)\Omega_1\Omega_2,
\end{aligned}
\end{equation}
or, equivalently, 
\[
{\mathbb I} \boldsymbol{\dot{\Omega}}
={\mathbb I} \boldsymbol{\Omega} \times \boldsymbol{\Omega},
\]
where $\boldsymbol{\Omega} = (\Omega_1, \Omega_2, \Omega_3)$ is the
body angular velocity vector and  $I_1,  I_2, I_3$  are the moments
of inertia of the rigid body.

\begin{exercise}
Can equations \eqref{rbe1} be cast into Lagrangian or
Hamiltonian form in any sense?  (Since there are an odd number of
equations, they cannot be put into canonical Hamiltonian form.)
\end{exercise}

\noindent
We could reformulate them as:
\begin{itemize}
\item Euler--Lagrange equations on $T\mathrm{SO(3)}$ or 
\item Canonical Hamiltonian equations on  $T^\ast\mathrm{SO(3)}$, 
\end{itemize}
by using Euler angles and their velocities, or their conjugate momenta.
However, these reformulations on $T{\mathrm SO(3)}$ or $T^\ast{\mathrm SO(3)}$
would answer a different question for a \textit{ six} dimensional system. We
are interested in these structures for the equations as given above.

\begin{answer}[Lagrangian formulation] \index{rigid body!Lagrangian formulation!action principle}
\index{variational principle!Euler--Poincar\'e}
The Lagrangian answer is this: These equations
may be expressed in Euler--Poincar\'e form on the Lie algebra $
\mathfrak{so}(3)\simeq\mathbb{R}^3$ using the Lagrangian
\begin{equation}\label{rbl-1}
l (\boldsymbol{\Omega}) = \tfrac{1}{2} (I _1 \Omega ^2 _1 + I _2
\Omega ^2 _2 + I _3 \Omega ^2 _3 )
=
\tfrac{1}{2}\boldsymbol{\Omega}^T \cdot \mathbb{I}\boldsymbol{\Omega}
,
\end{equation}
which is the (rotational) kinetic energy of the
rigid body.%
\footnote{The Hamiltonian answer to this question will be discussed later.}
\end{answer}

\begin{proposition}
The Euler rigid body equations are equivalent to the \emph{rigid body
action principle} for a \emph{reduced action} \index{reduced action}
\begin{equation}
\delta S_{\mathrm{red}} = \delta \int^b_a l (\boldsymbol{\Omega})\,dt 
= \delta \int^b_a \tfrac{1}{2}\boldsymbol{\Omega}^T \cdot \mathbb{I}\boldsymbol{\Omega}\,dt
= 0,
\end{equation}
where variations of $\boldsymbol{\Omega} $ are restricted to
be of the form
\begin{equation}
\delta \boldsymbol{\Omega} = \boldsymbol{\dot{ \Sigma}} +
\boldsymbol{\Omega} \times \boldsymbol{\Sigma},
\end{equation}
in which $\boldsymbol{\Sigma}(t)$ is a curve in $\mathbb{R}^3$ that
vanishes at the endpoints in time. 
\end{proposition}

\begin{proof}
Since
$ l (\boldsymbol{\Omega}) = \tfrac{1}{2} \langle  {\mathbb I}
\boldsymbol{\Omega}, \boldsymbol{\Omega} \rangle $, and ${\mathbb I}$
is symmetric, we obtain
\begin{align*}
\delta \int^b_a l(\boldsymbol{\Omega}) \, d t
& =  \int^b_a \langle  {\mathbb I} \boldsymbol{\Omega},\delta
\boldsymbol{\Omega}\rangle \, dt \\
& =  \int^b_a \langle  {\mathbb I} \boldsymbol{\Omega},
\boldsymbol{\dot{ \Sigma}}
      + \boldsymbol{\Omega} \times \boldsymbol{\Sigma}\rangle \, dt\\
& =  \int^b_a \left[ \left\langle
      - \frac{d }{d t}{\mathbb I} \boldsymbol{\Omega} ,
      \boldsymbol{\Sigma}\right\rangle 
      + \Big\langle {\mathbb I} \boldsymbol{\Omega}, \boldsymbol{\Omega} \times \boldsymbol{\Sigma}\Big\rangle\right] {d t}
+ \left\langle \mathbb{I} \boldsymbol{\Omega} \,,\, \boldsymbol{\Sigma} \right\rangle\Big|_a^b
\\& = \int^b_a \left\langle
- \frac{d }{d t} {\mathbb I} \boldsymbol{\Omega}
      + {\mathbb I} \boldsymbol{\Omega}\times \boldsymbol{\Omega} 
      \,\,,\,\boldsymbol{\Sigma}\right\rangle \,dt
\,,\end{align*}
upon integrating by parts and using the
endpoint conditions, $ \boldsymbol{\Sigma} (b) =
\boldsymbol{\Sigma} (a) = 0 $. Since $\boldsymbol{\Sigma}$ is
otherwise arbitrary,~\eqref{rbvp1} is equivalent to
\[ -\frac{d }{d t} ({\mathbb I} \boldsymbol{\Omega})
+ {\mathbb I} \boldsymbol{\Omega} \times \boldsymbol{\Omega}
= 0 , \]
which are Euler's rigid body equations \eqref{rbe1}.
\end{proof}

Let's derive this variational principle from the
\textit{standard} Hamilton's principle.

\subsection{Hamilton's principle for rigid body motion on $T{\mathrm SO}(3)$}

An element ${ O} \in {\mathrm SO}(3)$
gives the configuration of the body as a map of a \emph{reference configuration} 
\emph{reference configuration} 
${\mathcal B}\subset \mathbb{R}^3$ to the current
configuration ${ O}({\mathcal B})$. The map
${ O}$ takes a reference or label point $ \mathbf{X} \in
{\mathcal B} $ to a current point $ \mathbf{x} = { O}(\mathbf{X}) \in
{ O}({\mathcal B} )$. 

When the matrix ${ O}$ is time-dependent the rigid body undergoes motion
relative to the reference configuration. 
Thus, 
\[ \mathbf{x}(t) = {O}(t)\mathbf{X} \]
with ${O}(t)$ a curve parametrised by time in ${\mathrm SO}(3)$.
The relative velocity in fixed space of a point of the body is given by
\[\mathbf{\dot{ x}}(t) = {\dot{O}}(t) \mathbf{X} 
= {\dot{O}} { O} ^{-1}(t) \mathbf{x}(t).\]
As discussed earlier, since ${ O}$ is an
orthogonal matrix, ${ O} ^{-1} {\dot{O}}$ and
${\dot{O}}{ O}^{-1}$ are skew matrices. Consequently,
we can write (recall the hat map in Remark \ref{Remark-HatMap})
\begin{equation}\label{sav}
\mathbf{\dot{x}} = {\dot{O}}{ O} ^{-1} \mathbf{x} = \boldsymbol{\omega}
\times \mathbf{x}.
\end{equation} 
This formula defines the \emph{spatial angular velocity \index{angular velocity!spatial}
vector\/} $\boldsymbol{\omega}$.  Thus, the vector $\boldsymbol{\omega}
={\big({\dot{O}} { O} ^{-1}\big)}\boldsymbol{\widehat{\,}}$ is
associated with the {\em right\/} translation of ${\dot{O}}$ to the identity. 
The corresponding \emph{body angular velocity} vector is defined by \index{angular velocity!body}
\begin{equation}\label{bav}
\boldsymbol{\Omega} = { O} ^{ -1} \boldsymbol{\omega} ,
\end{equation}
so that $\boldsymbol{\Omega}\in\mathbb{R}^3$ is the angular velocity
relative to a body fixed frame. Notice that
\begin{align}\label{lefttranslation}
{ O} ^{ -1} {\dot{O}} \mathbf{X}
&= { O} ^{ -1} {\dot{O}}
        { O} ^{ -1} \mathbf{x} = { O} ^{ -1}
        (\boldsymbol{\omega} \times \mathbf{x}) \nonumber \\
&= { O}^{ -1} \boldsymbol{\omega} \times { O}^{ -1} \mathbf{x} 
= \boldsymbol{\Omega} \times \mathbf{X},
\end{align}
so that $\boldsymbol{\Omega}\times$ is given by {\em left\/} translation
of ${\dot{O}}$ to the identity. That is, the skew-symmetric matrix $\boldsymbol{\Omega}\times$
is given by
\[
\boldsymbol{\Omega}\times
=\big({ O} ^{-1}{\dot{O}}\big)\boldsymbol{\widehat{\,}}
.
\]
The \emph{kinetic energy} \index{kinetics energy!rigid body} is obtained by summing up $m | \mathbf{\dot{x}}
| ^2 / 2$ (where $|\,{\cdot }\,|$ denotes the Euclidean norm) over the
body. This yields
\begin{equation}\label{ke}
 K = \frac{1}{2} \int _{\mathcal B} \rho(\mathbf{X}) |
{\dot{O}} \mathbf{X} | ^2 \, d^3 X,
\end{equation}
in which $ \rho(\mathbf{X})$ is a given mass density in the reference
configuration. Since the action of $SO(3)$ on $\mathbb{R}^3$ preserves 
vector magnitudes, one finds
\[
| {\dot{O}} \mathbf{X} |  = | \boldsymbol{\omega} \times \mathbf{x} 
| = |
{ O} ^{-1}(\boldsymbol{\omega} \times \mathbf{x} ) 
| = |
({ O} ^{-1}\boldsymbol{\omega} \times { O} ^{-1} \mathbf{x} ) 
| = | 
\boldsymbol{\Omega} \times \mathbf{X} |,
\]
one sees that $K$ is a quadratic function of $\boldsymbol{\Omega}$. Writing
\begin{equation}\label{mit}
K = \tfrac{1}{2} \boldsymbol{\Omega} ^T \cdot{\mathbb I}\boldsymbol{\Omega}
\end{equation}
defines the \emph{moment of inertia tensor\/}, \index{moment of inertia! tensor} ${\mathbb I}$, which is a positive-definite
$(3\times 3)$ matrix (provided the body does not degenerate to a line interval) and the kinetic energy $K$ is a quadratic form.  
This quadratic form can be diagonalized by a change of basis; thereby defining the
principal axes and moments of inertia. In this basis, we write
$ {\mathbb I} = {\mathrm diag}(I _1, I _2, I_3)$.

The function $K$ is
taken to be the Lagrangian of the system on
$ T{\mathrm SO}(3) $. Notice that $K$ in equation \eqref{ke} is
{\em left\/} (not right) invariant on $T{\mathrm SO}(3)$, since
\[
\boldsymbol{\Omega}\times = { O} ^{-1}{\dot{O}}
\quad\Longleftrightarrow\quad 
\boldsymbol{\Omega} =\big({ O} ^{-1}{\dot{O}}\big)\boldsymbol{\widehat{\,}}
.
\]

In the framework of Hamilton's principle, the relation between
motion in $O\in SO(3)$ space and motion in body angular velocity $O^{-1}\dot{O} \in T_eSO(3)$
(or $\boldsymbol{\Omega}$) space is as follows.

\begin{proposition}
The curve $ {O} (t) \in {\mathrm SO}(3) $ satisfies the Euler--Lagrange
equations for
\begin{equation}\label{rbl-2}
 L ({ O}, {\dot{O}} )
= \frac{1}{2} \int_{{\mathcal B}}
\rho (\mathbf{X})
|  {\dot{O}} \mathbf{X} | ^2 \, d ^3 \mathbf{X},
\end{equation}
if and only if $\boldsymbol{\Omega} (t) $ defined by
${ O} ^{-1} {\dot{O}} \mathbf{v} = \boldsymbol{\Omega}
\times \mathbf{v}$ for all $\mathbf{v} \in \mathbb{R}^3$ satisfies
Euler's rigid body equations,
\begin{equation}\label{ee}
{\mathbb I} \boldsymbol{\dot{ \Omega}} = {\mathbb
I}\boldsymbol{\Omega}\times \boldsymbol{\Omega}.
\end{equation}
\end{proposition}
The proof of this relation will illustrate how
to reduce variational principles using their symmetry groups.
By Hamilton's principle, ${O}(t)$
satisfies the Euler--Lagrange equations, if and only if
\[
\delta\int L ({ O}, {\dot{O}} ) \, dt = 0 .
\]
 Let $ l (\boldsymbol{\Omega}) = \frac{1}{2} ( {\mathbb I}
\boldsymbol{\Omega})
\cdot \boldsymbol{\Omega} $, so that
$ l (\boldsymbol{\Omega}) = L ({ O}, {\dot{O}} ) $
where the matrix ${ O}$ and the vector $\boldsymbol{\Omega}$ are
related by the hat map, $\boldsymbol{\Omega} ={\big({ O}
^{-1}{\dot{O}}\big)}\boldsymbol{\widehat{\,}}$. Thus, the Lagrangian
$L$ is left SO(3)-invariant. That is, 
\[
l (\boldsymbol{\Omega}) = L ({ O}, {\dot{O}} )
= L (\mathbf{ e}, { O}^{-1}{\dot{O}} )
.
\]
To see how we should use this left-invariance to transform
Hamilton's principle, define the skew matrix
${\widehat{\Omega}}$  by
${\widehat{\Omega}}\mathbf{v} = \boldsymbol{\Omega} \times
\mathbf{v}$ for any $\mathbf{v} \in \mathbb{R}^3$.

We differentiate the relation
${ O} ^{-1} {\dot{O}} = {\widehat{\Omega}}$
with respect to ${ O}$ to get
\begin{equation}\label{rda}
- { O} ^{-1} (\delta { O}){ O}^{-1}
{\dot{O}}  + { O}^{-1} (\delta
{\dot{O}}) = \widehat{\delta \boldsymbol{\Omega}} \,.
\end{equation}
Let the skew matrix ${\widehat{\Sigma}}$ be defined by
\begin{equation}\label{dsig}
 {\widehat{\Sigma}} = {O}^{-1} \delta {O},
\end{equation}
and define the corresponding vector $\boldsymbol{\Sigma}$ by
\begin{equation}\label{dhat}
{\widehat{\Sigma}} \mathbf{v} = \boldsymbol{\Sigma} \times
\mathbf{v} .
\end{equation}
Note that
\[
{\dot{ \widehat{\Sigma}}}
= - { O}^{-1} {\dot{O}} { O} ^{-1}
\delta { O} + { O} ^{-1}
\delta{\dot{O}}
\,.\]
Consequently,
\begin{equation}\label{dela}
{ O} ^{-1} \delta {\dot{O}}
= {\dot{ \widehat{\Sigma}}} + { O} ^{-1}
{\dot{O}} {\widehat{\Sigma}}
\,.
\end{equation}
Substituting~\eqref{dela} and~\eqref{dsig}
into~\eqref{rda} gives
\[ -{\widehat{\Sigma}} {\widehat{\Omega}} 
+ {\dot{ \widehat{\Sigma}}} 
+ {\widehat{\Omega}} {\widehat{\Sigma}}
= \widehat{ \delta {\Omega}}, \]
that is, cf. relation \eqref{EP-Var-Id},
\begin{equation}\label{domh}
\widehat{ \delta {\Omega}} = 
{\dot{ \widehat{\Sigma}}} 
+ [
{\widehat{\Omega}},
{\widehat{\Sigma}} ] .
\end{equation}
The identity
$ [ {\widehat{\Omega}}, {\widehat{\Sigma}} ] =
(\boldsymbol{\Omega} \times
\boldsymbol{\Sigma}) \boldsymbol{\widehat{\ }}$ holds by the exercises in \eqref{hat-exercises}.
Hence, one has
\begin{equation}\label{dom}
\delta \boldsymbol{\Omega} = \boldsymbol{\dot{\Sigma}} +
\boldsymbol{\Omega}
\times \boldsymbol{\Sigma} \,.
\end{equation}


These calculations have proved the following:
\begin{theorem}  \label{red/variational/prpl}
For a Lagrangian that is left-invariant under SO(3),
Hamilton's variational principle
\begin{equation}\label{hprin}
\delta S = \delta \int^b_a L ({ O}, {\dot{O}} )\, d t = 0
\end{equation}   on $T{\mathrm SO}(3)$ is equivalent to the 
\emph{reduced variational principle\/} \index{variational principle!reduced}
\begin{equation}\label{rvprin}
\delta S_{\mathrm  red}= \delta \int^b_a l(\boldsymbol{\Omega}) \, d t = 0
\end{equation}  
with $\boldsymbol{\Omega} =\big({O}^{-1}
{\dot{O}}\big)\boldsymbol{\widehat{\,}}$ on $\mathbb{R}^3$ where the
variations $ \delta
\boldsymbol{\Omega} $ are of the form
\[
\delta \boldsymbol{\Omega} = \boldsymbol{\dot{\Sigma}} +
\boldsymbol{\Omega}
\times \boldsymbol{\Sigma},
\]
with
$\boldsymbol{\Sigma}(a) =\boldsymbol{\Sigma}(b) = 0 $.
\end{theorem}


\subsection{Reconstruction of ${ O}(t)\in SO(3)$} In 
\ref{red/variational/prpl}, Euler's equation for the rigid body
\begin{equation}
{\mathbb I} \boldsymbol{\dot{ \Omega}} =
{\mathbb I}\boldsymbol{\Omega}\times \boldsymbol{\Omega}
\,,
\label{RB-eqn}
\end{equation}
follows from the reduced variational principle \eqref{rvprin} for the
Lagrangian 
\begin{equation}\label{rb-redlag}
l(\boldsymbol{\Omega}) =  
 \tfrac{1}{2} ( {\mathbb I}
\boldsymbol{\Omega})
\cdot \boldsymbol{\Omega} 
,
\end{equation}
which is expressed in terms of the left-invariant time-dependent angular
velocity in the body, $\boldsymbol{\Omega}\in \mathfrak{so}(3)$. The
body angular velocity $\boldsymbol{\Omega}(t)$ yields the tangent vector
${\dot{O}}(t)\in T_{{ O}(t)}SO(3)$ along the integral curve
in the rotation group ${ O}(t)\in SO(3)$ by the relation,
\begin{equation}
{\dot{O}}(t)={ O}(t)\boldsymbol{\Omega}(t)
\,.
\label{recon-formula}
\end{equation}
This relation provides the \emph{reconstruction formula}. \index{reconstruction formula} Its solution as
a linear differential equation with time-dependent coefficients yields the
integral curve ${ O}(t)\in SO(3)$ for the orientation of the
rigid body, once the time dependence of $\boldsymbol{\Omega}(t)$ is
determined from the Euler equations.


\subsection{Hamilton--Pontryagin constrained rigid-body variational principle} \label{HamPont-EulerRB}
\index{constrained variational principle!Hamilton--Pontryagin} \index{Hamilton--Pontryagin principle!rigid
body!left action} Formula \eqref{domh} for the variation $\widehat{\Omega}$
of the skew-symmetric matrix  \index{rigid body!angular velocity!left-invariant}
\[\widehat{\Omega}= O^{-1} \dot{O}\] 
may be
imposed as a constraint in Hamilton's principle and thereby provide a
variational derivation of Euler's equations \eqref{RB-eqn} for rigid-body
motion in principal axis coordinates. This constraint is incorporated into
the matrix Euler equations, as follows. 
\index{rigid body!matrix Euler equations}

\begin{proposition}[Matrix Euler equations]
\label{MatrixEulerRBeqns}
Euler's rigid-body equation may be written in matrix form as
\begin{equation}
\frac{d{\Pi}}{dt}  
= 
-\,\big[\,{\widehat{\Omega}}\,,\,{\Pi}\,\big]
\quad\hbox{with}\quad
 {\Pi} = \mathbb{I}\widehat{\Omega}
 =
 \frac{\delta l }{\delta \widehat{\Omega}} 
  \,,
 \label{Euler's-eqnRB}
\end{equation} 
for the Lagrangian $l(\widehat{\Omega})$ given by
\begin{equation}
l = \frac{1}{2}\Big\langle \,\mathbb{I}\widehat{\Omega}
 \,,\, \widehat{\Omega}\, \Big\rangle
 \,,
  \label{Lag-RB}
\end{equation}
in which the bracket  
\begin{equation}
\big[\,\widehat{\Omega}\,,\,{\Pi} \big]
:=\widehat{\Omega}{\Pi}-{\Pi}\widehat{\Omega}
  \label{commutator-eqnRB}
\end{equation}
denotes the commutator and $\langle\,\cdot\,,\,\cdot\,\rangle$ denotes the \emph{trace pairing}, e.g.,
\begin{equation}
\Big\langle \,{\Pi}\,,\, \widehat{\Omega}\, \Big\rangle
=:
\frac{1}{2}\operatorname{trace} \Big({\Pi}^T\,\widehat{\Omega} \Big)
\,.
\label{tracepairing-RB}
\end{equation}
\end{proposition}

\begin{remark}
Note that the symmetric part of $\Pi$ does not contribute in the pairing and if set equal to zero initially, it will remain zero. 
\end{remark}

\begin{proposition}[Constrained variational principle]
\label{HamPont-VarPrinc-RBeqns}
The matrix Euler equations~(\ref{Euler's-eqnRB})
are equivalent to stationarity $\delta S = 0$ of the following \emph{constrained  action}:
\begin{eqnarray}
S(\widehat{\Omega},O,\dot{O},{\Pi})
&=&
\int^b_a l(\widehat{\Omega},O,\dot{O},{\Pi}) \, d t
\label{implicit-actionEulerRB}\\
&=&
\int^b_a \Big[l(\widehat{\Omega}) 
+ 
\langle\, {\Pi}
\,,\,
O^{-1}\dot{O} - \widehat{\Omega}\,\rangle
\Big]\, dt
\,.
\nonumber
\end{eqnarray}
\end{proposition}

\begin{remark}
The integrand of the constrained action in (\ref{implicit-actionEulerRB}) is similar to the formula for the Legendre transform, but its functional dependence is different. This variational approach is related to the classic \emph{Hamilton--Pontryagin principle} which is used in control theory.
\end{remark}
\begin{proof}
The variations of $S$ in formula (\ref{implicit-actionEulerRB}) are given by
\begin{eqnarray*}
\delta S 
&=&
\int^b_a \bigg\{
\Big\langle\,\frac{\delta l }{\delta \widehat{\Omega}} 
- {\Pi}\,,\, \delta \widehat{\Omega} \,\Big\rangle
\\&&\phantom{\int^b_a}+
\Big\langle\,\delta {\Pi}  
\,,\, (O^{-1}\dot{O} - \widehat{\Omega})\,\Big\rangle
+
\Big\langle\, {\Pi}  \,,\, \delta (O^{-1}\dot{O})
\,\Big\rangle
\bigg\}
\, dt
\,,
\end{eqnarray*}
where 
\begin{equation}
\delta(O^{-1}\dot{O})
=
\widehat{\Xi}\,\dot{\phantom{\,}}
+
[\,\widehat{\Omega} \,,\, \widehat{\Xi}\,]
\,,
\label{var-omegamatrix}
\end{equation}
and  $\widehat{\Xi}=(O^{-1}\delta{O})$ from equation \eqref{dsig}. \medskip

Substituting for $\delta(O^{-1}\dot{O})$ in \eqref{var-omegamatrix} into the last term of $\delta S$ produces
\begin{eqnarray}
\int^b_a
\Big\langle\, {\Pi}  \,,\, \delta (O^{-1}\dot{O})
\,\Big\rangle
\, dt
&=&
\int^b_a
\Big\langle\, {\Pi}  \,,\, \widehat{\Xi}\,\dot{\phantom{\,}}
+
[\,\widehat{\Omega} \,,\, \widehat{\Xi}\,]
\,\Big\rangle
\, dt
\nonumber\\
&=&
\int^b_a
\Big\langle\, 
-\,{\Pi}\,\dot{\phantom{\,}}  
-[\,\widehat{\Omega} \,,\, {\Pi}\,]
\,,\, \widehat{\Xi}
\,\Big\rangle
\, dt
\nonumber\\
&& +\
\Big\langle\, \Pi\,,\, \widehat{\Xi}
\,\Big\rangle\Big|_a^b
\,,
\label{endpt-terms-HP}
\end{eqnarray}
where one uses the cyclic properties of the trace operation for matrices,
\begin{equation}
\operatorname{trace}
\left({\Pi}^T\,\widehat{\Xi}\,\widehat{\Omega} \right)
=
\operatorname{trace}
\left(\widehat{\Omega}\,{\Pi}^T\,\widehat{\Xi} \right)
\,.
\label{cyclic-trace-property}
\end{equation}
Thus, stationarity  of the Hamilton--Pontryagin variational principle for vanishing 
endpoint conditions $\widehat{\Xi}(a)=0=\widehat{\Xi}(b)$ implies the following  set of equations:
\begin{eqnarray}
\frac{\delta l}{\delta \widehat{\Omega}} = {\Pi}
\,,\quad
O^{-1}\dot{O} = \widehat{\Omega}
\,,\quad
\frac{d\Pi}{dt} 
=  -
[\,\widehat{\Omega} \,,\, {\Pi}\,]
\,.
\label{symmHP-RBeqns}
\end{eqnarray}
\rem{
This Hamilton's principle is said to be \emph{implicit} because the definitions of the quantities describing the motion only emerge only \emph{after} the variations have been taken. 
}
\end{proof}

\begin{remark}[Interpreting the formulas in 
(\ref{symmHP-RBeqns})]
The first formula in (\ref{symmHP-RBeqns}) defines the angular momentum matrix ${\Pi}$ as the \emph{fibre derivative} of the Lagrangian with respect to the angular velocity matrix $\widehat{\Omega}$. The second formula is the reconstruction formula \eqref{recon-formula} for the solution curve $O(t)\in SO(3)$, given the solution $\widehat{\Omega}(t)=O^{-1}\dot{O}$. And the third formula is Euler's equation for rigid-body motion in matrix form.  
\index{fibre derivative} 
\end{remark}

\noindent
We transform the endpoint terms in (\ref{endpt-terms-HP}), arising on
integrating the variation $\delta S$ by parts in the proof of Theorem  
\ref{HamPont-VarPrinc-RBeqns}, into the spatial representation by setting $\widehat{\Xi}(t)=:O(t)\,\widehat{\xi}\,O^{-1}(t)$ and $\widehat{\Pi}(t)=:O(t)\widehat{\pi}(t)O^{-1}(t)$, as follows:
\begin{eqnarray}
\Big\langle\, \Pi\,,\, \widehat{\Xi}
\,\Big\rangle
=
\operatorname{trace}
\left({\Pi}^T\,\widehat{\Xi}\right)
=
\operatorname{trace}
\left({\pi}^T\,\widehat{\xi}\,\right)
=
\Big\langle\, \pi\,,\, \widehat{\xi}
\,\Big\rangle\,.
\label{endpt-terms-HP}
\end{eqnarray}
Thus, the vanishing of both endpoints for an arbitrary \emph{constant} infinitesimal spatial rotation $\widehat{\xi}=(\delta{O}O^{-1})=const$ implies constancy of the spatial angular momentum,
\begin{eqnarray}
\pi(a) = \pi(b)\,.
\label{Nthm-HP}
\end{eqnarray}
This is Noether's theorem for the rigid body. \index{rigid body!Noether's theorem}

\begin{theorem}[Noether's theorem for the rigid body]\label{Nthm-RB}
\index{Noether's theorem!rigid body}
Invariance of the constrained Hamilton--Pontryagin action under spatial rotations implies conservation of spatial angular momentum,
\begin{eqnarray}
\pi=O^{-1}(t)\Pi(t)O(t)=:{\rm Ad}^*_{O^{-1}(t)}\Pi(t)
.
\label{space-ang-mom}
\end{eqnarray}
\end{theorem}

\begin{proof}
\begin{eqnarray}
\frac{d}{dt}\Big\langle\, \pi\,,\, \widehat{\xi}
\,\Big\rangle
&=&
\frac{d}{dt}\Big\langle\, O^{-1}\Pi O\,,\, \widehat{\xi}
\,\Big\rangle
=
\frac{d}{dt}\operatorname{trace}
\left({\Pi}^T\,O^{-1}\widehat{\xi}O\right)
\nonumber\\&=&
\Big\langle\,\frac{d}{dt}\Pi + \big[\,\widehat{\Omega} \,,\, {\Pi}\, \big] \,,\, O^{-1}\widehat{\xi}O \,\Big\rangle
= 0
\nonumber\\&=:&
\Big\langle\,\frac{d}{dt}\Pi - {\rm ad}^*_{\widehat{\Omega}}{\Pi}\,,\, {\rm Ad}_{O^{-1}}\widehat{\xi}\,\Big\rangle,
\nonumber\\
\frac{d}{dt}\Big\langle\, {\rm Ad}^*_{O^{-1}}\Pi\,,\, \widehat{\xi}
\,\Big\rangle
&=&
\Big\langle\, {\rm Ad}^*_{O^{-1}}\Big(\,\frac{d}{dt}\Pi - {\rm ad}^*_{\widehat{\Omega}}{\Pi}\Big)\,,\, \widehat{\xi}\,\Big\rangle
.
\label{space-ang-mom-proof}
\end{eqnarray}
The proof of Noether's theorem for the rigid body is already on the second line. However, the last line gives a general result. 
\end{proof}

\begin{remark}\rm
The proof of Noether's theorem for the rigid body when the constrained Hamilton--Pontryagin action is invariant under spatial rotations also proves a general result in equation~(\ref{space-ang-mom-proof}),
with $\widehat{\Omega}=O^{-1}\dot{O}$ for a Lie group $O$, that as in the calculation in \eqref{coadjoint-calc},
\begin{equation}
\frac{d}{dt}\Big(\,{\rm Ad}^*_{O^{-1}}\Pi \,\Big)
=
{\rm Ad}^*_{O^{-1}}\Big(\,\frac{d}{dt}\Pi - {\rm ad}^*_{\widehat{\Omega}}{\Pi}\Big)
.
\label{Ad-dot-ad-id}
\end{equation}
This equation will be useful in the remainder of the text. In particular, it provides the solution of a differential equation defined on the dual of a Lie algebra. Namely, for a Lie group $O$ with Lie algebra $\mathfrak{o}$, the equation for $\Pi\in\mathfrak{o}^*$ and $\widehat{\Omega}=O^{-1}\dot{O}\in \mathfrak{o}$
\begin{equation}
\frac{d}{dt}\Pi - {\rm ad}^*_{\widehat{\Omega}}{\Pi} = 0
\quad\hbox{has solution}\quad
\Pi(t) =  {\rm Ad}^*_{O(t)}\pi 
\,,
\label{Ad-dot-ad-id}
\end{equation}
in which the constant $\pi\in\mathfrak{o}^*$ is obtained from the initial conditions.

\end{remark}

\begin{exercise}
Recover Euler's rigid body equations in \eqref{} from the Hamilton--Pontryagin variational principle for the 
following Lagrangian,

\end{exercise}
\begin{answer}
\end{answer}

\subsection{Hamiltonian form of rigid body motion in $\mathbb{R}^3$}
A dynamical system on a manifold $M$ 
\[
\mathbf{\dot{x}}(t)=\mathbf{F}(\mathbf{x})
,\quad \mathbf{x}\in M
\]
is said to be in \emph{Hamiltonian form}, if it can be expressed as
\[
\mathbf{\dot{x}}(t)=\{\mathbf{x},H\}
,\quad\hbox{for}\quad
H{:\ } M\mapsto \mathbb{R}
\,,
\]
in terms of a Poisson bracket operation, 
\[
\{\cdot, \cdot\}{:\ } 
\mathcal{F}(M)\times\mathcal{F}(M)\mapsto\mathcal{F}(M)
\,,
\]
that is bilinear, skew-symmetric and satisfies the Jacobi identity and
(usually) the Leibniz rule. 

As we shall explain, reduced equations arising from Lie group-invariant
Hamilton's principles are naturally Hamiltonian. If we
\emph{Legendre transform} our reduced Lagrangian for the $SO(3)$
left-invariant variational principle \eqref{rvprin} for
rigid body dynamics, then its simple, beautiful and well-known Hamiltonian
formulation emerges.  \index{Legendre transform}

\begin{definition}
The Legendre transformation $\mathbb{F}l
{:\ } \mathfrak{so}(3)\rightarrow\mathfrak{so}(3)^*$ is defined  by the fibre derivative 
$$
\mathbb{F} l (\Omega) = \frac{\delta l}{\delta \Omega} = \Pi \,.
$$
\end{definition}
The Legendre transformation defines the \emph{body angular momentum} by the
variations of the rigid-body's reduced Lagrangian with respect to the body
angular velocity. For the Lagrangian in \eqref{rb-redlag}, the
$\mathbb{R}^3$ components of the body angular momentum are
\begin{equation}\label{rbm}
\Pi_i  =  I_i\Omega_i = \frac{\partial l}{\partial \Omega _i } \,,
\quad i = 1, 2, 3\,.
\end{equation}
\index{rigid body!body angular momentum}

\subsection{Lie--Poisson Hamiltonian formulation of rigid body dynamics} \index{Lie--Poisson!Hamiltonian formulation} \index{Hamiltonian formulation!Lie--Poisson}
The rigid-body Hamiltonian is obtained from the \emph{reduced Legendre transform}, \index{reduced Legendre transform!rigid body} \index{rigid body!reduced Legendre transform}
\begin{equation}\label{RLT} 
h(\Pi) := \langle \Pi , \Omega\rangle - l(\Omega)\,,
\end{equation}
where the pairing $\langle\cdot, \cdot\rangle{:\ } 
\mathfrak{so}(3)^*\times\mathfrak{so}(3)\to\mathbb{R}$ is understood via the hat map in
components as the vector dot product on $\mathbb{R}^3$
\[ \langle \Pi , \Omega\rangle := \boldsymbol{\Pi}\cdot\boldsymbol{\Omega}. \]
Hence, one finds the expected expression for the rigid-body Hamiltonian
\begin{equation}\label{rbh} 
h = \tfrac{1}{2} \boldsymbol{\Pi}\cdot {\mathbb I}^{-1}\boldsymbol{\Pi}
:= \frac{\Pi^2_1}{2I_1} + \frac{\Pi^2_2}{2I_2} + \frac{\Pi^2_3}{2I_3}.
\end{equation}
The reduced Legendre transform $\mathbb{F}l$ for this case in \eqref{RLT} 
is a diffeomorphism, so we may take its differential to find 
\begin{align*}
dh(\Pi) = \scp{\frac{\partial h}{\partial \Pi} }{d\Pi} 
= \scp{ \Omega }{d\Pi}  + \scp{\Pi - \frac{\p L}{\p \Omega} }{d\Omega}
.\end{align*}
In $\mathbb{R}^3$ coordinates, this relation expresses the body angular
velocity as the derivative of the reduced Hamiltonian with respect to
the body angular momentum, namely (introducing grad-notation),
\[ \nabla_{\Pi}h := \frac{d h}{d\boldsymbol{\Pi}}
= \boldsymbol{\Omega} \,. \]
Hence, the reduced Euler--Lagrange equations for the Lagrangian $l(\boldsymbol{\Omega}) $ in \eqref{rb-redlag}, may be expressed equivalently in angular momentum vector components in $\mathbb{R}^3$ and
Hamiltonian
$h(\boldsymbol{\Pi})$ as:
\[ \frac{d }{d t} ({\mathbb I} \boldsymbol{\Omega})
= {\mathbb I} \boldsymbol{\Omega} \times \boldsymbol{\Omega}
\Longleftrightarrow 
\boldsymbol{\dot{\Pi}}
= 
\boldsymbol{\Pi}\times \frac{dh}{d \boldsymbol\Pi}
:=
\{\boldsymbol{\Pi},h\}. \]
This expression suggests we introduce the following rigid body Poisson
bracket on functions of the ${\boldsymbol \Pi}$'s as:
\begin{equation}\label{rbb}
 \{f,h\}({\boldsymbol  \Pi})
 =   -\,\Pi_i \frac{\p f}{\p \Pi_j} \epsilon_{ijk} \frac{\p h}{\p \Pi_k}
=: -\,{\boldsymbol  \Pi} \cdot \frac{df}{d \boldsymbol\Pi}\times \frac{dh}{d \boldsymbol\Pi}
\,.
 \end{equation} 
For the Hamiltonian \eqref{rbh}, one checks that 
the Euler equations in terms of the rigid-body angular
momenta follow as,
\begin{equation}\label{rbe2}
\begin{aligned}
\dot {\Pi}_1  &=  \frac{I_2 - I_3}{I_2I_3} \Pi_2\Pi_3, &
\dot {\Pi}_2  &=  \frac{I_3 - I_1}{I_3I_1} \Pi_3\Pi_1, &
\dot {\Pi}_3  &=  \frac{I_1 - I_2}{I_1I_2} \Pi_1\Pi_2
\,.
\end{aligned}
\end{equation}
That is, the equation 
\begin{equation}\label{rbe3}
\boldsymbol{\dot{\Pi }} = 
\boldsymbol{\Pi}\times \frac{dh}{d \boldsymbol\Pi}
\end{equation}
implies
\[
\frac{df}{dt} = \frac{df}{d \boldsymbol\Pi}\cdot \frac{d \boldsymbol\Pi}{dt} 
= \frac{df}{d \boldsymbol\Pi}\cdot \boldsymbol{\Pi}\times \frac{dh}{d \boldsymbol\Pi}
= -\, \boldsymbol{\Pi} \cdot \frac{df}{d \boldsymbol\Pi}\times \frac{dh}{d \boldsymbol\Pi}
= \{f, h\}
\,.
\]
The Poisson bracket proposed in \eqref{rbb} is an example of a \emph{Lie
Poisson bracket}, which we will show later satisfies the defining
relations to be a Poisson bracket. 
\begin{remark}\rm 
The bracket  in \eqref{rbb} may also be written equivalently as a differential form relation,
\begin{equation}\label{RBB}
 \{f,h\}d^3\Pi
:=  -\,d\big(|\bs{\Pi}|^2/2\big)\wedge df \wedge dh
\,.\end{equation}
Notice that this bracket is symplectic on level sets of $|\bs{\Pi}|^2$. It also vanishes 
if either $f$ or $h$ is a function of $|\bs{\Pi}|^2$. Hence, for $f(|\bs{\Pi}|^2)$, say,
the Poisson bracket proposed in \eqref{rbb} vanishes for every function $h(\bs{\Pi})$. Also,
the flow of either $f(\bs{\Pi})$ or $h(\bs{\Pi})$ takes place on level sets of $|\bs{\Pi}|^2$, where it reduces to a canonical Poisson bracket.
As we shall see, this is the hallmark of \emph{coadjoint motion}. 
\end{remark}
\subsection{Relation to the $\mathbb{R}^3$ Poisson bracket}
The rigid body Poisson bracket \eqref{rbb} is a special case of the
Poisson bracket for functions on $\mathbb{R}^3$,
\begin{equation}\label{r3pb}
\{f, h\} = -\nabla{c}\cdot\nabla{f}\times\nabla{h}
\,.\end{equation}
This bracket generates the motion 
\begin{equation}
\mathbf{\dot{x}} = \{\mathbf{x}, h\} =
\nabla{c}\times\nabla{h}
\,.\end{equation}
For this bracket the motion takes place along the
intersections of level surfaces of the functions $c$ and $h$ in
$\mathbb{R}^3$. In particular, for the rigid body, the motion takes place
along intersections of angular momentum spheres $c=\|\mathbf{x}\|^2/2$ and
energy ellipsoids $h=\mathbf{x}\cdot \mathbb{I}\mathbf{x}$. (See the cover
illustration of Marsden and Ratiu [2003].)

\begin{exercise}
Consider the $\mathbb{R}^3$ Poisson bracket
\begin{equation}
\{f, h\} = -\nabla{c}\cdot\nabla{f}\times\nabla{h}
\,.\end{equation}
Let  $c=\mathbf{x}^T\cdot\mathbb{C}\mathbf{x}$ be a quadratic form on
$\mathbb{R}^3$, and let $\mathbb{C}$ be the associated symmetric
$3\times3$ matrix. Determine the conditions on the quadratic function
$c(\mathbf{x})$ so that this Poisson bracket will satisfy the Jacobi
identity.
\end{exercise}

\begin{exercise}
Find the general conditions on the function $\mathbf{c}(\mathbf{x})$ so
that the $\mathbb{R}^3$ bracket
\[
\{f, h\} = -\nabla{c}\cdot\nabla{f}\times\nabla{h}
\] 
satisfies the defining properties of a Poisson bracket. Is this
$\mathbb{R}^3$ bracket also a derivation satisfying the Leibniz relation
for a product of functions on $\mathbb{R}^3$? If so, why?
\end{exercise}

\begin{exercise}
How is the $\mathbb{R}^3$ bracket related to the canonical Poisson
bracket? Hint: restrict to level surfaces of the function
$c(\mathbf{x})$. 
\end{exercise}

\begin{exercise}[Casimirs of the $\mathbb{R}^3$ bracket]
The Casimirs (or distinguished functions, as Lie called them) of a
Poisson bracket satisfy 
\[
\{c, h\}(\mathbf{x}) = 0
,\quad\forall h(\mathbf{x}).
\]
Suppose the function $\mathbf{c}(\mathbf{x})$ is chosen so that the
$\mathbb{R}^3$ bracket \eqref{r3pb} satisfies the defining properties of a
Poisson bracket. What are the Casimirs for the
$\mathbb{R}^3$ bracket \eqref{r3pb}? Why?
\end{exercise}

\begin{exercise}
Show that the motion equation
\[
\mathbf{\dot{x}} = \{\mathbf{x}, h\}
\]
for the $\mathbb{R}^3$ bracket \eqref{r3pb} is invariant under a certain
linear combination of the functions $c$ and $h$. Interpret this invariance
geometrically. 
\end{exercise}

\begin{exercise}

Use the Hamilton--Pontryagin approach to compute the dynamics arising from stationarity 
$\delta S = 0$ of the following \emph{right-invariant} constrained  action,
which is discussed in detail in \cite{gay2012reduced},
\begin{eqnarray}
S(\widehat{\omega},O,\dot{O},{\pi})
&=&
\int^b_a l(\widehat{\omega},O,\dot{O},{\pi}) \, d t
\label{implicit-actionEulerRB}\\
&=&
\int^b_a \Big[l(\widehat{\omega}) 
+ 
\langle\, {\pi}
\,,\,
\dot{O}O^{-1} - \widehat{\omega}\,\rangle
\Big]\, dt
\,.
\nonumber
\end{eqnarray}
\index{Hamilton--Pontryagin principle!rigid body!right action}
\end{exercise}

\begin{answer}
The variations of $S$ in formula (\ref{implicit-actionEulerRB}) are given by
\begin{align}
\begin{split}
\delta S 
&=
\int^b_a \bigg\{
\Big\langle\,\frac{\delta l }{\delta \widehat{\omega}} 
- {\pi}\,,\, \delta \widehat{\omega} \,\Big\rangle
\\&\phantom{\int^b_a}
+
\Big\langle\,\delta {\pi}  
\,,\, (\dot{O}O^{-1} - \widehat{\omega})\,\Big\rangle
+
\Big\langle\, {\pi}  \,,\, \delta (\dot{O}O^{-1})
\,\Big\rangle
\bigg\}
\, dt
\,,
\end{split}
\label{var-omegamatrix-right}
\end{align}
where 
\begin{equation}
\delta(\dot{O}O^{-1})
=
\widehat{\xi}\,\dot{\phantom{\,}}
-
[\,\widehat{\omega} \,,\, \widehat{\xi}\,]
\,,
\end{equation}
and one defines $\widehat{\xi}:=\delta{O}O^{-1}$. \medskip

Substituting for $\delta(\dot{O}O^{-1})$ in \eqref{var-omegamatrix} into the last term of $\delta S$ produces
\begin{eqnarray}
\int^b_a
\Big\langle\, {\pi}  \,,\, \delta (\dot{O}O^{-1})
\,\Big\rangle
\, dt
&=&
\int^b_a
\Big\langle\, {\pi}  \,,\, \widehat{\xi}\,\dot{\phantom{\,}}
-
[\,\widehat{\omega} \,,\, \widehat{\xi}\,]
\,\Big\rangle
\, dt
\nonumber\\
&=&
\int^b_a
\Big\langle\, 
-\,\dot{\pi}
-[\,\widehat{\omega} \,,\, {\pi}\,]
\,,\, \widehat{\xi}
\,\Big\rangle
\, dt
\nonumber\\
&& +\
\Big\langle\, \pi\,,\, \widehat{\xi}
\,\Big\rangle\Big|_a^b
\,.
\label{endpt-terms-HP-right}
\end{eqnarray}
Here, the square brackets $[\,\cdot\,,\,\cdot\,]$ denote matrix commutator and the second line in \eqref{endpt-terms-HP-right} 
applies integration by parts and uses the cyclic property of the trace operation for matrices,
\begin{equation}
\operatorname{trace}
\left({\pi}^T\,\widehat{\xi}\,\widehat{\omega} \right)
=
\operatorname{trace}
\left(\widehat{\omega}\,{\pi}^T\,\widehat{\xi} \right)
\,.
\label{cyclic-trace-property}
\end{equation}
Thus, stationarity  of the Hamilton--Pontryagin variational principle  in \eqref{var-omegamatrix-right} for vanishing 
endpoint conditions $\widehat{\xi}(a)=0=\widehat{\xi}(b)$ implies the following  set of equations:
\begin{eqnarray}
\frac{\delta l}{\delta \widehat{\omega}} = {\pi}
\,,\quad
\dot{O}O^{-1} = \widehat{\omega}
\,,\quad
\frac{d\pi}{dt} 
=  
[\,\widehat{\omega} \,,\, {\pi}\,]
\,.
\label{symmHP-RBeqns-right}
\end{eqnarray}
\end{answer}

\begin{exercise}
Recall that the Lagrangian $l(\widehat{\Omega})$ for $\widehat{\Omega}=O^{-1}\dot{O} $ is given in equation \eqref{Lag-RB} for rigid body motion in the body frame. 
Derive the equations of rigid body dynamics in the spatial frame by using the Hamilton--Pontryagin approach 
after writing the Lagrangian from \eqref{Lag-RB} in terms of the spatial angular velocity $\widehat{\omega}=\dot{O} O^{-1}$.
\end{exercise}

\index{rigid body!angular velocity!right-invariant}

\begin{answer}
The Euler's rigid-body Lagrangian from \eqref{Lag-RB} is written in terms of the spatial angular velocity $\widehat{\omega}=\dot{O} O^{-1}$
by  substituting relation the relation between body and spatial angular velocities in \eqref{BodySpaceAngVel-Ad}. Hence,
the rigid body Lagrangian  in the spatial frame transforms into
\begin{equation}
l(\widehat{\Omega}):=
\tfrac12\Big\langle \,\widehat{\Omega}\,,\, \mathbb{I} \widehat{\Omega}\, \Big\rangle
= \tfrac12\Big\langle \,\widehat{\omega} \,,\mathbb{I}_{spat}(t)\, \widehat{\omega}\,\Big\rangle
=: \ell(\widehat{\omega},\mathbb{I}_{spat}) 
\,.
\label{tracepairing-RB-right}
\end{equation}
In terms of the \emph{trace pairing} in \eqref{tracepairing-RB} one may verify this formula directly,
\begin{equation}
\frac{-1}{2}\text{Tr} \big(\,\widehat{\Omega}\,\mathbb{I}\,\widehat{\Omega} \,\big)
 = \frac{-1}{2}\text{Tr} \big(\,\widehat{\omega} \,O(t)\mathbb{I}O^{-1}(t)\, \widehat{\omega}\, \big)
 =: \frac{-1}{2}\text{Tr} \big(\,\widehat{\omega} \,\mathbb{I}_{spat}(t)\, \widehat{\omega}\, \big)
 \,.
  \label{Lag-RB-right}
\end{equation}
Thus, in the spatial frame the moment of inertia becomes a dynamical variable, as
\begin{equation}
\mathbb{I}_{spat}(t) := O(t)\mathbb{I}O^{-1}(t) 
\,.  
\label{def: I-spat}
\end{equation}
From its definition, the time-dependent moment of inertia $\mathbb{I}_{spat}(t)$ satisfies the auxiliary equations,
\begin{align}
\begin{split}
\frac{d\mathbb{I}_{spat}}{dt} &= \widehat{\omega} \,\mathbb{I}_{spat} -  \mathbb{I}_{spat}\,\widehat{\omega} =: \big[\widehat{\omega}\,, \,\mathbb{I}_{spat}\big]
\,,\\
\delta \mathbb{I}_{spat} &= \big[\widehat{\xi}\,, \,\mathbb{I}_{spat}\big]
\quad \hbox{with}\quad 
\widehat{\xi} : = \delta O \, O^{-1}
\,.  
\end{split}
\label{dyn: I-spat}
\end{align}
Likewise, the variational relation for the spatial angular velocity becomes
\begin{equation}
\delta(\dot{O}O^{-1})
=
\widehat{\xi}\,\dot{\phantom{\,}}
-
[\,\widehat{\omega} \,,\, \widehat{\xi} \,]
\quad \hbox{with}\quad 
\widehat{\xi} : = \delta O \, O^{-1}
\,.
\label{var-omegamatrix-right1}
\end{equation}
Thus, in the spatial frame one is dealing with a rigid body Lagrangian that is not right invariant and 
depends on both $\widehat{\omega}(t) $ and $\mathbb{I}_{spat}(t)$,
\begin{eqnarray}
S(\widehat{\omega},O,\dot{O},{\pi})
&=&
\int^b_a \Big[ \ell(\widehat{\omega},\mathbb{I}_{spat}) 
+ 
\langle\, {\pi}
\,,\,
\dot{O}O^{-1} - \widehat{\omega}\,\rangle
\Big]\, dt
\,.\label{implicit-actionEulerRB-right}\\
\,.
\nonumber
\end{eqnarray}
The variations of the action $S$ in formula (\ref{implicit-actionEulerRB-right}) are now given by
\begin{align}
\begin{split}
\delta S 
&=
\int^b_a \bigg\{
\Big\langle\,\frac{\delta \ell }{\delta \widehat{\omega}} 
- {\pi}\,,\, \delta \widehat{\omega} \,\Big\rangle
+ \Big\langle\,\delta {\pi}  
\,,\, (\dot{O}O^{-1} - \widehat{\omega})\,\Big\rangle
\\&\phantom{\int^b_a}
+
\Big\langle\, {\pi}  \,,\, \delta (\dot{O}O^{-1})
\,\Big\rangle
+ \Big\langle\,\frac{\delta \ell }{\delta \mathbb{I}_{spat} } 
\,,\, \delta \mathbb{I}_{spat}  \,\Big\rangle
\bigg\}
\, dt
\\&=
\int^b_a \bigg\{
\Big\langle\,\frac{\delta \ell }{\delta \widehat{\omega}} 
- {\pi}\,,\, \delta \widehat{\omega} \,\Big\rangle
+ \Big\langle\,\delta {\pi}  
\,,\, (\dot{O}O^{-1} - \widehat{\omega})\,\Big\rangle
\\&\phantom{\int^b_a}
+ 
\Big\langle\,- \dot{\pi}  - \big[\pi\,,\,\widehat{\omega} \,\big] 
+ \Big[\mathbb{I}_{spat}\,,\,\frac{\delta \ell }{\delta \mathbb{I}_{spat} } \Big]
\,,\, \widehat{\xi} \,\Big\rangle
\bigg\}
\, dt
\,,
\end{split}
\label{spatialRBvars}
\end{align}
where we have applied formula \eqref{implicit-actionEulerRB-right} for $\delta(\dot{O}O^{-1})$ 
and then integrated by parts in time. A direct calculation with the Lagrangian 
$\ell(\widehat{\omega},\mathbb{I}_{spat})$ in equation \eqref{tracepairing-RB-right} reveals that the 
last two terms in the variations paired with $\widehat{\xi}$ in  \eqref{spatialRBvars} cancel each other, 
because for this Lagrangian the equality holds that
\[
\big[\pi\,,\,\widehat{\omega} \,\big] = \Big[\mathbb{I}_{spat}\,,\,\frac{\delta \ell }{\delta \mathbb{I}_{spat} } \Big]
\,.\]

Consequently, the dynamics of the free rigid body in spatial variables satisfies
\begin{align*}
\frac{d\pi}{dt} = 0
\quad\hbox{and}\quad
\frac{d\mathbb{I}_{spat}}{dt} = \big[\widehat{\omega}\,, \,\mathbb{I}_{spat}\big]
\quad\hbox{with}\quad
\pi = \mathbb{I}_{spat}(t)\, \widehat{\omega}
\,,
\end{align*}
for spatial angular velocity $\widehat{\omega}=\dot{O} O^{-1}$ and
dynamical moment of inertia $\mathbb{I}_{spat}(t) := O(t)\mathbb{I}O^{-1}(t)$.

\end{answer}

\begin{exercise}
Show that the Hamiltonian equations for the spatial dynamics of the rigid body in \eqref{spatialRBvars} 
may be expressed via the following Lie--Poisson matrix operator 
\begin{align}
\frac{d}{dt}
\begin{pmatrix}
\pi \\ \mathbb{I}_{spat}
\end{pmatrix}
= -
\begin{pmatrix}
\big[\pi\,,\,\Box \big]  & \big[\mathbb{I}_{spat} \,,\,\Box \big]
\\ \big[\mathbb{I}_{spat} \,,\,\Box \big]  & 0
\end{pmatrix}
\begin{pmatrix}
\delta \mathrm{h}/\delta \pi = \widehat{\omega} 
\\ 
\delta \mathrm{h}/\delta \mathbb{I}_{spat} = -\,\delta \ell/\delta \mathbb{I}_{spat}
\end{pmatrix}
,\label{spatialRB-LPB}
\end{align}
with $(\pi ,\mathbb{I}_{spat})\in [\mathfrak{so}(3)\circledS {\rm Sym}^3_+(\mathbb{R})]^*$  
where $\circledS$ denotes semidirect product action of the Lie algebra 
$\mathfrak{so}(3)$ represented as $3\times3$ antisymmetric real matrices acting on  
the vector space of $3\times3$ positive symmetric real matrices, ${\rm Sym}^3_+(\mathbb{R})$,
and superscript $\phantom{\,}^*$ denotes the dual semidirect product Lie algebra under matrix trace pairing.
\index{Lie algebra!semidirect product}
\end{exercise}

\begin{remark}\rm
Semidirect product action by matrix multiplication will be discussed further for the example of the 
heavy top in body coordinates in the next lecture. 
For now, one should keep in mind that semidirect product action arose here when the left-invariant 
angular velocity producing $SO(3)$ rotation symmetry of the rigid body Lagrangian in body coordinates 
was broken by transforming to  the right-invariant angular velocity in spatial coordinates. 
\index{broken symmetry!semidirect product}
\end{remark}

\begin{exercise}
Formulate and analyse the equations of motion on $\mathfrak{so}()3^*\times T^*Q$ 
for a rigid body that has a flywheel attached whose axis of rotation is aligned with the intermediate principal axis of
the rigid body. The flywheel's rotation angle $\alpha$ has a harmonic restoring force with spring constant $k$. 
The kinetic energy of this system is given by
\begin{eqnarray*}
KE
=
\frac{1}{2}I_1\Omega_1^2 
+
\frac{1}{2}I_2\Omega_2^2 
+
\frac{1}{2}I_3\Omega_3^2 
+
\frac{1}{2}J_2(\dot{\alpha}+\Omega_2)^2 
\,,
\end{eqnarray*}
where 
${\bs{\Omega}}=(\Omega_1,\Omega_2,\Omega_3)$ is the angular
velocity vector, $\dot{\alpha}$ is the angular rotational rate of the flywheel about the 
intermediate principal axis of the rigid body, and 
$I_1$, $I_2$, $I_3$ and $J_2$ are positive constants.
\begin{itemize}
\item
Explain why Hamilton's principle for this system may be written as 
\begin{align}
\begin{split}
0=\delta S &= \delta \int_0^T L(\bs{\Omega},\alpha , p_\alpha) \,dt  
\\&= \delta \int_0^T \frac12  \bs{\Omega}\cdot \mathbb{I}\bs{\Omega} + p_\alpha(\dot{\alpha}+\Omega_2) 
- \Big(\frac{1}{2J_2} p_\alpha^2 + \frac12 k\alpha^2\Big)\,dt
\end{split}
\label{gyro-action}
\end{align}
\item
Find the angular momenta ${\boldsymbol{\Pi}}\in\mathbb{R}^3$ and $p_\alpha\in\mathbb{R}^1$.
\item
Legendre transform to obtain the Hamiltonian in the variables 
${\boldsymbol{\Pi}},p_\alpha,\alpha$.
\item
Write the equations of motion for this system in Lie--Poisson bracket form.
\end{itemize}

\end{exercise}

\begin{answer}
\begin{itemize}
\item
The variations of the Lagrangian in Hamilton's principle \eqref{gyro-action} with respect to to $\bs{\Omega}$ and $p_\alpha$
yield
\begin{align}
\begin{split}
\delta \bs{\Omega}:& \quad \bs{\Pi}=\big(I_1\Omega_1 \,,\,I_2\Omega_2 + p_\alpha\,,\, I_3\Omega_3\big)^T
\,,\\
\delta  p_\alpha:& \quad  p_\alpha = J_2( \dot{\alpha} + \Omega_2 )
\,.
\end{split}
\label{gyro-vars}
\end{align}

\item
The Legendre transform 
\begin{align}
H(\bs{\Pi},\alpha,p_\alpha) = \bs{\Pi }\cdot \bs{\Omega} +  p_\alpha \dot{\alpha} - L(\bs{\Omega},\alpha , p_\alpha) 
\,,\label{gyro-LegTrans}
\end{align}
produces the Hamiltonian for this system
\begin{align}
H(\bs{\Pi},\alpha,p_\alpha) = \frac{\Pi_1^2}{2I_1}  + \frac{1}{2I_2} (\Pi_2 - p_\alpha)^2 
+ \frac{\Pi_3^2}{2I_3} + \frac{1}{2J_2} p_\alpha^2 + \frac{1}{2}k\alpha^2
.\label{gyro-Ham}
\end{align}
Notice that the angular momentum shift $(\Pi_2 - p_\alpha)$ along the intermediate axis couples the dynamics of the 
two degrees of freedom. 
\item
The equations of motion for this system in Lie--Poisson bracket form are
\begin{align}
\frac{d}{dt} \begin{pmatrix} \bs{\Pi} \\ p_\alpha \\ \alpha \end{pmatrix} 
= \begin{bmatrix} 
\bs{\Pi} \,\times & 0 & 0 
\\ 
0 & 0 & -1
\\
0 & 1 & 0
\end{bmatrix} 
\begin{pmatrix} 
\p H/\p \bs{\Pi} = \bs{\Omega} \\ \p H/ \p p_\alpha =  p_\alpha/J_2 -  \Omega_2 = \dot{\alpha} \\ \p H/ \p \alpha = k \alpha
\end{pmatrix}
.\label{gyro-EP-brkt}
\end{align}
\end{itemize}
\end{answer}

\begin{exercise}
Formulate and analyse the equations of motion on $\mathfrak{so}(3)^*\times T^*\mathbb{R}^3$ for a rigid body 
whose angular velocity $\bs{\Omega}\in \mathfrak{so}(3)$ acts on $\bs{q}\in \mathbb{R}^3$  as
\[
\bs{\dot{q}} + (\bs{\Omega}+\bs{\Upsilon})\times \bs{q} = 0
\,,
\]
where $\bs{\Upsilon}\in\mathbb{R}^3$ is constant. Hamilton's principle for the corresponding constrained  rigid body motion is given 
via the Clebsch approach as
\begin{align}
\begin{split}
0=\delta S &= \delta \int_0^T L(\bs{\Omega}) 
+ \bs{p}\cdot(\bs{\dot{q}} + (\bs{\Omega}+\bs{\Upsilon})\times \bs{q}) \,dt
\\&= \delta \int_0^T \frac12  \bs{\Omega}\cdot \mathbb{I}\bs{\Omega} 
+ \bs{p}\cdot(\bs{\dot{q}}+(\bs{\Omega}+\bs{\Upsilon})\times \bs{q}) \,dt
\,.\end{split}
\label{ALT-action}
\end{align}
\begin{itemize}
\item
Take the variation of the action integral $S$ in $\bs{\Omega}$ and determine the Euler--Poincar\'e equation
for $ \bs{\Pi}:={\p L}/{\p \bs{\Omega}}$, by using the equations arising from 
 the other variations in $\bs{q}$ and $\bs{p}$ as constraints. This is the Clebsch variational approach.
 \index{Clebsch variational approach}
\item
Write the equations of motion for this system in Lie--Poisson bracket form.
\end{itemize}

\end{exercise}

\begin{answer}
\begin{itemize}
\item
The stationary variation in $\bs{\Omega}$ of the action integral $S$ in \eqref{ALT-action} implies 
\[
 \frac{\p L}{\p \bs{\Omega}} - (\bs{q}\times \bs{p}) = 0\,,
\]
and the variations of the Clebsch constraint  in $\bs{q}$ and $\bs{p}$ imply that
\[
\frac{d}{dt} (\bs{q}\times \bs{p}) = (\bs{q}\times \bs{p})\times (\bs{\Omega}+\bs{\Upsilon})
\,.\]
Hence, 
\begin{align}
\frac{d}{dt}  \frac{\p L}{\p \bs{\Omega}} = \frac{\p L}{\p \bs{\Omega}} \times(\bs{\Omega}+\bs{\Upsilon})
\,.
\label{ALT-action1}
\end{align}
\item
To write the equations of motion for this system in Lie--Poisson bracket form, we Legendre transform to
derive the Hamiltonian as 
\begin{align}
\begin{split}
H(\bs{\Pi}) &= \bs{\Pi}\cdot\bs{\Omega} - L(\bs{\Omega})
\,,\\
dH(\bs{\Pi}) &=  \frac{\p H}{\p \bs{\Pi}} \cdot d\bs{\Pi} =  \bs{\Omega} \cdot d\bs{\Pi} + \Big( \bs{\Pi} -  \frac{\p L}{\p \bs{\Omega}}\Big) \cdot d \bs{\Omega}
\,.\end{split}
\label{ALT-action2}
\end{align}
Then, upon rewriting equation \eqref{ALT-action1} by using \eqref{ALT-action2} we find 
\begin{align}
\begin{split}
\frac{d}{dt} \bs{\Pi}  &=  \bs{\Pi} \times(\bs{\Omega}+\bs{\Upsilon}) =  \bs{\Pi} \times  \frac{\p H}{\p \bs{\Pi}} 
\,,\\
 \Longrightarrow \quad H(\bs{\Pi}) &=  \frac12 \bs{\Pi} \cdot \mathbb{I}^{-1}\bs{\Pi}  +  \bs{\Pi} \cdot \bs{\Upsilon}
\,,\end{split}
\label{ALT-action3}
\end {align}
where the Lie--Poisson bracket is the same as for standard rigid body dynamics. However, the rigid body Hamiltonian 
has acquired an additional term which in the motion equation \eqref{ALT-action3} shifts the angular velocity, as 
$\bs{\Omega}\to (\bs{\Omega}+\bs{\Upsilon})$. For more discussion, see, e.g., \cite{arnaudon2018noise}.

\end{itemize}

\end{answer}

\newpage
\vspace{4mm}\centerline{\textcolor{shadecolor}{\rule[0mm]{6.75in}{-2mm}}\vspace{-4mm}}
\section{Broken Symmetry: Heavy top equations}\label{sec-BrokenSym}

\secttoc

\textbf{What is this lecture about?} This lecture reformulates Euler's equations for 
heavy top dynamics as breaking of the $SO(3)$ symmetry of the invariant 
variational principles of Lagrange and Hamilton for rigid body dynamics. 
\index{broken symmetry!semidirect product}

\subsection{Introduction and definitions}
A top is a rigid body of mass $m$ rotating with a fixed point of support
in a constant gravitational field of acceleration $-g\mathbf{\hat{z}}$
pointing vertically downward. The orientation of  the body relative to the
vertical axis $\mathbf{\hat{z}}$ is defined by the unit vector
$\boldsymbol{\Gamma}={ O} ^{-1}(t)\mathbf{\hat{z}}$ for a curve
${O}(t)\in SO(3)$. \smallskip

According to its definition, the unit vector
$\boldsymbol{\Gamma}$ represents the motion of the vertical direction as
seen from the rotating body. Consequently, it satisfies the auxiliary
motion equation,
\[
\boldsymbol{\dot{\Gamma}}
=
-\,\big({O}^{-1} {\dot{O}}(t)\big)\boldsymbol{\Gamma}
=
\boldsymbol{\Gamma}\times\boldsymbol{\Omega}
\,.
\]
Here the rotation matrix ${O}(t)\in SO(3)$, the skew matrix
${\widehat{\Omega}}={ O} ^{-1} {\dot{O}}\in \mathfrak{so}(3)$
and the body angular frequency vector $\boldsymbol{\Omega}\in\mathbb{R}^3$
are related by the hat map,  $\boldsymbol{\Omega} =\big({ O}
^{-1}{\dot{O}}\big)\boldsymbol{\widehat{~}}$, where 
${\widehat{~}}\,{:\ } (\mathfrak{so}(3), [\cdot, \cdot])
\to (\mathbb{R}^3, \times)$ with 
${\widehat{\Omega}}\mathbf{v} = \boldsymbol{\Omega} \times
\mathbf{v}$ for any $\mathbf{v} \in \mathbb{R}^3$.

The motion of a heavy top of weight $m{\rm g}$  is given by Euler's equations in vector form,
\begin{align}\label{top-eqns-vector}
\begin{split}
{\mathbb I} \boldsymbol{\dot{\Omega}}
&={\mathbb I} \boldsymbol{\Omega} \times \boldsymbol{\Omega}
+ m{\rm g}\, \boldsymbol{\Gamma}\times\boldsymbol{\chi}
,\\
\boldsymbol{\dot{\Gamma}}
&=
\boldsymbol{\Gamma}\times\boldsymbol{\Omega}
\,,
\end{split}
\end{align}
where $\boldsymbol{\Omega}, 
\boldsymbol{\Gamma}, \boldsymbol{\chi}\in\mathbb{R}^3$ are vectors in the
rotating body frame. 

Here 
\begin{itemize}
\item
$\boldsymbol{\Omega} = (\Omega_1,\Omega_2, \Omega_3)$ is the body angular
velocity vector,
\item
${\mathbb I}={\mathrm diag}(I_1,  I_2, I_3)$  is the
moment of inertia tensor, diagonalized in the body principle axes,
\item
$\boldsymbol{\Gamma}=O^{-1}(t)\mathbf{\hat{z}}$ represents the motion of
the unit vector along the vertical axis, as seen from the body,
\item
$\boldsymbol{\chi}$ is the constant vector in the body from the point of
support to the body's center of mass,
\item
$m$ is the total mass of the body
and ${\rm g}$ is the constant acceleration of gravity.
\end{itemize}

\subsection{Heavy top action principles}

\emph{(A) Euler--Poincar\'e reduced heavy top action integral.}

\index{heavy top!geometric treatment} \index{heavy top!body representation} 

\begin{proposition}\label{ht-EPactprinc}
The heavy top equations are equivalent to the \emph{heavy top
action principle} for an Euler--Poincar\'e \emph{reduced action integral}
\index{variational principle!Euler--Poincar\'e}
\begin{equation}\label{rbvp1}
\delta S_{\mathrm red}=0
,\quad\hbox{with}\quad
S_{\mathrm red}= \int^b_a l(\boldsymbol{\Omega},\boldsymbol{\Gamma}) \, d t
= \int^b_a \tfrac{1}{2}
\langle\mathbb{I}\boldsymbol{\Omega},\boldsymbol{\Omega}\rangle
- \langle m{\rm g}\boldsymbol{\chi}, \boldsymbol{\Gamma}\rangle \, dt
\,,
\end{equation}
where variations of $\boldsymbol{\Omega}$ and $\boldsymbol{\Gamma}$ are
restricted to be of the form
\begin{equation}
\delta \boldsymbol{\Omega} = \boldsymbol{\dot{ \Sigma}} +
\boldsymbol{\Omega} \times \boldsymbol{\Sigma}
\quad\hbox{and}\quad
\delta \boldsymbol{\Gamma} = 
\boldsymbol{\Gamma} \times \boldsymbol{\Sigma}
\,,
\end{equation}
arising from variations of the definitions $\boldsymbol{\Omega}
=\big({ O} ^{-1}{\dot{O}}\big)\boldsymbol{\widehat{~}}$ and
$\boldsymbol{\Gamma}={ O} ^{-1}(t)\mathbf{\hat{z}}$  in which
$\boldsymbol{\Sigma}(t) =\big({O}^{-1}
\delta{O}\big)\boldsymbol{\widehat{~}}$ is a curve in
$\mathbb{R}^3$ that vanishes at the endpoints in time. 
\end{proposition}

\begin{proof}
Since ${\mathbb I}$ is symmetric and $\boldsymbol{\chi}$ is constant, we 
obtain the variation,
\begin{align*} 0 =
\delta \int^b_a l(\boldsymbol{\Omega},\boldsymbol{\Gamma}) \, d t
& =  \int^b_a 
\langle\,  {\mathbb I} \boldsymbol{\Omega}
, 
\delta\boldsymbol{\Omega}\rangle 
-
\langle  m{\rm g}\,\boldsymbol{\chi}
, 
\delta\boldsymbol{\Gamma}\rangle \, dt \\
& =  \int^b_a 
\langle\,  {\mathbb I} \boldsymbol{\Omega}
, 
\boldsymbol{\dot{ \Sigma}}
      + \boldsymbol{\Omega} \times \boldsymbol{\Sigma}\rangle
-
\langle  m{\rm g}\,\boldsymbol{\chi}, 
\boldsymbol{\Gamma} \times \boldsymbol{\Sigma}
\rangle 
\, dt\\
& =  \int^b_a  \left\langle
      - \frac{d }{d t}{\mathbb I} \boldsymbol{\Omega}
, 
      \boldsymbol{\Sigma}\right\rangle + \left\langle\, {\mathbb I}
\boldsymbol{\Omega}
, 
      \boldsymbol{\Omega} \times
\boldsymbol{\Sigma}\right\rangle 
-
\langle  m{\rm g}\,\boldsymbol{\chi}, 
\boldsymbol{\Gamma} \times \boldsymbol{\Sigma}
\rangle 
{d t}\\
& = \int^b_a \left\langle
- \frac{d }{d t} {\mathbb I} \boldsymbol{\Omega}
      + {\mathbb I} \boldsymbol{\Omega}\times \boldsymbol{\Omega}
+
m{\rm g}\, \boldsymbol{\Gamma}\times\boldsymbol{\chi}
\,, 
\boldsymbol{\Sigma} \right\rangle d t
+ 
\scp{{\mathbb I} \boldsymbol{\Omega}} {\boldsymbol{\Sigma}}\Big|_a^b
\,.
\end{align*}
upon integrating by parts and using the
endpoint conditions, $ \boldsymbol{\Sigma} (b) =
\boldsymbol{\Sigma} (a) = 0 $. 

Since $\boldsymbol{\Sigma}$ is
otherwise arbitrary, vanishing in the variation of the reduced action integral 
of $S_{\mathrm red}$ in \eqref{rbvp1} is equivalent to
\[- \frac{d }{d t} {\mathbb I} \boldsymbol{\Omega}
      + {\mathbb I} \boldsymbol{\Omega}\times \boldsymbol{\Omega}
+
m{\rm g}\, \boldsymbol{\Gamma}\times\boldsymbol{\chi}
= 0 \,, \]
which is Euler's motion equation for the heavy top
\eqref{top-eqns-vector}. This motion equation is completed by the
auxiliary equation
$\boldsymbol{\dot{\Gamma}}=\boldsymbol{\Gamma}\times\boldsymbol{\Omega}$
in \eqref{top-eqns-vector} arising from the definition of $\boldsymbol{\Gamma}$. 
\end{proof}

\emph{(B) Hamilton--Pontryagin constrained heavy top action integral.}

\index{heavy top!Hamilton--Pontryagin} \index{Hamilton--Pontryagin!heavy top} 

Heavy top dynamics can also be derived by using Lagrange multipliers to impose the constraints on the reduced Lagrangian, just as was done in the Hamilton--Pontryagin principle for the matrix Euler equations in Section \ref{HamPont-EulerRB}. In particular, we apply the Hamilton--Pontryagin approach to the following class of constrained action integrals,
\begin{eqnarray}
S &=& \int L(g,\,\dot{g}\,,\,\,\hat{e}_3) \,dt
\\
&=& \int  \bigg\{
l(\Omega, \Gamma ) 
+ \Big\langle\Pi\,,\,g^{-1}\dot{g}-\Omega \Big\rangle
+
 \Big\langle\chi\,,\,g^{-1}\hat{e}_3-\Gamma \Big\rangle
\bigg\}\,dt
\,,
\label{reduced-constrained-lag1}
\end{eqnarray}
where one identifies
\begin{equation}
l(\Omega, \Gamma ) =  
L(e,\,g^{-1}\dot{g},\,g^{-1}\hat{e}_3)
\,.
\label{reduced-constrained-lag2}
\end{equation}
This class of action integrals produces constrained equations of motion determined from the Hamilton--Pontryagin principle. 

\begin{remark}\label{HP-feature}\rm
A feature of the Hamilton--Pontryagin principle is the freedom to modify the constraint relations in \eqref{reduced-constrained-lag1} and \eqref{reduced-constrained-lag2} and thereby accommodate alternative  types of dynamical interpretations between the physical quantities $(\Omega, \Gamma )$ and the transport maps $(g^{-1}\dot{g},g^{-1}\hat{e}_3)$. For an example of the use of this feature, see section \ref{BKeqn-sec} on nonlinear shallow water equations.
\end{remark}

\begin{theorem}[Hamilton--Pontryagin action principle]\label{HeavyTop-motion-thm}
\index{variational principle!Hamilton--Pontryagin}\hfill

The stationarity condition for the constrained Hamilton--Pontryagin principle defined in Equations \eqref{reduced-constrained-lag1} and \eqref{reduced-constrained-lag2} implies the following equation of motion,
\begin{equation}
\Big(\frac{d}{dt}-{\rm ad}^*_\Omega\Big)\Pi
= \chi \diamond\Gamma
\,,
\label{TopHamPont-eqn}
\end{equation}
with $\Pi=\delta l / \delta\Omega$ and $\chi = \delta l/\delta \Gamma$,  where $\chi$ defines the vector in the body directed from the point of support to the centre of mass. 
\end{theorem}

\begin{exercise}
Prove that Theorem \ref{HeavyTop-motion-thm} for the Hamilton--Pontryagin action principle yields the same $\mathbb{R}^3$ vector equations \eqref{top-eqns-vector} as determined from the Euler--Poincar\'e action principle, when one sets $g=O\in SO(3)$, $\Gamma=\bs{\Gamma}\in \mathbb{R}^3$ and $\chi=m{\rm g}\bs{\chi}\in \mathbb{R}^3$.
\end{exercise}

\begin{answer}
The result here hinges on the variational relations with $\xi := g^{-1}\delta g\in \mathfrak{g}$,
\begin{equation}
\delta (g^{-1}\dot{g}) = \frac{d\xi}{dt} + {\rm ad}_{g^{-1}\dot{g}}\xi 
\quad\hbox{and}\quad 
\delta (g^{-1}\hat{e}_3) = -\, \pounds_\xi \hat{e}_3
\label{HamPont-relations}
\end{equation}
The variations then yield the constraints $g^{-1}\dot{g}=\Omega$ and $g^{-1}\hat{e}_3=\Gamma$ as well as 
\begin{equation}
\frac{\delta l }{\delta \Omega} = \Pi\,,\quad
\frac{\delta l }{\delta \Gamma} = \chi \,,\quad
\label{HamPont-variations}
\end{equation}
and after defining dual actions and the diamond operator, then  integrating by parts in time one has
the Hamilton--Pontryagin equations,
\begin{equation}
\Big(\frac{d}{dt}-{\rm ad}^*_{g^{-1}\dot{g}}\Big)\frac{\delta l }{\delta \Omega} 
= \frac{\delta l }{\delta \Gamma} \diamond\Gamma
\quad\hbox{and}\quad
\frac{d\Gamma}{dt} = - \pounds_{g^{-1}\dot{g}} \Gamma
\,,
\label{TopHamPont-motion}
\end{equation}
which completes the proof of the motion equation in \eqref{TopHamPont-eqn}
and the advection equation for $\Gamma:= g^{-1}\hat{e}_3$.
\end{answer}

\smallskip

\emph{Legendre transformation for the heavy top.}

The Legendre transformation for
$l(\boldsymbol{\Omega},\boldsymbol{\Gamma})$ gives the body angular
momenta,
\begin{eqnarray*}
\label{ht Legendre}
\boldsymbol{\Pi} 
= \frac{\partial l}{\partial\boldsymbol{\Omega}}
= {\mathbb I} \boldsymbol{\Omega}
\,.
\end{eqnarray*}
The well known energy Hamiltonian for the heavy top then emerges from the 
\emph{reduced Legendre transformation} \index{reduced Legendre transform}
\begin{eqnarray}
\label{ht Hamiltonian}
h(\boldsymbol{\Pi},\boldsymbol{\Gamma})
=
\boldsymbol{\Pi}\cdot\boldsymbol{\Omega}
-
l(\boldsymbol{\Omega},\boldsymbol{\Gamma})
=
\tfrac{1}{2}\langle\boldsymbol{\Pi}, 
\mathbb{I}^{-1}\boldsymbol{\Pi}\rangle 
+
\langle  m{\rm g}\,\boldsymbol{\chi}
, 
\boldsymbol{\Gamma}\,\rangle 
,
\end{eqnarray}
which is the sum of the kinetic and potential energies of the heavy top.

\begin{definition}[Functional variational derivative]
Let $f,h{:\ } \mathfrak{g} ^{\ast}\to\mathbb{R}$ be two real-valued functions on the
dual space 
$\mathfrak{g} ^{\ast}$.  Upon denoting elements of $\mathfrak{g} ^{\ast}$ by
$\mu$, the functional variational derivative of $f$   at $\mu$ is defined as the
unique element $\delta  f/ \delta \mu$ of $\mathfrak{g}$ emerging in the following limit
\begin{equation}\label{fd}
\lim_{\varepsilon \rightarrow 0}
\frac{1}{\varepsilon }[f(\mu
+ \varepsilon \delta \mu) - f(\mu )]
=  \left\langle \delta  \mu , \frac{
\delta  f}{\delta  \mu } \right\rangle ,
\end{equation}
for all $\delta\mu \in \mathfrak{g}^*$, where 
$\left\langle\cdot , \cdot\right\rangle$ denotes the pairing between
$\mathfrak{g} ^{\ast} $ and  $\mathfrak{g}$.%
\footnote{For fluid dynamics, $\scp{\,\cdot\,}{\,\cdot\,}$ 
is the $L^2$ integral pairing between two functions. See equation \eqref{l2p}.}
\end{definition}

\begin{definition}[Lie--Poisson brackets and Lie--Poisson equations]
The \emph{$(\pm)$  Lie--Poisson brackets} for the Euler--Poincar\'e 
equations for a kinetic energy Lagrangian are defined by
\begin{equation}
\{f, h\}_\pm (\mu )  
=  \pm \left\langle \mu , \left[ \frac{ \delta f}{\delta  \mu},
\frac{\delta  h}{\delta \mu } \right] \right\rangle 
=  \mp \left\langle \mu , 
\operatorname{ad}_{\delta h/\delta \mu}
\frac{ \delta f}{\delta  \mu}\right\rangle
.
\end{equation}
The corresponding \emph{Lie--Poisson equations}, determined by $\dot f =
\{ f,h\}
$ read
\begin{equation}\label{alpe}
\dot \mu  = \{ \mu,h\}
= \mp \operatorname{ad} ^{* }_{\delta h/\delta \mu} \mu
,
\end{equation}
where one defines the $\ad^*$ operation in terms of the pairing
$\left\langle\cdot , \cdot\right\rangle$, by
\[
\{ f,h\}
=
\left\langle \mu , 
\operatorname{ad}_{\delta h/\delta \mu}
\frac{ \delta f}{\delta  \mu}\right\rangle
=
\left\langle \operatorname{ad}^*_{\delta h/\delta \mu}\mu , 
\frac{ \delta f}{\delta  \mu}\right\rangle
.
\]
\end{definition}
The Lie--Poisson setting of mechanics is a special case of the general
theory of systems on Poisson manifolds, for which there is now
an extensive theoretical development. (See Marsden and Ratiu \cite{MaRa1994} for a
start on this literature.) 

\subsection{Lie--Poisson brackets and momentum maps}  An important 
feature of the rigid body bracket carries  over to general Lie algebras.
Namely, {\em Lie--Poisson brackets on $\mathfrak{g}^*$ arise from canonical
brackets on the cotangent bundle\/} (phase space) $T^\ast G$ associated
with a Lie group $G$ which has $\mathfrak{g}$ as its associated Lie
algebra. Thus, the process by which the Lie--Poisson brackets arise is the
\emph{momentum map}
\[
T^\ast G \mapsto \mathfrak{g}^\ast
.
\]
For example, a rigid body is free to rotate about its center of
mass and $G$ is the (proper) rotation group ${\mathrm SO}(3)$. The choice of
$T^\ast G$ as the primitive phase space is made according to
the classical procedures of mechanics described earlier.  For
the description using Lagrangian mechanics, one forms the
velocity phase space $TG$.  The Hamiltonian description on 
$T^\ast G$ is then obtained
by standard procedures: Legendre transforms, etc.

The passage from $T^\ast G$ to the space of ${\boldsymbol
\Pi}$'s (body angular momentum space) is determined by {\em
left\/} translation on the group.  The mapping $T^\ast SO(3)\to{\boldsymbol \Pi}$ 
is  an example of a {\em momentum map\/}; that is, a
mapping whose components are the ``Noether  quantities''
associated with a symmetry group.  The map from $T^\ast G$ to
$\mathfrak{g}^\ast$ being a Poisson map {\em is a
general fact about momentum maps\/}.  The Hamiltonian point of
view of all this is a standard subject.

\begin{remark}\rm [Lie--Poisson description of the heavy top]

As it turns out, the underlying Lie algebra for the Lie--Poisson
description of the heavy top consists of the Lie algebra $\mathfrak{se}(3)$
of infinitesimal Euclidean motions in $\mathbb{R}^3$. This is a bit
surprising, because heavy top motion itself does {\em not\/} actually
arise through actions of the Euclidean group of rotations and translations
on the body, since the body has a fixed point! 

Instead, the Lie algebra
$\mathfrak{se}(3)$ arises for another reason associated with the 
\emph{breaking} of the ${\mathrm SO}(3)$ isotropy by the presence of the
gravitational field. This symmetry breaking introduces a semidirect-product
Lie--Poisson structure which happens to coincide with the dual of the Lie algebra
$\mathfrak{se}(3)$ in the case of the heavy top. As we shall see later, a close
parallel exists between this case and the Lie--Poisson structure for compressible
fluids. \index{Lie algebra!semidirect product}
\end{remark}

\subsection{Lie--Poisson brackets for the heavy top}
The \emph{Lie algebra of the special Euclidean group in 3D} is
${se}(3) = \mathbb{R}^3  \times \mathbb{R}^3$ with the  Lie bracket
\begin{equation}\label{htla} 
\big[(\boldsymbol{\xi}, \mathbf{u} )\,,\,
(\boldsymbol{\eta} , \mathbf{ v} )\big]  
=  
(\boldsymbol{\xi} \times \boldsymbol{\eta}\,,\,
\boldsymbol{\xi} \times \mathbf{ v} - \boldsymbol{\eta}
\times \mathbf{u})  
\,.
\end{equation}
%
We identify the dual space of $\mathfrak{se}(3)$ with pairs $({\boldsymbol
\Pi},\boldsymbol{\Gamma} )$; the corresponding
$  (-)  $ Lie--Poisson bracket called the \emph{heavy top
bracket\/} is
\begin{eqnarray}\label{ht-LPB}
\{f, h\}(\boldsymbol{\Pi},\boldsymbol{\Gamma})
=
-\boldsymbol{\Pi}\cdot\nabla_{\Pi}f\times\nabla_{\Pi}h
\,-\boldsymbol{\Gamma}\cdot\big(\nabla_{\Pi}f\times\nabla_{\Gamma}h
-\nabla_{\Pi}h\times\nabla_{\Gamma}f\big)
\,.
\end{eqnarray}
This Lie--Poisson bracket and the Hamiltonian \eqref{ht
Hamiltonian} recover the equations \eqref{top-eqns-vector} and
\eqref{chi-eqns-aux} for the heavy top, as
\begin{eqnarray}
\label{ht-eqns}
\boldsymbol{\dot{\Pi}}
=
\{\boldsymbol{\Pi}, h\}
&=&
\boldsymbol{\Pi}\times\nabla_{\Pi}h
+\boldsymbol{\Gamma}\times\nabla_{\Gamma}h
=
\boldsymbol{\Pi}\times\mathbb{I}^{-1}\boldsymbol{\Pi}
+\boldsymbol{\Gamma}\times m{\rm g}\,\boldsymbol{\chi}
\,,\\
\boldsymbol{\dot{\Gamma}}
=
\{\boldsymbol{\Gamma}, h\}
&=&
\boldsymbol{\Gamma}\times\nabla_{\Pi}h
=
\boldsymbol{\Gamma}\times\mathbb{I}^{-1}\boldsymbol{\Pi}
\,.
\end{eqnarray}

\index{Lie algebra!semidirect product}\index{broken symmetry!semidirect product}

\begin{remark}\rm [Semidirect products and symmetry breaking -- \emph{General Picture}]
The Lie algebra of the Euclidean group has a structure which is
a special case of what is called a \emph{semidirect product\/}, denoted $\circledS$.
Here, it is the semidirect product action $\mathfrak{so}(3)\circledS\mathbb{R}^3$ of the Lie
algebra of rotations $\mathfrak{so}(3)$ acting on the infinitesimal translations $\mathbb{R}^3$,
which happens to coincide with $se(3,\mathbb{R})$.  In general, the Lie bracket for
semidirect product action $\mathfrak{g}\,\circledS\,V$ of Lie algebra $\mathfrak{g}$
on vector space $V$ is given by 
\begin{align}
\big[(X,a), (\overline{X},\overline{a})\big]
=
\big([X,\overline{X}\,],\overline{X}(a)-X(\overline{a})\big)
\,,\label{SDPaction}
\end{align}
in which $X,\overline{X}\in\mathfrak{g}$ and $a,\overline{a}\in V$.
The action of the Lie algebra $\mathfrak{g}$ on the vector space $V$ is denoted, for
example, $X(\overline{a})$. Usually, the action $\mathfrak{g}\times V\to V$ would be the Lie derivative,
$\pounds$, or its transpose  $\pounds^T$. Let variables $\mu\in\mathfrak{g}^*$ and $b\in V^*$ be dual, 
respectively, to $X,\overline{X}\in\mathfrak{g}$ and $a,\overline{a}\in V$ as
\begin{align}
\left[\left(\frac{\p F}{\p \mu} , \frac{\p F}{\p b}\right), \left(\frac{\p H}{\p \mu},\frac{\p H}{\p b}\right)\right]
= \mp
\left( \left[\frac{\p F}{\p \mu},\frac{\p H}{\p \mu}\,\right] \,,\,
\pounds^T_{\frac{\p H}{\p \mu}} \frac{\p F}{\p b} - \pounds^T_{\frac{\p F}{\p \mu}} \frac{\p H}{\p b}\right)
\label{Dual-vars}
\end{align}
where in Lie symmetry reduction of the Lagrangian in Hamilton's principle one chooses $(-)$ resp. $(+)$ 
signs for Lie algebras which are invariant under right (resp. left) Lie-group action.

Consequently, one may write the Lie-Poisson bracket dual to the semidirect-product 
action of vector fields on vector spaces as, \index{broken symmetry!semidirect product}
\begin{align}
\begin{split}
\Big\{ F,H \Big\} :&= \mp\scp{(\mu,b)}{\left[\left(\frac{\p F}{\p \mu} , \frac{\p F}{\p b}\right), \left(\frac{\p H}{\p \mu},\frac{\p H}{\p b}\right)\right]}
\\&= \mp
\scp{(\mu,b)} { \left( \left[\frac{\p F}{\p \mu},\frac{\p H}{\p \mu}\,\right] ,
\pounds^T_{\frac{\p H}{\p \mu}}\frac{\p F}{\p b} - \pounds^T_{\frac{\p F}{\p \mu}}\frac{\p H}{\p b}\right) }
\\&= 
\pm\scp{\mu} {\ad_{\frac{\p F}{\p \mu}} \frac{\p H}{\p \mu} }
\pm\scp{\pounds^T_{\frac{\p H}{\p \mu} }\frac{\p F}{\p b} - \pounds^T_{\frac{\p F}{\p \mu}}\frac{\p H}{\p b}}{b}
\\&=
\pm\scp{\ad^*_{\frac{\p H}{\p \mu}}\mu} {\frac{\p F}{\p \mu}}
\pm \scp{\frac{\p F}{\p b}} {-\pounds_{\frac{\p H}{\p \mu} } b}
\mp \scp{\frac{\p H}{\p b}} {-\pounds_{\frac{\p F}{\p \mu}}b}
\\&=
\mp\scp{\ad^*_{\frac{\p H}{\p \mu}}\mu} {\frac{\p F}{\p \mu}}
\mp \scp {\pounds_{\frac{\p H}{\p \mu} } b} {\frac{\p F}{\p b}}
\mp \scp{\frac{\p H}{\p b}\diamond b} {\frac{\p F}{\p \mu} }
.\end{split}
\label{SDP-LPB}
\end{align}
Hence, the \index{Lie algebra!semidirect product} \index{Lie--Poisson!semidirect product}
Lie-Poisson equations for semidirect-product action may be written in matrix operator form as
\begin{align}
\frac{d}{dt} 
\begin{bmatrix}
\mu \\ b
\end{bmatrix}
=
\mp
\begin{bmatrix}
\ad^*_{\Box}\mu & \Box \diamond b 
\\
\pounds_\Box b & 0
\end{bmatrix}
\begin{bmatrix}
{\p H}/{\p \mu} \\  {\p H}/{\p b}
\end{bmatrix}
=
\mp
\begin{bmatrix}
\ad^*_{\frac{\p H}{\p \mu}}\mu + \frac{\p H}{\p b}\diamond b
\\ \pounds_{\frac{\p H}{\p \mu} } b
\end{bmatrix}
,
\label{SDP-LPB-eqns}
\end{align}
with $(-)$ resp. $(+)$ signs for Lie algebras invariant under right (resp. left) Lie-group action.
\index{broken symmetry!semidirect product}

For the left-invariant variables in the heavy top Hamiltonian in \eqref{ht Hamiltonian} 
the motion equations in \eqref{SDP-LPB-eqns} with the $(+)$ sign become
\begin{align}
\frac{d}{dt} 
\begin{bmatrix}
\boldsymbol{\Pi} \\ \boldsymbol{\Gamma}
\end{bmatrix}
=
+
\begin{bmatrix}
\boldsymbol{\Pi}\times & \boldsymbol{\Gamma}\times
\\
\boldsymbol{\Gamma}\times & 0
\end{bmatrix}
\begin{bmatrix}
{\p H}/{\p \boldsymbol{\Pi}} = \mathbb{I}^{-1}\boldsymbol{\Pi}
\\  
{\p H}/{\p \boldsymbol{\Gamma}} = m{\rm g}\,\boldsymbol{\chi}
\end{bmatrix}
=+
\begin{bmatrix}
\boldsymbol{\Pi}\times  \frac{\p H}{\p \boldsymbol{\Pi}}
+ \boldsymbol{\Gamma}\times \frac{\p H}{\p \boldsymbol{\Gamma}}
\\ \boldsymbol{\Gamma}\times \frac{\p H}{\p \boldsymbol{\Pi}}
\end{bmatrix}
.
\label{SDP-HT-eqns}
\end{align}
This recovers the heavy top equations in \eqref{ht-eqns} for the Hamiltonian in \eqref{ht Hamiltonian},
\begin{eqnarray}
\label{ht Hamiltonian2}
h(\boldsymbol{\Pi},\boldsymbol{\Gamma})
=
\boldsymbol{\Pi}\cdot\boldsymbol{\Omega}
-
l(\boldsymbol{\Omega},\boldsymbol{\Gamma})
=
\tfrac{1}{2}\langle\boldsymbol{\Pi}, 
\mathbb{I}^{-1}\boldsymbol{\Pi}\rangle 
+
\langle  m{\rm g}\,\boldsymbol{\chi}
, 
\boldsymbol{\Gamma}\,\rangle 
,
\end{eqnarray}
which is the sum of the kinetic and potential energies of the heavy top.
\end{remark}

\begin{exercise} 

Determine the Hamiltonian equations via Hamilton's 
principle for a  heavy top whose body variables and parameters are made \textit{complex}, 
$\bs{\Omega}, \bs{\Gamma}, \bs{\chi}\in\mathbb{C}^3$, 
\[
\bs{\Omega} = \bs{\Omega}_\Re + i \bs{\Omega}_\Im
\,,\quad
\bs{\Gamma} = \bs{\Gamma}_\Re + i \bs{\Gamma}_\Im
\,,\quad
\bs{\chi} = \bs{\chi}_\Re + i \bs{\chi}_\Im
\,.\]
The imaginary part of its Lagrangian would be, 
\begin{align}\begin{split}
\Im\ell (\bs{\Omega}_\Re ,\bs{\Omega}_\Im)
= 
\bs{\Omega}_\Re \cdot \mathbb{I} \bs{\Omega}_\Im 
- m{\rm g} \big( \bs{\chi}_\Re\bs{\Gamma}_\Im +  \bs{\chi}_\Im\bs{\Gamma}_\Re \big)
\label{CHT-Lag}
\end{split}\end{align}
for a real moment of inertia $\mathbb{I}={\rm diag} (I_1,I_2,I_3)$.

As one would expect, the corresponding Hamiltonian arising after a Legendre transform 
is
\begin{align}\begin{split}
\Im h( \bs{\Pi}_\Re , \bs{\Gamma}_\Re ,\bs{\Pi}_\Im , \bs{\Gamma}_\Im )
= 
\bs{\Pi}_\Re \cdot \mathbb{I}^{-1} \bs{\Pi}_\Im 
+ m{\rm g} \big( \bs{\chi}_\Re\bs{\Gamma}_\Im +  \bs{\chi}_\Im\bs{\Gamma}_\Re \big)
\,.\label{CHT-Lag}
\end{split}\end{align}
Derive the equations of motion for this complex heavy top via its
Euler--Poincar\'e equations, then find the $\mathfrak{so}(3)^*\times \mathfrak{so}(3)^*$ 
Lie-Poisson bracket for this system. What are the Casimirs for this Lie-Poisson  system?

\end{exercise}

\begin{answer}
The $\mathfrak{so}(3)^*\times \mathfrak{so}(3)^*$ 
block diagonal Lie-Poisson bracket for this system is given by
\begin{equation}
\frac{d}{d t}\left(\begin{array}{l}
\bs{\Pi}_\Re \\
\bs{\Gamma}_\Re \\
\bs{\Pi}_\Im \\
\bs{\Gamma}_\Im
\end{array}\right)
=
\left(\begin{array}{cccc}
\bs{\Pi}_\Re \times & \bs{\Gamma}_\Re  \times & 0 & 0  \\
\bs{\Gamma}_\Re \times &0 & 0 & 0  \\
0 & 0 & \bs{\Pi}_\Im \times & \bs{\Gamma}_\Im \times  \\
0 & 0 & \bs{\Gamma}_\Im \times & 0  
\end{array}\right)
\left(\begin{array}{l}
\partial H / \partial \bs{\Pi}_\Re:=\mathbb{I}^{-1} \bs{\Pi}_\Im \\
\partial H / \partial \bs{\Gamma}_\Re := m g \bs{\chi}_\Im \\
\partial H / \partial \bs{\Pi}_\Im := \mathbb{I}^{-1} \bs{\Pi}_\Re \\
\partial H  / \partial \bs{\Gamma}_\Im := m g \bs{\chi}_\Re
\end{array}\right) .
\end{equation}
with intertwined equations of motion 
\begin{equation}
\frac{d}{d t}\left(\begin{array}{l}
\bs{\Pi}_\Re \\
\bs{\Gamma}_\Re \\
\bs{\Pi}_\Im\\
\bs{\Gamma}_\Im
\end{array}\right)=\left(\begin{array}{l}
\bs{\Pi}_\Re \times \bs{\Omega}_\Im + m g \bs{\Gamma}_\Re \times \bs{\chi}_\Im \\
\bs{\Gamma}_\Re \times \bs{\Omega}_\Im\\
\bs{\Pi}_\Im\times \bs{\Omega}_\Re + m g \bs{\Gamma}_\Im \times \bs{\chi}_\Re\\
\bs{\Gamma}_\Im \times \bs{\Omega}_\Re
\end{array}\right)
\end{equation}
whose Casimirs are $\bs{\Pi}_\Re\cdot\bs{\Gamma}_\Re$, $|\bs{\Gamma}_\Re|^2$, and 
$\bs{\Pi}_\Im\cdot\bs{\Gamma}_\Im$, $|\bs{\Gamma}_\Im|^2$.
\end{answer}

\subsection{The heavy top equations  via the Kaluza--Klein construction} 
The Lagrangian in the heavy top action principle \eqref{rbvp1} may be transformed into
a quadratic form. This is accomplished by suspending the system in a higher dimensional
space via the \emph{Kaluza--Klein construction\/}. This construction proceeds for the
heavy top as a modification of the well-known Kaluza--Klein construction for a
charged particle in a prescribed magnetic field \cite{HoMaRa1998a}.

Let $Q_{KK}$ be the manifold $SO(3)\times\mathbb{R}^3$ with
variables
$(O,\mathbf{q})$. On $Q_{KK} $ introduce the \emph{Kaluza--Klein
Lagrangian} $L_{KK}{:\ } TQ_{KK}\simeq TSO(3)\times T\mathbb{R}^3\mapsto
\mathbb{R}$ in which one observes that the coordinate $\mathbf{q},\in\mathbb{R}^3$ is absent, as
\begin{equation}
L_{KK}(O, {\dot{O}},\mathbf{\dot{q}}
;\mathbf{\hat{z}}) 
=
L_{KK}(\boldsymbol{\Omega},\boldsymbol{\Gamma},\mathbf{\dot{q}}) 
=
\tfrac{1}{2} \langle\,  \mathbb{I} \boldsymbol{\Omega}
, 
\boldsymbol{\Omega}\,\rangle 
+
\tfrac{1}{2}|\boldsymbol{\Gamma}+\mathbf{\dot{q}}|^2
,
\label{KK-metric}
\end{equation}
with $\boldsymbol{\Omega} =\big(O^{-1}
{\dot{O}}\big)\boldsymbol{\widehat{\,}}$
and $\boldsymbol{\Gamma}={ O}^{-1}\mathbf{\hat{z}}$.
The Lagrangian $L_{KK} $ is positive definite in
$(\boldsymbol{\Omega},\boldsymbol{\Gamma},\mathbf{\dot{q}})$;
so it may be regarded as the kinetic energy of a metric, the
\emph{Kaluza--Klein metric\/} on $TQ_{KK} $.

The Legendre transformation for $L_{KK}$ gives the momenta
\begin{equation}
\label{KK Legendre2}
\frac{\p L_{KK}}{\p \boldsymbol{\Omega}} =
{\mathbb I} \boldsymbol{\Omega} =: \boldsymbol{\Pi} 
\qquad \text{and} \qquad 
\frac{\p L_{KK}}{\p \mathbf{\dot{q}}} =
\boldsymbol{\Gamma}+\mathbf{\dot{q}} =: \mathbf{p} 
\,.
\end{equation}
Since $L_{KK}$ does not depend on $\mathbf{q}$, the Euler--Lagrange
equation
\[
\frac{d}{dt}\frac{\partial L_{KK}}{\partial \mathbf{\dot{q}}}  
=
\frac{\partial L_{KK}}{\partial \mathbf{q}} = 0
,
\]
shows that the $\mathbf{q} $-momentum given by  $\mathbf{p} = \partial L_{KK}/\partial\mathbf{\dot{q}}$
is conserved. We now identify the \emph{constant vector\/} $\mathbf{p}$
 as the following constant vector in the body,
\[
\mathbf{p} = \boldsymbol{\Gamma}+\mathbf{\dot{q}}
=
-\,m{\rm g}\,\boldsymbol{\chi}
.\]
After this identification, the heavy top action principle in
\ref{ht-actprinc} with the Kaluza--Klein Lagrangian returns Euler's
motion equation for the heavy top \eqref{top-eqns-vector}. 

The Hamiltonian
$H_{KK}$ associated to $L_{KK}$ by the Legendre transformation
\eqref{KK Legendre2} is
\begin{eqnarray*}
H_{KK}(\boldsymbol{\Pi},\boldsymbol{\Gamma}, \mathbf{p}) 
&=& 
\boldsymbol{\Pi}\cdot\boldsymbol{\Omega}
 + 
\mathbf{p}\cdot\mathbf{\dot{q}}
- 
L_{KK}(\boldsymbol{\Omega},\boldsymbol{\Gamma},\mathbf{\dot{q}})  
\nonumber \\
&=& 
\tfrac{1}{2}\boldsymbol{\Pi}\cdot\mathbb{I}^{-1}\boldsymbol{\Pi}
-
\mathbf{p}\cdot\boldsymbol{\Gamma}
+
\tfrac{1}{2}|\mathbf{p}|^2
\nonumber \\
&=& 
\tfrac{1}{2}\boldsymbol{\Pi}\cdot\mathbb{I}^{-1}\boldsymbol{\Pi}
+
\tfrac{1}{2}|\mathbf{p}-\boldsymbol{\Gamma}|^2
-
\tfrac{1}{2}|\boldsymbol{\Gamma}|^2
.
\end{eqnarray*}
Recall that $\boldsymbol{\Gamma}$ is a unit vector. 
On the constant level set $|\boldsymbol{\Gamma}|^2=1$, the
Kaluza--Klein Hamiltonian $H_{KK}$ is a positive quadratic function,
shifted by a constant. Likewise, on the constant level set $\mathbf{p} =
-m{\rm g}\,\boldsymbol{\chi}$, the Kaluza--Klein Hamiltonian $H_{KK}$ is a
function of only the variables
$(\boldsymbol{\Pi},\boldsymbol{\Gamma})$ and is equal to the Hamiltonian
\eqref{ht Hamiltonian} for the heavy top up to an additive constant.
Consequently, the Lie--Poisson equations for the Kaluza--Klein Hamiltonian
$H_{KK}$ now reproduce Euler's motion equation for the heavy top
\eqref{top-eqns-vector}.

\index{heavy top!Hamilton--Pontryagin} \index{Hamilton--Pontryagin!heavy top} 
\begin{exercise}
Use the Hamilton--Pontryagin principle to calculate the equations for heavy top dynamics in the spatial frame.
\end{exercise}

\begin{answer}
In the spatial frame one is dealing with a rigid body Lagrangian that is not right invariant and 
depends on both $\widehat{\omega}(t) $, $\mathbb{I}_{spat}(t)$, and $\chi_{spat}(t):= \chi O^{-1}(t)$,
cf. equation \eqref{reduced-constrained-lag1},
where its definition implies the dynamics of $\chi_{spat}$; namely,
\[
\frac{\chi_{spat}}{dt} = - [\chi_{spat} \,,\,\widehat{\omega}]
\,.
\]
\begin{eqnarray}
S(\widehat{\omega},O,\dot{O},{\pi})
&=&
\int^b_a \Big[ \ell(\widehat{\omega},\mathbb{I}_{spat},\chi_{spat}) 
+ 
\langle\, {\pi}
\,,\,
\dot{O}O^{-1} - \widehat{\omega}\,\rangle
\Big]\, dt
\,.\label{implicit-actionEulerHT-right}\\
\,.
\nonumber
\end{eqnarray}
The variations of the action $S$ in formula (\ref{implicit-actionEulerRB-right}) are now given by
\begin{align}
\begin{split}
\delta S 
&=
\int^b_a \bigg\{
\Big\langle\,\frac{\delta \ell }{\delta \widehat{\omega}} 
- {\pi}\,,\, \delta \widehat{\omega} \,\Big\rangle
+ \Big\langle\,\delta {\pi}  
\,,\, (\dot{O}O^{-1} - \widehat{\omega})\,\Big\rangle
+
\Big\langle\, {\pi}  \,,\, \delta (\dot{O}O^{-1})
\,\Big\rangle
\\&\phantom{\int^b_a}
+ \Big\langle\,\frac{\delta \ell }{\delta \mathbb{I}_{spat} } 
\,,\, \delta \mathbb{I}_{spat}  \,\Big\rangle
+ \Big\langle\,\frac{\delta \ell }{\delta \chi_{spat} } 
\,,\, \delta \chi_{spat}  \,\Big\rangle
\bigg\}
\, dt
\\&=
\int^b_a \bigg\{
\Big\langle\,\frac{\delta \ell }{\delta \widehat{\omega}} 
- {\pi}\,,\, \delta \widehat{\omega} \,\Big\rangle
+ \Big\langle\,\delta {\pi}  
\,,\, (\dot{O}O^{-1} - \widehat{\omega})\,\Big\rangle
\\&\phantom{\int^b_a}
+ 
\Big\langle\,- \dot{\pi}  - \big[\pi\,,\,\widehat{\omega} \,\big] 
+ \Big[\mathbb{I}_{spat}\,,\,\frac{\delta \ell }{\delta \mathbb{I}_{spat} } \Big]
+ \Big[\chi_{spat} \,,\, \frac{\delta \ell }{\delta \chi_{spat} }  \,\Big] 
\,,\, \widehat{\xi} \,\Big\rangle
\bigg\}
\, dt
\,,
\end{split}
\label{spatialHTvars}
\end{align}
where we have applied formula \eqref{implicit-actionEulerHT-right} for $\delta(\dot{O}O^{-1})$ 
and then integrated by parts in time. 
\end{answer}

\begin{exercise}
Show that the Hamiltonian equations for the spatial dynamics of the heavy top in \eqref{spatialHTvars} 
may be expressed via the following Lie--Poisson matrix operator 
\begin{align}
\frac{d}{dt}
\begin{pmatrix}
\pi \\ \mathbb{I}_{spat} \\ \chi_{spat}
\end{pmatrix}
= -
\begin{pmatrix}
\big[\pi\,,\,\Box \big]  & \big[\mathbb{I}_{spat} \,,\,\Box \big] & \big[\chi_{spat} \,,\,\Box \big]
\\ \big[\mathbb{I}_{spat} \,,\,\Box \big]  & 0 & 0
\\ \big[\chi_{spat} \,,\,\Box \big]  & 0 & 0
\end{pmatrix}
\begin{pmatrix}
\delta \mathrm{h}/\delta \pi = \widehat{\omega} 
\\ 
\delta \mathrm{h}/\delta \mathbb{I}_{spat} = -\,\delta \ell/\delta \mathbb{I}_{spat}
\\ 
\delta \mathrm{h}/\delta \chi_{spat} = -\,\delta \ell/\delta \chi_{spat}
\end{pmatrix}
\label{spatialRB-LPB}
\end{align}
with $(\pi ,\mathbb{I}_{spat},\chi_{spat})\in [\mathfrak{so}(3)\circledS ({\rm Sym}^3_+(\mathbb{R}) \oplus \mathbb{R}^3_+)]^*$  
where $\circledS$ denotes semidirect product action of the Lie algebra 
$\mathfrak{so}(3)$ represented as $3\times3$ antisymmetric real matrices acting on  
the direct sum of vector spaces comprising  $3\times3$ positive symmetric real matrices and 3D vectors, $({\rm Sym}^3_+(\mathbb{R})\oplus \mathbb{R}^3_+)$,
and superscript $\phantom{\,}^*$ denotes the dual semidirect product Lie algebra under matrix trace pairing.
See \cite{lewis1992heavy} for a complete a geometric treatment of the heavy top in both the body and spatial representations. 
  title={The heavy top: a geometric treatment} 
  \index{heavy top!geometric treatment} \index{heavy top!spatial representation}
\index{Lie algebra!semidirect product}\index{semidirect product!broken symmetry} \index{broken symmetry!semidirect product}
\end{exercise}

\emph{Summary.}
Geometric mechanics deals with dynamical systems
defined by Lie group invariant variational principles (VP/G),
such as geodesic motion on a Lie group $G$ whose metric 
is invariant under the action of the Lie group $G$. An example is 
the variational formulation of Euler's rigid body equations 
in three dimensions whose solutions are then seen to be 
geodesics on the rotation group  $SO(3)$. 

\index{symmetry!breaking}\index{Lie group!semidirect product action}
Geometric mechanics also deals with the effects of breaking the symmetry of a VP/G 
from $G$ to a subgroup $G_0\subset G$ on whose coset $G/G_0$ the Lie group $G$ may act
by the semidirect product, denoted as $G\circledS(G/G_0)$.
An example is Euler's heavy top equations, in which the fixed vertical direction $\zh$ of gravity reduces the 
Lie symmetry of the VP/G for the rigid body motion from $SO(3)$ to $SO(2)$. Namely, 
the remaining Lie group symmetry $SO(2)$ corresponds only to rotations which leave the direction of gravity $\zh$ invariant. 
For the heavy top, the (left) action $O(t)^{-1}\zh$ of $O(t)\in SO(3)$ on its coset $SO(3)/SO(2)$ comprises the 
motion in time $t$ of the fixed direction of gravity $\zh$ as seen by an observer in the reference frame of the rotating top. 
The variational formulation of Euler's heavy top equations then leads to motion on the
semidirect-product Lie group $SO(3)\circledS (SO(3)/SO(2))$, which is isomorphic to the Euclidean group
of rotations and translations in three dimensions, denoted $SE(3)$. \index{Lie group!semidirect product!$SE(3)$}

\begin{remark}\rm \index{Lie--Poisson brackets!dual Lie algebra}\index{Lie algebra!semidirect product}
Lie--Poisson brackets defined on the dual spaces of semidirect product Lie algebras
tend to occur under rather general circumstances when the symmetry in $T^\ast G $ is
broken, for example, reduced to an isotropy subgroup of a set of parameters.  In
particular, there are similarities in structure between the Poisson bracket for
compressible flow and that for the heavy top. In the latter case, the vertical
direction of gravity breaks isotropy of $\mathbb{R}^3$ from
$SO(3)$ to $SO(2)$ and the dynamics of the $SO(3)$ flow acts on $\boldsymbol{\Gamma}\in SO(3)/SO(2)$.
In the case of compressible fluid flow breaks the initial density configuration breaks the 
allowed diffeomorphisms from ${\rm Diff}(M)$ to ${\rm Diff}_{\rho_0}(M)$ the isotropy transformations
of the initial density distribution, $\rho_0$. Thus, the dynamics of the ${\rm Diff}(M)$ flow acts on 
$\rho\in {\rm Diff}(M)/{\rm Diff}_{\rho_0}(M)$. \index{broken symmetry!semidirect product}
\index{semidirect product!broken symmetry}

The general theory for semidirect products has been reviewed in a
variety of places, including Marsden, Ratiu and Weinstein~\cite{MaRaWe1984a,MaRaWe1984b}. 
Many interesting examples of Lie--Poisson brackets on semidirect products
exist for fluid dynamics. \index{Lie algebra!semidirect product}\index{Lie--Poisson brackets!semidirect product}

These semidirect-product Lie--Poisson Hamiltonian fluid theories range from
simple fluids, to charged fluid plasmas, to magnetized fluids, to
multiphase fluids, to super fluids, to Yang--Mills plasmas, relativistic, or
not, and to liquid crystals, as well as to other complex fluids. See, for example, the papers by Gibbons, Holm and
Kupershmidt~\cite{GiHoKu1982,HoKu1982,HoKu1983,HoKu1988}.
For discussions of many of these theories from the Euler--Poincar\'e viewpoint, see
Holm, Marsden and Ratiu~\cite{HoMaRa1998a} and Holm~\cite{Ho2002a}.
\end{remark}

\newpage
\vspace{4mm}\centerline{\textcolor{shadecolor}{\rule[0mm]{6.75in}{-2mm}}\vspace{-4mm}}
\section[Geometric ray optics in $\mathbb{R}^3$]{Lagrangian \& Hamiltonian methods for geometric ray optics in translation invariant, axisymmetric material}
\label{geometric-optics-app}

\secttoc

\textbf{What is this lecture about?} This lecture reformulates Fermat's principle for 
geometrical optics with axial symmetry to illustrate the modern geometrical mechanics 
versions of the variational methods of Lagrange and Hamilton. 

\subsection{Fermat's principle: Rays take paths of least optical length}
In geometrical optics, the ray path is determined by Fermat's principle
of least optical length,
\[
\delta \int n(x,y,z)\,ds = 0
\,.
\]
Here $n(x, y, z)$ is the index of refraction at the
spatial point $(x, y, z)$ and $ds$ is the element of arc
length along the ray path through that point.
Choosing coordinates so that the $z$--axis coincides
with the optical axis (the general direction of propagation), gives
\[
ds
= [(dx)^2 + (dy)^2 + (dz)^2]^{1/2}
= [1+\dot{x}^2 + \dot{y}^2 ]^{1/2}\,dz
,
\]
with $\dot{x} = dx/dz$ and $\dot{y}= d y / d z $. Thus, Fermat's
principle can be written in Lagrangian form, with
$z$ playing the role of time, 
\[
\delta \int L(x,y,\dot{x},\dot{y},z)\,dz = 0
.
\]
Here, the optical Lagrangian is,
\[
L(x,y,\dot{x},\dot{y},z) 
=  
n(x,y,z)[1+\dot{x}^2 + \dot{y}^2 ]^{1/2}
=:
n/\gamma
,
\]
or, equivalently, in two-dimensional vector notation with
$\mathbf{q}=(x,y)$,
\[
L(\mathbf{q},\mathbf{\dot{q}},z) 
=  
n(\mathbf{q},z)[1+|\mathbf{\dot{q}}|^2 ]^{1/2}
=:
n/\gamma
\quad\hbox{with}\quad
\gamma = [1+|\mathbf{\dot{q}}|^2 ]^{-1/2}
\le1
.
\]
Consequently, the vector Euler--Lagrange equation
of the light rays is
\[
\frac{d}{ds}\Big(n\frac{d\mathbf{q}}{ds}\Big)
=
\gamma\frac{d}{dz}\Big(n\gamma\frac{d\mathbf{q}}{dz}\Big)
=
\frac{\partial n}{\partial \mathbf{q}}
.
\]
The momentum $p$ canonically conjugate to the
ray path position $q$ in an ``image plane'', or on an
``image screen'', at a fixed value of $z$ is given by
$$\mathbf{p}=\frac{\partial L}{\partial \mathbf{\dot{q}}}
= n\gamma\mathbf{\dot{q}}$$
which satisfies $|\mathbf{p}|^2=n^2(1-\gamma^2)$.  This implies the velocity
\[\mathbf{\dot{q}}=\mathbf{p}/(n^2-|\mathbf{p}|^2)^{1/2}.\]

Hence, the momentum is real-valued and the Lagrangian is hyperregular,
provided $n^2-|\mathbf{p}|^2>0$. When $n^2=|\mathbf{p}|^2$, the ray
trajectory is vertical and has \textit{ grazing incidence} with the image
screen.

Defining $\sin \theta = d z / d s = \gamma$ leads to
$|\mathbf{p}| = n \cos \theta$, and gives the following geometrical
picture of the ray path. Along the optical axis (the $z$--axis)
each image plane normal to the axis is pierced at
a point $\mathbf{q} = (x, y)$ by a vector of magnitude $n(\mathbf{q}, z)$
tangent to the ray path. This vector makes an angle $\theta$ 
to the plane. The projection of this vector onto
the image plane is the canonical momentum $\mathbf{p}$.
This picture of the ray paths captures all but the
rays of grazing incidence to the image planes.
Such grazing rays are ignored in what follows.

\begin{figure} [h]

\centering 

\includegraphics[width=0.6\textwidth]{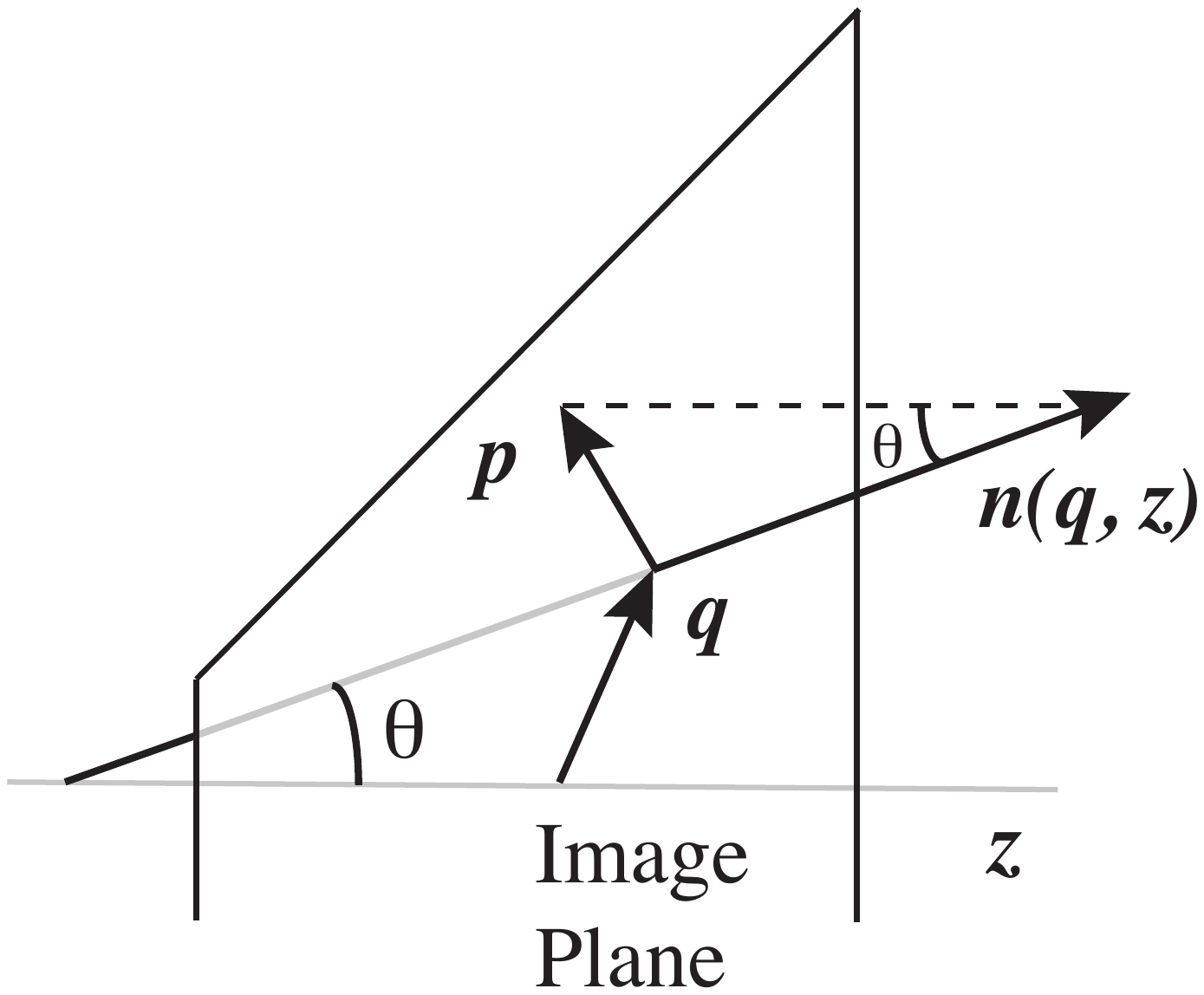}
%
 \caption{
 The canonical momentum $\mathbf{p}$ associated with the coordinate
 $\mathbf{q}$ on the image plane at $z$ has
 magnitude $|\mathbf{p}| = n(\mathbf{q}, z)\, {\sin}\, \theta$, where ${\cos}\, \theta = dz/ds$ is the direction cosine of the ray with respect to the optical $z$-axis.} 
 \label{fig:canonical-screen-optics-expand} 
 \end{figure} 

Passing now via the usual Legendre transformation from the Lagrangian to
the Hamiltonian description gives
\[ H =\mathbf{p}\cdot\mathbf{\dot{q}} - L =
n\gamma|\mathbf{\dot{q}}|^2 -n/\gamma
= - n \gamma = -\big[n(\mathbf{q},z)^2 - |\mathbf{p}|^2\big]^{1/2}
.\]
Thus, in the geometrical picture, the component
of the tangent vector of the ray-path along the optical
axis is (minus) the Hamiltonian, that is, $n(\mathbf{q},z)\sin \theta = - H$.

The phase space description of the geometric optics ray paths now follows from Hamilton's
equations,
\[
\mathbf{\dot{q}}
=\frac{\partial H}{\partial \mathbf{p}}
=\frac{-1}{H}\mathbf{p}
,\qquad
\mathbf{\dot{p}}
=
-\frac{\partial H}{\partial \mathbf{q}}
=
\frac{-1}{2H}\frac{\partial n^2}{\partial \mathbf{q}}
.
\]

\begin{remark}\rm [Translation invariant media]
If $n = n(\mathbf{q})$, so
that the medium is translation invariant along the optical axis, $z$, then
$H = - n\sin \theta$ is conserved. (Conservation of $H$ at an interface is
Snell's law.) For translation-invariant media, the vector ray-path equation
simplifies to
$$\mathbf{\ddot{q}}=-
\frac{1}{2H^2}
\frac{\partial n^2}{\partial \mathbf{q}}.$$
This is Newtonian dynamics for $\mathbf{q}\in\mathbb{R}^2$.

Thus, in this case geometrical ray tracing reduces
to ``Newtonian dynamics'' in $z$, with potential
$-n^2(\mathbf{q})$ and with ``time'' rescaled along each path
by the constant value of $\sqrt{2}H$ determined from the initial
conditions for each ray.
\end{remark}

\subsection{Axisymmetric, translation invariant optical materials}
In axisymmetric, translation invariant media, the index of refraction is
a function of the radius alone. Axisymmetry implies an additional constant
of motion and, hence, reduction of the Hamiltonian system for the light
rays to phase plane analysis. For such media, the index of refraction
satisfies
\[
n ( \mathbf{q} , z ) = n ( r )
,\qquad
r = |\mathbf{q}|
.
\]
Passing to polar coordinates $(r, \phi)$ with $\mathbf{q} = (x, y)
= r(\cos \phi, \sin \phi)$ leads in the usual way to
\[
|\mathbf{p}|^2
=
p_r^2
+  p_{\phi}^2/r^2
.
\]
Consequently, the optical Hamiltonian,
\[
H
= -\big[n(r)^2 - p_r^2 -  p_{\phi}^2/r^2\big]^{1/2}
\]
is independent of the azimuthal angle $\phi$; so its
canonically conjugate ``angular momentum'' $p_{\phi}$ is
conserved.

Using the relation $\mathbf{q}\cdot\mathbf{p} = rp_r$ leads to an
interpretation of $p_{\phi}$ in terms of the image-screen
phase space variables $\mathbf{p}$ and $\mathbf{q}$. Namely,
\[
|\mathbf{p}\times\mathbf{q}|^2
=
|\mathbf{p}|^2|\mathbf{q}|^2
-
(\mathbf{p}\cdot\mathbf{q})^2
=
p_{\phi}^2
\,.\]
The conserved quantity $p_{\phi} = \mathbf{p}\times\mathbf{q}
= yp_x - xp_y$ is called the skewness function, or the \emph{Petzval invariant}
for axisymmetric media. 

Vanishing of $p_{\phi}$ occurs for \textit{ meridional
rays}, for which $\mathbf{p}$ and $\mathbf{q}$ are collinear in the image
plane. On the other hand, $p_{\phi}$ takes its maximum value for \textit{
sagittal rays}, for which $\mathbf{p}\cdot\mathbf{q}=0$, so that
$\mathbf{p}$ and
$\mathbf{q}$ are orthogonal in the image plane. 

\begin{exercise}[Exercise: Axisymmetric, translation invariant materials]$\,$

Write Hamilton's canonical equations for axisymmetric, translation
invariant media. 

Solve these equations for the case of an optical fibre
with radially graded index of refraction in the following form:
\[
n^2(r) = \lambda^2 + (\mu-\nu r^2)^2
,\quad\lambda, \mu, \nu=\mathrm{constants,}
\]
by reducing the problem to phase plane analysis. 

How does the corresponding phase space
portrait differ between $p_{\phi}=0$ and $p_{\phi}\ne0$? 

Show that for
$p_{\phi}\ne0$ the problem reduces to a Duffing oscillator in a rotating
frame, up to a rescaling of time by the value of the Hamiltonian on each
ray ``orbit.''
\end{exercise}

\subsection{The Petzval invariant and its Poisson bracket relations}
The skewness function in the image plane
\[
S = p_{\phi} = \mathbf{p}\times\mathbf{q}\cdot  \mathbf{\hat{z}} = yp_x - xp_y
\]
generates rotations of phase space, of $\mathbf{q}$ and $\mathbf{p}$
jointly, each in its plane, around the optical axis $ \mathbf{\hat{z}}$.
Its square, $S^2$ (called the Petzval invariant) is conserved for ray
optics in axisymmetric media. That is, $\{S^2,H\}=0$ for optical
Hamiltonians of the form,
\[
H
= -\big[n(|\mathbf{q}|^2)^2 - |\mathbf{p}|^2\big]^{1/2}
.
\]
We define the axisymmetric invariant coordinates by the map
$T^*\mathbb{R}^2\mapsto\mathbb{R}^3$ $(\mathbf{q},\mathbf{p})\mapsto
(X_1,X_2,X_3)$, 
\begin{align}
X_1=|\mathbf{q}|^2\ge0
,\quad
X_2=|\mathbf{p}|^2\ge0
,\quad
X_3=\mathbf{p}\cdot\mathbf{q}
.
\label{Sp-momap}
\end{align}
The following Poisson bracket relations hold
\[
\{S^2,X_1\}=0
,\quad
\{S^2,X_2\}=0
,\quad
\{S^2,X_3\}=0
\,,
\]
since rotations preserve dot products.
In terms of these invariant coordinates,
the Petzval invariant and optical Hamiltonian satisfy
\begin{align}
S^2=X_1X_2-X_3^2\ge0
,\quad\hbox{and}\quad
H^2=n^2(X_1)-X_2\ge0
\,.
\label{S+H}
\end{align}
The level sets of $S^2$ are hyperboloids of revolution around 
the $X_1 = X_2$ axis, extending up through the interior of the $S = 0$
cone, and lying between the $X_1$-- and $X_2$--axes. The level sets of $H^2$
in \eqref{S+H} depend on the functional form of the index of refraction, but they are
$X_3$--independent.

\subsection{Hamilton's characteristic function for optics in 3D}
The tangents to Fermat's light rays in an isotropic medium are normal to Huygens' wave fronts. The phase of such a wave front is given by \cite{born2013principles}
\begin{eqnarray}
\phi = \int \mathbf{k}\cdot d \mathbf{r} - \omega(\mathbf{k},\mathbf{r}) \,dt
\,.
\label{phase-reln}
\end{eqnarray}
The Huygens wave front is a travelling wave, for which the phase $\phi$ is constant. Consequently, the phase shift 
\[
\int \mathbf{k}\cdot d \mathbf{r} = \int \frac{d \phi}{d \mathbf{r}}\cdot d \mathbf{r} = \int d\phi
= \int\omega(\mathbf{k},\mathbf{r}) \,dt
\]
along a ray trajectory for a travelling wave is given by the integral $ \int\omega(\mathbf{k},\mathbf{r}) \,dt$. 

Physically, the index of refraction $n(\mathbf{r})$ of the medium at position $\mathbf{r}$ enters the travelling wave phase speed $\omega/k$ as 
\begin{eqnarray*}
\frac{\omega}{k} = \frac{c}{n(\mathbf{r})} 
\,,\qquad
k = |\mathbf{k}|
\,,
\end{eqnarray*}
where $c$ is the speed of light in a vacuum and the index of refraction in a material medium always  satisfies $n(\mathbf{r})>1$. 

As it turns out, the frequency $ \omega $ of the travelling wave plays the role of the Hamiltonian and the wave vector $\mathbf{k}$ corresponds to the canonical momentum.  Consequently, we may write Hamilton's canonical equations for a wave front as 
 \begin{eqnarray}
\frac{d\mathbf{r}}{dt}
&=&\frac{\partial \omega}{\partial \mathbf{k}}
= \frac{c}{n(\mathbf{r})} \frac{\mathbf{k}}{k}
= \frac{c^2}{n^2\omega}\mathbf{k}
\,,\label{Ham-canon-position}
\\
\frac{d\mathbf{k}}{dt}
&=&
-\,\frac{\partial \omega}{\partial \mathbf{r}}
= \frac{ck}{2n^3}
\frac{\partial n^2}{\partial \mathbf{r} }
= \frac{\omega}{n}
\frac{\partial n}{\partial \mathbf{r} }
\,.
\label{Ham-canon-mom}
\end{eqnarray}
After a short manipulation, these canonical equations combine into
\begin{equation}
\frac{n^2}{c}\frac{d}{dt}\bigg(
\frac{n^2}{c}\frac{d\mathbf{r}}{dt}
\bigg)
=
\frac12\frac{\partial n^2}{\partial \mathbf{r} }
\,.
\label{eikonal-eqn3D-halfstep}
\end{equation} 
Equation~(\ref{eikonal-eqn3D-halfstep}) may also be expressed in terms of a \textit{different} variable time increment $cdt = n^2d\tau$ in the form of Newton's second law
\index{Newton's second law}
\begin{equation}
\frac{d^2\mathbf{r}}{d\tau^2}
=
\frac{1}{2}
\frac{\partial n^2}{\partial \mathbf{r} }
\hspace{1cm}\hbox{(Newton's $2^{nd}$ law)}\, .
\label{eikonal-eqn3D-Newton1}
\end{equation}
If instead of $\tau$ we define the variable time increment $cdt = nd \sigma$, then equation~(\ref{eikonal-eqn3D-halfstep}) yields the following \emph{eikonal equation} for the paths of light rays in geometric optics, $\mathbf{r}(\sigma)\in\mathbb{R}^3$, as
\index{eikonal equation}
\begin{equation}
\frac{d}{d\sigma}\bigg(
n(\mathbf{r})\frac{d\mathbf{r}}{d \sigma}
\bigg)
=
\frac{\partial n}{\partial \mathbf{r} }
\hspace{1cm}\hbox{(Eikonal equation)}\, .
\label{eikonal-eqn3D-HJ}
\end{equation}
This equation also follows from Fermat's principle of 
stationarity of the optical length under variations of the ray paths,
\begin{equation}
\delta \int_A^B 
n(\mathbf{r}(\sigma))\,d \sigma = 0
\hspace{1cm}\hbox{(Fermat's principle)}\, ,
\label{Fermat-sigma}
\end{equation}
with arc-length parameter $\sigma $, satisfying $d \sigma ^2=d\mathbf{r}(\sigma)\cdot d\mathbf{r}(\sigma)$ and, hence, $|d\mathbf{r}/d \sigma |=1$. 

From this vantage point, one sees that replacing $\mathbf{k}\to \frac{\omega}{c} \nabla {\mathbb S}$ in the first Hamilton equation in (\ref{Ham-canon-position}) yields 
\begin{equation}
n(\mathbf{r})\frac{d\mathbf{r}}{d \sigma}
=
\nabla {\mathbb S}(\mathbf{r})
\hspace{1cm}\hbox{(Huygens' equation)}\, ,
\label{wave-raypath-eqn1}
\end{equation}\index{Huygens!equation}%
from which the eikonal equation~(\ref{eikonal-eqn3D-HJ}) may be recovered by differentiating and using
\begin{equation}
d/d \sigma =n^{-1}\nabla{{\mathbb S}}\cdot\nabla 
\quad\hbox{and}\quad 
|\nabla {\mathbb S}|^2=n^2
\,.
\label{Xtrans-sigma}
\end{equation}
\begin{exercise}
Derive the  eikonal equation~(\ref{eikonal-eqn3D-HJ}) by differentiating Huygens' equation \eqref{wave-raypath-eqn1} and 
using the transformation relations in equation \eqref{Xtrans-sigma}.
\end{exercise}

\subsection{$\mathbb{R}^3$ Poisson bracket for ray optics}
The Poisson brackets among the axisymmetric
variables $X_1$, $X_2$ and $X_3$ in \eqref{Sp-momap} close among themselves,
\begin{align}
\{X_1,X_2\}=4X_3
,\quad 
\{ X_2 , X_3 \} = - 2 X_2 
,\quad 
\{ X_3 , X_1 \} = - 2 X_1 
.
\label{PBrelXYZ}
\end{align}
These Poisson brackets derive from a single $\mathbb{R}^3$  
Poisson bracket for $\mathbf{X}=(X_1,X_2,X_3)$ given by
\[
\{F,H\}=-\nabla{S^2}\cdot\nabla{F}\times\nabla{H}
\,.\]
Consequently, we may re-express the equations of
Hamiltonian ray optics in axisymmetric media
with $H = H(X_1, X_2 )$ as
\[
\mathbf{\dot{X}}=\nabla{S^2}\times\nabla{H}
.
\]
with Casimir $S^2$, for which $\{ S^2 , H \} = 0$, for every $H$.
Thus,  the flow preserves volume ($\div\mathbf{\dot{X}}=0$) and the
evolution takes place on intersections of level surfaces of the
axisymmetric media invariants $S^2$ and $H(X_1,X_2)$. 

\subsection{Recognition of the Lie--Poisson bracket for ray optics}

The Casimir invariant $S^2=X_1X_2-X_3^2$ is quadratic. In such cases, one may
write the $\mathbb{R}^3$ Poisson bracket in the suggestive form
\[
\{F,H\}
=
-\,C^k_{ij}X_k
\frac{\partial F}{\partial X_i}
\frac{\partial H}{\partial X_j}.
\]
In this particular case, $\smash{C^3_{12}=4}$, $\smash{C^2_{23}=2}$ and
$\smash{C^1_{31}=2}$ and
the rest either vanish, or are obtained from antisymmetry of
$\smash{C^k_{ij}}$
under exchange of any pair of its indices. These values are the structure
constants of any of the Lie algebras $\mathfrak{sp}(2,\mathbb{R})$, $\mathfrak{so}(2,1)$,
$\mathfrak{su}(1,1)$, or $\mathfrak{sl}(2,\mathbb{R})$. Thus, the reduced description of
Hamiltonian ray optics in terms of axisymmetric $\mathbb{R}^3$ variables
is said to be ``Lie--Poisson'' on the dual space of any of these Lie
algebras; say, $\mathfrak{sp}(2,\mathbb{R})^*$ for definiteness. We will have more to
say about Lie--Poisson brackets later, when we reach the Euler--Poincar\'e
reduction theorem. 

\begin{remark}\rm [Coadjoint orbits]
As one might expect, the coadjoint orbits of the symplectic group $SP(2,\mathbb{R})$ lie on the level sets of 
the Petzval invariant $S^2=X_1X_2-X_3^2$ in $\mathbb{R}^3$, in contrast to the level sets of spheres $X_1^2+X_2^2+X_3^2$
 for the orthogonal group $SO(3)$.

Level sets of the Petzval invariant $S^2=X_1X_2-X_3^2$ comprise hyperboloids of revolution $\mathsf{H}^2\otimes S^1$ around the $X_1 = X_2$ axis in the $\pi/4$ direction on the horizontal plane, $X_3 = 0$. Level sets of the Hamiltonian $H$ in \eqref{S+H} are independent of the vertical coordinate. The axisymmetric invariants $\mathbf{X}=(X_1,X_2,X_3)\in\mathbb{R}^3$ evolve along the intersections of these level sets by $\mathbf{\dot{X}}=\nabla{S^2}\times\nabla{H}$, as the vertical Hamiltonian knife $H=constant$ slices through the hyperbolic onion of level sets of the Petzval invariant, $S^2$. To visualise the hyperbolic onion in $\mathbb{R}^3$ notice that in the coordinates
  $
 Y_1=(X_1+X_2)/2
\,,
Y_2=(X_2-X_1)/2
\,,
Y_3=X_3
,$
one has $S^2=Y_1^2-Y_2^2-Y_3^2$. 
%
\begin{figure} [h!]

\centering 

\includegraphics[width=.6\textwidth]{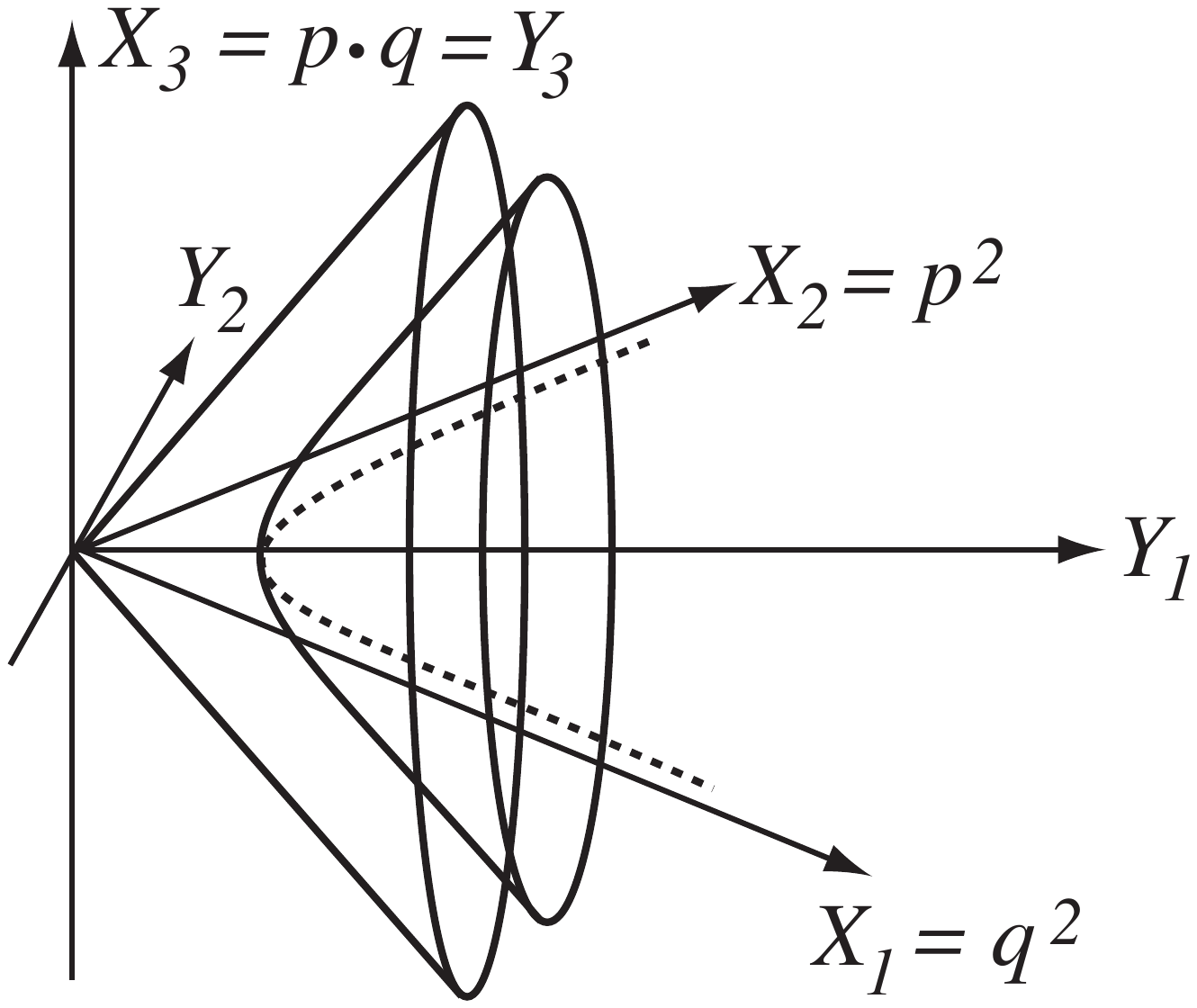} 
 \caption{The hyperbolic onion of level sets of the Petzval invariant $\mathsf{H}^2$ given in $\mathbb{R}^3$ coordinates as $S^2=X_1X_2-X_3^2=const$.} 
 \label{fig:hyperboliconion} 

 \end{figure} 

Being invariant under the flow (integral curves on $\mathsf{H}^2$) of the Hamiltonian vector field given by
\begin{align}
X_S = \{\cdot, S\}
= \frac{\partial S}{\partial p}\frac{\partial }{\partial q}
-
\frac{\partial S}{\partial q}\frac{\partial }{\partial p}
\,,\end{align}
each point on any layer $\mathsf{H}^2$ of the \emph{hyperbolic onion} \index{hyperbolic onion} $\mathsf{H}^2\otimes S^1$ in Figure \ref{fig:hyperboliconion} comprises an $S^1$ orbit in phase space under rotation by $p_\phi$, since the quantities $(X_1,X_2,X_3)$ are all invariant under $S^1$ rotations about the optical axis. In phase space, this orbit is a circular rotation by an angle $\phi$ about the optical axis of both $\mathbf{q}$ and $\mathbf{p}$ on an $\mathbb{R}^2$ image screen at position $z$.
\end{remark}

\begin{exercise}[Potential relation to the Hopf fibration]
The Petzval fibration $R^4\to \mathsf{H}^2\otimes S^1$ (locally) seems to be a hyperbolic 
analog of the spherical Hopf fibration: $S^3\to S^2\otimes S^1$ (locally). 
This observation may be worth further investigation.
\end{exercise}

\begin{definition}
\emph{Hamiltonian matrices} \index{Hamiltonian matrices} $m_i$ with $i=1,2,3$ 
each satisfy 
\begin{equation}
   Jm_i+m_i^TJ=0
   \,, \quad \hbox{where}\quad
J=\left( 
\begin{array}{cc} 
0 & -1 \\ 
1 & 0
\end{array} 
\right)
\,.
\label{symplectic-Lie-algebra}
\end{equation}   
That is, $Jm_i = (Jm_i)^T$ is a symmetric matrix.

The following traceless constant Hamiltonian matrices provide a representation the symplectic 
Lie algebra $sp(2,\mathbb{R})$
\begin{equation}
m_1= 
\left( 
\begin{array}{cc} 
0 & 0 \\ 
-2 & 0
\end{array} 
\right) 
,\ 
m_2= 
\left( 
\begin{array}{cc} 
0 & 2 \\ 
0 & 0
\end{array} 
\right) 
,\ 
m_3
=
\left( 
\begin{array}{cc} 
1 & 0 \\ 
0 & -1
\end{array} 
\right) 
,
\label{Ham-matrices}
\end{equation}
\end{definition}

\begin{exercise}
Show that the corresponding matrix commutation relations 
of the Hamiltonian matrices are
\[
[m_1,m_2] = 4m_3
\,,\quad
[m_2, m_3] = - \,2 m_2
\,,\quad
[m_3,m_1] = -\,2m_1
\,.\]
\end{exercise}

\begin{definition}
The map ${\cal J}:\, T^*\mathbb{R}^2\simeq\mathbb{R}^2\times\mathbb{R}^2\to sp(2,\mathbb{R})^*\,$ is defined by
\begin{eqnarray}
{\cal J}^\xi(\mathbf{z})
&:=&
\Big\langle 
{\cal J}(\mathbf{z}),\,\xi
\Big\rangle_{{sp}(2,\mathbb{R})^*\times {sp}(2,\mathbb{R})}
\nonumber \\&=&
\Big( \mathbf{z},\,J\xi\mathbf{z}\Big)_{\mathbb{R}^2\times \mathbb{R}^2}
\nonumber \\&:=&
z_k(J\xi)_{kl}z_l
\nonumber \\&=&
\mathbf{z}^T\cdot J\xi \mathbf{z}
\nonumber \\&=&
{\rm tr} \Big((\mathbf{z}\otimes \mathbf{z}^TJ) \xi \Big)\,,
\label{sp2-Poisson-map}
\end{eqnarray}
where $\xi\in {sp}(2,\mathbb{R})$ and $\mathbf{z}=(\mathbf{q},\,\mathbf{p})^T\in\mathbb{R}^2\times\mathbb{R}^2$.
\end{definition}
\begin{remark}\rm 
The map ${\cal J}(\mathbf{z})$ obtained in the last line of  \eqref{sp2-Poisson-map}
\begin{eqnarray}
{\cal J}(\mathbf{z}) = 
(\mathbf{z}\otimes \mathbf{z}^TJ)
\in sp(2,\mathbb{R})^*
\label{sp2-mommap}
\end{eqnarray}
sends $\mathbf{z}=(\mathbf{q},\,\mathbf{p})^T\in\mathbb{R}^2\times\mathbb{R}^2$ to ${\cal J}(\mathbf{z})=(\mathbf{z}\otimes \mathbf{z}^TJ)$, which is an element of $\mathfrak{sp}(2,\mathbb{R})^*$, the dual space to $\mathfrak{sp}(2,\mathbb{R})$. Under the pairing $\langle\,\cdot\,,\,\cdot\,\rangle:\,\mathfrak{sp}(2,\mathbb{R})^*\times \mathfrak{sp}(2,\mathbb{R})\to \mathbb{R}$ given by the trace of the matrix product, one finds the Hamiltonian, or phase-space function,
\begin{eqnarray}
\Big\langle {\cal J}(\mathbf{z}),\,\xi\Big\rangle
=
{\rm tr}\,\big( {\cal J}(\mathbf{z})\, \xi \big)
\,,
\label{mommap-formula}
\end{eqnarray}
in which $\xi\in \mathfrak{sp}(2,\mathbb{R})$ and ${\cal J}(\mathbf{z}) = (\mathbf{z}\otimes \mathbf{z}^TJ)
 \in \mathfrak{sp}(2,\mathbb{R})^*$ is a cotangent lift \emph{momentum map}. \index{momentum map!cotangent lift}
  \index{cotangent lift!momentum map}
\end{remark}

\begin{remark}\rm [Map to axisymmetric invariant variables]

The map 
\[{\cal J}:\, T^*\mathbb{R}^2\simeq\mathbb{R}^2\times\mathbb{R}^2\to sp(2,\mathbb{R})^*\,
\] 
in \eqref{sp2-Poisson-map} for $Sp(2,\mathbb{R})$ acting diagonally on $\mathbb{R}^2\times\mathbb{R}^2$ in equation \eqref{sp2-mommap} may be expressed in matrix form as
\begin{eqnarray}
{\cal J} 
&=& (\mathbf{z}\otimes \mathbf{z}^TJ)
\nonumber\\
&=& 2
\left(
\begin{array}{cc}
\mathbf{p}\cdot\mathbf{q} & -\,|\mathbf{q}|^2 \\
|\mathbf{p}|^2 & -\,\mathbf{p}\cdot\mathbf{q}
\end{array}
\right)
\nonumber\\
&=&
2
\left(
\begin{array}{cc}
X_3 & -\,X_1 \\
X_2 & -\,X_3
\end{array}
\right)
\,.
\label{mommap-matriX}
\end{eqnarray}
This is none other than the matrix form of the map
\eqref{Sp-momap} to axisymmetric invariant variables, 
\[
T^*\mathbb{R}^2\to\mathbb{R}^3:\,
(\mathbf{q},\mathbf{p})^T
\to
\mathbf{X}=(X_1,X_2,X_3)
\,,
\]
defined as 
\begin{equation}
X_1=|\mathbf{q}|^2\ge0
\,,\quad
X_2=|\mathbf{p}|^2\ge0
\,,\quad
X_3=\mathbf{p}\cdot\mathbf{q}
\,.
\label{axisymmetric-vars}
\end{equation}
Applying the momentum map ${\cal J}$ to the vector of Hamiltonian matrices $\mathbf{m}=(m_1,\,m_2,\,m_3)$ in equation \eqref{Ham-matrices} yields the individual components
\begin{eqnarray}
{\cal J}\cdot \mathbf{m}
=
4\mathbf{X}
\quad\Longleftrightarrow\quad
\mathbf{X}
=
\frac{1}{4}\,z_k(J\mathbf{m})_{kl}z_l
\,.
\label{axi-decomp}
\end{eqnarray}
Thus, the momentum map ${\cal J}:\, T^*\mathbb{R}^2\simeq\mathbb{R}^2\times\mathbb{R}^2\to sp(2,\mathbb{R})^*\,$ recovers the components of the vector $\mathbf{X}=(X_1,\,X_2,\,X_3)$ at any point on a level set of the Petzval invariant $S^2=X_1X_2-X_3^2$. 

\end{remark}

\begin{exercise}
Consider the $\mathbb{R}^3$ Poisson bracket
\begin{equation}
\{f, h\} = -\nabla{c}\cdot\nabla{f}\times\nabla{h}
\end{equation}
Let  $c=\mathbf{x}^T\cdot\mathbb{C}\mathbf{x}$ be a quadratic form on
$\mathbb{R}^3$, and let $\mathbb{C}$ be the associated symmetric
$3\times3$ matrix. Show that this is the Lie--Poisson bracket for the Lie
algebra structure
\[
[\mathbf{u},\mathbf{v}]_{\mathbb{C}}
=
\mathbb{C}(\mathbf{u}\times\mathbf{v})
\]
What is the underlying matrix Lie algebra? What are the coadjoint orbits
of this Lie algebra? What are its structure constants? What quadratic form 
provides the pairing of this Lie algebra with its dual Lie algebra? 
\end{exercise}

For more discussion of Lagrangian and Hamiltonian methods for geometric ray optics,
see \cite{holm2011geometricI,holm2011geometricII}.

\newpage
\vspace{4mm}\centerline{\textcolor{shadecolor}{\rule[0mm]{6.75in}{-2mm}}\vspace{-4mm}}
\section{Introduction to vector fields}

\secttoc

\textbf{What is this lecture about?} This lecture defines vector fields as tangents to the curves obtained from actions of 
smooth invertible maps on smooth manifolds; it then explains how this definition determines the properties of the vector fields.

\begin{definition}[Vector fields] \index{vector field!definition} 
A \emph{vector field} $X$ on a manifold $M$ is a map $X:M\to TM$ that assigns a vector $X(q)$ at every point $q\in M$. 
The real vector space of vector fields on a manifold $M$ is denoted by $\mathfrak{X}(M)$. 
\end{definition}

\begin{definition}
A \emph{time-dependent vector field} \index{time-dependent vector field} is a map 
\[
X:M\times \mathbb{R}\to TM
\]
such that $X(q,t)\in T_qM$ for each $q\in M$ and $t\in\mathbb{R}$. 
\end{definition}

\begin{definition}
[Integral curves] 
An \emph{integral curve} \index{integral curve} of vector field $X(q)$ with initial condition $q_0$ is a differentiable map $q:\,]t_1,t_2[\to M$ such that the open interval $]t_1,t_2[$ contains the initial time $t=0$, at which $q(0)=q_0$ and the tangent vector coincides with the vector field
\[
\dot{q}= X(q(t))
\]
for all $t\in\,]t_1,t_2[$.
\end{definition}

\begin{remark}\rm 
In what follows we shall always assume we are dealing with vector fields that
satisfy the conditions required for their integral curves to exist and be  
unique.
\end{remark}

\begin{definition}[Vector field basis]
\index{vector field!basis} \index{basis!vector} \index{dual basis}\label{basis!dual}
The components of a vector field $\dot{q}$ are defined by its directional derivatives in the chosen coordinate basis.
\end{definition}
\begin{example}\rm [Vector field basis]
An example of a vector basis for the components of a vector field is given by
\begin{equation}
 \dot{q} =  \dot{q}^a\,\frac{\pa}{\pa q^a}
 \quad\hbox{\bfi(Vector basis)}.
\end{equation}
That is, in the vector basis defined by the coordinate derivatives ${\pa}/{\pa q^a}$, 
the vector field $\dot{q}$ has components $\dot{q}^a$, $a=1,\dots,K$.
\end{example}

\subsection{Motions, pull-backs, push-forwards \&  commutators}
\begin{itemize}

\item
A \emph{motion} is defined as a smooth curve $q(t)\in M$ parameterised by $t\in\mathbb{R}$ that solves the \emph{motion equation}, which is a system of differential equations
\begin{equation}
\dot{q}(t) = \frac{dq}{dt} = f(q) \in TM
\,,
\label{motion-eqn}
\end{equation}
or in components
\begin{equation}
\dot{q}^i(t) = \frac{dq^i}{dt} = f^i(q)
\quad i=1,2,\dots,n\,.
\label{motion-eqn-compon}
\end{equation}

\item
The map $f: q\in M\to f(q)\in T_qM$ is a \emph{vector field}. 

According to standard theorems about differential equations that are not proven in this course, the solution, or integral curve, $q(t)$ exists, provided $f$ is sufficiently smooth, which will always be assumed to hold. 

\item
Vector fields can also be defined as \emph{differential operators} that act on functions, as 
\begin{equation}
\frac{d}{dt} G(q) = \dot{q}^i(t)\frac{\partial G}{ \partial q^i} = f^i(q)\frac{\partial G}{ \partial q^i}
\quad i=1,2,\dots,n,
\quad\hbox{(sum on repeated indices)}
\label{vector-fieldQ-time}
\end{equation}
for any smooth function $G(q): M\to \mathbb{R}$.

\item
To indicate the dependence of the solution of its initial condition $q(0)=q_0$, we write the motion as a smooth transformation
\[
q(t) = \phi_t(q_0)
\,.
\]
Because the vector field $f$ is independent of time $t$, for any fixed value of $t$ we may regard 
$\phi_t$ as  mapping from $M$ into itself that satisfies the \emph{composition law} \index{composition law}
\[
\phi_t\circ\phi_s=\phi_{t+s}
\]
and 
\[
\phi_0 = {\rm Id}
\,.
\]
Setting $s=-\,t$ shows that $\phi_t$ has a smooth inverse. A smooth mapping that has a smooth inverse is called a \emph{diffeomorphism}. \index{diffeomorphism} Geometric mechanics deals with diffeomorphisms. 

\item
The smooth mapping $\phi_t:\mathbb{R}\times M\to M$ that determines the solution $\phi_t\circ q_0=q(t)\in M$ of the motion equation (\ref{motion-eqn}) with initial condition $q(0)=q_0$ is called the \emph{flow} \index{flow of vector field!definition} of the vector field $Q$. 

A point $q^\star\in M$ at which $f(q^\star)=0$ is called a \emph{fixed point} of the flow $\phi_t$, or an \emph{equilibrium}.

Vice versa, the vector field $f$ is called the \emph{infinitesimal transformation} \index{infinitesimal transformation} 
of the mapping $\phi_t$, since
\[
\frac{d}{dt}\bigg|_{t=0}(\phi_t\circ q_0) = f(q)
\,.
\]
That is, $f(q)$ is the \emph{linearisation} of the flow map $\phi_t$ at the point $q\in M$.

More generally, the \emph{directional derivative} of the function $h$ along the vector field $f$ is given by the action of a differential operator, as
\[
\frac{d}{dt}\bigg|_{t=0}h\circ\phi_t  
= \left[\frac{\partial h}{ \partial \phi_t}
\frac{d}{dt}(\phi_t\circ q_0)
\right]_{t=0}
= \frac{\partial h}{ \partial q^i}\dot{q}^i
= \frac{\partial h}{ \partial q^i}f^i(q)
=: Qh
\,.\]

\item
Under a smooth change of variables $q=c(r)$ the vector field $Q$ in the expression $Qh$   transforms as 
\begin{equation}
Q = f^i(q)\frac{\partial}{ \partial q^i}
\quad\mapsto\quad
R = g^j(r) \frac{\partial}{ \partial r^j}
\label{vector-fieldR1}
\end{equation}
with
\begin{equation}
g^j(r) \frac{\partial c^i}{\partial r^j} = f^i(c(r))
\quad\hbox{or}\quad
g = c_r^{-1}f\circ c
\,,
\label{vector-fieldR2}
\end{equation}
where $c_r$ is the \emph{Jacobian matrix} \index{Jacobian matrix}  of the transformation. 
That is, since $h(q)$ is a function of $q$,
\[(Qh)\circ c= R(h\circ c)\,.\] 

We express the transformation between the vector fields as $R=c^*Q$ and write this relation as
\begin{equation}
(Qh)\circ c =: c^*Q (h\circ c) 
\,.
\label{vector-field-pull-back}
\end{equation}
The expression $c^*Q $ is called the \emph{pull-back} \index{pull-back} of the vector field $Q$ by the map $c$. Two vector fields are equivalent under a map $c$, if one is the pull-back of the other, and fixed points are mapped into fixed points.

The inverse of the pull-back is called the \emph{push-forward}. \index{push-forward} Namely, the push-forward by the map $c$ is the pull-back by the inverse map, $c^{-1}$. 

\item
The \emph{commutator} \index{commutator!vector fields} 
\[
QR-RQ =: \big[Q,\,R\big] 
\]
of two vector fields $Q$ and $R$ defines another vector field. Indeed, if 
\[
Q = f^i(q)\frac{\partial}{ \partial q^i}
\quad\hbox{and}\quad
R = g^j(q)\frac{\partial}{ \partial q^j}
\,,\]
then
\[
\big[Q,\,R\big] =  \left(f^i(q)\frac{\partial g^j(q)}{ \partial q^i} - 
g^i(q)\frac{\partial  f^j(q)}{ \partial q^i}\right)\frac{\partial}{ \partial q^j}
\,,\]
because the second-order derivative terms cancel.  By the pull-back relation (\ref{vector-field-pull-back}) we have
\begin{equation}
c^*\big[Q,\,R\big] = \big[c^*Q,\,c^*R\big]
\label{commutator-pull-back}
\end{equation}
under a change of variables defined by a smooth map, $c$.
This means the definition of the vector field commutator is independent of the choice of coordinates. As we shall see in Corollary \ref{JacobiID}, the \emph{tangent} to the relation $c_t^*\big[Q,\,R\big] = \big[c_t^*Q,\,c_t^*R\big]$ at the identity $t=0$ is the \emph{Jacobi condition} \index{Jacobi condition} for the vector fields to form an algebra.

\item
Pullbacks of vector fields lead to Lie derivative expressions. 
\begin{definition}[Lie derivative of a vector field]
The \emph{Lie derivative} \index{Lie derivative} of a vector field $Y\in\mathfrak{X}$ 
by another vector field $X\in\mathfrak{X}$ is defined by linearising the flow $\phi_t$ 
of $X$ around the identity $t=0$,
\begin{eqnarray*}
\pounds_XY = \frac{d}{dt}\bigg|_{t=0}\phi^*_tY
\quad\hbox{maps}\quad
\pounds_X\in\mathfrak{X} \mapsto  \mathfrak{X}
\,.
\end{eqnarray*}
\end{definition}
\begin{theorem}
The Lie derivative $\pounds_XY$ of a vector field $Y$ by a vector field $X$ satisfies
\begin{equation}
\pounds_XY = \frac{d}{dt}\bigg|_{t=0}\phi^*_tY
= [X,\,Y]
\,,
\label{LieBrkt}
\end{equation}
where $[X,\,Y]=XY-YX$ is the commutator of the vector fields $X$ and $Y$.
\end{theorem}

\begin{proof}
Denote the vector fields in components as
\begin{eqnarray*}
X = X^i(q)\frac{\partial}{ \partial q^i} = \frac{d}{dt}\bigg|_{t=0}\phi^*_t
\quad\hbox{and}\quad
Y = Y^j(q)\frac{\partial}{ \partial q^j}
\,.
\end{eqnarray*}
Then, by the pull-back relation (\ref{vector-field-pull-back}) a direct computation yields, on using the matrix identity $dM^{-1}=-\,M^{-1}dM M^{-1}$,
\begin{align*}
\pounds_XY 
&= \frac{d}{dt}\bigg|_{t=0}\phi^*_tY
= \frac{d}{dt}\bigg|_{t=0} \left(Y^j(\phi_tq)\frac{\partial}{ \partial (\phi_tq)^j}\right)
\\
&= \frac{d}{dt}\bigg|_{t=0} \left(Y^j(\phi_tq)
\left[\frac{ \partial (\phi_tq)}{\partial q}^{-1}\right]^k_j
\frac{\partial}{ \partial q^k}\right)
\\
&= 
\left(
X^j \frac{ \partial Y^k}{\partial q^j}
-
Y^j \frac{ \partial X^k}{\partial q^j}
\right)\frac{\partial}{ \partial q^k}
\\
&= 
[X,\,Y]
\,.
\end{align*}
\end{proof}

\begin{corollary}\label{JacobiID}
The Lie derivative of the relation (\ref{commutator-pull-back}) for the pull-back of the commutator 
$c_t^*\big[Y,\,Z\big] = \big[c_t^*Y,\,c_t^*Z\big]$ yields the \emph{Jacobi condition} for the vector fields to form an algebra.
\end{corollary}

\begin{proof}
By the product rule and the definition of the Lie bracket (\ref{LieBrkt}) we have
\begin{align*}
\frac{d}{dt}\bigg|_{t=0}\phi^*_t\big[Y,\,Z\big]
=
\big[X,\big[Y,\,Z\big]\big]
=
\big[ [X,Y],\,Z\big] + \big[ Y,\,[X,Z]\big]
=
\frac{d}{dt}\bigg|_{t=0}\big[\phi^*_tY,\,\phi^*_t Z\big]
\end{align*}
This is the \emph{Jacobi identity} for vector fields.
\end{proof}

\end{itemize}

\subsection{Lie algebras of vector fields}

\begin{definition}[The $\ad$--operation]
For $A \in \mathfrak{g}$ we define the operator $\ad_A$ to be the operator
$\ad{:\ } \mathfrak{g}\times\mathfrak{g}\to\mathfrak{g}$ that maps $B \in
\mathfrak{g}$ to $[A,B]$. We write $\ad_A B = [A,B]$.
\end{definition} 

\begin{definition} A \emph{representation} of a Lie algebra $\mathfrak{g}$ 
\index{representation!Lie algebra!faithful} 
\index{Lie algebra!representation!adjoint}
on a vector space $V$ is a mapping $\rho$ from $\mathfrak{g}$ to the
linear transformations of $V$ such that for $A,B \in \mathfrak{g}$ and any
constant scalar $c$,
\begin{enumerate}
\item[(i)] $\rho(A + cB) =\rho(A) + c\rho(B)$,
\item[(ii)] $\rho([A,B]) = \rho(A)\rho(B) - \rho(B)\rho(A)$.
\end{enumerate}
If the map $\rho$ is one-to-one, the representation is said to \emph{faithful}.
\end{definition}

\begin{exercise}
For a Lie algebra $\mathfrak{g}$, show that the map $A \to (\ad A)$ is a
representation of the Lie algebra $\mathfrak{g}$, with $\mathfrak{g}$
itself the vector space of the representation. This is called the
\emph{adjoint representation}. \index{adjoint!representation}
\end{exercise}

\begin{example}\rm [Vector field representations of Lie algebras] 
The Jacobi--Lie bracket of the vector fields $\xi$ and $\eta$ in
\ref{JLbrkt-def} may be represented in coordinate
charts as
\[
\eta=\frac{dx}{ds}\Big|_{s=0}=v(x)
,\quad\hbox{and}\quad
\xi=\frac{dx}{dt}\Big|_{t=0}=u(x)
.
\]
The Jacobi--Lie bracket of these two vector fields yields a third vector
field,
\begin{align*}
\xi\eta-\eta\xi &=
\frac{d\eta}{dt}\Big|_{t=0}-\frac{d\xi}{ds}\Big|_{s=0} \\
&= \frac{dv}{dx}\frac{dx}{dt}\Big|_{t=0} -\frac{du}{dx}\frac{dx}{ds}\Big|_{s=0}
= \frac{dv}{dx}\cdot u -\frac{du}{dx}\cdot v
= u\cdot\nabla v - v \cdot\nabla u.
\end{align*}
Thus, the Jacobi--Lie bracket of vector fields at the tangent space of the
identity $T_eG$ is closed and may be represented in coordinate charts by
the Lie bracket (commutator of vector fields)
\[
[\xi,\eta] :=\xi\eta-\eta\xi = u\cdot\nabla v - v \cdot\nabla u =: [u,v].
\]
\end{example}

This example also proves the following 
\begin{proposition}
Let $\mathfrak{X}(\mathbb{R}^n)$ be the set of vector fields defined on
$\mathbb{R}^n$. A Lie algebra $\mathfrak{g}$ may be
represented on coordinate charts by vector fields 
$\smash{X_{\xi}=X_{\xi}^i\frac{\partial}{\partial x^i}}\in
\mathfrak{X}(\mathbb{R}^n)$ for each element $\xi\in\mathfrak{g}$. This
vector field representation satisfies
\[
X_{[\xi,\eta]}=[X_{\xi},X_{\eta}]
\]
where $[\xi,\eta]\in\mathfrak{g}$ is the Lie algebra product and
$[X_{\xi},X_{\eta}]$ is the vector field commutator.
\end{proposition}
\bigskip

\begin{exercise}
\textbf{Integral curves of vector fields on the real line.}
Calculate the integral curves and identify the group action generated by the two vector fields
on the real line, $v_1=\p_x$ and $v_2= x\p_x$. Find the matrix Lie group isomorphic to the Lie group 
generated by the integral curves of vector fields $v_1$ and $v_2$. 
\end{exercise}
\begin{answer}$\,$\\
The commutator relation for $v_1$ and $v_2$ is
\[
\big[v_1,v_2\big] = \big[\p_x\,,\,x\p_x\big] = \p_x = v_1
\,.
\]
The integral curves for $v_1$ and $v_2$ are computed from 
\[
\frac{dx}{d \ep_i} = X_i(x)\quad i=1,2,
\quad \hbox{with} \quad X_1=1 \quad \hbox{and} \quad X_2 = x.
\]
by integrating to find translations in $\ep_1$
\[
\int d\ep_1 = \ep_1 =  \int \frac{dx}{1} = x_{\ep_1} - x_0 
\quad\Longrightarrow \quad 
x_{\ep_1}  = g_{\ep_1}x_0 = x_0 + \ep_1
\]
and scaling transformations in $\ep_2$,
\[
\int d\ep_2 = \ep_2 =  \int \frac{dx}{x} = \log({x_{\ep_2}/x_0})
\quad\Longrightarrow \quad 
x_{\ep_2} = g_{\ep_2}x_0 = x_0 e^{\ep_2}
\,.\]
Combining these transformations produces the affine group action
\[
g_{\ep_1}g_{\ep_2}x_0 = e^{\ep_2}x_0 + \ep_1
\,.\]
This group action can be obtained as a left action 
of the upper triangular $2\times 2$ matrices on the column vector $[x_0 \,, 1]^T$,
\[
\begin{bmatrix}
e^{\ep_2} & \ep_1 \\
0 & 1
\end{bmatrix}
\begin{bmatrix}
x_0 \\ 1
\end{bmatrix}
=
\begin{bmatrix}
e^{\ep_2}x_0 + \ep_1
\\
1
\end{bmatrix}
\,.\]
It can also be written as a right action of lower triangular $2\times 2$ matrices 
on the row vector $[x_0\,, 1]$,
\[
\begin{bmatrix}
x_0 & 1
\end{bmatrix}
\begin{bmatrix}
e^{\ep_2} & 0 \\
 \ep_1 & 1
\end{bmatrix}
=
\begin{bmatrix}
e^{\ep_2}x_0 + \ep_1 & 1 
\end{bmatrix}
\,.\]

\end{answer} 

\begin{exercise}
\textbf{Integral curves of vector fields on the real line (continued).}
Calculate the integral curves and identify the group action generated by the three vector fields
on the real line, $v_1=\p_x$, $v_2= x\p_x$ and $v_3= -x^2\p_x$. Find the matrix Lie group isomorphic to the Lie group 
generated by the Integral curves of vector fields $v_1$, $v_2$ and $v_3$. 
\end{exercise}

\begin{answer} 
The commutator relation for $v_1$, $v_2$ and $v_3$ is given by
\begin{equation}
\hbox{
$
[v_i,\,v_j]
=
c_{ij}^k
v_k
=
$
\begin{tabular}{| c | c |  }	\hline	
$[\,\cdot\,,\,\cdot\,]$	& 	
$
\begin{array}{ccc}
\ v_1\quad &  v_2\quad & v_3
\end{array}
$
\\ \hline
$
\begin{array}{c}
v_1 \\  v_2 \\ v_3
\end{array}
$
&	
$
\begin{array}{ccc}
0   & v_1  & -2v_2
\\
-v_1 &  0  & -v_3
\\
2v_2 & v_3 &   0
\end{array}
$\\
\hline
\end{tabular}
}\,.
\label{sp2R-commutable}
\end{equation}

The transformations in $\ep_3$, is given by
\[
\int d\ep_3 = \ep_3 =  \int \frac{dx}{-x^2} = \frac{1}{x_{\ep_3}} - \frac{1}{x_0}
\quad\Longrightarrow \quad 
x_{\ep_3} = g_{\ep_3}x_0 = \frac{x_0}{\ep_3x_0 + 1}
\,.\]
Combining these transformations produces the projective group action
\[
g_{\ep_1}g_{\ep_2}g_{\ep_3}x_0 =  \frac{e^{\ep_2}x_0}{1+\ep_3x_0}  + \ep_1
=  \frac{ \big(e^{\ep_2} + \ep_1\ep_3\big)x_0 + \ep_1 }{\ep_3x_0 + 1} 
\,,\]
which can be written as a right action of $2\times 2$ matrices 
on the row vector $[x_0\,, 1]$,
\[
\begin{bmatrix}
x_0 & 1
\end{bmatrix}
\begin{bmatrix}
e^{\ep_2} +  \ep_1\ep_3 & \ep_3 \\
 \ep_1  & 1
\end{bmatrix}
=
\begin{bmatrix}
 \big(e^{\ep_2}+ \ep_1\ep_3\big)x_0 + \ep_1 &  \ep_3x_0 + 1
\end{bmatrix}
\simeq 
\begin{bmatrix}
\frac{ \big(e^{\ep_2} + \ep_1\ep_3\big)x_0 + \ep_1 } {\ep_3x_0 + 1} & 1
\end{bmatrix}
\,.\]
\end{answer} 


\newpage
\vspace{4mm}\centerline{\textcolor{shadecolor}{\rule[0mm]{6.75in}{-2mm}}\vspace{-4mm}}
\section{Matrix Lie groups and Lie algebras}

\secttoc

\textbf{What is this lecture about?} This lecture lays out the definitions and basic properties of matrix Lie groups and Lie algebras,  
as well as their actions on finite dimensional manifolds.

\subsection{Matrix Lie groups}

\begin{definition}
A \emph{group} \index{Lie group!definition} is a set of elements with:
\begin{enumerate}
\item A binary product (multiplication), $G \times G \to G$, such that
\begin{itemize}
\item the product of $g$ and $h$ is written $gh$, and
\item the product is associative: $(gh)k=g(hk)$.
\end{itemize}
\item An identity element $e$ such that $eg=g$ and $ge=g$, $\forall g\in G$
\item An inverse operation $G \to G$, such that $gg^{-1}=g^{-1}g=e$
\end{enumerate}
\end{definition}

\begin{definition}
A \emph{Lie group} is a smooth manifold $G$ which is a group
and for which the group operations of multiplication, 
$(g,h) \to gh$ for $g,h \in G$, and inversion, $g \to g^{-1}$ with
$gg^{-1}=g^{-1}g=e$, are smooth.
\end{definition}

\begin{definition}\label{MatrixLG-def}
A \emph{matrix Lie group} is a set of invertible $n\times n$ matrices
which is closed under matrix multiplication and which is a submanifold of
$\mathbb{R}^{n\times n}$.
The conditions showing that a matrix Lie group is a Lie group are easily
checked:
\begin{itemize}
\item A matrix Lie group is a manifold, because it is a submanifold of
$\mathbb{R}^{n\times n}$
\item Its group operations are smooth, since they are algebraic operations on
the matrix entries.
\end{itemize}
\end{definition}

\begin{example}\rm [The general linear group $GL(n,\mathbb{R})$]
The matrix Lie group $GL(n,\mathbb{R})$ is the group of linear
isomorphisms of $\mathbb{R}^n$ to itself.  The dimension of the matrices
in $GL(n,\mathbb{R})$ is $n^2$.
\end{example}

\begin{proposition}\label{Smap-prop}
Let $K \in GL(n,\mathbb{R})$ be a symmetric matrix,
$K^T=K$. Then the subgroup $S$ of $GL(n,\mathbb{R})$ defined by the
mapping 
\[S = \{U \in GL(n,\mathbb{R})|U^TKU = K\}\]
is a submanifold of $\mathbb{R}^{n\times n}$ of dimension 
$n(n-1)/2$.
\end{proposition}

\begin{remark}\rm 
The subgroup $S$ leaves invariant a certain symmetric quadratic form
under linear transformations,
$S\times\mathbb{R}^n\to\mathbb{R}^n$ given by
$\mathbf{x}\to U\mathbf{x}$, since
\[
\mathbf{x}^T K\mathbf{x}
=
\mathbf{x}^T U^TKU\mathbf{x}
.
\]
So the matrices $U \in S$ change the basis for this
quadratic form, but they leave its value unchanged. Thus, $S$ is the
\emph{isotropy subgroup} \index{isotropy subgroup!quadratic form} of the quadratic form associated with $K$. 
\end{remark} 

\begin{proof}
\begin{itemize}
\item Is $S$ a subgroup? We check the following three defining properties
\begin{enumerate}
\item
Identity: $I\in S$ because $I^TKI=K$.
\item
Inverse:
$U\in S\Longrightarrow U^{-1}\in S$, because
$$K=U^{-T}(U^TKU)U^{-1}=U^{-T}(K)U^{-1}.$$
\item
Closed under multiplication:
$U, V\in S\Longrightarrow UV\in S$, because
$$(UV)^TKUV=V^T(U^TKU)V=V^T(K)V=K.$$
\end{enumerate}
\item
Hence, $S$ is a subgroup of $GL(n,\mathbb{R})$. 

\item
Is $S$ is a submanifold of $\mathbb{R}^{n\times n}$  of dimension 
$n(n-1)/2$? 
\begin{itemize}
\item
Indeed, $S$ is the zero locus of the mapping $UKU^T - K$. This makes
it a submanifold, because it turns out to be a submersion. 
\item
For a submersion, the dimension of the level set is the dimension of the
domain minus the dimension of the range space. In this case, this
dimension is $n^2-n(n+1)/2=n(n-1)/2$.  $\Box$
\end{itemize}
\end{itemize}
\end{proof}

\begin{exercise}
Explain why one can conclude that the zero locus map for $S$ is a
submersion. In particular, pay close attention to establishing the
constant rank condition for the linearization of this map.
\end{exercise}

\begin{answer}
Here is why $S$ is a submanifold of $R^{n\times n}$.

First, $S$ is the zero locus of the mapping 
\[
U\to U^TKU - K
\qquad\hbox{(locus map)}.
\] 
Let $U\in S$, and let $\delta{U}$ be an arbitrary element of $R^{n\times
n}$. Then linearize to find
\[
(U + \delta{U})^TK(U + \delta{U}) - K =
U^TKU - K + \delta{U}^TKU + U^TK\delta{U} + O(\delta{U})^2.
\]
We may conclude that $S$ is a submanifold of $R^{n\times n}$ if we can
show that the linearization of the locus map, namely the linear mapping 
defined by
\[
L\equiv \delta{U}\to \delta{U}^TKU + U^TK\delta{U}
,\qquad R^{n\times n}\to R^{n\times n}
\] 
has constant rank for all $U\in S$.
\end{answer}

\begin{lemma} The linearization map $L$ is onto the space of ${n\times n}$
of symmetric matrices and hence the original map is a submersion.
\end{lemma}

\begin{proof}[Proof that $L$ is onto]$\,$

\begin{itemize}
\item
Both the original locus map and the image of $L$ lie in the subspace
of ${n\times n}$ symmetric matrices. 
\item
Indeed, given $U$ and any symmetric
matrix
$S$ we can find $\delta{U}$ such that 
\[
 \delta{U}^TKU + U^TK\delta{U} = S
.
\] 
Namely 
\[
 \delta{U}=K^{-1}U^{-T}S/2
.
\] 
\item
Thus, the linearization map $L$ is onto the space of ${n\times n}$
of symmetric matrices and the original locus map $U\to UKU^T - K$ to the
space of symmetric matrices is a submersion.  $\Box$
\end{itemize}
\end{proof}

For a submersion $S$,
the dimension of the level set is the dimension of the domain minus the
dimension of the range space. Here, this dimension is $n^2 - n(n +
1)/2 = n(n - 1)/2$.

\begin{corollary}[$S$ is a matrix Lie group]
$S$ is both a subgroup and a submanifold of the general linear group
$GL(n,\mathbb{R})$. Thus, by \ref{MatrixLG-def}, $S$ is
a matrix Lie group.
\end{corollary}

\begin{exercise}
What is the tangent space to $S$ at the identity, $T_IS$? 
\end{exercise}

\begin{exercise}
Show that for any pair of matrices $A,B\in T_IS$, the matrix
commutator $[A,B]\equiv AB-BA \in T_IS$.
\end{exercise}

\begin{proposition} 
The linear space of matrices $A$ satisfying 
\[
A^TK+KA=0
\]
defines $T_IS$, the tangent space at the identity
of the matrix Lie group $S$ defined in \ref{Smap-prop}.
\end{proposition}

\begin{proof}
Near the identity the defining condition for $S$
expands to 
\[
(I+\epsilon A^T + O(\epsilon^2))K
(I+\epsilon A + O(\epsilon^2)) = K
,\quad\hbox{for}\quad \epsilon\ll1
.
\]
At linear order $O(\epsilon)$ one finds,
\[
A^TK + KA =0
.\]
This relation defines the linear space of matrices $A\in T_IS$.
\end{proof}

If $A,B\in T_IS$, does it follow that $[A,B]\in T_IS$?

Using $[A,B]^T=[B^T,A^T]$, we check \emph{closure} by a direct computation,
\begin{eqnarray*}
[B^T,A^T]K+K[A,B]&=&B^TA^TK-A^TB^TK+KAB-KBA \\
&=&B^TA^TK-A^TB^TK-A^TKB+B^TKA=0.
\end{eqnarray*}
Hence, the tangent space of $S$ at the identity $T_IS$ is closed under the
matrix commutator $[\,\cdot\,,\, \cdot\,]$. 

\begin{remark}\rm 
In a moment, we will show that the matrix commutator for $T_IS$ also
satisfies the Jacobi identity. This will imply that the condition $A^TK +
KA =0$ defines a matrix Lie algebra.
\end{remark}

\subsection{Defining matrix Lie algebras}
We are ready to prove the following proposition, in preparation for defining matrix
Lie algebras. 

\begin{proposition} \label{matrixLG-comprop}
Let $S$ be a matrix Lie group, and let $A,B \in T_IS$
(the tangent space to $S$ at the identity element). Then $AB - BA \in
T_IS$.
\end{proposition}
The proof makes use of a lemma.
\begin{lemma} Let $R$ be an arbitrary element of a matrix Lie group $S$,
and let $B \in T_IS$. Then $RBR^{-1}\in T_IS$.
\end{lemma}

\begin{proof}
Let $R_B(t)$ be a curve in $S$ such that $R_B(0) = I$ and $R^\prime(0)
= B$. Define $S(t) = RR_B(t)R^{-1}\in T_IS$ for all $t$. Then $S(0) = I$
 and $S^\prime(0)=RBR^{-1}$. Hence, $S^\prime(0) \in T_IS$, thereby proving
the lemma. 
\end{proof}

\begin{proof}[Proof of \ref{matrixLG-comprop}] Let $R_A(s)$ be a
curve in $S$ such that
$R_A(0) = I$ and $R_A^\prime(0) = A$. 
Define $S(t) = R_A(t)BR_A(t)^{-1}\in T_IS$.
Then the lemma implies  that $S(t)\in T_IS$ for every $t$.
Hence, $S^\prime(t)\in T_IS$, and in particular, $S^\prime(0) = AB - BA\in
T_IS$.
\end{proof}

\begin{definition}[Matrix commutator]
For any pair of $n \times n$ matrices $A,B$, the \emph{matrix
commutator} is defined as  $[A,B] = AB - BA$.
\end{definition} 

\begin{proposition}[Properties of the matrix commutator]
The matrix commutator has the following
two properties:
\begin{enumerate}
\item[(i)] Any two $n \times n$ matrices $A$ and $B$ satisfy
\[[B,A] = -[A,B].\] (This is
the property of skew-symmetry.)
\item[(ii)] Any three $n \times n$ matrices $A$, $B$ and $C$ satisfy
\[[[A,B],C] + [[B,C],A] + [[C,A],B] = 0.\]
(This is known as the \emph{Jacobi identity}.)
\end{enumerate}
\end{proposition}

\begin{definition}[Matrix Lie algebra]
A matrix Lie algebra $\mathfrak{g}$ is a set of $n \times n$ matrices
which is a vector space with respect to the usual operations of matrix
addition and multiplication by real numbers (scalars) and which is closed
under the matrix commutator $[\,\cdot\,,\, \cdot\,]$.
\end{definition}

\begin{proposition} For any matrix Lie group $S$, the tangent space at
the identity $T_IS$ is a matrix Lie algebra.
\end{proposition}
\begin{proof}
This follows by \ref{matrixLG-comprop} and because $T_IS$ is a vector space.
\end{proof}

\subsection{Examples of matrix Lie groups}

\begin{example}\rm [The Orthogonal Group $O(n)$]
The mapping condition $U^TKU=K$ in \ref{Smap-prop}
specializes for $K = I$ to $U^TU=I$, which defines the orthogonal group.
Thus, in this case, $S$ specializes to $O(n)$, the group of ${n\times n}$
orthogonal matrices. The orthogonal group is of special interest in the
dynamics of rotating rigid bodies.
\end{example}

\begin{corollary}[$O(n)$ is a matrix Lie group]
By \ref{Smap-prop} the orthogonal group $O(n)$ 
is both a subgroup and a submanifold of the general linear group
$GL(n,\mathbb{R})$. Thus, by \ref{MatrixLG-def}, the
orthogonal group $O(n)$ is a matrix Lie group.
\end{corollary}

\begin{example}\rm [The Special Linear Group $SL(n,\mathbb{R})$]
The subgroup of $GL(n,\mathbb{R})$ with $\det(U)=1$ is called
$SL(n,\mathbb{R})$. 
\end{example}

\begin{example}\rm [The Special Orthogonal Group $SO(n)$]
The special case of $S$ with $\det(U)=1$ and $K=I$ is called
$SO(n)$. In this case, the mapping condition $U^TKU=K$ specializes
to $U^TU=I$ with the extra condition $\det(U)=1$. 
\end{example}

\begin{example}\rm [The tangent space of $SO(n)$ at the identity
$T_ISO(n)$] The special case with $K=I$ of $T_ISO(n)$
yields, 
\[
A^T+A=0
.
\]
These are antisymmetric matrices. Lying in the tangent space at the
identity of a matrix Lie group, this linear vector space forms a matrix
Lie algebra. 
\end{example}

\begin{example}\rm [The Symplectic Group] Suppose $n = 2l$ (that is, let $n$ be 
even) and consider the nonsingular skew-symmetric matrix
\[
J=
\left[
\begin{matrix} 0 &I \\ -I&0 \end{matrix}
\right]
\]
where $I$ is the $l\times l$ identity matrix. One may verify that
\[Sp(l) = \{U \in GL(2l,\mathbb{R})|U^TJU = J\}\]
is a group. This is called the symplectic group. Reasoning as before, the matrix
algebra $T_ISp(l)$ is defined as the set of $n\times n$ matrices $A$ satisfying
$JA^T+AJ=0$. This algebra is denoted as $sp(l)$.
\end{example}

\begin{example}\rm [The Special Euclidean Group]
Consider the set of $4\times4$ matrices of the form
\[ E(R,v) = \left[ \begin{matrix} R &v \\ 0&1 \end{matrix} \right] \]
where $R\in SO(3)$ and $v\in\mathbb{R}^3$. This set of $4\times4$ matrices forms 
a \textit{representation} of the special Euclidean group in three dimensions, denoted $SE(3)$. 
\index{semidirect product!broken symmetry} \index{broken symmetry!semidirect product}
The special Euclidean group $SE(3)$ is of
central interest in mechanics since it describes the set of rigid motions
and linear coordinate transformations of three-dimensional space. 
\end{example}

\begin{exercise}
A point $P$ in $\mathbb{R}^3$ undergoes a rigid motion associated with
$E(R_1, v_1)$ followed by a rigid motion associated with $E(R_2, v_2)$.
What matrix element of $SE(3)$ is associated with the composition of these
motions in the given order?
\end{exercise}

\begin{exercise}
Multiply the special Euclidean matrices of $SE(3)$. Investigate their
matrix commutators in their tangent space at the identity. (This is an
example of a semidirect product Lie group.)
\end{exercise}

\begin{exercise}[Tripos question]
When does a stone at the equator of the Earth weigh the most? Two hints:
(a) Assume the Earth's orbit is a circle around the Sun and ignore the
declination of the Earth's axis of rotation. (b) This is an exercise in
using $SE(2)$. 
\end{exercise}

\begin{exercise}
Suppose the $n\times n$ matrices $A$ and $M$ satisfy \[AM+MA^T = 0\,.\]
Show that $\exp(At)M \exp(A^T t) = M$ for all $t$. Hint: $A^nM=M(-A^T)^n$. This
direct calculation shows that for $A \in so(n)$ or $A \in sp(l)$, we have $\exp(At)\in
SO(n)$ or
$\exp(At) \in Sp(l)$, respectively.
\end{exercise}

\subsection{Lie group actions} The action of a Lie group $G$ on a
manifold $M$ is a group of transformations of $M$ associated to 
elements of the group $G$, whose composition acting on $M$ is corresponds
to group multiplication in $G$.

\begin{definition}[Left and Right actions] Let $M$ be a manifold and let $G$ be a Lie group. A
\emph{left action} of a Lie group \index{left action!Lie group} $G$ on $M$ is a smooth mapping 
$\Phi{:\ }  G\times M\to M$ such that 
\begin{enumerate}
\item[(i)] $\Phi(e, x) = x \hbox{ for all } x \in M$, 
\item[(ii)] $\Phi(g, \Phi(h, x)) = \Phi(gh, x)$ for all $g, h \in G$ and $x \in
M$, and 
\item[(iii)] $\Phi(g, \cdot)$ is a diffeomorphism on $M$ 
for each $g \in G$. 
\end{enumerate}
We often use the convenient notation $gx$ for $\Phi(g, x)$ and
think of the group element $g$ acting on the point $x \in M$. The
associativity condition (ii) above then simply reads $(gh)x = g(hx)$.
\end{definition}
Similarly, one can define a \emph{right action}, \index{right action!Lie group}
which is a map $\Psi{:\ } 
M\times G\to M$  satisfying $\Psi(x,e) = x$ and $\Psi(\Psi(x,g),h) =
\Psi(x,gh)$. The convenient notation for right action is $xg$ for
$\Psi(x, g)$, the right action of a group element $g$ on the point
$x\in M$. Associativity $\Psi(\Psi(x,g),h) = \Psi(x,gh)$ is
then be expressed conveniently as $(xg)h=x(gh)$.

\begin{example}\rm [Properties of Lie group actions] \index{Lie group actions!properties}
The action $\Phi{:\ }  G\times M\to M$ of a group $G$ on a manifold $M$ is said
to be
\begin{enumerate} 
\item
\emph{Transitive}, if for every $x,y\in M$ there exists a $g\in G$, such
that $gx=y$;
\item
\emph{Free}, if it has no fixed points, that is, $\Phi_g(x)=x$ implies
$g=e$; and 
\item
\emph{Proper}, if whenever a convergent subsequence $\{x_n\}$ in $M$
exists, and the mapping $g_nx_n$ converges in $M$, then $\{g_n\}$ has
a convergent subsequence in $G$.
\end{enumerate}
\end{example}

\subsection*{Orbits} Given a group action of $G$ on $M$, for a
given point $x \in M$, the subset
\[{\cal O}(x)= \{gx ~|~ g \in G\}\subset M ,\] 
is called the \emph{group orbit} \index{group orbit} through $x$. In finite dimensions, it can
be shown that group orbits are always smooth (possibly immersed) manifolds.
Group orbits generalise the notion of orbits of a dynamical system. 

\begin{exercise}
The flow of a vector field on $M$ can be thought of as an action of
$\mathbb{R}$ on $M$. Show that in this case the general notion of group
orbit reduces to the familiar notion of orbit used in dynamical systems.
\end{exercise}

\begin{exercise}
Compute the group orbit of the action of unitary transformations $SU(2)$ on complex space
$\mathbb{C}^2$. For details of the formulation, see \cite{holm2011geometricI}.
\end{exercise}

\begin{theorem}
Orbits of proper group actions are embedded submanifolds.
\end{theorem}
This theorem is stated by Marsden and Ratiu~\cite[Chapter~9]{MaRa1994},
who refer in turn to Abraham and Marsden~\cite{AbMa1978} for the proof.

\begin{example}\rm [Orbits of $SO(3)$]\index{orbits of $SO(3)$}
A simple example of a group orbit is the action of $SO(3)$ on
$\mathbb{R}^3$ given by matrix multiplication: The action of $A\in SO(3)$
on a point $\mathbf{x}\in\mathbb{R}^3$ is simply the product
$A\mathbf{x}$. In this case, the orbit of the origin is a single point
(the origin itself), while the orbit of any other point is the sphere
through that point.
\end{example}

\begin{example}\rm [Orbits of a Lie group acting on itself]
The action of a group $G$ on itself from either the left, or the right,
also produces group orbits. This action sets the stage for
discussing the tangent lifted action of a Lie group on its tangent bundle.

Left and right translations on the group are denoted \index{right translations}
$L_g$ and $R_g$, respectively. For example, $L_g {:\ }  G \to G$ is the map
given by $h\to gh$, while $R_g {:\ }  G \to G$ is the map given
by $h\to hg$, for $g,h\in G$.
\begin{enumerate}
\item[(a)]
\emph{Left translation} \index{Left translation} $L_g{:\ } G\to G;\,h\to gh$ defines a transitive and
free action of $G$ on itself. Right multiplication 
$R_g{:\ } G\to G;\,h\to hg$ defines a right action,  while $h\to hg^{-1}$
defines a left action of $G$ on itself. Thus, a right action by the inverse is a left action. 
\item[(b)]
$G$ acts on $G$ by conjugation, $g\to I_g=R_{g^{-1}}\circ L_g$. The map
$I_g{:\ } G\to G$ given by $h\to ghg^{-1}$ is the \emph{inner automorphism}
\index{inner automorphism!conjugacy classes}
associated with $g$. Orbits of this action are called \emph{conjugacy classes}. 
\item[(c)]
Differentiating conjugation at $e$ gives the \emph{adjoint action} \index{adjoint action} 
of $G$ on $\mathfrak{g}$: 
\[
\Ad_g:=T_eI_g\,:\,T_eG=\mathfrak{g}\to T_eG=\mathfrak{g}.
\]
Explicitly, the \emph{adjoint action} of $G$ on $\mathfrak{g}$ is given by
\[
\Ad{:\ } G\times \mathfrak{g}\to \mathfrak{g}
,\quad
\Ad_g(\xi)=T_e(R_{g^{-1}}\circ L_g)\xi
\,.\]
We have already seen an example of adjoint action for matrix Lie groups
acting on matrix Lie algebras, when we defined 
$S(t) = R_A(t)BR_A(t)^{-1}\in T_IS$ as a key step in the proof of Proposition
\ref{matrixLG-comprop}.  
\item[(d)]
The \emph{co-Adjoint action} \index{coadjoint action}	
of $G$ on $\mathfrak{g}^*$, the dual of the Lie
algebra $\mathfrak{g}$ of $G$, is defined as follows. Let
$\Ad^*_g{:\ } \mathfrak{g}^*\to\mathfrak{g}^*$ be the dual of $\Ad_g$,
defined by 
\[
\langle\Ad^*_g\alpha,\xi\rangle
=
\langle\alpha,\Ad_g\xi\rangle
\]
for $\alpha\in \mathfrak{g}^*$, $\xi\in \mathfrak{g}$ and pairing
$\langle\cdot, \cdot\rangle{:\ } \mathfrak{g}^*\times\mathfrak{g}
\to\mathbb{R}$. Then the map
\[
\Phi^*{:\ } G\times \mathfrak{g}^*\to\mathfrak{g}^*
\quad\hbox{given by}\quad
(g,\alpha)\mapsto \Ad^*_{g^{-1}}\alpha
\]
is the co-Adjoint action of $G$ on $\mathfrak{g}^*$.
\end{enumerate}
\end{example}

\subsection{Examples: $SO(3)$, $SE(3)$, etc}

\subsubsection{A basis for the matrix Lie algebra $\mathfrak{so}(3)$ and a map to
$\mathbb{R}^3$} 
The Lie algebra of $SO(n)$ is called $\mathfrak{so}(n)$. A basis 
$(e_1,e_2,e_3)$ for $\mathfrak{so}(3)$ when
$n=3$ is given by 
\[
\mathbf{\hat{x}} =
\left[ \begin{matrix} 0 &-z&y \\ z&0&-x \\ -y&x&0 \end{matrix} \right]
= xe_1+ye_2+ze_3
\]
\begin{exercise}
Show that $[e_1,e_2]=e_3$ and cyclic permutations, while all other
matrix commutators among the basis elements vanish. 
\end{exercise}

\begin{example}\rm [The isomorphism between $\mathfrak{so}(3)$ and $\mathbb{R}^3$]
The previous equation may be written equivalently by defining the
hat-operation $\widehat{\,\cdot\,}$ as 
\[
\mathbf{\hat{x}}_{ij}
=
\epsilon_{ijk}x^k
,\quad\hbox{where}\quad
(x^1,x^2,x^3)=(x,y,z)
.
\]
Here $\epsilon_{123}=1$ and $\epsilon_{213}=-1$, with cyclic permutations.
The totally antisymmetric tensor
$\epsilon_{ijk}=-\epsilon_{jik}=-\epsilon_{ikj}$ also defines the cross
product of vectors in $\mathbb{R}^3$. Consequently, we may write,
\[
(\mathbf{x}\times\mathbf{y})_i
=
\epsilon_{ijk}x^jy^k
=
\mathbf{\hat{x}}_{ij}y^j
,\quad\hbox{that is,}\quad
\mathbf{x}\times\mathbf{y}
=
\mathbf{\hat{x}}\mathbf{y}
\,.\]
\end{example}

\begin{exercise}
What is the analog of the hat map $\mathfrak{so}(3)\mapsto\mathbb{R}^3$ for the three
dimensional Lie algebras $\mathfrak{sp}(2,\mathbb{R})$, $\mathfrak{so}(2,1)$,
$\mathfrak{su}(1,1)$, or $\mathfrak{sl}(2,\mathbb{R})$?
\end{exercise}

Background reading for this lecture is Marsden and Ratiu~\cite[Chapter 9]{MaRa1994}.

\subsubsection{Compute the Adjoint and adjoint operations by differentiation}

\begin{enumerate}
\item
Differentiate $I_g(h)$ with respect to $h$ at $h=e$ to produce the 
\emph{Adjoint operation}\index{Adjoint operation}
\[
\Ad{:\ } G\times\mathfrak{g}\to\mathfrak{g}
\,:\quad
\Ad_g\eta=T_eI_g\eta
\]
\item
Differentiate $\Ad_g\eta$ with respect to $g$ at $g=e$ in the direction $\xi$
to get the \emph{Lie bracket} \index{Lie bracket!adjoint operation}
$[\xi,\eta]{:\ } \mathfrak{g}\times\mathfrak{g}\to\mathfrak{g}$ and thereby
to produce the \emph{adjoint operation}
\[
T_e(\Ad_g\eta)\xi=[\xi,\eta]=\ad_{\xi}\eta
\]
\end{enumerate}

\subsubsection{Compute the co-Adjoint and coadjoint operations by taking duals}

\begin{enumerate}
\item
$\Ad^*_g{:\ } \mathfrak{g}^*\to\mathfrak{g}^*$, the dual of $\Ad_g$,
is defined by 
\[
\langle\Ad^*_g\alpha,\xi\rangle
=
\langle\alpha,\Ad_g\xi\rangle
\]
for $\alpha\in \mathfrak{g}^*$, $\xi\in \mathfrak{g}$ and nondegenerate pairing
$\langle\cdot, \cdot\rangle{:\ } \mathfrak{g}^*\times\mathfrak{g}
\to\mathbb{R}$. The map
\[
\Phi^*{:\ } G\times \mathfrak{g}^*\to\mathfrak{g}^*
\quad\hbox{given by}\quad
(g,\alpha)\mapsto \Ad^*_{g^{-1}}\alpha
\]
defines the \emph{co-Adjoint action} \index{co-Adjoint action} of $G$ on $\mathfrak{g}^*$.
\item
The pairing 
\[
\langle\ad^*_{\xi}\alpha,\eta\rangle
=
\langle\alpha,\ad_{\xi}\eta\rangle
\]
defines the \emph{coadjoint action} \index{coadjoint action} of $\mathfrak{g}$ on $\mathfrak{g}^*$,
for $\alpha\in \mathfrak{g}^*$ and $\xi,\eta\in \mathfrak{g}$.
\end{enumerate}

See \cite[Chapter~9]{MaRa1994} for more discussion of the $\Ad$ and $\ad$
operations.


\subsubsection{The Lie algebra $\mathfrak{so}(3)$ and its dual}

The special orthogonal group is defined by
\[
SO(3): = \{A \mid A \text{~a~} 3 \times 3
\text{~orthogonal matrix}, \operatorname{\det}(A) = 1\}
.
\]
Its Lie algebra $\mathfrak{so}(3)$ is formed by $3\times 3 $ skew
symmetric matrices, and its dual is denoted $\mathfrak{so}(3)^\ast $. 

\subsubsection{The isomorphism $(\widehat{\,\cdot\,})\,{:\ } ({so}(3),
[\cdot, \cdot]) \to (\mathbb{R}^3, \times)$}

The Lie algebra $({so}(3), [\cdot, \cdot])$, where $[\cdot,
\cdot]$ is the commutator bracket of matrices, is isomorphic to the Lie
algebra $(\mathbb{R}^3, \times) $, where $\times $ denotes the vector
product in $\mathbb{R}^3$, by the isomorphism
$$\label{so three isomorphism in coordinates}
\mathbf{u}: =(u^1, u^2, u ^3) \in \mathbb{R}^3 \mapsto 
\mathbf{\hat{u}}: =
\left[ \begin{matrix} 0&-u^3&u^2\\ u^3&0&-u^1\\ -u^2&u^1&0 \end{matrix}
\right] \in {so}(3),$$
that is, $\mathbf{\hat{u}}_{ij}: =-\epsilon_{ijk}u^k$.
Equivalently, this isomorphism is given by
\begin{eqnarray*}
\label{so three isomorphism}
\mathbf{\hat{u}} \mathbf{v} = \mathbf{u}\times \mathbf{v} \quad
\text{for all} \quad \mathbf{u}, \mathbf{v}\in \mathbb{R}^3.
\end{eqnarray*}
The following formulas for $\mathbf{u}, \mathbf{v}, \mathbf{w} \in
\mathbb{R}^3$ may be easily verified:
\begin{eqnarray*}
\label{bracket relation}
(\mathbf{u} \times \mathbf{v})\widehat{\phantom{u}} 
&=& [\mathbf{\hat{u}},
\mathbf{\hat{v}}]\\
\label{triple product}
[\mathbf{\hat{u}},\mathbf{\hat{v}}]\mathbf{w} 
&=&
(\mathbf{u} \times \mathbf{v}) \times \mathbf{w}\\
\label{dot product}
\mathbf{u}\cdot \mathbf{v} 
&=& - \tfrac12 \operatorname{trace}(\mathbf{\hat{u}} \mathbf{\hat{v}}).
\end{eqnarray*}

\subsubsection{The $\Ad$ action of $SO(3)$ on ${so}(3)$}

The corresponding
adjoint action of $SO(3)$ on
$\mathfrak{so}(3)$ may be obtained as follows. For $SO(3)$ we have $I_A(B)=ABA^{-1}$.
Differentiating $B(t)$ at $B(0)=\Id$ gives
\[
\Ad_A\mathbf{\hat{v}}
=\frac{d}{dt}\Big|_{t=0}AB(t)A^{-1}
=A\mathbf{\hat{v}}A^{-1}
,\quad\hbox{with}\quad
\mathbf{\hat{v}}=B^\prime(0).
\] 
One calculates the pairing with a vector $\mathbf{w}\in\mathbb{R}^3$ as
\[
\Ad_A\mathbf{\hat{v}}(\mathbf{w})
=
A\mathbf{\hat{v}}(A^{-1}\mathbf{w})
=
A(\mathbf{v}\times A^{-1}\mathbf{w})
=
A\mathbf{v}\times\mathbf{w}
=
(A\mathbf{v})\,\widehat{~}\,\mathbf{w}
\]
where we have used a relation
\begin{eqnarray*}
\label{times relation}
A(\mathbf{u} \times \mathbf{v}) = A \mathbf{u} \times A \mathbf{v}
\end{eqnarray*}
which holds for any $\mathbf{u}, \mathbf{v} \in \mathbb{R}^3$ and $A \in
SO(3)$.

Consequently,
\begin{equation}
\Ad_A\mathbf{\hat{v}}=(A\mathbf{v})\,\widehat{~}
\label{Ad-Av-SO3a}
\end{equation}
Identifying $\mathfrak{so}(3)\simeq\mathbb{R}^3$ then gives 
\begin{equation}
\Ad_A\mathbf{v}=A\mathbf{v}.
\label{Ad-Av-SO3b}
\end{equation}
So (speaking prose all our lives) the adjoint action of $SO(3)$ on $\mathfrak{so}(3)$
may be identified with  multiplication of a matrix in $SO(3)$ times a
vector in $\mathbb{R}^3$.

\subsubsection{The $\ad$ action of $\mathfrak{so}(3)$ on ${so}(3)$}
Differentiating again gives the $\ad$--action of the Lie algebra $\mathfrak{so}(3)$ on
itself:
\[
[\mathbf{\hat{u}}, \mathbf{\hat{v}}] =
\operatorname{ad}_{\mathbf{\hat{u}}}
\mathbf{\hat{v}} =
\left.\frac{d}{dt}\right|_{t = 0} \left(e^{t
\mathbf{\hat{u}}}\mathbf{v}\right)
\!\!\!\widehat{\phantom{A}} = (\mathbf{\hat{u}}
\mathbf{v})\!\widehat{\phantom{A}} = (\mathbf{u}
\times \mathbf{v})\!\widehat{\phantom{A}}.
\]
So in this isomorphism the vector cross product is identified with the
matrix commutator of skew symmetric matrices.

\subsubsection{Infinitesimal generator}
Likewise, the \emph{infinitesimal generator} \index{infinitesimal generator} corresponding to
$\mathbf{u}\in\mathbb{R}^3$ has the expression
\begin{eqnarray*}
\label{inf generator for rotations}
\mathbf{u}_{\mathbb{R}^3} (\mathbf{v}) := \left.\frac{d}{dt}\right|_{t =
0} e^{t \mathbf{\hat{u}}} \mathbf{v} = \mathbf{\hat{u}}\, \mathbf{v} = \mathbf{u}
\times \mathbf{v}.
\end{eqnarray*}

\begin{exercise}
What is the analog of the hat map $\mathfrak{so}(3)\mapsto\mathbb{R}^3$ for the three
dimensional Lie algebras $\mathfrak{sp}(2,\mathbb{R})$, $\mathfrak{so}(2,1)$,
$\mathfrak{su}(1,1)$, or $\mathfrak{sl}(2,\mathbb{R})$?
\end{exercise}

\subsection*{The dual Lie algebra isomorphism
$\mathbf{\breve{~}}\,{:\ } \mathfrak{so}(3)^\ast\to
\mathbb{R}^3$}

\subsubsection{Coadjoint actions} 
The dual ${so}(3) ^\ast$ is identified with $\mathbb{R}^3$
by the breve $(\,\breve{\phantom{~}}\,)$ isomorphism 
\[
\mathbf{\Pi} \in \mathbb{R}^3  \mapsto
\mathbf{\breve{\Pi}} \in {so}(3) ^\ast
{:\ } \mathbf{\breve{\Pi}}(\mathbf{\hat{u}}): = \mathbf{\Pi} \cdot
\mathbf{u}
\quad\hbox{for any}\quad
\mathbf{u}\in \mathbb{R}^3
.\]
In terms of this isomorphism, the co-Adjoint action of $SO(3)$ on
$\mathfrak{so}(3)^\ast$ is given by
\begin{eqnarray}
\label{coadjoint so three action}
\operatorname{Ad}^\ast _{A^{-1}} \mathbf{\breve{\Pi}} = (A\boldsymbol{\Pi})\!\breve{\phantom{A}}
\end{eqnarray}
and the coadjoint action of $\mathfrak{so}(3) $ on $\mathfrak{so}(3) ^\ast$ is
given by
\begin{eqnarray}
\label{coadjoint so three lie algebra action}
\operatorname{ad}^\ast_{\widehat{\mathbf{u}}}
\breve{\boldsymbol{\Pi}} = (\boldsymbol{\Pi}\times
\mathbf{u})\breve{\phantom{u}}.
\end{eqnarray}

\subsubsection{Computing the co-Adjoint action of $SO(3)$ on $\mathfrak{so}(3)^\ast$}
For $A\in SO(3)$ and $\mathbf{\hat{u}}\in \mathfrak{so}(3)$, 
the co-Adjoint action of $SO(3)$ on $\mathfrak{so}(3)$ is obtained from
\begin{align*}
\left(\operatorname{Ad}^\ast _{A^{-1}} \mathbf{\breve{\Pi}}\right) (\mathbf{\hat{u}}) 
&= \mathbf{\breve{\Pi}} (\operatorname{Ad}_{A^{-1}}\mathbf{\hat{u}})
= \mathbf{\breve{\Pi}}  \big((A^{-1}\mathbf{u})\!\widehat{\phantom{u}} \big)
= \boldsymbol{\Pi}\cdot A^{T}\mathbf{u} \\ &= A \boldsymbol{\Pi} \cdot \mathbf{u}
=(A \boldsymbol{\Pi})\!\breve{\phantom{A}}(\mathbf{\hat{u}})
\,,
\end{align*}
where the second step applies equation \eqref{Ad-Av-SO3a}.
Finally, the co-Adjoint action of $SO(3) $ on $\mathfrak{so}(3) ^\ast$ is
expressed as in \eqref{coadjoint so three action}
\[
\operatorname{Ad}^\ast _{A^{-1}} \mathbf{\breve{\Pi}} = (A
\boldsymbol{\Pi})\!\breve{\phantom{A}}
.
\]

Consequently, the \emph{co-Adjoint orbit} \index{co-Adjoint orbit} 
$ \mathcal{O} = \left\{A \boldsymbol{\Pi} \mid
 A\in SO(3)\right\} \subset \mathbb{R}^3$ of $SO(3)$ through
$\boldsymbol{\Pi} \in \mathbb{R}^3$ is a $2$--sphere  of
radius $\|\boldsymbol{\Pi}\|$.

\subsubsection{Computing the coadjoint action of $\mathfrak{so}(3)$ on $\mathfrak{so}(3)^\ast$} Let
$\mathbf{u}, \mathbf{v} \in \mathbb{R}^3$ and note that
\begin{align*}
\big\langle\operatorname{ad}^\ast_{\mathbf{\hat{u}}}
\breve{\boldsymbol{\Pi}},
\mathbf{\hat{v}}\big\rangle
&= \big\langle \breve{\boldsymbol{\Pi}}, \big[ \mathbf{\hat{u}},
\mathbf{\hat{v}} \big] \big\rangle
= \big\langle \breve{\boldsymbol{\Pi}}, (\mathbf{u} \times \mathbf{v})
\widehat{\phantom{u}} \big\rangle
= \boldsymbol{\Pi}\cdot(\mathbf{u}\times \mathbf{v}) \\
&= (\boldsymbol{\Pi}\times \mathbf{u})\cdot \mathbf{v}
= \big\langle \boldsymbol{\Pi}\times \mathbf{u})\breve{\phantom{u}},
\mathbf{\hat{v}}\big\rangle,
\end{align*}
which shows that $\operatorname{ad}^\ast_{\mathbf{\hat{u}}}
\breve{\boldsymbol{\Pi}} = (\boldsymbol{\Pi}\times
\mathbf{u})\breve{\phantom{u}}$, thereby proving \eqref{coadjoint so three
lie algebra action}. 

Consequently, the tangent space of the \emph{co-Adjoint orbit} $ \mathcal{O}$
$$T_{\boldsymbol{\Pi}}{\mathcal{O}} = \big\{\boldsymbol{\Pi} \times
\mathbf{u}~\mid~\mathbf{u}\in \mathbb{R}^3 \big\},$$
since the plane perpendicular to
$\boldsymbol{\Pi}$, that is, the tangent space to the sphere centered
at the origin of radius $\| \boldsymbol{\Pi}\|$, is given by
$\big\{\boldsymbol{\Pi} \times \mathbf{u} \mid
\mathbf{u}\in \mathbb{R}^3 \big\}$.

\newpage
\vspace{4mm}\centerline{\textcolor{shadecolor}{\rule[0mm]{6.75in}{-2mm}}\vspace{-4mm}}
\section{Rigid body equations on $\operatorname{SO}(n)$}

\secttoc

\textbf{What is this lecture about?} This lecture outlines Manakov's approach to proving 
algebraic integrability of the $SO(n)$ rigid body dynamics via its isospectral eigenvalue problem.

\subsection{Manakov's formulation of the $SO(n)$ rigid body}
\label{ManakovFormRB}
\index{Manakov!commutator form}
\index{rigid body!Manakov's formulation}
\begin{proposition}[Manakov \cite{Man1976}]
Euler's equations for a rigid body on $SO(n)$ take the matrix commutator form,
\begin{equation}
\frac{dM}{dt}
=
[\,M\,,\, {\Omega}\,]
\quad\hbox{with}\quad
M = \mathbb{A}{\Omega}+{\Omega}\mathbb{A}
\,,
\label{commutator-form}
\end{equation}
where the $n\times n$ matrices $M,\,\Omega$ are skew-symmetric (forgoing superfluous hats) and $\mathbb{A}$ is symmetric.
\end{proposition}
\begin{proof}
Manakov's commutator form of the $SO(n)$ rigid-body equations~(\ref{commutator-form}) follows as the Euler--Lagrange equations for Hamilton's principle $\delta S =0$ with $S=\int l\,dt$ for the Lagrangian 
\[l(\Omega) = - \frac{1}{2} {\rm tr}({\Omega} \mathbb{A}{\Omega})\,,\] 
where ${\Omega} = O^{-1}\dot{O}\in \mathfrak{so}(n)$ and the $n\times n$ matrix $\mathbb{A}$ is symmetric. Taking matrix variations in Hamilton's principle yields
\begin{equation*}
\delta S 
=
- \frac{1}{2} \int^b_a
{\rm tr}\big(
\delta {\Omega} \,
(\mathbb{A}{\Omega}+{\Omega}\mathbb{A})
\big)\, dt
=
- \frac{1}{2} \int^b_a
{\rm tr}\big(
\delta {\Omega} \,
M
\big)\, dt
\,,
\end{equation*}
after cyclically permuting the order of matrix multiplication under the trace and substituting $M := \mathbb{A}{\Omega}+{\Omega}\mathbb{A}$. Using the Euler--Poincar\'e variational formula for $\delta {\Omega}$ now leads to
\begin{eqnarray*}
\delta S 
&=&
- \frac{1}{2} \int^b_a
{\rm tr}\big(
({\Xi}\,\dot{\phantom{\,}}
+
{\Omega}{\Xi} 
-
{\Xi} {\Omega}) 
M
\big)\, dt
\,.
\end{eqnarray*}
Integrating by parts and permuting under the trace then yields the equation
\begin{eqnarray*}
\delta S 
&=&
\frac{1}{2} \int^b_a
{\rm tr}\big({\Xi}\,
(\,\dot{M}
+
{\Omega}M
-
M\Omega\,)
\big)\, dt
\,.
\end{eqnarray*}
Finally, invoking stationarity for arbitrary $\Xi$ implies the commutator form (\ref{commutator-form}). 
\end{proof}

\subsection{Matrix Euler--Poincar\'e equations}
Manakov's commutator form of the rigid-body equations recalls much earlier work by Poincar\'e \cite{poincare1901forme}, who also noticed that the matrix commutator form of Euler's rigid-body equations suggests an  additional mathematical structure going back to Lie's theory of groups of transformations depending continuously on parameters. In particular, Poincar\'e  \cite{poincare1901forme} remarked that the commutator form of Euler's rigid-body equations would make sense for any Lie algebra, not just for $\mathfrak{so}(3)$. The proof of Manakov's commutator form (\ref{commutator-form}) by Hamilton's principle is essentially the same as Poincar\'e's proof in \cite{poincare1901forme}.
\index{Poincar\'e!1901 paper}

\vskip1sp
\begin{theorem}[Matrix Euler--Poincar\'e equations] \index{Euler--Poincar\'e equations!matrices}
\label{EPleft-Po1901}
The Euler--Lagrange
equations for Hamilton's principle $\delta S =0$ with $S=\int l(\Omega)\,dt$ may be expressed in matrix commutator form,
\begin{equation}
\frac{dM}{dt}
=
[\,M\,,\, {\Omega}\,]
\quad\hbox{with}\quad
M = \frac{\delta l}{\delta\Omega}
\,,
\label{EP-form}
\end{equation} 
for any Lagrangian $l(\Omega)$, where ${\Omega} = g^{-1}\dot{g}\in \mathfrak{g}$ and $\mathfrak{g}$ is the matrix Lie algebra of any matrix Lie group $G$. 
\end{theorem}

\begin{proof}
The proof here is the same as the proof of Manakov's commutator formula via Hamilton's principle, modulo replacing $O^{-1}\dot{O}\in so(n)$ with $g^{-1}\dot{g}\in \mathfrak{g}$.  
\end{proof}

\begin{remark}\rm 
Poincar\'e's observation leading to the matrix Euler--Poincar\'e equation~(\ref{EP-form}) was reported in two pages with no references, also using a variational approach \cite{poincare1901forme}. Hence, the name Euler--Poincar\'e equations. Note that if ${\Omega} = g^{-1}\dot{g}\in \mathfrak{g}$, then $M=\delta l/\delta {\Omega} \in \mathfrak{g}^*$, where the dual is defined in terms of the matrix trace pairing. \index{Euler--Poincar\'e equations!matrices}
\end{remark}

\begin{exercise}
Retrace the proof of the variational principle for the Euler--Poincar\'e
equation, replacing the left-invariant quantity $g^{-1}\dot{g}$ with 
the right-invariant quantity $\dot{g}g^{-1}$.
\vskip-19pt
\end{exercise}

\index{rigid body!eigenvalue problem}

\vskip-12pt\vskip1sp
\subsection{An isospectral eigenvalue problem for the $SO(n)$ rigid body}
\label{RB evprob}\index{rigid body!isospectral problem}

The solution of the $SO(n)$ rigid-body dynamics
\begin{equation*}
\frac{dM}{dt}
=
[\,M\,,\, {\Omega}\,]
\quad\hbox{with}\quad
M = \mathbb{A}{\Omega}+{\Omega}\mathbb{A}
\,,
\end{equation*}
for the evolution of the $n\times n$ skew-symmetric matrices $M,\,\Omega$, with constant symmetric $\mathbb{A}$,
is given by a similarity transformation (later to be identified as coadjoint motion), \index{coadjoint motion}
\begin{eqnarray*}
M(t)=O(t)^{-1}M(0)O(t)=:{\rm Ad}^*_{O(t)}M(0)
\,,
\end{eqnarray*}
with $O(t)\in SO(n)$ and $\Omega := O^{-1}\dot{O}(t)$.
Consequently, the evolution of $M(t)$ is \emph{isospectral}. \index{isospectral}
This means that 
\begin{itemize}
\item
The initial eigenvalues of the matrix $M(0)$ are preserved by the motion; that is, $d\lambda/dt=0$ in 
\[
M(t)\psi(t)=\lambda\psi(t)
\,,
\] 
provided its eigenvectors $\psi\in\mathbb{R}^n$ evolve according to
\begin{eqnarray*}
\psi(t)=O(t)^{-1}\psi(0)
\,.
\end{eqnarray*}
The proof of this statement follows from the corresponding property of similarity transformations.
\item
Its matrix invariants are preserved:
\[
\frac{d}{dt}{\rm tr}(M-\lambda {\rm Id})^K=0
\,,
\]
for every non-negative integer power $K$.

This is clear because the invariants of the matrix $M$ may be expressed in terms of its eigenvalues; but these are invariant under a similarity transformation. 
\end{itemize}
\begin{proposition}
Isospectrality allows the quadratic rigid-body dynamics (\ref{commutator-form}) on $SO(n)$ to be rephrased as a system of two coupled linear equations: the eigenvalue problem for $M$ and an evolution equation for its eigenvectors $\psi$, as follows:
\begin{eqnarray*}
M\psi=\lambda\psi
\quad\hbox{and}\quad
\dot{\psi}=-\,\Omega\psi
\,,
\quad\hbox{with}\quad
\Omega=O^{-1}\dot{O}(t)
\,.
\end{eqnarray*}
\end{proposition}
\begin{proof} Applying isospectrality  in the time derivative of the first equation yields 
\[
(\,\dot{M}+[\,\Omega,M\,]\,)\psi
+
(M-\lambda {\rm Id})(\dot{\psi}+\Omega\psi)=0
\,.
\]
Now substitute the second equation to recover (\ref{commutator-form}).
\end{proof}

\subsection{Manakov's proof of algebraic integrability of the $SO(4)$ rigid body}

The Euler equations on $SO(4)$ are
\[ \frac{dM}{dt}= M\Omega - \Omega M =[M,\Omega]  \]
where $\Omega$ and $M$ are skew symmetric $4\times4$ matrices. The angular
frequency $\Omega$ is a linear function of the angular momentum, $M$.
Manakov~\cite{Man1976} ``deformed'' these equations into
\[ 
\frac{d}{dt}(M+\lambda A)=[(M+\lambda A),(\Omega+\lambda B)] , 
\]
where $A$, $B$ are also skew symmetric $4\times4$ matrices and $\lambda$
is a scalar constant parameter. For these equations to hold for any value
of $\lambda$, the coefficent of each power must vanish. 

\begin{itemize}
\item
The coefficent of $\lambda^2$ is
\[ 0=[A,B] \]
So $A$ and $B$ must commute. So, let them be constant and diagonal:
\[ A_{ij}={\mathrm diag}(a_i)\delta_{ij} ,\quad B_{ij}={\mathrm diag}(b_i)\delta_{ij}
\tag{no sum} \]
\item
The coefficent of $\lambda$ is
\[ 0=\frac{dA}{dt}=[A,\Omega]+[M,B] \]
Therefore, by antisymmetry of $M$ and $\Omega$,
\[ (a_i-a_j)\Omega_{ij}=(b_i-b_j)M_{ij} \qquad\Longleftrightarrow\qquad
\Omega_{ij} = \frac{b_i-b_j}{a_i-a_j}M_{ij} \tag{no sum} \]
\item
Finally, the coefficent of $\lambda^0$ is the Euler equation,
\[ \frac{dM}{dt}=[M,\Omega] , \]
but now with the restriction that the moments of inertia are of the form,
\[ \Omega_{ij} = \frac{b_i-b_j}{a_i-a_j}M_{ij} \tag{no sum} \]
which turns out to possess only 5 free parameters.
\end{itemize}
With these conditions, Manakov's deformation of the $SO(4)$ rigid body
implies for every power $n$ that
\[ \frac{d}{dt}(M+\lambda A)^n=[(M+\lambda A)^n,(\Omega+\lambda B)] , \]
Since the commutator is antisymmetric, its trace vanishes and one has
\[ \frac{d}{dt}\trace (M+\lambda A)^n=0 \]
after commuting the trace operation with time derivative. Consequently, 
\[ \trace (M+\lambda A)^n=\text{ constant} \]
for each power of $\lambda$. That is, all the coefficients of each power
of $\lambda$ are constant in time for the $SO(4)$ rigid body.
Manakov~\cite{Man1976} proved that these constants of motion are sufficient to
completely determine the solution. 

\begin{remark}\rm 
This result generalizes considerably. First, it holds for $SO(n)$. Indeed, as as proven
using the theory of algebraic varieties by Haine~\cite{Ha1984}, Manakov's method captures all
the algebraically integrable rigid bodies on $SO(n)$ and the moments of inertia of
these bodies possess only $2n-3$ parameters. (Recall that in Manakov's case for $SO(4)$
the moment of inertia possesses only five parameters.) Moreover,
Mi\v{s}\v{c}enko and Fomenko~\cite{MiFo1978} prove that every compact Lie group admits a family of left-invariant metrics with completely integrable geodesic flows.
\end{remark}

\begin{exercise}
Try computing the constants of motion $\trace (M+\lambda A)^n$ for
the values $n=2, 3, 4$. 
How many additional constants of motion are needed for integrability for
these cases? How many for general $n$? Hint: keep in mind that $M$ is a skew symmetric
matrix, $M^T=-M$, so the trace of the product of any diagonal matrix times an odd power
of $M$ vanishes.
\end{exercise}

\begin{answer}
The traces of the powers $\trace (M+\lambda A)^n$ are given by
\begin{eqnarray*}
n{=}2:&~& \tr M^2 {+} 2\lambda\tr (AM) {+} \lambda^2\tr A^2 \\
n{=}3:&~& \tr M^3 {+} 3\lambda\tr (AM^2) {+} 3\lambda^2\tr A^2M
{+} \lambda^3\tr A^3 \\
n{=}4:&~& \tr M^4 {+} 4\lambda\tr (AM^3) 
{+} \lambda^2(2\tr A^2M^2 {+} 4\tr AMAM) 
\\ &&   {+} \lambda^3\tr A^3M
{+} \lambda^4\tr A^4
\end{eqnarray*}
The number of conserved quantities for $n=2,3,4$ are, respectively, 
one ($C_1=\tr M^2$), one ($I_1=\tr AM^2$) and two ($C_2=\tr M^4$ and
$I_2=2\tr A^2M^2 + 4\tr AMAM$). The quantities $C_1$ and $C_2$ are Casimirs
for the Lie--Poisson bracket for the rigid body. Thus, $\{C_1,H\}=0=\{C_2,H\}$ for
any Hamiltonian $H(M)$; so of course $C_1$ and $C_2$ are conserved. However, each
Casimir only reduces the dimension of the system by one. The dimension of the original
phase space is dim$\,T^*SO(n)=n(n-1)$. This is reduced in half by left invariance of the
Hamiltonian to the dimension of the dual Lie algebra dim$\,so(n)^*=n(n-1)/2$. For $n=4$,
dim$\,\mathfrak{so}(4)^*=6$. One then subtracts the number of Casimirs (two) by passing to
their level surfaces, which leaves four dimensions remaining in this case. The other two
constants of motion $I_1$ and $I_2$ turn out to be sufficient for integrability, because
they are in involution $\{I_1,I_2\}=0$ and because the level surfaces of the Casimirs
are symplectic manifolds, by the Marsden--Weinstein reduction theorem~\cite{MaWe1974}. 
For more details, see Ratiu~\cite{Ra1980}.
\end{answer}

\begin{exercise}
How do the Euler equations look on $\mathfrak{so}(n)^*$ as a matrix equation? Is there
an analog of the hat map for $\mathfrak{so}(3)^*$? Hint: the Lie algebra $\mathfrak{so}(4)$ is
locally isomorphic to $\mathfrak{so}(3)\times \mathfrak{so}(3)$.
\end{exercise}

\subsection{Implications of left invariance}

This Hamiltonian $H(M)$ for the $SO(n)$ rigid body equations is invariant 
under the action of $SO(n)$ from the left. 
The corresponding conserved momentum map under this symmetry is
known from the previous section as
\[ J_L{:\ } T^*SO(n)\mapsto \mathfrak{so}(n)^* \quad\hbox{is}\quad J_L(Q,P)=PQ^T \]
On the other hand, we know (from section~\ref{sec:16}) that the momentum map
for right action is 
\[ J_R{:\ } T^*SO(n)\mapsto \mathfrak{so}(n)^* ,\quad J_R(Q,P)=Q^TP \]
Hence $M=Q^TP=J_R$. Therefore, one computes
\begin{eqnarray*}
H(Q,P)&=&H(Q,Q\cdot M)=H(\Id ,M) \quad\hbox{(by left invariance)} \\
&=& H(M)=\tfrac{1}{2}\langle M, \mathbb{I}^{-1}(M)\rangle \\
&=&\tfrac{1}{2}\langle Q^TP, \mathbb{I}^{-1}(Q^TP)\rangle
\end{eqnarray*}
Hence, we may write the $SO(n)$ rigid body Hamiltonian as 
\[ H(Q,P) =\tfrac{1}{2}\big\langle Q^TP, \Omega(Q,P)\big\rangle \]
Consequently, the variational derivatives of $H(Q,P)=\frac{1}{2}\langle
Q^TP, \Omega(Q,P)\rangle$ are
\begin{eqnarray*}
\delta H 
&=&  \left\langle Q^T\delta{P}+\delta{Q}^TP, \Omega(Q,P) \right\rangle \\
&=&  \tr(\delta{P}^TQ\Omega) + \tr(P^T\delta{Q}\Omega) \\
&=&  \tr(\delta{P}^TQ\Omega) + \tr(\delta{Q}\Omega P^T) \\
&=&  \tr(\delta{P}^TQ\Omega) + \tr(\delta{Q}^T P\Omega^T) \\
&=&  \left\langle \delta{P}, Q\Omega \right\rangle -
  \left\langle \delta{Q}, P\Omega \right\rangle 
\end{eqnarray*}
where skew symmetry of $\Omega$ is used in the last step, that is,
$\Omega^T=-\,\Omega$. Thus, Hamilton's
canonical equations take the symmetric form,
\begin{equation}\label{HamCan-eqns}
\begin{aligned}
\dot Q&= \frac{\delta H}{\delta P}= Q\Omega , \\
\dot P&= -\frac{\delta H}{\delta Q}= P\Omega .
\end{aligned}
\end{equation}
Equations \eqref{HamCan-eqns} are the \emph{symmetric generalized rigid body
equations}, derived earlier by Bloch, Brockett and Crouch~\cite{BlCr1996,BlBrCr1997}
from the viewpoint of optimal control. Combining them yields the left-invariant relations,
\[
Q^{-1}\dot{Q}=\Omega =P^{-1}\dot{P}
\Longleftrightarrow
\frac{d}{dt}(PQ^T)=0
,
\]
in agreement with conservation of the momentum map $J_L(Q,P)=PQ^T$
corresponding to symmetry of the Hamiltonian under left action of $SO(n)$.
This momentum map is the angular momentum in space, which is related to
the angular momentum in the body by $PQ^T=m=QMQ^T$. Thus, we recognize the
canonical momentum as $P=QM$ (see \ref{ex-ad-star}), and the
momentum maps for left and right actions as
\begin{align*}
J_L&=m=PQ^T &&\text{(spatial angular momentum)} \\
J_R&=M=Q^TP &&\text{(body angular momentum)}
\end{align*} \index{angular momentum!spatial} \index{angular momentum!body}
Thus, momentum maps $T^*G\mapsto\mathfrak{g}^*$ corresponding to
symmetries of the Hamiltonian produce conservation laws; while momentum
maps $T^*G\mapsto\mathfrak{g}^*$ which do \textit{ not} correspond to
symmetries may be used to re-express the equations on $\mathfrak{g}^*$,
in terms of variables on $T^*G$.

\begin{exercise}
Write Manakov's deformation of the rigid body equations in the symmetric form
\eqref{HamCan-eqns}.
\end{exercise}

\newpage
\vspace{4mm}\centerline{\textcolor{shadecolor}{\rule[0mm]{6.75in}{-2mm}}\vspace{-4mm}}
\section{Exercises: Inside the Geometric Mechanics Cube} \index{Exercises}
\label{sec-Exercises}

\secttoc

\textbf{What is this lecture about?} This lecture provides a plethora of instructive worked examples of 
geometric formulations of finite dimensional dynamics based on Lie symmetry of Hamilton's principle.

\subsection{Introduction}
Following Noether \cite{noether1918invariante}, geometric mechanics deals with Lie group invariant variational principles.

This section applies the framework of Geometric Mechanics to Hamilton's principle (HP) for a dynamical system with two degrees of freedom (dof$_1$ and dof$_2$) taking values on $(\mathbb{R}_1^n \times \mathbb{R}_2^d)$, respectively. The Lagrangian $L: T(\mathbb{R}_1^n \times \mathbb{R}_2^d)\to \mathbb{R}$ is invariant under the transitive action of a Lie group $G$ on $\mathbb{R}_1^n$, given by $G\times \mathbb{R}_1^n\to \mathbb{R}_1^n$. Geometric mechanics uses this framework to gain insight into the similarities in structure among many different topics ranging from geometric optics, to classical mechanics, to quantum mechanics, and onward to ideal fluid mechanics. The main goal of the lecture is to gain insight into how the framework of  Geometric Mechanics applies the transformation theory of smooth invertible flows of Lie groups acting on configuration manifolds to unify our perception of physically disparate topics. \smallskip

\noindent
$\bullet$ Hamilton's principle (HP): \\$\delta S=0$ with $S=\int_a^b L(q,v) + \langle p, \dot{q} - v \rangle_{TM}\,dt$ and $S = \int_a^b  \langle p, \dot{q}\rangle_{TM} - H(q,p)  dt $
\smallskip

\noindent
$\bullet$ Reduction by Lie symmetry $TM$:\\ $q_t=g_tq_0,\,\dot{q_t} = \dot{g_t}q_0$. Let $L(g,v)=L(kg,kv)$, $k\in G$, set $L(e,g^{-1}v)=:l(\xi)$, apply HP.
\smallskip

\noindent
$\bullet$ Noether's theorem: 1-parameter Lie $G$-symmetry of HP for $\xi\in\mathfrak{g}\simeq T_eG$ implies conservation of \\
\centerline{$\langle \frac{\partial L}{\partial \dot{q}}\,,\, \delta q \rangle_{TM} =\langle p\,,\, \delta q \rangle_{TM}
=\langle p\,,\, -\pounds_\xi q \rangle_{TM} =: \langle \,p\diamond q, \xi\rangle_\mathfrak{g} 
= \langle J(q,p), \xi\rangle_\mathfrak{g} $,} which is the Hamiltonian defined by the $\xi$-component of the momentum map 
$J(q,p)= p\diamond q\in \mathfrak{g}^*$.
\smallskip

\noindent
$\bullet$ Legendre transformation $(LT)$:\\ $p:={\partial L}/{\partial \dot{q}}$, 
\quad $H(q,p) := \langle p, v\rangle_{TM} - L(q,v)$,
\ and\ $J:={\partial l}/{\partial \xi}$, \quad $h(J) := \langle J, \xi\rangle_\mathfrak{g} - l(\xi)$
\smallskip

\noindent
$\bullet$ Reduced (left-invariant) Hamilton's principle:\\ $S_{red}=\int_a^b l(\xi) + \langle J, g^{-1}\dot{g} - \xi \rangle_\mathfrak{g}\,dt$ and $S_{red} = \int_a^b  \langle J, g^{-1}\dot{g}\rangle_\mathfrak{g} - h(J)  dt $
\smallskip


\noindent
In studying these notes, the reader might keep in mind the equivariance of the following diagram. 

\bfi{Six commuting diagrams comprise the box of GM relations.}

\begin{figure}[h!]
\begin{center}
\includegraphics*[width = \textwidth]{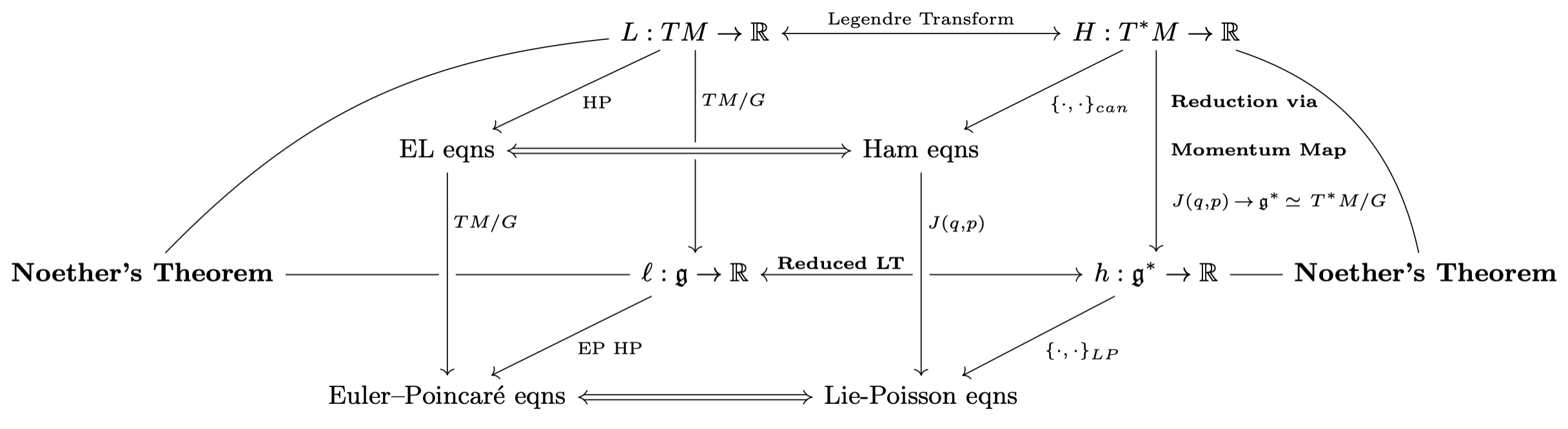}
\end{center}\vspace{-2mm}

\caption{ \textbf{\textsf{Geometric Mechanics as a cube of  six commuting diagrams.}}}
\label{GMcube1-redux}
\end{figure}


\begin{figure}[h!]
\tiny
\begin{center}\hspace*{-1.4cm}
\begin{tikzcd}[column sep=1.6em,arrows=-latex]
{\left.  \begin{array}{c} 
\frac{d}{dt}\frac{\partial L}{\partial \dot{q}}-\frac{\partial L}{\partial q}=0 , 
\\[6pt] \big\langle \frac{\partial L}{\partial \dot{q}}\,,\, \delta q \big\rangle_{TM}\big|_{t=a}^{t=b}=0
\\ \end{array} \right\} }
\arrow[dd, dotted] 
&  & L(q,v):TM \to \mathbb{R} \arrow[dd, "\text{Reduction by Lie symmetry}" description] 
\arrow[rr, "\mathcal{L}T"] \arrow[ll, "\delta S=0"'] 
&  & H(q,p):T^*M \to \mathbb{R} 
\arrow[dd, "\langle p {,\,} \delta q \rangle_{TM} = \langle J(q{,}p){,\,} \xi\rangle_\mathfrak{g}\,\text{(momentum map)}" description] 
\arrow[rr, "\delta S=0"] 
&  & {\left \{  \begin{array}{c} \frac{\partial H}{\partial p} =\dot{q},\quad  \frac{\partial H}{\partial q} = -\,\dot{p}, 
\\[6pt] 
{ \Big( \iota_{X_H} (dq\wedge dp) = dH \Big) } \\ \end{array} \right.} 
\arrow[dd, dotted] \\
&  &  &  &  &  &  \\
\begin{array}{c} 
\frac{d}{dt}\frac{\partial l}{\partial \xi^\alpha} 
= \frac{\partial l}{\partial \xi^\gamma} c_{\alpha\beta}^\gamma  \xi^\beta
\\[6pt] 
\Big(\frac{d}{dt}\frac{\partial l}{\partial \xi}=\text{ad}^*_{\xi}\frac{\partial l}{\partial \xi} \Big)
\end{array}
\arrow[uu, dotted] 
&  & l(\xi):\mathfrak{g}\to \mathbb{R} \arrow[ll, "\delta S_{red}=0"'] \arrow[rr] 
&  & h(J):\mathfrak{g}^*\to\mathbb{R} 
\arrow[rr, "\delta S_{red}=0"] 
\arrow[ll, "\text{Reduced }\mathcal{L}T"'] 
&  & \begin{array}{c} 
\frac{dJ_\alpha}{dt} = \{J_\alpha,J_\beta\} \frac{\partial h}{\partial J_\beta}
= J_\gamma c_{\alpha\beta}^\gamma 
\frac{\partial h}{\partial J_\beta}\arrow[uu, dotted]
\\[6pt] 
\Big(\frac{dJ}{dt}=\text{ad}^*_{\partial h/\partial J}J \Big)
\end{array}
\end{tikzcd}
\end{center}
\caption{ \ \textbf{\textsf{Unfolding the cube of commuting diagrams for Geometric Mechanics }}}
\end{figure}
\vspace{-5mm}
\noindent \normalsize 
\newpage

\begin{exercise}
Write Hamilton's principle for two degrees of freedom in the configuration manifold $(q_1,q_2)\in M= \mathbb{R}_1^n \times \mathbb{R}_2^d$.
\begin{enumerate}[(a)] 
	\item Formulate the Lagrangian $L\in C^\infty(TM)$ as $L(q_1,u_1;q_2,u_2): T(\mathbb{R}_1^n \times \mathbb{R}_2^d)\to \mathbb{R}$, where, e.g., $u_1\in T_{q_1}\mathbb{R}_1^n$ lies in the tangent fibre over the point $q_1\in \mathbb{R}_1^n$.
	\item For a family of curves $(q_1(t),q_2(t))$ with tangents $(\dot{q}_1(t),\dot{q}_2(t))$, define Hamilton's variational principle as:  
	\[
	0=\delta S=\delta \int_a^b L(q_1,u_1;q_2,u_2) 
	+ \langle p_1\,,\,\tfrac{dq_1}{dt} - u_1\rangle + \langle p_2 \,,\tfrac{dq_2}{dt} - u_2\,	\rangle \,dt
	\,,\]
	with natural pairing $\langle\,\cdot\,,\,\cdot\,\rangle: T^*M\times TM\to \mathbb{R}$ for $p_i\in T^*M$ and $u_i\in TM$, $i=1,2$.
	\item Set $\delta S=0$ under variations $\delta q_i$, $\delta u_i$ and $\delta p_i$ for $i=1,2$ to derive the  Euler-Lagrange (EL) equations: 
	\[
	p_1 = \frac{\partial L}{\partial u_1}\Big|_{u_1=\dot{q}_1 }, \quad
	\frac{dp_1}{dt}
	=  \frac{\partial L}{\partial q_1}
	\quad\hbox{and} \quad
	p_2 = \frac{\partial L}{\partial u_2}\Big|_{u_1=\dot{q}_2 }
	\,,\quad
	\frac{dp_2}{dt}=  \frac{\partial L}{\partial q_2}
	\,,\]
	for endpoint conditions $\langle p_1,\delta q_1\rangle|_a^b=0$ 
	and $\langle p_2,\delta q_2\rangle|_a^b=0$.
	\end{enumerate}
	\end{exercise}
\smallskip

%
\begin{exercise}
	Recall that a Lie group $G$ is a group whose transformations $g_\epsilon$ depend differentiably on a set of parameters, $\epsilon\in\mathbb{R}$. 
	
	Show that a Lie group is also a manifold. (Hint: A manifold is a space on which the rules of calculus apply.) 
	 \end{exercise}
	
\begin{exercise}
	The vector space $T_eG=\mathfrak{g}$ (i.e., the tangent space of $G$ evaluated at the identity of $G$) is called the Lie algebra of the Lie group $G$. 
	
	Show that the differential $\frac{d}{d\epsilon}\big|_{\epsilon=0}$ of the composition law of the \textit{left action} of a Lie group $G$ evaluated at the identity $\epsilon=0$  yields its Lie algebra bracket operation $[\,\cdot\,,\,\cdot\,]: \mathfrak{g}\times \mathfrak{g}\to \mathfrak{g}$, also written as the adjoint action $\mathrm{ad}_\xi \theta = [\xi,\theta]=- [\theta,\xi]$ for $\xi,\theta \in \mathfrak{g}$.
	\end{exercise}

\begin{answer}
Consult chapters 4, 5, and 6 of \cite{Holm_Book2_2008}
\end{answer}

\begin{exercise}
	Suppose Lie group $G$ acts transitively on submanifold $\mathbb{R}_1^n$ of the configuration manifold $M=\mathbb{R}_1^n \times \mathbb{R}_2^d$, so that $q_1(\epsilon)=g_\epsilon q_1(0)$, with $\epsilon=0$ at the identity element of $G$. Assume this $G$-action only affects submanifold $\mathbb{R}_1^n$ and does not act on the remaining submanifold $\mathbb{R}_2^d$ of the configuration manifold $M$.

	Prove the Noether theorem \cite{noether1918invariante}, that each 1-parameter Lie symmetry of the Lagrangian in Hamilton's principle implies a conserved quantity under the dynamics of the corresponding EL equations for this composite system on $M=\mathbb{R}_1^n \times \mathbb{R}_2^d$.
	\end{exercise}

\begin{answer}$\,$

Proof: 
	
	Invariance of the Lagrangian $\delta L = \frac{dL}{d\epsilon}|_{\epsilon=0} = 0$ for an infinitesimal Lie symmetry transform $(\delta q_1=\frac{dq_1}{d\epsilon}|_{\epsilon=0}$ and $\delta u_1=\frac{du_1}{d\epsilon}|_{\epsilon=0})$ implies (via the endpoint condition $\langle p_1,\delta q_1\rangle|_a^b=0$) that if the EL equations hold, then the pairing $\langle p_1,\delta q_1\rangle$ must be conserved. Here, $\delta u_1\in T_{q_1}\mathbb{R}_1^n$ is the tangent lift at $q_1$ of the action 
	\[
	\delta q_1=\frac{d}{d\epsilon}\Big|_{\epsilon=0}q_1(\epsilon)=:-\pounds_\xi q_1
	\]
	of Lie algebra element $\xi\in\mathfrak{g}$ on position $q_1\in \mathbb{R}_1^n\subset M$. The operation $\pounds_\xi q_1$ is the Lie derivative action of $\xi\in\mathfrak{g}$ on the position $q_1\in \mathbb{R}_1^n$.

	Define the diamond operation to relate vector space pairings 
	\[
	\langle p_1,\delta q_1\rangle = \langle p_1\,,\, -\pounds_\xi  q_1\rangle =: \langle p_1 \diamond q_1\,,\, \xi \rangle_\mathfrak{g}
	\]
	In pairing $\langle\,\cdot\,,\,\cdot\,\rangle_\mathfrak{g}: \mathfrak{g}^*\times \mathfrak{g}\to \mathbb{R}$ we have $p_1 \diamond q_1\in \mathfrak{g}^* := T_e^*G$ and $\xi\in \mathfrak{g}:=T_eG$. 
	
	The map $\mu=p_1 \diamond q_1$ is called a cotangent-lift momentum map $\mu:T^*\mathbb{R}_1^n \to  \mathfrak{g}^*$. \\ Cotangent-lift momentum maps also figure in the definitions of Lie-Poisson brackets in Hamiltonian dynamics.
\end{answer}	
 
\subsection{Reduction by Lie symmetry}
\begin{exercise}
	Reduction by Lie symmetry of Lagrangian:
	\[T(\mathbb{R}_1^n \times \mathbb{R}_2^d) \to ((T\mathbb{R}_1^n)/G) \times T\mathbb{R}_2^d \simeq \mathfrak{g} \times 		T\mathbb{R}_2^d\] for a Lagrangian that is left-invariant under the Lie group action $G\times T\mathbb{R}_1^n \to T\mathbb{R}_1^n$.
	
	The left-invariant $G$-reduced Lagrangian appears in the action integral of Hamilton's principle as
	\[
	S =\int_a^bL(\xi;q_2,u_2) + \langle \mu\,,\,g^{-1}\dot{g}-\xi\rangle_\mathfrak{g} + \langle p_2 \,,\dot{q}_2-u_2\,\rangle dt
	\,.\]
	Assuming that $\dot{q}_2$ (resp. $\xi$) may be obtained from $p_2$ and $q_2$ (resp. from $\mu = \frac{\partial L}{\partial \xi}$), show that Hamilton's principle $\delta S=0$ implies: 
	\[
	\mu = \frac{\partial L}{\partial \xi}
	\,,\quad
	\frac{d\mu}{dt}=\mathrm{ad}^*_\xi 	\mu
	\quad\hbox{and}\quad
	p_2 = \frac{\partial L}{\partial u_2}\Big|_{u_2=\dot{q}_2}
	\,,\quad
	\frac{dp_2}{dt} = \frac{\partial L}{\partial q_2}
	\,,\] 
	where $\mathrm{ad}^*_\xi \mu$ is defined by $\langle \mathrm{ad}^*_\xi \mu \,,\, \theta \rangle_\mathfrak{g} :=  \langle \mu\,,\, \mathrm{ad}_\xi \theta  \rangle_\mathfrak{g}$, with $\theta:=g^{-1}\delta{g}$ and $\delta{g}:= \frac{dg}{d\epsilon}|_{\epsilon=0}$.
	These are the Euler--Poincar\'e (EP) equations for $\mu = \frac{\partial L}{\partial \xi} \in \mathfrak{g}^*$ and the EL equations for $(q_2,p_2)\in T^*\mathbb{R}^d$.
	
	Hint: the derivation of the EP equations from Hamilton's principle follows by defining $\theta=g^{-1}\delta{g}$ with $\delta{g}:= \frac{dg}{d\epsilon}|_{\epsilon=0}$ and proving by equality of cross derivatives in $\frac{d}{dt}$ and $\frac{d}{d\epsilon}|_{\epsilon=0}$ that 
	\[
	\frac{d\xi}{d\epsilon}\Big|_{\epsilon=0} - \frac{d\theta}{dt} = - \mathrm{ad}_\theta \xi 
	\quad\hbox{with}\quad
	\mathrm{ad}_\theta \xi=[\theta,\xi]
	\quad\hbox{for left Lie group action}
	\,.\]
	\end{exercise}

\begin{answer}$\,$

Proof: 
	
To show that the combination of EP and EL equations above follows from Hamilton's principle, one first shows by direct computation that,  
	\begin{align*}
	0 = \delta S &= \int_a^b \left\langle\frac{\partial L}{\partial \xi} - \mu\,,\,\delta \xi\right\rangle_\mathfrak{g}
	+  \left\langle \mu\,,\,\frac{d\theta}{dt} + \mathrm{ad}_\xi\theta \right\rangle_\mathfrak{g}
	\\&\qquad + \left\langle\frac{\partial L}{\partial u_2} - p_2\,,\,\delta u_2 \right\rangle_{T\mathbb{R}_2}
	+ \left\langle\frac{\partial L}{\partial q_2} - \dot{p}_2\,,\,\delta q_2 \right\rangle_{T\mathbb{R}_2}
	\hspace{-4mm}dt
	\,,
	\end{align*}
	with the vanishing endpoint term $\langle p_2,\delta q_2\rangle|_a^b=0$. 
	Then, upon integration by parts in time and use of the definition of $\mathrm{ad}^*_\xi \mu$, show that
	\begin{align*}
	0 = \delta S &= \int_a^b \left\langle\frac{\partial L}{\partial \xi} - \mu\,,\,\delta \xi\right\rangle_\mathfrak{g}
	- \left\langle \frac{d\mu}{dt} - \mathrm{ad}^*_\xi\mu \,,\, \theta\right\rangle_\mathfrak{g}
	\\&\qquad + \left\langle\frac{\partial L}{\partial u_2} - p_2\,,\,\delta u_2 \right\rangle_{T\mathbb{R}_2}
	+ \left\langle\frac{\partial L}{\partial q_2} - \dot{p}_2\,,\,\delta q_2 \right\rangle_{T\mathbb{R}_2}
	\hspace{-4mm}dt
	\,,
	\end{align*}
	by imposing another vanishing endpoint term $\langle\mu,\theta\rangle_\mathfrak{g}|_a^b=0$. 
\end{answer}

\smallskip

\begin{exercise}
	Derive canonical Hamiltonian dynamics for two degrees of freedom, as follows. 
	\begin{enumerate}
	\item For $M=(\mathbb{R}_1^n \times \mathbb{R}_2^d)$, pass from the Lagrangian $L\in C^\infty(TM)$ to the Hamiltonian $H\in C^\infty(T^*M)$  via the Legendre transformation
	\[ 
	H(q_1,p_1;q_2,p_2) := \langle p_1\,,\,u_1\rangle + \langle p_2 \,,u_2\,\rangle - L(q_1,u_1;q_2,u_2)
	\,.\]
	Then equate differentials of left and right sides of the Legendre transformation and use EL equations to find Hamilton's canonical equations:  
	\begin{align*}
	H_{q_1}&=-L_{q_1} = - \,\dot{p}_1 \,, \quad H_{p_1}=u_1=\dot{q}_1
	\\
	H_{q_2}&=-L_{q_2} = - \,\dot{p}_2 \,, \quad H_{p_2}=u_2=\dot{q}_2\,.
	\end{align*}
	\item Via the chain rule for $\frac{dF}{dt}$ with $F(q_1,p_1;q_2,p_2)$, show that Hamilton's  equations follow from a bracket operation $\{\,\cdot\,,\,\cdot\, \}$ by $\{q_1,p_1\}=1$, $\{q_2,p_2\}=1$ and zero otherwise. 
	
	Verify that the bracket operation $\{F,H\}$ defined by 
	\begin{align*}
	\frac{dF}{dt} &= Tr \left(\begin{bmatrix}\frac{\partial F}{\partial q} \\ \frac{\partial F}{\partial p} \end{bmatrix}^T
	\begin{bmatrix} 0 & Id \\ -Id & 0 \end{bmatrix} 
	\begin{bmatrix}\frac{\partial H}{\partial q} \\ \frac{\partial H}{\partial p} \end{bmatrix}\right)
	\\&=
	\sum_{i=1}^2 \frac{\partial H}{\partial p_i}\frac{\partial F}{\partial q_i}-\frac{\partial H}{\partial q_i}\frac{\partial F}{\partial p_i}
	=: \{F,H\}
	\,,\end{align*}
	is skew, and satisfies both the Leibnitz product rule and the Jacobi identity, $\{F,\{G,H\}\} + c.p. = 0$. 
	\end{enumerate}
	\end{exercise}
\smallskip

\begin{exercise}  
	Show that \emph{Hamiltonian vector fields} defined via the canonical Poisson bracket $\{\,\cdot\,,\,\cdot\, \}$ as 
	\[X_H=\{\,\cdot\,,H\}\in \mathfrak{X}(T^*(M))\]
	satisfy the commutation relation 
	\[
	[X_F,X_H] = - X_{\{F,H\}}
	\]
	for phase-space functions $F,H\in C^\infty(T^*(M))$.  
\end{exercise}

\begin{answer} 
This follows from Jacobi's identity for the canonical Poisson bracket.
\end{answer} 
 
\smallskip

 \begin{exercise}  
 Explain how the Lie derivative of a phase space function $F$ by a Hamiltonian vector field $X_H$ is related to the canonical Poisson bracket. 
	\end{exercise}  

\begin{answer} 

\centerline{$\pounds_{X_F} H = X_F (H) = \{F\,,\, H\} $.}
\end{answer} 
 
 \begin{exercise}   Prove that:
	\\(a) Invariance of a scalar function $H$ under the flow $\phi^{\sss{X}}_\epsilon$ of a vector field $X$ means that the Lie derivative $\pounds_X H$ vanishes. That is, ${\phi^{\sss{X}}_\epsilon}^*H(z)=H(\phi^{\sss{X}}_\epsilon(z)) = H(z) \Leftrightarrow  \pounds_X H = 0$. 
	 \\(b)The Lie derivative action $\pounds_XF$ on a scalar phase space function $F(z)$ is the familiar directional derivative.
	 \\(c) A smooth vector field $X$ generates a flow $\phi^{\sss{X}}_\epsilon$ which takes place along its integral curves.  
	\end{exercise}  

\begin{answer} 

(a) Invariance of a scalar function $H$ under the flow $\phi^{\sss{X}}_\epsilon$ of a vector field $X$ means that the Lie derivative $\pounds_X H$ vanishes. This property follows from the \emph{Lie chain rule}, given by \index{Lie chain rule}
\begin{align}
\frac{d}{d \epsilon}\big(\phi^{{\sss{X}}*}_\epsilon H(z)\big) = \frac{d}{d \epsilon}H\big(\phi^{\sss{X}}_\epsilon(z) \big)
= \phi^{{\sss{X}}*}_\epsilon \big( \pounds_X H \big) \Leftrightarrow  \pounds_X H = 0
\,,
\label{LieChainRule}
\end{align}
by evaluating at the identity $(\epsilon=0)$ with $\phi^{{\sss{X}}*}_0=Id$.

(b) The directional derivative $X(H)=\frac{\partial H}{\partial z^i} X^i(z)$ governs how a function such as the Hamiltonian $H(z)$ changes along the flow of the vector field $X$. The directional derivative of a scalar function is a special case of the Lie derivative $\pounds_X$, since $X(H)=\pounds_XH$ for scalar functions. To see this, one computes 
via the converse of the Lie chain rule, \index{Lie chain rule}
\begin{align*}
\pounds_X H &= \frac{d}{d \epsilon}(\phi^{{\sss{X}}*}_\epsilon H(z)\Big|_{\epsilon=0} 
\\&= \frac{d}{d \epsilon}\Big|_{\epsilon=0}H (\phi^{\sss{X}}_\epsilon z)= \frac{\partial H}{\partial z}\cdot X(z)=X(H)
\,.\end{align*}
 	
(c) The integral curves of a vector field $X$ are given by the unique solution of the system of ODEs $\frac{dz^i}{d\epsilon}=X^i(z)$ with initial condition $z^i(0)$. Consequently, the integral curve $\phi^{\sss{X}}_\epsilon$ of a vector field $X$ satisfies the flow condition, $\phi^{\sss{X}}_\epsilon\circ \phi^{\sss{X}}_\tau=\phi^{\sss{X}}_{\epsilon+\tau}$. 
\end{answer}

 \begin{exercise}
	Derive a coordinate-free definition of the Poisson bracket in terms of the operation of  insertion 
	\[
	\iota: \mathfrak{X} \times \Lambda^n  \to \Lambda^{n-1}
	\]
	of Hamiltonian vector fields $X_F$ and $X_H$ into the closed symplectic 2-form $\omega\in \Lambda^2$ with $d\omega=0$. 
	\end{exercise}
	
\begin{answer}

	The required formula maybe verified to be
	\[
	\{F, H\} = \omega(X_F,X_H) = \iota_{X_H} (\iota_{X_F} \omega)
	\,,\]
	which is related to the original phase space coordinates by, 
	\begin{align*}
	\omega &= \sum_{i=1}^2dq_i\wedge dp_i = - \sum_{i=1}^2 dp_i\wedge dq_i 
	\,,\\X_H &=\{\,\cdot\,,H\}
	=\sum_{i=1}^2 \frac{\partial H}{\partial p_i}\frac{\partial }{\partial q_i}-\frac{\partial H}{\partial q_i}\frac{\partial }{\partial p_i}
	\,,\end{align*}
	which is familiar from earlier lectures.
	\end{answer}
 
 \begin{exercise}
	Explicitly calculate out the verification of the coordinate-free definition of the Poisson bracket in terms of the operation of  insertion 
	\[
	\iota: \mathfrak{X} \times \Lambda^n  \to \Lambda^{n-1}
	\]
	for the specific Hamiltonian vector fields $X_F$ and $X_H$ into the exact symplectic 2-form $\omega\in \Lambda^2$ with $d\omega=0$. 
	
The dynamics along each integral curves of $X_H$ is determined by \[dH =\omega(X_H\,,\,\cdot\,) = \iota_{X_H} \omega\,,\] in which the vector field $X_H$ is inserted ($\iota$) into the symplectic 2-form $\omega$ to create the exact 1-form $dH$. 
	
In the original coordinates, this is
\[
dH  = \iota_{(\frac{\partial H}{\partial p}\frac{\partial }{\partial q}-\frac{\partial H}{\partial q}\frac{\partial }{\partial p})}(dq\wedge dp)
= \frac{\partial H}{\partial q}dq + \frac{\partial H}{\partial p}dp
\,.\] 
Show that $\omega(X_F,X_H) = \{F, H\}$ in the original coordinates.
\end{exercise}

\begin{answer}

\begin{align*}
\frac{dF}{dt} &= \omega(X_F,X_H) 
\\&= \iota_{X_H} (\iota_{X_F} \omega) = \iota_{X_H} dF
\\&=\iota_{(\frac{\partial H}{\partial p}\frac{\partial }{\partial q}
-\frac{\partial H}{\partial q}\frac{\partial }{\partial p})}\Big(\frac{\partial F}{\partial q}dq + \frac{\partial F}{\partial p}dp\Big) 
\\&= \frac{\partial H}{\partial p}\frac{\partial F}{\partial q} - \frac{\partial H}{\partial q}\frac{\partial F}{\partial p}
\\&=\{F, H\} \,.
\end{align*}
	The opposite calculation was used to motivate introducing the insertion operation in previous lectures.
\end{answer}

 \begin{exercise}
	Use Cartan's geometric definition of the \emph{Lie derivative} of the symplectic 2-form $\omega$ with respect to the Hamiltonian vector field $X_F=\{\,\cdot\,,\,F\,\}$ to show that the symplectic form $\omega$ is invariant under the Lie algebra actions of Hamiltonian vector fields.  \index{Lie derivative!Hamiltonian vector field}
	\end{exercise}

	\begin{answer}The coordinate-free expression 
	\[
	\pounds_{X_F}\omega
	=d(\iota_{X_F}\omega)+ \iota_{X_F}d\omega
	= (d\,\iota_{X_F}+ \iota_{X_F}d)\,\omega
	\]
	is Cartan's geometric definition of the \emph{Lie derivative} of the symplectic 2-form $\omega$ with respect to the Hamiltonian vector field $X_F=\{\,\cdot\,,\,F\,\}$. 
	
	Since the symplectic form $\omega$ is closed ($d\omega=0$) and $\iota_{X_F} \omega=dF$ for a Hamiltonian vector field $X_F$, we have 
	\[
	\pounds_{X_F}\omega=d(\iota_{X_F}\omega)=d^2F=0
	\,.\]
	The finite transformation $\phi_\epsilon$ generated by the left-invariant Hamiltonian vector field $X_F=\phi_\epsilon^{-1}\phi'_\epsilon|_{\epsilon=0}$ is called a \emph{symplectic flow}. \index{symplectic flow}
\end{answer}

\subsection{Lie chain rule and Noether's theorem}\index{Lie chain rule}

\begin{exercise}
	Use the Lie chain rule and the equivalence of the dynamic and Cartan definitions of the Lie derivative of a differential form to prove that a smooth symplectic flow ${\phi_\epsilon^{\sss X_F}}$ generated by a Hamiltonian vector field given by $X_F=\frac{d}{d\epsilon}\phi_\epsilon\big|_{\epsilon=0}=\{\,\cdot\,,\,F\,\}$ with $\iota_{X_F} \omega=dF$ preserves the symplectic 2-form $\omega$ under the pull-back ${\phi_{\scaleto{\epsilon}{4pt}}}^*\omega(q,p) := \omega(\phi_\epsilon q,\phi_\epsilon p)$. 
 \end{exercise}
  \begin{answer}
	In the context of flows here, the dynamic definition of the Lie derivative is natural, 
	\[
	\pounds_{X_F}\omega=\frac{d}{d \epsilon}({\phi_{\scaleto{\epsilon}{4pt}}}^*\omega)\big|_{\epsilon=0}
	\quad\hbox{with}\quad
	X_F=\phi_\epsilon^{-1}\phi'_\epsilon\big|_{\epsilon=0}\,.
	\]
	One immediately finds, 
	\begin{align*}
	\frac{d}{d \epsilon}(\phi^*_\epsilon\omega)
	&=\phi^*_\epsilon(\pounds_{X_F}\omega)
	\overset{d\omega=0}= \phi^*_\epsilon d(\iota_{X_F}\omega)
	\overset{\iota_{X_F}\omega=dF}=
	\phi^*_\epsilon d(dF)=0
	\,,\end{align*} 
	since $d^2F=0$. 

The first step invokes the Lie chain rule. The second step invokes the equivalence of the dynamic and Cartan definitions of the Lie derivative of a differential form.
\end{answer} \index{Lie chain rule}
 
\smallskip

\begin{exercise}
Show that the diamond operation $(\diamond)$ is natural under $\mathrm{Ad}^*$. That is,
\begin{align}
\operatorname{Ad}_g^*(b \diamond a)=(b g) \diamond(a g)
=\langle g^*b \diamond g^*a, w\rangle_V 
\,,
\label{PushForward-ID}
\end{align}
 for $g \in \operatorname{Diff}(\mathcal{D}), b \in V$ and $a \in V^*$.
\end{exercise}

\begin{answer}
Let $V$ be a vector space and let $V^*$ be its dual under a pairing $\langle b, a \rangle_V$ for $b\in V$ and $a\in V^*$. Assume that the smooth invertible maps (diffeomorphisms) $\operatorname{Diff}(\mathcal{D})$ defined in domain $\mathcal{D}$ has a right representation on $V$ and an induced right representation on $V^*$, both of which are denoted by composition. Also, let $w \in \mathfrak{X}(\mathcal{D})$ be an arbitrary fixed vector field and let $\langle\cdot, \cdot\rangle_\mathfrak{X}$ be the $L^2$ pairing between the Lie algebra of vector fields and its dual $\mathfrak{X}^*(\mathcal{D})$, the 1-form densities. The diamond operation $(\diamond)$ in this situation is defined as 
\[
 \langle b \diamond a\,,\, w \rangle_\mathfrak{X} 
 := \langle b\,,\, -\pounds_w a \rangle_V
\,.\]
Next, recall the proof that the Lie derivative by vector fields is natural under Adjoint action of a Lie group element $g\in G$ on its Lie algebra $\mathfrak{g}$, which in this case becomes for $g\in \operatorname{Diff}(\mathcal{D})$, $w \in \mathfrak{X}(\mathcal{D})$,  $b\in V$ and $a\in V^*$,
\begin{align}
\mathcal{L}_{\operatorname{Ad}_g w} a=g_*\left(\mathcal{L}_w g^* a\right)=\left(\mathcal{L}_w a g\right) g^{-1}
.
\label{PushForward-nat}
\end{align}
Upon recalling that
the differential is natural under the push-forward $g_*$, the identity $\eqref{PushForward-nat}$ can be shown by via the Cartan form of the Lie derivative  as,
$$
\left.\left.\left.\left.\mathcal{L}_{\mathrm{Ad}_g w} a=\mathbf{d}\left(g_* w\right\lrcorner a\right)+g_* w\right\lrcorner \mathbf{d} a=g_*\left(\mathbf{d}(w\lrcorner g^* a\right)+w\right\lrcorner \mathbf{d} g^* a\right)=g_*\left(\mathcal{L}_w g^* a\right) .
$$
Since vector field $w$ is fixed, equation \eqref{PushForward-nat}, implies that the diamond operation $(\diamond)$ is natural under $\mathrm{Ad}^*$, for $\operatorname{Diff}(\mathcal{D})$ acting by push-forward on arbitrary vector spaces $V$ and $V^*$ defined in domain $\mathcal{D}$,
\begin{align*}
\left\langle\operatorname{Ad}_g^*(b \diamond a), w\right\rangle_\mathfrak{X}
& =\left\langle b \diamond a, \operatorname{Ad}_g w\right\rangle_\mathfrak{X} 
 =\left\langle-b, \mathcal{L}_{\left(\operatorname{Ad}_g w\right)} a\right\rangle_V 
 =\left\langle-b,\left(\mathcal{L}_w a g\right) g^{-1}\right\rangle_V \\
& =\left\langle-b g, \mathcal{L}_w a g\right\rangle_V
=\langle(b g) \diamond(a g), w\rangle_V
=\langle g^*b \diamond g^*a, w\rangle_V
\,.
\end{align*}

\end{answer}

\smallskip

\begin{exercise} State and prove Noether's Theorem on the Hamiltonian side. \end{exercise}

  \begin{answer}
	On the Hamiltonian side, Noether's theorem states that if the 1-parameter flow $\phi^{\sss{X_F}}_\epsilon$ is a symmetry of $H$, then skew symmetry of the Poisson bracket $  \{F\,,\, H\} = 0 = -\, \{H\,,\, F\}$ implies that the Hamiltonians $F(z(t))$ and $H(z(t))$ in the notation $z=(q,p)\in T^*M$ are preserved under each other's dynamics. 

	Proof: For Hamiltonian vector fields $X_F:=\{\,\cdot\,,\,F\}$,  the antisymmetry of the Poisson bracket implies
\begin{align*}
\pounds_{X_F} H  &= X_F (H) = \{F\,,\, H\} 
\\&= - \{H\,,\, F\} = -X_H (F) = - \pounds_{X_H} F
\,.\end{align*}
In this situation, if the flow $\phi^{\sss{X_F}}_\epsilon$ is a symmetry of $H$, then both $\phi^{\sss{X_F}*}_\epsilon H(z)=H(\phi^{\sss{X_F}}_\epsilon z)=H(z)$ and  $\phi^{\sss{X_H}*}_\epsilon F(z)=F(\phi^{\sss{X_H}}_t z)=F(z)$ are conserved, so that the flow $\phi^{\sss{X_H}}_t$ is a symmetry of $F$. Consequently, $  \{F\,,\, H\} = 0 = -\, \{H\,,\, F\}$ implies that the Hamiltonians $F(z(t))$ and $H(z(t))$ are preserved under each other's dynamics. 

This is Noether's Theorem on the Hamiltonian side.
 \end{answer}

\medskip

\begin{exercise} What are the Hamiltonian actions of Noether's cotangent-lift momentum map? Provide proofs. \end{exercise} 
  \begin{answer}
Suppose the variation $\delta q_1$ in Noether's theorem is defined by the infinitesimal action (i.e., the action tangent to the identity) of the Lie group element $g_\epsilon\in G$ parameterised by $\epsilon\in\mathbb{R}$  acting on the configuration space coordinate $q_1\in\mathbb{R}_1^n$ by push-forward $q_1(\epsilon)=q_1(0)g_\epsilon^{-1}$ (i.e., pull-back by the inverse). 

In this case Noether's theorem on the Hamiltonian side introduces a quantity $N^\xi(q_1,p_1)$ given by
	\begin{equation}
	N^\xi(q_1,p_1) := \langle p_1,\delta q_1\rangle_{TM} = \langle p_1\,,\, -\pounds_\xi  q_1\rangle_{TM} 
	\,,
	\end{equation}
	where the variation $\delta q_1$ is given by the Lie algebra action
	\begin{equation}
	\delta q_1=-\pounds_\xi  q_1 = \frac{d}{d\epsilon} \Big|_{\epsilon=0} [q_1(0)g_\epsilon^{-1}]
	= \frac{d}{d\epsilon} \Big|_{\epsilon=0} q_1(\epsilon)
	\,,
	\label{push-forward-chain}
	\end{equation}
	with $q_1(\epsilon):= q_1(0)g_\epsilon^{-1} = g_{\epsilon\,*}q_1(0)$.

\end{answer}

\begin{exercise} To prove the statement \eqref{push-forward-chain} above, show that the tangent to the push-forward evaluated at the identity $(\epsilon=0)$ is given by (minus) the Lie derivative with respect to $\xi= [\frac{dg}{d\epsilon}  g^{-1}\big]_{\epsilon=0}$, by first proving the Lie chain rule for the push-forward, \index{Lie chain rule}
	\[\frac{d}{d\epsilon} (g_{\epsilon\,*}q_1(0)) 
	= - \,\pounds_{ \frac{dg}{d\epsilon}  g_\epsilon^{-1} }   (g_{\epsilon\,*}q_1(0))\,,\]
	then evaluating at the identity.
\end{exercise} 
	
\begin{answer}
	To prove the Lie chain rule $\frac{d}{d\epsilon} (g_{\epsilon\,*}q_1(0)) 
	= - \pounds_{\frac{dg}{d\epsilon}  g_\epsilon^{-1}}  \,g_{\epsilon\,*}q_1(0)$, one computes
	\begin{align*}
	\frac{d}{d\epsilon} q_1(0) g_\epsilon^{-1} 
	&= - \Big[ q_1(0)g_\epsilon^{-1} \frac{dg}{d\epsilon}  g_\epsilon^{-1} \Big]
	= -\Big[ \pounds_{ \frac{dg}{d\epsilon}  g_\epsilon^{-1} }  q_1(0)g_\epsilon^{-1}\Big]
	= - \pounds_{\frac{dg}{d\epsilon}  g_\epsilon^{-1}}  \,g_{\epsilon\,*}q_1(0)
	\,.
	\end{align*}
(The statement above then holds at $\epsilon=0)$.
\end{answer}

\begin{exercise} Write the Noether Hamiltonian in terms of the diamond operator and identify the momentum map.
	\end{exercise} 
	
\begin{answer}	
	By writing
	\[
	N^\xi(q_1,p_1)  = \langle p_1\,,\, -\pounds_\xi  q_1\rangle_{TM} 
	= \langle p_1\diamond q_1\,,\, \xi  \rangle_{\mathfrak{g}} 
	\]
	one identifies the momentum map as $N(q_1,p_1)=p_1\diamond q_1$. 
\end{answer}

\begin{exercise} 
	Compute the canonical transformation generated by the corresponding phase-space Hamiltonian $N^\xi(q_1,p_1)$. 
	\end{exercise} 
		
	\begin{answer}
The canonical transformations may be calculated as 
	\begin{align*}
	\delta q_1 &= \{q_1, N^\xi\} = \frac{\partial N^\xi}{\partial p_1} = -\pounds_\xi  q_1 \,,
	\\
	\delta p_1 &= \{p_1, N^\xi\} = - \,\frac{\partial N^\xi}{\partial q_1} = \pounds_\xi^T  p_1 	\,.
	\end{align*}
	This is the infinitesimal action of the Lie group $G$ on the cotangent space $T^*\mathbb{R}_1^n$ for the first degree of freedom with phase-space coordinates $(q_1,p_1)$.
\end{answer}

\begin{exercise} 
	An infinitesimal left action  of $SO(3)$ on $(\mathbf{q},\mathbf{p})\in\mathbb{R}^3\times \mathbb{R}^3$ is given by
	\[ 
	(\delta \mathbf{q},\delta \mathbf{p}) = (\widehat{\xi}\mathbf{q},\widehat{\xi}\mathbf{p})
	\,.\]
	where $\widehat{\xi}\in \mathfrak{so}(3)$ is a $3\times3$ skew-symmetric matrix. These infinitesimal actions of the Lie algebra 		$\mathfrak{so}(3)$ are given by the vector cross product, as $\mathfrak{so}(3)\simeq \mathbb{R}^3$ via the hat map $\widehat{N}_{ij}=-\,\epsilon_{ijk}{N^k}$ for $k=1,2,3$, as
	\[ 
	(\delta \mathbf{q},\delta \mathbf{p}) 
	= (\widehat{\xi}_{ij}q_j ,\widehat{\xi}_{ij}p_j) 
	= (- \epsilon_{ijk}\xi^kq^j - \epsilon_{ijk}\xi^kp^j)  
	= (\boldsymbol{\xi} \times\mathbf{q}, \boldsymbol{\xi} \times\mathbf{p})
	\,,\]
	Show that the Noether quantity $N^\xi = \mathbf{N}\cdot\bs{\xi}$ with \[\mathbf{N}=\mathbf{q}\times\mathbf{p}\in \mathfrak{so}(3)^*\simeq \mathbb{R}^3\] provides this infinitesimal left action as a symplectic transformation of $(\mathbf{q},\mathbf{p})\in T^*\mathbb{R}^3$. Identify the corresponding conserved quantity. 
 \end{exercise} 
	 
\begin{answer}	
	\begin{align*}
	N^\xi(q,p) := \langle p,\delta q\rangle_{\mathbb{R}^3} &= \mathbf{p}\cdot\delta \mathbf{q}
	\\
	= \mathbf{p}\cdot  \boldsymbol{\xi}  \times \mathbf{q} &= \mathbf{q}\times\mathbf{p}\cdot  \boldsymbol{\xi}   
	=: \mathbf{N}(\mathbf{q},\mathbf{p})\cdot \boldsymbol{\xi}
	\,.
	\end{align*}
The corresponding conserved quantity is the spatial angular momentum. See \dots
\end{answer}

\medskip

\begin{exercise} More about Noether's theorem and Hamiltonian dynamics for two degrees of freedom: 

Previously in an earlier part of this lecture we transformed to the reduced Lagrangian $L(\xi,q_2 , u_2 )$ defined on 
\[\big(T\mathbb{R}_1^n/G\big) \times T\mathbb{R}_2^d \simeq \mathfrak{g} \times T\mathbb{R}_2^d\] and found the EP and EL equations, 
	\[
	\frac{d\mu}{dt}=\mathrm{ad}^*_\xi 	\mu
	\quad\hbox{with}\quad
	\mu := \frac{\partial L}{\partial \xi}
	\]
	and
	\[
	\frac{d p_2}{dt} := \frac{\partial L}{\partial q_2}
	\quad\hbox{with}\quad
	p_2 = \frac{d}{dt}\frac{\partial L}{\partial u_2} \Big|_{u_2=\dot{q}_2}
	\,.
	\] 
	\end{exercise} 
	 

\begin{exercise}
	Compute the partial derivatives of the corresponding Hamiltonian by Legendre transforming to
	$(T^*\mathbb{R}_1^n/G\big) \times T^*\mathbb{R}_2^d \simeq \mathfrak{g}^* \times T\mathbb{R}_2^d$
	via 
	\[
	H(\mu;q_2,p_2) = \langle \mu,\xi\rangle_\mathfrak{g} + \langle p_2 , u_2 \rangle - L(\xi,q_2 , u_2 )
	\,.\] 
	\end{exercise} 
	 
\begin{answer}
	The Legendre transform yields the following partial derivatives of the corresponding Hamiltonian
	\begin{align*}
	dH(\mu;q_2,p_2) &= \Big\langle d\mu,\xi \Big\rangle_\mathfrak{g}
	+ \Big\langle \mu - \frac{\partial L}{\partial \xi} , d\xi \Big\rangle_\mathfrak{g}
	\\&+ \langle dp_2 , u_2 \rangle + \Big\langle p_2 - \frac{\partial L}{\partial u_2} , du_2 \Big\rangle
	- \Big\langle   \frac{\partial L}{\partial q_2} , dq_2 \Big\rangle
	\\&= \Big\langle d\mu, \frac{\partial H}{\partial \mu} \Big\rangle_\mathfrak{g} 
	+ \Big\langle dp_2 , \frac{\partial H}{\partial p_2}  \Big\rangle
	+ \Big\langle dq_2 , \frac{\partial H}{\partial q_2}  \Big\rangle
	\,.
	\end{align*}
\end{answer}

\begin{exercise}
	Show the following relations by equating like terms in the partial derivatives of the Lagrangian and the corresponding Hamiltonian, 
	\[
	\frac{\partial H}{\partial \mu} = \xi \,,\quad
	\frac{\partial H}{\partial p_2} = u_2\,,\quad
	\frac{\partial H}{\partial q_2} = -\,\frac{\partial L}{\partial q_2}\,,\quad
	\]
	with
	\[p_2 = \frac{\partial L}{\partial u_2} 
	\quad\hbox{and}\quad
	\mu = \frac{\partial L}{\partial \xi} 
	\,,\]
	so that the EL and EP equations imply 
	\[
	EL:\ \frac{\partial L}{\partial q_2} = \frac{dp_2}{dt} = -\, \frac{\partial H}{\partial q_2}
	\,,\quad
	u_2 = \frac{dq_2}{dt} = \frac{\partial H}{\partial p_2}
	\,,\]
	and 
	\[
	\frac{d\mu}{dt}=\mathrm{ad}^*_{\frac{\partial H}{\partial \mu}} \mu	
	\,.\]
	\end{exercise}

\begin{exercise}
	Compute the following sum of a Lie-Poisson (LP) bracket and a symplectic Poisson bracket by expanding out 
	the time derivative \\
	$dF(\mu,q_2,p_2)/dt$ as
	\begin{align*}
	\frac{dF}{dt}  
	&= \left\langle  \mathrm{ad}^*_{\frac{\partial H}{\partial \mu}}\mu ,\frac{\partial F}{\partial \mu} \right\rangle_\mathfrak{g} 
	+ \left\langle \frac{\partial H}{\partial p_2}  ,\frac{\partial F}{\partial q_2} \right\rangle_{T\mathbb{R}_2}
	- \left\langle \frac{\partial H}{\partial q_2}  ,\frac{\partial F}{\partial p_2} \right\rangle_{T\mathbb{R}_2}
	\\&=: \{F,H\}\,.
	\end{align*}
	\end{exercise}

\begin{exercise}
	Write the combined Poisson bracket as a block diagonal combination of Lie-Poisson and symplectic Poisson brackets,
	\begin{align*}
	\frac{dF}{dt} &= Tr \left(\begin{bmatrix}\frac{\partial F}{\partial \mu} \\
	\\ \frac{\partial F}{\partial q_2} \\ \\ \frac{\partial F}{\partial p_2} \end{bmatrix}^T
	\begin{bmatrix} \mathrm{ad}^*_{\Box} \mu &  0  & 0 
	\\ \\
	0 & 0 & 1
	\\ \\
	0 & -1 & 0
	\end{bmatrix} 
	\begin{bmatrix}\frac{\partial H}{\partial \mu} \\
	\\ \frac{\partial H}{\partial q_2} \\ \\ \frac{\partial H}{\partial p_2} \end{bmatrix}\right)
	\\&=
	-\left\langle \mu , \left[ \frac{\partial F}{\partial \mu}\,,\,\frac{\partial H}{\partial \mu}\right] \right\rangle_\mathfrak{g} 
	+
	\frac{\partial H}{\partial p_2}\frac{\partial F}{\partial q_2}-\frac{\partial H}{\partial q_2}\frac{\partial F}{\partial p_2}
	=: \{F,H\}
	\,.\end{align*}
	\end{exercise}
	 
\begin{exercise}
Prove that the Jacobi identity holds for this block-diagonal Poisson operator.  
\end{exercise}
	 \begin{answer}	
The Lie-Poisson bracket holds for the first term, because it is a linear functional of the Lie bracket of Hamiltonian vector fields. The second part of the bracket is canonical and is independent from the first part. Thus, the sum of the two Poisson brackets also satisfies the Jacobi identity. 
\end{answer}	


 \begin{exercise}
	Show by direct calculation that Hamilton's canonical equations yield the block-diagonal Poisson operator above when the Hamiltonian is given by $H(\mu = p_1 \diamond q_1;q_2,p_2)$.
	
That is, when the $(q_1,p_1)\in T^*\mathbb{R}_1^n$ dependence of the original Hamiltonian depends only on the Noether quantity $\mu = p_1 \diamond q_1\in \mathfrak{g}^*$, show that its evolution is given by the LP equation, 
	\[
	\frac{d\mu}{dt}=\mathrm{ad}^*_\xi 	\mu
	\quad\hbox{with}\quad
	\mu = p_1 \diamond q_1
	\quad\hbox{and}\quad
	\xi = \frac{\partial H(\mu;q_2,p_2)}{\partial \mu} 
	\,.\]
	In this case, the Hamiltonian $H(\mu = p_1 \diamond q_1;q_2,p_2): \mathfrak{g}^*\times T^*\mathbb{R}_2^d$ is said to have \emph{collectivised} in the $T^*\mathbb{R}_1^n$ phase-space variables and remained symplectic in the $T^*\mathbb{R}_2^d$ phase-space variables. 
	\end{exercise}
	
	 
\begin{answer}	 
\begin{proof}
For an arbitrary fixed $\eta\in \mathfrak{X}$, one computes the pairing
\begin{align}
\begin{split}
\left\langle 
\p_t \mu \,,\, \eta 
\right\rangle_{\mathfrak{X}}
&=   
\left\langle 
\p_t p\diamond q + p\diamond \p_t q\,,\, \eta 
\right\rangle_{\mathfrak{X}}
\\
&=   
\left\langle 
(\pounds_{\xi}^Tp) \diamond q 
- p\diamond \pounds_{\xi} q\,,\, \eta 
\right\rangle_{\mathfrak{X}}
\\&=   
\left\langle 
p\,,\, (-\pounds_{\xi} \pounds_{\eta} + \pounds_{\eta} \pounds_{\xi} )q\,
\right\rangle_{V}
\\&=   
\left\langle 
p\,,\, -\,\big({\rm ad}_{\xi}{\eta}\big)q\,
\right\rangle_{V}
=
\left\langle 
p\diamond q\,,\, {\rm ad}_{\xi}{\eta}\,
\right\rangle_{\mathfrak{X}}
\\&=
\left\langle 
 {\rm ad}^*_{\xi}(p\diamond q)\,,\,{\eta}\,
\right\rangle_{\mathfrak{X}}
=
\Big\langle 
 \pounds_{\xi}\mu\,,\,{\eta}\,
\Big\rangle_{\mathfrak{X}}\,.
\end{split}
\label{calc-lem}
\end{align}
Since $\eta\in \mathfrak{X}$ was arbitrary, the last line completes the proof of the Lemma. 
In the last step we have also used the fact that coadjoint action is identical to Lie-derivative action 
for vector fields acting on 1-form densities such as $m=p\diamond q$. 
\end{proof}

\end{answer}


\begin{exercise} 

{\bf $G=GL(n,\mathbb{R})\circledS\mathbb{R}^n$ affine invariant motions}\smallskip

Begin with the Lagrangian given by the kinetic energy of the Fisher-Rao metric 
$g(TG)$  for multivariate Gaussian probability densities \cite{barbaresco2020lie}
\[
L(S,\dot{S},\mathbf{\dot{q}})
=\frac{1}{2}\,{\rm tr}\Big(\dot{S}S^{-1}\dot{S}S^{-1}\Big)
+
\frac{1}{2}\,\mathbf{\dot{q}}^T S^{-1}\mathbf{\dot{q}}
\,.\]
Here $S$ is an $n\times n$ symmetric matrix and
$\mathbf{q}\in\mathbb{R}^n$ is an
$n-$component column vector. The quantities $S$ and $\mathbf{q}$ represent, respectively, the covariance matrix and 
the mean of the Gaussian probability distribution. 
Conveniently, the Lagrangian $L(S,\dot{S},\mathbf{\dot{q}})$ is independent of the coordinate $\mathbf{q}$.
\begin{description}
\item [(a)] 
Calculate the Euler-Lagrange equations for this Lagrangian. 
\item [(b)]
Legendre transform to find the Hamiltonian for this system and write its \emph{canonical} equations.
\item [(c)]
Show that the Lagrangian and Hamiltonian for this system are both  invariant under the group action 
of the general affine group $G=GL(n)\circledS\mathbb{R}^n$
\[
\mathbf{q}\to G\mathbf{q}
\quad\hbox{and}\quad
S\to GSG^T
\]
for any constant invertible $n\times n$ matrix, $G$.
\item [(d)]
\begin{description}
\item [(i)] 
Linearise this group action around the identity in terms of $A=G'G^{-1}$ and construct the infinitesimal transformations $X_A\mathbf{q}$ and $X_AS$ for the linearised action of $G$ on the configuration space $(\mathbf{q},S)$. 
\item [(ii)] 
Find the phase space function (\emph{infinitesimal generator}) whose canonical Poisson brackets produce these infinitesimal transformations by pairing $X_A\mathbf{q}$ and $X_AS$ with the corresponding canonical momenta and summing.
\item [(iii)] 
Compute the Poisson bracket of the canonical momenta with the infinitesimal generator. (This is the \emph{cotangent lift} to the full phase space of the infinitesimal action of $G$ on the configuration space.)
\end{description}

\item [(e)]
\begin{description}
\item [(i)] 
Verify directly that the infinitesimal generator of the $G$-action is a conserved $n\times n$ matrix quantity by using the equations of motion. 
\item [(ii)] Determine whether this Hamiltonian system has sufficiently many conservation laws in involution to be completely integrable, for any dimension $n$. 
\end{description}
\end{description}

\end{exercise} 



\begin{answer}
\begin{enumerate}[(a)] 
\item
The Euler-Lagrange equations for this Lagrangian are
\[
\ddot{S} + \mathbf{q}\otimes \mathbf{q}^T  - \dot{S}S^{-1}\dot{S} = 0
\,,\quad\hbox{and}\quad
\mathbf{\ddot{q}} + \dot{S}S^{-1}\mathbf{\dot{q}} = 0
\,.\]
\item The Legendre transform yields fibre derivatives,
\[
P=\frac{\partial L}{\partial \dot{S}}
=S^{-1}\dot{S}S^{-1}
\quad\hbox{and}\quad
\mathbf{p}
=\frac{\partial L}{\partial \mathbf{\dot{q}}}
=S^{-1}\mathbf{\dot{q}}
\]
Thus, the Hamiltonian $H(\mathbf{q},\mathbf{p},S,P)$ is
\[
H(\mathbf{q},\mathbf{p},S,P)
=\frac{1}{2}{\rm tr}\,\big(PS\cdot PS\big) 
+ \frac{1}{2}\mathbf{p}\cdot S\mathbf{p}
\,,
\]
 and its canonical Hamilton equations are:
\[
\dot{S}=\frac{\partial H}{\partial P}=SPS
\,,\quad
\dot{P}
=-\,\frac{\partial H}{\partial S}
=-\Big(
PSP+\frac{1}{2}\mathbf{p}\otimes\mathbf{p}
\Big),
\]
\[
\mathbf{\dot{q}}=\frac{\partial H}{\partial \mathbf{p}}=S\mathbf{p}
\,,\quad
\mathbf{\dot{p}}=\frac{\partial H}{\partial \mathbf{q}}=0
\,.
\]
\item  
Under the $G=GL(n,\mathbb{R})\circledS\mathbb{R}^n$ affine group action 
$
\mathbf{q}\to G\mathbf{q}
\quad\hbox{and}\quad
S\to GSG^T
$
for any constant invertible $n\times n$ matrix, $G$,
one finds $\dot{S}S^{-1}\to G\dot{S}S^{-1}G^{-1}$ and 
$\mathbf{\dot{q}}\cdot S^{-1}\mathbf{\dot{q}}
\to \mathbf{\dot{q}}\cdot S^{-1}\mathbf{\dot{q}}$. Hence, $L\to L$ and this Lagrangian is invariant.

Likewise, $P\to G^{-T}PG^{-1}$ so $PS\to G^{-T}PSG^{T}$ and
$\mathbf{p}\to G^{-T}\mathbf{p}$ so that $S\mathbf{p}\to GS\mathbf{p}$.
Hence, $H\to H$, as well; so both $L$ and $H$ for this system are
invariant under the $G$ affine group action. 
\item
\begin{enumerate}[(i)]
\item  
The infinitesimal actions for 
$G(\epsilon)=Id+\epsilon{A}+O(\epsilon^2)$,
where $A\in \mathfrak{g}(n)$ are 
\[
X_A\mathbf{q}
=\frac{d}{d\epsilon}\Big|_{\epsilon=0}
G(\epsilon)\mathbf{q}
=A\mathbf{q}
\quad\hbox{and}\quad
X_AS
=\frac{d}{d\epsilon}\Big|_{\epsilon=0}
\Big(G(\epsilon)SG(\epsilon)^T\Big)
=AS+SA^T
\]
\item 
Pairing $X_A\mathbf{q}$ and $X_AS$ with their corresponding canonical momenta and 
summing using $P^T=P$ yields
\[
\langle J,A\rangle
:=
{\rm tr}\,(PX_AS)+\mathbf{p}\cdot X_A\mathbf{q}
=
{\rm tr}\,\big(P(AS+SA^T)\big)+\mathbf{p}\cdot A\mathbf{q}
\]
Hence,
\[
\langle J,A\rangle
:=
{\rm tr}\,\big(JA^T\,\big)
=
{\rm tr}\,\big((2SP+\mathbf{q}\otimes\mathbf{p})A\big)
\,,\quad\hbox{so}\quad
J=(2PS+\mathbf{p}\otimes\mathbf{q})
\]
where $J: T^*(S, \mathbf{q})\to \mathfrak{g}^*$
is the momentum map of the cotangent lifted action of $G(n)$ 
relative to the canonical symplectic form.
\item
For any choice of the matrix $A$, the Poisson bracket with $\langle J,A\rangle$ generates the Hamiltonian vector field
\begin{eqnarray*}
\Big\{\,\cdot\,,\langle J,A\rangle \Big\}
&=&
{\rm tr}\,\Big(\frac{\partial \langle J,A\rangle}{ \partial P}
\frac{\partial }{ \partial S}\Big)
+
\frac{\partial \langle J,A\rangle}{ \partial \mathbf{p}}
\cdot\frac{\partial }{ \partial \mathbf{q}}
\\&&
-\
{\rm tr}\,\Big(\frac{\partial \langle J,A\rangle}{ \partial S}
\frac{\partial }{ \partial P}\Big)
-
\frac{\partial \langle J,A\rangle}{ \partial \mathbf{q}}
\cdot\frac{\partial }{ \partial \mathbf{p}}
\\
&=&
{\rm tr}\,\Big((AS+SA^T)
\frac{\partial }{ \partial S}\Big)
+
A\mathbf{q}
\cdot\frac{\partial }{ \partial \mathbf{q}}
\\&&
-\
{\rm tr}\,\Big((PA+A^TP)
\frac{\partial }{ \partial P}\Big)
-
A^T\mathbf{p}
\cdot\frac{\partial }{ \partial \mathbf{p}}
\end{eqnarray*}
which recovers the infinitesimal action on $(S,\mathbf{q})$ and provides the cotangent-lifted 
infinitesimal action on the canonical momenta $(P,\mathbf{p})$.
\end{enumerate}
\item
\begin{enumerate}[(i)] 
\item 
Conservation of $\langle J,A\rangle$ is verified directly in
\[
\frac{d}{dt}\langle J,A\rangle=\langle \dot{J},A\rangle
\]
by computing
\begin{eqnarray*}
\dot{J} &=& 
\Big(2\dot{P}S+2P\dot{S}
+\mathbf{\dot{p}} \otimes \mathbf{q}
+\mathbf{p}\otimes\mathbf{\dot{q}} \Big)
\\
&=&
-\Big(2PSP+(\mathbf{p}\otimes\mathbf{p})\Big)S
+2P\Big(SPS\Big)
+\mathbf{0} \otimes \mathbf{q}
+\mathbf{p}\otimes S\mathbf{p}
\\
&=&
0
\,.
\end{eqnarray*}
\item
The system has $n(n+1)/2 + n=n(n+3)/2$ degrees of freedom. It conserves the $n$ components of linear momentum $\mathbf{p}$ and the $n(n+1)/2$ components of $J$. Thus, there is one constant of motion for each degree of freedom. 

However, these two sets of independent conservation laws do not Poisson commute, since
\[
\Big\{\,\mathbf{p}\,,\langle J,A\rangle \Big\}
=
-
A^T\mathbf{p}
\,.
\]
This means that the naive count of degrees of freedom will not produce  complete integrability, because the momentum map constants of motion arising from Noether's theorem are not in involution. In general, something more would be needed for complete integrability of this system to hold.  This is a potential research question. For more discussion of the geometric dynamics based on the Fisher-Rao metric in probability theory, see \cite{barbaresco2020lie}.
	\end{enumerate}
\end{enumerate}
\end{answer}


\begin{exercise}\label{ex-28}
\text{Hamiltonian symmetry reduction by stages}\\
\begin{enumerate}[(a)]
\item
Write Hamilton's equations on $\mathfrak{so}^*(4)\simeq \mathfrak{so}(3)^*\times \mathfrak{so}(3)^*$ using the Poisson brackets 
\begin{eqnarray*}
\{M_i,\,M_j\}  = \epsilon_{ijk}M_k
\,,\quad
\{\,N_i,\,N_j\,\}  = \epsilon_{ijk}N_k
\,, \quad
\{M_i,\,N_j\}  = 0
\,,
\end{eqnarray*}

among the components of the $\mathbb{R}^3$ vectors $\mathbf{M}$ and $\mathbf{N}$.
\item 
Compute the equations of motion and identify the functionally independent conserved quantities for the following two Hamiltonians
\begin{eqnarray}
H_1=   \mathbf{\hat{z}}\cdot(\mathbf{M}\times\mathbf{N}) 
\quad\hbox{and}\quad
H_2= \mathbf{M}\cdot\mathbf{N}
\,.
\label{LJ-Ham}
\end{eqnarray}
\item 
Determine whether these Hamiltonians have sufficiently many symmetries and associated conservation laws to be completely integrable (i.e., reducible to Hamilton's canonical equations for a single degree of freedom) and explain why. 
\item 
Transform the Hamiltonians in (\ref{LJ-Ham}) from Cartesian components of the vectors $(\mathbf{M},\mathbf{N})\in\mathbb{R}^3\times\mathbb{R}^3$ into spherical coordinates $(\theta,\phi)\in S^2$ and $(\bar\theta,\bar\phi)\in S^2$, respectively.
\item 
Use the $S^1$ symmetries and their associated  conservation laws to reduce the dynamics in $\mathbb{R}^3\times\mathbb{R}^3$ to canonical Hamiltonian equations first on $S^2\times S^2$ and then on $S^2$ by a two-stage sequence of canonical transformations.
\end{enumerate}
\end{exercise}

\begin{answer}
\rm
\begin{enumerate}[(a)]
\item
The Hamiltonian equations for this system are 
\[
\mathbf{\dot{M}}=\mathbf{M}\times \frac{\partial H}{\partial \mathbf{M}}
\quad\hbox{and}\quad
\mathbf{\dot{N}}=\mathbf{N}\times \frac{\partial H}{\partial \mathbf{N}}
\]
\item
The system with Hamiltonian $H_1=   \mathbf{\hat{z}}\cdot(\mathbf{M}\times\mathbf{N})$ conserves 
\[
|\mathbf{M}|^2\,,\quad |\mathbf{N}|^2\quad\hbox{and}\quad L_3=M_3+N_3
\,.
\]
The first two conservation laws reduce the problem to $S^2\times S^2$ and the last one provides a further $SO(2)$ symmetry under simultaneous rotation of each of the spheres about its vertical 3-axis. As we shall see, this symmetry and its conservation law are enough to reduce $S^2\times S^2$ to $S^2$ and thereby make the system completely integrable. 
\item
The system with Hamiltonian $H_2= \mathbf{M}\cdot\mathbf{N}$ conserves $|\mathbf{M}|^2$, $|\mathbf{N}|^2$  and all the components of  $\mathbf{L}=\mathbf{M}+\mathbf{N}$. These conserved quantities are \textit{not} all functionally independent, since
\[
|\mathbf{L}|^2=|\mathbf{M}+\mathbf{N}|^2
= 
 |\mathbf{M}|^2 + |\mathbf{N}|^2 + 2\,\mathbf{M}\cdot\mathbf{N}
\,.
\]
However, enough symmetry still remains for these equations to be integrated by employing $L_3=M_3+N_3$ and its associated $SO(2)$ symmetry for simultaneous rotation of each of the spheres about its vertical 3-axis. This symmetry reduces its $S^2\times S^2$ phase space to $S^2$ and thereby allows it to be integrated as before. 
\item
The vectors $\mathbf{M}$ and $\mathbf{N}$ may be written in spherical coordinates $(\theta,\phi)$ and $(\bar\theta,\bar\phi)$, respectively, as
\begin{eqnarray*}
\mathbf{M} &=& (M_1,M_2,M_3)^T
\\
 &=& M(\sin\theta\cos\phi,\sin\theta\sin\phi,\cos\theta)^T
\,,\\
\mathbf{N} &=&  (N_1,N_2,N_3)^T
\\
 &=& N(\sin\bar\theta\cos \bar\phi,\sin \bar\theta\sin \bar\phi,\cos \bar\theta)^T
\,.\end{eqnarray*}
In terms of these variables we may write 
\[
H_1=M_1N_2-M_2N_1
\hbox{ and }
H_2= M_1N_1+M_2N_2+M_3N_3
\,.
\]
\item
Reduction $S^2\times S^2\to S^2$ may be accomplished by a canonical transformation using conservation of $|\mathbf{M}|^2$, $|\mathbf{N}|^2$ and $L_3=M_3+N_3$. The symplectic form on $S^2\times S^2$ is given in spherical coordinates by
\begin{eqnarray}
\omega = M^2 d\cos\theta\wedge d\phi + N^2 d\cos\bar\theta\wedge d\bar\phi 
\,.
\label{LJ-sphere}
\end{eqnarray}
We transform to weighted sum and difference variables by
\begin{eqnarray*}
\sqrt{2}\lambda &=& M_3+N_3 = M\cos\theta + N\cos\bar\theta
\,,\qquad
\sqrt{2}\alpha = M\phi + N\bar\phi
\,,\\
\sqrt{2}\kappa &=& M_3-N_3 = M\cos\theta - N\cos\bar\theta
\,,\qquad
\sqrt{2}\beta = M\phi - N\bar\phi
\,.
\label{sum-diff}
\end{eqnarray*}
This transformation is canonical and yields the new symplectic form,
\begin{eqnarray}
\omega = d\kappa\wedge d\beta + d\lambda\wedge d\alpha  
\,.
\label{LJ-symp}
\end{eqnarray}
Expressing the Hamiltonians $H_1$ and $H_2$ in terms of these new canonical variables reduces the problem to the $(\kappa, \beta )$ phase plane, with motion parameterised by the 3rd component of total angular momentum $\lambda$ and independent of its canonically conjugate angle, $\alpha$. In each case Hamilton's canonical equations separate into reduced dynamics on $S^2$, plus reconstruction of the phase, $\alpha \in S^1$,
\[
\underbrace{\
\dot{\kappa} = -\, \frac{\partial H}{\partial \beta}
\,,\quad
\dot{\beta} =  \frac{\partial H}{\partial \kappa}
\,,\
}_{\hbox{\rm Reduced dynamics on $S^2$}}
\qquad
\underbrace{\
\dot{\lambda} = -\, \frac{\partial H}{\partial \alpha} =0
\,,\quad
\dot{\alpha} =  \frac{\partial H}{\partial \lambda}\
}_{\hbox{\rm Reconstruction of the phase, $\alpha $}}
\]
\end{enumerate} 
\end{answer}


\subsection{Composition of maps and Lagrangian reduction by stages}\label{subsec-CoM-LagRedux}


\begin{exercise}
 Reduction by stages of the composition of two non-commutative left group actions $G_1\times G_2$.

Consider the following Hamilton's principle on the product space ($TG_1\times TG_2)$ and
verify the symmetry reduction by stages  $TG_1/G_1\times (TG_2/G_2)/G_1$.
\begin{align}
\begin{split}
S &= \int_{t_1}^{t_2}
L(g_1,\dot{g}_1;  g_2,\dot{g}_2)\,dt
\\
 &= \int_{t_1}^{t_2}
L\big(g_1^{-1}\dot{g}_1, \Ad_{g_1^{-1}}( g_2^{-1}\dot{g}_2)\big)\,dt
\\
 &=: \int_{t_1}^{t_2}
L( \Omega_1,  \Omega_2 )\,dt
\end{split}
\label{CoM-var}
\end{align}
where one introduces the following two velocities \cite{holm2023lagrangian}
\[\Omega_1=g_1^{-1}\dot{g}_1
\,,\ \ 
\Omega_2 = g_1^{-1}(\dot{g}_2g_2^{-1})g_1 =: \Ad_{g_1^{-1}}( g_2^{-1}\dot{g}_2)
\,,
\]
for the nested symmetries of ($TG_1\times TG_2)$ under the respective left actions of $G_1\times G_2$. 
\end{exercise}



\begin{exercise}
After verifying the line of reasoning in deriving the Lagrangian shown above, 
prove the following equations for iterated left actions leading to semidirect-product Lie algebras,
with \cite{holm2023lagrangian} \index{Lie algebra!semidirect product}
\index{semidirect product!broken symmetry} \index{broken symmetry!semidirect product}
\[
\omega_1= g_1^{-1} \delta g_1 =: g_1^{-1}g'_1 
\,,\quad
\omega_2 = \Ad_{g_1^{-1}}( g_2^{-1}g'_2)
\,.\]
\begin{align}
\begin{split}
\Omega'_1 - \dot{\omega}_1 &=    \ad_{\Omega_1} \omega_1 := [\Omega_1 , \omega_1]
\,,\\
\Omega'_2 -\dot{\omega}_2 
  &=  \ad_{\Omega_2}\omega_2 + \ad_{\Omega_2} \omega_1 + \ad_{\Omega_1} \omega_2
\\&= \ad_{\Omega_1 + \Omega_2}\omega_2 + \ad_{\Omega_2} \omega_1
\,.
\end{split} 
\label{SDP-Iterates12}
\end{align} 
\end{exercise}


\begin{answer}
Recall the definitions,
\[\Omega_1=g_1^{-1}\dot{g}_1 
\,,\ \ 
\Omega_2 =:: \Ad_{g_1^{-1}}( g_2^{-1}\dot{g}_2)
\,.
\]
The first relation has already been derived in equation \eqref{EP-Var-Id}.
\[
\Omega'_1  =  (g_1'g_1^{-1})\dot{\phantom{\,}} - \ad_{\dot{g}_1g_1^{-1}} g'_1g_1^{-1}
= \dot{\omega}_1 - \ad_{\Omega_1} \omega_1
\,.
\]
One derives the second formula in \eqref{SDP-Iterates12} by direct calculation of $\Omega'_2$ and $\dot{\omega}_2 $ , as
 follows. 
\begin{align*}
\begin{split}
\Omega'_2   &= \Ad_{g_1^{-1}}( g_2^{-1}\dot{g}_2)'
+ \big[\omega_2,\Omega_1\big]
\\
\dot{\omega}_2   &= \Ad_{g_1^{-1}}( g_2^{-1}g'_2)\dot{\phantom{\,}}
+ \big[\omega_1,\Omega_2\big]
\,.
\end{split} 
\end{align*} 
Consequently, the difference is given by
\begin{align*}
\Omega'_2 - \dot{\omega}_2 &= 
\Ad_{g_1^{-1}} \big( ( g_2^{-1}\dot{g}_2)'  - ( g_2^{-1}g'_2)\dot{\phantom{\,}} \big)
+ \big[\omega_2,\Omega_1\big] - \big[\omega_1,\Omega_2\big]
\\&= [\Omega_2,\omega_2]
+ \big[\Omega_2,\omega_1\big] + \big[\Omega_1,\omega_2\big]
=  \ad_{\Omega_1 + \Omega_2}\omega_2 + \ad_{\Omega_2} \omega_1
\,.\end{align*} 
This is how infinitesimal transformations compose under iteration of left semidirect-product 
Lie algebra action,  \cite{bruveris2011momentum,Holm_Book2_2008,holm2023lagrangian}. 
\end{answer}

\index{Lie algebra!semidirect product}



\begin{exercise}
 Calculate the Euler--Poincar\'e equations of motion resulting from 
 \[
 0 = \delta S = \delta  \int_{t_1}^{t_2}
L( \Omega_1,  \Omega_2 )\,dt
 \]
Write these Euler--Poincar\'e equations of motion in Lie-Poisson form.
\end{exercise} 


\begin{answer} 
The Euler--Poincar\'e equations of motion resulting from Hamilton's principle 
can be read off from the following variational calculation, 
\begin{align*}
 0 &= \delta S = \delta \!\! \int_{t_1}^{t_2}  \!\! L( \Omega_1,  \Omega_2 )\,dt
 = \int_{t_1}^{t_2}  \!\! \scp{\frac{\delta L}{\delta \Omega_1}}{\delta\Omega_1}
 +  \scp{\frac{\delta L}{\delta \Omega_2}}{\delta\Omega_2}\,dt
 \\& 
=  \int_{t_1}^{t_2} \scp{\frac{\delta L}{\delta \Omega_1}}{\dot{\omega}_1 + \ad_{\Omega_1} \omega_1}
 +  \scp{\frac{\delta L}{\delta \Omega_2}}{\dot{\omega}_2 + \ad_{\Omega_1 + \Omega_2}\omega_2 + \ad_{\Omega_2} \omega_1}\,dt
 \\& 
 =  \int_{t_1}^{t_2} \scp{ -\,\big(\p_t -  \ad^*_{\Omega_1} \big)\frac{\delta L}{\delta \Omega_1}
+ \ad^*_{\Omega_2}\frac{\delta L}{\delta \Omega_2}}{ \omega_1 }
+ \scp{-\big( \p_t - \ad^*_{\Omega_1 + \Omega_2}\big) \frac{\delta L}{\delta \Omega_2} } {\omega_2} \,dt
\,.\end{align*} 
The Lie-Poisson form of these Euler--Poincar\'e equations of motion is given in terms of angular moments 
defined as 
\[
\Pi_1 := \frac{\delta L}{\delta \Omega_1}
\quad\hbox{and}\quad 
\Pi_2 := \frac{\delta L}{\delta \Omega_2}
\,.\]
Namely, the Lie-Poisson form is the following.
\begin{align}
\p_t\begin{pmatrix} \Pi_1 \\   \Pi_2  \end{pmatrix}
&= 
	\begin{bmatrix} 
	\mathrm{ad}^*_{\Box} \Pi_1 &  \mathrm{ad}^*_{\Box} \Pi_2  
	\\ 
	\mathrm{ad}^*_{\Box} \Pi_2 & \mathrm{ad}^*_{\Box} \Pi_2 ]
	\end{bmatrix} 
  \begin{pmatrix}
    {\delta h}/{\delta \Pi_1} = \Omega_1 \\ 
    {\delta h}/{\delta \Pi_2} = \Omega_2 
    \end{pmatrix}
.
\label{2rotors-LPHam}
\end{align} 
If one transforms variables by the linear transformation $(\Pi_1,\Pi_2)\to (\Pi_1-\Pi_2,\Pi_2)$
them the Poisson matrix in \eqref{2rotors-LPHam} transforms to
\[
\begin{bmatrix}
1 & -1 \\ 0 & 1
\end{bmatrix}
	\begin{bmatrix} 
	\mathrm{ad}^*_{\Box} \Pi_1 &  \mathrm{ad}^*_{\Box} \Pi_2  
	\\ 
	\mathrm{ad}^*_{\Box} \Pi_2 & \mathrm{ad}^*_{\Box} \Pi_2 ]
	\end{bmatrix} 
\begin{bmatrix}
1 & 0 \\ -1 & 1
\end{bmatrix}
=
	\begin{bmatrix} 
	\mathrm{ad}^*_{\Box} (\Pi_1-\Pi_2) & 0 
	\\ 
						0  & \mathrm{ad}^*_{\Box} \Pi_2 ]
	\end{bmatrix} ,
\]
which is the same Lie-Poisson bracket as appears in Exercise \ref{ex-28}, upon identifying 
$G_1\times G_2 \to SO(3)\times SO(3)$,  $(\Pi_1-\Pi_2)\to M$, and $\Pi_2\to N$.
\end{answer}



\begin{exercise}
 Reduction by stages of the composition of three non-commutative \emph{left} group actions $G_1\times G_2\times G_3$.

Consider the following Hamilton's principle on the product space ($TG_1\times TG_2\times TG_3)$ and
verify the symmetry reduction by stages  $TG_1/G_1\times (TG_2/G_2)/G_1\times ((TG_3/G_3)/G_2)/G_1)$ 
\begin{align}
\begin{split}
S &= \int_{t_1}^{t_2}
L(g_1,\dot{g}_1;  g_2,\dot{g}_2;  g_3,\dot{g}_3)\,dt
\\
 &= \int_{t_1}^{t_2}
L( g_1^{-1}\dot{g}_1 \,; \Ad_{g_1^{-1}}( g_2^{-1}\dot{g}_2) \,;  \Ad_{g_1^{-1}}\Ad_{g_2^{-1}}(\dot{g}_3g_3^{-1})\,dt
\\
 &=: \int_{t_1}^{t_2}
L( \Omega_1; \Omega_2;  \Omega_3 )\,dt
\end{split}
\label{CoM-angvar}
\end{align}
where one introduces sequential $\Ad$ actions to define three velocities 
\begin{align}
\Omega_1 := g_1^{-1}\dot{g}_1  
\,,\ \ 
\Omega_2 :=  \Ad_{g_1^{-1}}( g_2^{-1}\dot{g}_2)
\,,\ \
\Omega_3 := \Ad_{g_1^{-1}}\Ad_{g_2^{-1}}(\dot{g}_3g_3^{-1})
\,,
\label{CoM-angveloc}
\end{align}
for the nested symmetries of ($TG_1\times TG_2\times TG_3)$ under the respective left actions of $G_1\times G_2\times G_3$. 
\end{exercise}


\begin{exercise}
After verifying the line of reasoning in deriving the Lagrangian shown above, 
introduce variations with sequential $\Ad$ actions, 
\begin{align}
\omega_1 := g_1^{-1}{g'}_1  
\,,\ \ 
\omega_2 :=  \Ad_{g_1^{-1}}( g_2^{-1}{g'}_2)
\,,\ \
\omega_3 := \Ad_{g_1^{-1}}\Ad_{g_2^{-1}}(g_3^{-1}{g'}_3)
\,.
\label{CoM-angvar}
\end{align}
Then prove the following equations for iterated left actions leading to semidirect-product Lie algebras,
\index{semidirect product!broken symmetry} \index{broken symmetry!semidirect product}
\begin{align}
\begin{split}
\Omega'_1 - \dot{\omega}_1 &=   \ad_{\Omega_1} \omega_1
\,,\\
\Omega'_2 - \dot{\omega}_2 
  &=  \ad_{\Omega_2}\omega_2 + \ad_{\Omega_2} \omega_1 + \ad_{\Omega_1} \omega_2
\\&= \ad_{\Omega_1 + \Omega_2}\omega_2 + \, \ad_{\Omega_2} \omega_1
\,,\\
\Omega'_3 - \dot{\omega}_3 
  &= \ad_{\Omega_3}\omega_3 + \ad_{\Omega_3} (\omega_1 + \omega_2) + \ad_{\Omega_1 + \Omega_2} \omega_3
\\&= \ad_{\Omega_1 + \Omega_2 + \Omega_3}\omega_3 + \ad_{\Omega_3} (\omega_1 + \omega_2)
\,.
\end{split} 
\label{SDP-angIterates123}
\end{align} 
The sequence of EP variational relations for compositions of group actions follows a clear pattern \cite{holm2023lagrangian}.
\index{composition of maps}
\end{exercise}


\begin{answer}
The definitions in equations \eqref{CoM-angveloc} and \eqref{CoM-angvar} have already implied 
the first two equations in the set \eqref{SDP-angIterates123} in the previous exercise.

The last equation in \eqref{SDP-angIterates123} is obtained from the following intermediate relations, 
\begin{align}
\begin{split}
\Omega'_3 &= -(\omega_1 + \omega_2 + \omega_3)\Omega_3\,
+ \Ad_{g_1^{-1}}\Ad_{g_2^{-1}}( g_3^{-1}\dot{g}'_3)
+ (\Omega_1 + \Omega_2 )\omega_3
\,,\\
\dot{\omega}_3 &= -(\Omega_1 + \Omega_2 + \Omega_3)\omega_3\,
+ \Ad_{g_1^{-1}}\Ad_{g_2^{-1}}( g_3^{-1}\dot{g}'_3)
+ (\omega_1 + \omega_2 )\Omega_3
\end{split} 
\label{SDP-angIterates3}
\end{align} 
Taking the difference of the previous two equations yields
\begin{align}
\begin{split}
\Omega'_3 - \dot{\omega}_3
&=
\big[ \Omega_1 + \Omega_2 + \Omega_3 \big]\omega_3 + \big[\Omega_3 \,,\, \omega_1 + \omega_2 \big]
\\
&=
\ad_{(\Omega_1 + \Omega_2 + \Omega_3)}\omega_3 + \ad_{\Omega_3 }(\omega_1 + \omega_2)
\end{split}
\label{SDP-angDiff3}
\end{align}

\end{answer}

\begin{exercise}
 Reduction by stages of the composition of three non-commutative right group actions $G_1\times G_2\times G_3$.

Consider the following Hamilton's principle on the product space ($TG_1\times TG_2\times TG_3)$ and
verify the symmetry reduction by stages  $TG_1/G_1\times (TG_2/G_2)/G_1\times ((TG_3/G_3)/G_2)/G_1)$ 
\begin{align}
\begin{split}
S &= \int_{t_1}^{t_2}
L(g_1,\dot{g}_1;  g_2,\dot{g}_2;  g_3,\dot{g}_3)\,dt
\\
 &= \int_{t_1}^{t_2}
L(\dot{g}_1g_1^{-1}; ( \dot{g}_2g_2^{-1})g_1^{-1};  (\dot{g}_3g_3^{-1})g_2^{-1}g_1^{-1})\,dt
\\
 &=: \int_{t_1}^{t_2}
L( u_1; u_2;  u_3 )\,dt
\end{split}
\label{CoM-var}
\end{align}
where $g_{1*}$, resp. $(g_1g_2 )_{*}$ denotes push-forward by $g_1$, resp.  $g_1g_2$, and one introduces the following three velocities 
\[u_1=\dot{g}_1g_1^{-1}
\,,\ \ 
u_2=(\dot{g}_2g_2^{-1})g_1^{-1} =: g_{1*} (\dot{g}_2g_2^{-1})
\,,\ \
u_3=(\dot{g}_3g_3^{-1})g_2^{-1}g_1^{-1} =: g_{1*} g_{2*}  (\dot{g}_3g_3^{-1})
\,,
\]
for the nested symmetries of ($TG_1\times TG_2\times TG_3)$ under the respective right actions of $G_1\times G_2\times G_3$. 
\end{exercise}


\begin{exercise}
After verifying the line of reasoning in deriving the Lagrangian shown above, 
prove the following equations for iterated right actions leading to semidirect-product Lie algebras,
\index{semidirect product!broken symmetry} \index{broken symmetry!semidirect product}
with $w_1=\delta g_1 g_1^{-1}=:g'_1 g_1^{-1}$, $w_2 = g_{1*}g'_2 g_2^{-1}$, $w_3 = g_{1*}g_{2*}g'_3 g_3^{-1}$.
\begin{align}
\begin{split}
u'_1 - \dot{w}_1 &=   - \ad_{u_1} w_1
\,,\\
u'_2 - \dot{w}_2 
  &=  -\,\ad_{u_2}w_2 - \ad_{u_2} w_1 - \ad_{u_1} w_2
\\&= -\,\ad_{u_1 + u_2}w_2 - \, \ad_{u_2} w_1
\,,\\
u'_3 - \dot{w}_3 
  &= -\,\ad_{u_3}w_3 - \ad_{u_3} (w_1 + w_2) - \ad_{u_1 + u_2} w_3
\\&= -\,\ad_{u_1 + u_2 + u_3}w_3 - \ad_{u_3} (w_1 + w_2)
\,.
\end{split} 
\label{SDP-Iterates123}
\end{align} 
The sequence of EP variational relations for compositions of group actions follows a clear pattern.
Note the changes of sign for right action in \eqref{SDP-Iterates123}, relative to equations \eqref{SDP-angIterates3} 
 for left action \cite{holm2023lagrangian}.
\end{exercise}


\begin{answer}
Recalling the definitions,
\[u_1=\dot{g}_1g_1^{-1}
\,,\ \ 
u_2 =: g_{1*} (\dot{g}_2g_2^{-1})
\,,\ \
u_3 =: g_{1*} g_{2*}  (\dot{g}_3g_3^{-1})
\,,
\]
The first relation has already been derived in equation \eqref{EP-Var-Id}.
\[
u'_1  =  (g_1'g_1^{-1})\dot{\phantom{\,}} - \ad_{\dot{g}_1g_1^{-1}} g'_1g_1^{-1}
= \dot{w}_1 - \ad_{u_1} w_1
\,.
\]
One derives the second formula in \eqref{SDP-Iterates123} for $u'_2 - \dot{w}_2 $ by direct calculation. as
 follows. 
\begin{align}
\begin{split}
u'_2  &= (\dot{g}_2g_2^{-1})' g_1^{-1} - \dot{g}_2g_2^{-1}g_1^{-1} g'_1g_1^{-1}
\\&= \big[(g'_2g_2^{-1})\dot{\phantom{\,}} - \ad_{\dot{g}_2g_2^{-1}}g'_2g_2^{-1}\big] g_1^{-1}  
- \pounds_{w_1}u_2
\\&= g_{1*}\big[(g'_2g_2^{-1})\dot{\phantom{\,}} - \ad_{\dot{g}_2g_2^{-1}}g'_2g_2^{-1}\big]   
- \pounds_{w_1}u_2
\\&= \p_t \big( g_{1*}(g'_2g_2^{-1})\big) + \pounds_{\dot{g}_1g_1^{-1}} \big( g_{1*}(g'_2g_2^{-1})\big) 
+  \big(\pounds_{g_{1*}\dot{g}_2g_2^{-1}} \Big)g_{1*} \dot{g}_2g_2^{-1}
- \pounds_{w_1}u_2
\\&= \dot{w}_2  - \ad_{u_1} w_2  -\,\ad_{u_2}w_2 - \ad_{u_2} w_1 
\\&= \dot{w}_2 -\,\ad_{u_1 + u_2}w_2 - \, \ad_{u_2} w_1
\,.
\end{split} 
\label{SDP-Iterates12}
\end{align} 
\begin{remark}\rm 
After proving the first two formulas in \eqref{SDP-Iterates12} for right action
and comparing with \eqref{SDP-angIterates123} for left action the following patterns emerge
for higher iterates,
\begin{align}
\begin{split}
\Omega'_m - \dot{\omega}_m
&=
{\rm ad}_{\sum_{k=1}^{m}\Omega_k}\omega_m
+
{\rm ad}_{\Omega_m} \sum_{k=1}^{m-1}\omega_k
\quad\hbox{for left action}
\,,\\
u'_m - \dot{w}_m
&=
-\,
{\rm ad}_{\sum_{k=1}^{m}u_k}w_m
-\,
{\rm ad}_{u_m} \sum_{k=1}^{m-1}w_k
\quad\hbox{for right action}
\,.
\end{split} 
\label{SDP-rightleft}
\end{align} 
This is how Euler-Poicar\'e variations compose under iteration of semidirect-product 
Lie algebra action,  \cite{bruveris2011momentum,Holm_Book2_2008,holm2023lagrangian}. 
\index{composition of maps!semidirect product!}
\index{semidirect product!broken symmetry} \index{broken symmetry!semidirect product}
\end{remark}
\end{answer}


\begin{exercise}
Show by direct calculation that Hamilton's principle for C$\circ$M variations $\delta u_1, \delta (g_{1*}u_2), \delta ((g_1g_2 )_{*}u_3)$ of the frame-shifted velocities $u_1, g_{1*}u_2;  (g_1g_2 )_{*}u_3$ in the last line of the system \eqref{CoM-var}  yields
\begin{align}
\begin{split}
0 = \delta S_{red} &= \int_{t_1}^{t_2} 
\scp{\frac{\delta \ell}{\delta u_1}}{ (\p_t   - \ad_{u_1}) w_1} 
\\&\qquad 
+ \scp{ \frac{\delta \ell}{\delta \big(g_{1*}u_2\big)}}{ (\p_t   - \ad_{u_1+u_2} ) w_2 - \pounds_{w_1} u_2 } 
\\&\qquad 
+ \scp{ \frac{\delta \ell}{\delta \big((g_1g_2 )_{*}u_3\big)} }{ \big(\p_t  - \ad_{u_1+u_2 + u_3}\big)  w_3 
+ \pounds_{(w_1 + w_2)} u_3 }
\end{split}
\label{CoM-LagVars}
\end{align}
\end{exercise} 


\begin{exercise}
Upon defining the momentum variables
\[
\frac{\delta \ell}{\delta u_1} = m_1
\,,\quad
\frac{\delta \ell}{\delta  \big(g_{1*}u_2\big)} = m_2
\,,\quad
\frac{\delta \ell}{\delta  \big((g_1g_2 )_{*}u_3\big)} = m_3
\,,\]
show that integrating by parts in \eqref{CoM-LagVars} and collecting coefficients of the variations $w_1$, $w_2$, and $w_3$ 
leads to the following three EP equations upon separately setting to zero each coefficient of the variations $w_1$, $ w_2$ and $w_3$,
\begin{align}
\p_t\begin{pmatrix} m_1 \\  \\ m_2 \\  \\ m_3 \end{pmatrix}
&= -
	\begin{bmatrix} 
	\mathrm{ad}^*_{\Box} m_1 &  \mathrm{ad}^*_{\Box} m_2  & \mathrm{ad}^*_{\Box} m_3
	\\ \\
	\mathrm{ad}^*_{\Box} m_2 & \mathrm{ad}^*_{\Box} m_2 & \mathrm{ad}^*_{\Box} m_3
	\\ \\
	\mathrm{ad}^*_{\Box} m_3 & \mathrm{ad}^*_{\Box} m_3 & \mathrm{ad}^*_{\Box} m_3
	\end{bmatrix} 
  \begin{pmatrix}
    \frac{\delta h}{\delta m_1} = u_1 \\ \\
    \frac{\delta h}{\delta m_2} = u_2 \\ \\
    \frac{\delta h}{\delta m_3} = u_3
    \end{pmatrix}
\label{Poisson-Part1}
    \\&= -
    \begin{pmatrix} \mathrm{ad}^*_{u_1} m_1 +  \mathrm{ad}^*_{u_2} m_2 +  \mathrm{ad}^*_{u_3} m_3 
    \\ \\
    \mathrm{ad}^*_{u_1 + u_2} m_2 +  \mathrm{ad}^*_{u_3} m_3 
    \\ \\
    \mathrm{ad}^*_{u_1 + u_2 + u_3} m_3
    \end{pmatrix}.
\label{Poisson-Part2}
\end{align}
The matrix operator in square brackets in \eqref{Poisson-Part1} defines a Lie-Poisson bracket $\{f,h\}=\langle\mu,[df,dh]\rangle$
on the dual of the following \emph{nested} semidirect product Lie algebra \index{semidirect product!nested}
\index{Lie algebra!nested semidirect product!$\mathfrak{g}_1 \circledS  (\mathfrak{g}_2 \circledS \mathfrak{g}_3 )$}
\[
\mathfrak{s} = \mathfrak{g}_1\ \circledS  \big(\mathfrak{g}_2\ \circledS\ \mathfrak{g}_3 \ \big)
\,.\]
Physically, the Lie-Poisson bracket in \eqref{Poisson-Part1} refers to three types of fluid flow.
It thus mimics L. F. Richardson's famous ``whorls within whorls'' reference in characterising fluid dynamics.
\begin{quote}
Big whorls have little whorls, that feed on their velocity, and little whorls have lesser whorls, and so on to viscosity

-- L. F. Richardson (1922)
\end{quote}
This is `Richardson's triple'.  In the composition of three non-commutative right group actions $G_1\times G_2\times G_3$
with velocities 
\[TG_1/G_1\times (TG_2/G_2)/G_1\times ((TG_3/G_3)/G_2)/G_1)\,,\]
the flow of $G_1$ can represent the big whorls;  the flow of $G_2$ can represent the little whorls carried in the frame of motion of  the big whorls; and $G_3$ can represent the lesser whorls carried along successively by the two other whorls. 

\end{exercise}


\begin{exercise}
Show that the linear transformation of variables to 
\[\mu_1= m_1 -m_2\,,\quad\mu_2= m_2 -m_3\,,\quad \mu_3 = m_3\,,\]
diagonalises the \textit{entangled Poisson matrix} in \eqref{Poisson-Part1} and leads to 
equivalent equations with an \textit{untangled Poisson matrix},
\begin{align}
\begin{split}
\p_t\begin{pmatrix} \mu_1  \\ \mu_2  \\ \mu_3 \end{pmatrix}
&= -
	\begin{bmatrix}
	\mathrm{ad}^*_{\Box} \mu_1 & 0 & 0
	\\
	0 & \mathrm{ad}^*_{\Box} \mu_2 & 0
	\\
	0 & 0 & \mathrm{ad}^*_{\Box} \mu_3
	\end{bmatrix}  
 \begin{pmatrix}
    \frac{\delta h}{\delta \mu_1} = u_1 \\ 
    \frac{\delta h}{\delta \mu_2} = u_1 + u_2 \\ 
    \frac{\delta h}{\delta \mu_3} = u_1 + u_2 + u_3
    \end{pmatrix}
\\&= -
    \begin{pmatrix} \mathrm{ad}^*_{u_1} \mu_1 
    \\
    \mathrm{ad}^*_{(u_1+u_2)} \mu_2
    \\
    \mathrm{ad}^*_{(u_1+u_2+u_3)} \mu_3    \end{pmatrix} 
\,.\end{split}
\label{Poisson_untangled}
\end{align}
The untangled diagonal matrix operator in equation \eqref{Poisson_untangled} defines a Lie-Poisson bracket $\{f,h\}=\langle\mu,[df,dh]\rangle$ on the dual of the following direct product Lie algebra
\[
\mathfrak{s}_{diag} = \mathfrak{g}_1\ \otimes  \mathfrak{g}_2\ \otimes\ \mathfrak{g}_3 
\,.\]
This untangling property upon applying momentum shifts is typical in Lagrangian reduction by stages 
based on the composition of Lie group actions.
\end{exercise}


\begin{exercise}

Consider the following Hamilton's principle on the product space ($TG_1\times TG_2\times V)$ with left actions 
of Lie groups $G_1$ and $G_2$ on a vector space $V$ and verify its symmetry reduction by stages 
for the following nested sum of left semi-direct product Lie algebra actions
\begin{align}
\mathfrak{g}_1\,\circledS\, (\mathfrak{g}_2\, \oplus \,V) 
\oplus  (\mathfrak{g}_2\, \circledS\, V) 
\,.\label{NestedSDP-left}
\end{align}
Symmetry reduction by stages of Hamilton's principle in this nested semidirect-product dynamics may be written 
for a \emph{fixed} element $\Theta_0\in V$ as\index{semidirect product!nested}
\begin{align}
\begin{split}
S &= \int_{t_1}^{t_2}
L(g_1,\dot{g}_1;  g_2,\dot{g}_2; \Theta_0)\,dt
\\
 &= \int_{t_1}^{t_2}
L( g_1^{-1}\dot{g}_1 \,; \Ad_{g_1^{-1}}( g_2^{-1}\dot{g}_2) \,;  g_1^{-1}g_2^{-1}\Theta_0 )\,dt
\\
 &=: \int_{t_1}^{t_2}
L( \Omega_1; \Omega_2; \Theta)\,dt
\end{split}
\label{RBS-HamPrinc}
\end{align}
where the sequential $\Ad$ actions define two velocities and a curve in $V$ 
parameterised by time $t$ as
\begin{align}
\Omega_1 := g_1^{-1}\dot{g}_1  
\,,\ \ 
\Omega_2 :=  \Ad_{g_1^{-1}}( g_2^{-1}\dot{g}_2)
\,,\ \
{\Theta} :=  g_1^{-1}g_2^{-1}\Theta_0 
\,,
\label{RBS-veloc}
\end{align}
for the nested symmetries of ($TG_1\times TG_2\times V)$ under the respective left actions of $G_1\times G_2$ 
parametrised by time $t$ on a fixed element $\Theta_0\in V$. 

Derive the following evolutionary equation for $\Theta$ from its definition,
\begin{align}
\frac{d}{dt}{\Theta} =  - \mathcal{L}_{\Omega_1 + \Omega_2} \Theta 
\,.\label{theta-eqn}
\end{align}
\end{exercise}


\begin{answer}
By a direct calculation one finds the auxiliary evolutionary equation for $\Theta$ as
\begin{align}
\begin{split}
\frac{d}{dt} \Theta
=& \frac{d}{dt} \big( g_1^{-1}g_2^{-1}\Theta_0 \big)
= 
- \,(g_1^{-1}\dot{g}_1) \Theta - \Ad_{g_1^{-1}}(g_2^{-1}\dot{g}_2) \Theta 
\\&= 
-\mathcal{L}_{\Omega_1} \Theta - \mathcal{L}_{\Omega_2} \Theta 
=  - \mathcal{L}_{\Omega_1 + \Omega_2} \Theta 
\,.\end{split}
\label{theta-eqn-proof}
\end{align}

\end{answer}


\begin{exercise}\label{Ex41}
Verify the line of reasoning in deriving the Lagrangian shown above, 
introduce variations with the sequential $\Ad$ actions of $G_1\times G_2$, 
\begin{align}
\omega_1 := g_1^{-1}{g'}_1  
\,,\ \ 
\omega_2 :=  \Ad_{g_1^{-1}}( g_2^{-1}{g'}_2)
\,,\ \
\Theta' :=    - \mathcal{L}_{\omega_1 + \omega_2} \Theta 
\,.
\label{RBS-vars}
\end{align}
Then use the following equations for iterated left actions,
\begin{align}
\begin{split}
\Omega'_1 - \dot{\omega}_1 &=   \ad_{\Omega_1} \omega_1
\,,\\
\Omega'_2 - \dot{\omega}_2 
  &= \ad_{\Omega_1 + \Omega_2}\omega_2 + \, \ad_{\Omega_2} \omega_1
\,,\\
\Theta' &=    - \mathcal{L}_{\omega_1 + \omega_2} \Theta \,,
\end{split} 
\label{RBS-Iterates12}
\end{align} 
to derive the equations of motion which follow from Hamilton's principle with the 
Lagrangian in the last line of \eqref{RBS-HamPrinc}.
\end{exercise}


\begin{answer}
The first two equations for variations in the set \eqref{RBS-Iterates12} have already been determined 
in previous exercises via their definitions in equations \eqref{CoM-angveloc} and \eqref{CoM-angvar}.
Consequently, one may calculate the corresponding Euler-Poicar\'e equations, as follows,  
\begin{align}
\begin{split}
0 = \delta S &= \delta \int_{t_1}^{t_2} L( \Omega_1; \Omega_2; \Theta)\,dt
\\&=
\int_{t_1}^{t_2} \scp{ \frac{\delta L}{\delta \Omega_1} }{\delta \Omega_1} 
+ \scp{ \frac{\delta L}{\delta \Omega_2} }{\delta \Omega_2} 
+ \scp{ \frac{\delta L}{\delta \Theta} }{\delta \Theta} 
\\&=
\int_{t_1}^{t_2} \scp{ \frac{\delta L}{\delta \Omega_1} }{\dot{\omega}_1 +   \ad_{\Omega_1} \omega_1} 
+ \scp{ \frac{\delta L}{\delta \Theta} }{  - \mathcal{L}_{\omega_1 + \omega_2} \Theta } 
\\&\qquad
+ \scp{ \frac{\delta L}{\delta \Omega_2} }
{\dot{\omega}_2 + \ad_{\Omega_1 + \Omega_2}\omega_2 + \, \ad_{\Omega_2} \omega_1} 
\,dt
\\&=
\int_{t_1}^{t_2} 
\scp{ -\,\frac{d}{dt} \frac{\delta L}{\delta \Omega_1} 
+   \ad^*_{\Omega_1} \frac{\delta L}{\delta \Omega_1} 
+ \ad^*_{\Omega_2} \frac{\delta L}{\delta \Omega_2}
  +  \frac{\delta L}{\delta \Theta} \diamond \Theta } {\omega_1 }
\\&\qquad
+ \scp{ -\,\frac{d}{dt}\frac{\delta L}{\delta \Omega_2} 
+ \ad^*_{\Omega_1 + \Omega_2} \frac{\delta L}{\delta \Omega_2}
+  \frac{\delta L}{\delta \Theta} \diamond \Theta}{\omega_2 }
\,dt
\,.\end{split}
\label{RBS-HamPrincvars}
\end{align}

\end{answer}


\begin{exercise}\label{Ex42}

Use the symmetry-reduced Legendre transform to obtain the Hamiltonian formulation of 
the dynamics for nested left semidirect-product action. \index{semidirect product!nested}

\end{exercise}


\begin{answer}
The symmetry-reduced Legendre transform for this case is
\[
H(\Pi_1,\Pi_2,\Theta) = \scp{\Pi_1}{\Omega_1} + \scp{\Pi_2}{\Omega_2} - L(\Omega_1,\Omega_2,\Theta)
\]
It yields the following variational derivatives of the Hamiltonian
\[
\frac{\delta H}{\delta \Pi_1} = \Omega_1\,,\quad
\frac{\delta H}{\delta \Pi_2} = \Omega_2\,,\quad
\frac{\delta H}{\delta \Theta} = -\, \frac{\delta L}{\delta \Theta} 
\,.
\]
Rearranging the equations into Hamiltonian form yields
\begin{align}
\p_t\begin{pmatrix} \Pi_1 \\  \\ \Pi_2 \\  \\ \Theta \end{pmatrix}
&= 
	\begin{bmatrix} 
	\mathrm{ad}^*_{\Box} \Pi_1 &  \mathrm{ad}^*_{\Box} \Pi_2  & -\,{\Box} \diamond \Theta
	\\ \\
	\mathrm{ad}^*_{\Box} \Pi_2 & \mathrm{ad}^*_{\Box} \Pi_2 &  -\,{\Box} \diamond \Theta
	\\ \\
	- \mathcal{L}_{\Box}\Theta & - \mathcal{L}_{\Box}\Theta & 0
	\end{bmatrix} 
  \begin{pmatrix}
    \frac{\delta H}{\delta \Pi_1} = \Omega_1 \\ \\
    \frac{\delta H}{\delta \Pi_2} = \Omega_2 \\ \\
    \frac{\delta H}{\delta \Theta} = -\, \frac{\delta L}{\delta \Theta}
    \end{pmatrix}
\label{Poisson-NestedSDP-Part1}
    \\&= 
    \begin{pmatrix} 
    \mathrm{ad}^*_{\Omega_1} \Pi_1 +  \mathrm{ad}^*_{\Omega_2} \Pi_2 
    -  \frac{\delta H}{\delta \Theta} \diamond \Theta
    \\ \\
    \mathrm{ad}^*_{\Omega_1+\Omega_2} \Pi_2    -  \frac{\delta H}{\delta \Theta} \diamond \Theta
    \\ \\
    - \mathcal{L}_{\Omega_1+\Omega_2}\Theta
    \end{pmatrix}.
\label{Poisson-NestedSDP-Part2}
\end{align}
The matrix operator in square brackets in \eqref{Poisson-NestedSDP-Part1} defines 
a Lie-Poisson bracket $\{f,h\}=\langle\mu,[df,dh]\rangle$
on the dual of the following \emph{nested} semidirect product Lie algebra
\begin{align}
\mathfrak{g}_1\,\circledS\, (\mathfrak{g}_2\, \oplus \,V) 
\oplus  (\mathfrak{g}_2\, \circledS\, V) 
\,,\label{NestedSDP-LP}
\end{align}
with dual coordinates $\Pi_1\in \mathfrak{g}^*_1$, $\Pi_2\in \mathfrak{g}^*_2$ and $\Theta\in V$.

\end{answer}



\begin{exercise}

Write the Lie-Poisson bracket dual to $\mathfrak{g}_1\,\circledS\, (\mathfrak{g}_2\, \oplus \,V) 
\oplus  (\mathfrak{g}_2\, \circledS\, V)$ for invariance of the Lagrangian in \eqref{RBS-HamPrincvars} 
under \emph{right} Lie group action. 

\end{exercise}


\begin{answer}

Relative to left invariance $g^{-1}\dot{g}$ the right invariant case $\dot{g}g^{-1}$ simply reverses the signs 
in adjoint and coadjoint action. Consequently, one may read off the correct Lie-Poisson operator for right action 
by reversing the signs of the $\ad^*$ operations. 

In particular, the Lie-Poisson bracket for right action in this case becomes,
\begin{align}
\p_t\begin{pmatrix}m_1 \\  \\m_2 \\  \\ \rho \end{pmatrix}
&= -
	\begin{bmatrix} 
	\mathrm{ad}^*_{\Box} m_1 &  \mathrm{ad}^*_{\Box}m_2  & {\Box} \diamond \rho
	\\ \\
	\mathrm{ad}^*_{\Box}m_2 & \mathrm{ad}^*_{\Box}m_2 &  {\Box} \diamond \rho
	\\ \\
	\mathcal{L}_{\Box}\rho & \mathcal{L}_{\Box}\rho & 0
	\end{bmatrix} 
  \begin{pmatrix}
    \frac{\delta H}{\delta m_1} =  u_1 \\ \\
    \frac{\delta H}{\delta  m_2} = u_2 \\ \\
    \frac{\delta H}{\delta  \rho} = -\, \frac{\delta L}{\delta \rho}
    \end{pmatrix},
\label{Poisson-NestedSDP-right}
\end{align}
where one regards $m_1$ and $m_2$ as two fluid momenta whose corresponding velocities $u_1$ and $u_2$ which both transport the other variable, $\rho$, which for fluid dynamics would be the mass density, or charge density in the fluid plasma application.\footnote{In the case of stochastic transport, the velocities $u_1$ and $u_2$ 
in \eqref{Poisson-NestedSDP-right} would become stochastic processes. For a discussion of stochastic fluid transport, see \cite{holm2015variational}.}

\end{answer}

\subsection{Plasma physics applications of Lie-Poisson brackets} \label{subsec-5.4}
The result in equation \eqref{Poisson-NestedSDP-right} applies to fluid dynamics in any dimension.
In the fluid context, the Poisson bracket \eqref{Poisson-NestedSDP-right} can be compared with the 
Lie-Poisson bracket in \eqref{Poisson-Part1}.  
The apparently slight difference between these two Lie-Poisson brackets turns out to matter significantly,
both geometrically and physically. 	
In a fluids interpretation, the Lie-Poisson bracket in \eqref{Poisson-Part1} refers to three types of fluid flow, each carried along by the previous group action. In contrast, the Lie-Poisson bracket in \eqref{Poisson-NestedSDP-right} can involve physical quantities which may be carried by two different vector fields which in turn influence each other.

Remarkably, this Lie-Poisson bracket for right action applies to a reduced model of Alfv\'en wave turbulence equations for quasi-neutral plasma flow with the magnetic field in the plane of flow. In addition, the two vector fields in this case both interact with the charge density which is carried along by the two different types of transport. 
For more information and explanation of the physical meaning of these variables, see 
\cite{hazeltine1985shear,hazeltine1985electromagnetic}.


\begin{exercise}

Find the functionals whose variational derivatives are in the kernel of the 
Lie-Poisson operator in equation  \eqref{Poisson-NestedSDP-right}. 

These functionals are conserved for any Hamiltonian written in terms of these variables. 

Hint: One efficient way to find them would be to transform variables to make the 
Lie-Poisson operator as diagonal as possible. 

\end{exercise}

\begin{answer}

The change of variables in the Poisson matrix of \eqref{Poisson-NestedSDP-right} 
from $(m_1,m_2,\rho)$ to $(\mu,m_2,\rho)$  with $\mu=m_1-m_2$ yields
\begin{align}
\p_t\begin{pmatrix}\mu \\  \\m_2 \\  \\ \rho \end{pmatrix}
&= -
	\begin{bmatrix} 
	\mathrm{ad}^*_{\Box} \mu &  0  & 0
	\\ \\
	0 & \mathrm{ad}^*_{\Box}m_2 &  {\Box} \diamond \rho
	\\ \\
	0 & \mathcal{L}_{\Box}\rho & 0
	\end{bmatrix} 
  \begin{pmatrix}
    \frac{\delta C}{\delta \mu}  \\ \\
    \frac{\delta C}{\delta  m_2}  \\ \\
    \frac{\delta C}{\delta  \rho} 
    \end{pmatrix},
\label{Poisson-NestedSDP-right-diag}
\end{align}
To prove this statement, notice that the Jacobian matrix for this transformation is given by
\begin{align*}
J = 
	\begin{bmatrix} 
	1 &  -1\  & 0
	\\ 
	0 & \ 1\ &  0
	\\ 
	0 & \ 0\ &  1
	\end{bmatrix}.
\end{align*}
Multiplying the Poisson matrix in \eqref{Poisson-NestedSDP-right} by Jacobian $J$ from the left
 and by its transpose $J^T$ from the right produces the transformed Poisson matrix in 
 \eqref{Poisson-NestedSDP-right-diag}. This linear change of variables preserves the eigenvalues 
 of the the Poisson matrix. In particular, such linear transformations preserve the matrix null eigenvectors.

Consequently, one may check that the variational derivatives of the following functions 
$C_1,C_2,C_3$ are Casimirs, 
\begin{align}
C_1 = F_1(\mu)\,,\quad C_2 = m_2 F_2(\rho)\quad \hbox{and}\quad C_3 = F_3(\rho),
\label{Poisson-NestedSDP-Casimirs}
\end{align}
where $F_1$, $F_2$ and $F_3$ are arbitrary differentiable functions of their arguments. 
That is, their variational derivatives are null eigenvectors of  Lie-Poisson bracket in 
\eqref{Poisson-NestedSDP-right-diag} as well as the transformed Lie-Poisson bracket 
in \eqref{Poisson-NestedSDP-right-diag}.

\end{answer}


\begin{exercise}\label{Ex3NestedSDP}

Write the Lie-Poisson bracket dual to the following \emph{twice nested} semidirect product Lie algebra 
\begin{align}
\mathfrak{s} = \mathfrak{g}_1\ \circledS\ \Big(V_1 \oplus  \big(\mathfrak{g}_2\ \circledS\ (V_2 \oplus (\mathfrak{g}_3\ \circledS\ V_3) \ \big) \ \Big).
\label{2xNestedAlgebras}
\end{align}
for invariance of a Lagrangian under twice nested \emph{right} Lie group actions
on the product space ($TG_1\times V_1) \times (TG_2\times V_2)\times (TG_3\times V_3)$. 
\index{semidirect product!nested}
\end{exercise}


\begin{answer}

Legendre transforming such a Lagrangian invariant under twice nested right Lie group actions leads to 
Euler--Poincar\'e equations whose twice nested Lie-Poisson Hamiltonian formulation
may be displayed in the following matrix form with reduced Hamiltonian $h(m_k,a_k):\Pi_k (\mathfrak{X}_k^*\times V^*_k)\rightarrow\mathbb{R}$ with $m_k := {\delta \ell}/{\delta u_k}$.
\small
\begin{align*}
    \p_t\begin{pmatrix} m_1\\  a_1 \\   m_2  \\  a_2  
    \\   m_3  \\  a_3 \end{pmatrix} 
    =
    -\begin{pmatrix} 
    \ad^*_{\fbox{}}m_1 &  \Box{}\diamond a_1 
    & { {  \Box{}\diamond m_2 } } 
    &{ {  \Box{}\diamond a_2 } } 
    & {  {  \Box{}\diamond m_3} } 
    & {  {  \Box{}\diamond a_3} } 
    \\
     \mathcal{L}_{\fbox{}}a_1 & 0 & 0 & 0 & 0 & 0\\
     \mathcal{L}_{\fbox{}} m_2 & 0
    &  \ad^*_{\fbox{}} m_2 
    &  \Box{}\,\diamond \,a_2 
    & {  \Box{}\diamond m_3}   
    & {  \Box{}\diamond a_3}  \\
     \mathcal{L}_{\fbox{}} a_2 & 0 &  \mathcal{L}_{\fbox{}}a_2 & 0 & 0 & 0 \\
     \mathcal{L}_{\fbox{}} m_3 & 0&   \mathcal{L}_{\fbox{}} m_3 & 0  
    &  \ad^*_{\fbox{}}m_3 &  \Box{}\,\diamond\,a_3 \\
     \mathcal{L}_{\fbox{}} a_3 & 0 &  \mathcal{L}_{\fbox{}} a_3 & 0 
    &  \mathcal{L}_{\fbox{}}a_3 & 0  \\
    \end{pmatrix}
    \begin{pmatrix}
     \frac{\delta h}{\delta m_1} = u_1\\  \frac{\delta h}{\delta a_1} = -\frac{\delta \ell}{\delta a_1} \\ 
     \frac{\delta h}{\delta m_2} = u_2 \\  \frac{\delta h}{\delta a_2} = -\frac{\delta \ell}{\delta a_2} \\
     \frac{\delta h}{\delta m_3} = u_3 \\  \frac{\delta h}{\delta a_3} = -\frac{\delta \ell}{\delta a_3} 		   
\end{pmatrix}
\,. 
\end{align*}
\normalsize
 Again, the nested Lie-Poisson bracket may be `untangled' via momentum shifts into block diagonal form. 

The pattern of Lie algebraic self similarity for extension to further nested Lie group actions on additional degrees of freedom should now be clear. 

\end{answer}




\newpage
\part{Geometric Mechanics on Manifolds}

\begin{figure}[H]
\begin{center}
\includegraphics[width=0.75\textwidth]{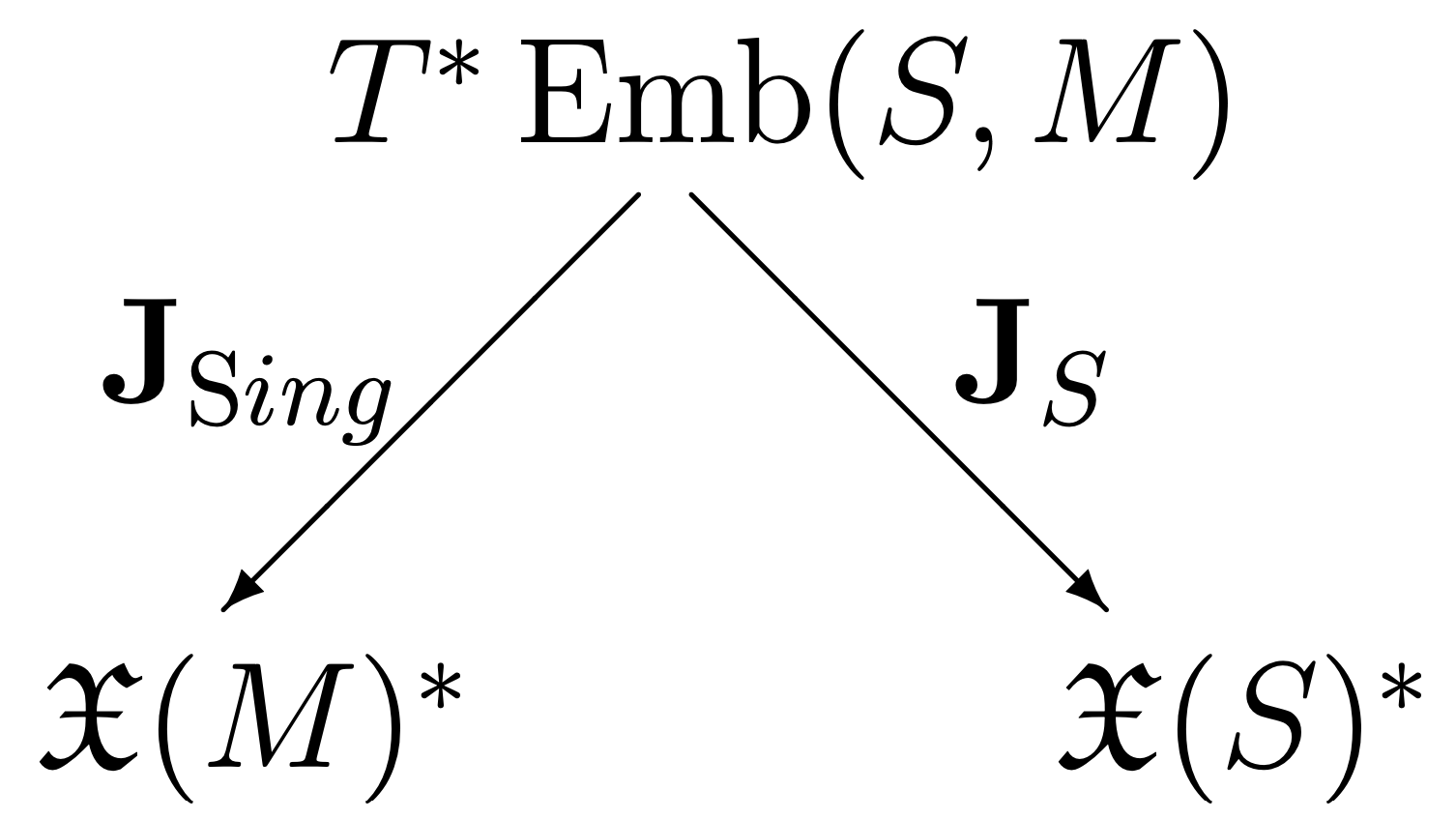}
\caption{The dual pair of momentum maps for continuum mechanics \cite{HoMa2004}}
\end{center}
\end{figure}\vspace{-5mm}
\vspace{4mm}\centerline{\textcolor{shadecolor}{\rule[0mm]{6.75in}{-2mm}}\vspace{-4mm}}

\newpage

\newpage
\vspace{4mm}\centerline{\textcolor{shadecolor}{\rule[0mm]{6.75in}{-2mm}}\vspace{-4mm}}
\section{Geometric structure of classical mechanics}\label{smooth manifold}

\secttoc

\textbf{What is this lecture about?} This lecture introduces basic vocabulary and notation 
for discussing the geometric mechanics of flows on manifolds.

\subsection{Manifolds}

Configuration space: coordinates $q\in M$, where $M$ is a smooth manifold.

The composition $\phi_\beta \circ \phi^{-1}_\alpha$ is a smooth change
of variables.

For later, smooth coordinate transformations: $q\to Q$ with
$dQ=\frac{\partial Q}{\partial q} dq$.

\begin{definition}
A smooth manifold $M$ is a set of points together with a finite (or
perhaps countable) set of subsets  $U_\alpha\subset M$ and one-to-one mappings
$\phi_\alpha{:\ } U_\alpha \to \mathbb{R}^n$ such that
\begin{enumerate}
\item \ $\bigcup_\alpha U_\alpha = M$
\item For every nonempty intersection $U_\alpha \cap U_\beta$, the set 
$\phi_\alpha \left(U_\alpha \cap U_\beta\right)$ is an open subset of
$\mathbb{R}^n$ and the one-to-one mapping $\phi_\beta \circ \phi_\alpha^{-1}$
is a smooth function on  $\phi_\alpha \left(U_\alpha \cap U_\beta\right)$. 
\end{enumerate}
\end{definition}

\begin{remark}\rm 
The sets $U_\alpha$ in the definition are called \emph{coordinate charts}. \index{manifold!coordinate charts}
The mappings $\phi_\alpha$ are called \emph{coordinate functions} or
\emph{local coordinates}. A collection of charts satisfying 1 and 2 is called
an \emph{atlas}. Condition 3 allows the definition of manifold to be made
independently of a choice of atlas. A set of charts satisfying 1 and 2 can
always be extended to a maximal set; so, in practice, conditions 1 and 2
define the manifold.
\end{remark}

\begin{example}\rm 
Manifolds often arise as intersections of zero level sets 
\[M=\left\{x \big| f_i(x) = 0, \ i=1,\dots,k\right\},\] for a given
set of functions
$f_i{:\ }  \mathbb{R}^n\to\mathbb{R}$, $i=1,\dots,k$.
If the gradients
$\nabla f_i$ are linearly independent, or more generally if the rank of
$\left\{\nabla f(x)\right\}$ is a constant $(r)$ for all $x$,  then $M$ is
a smooth manifold of dimension $n-r$. The proof uses the Implicit Function
Theorem to show that an $(n{-}r)$--dimensional coordinate chart may be
defined in a neighborhood of each point on $M$. In this situation, the
set $M$ is called a \emph{submanifold} of $\mathbb{R}^n$ (see
Lee~\cite{Le2003}).
\end{example}

\begin{definition} If $r=k$, then the map $\left\{f_i\right\}$ is called a
\emph{submersion}. \index{manifold!submersion}
\end{definition}

\begin{exercise} 
Prove that the zero sets of all submersions are submanifolds (see Lee~\cite{Le2003}).
\end{exercise}

\begin{definition}[Tangent space to level sets]
Let \[M=\left\{x \big| f_i(x) = 0, \ i=1,\dots,k\right\}\] be a manifold in
$\mathbb{R}^n$. (Note that the zero set is a manifold because the map has constant rank.)

The \emph{tangent space} \index{manifold!tangent space} at each $x\in M,\ $ is defined by 
\[T_xM=\bigg\{ v\in \mathbb{R}^n \ \bigg| \ \frac{\partial f_i}{\partial
x^a} (x) v^a = 0,\ i=1,\dots,k\bigg\}.\]  Note: we use the 
\emph{summation convention}. \index{summation convention}
That is, repeated indices are summed over their range.
\end{definition}

\begin{remark}\rm  
The tangent space is a linear vector space.
\end{remark} 


\begin{example}\rm [Tangent space to the sphere in $\mathbb{R}^3$]
The sphere $S^2$ is the set of points $(x,y,z)\in\mathbb{R}^3$ solving
$x^2 + y^2 + z^2 = 1$.  The tangent space to the unit sphere at such a point
$(x,y,z)$ is the plane containing vectors
$(u,v,w)$ satisfying $xu + yv + zw = 0$.

\end{example}

\begin{definition}[Tangent bundle] \label{tanbun-def} \index{manifold!tangent bundle}
The \emph{tangent bundle} of a manifold $M$, denoted by $TM$, is the
smooth manifold whose underlying set is the disjoint union of the tangent
spaces to $M$ at the points $x\in M$; that is, 
\[TM = \bigcup_{x\in M}T_xM\,.\]
Thus, a point of $TM$ is a vector $v$ which is tangent to $M$ at some point
$x\in M$.
\end{definition}

\begin{example}\rm [Tangent bundle $TS^2$ of $S^2$]
The tangent bundle $TS^2$ of $S^2\in \mathbb{R}^3$ is the union of the
tangent spaces of $S^2$:
$$TS^2 = \big\{ (x,y,z;u,v,w) \in \mathbb{R}^6 ~\big|~ x^2 + y^2 + z^2 = 1 
\text{ and } xu + yv + zw = 0 \big\}.$$
\end{example}

\begin{remark}\rm [Dimension of tangent bundle $TS^2$]
Defining $TS^2$ requires two independent conditions in
$\mathbb{R}^6$; so \textrm{dim}$TS^2=4$.
\end{remark} 

\begin{exercise}
Define the sphere $S^{n-1}$ in $\mathbb{R}^n$. What is the dimension of
its tangent space $TS^{n-1}$?
\end{exercise}

\begin{example}\rm [The two stereographic projections of $S^2\to\mathbb{R}^2$]
\index{stereographic projection} \index{Riemann sphere} \label{2StereoProj-eg}
The unit sphere 
\[
S^2=\{(x,y,z): x^2+y^2+z^2=1\}
\]
is a smooth two-dimensional manifold realised as the level set of a submersion in $\mathbb{R}^3$. Let
\[
U_N = S^2\backslash \{0,0,1\}
,\quad\hbox{and}\quad
U_S = S^2\backslash \{0,0,-1\}
\]
be the subsets obtained by deleting the North and South poles of $S^2$, respectively. Let 
\[
\chi_N{:\ } U_N\to (\xi_N,\eta_N)\in \mathbb{R}^2
,\quad\hbox{and}\quad
\chi_S{:\ } U_S\to (\xi_S,\eta_S)\in \mathbb{R}^2
\]
be stereographic projections from the North and South poles onto the equatorial plane, $z=0$. 

Thus, one may place two different coordinate patches in $S^2$ intersecting
everywhere except at the points along the $z$--axis at $z=1$ (North pole) and $z=-1$ (South
pole).

In the equatorial plane $z=0$, one may define two sets of
(right-handed) coordinates, 
\[
\phi_\alpha{:\ } U_\alpha\to\mathbb{R}^2\backslash\{0\}
,\quad
\alpha=N,S,
\]
obtained by the following two stereographic projections from the North and South poles:
\begin{enumerate}
\item  (valid everywhere except $z=1$)
\[
\phi_N(x,y,z)=(\xi_N, \eta_N)
=\left(\frac{x}{1-z}, \frac{y}{1-z}\right)
,\]
\item (valid everywhere except $z=-1$)
\[
\phi_S(x,y,z)=(\xi_S, \eta_S)
=\left(\frac{x}{1+z}, \frac{-y}{1+z}\right)
.\]
\end{enumerate}
(The two complex planes are identified differently with the plane $z = 0$. An orientation-reversal is necessary to maintain consistent coordinates on the sphere.) 

One may check directly that on the overlap $U_N\cap U_S$ the map,
\[
\phi_N\circ\phi_S^{-1}{:\ } \mathbb{R}^2\backslash\{0\}
\to \mathbb{R}^2\backslash\{0\}
\]
is a smooth diffeomorphism, given by the inversion
\[
\phi_N\circ\phi_S^{-1} (x, y)
=
\Big(\frac{x}{x^2+y^2}, \frac{y}{x^2+y^2}\Big)
.
\] 
\end{example}

\begin{exercise}
Construct the mapping from
$(\xi_N,\eta_N)\to(\xi_S,\eta_S)$ and verify that it is a diffeomorphism in  $\mathbb{R}^2\backslash\{0\}$.
Hint: $(1+z)(1-z)=1-z^2=x^2+y^2$.
\end{exercise}

\begin{answer}
\[
(\xi_S,-\eta_S)=\frac{1-z}{1+z}(\xi_N,\eta_N)
=\frac{1}{\xi_N^2+\eta_N^2}(\xi_N,\eta_N)
.
\]
The map $(\xi_N,\eta_N)\to(\xi_S,\eta_S)$ is smooth and invertible
except at $(\xi_N,\eta_N)=(0,0)$.
\end{answer}
\begin{figure}
 
\centering
 
\begin{minipage}{0.5\textwidth}
    \centering
       \hspace{0.1\textwidth}
    \includegraphics[width=\textwidth]{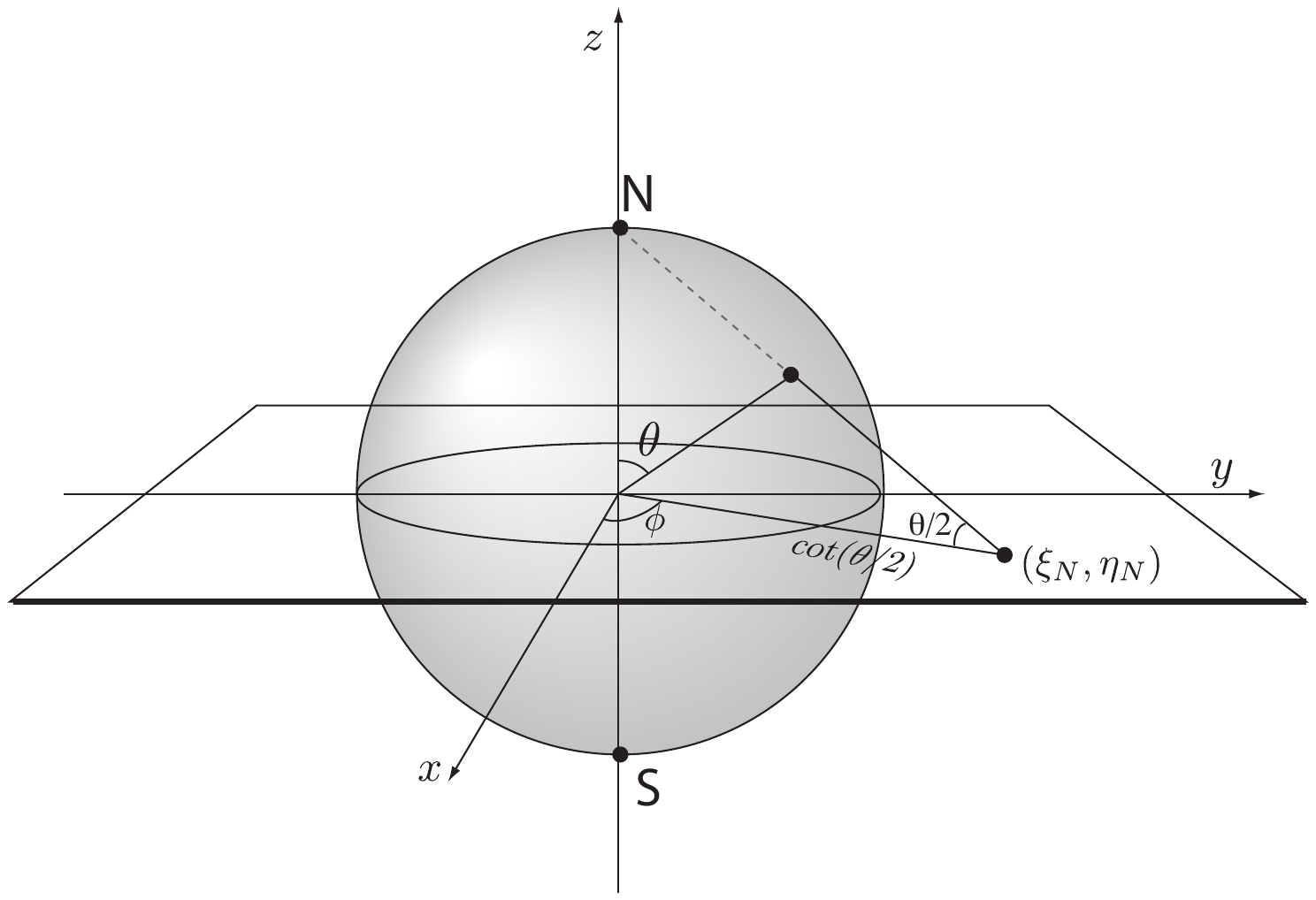}
\end{minipage}\hfill
\begin{minipage}{0.5\textwidth}
    \centering
    \hspace{-0.1\textwidth}
    \includegraphics[width=\textwidth]{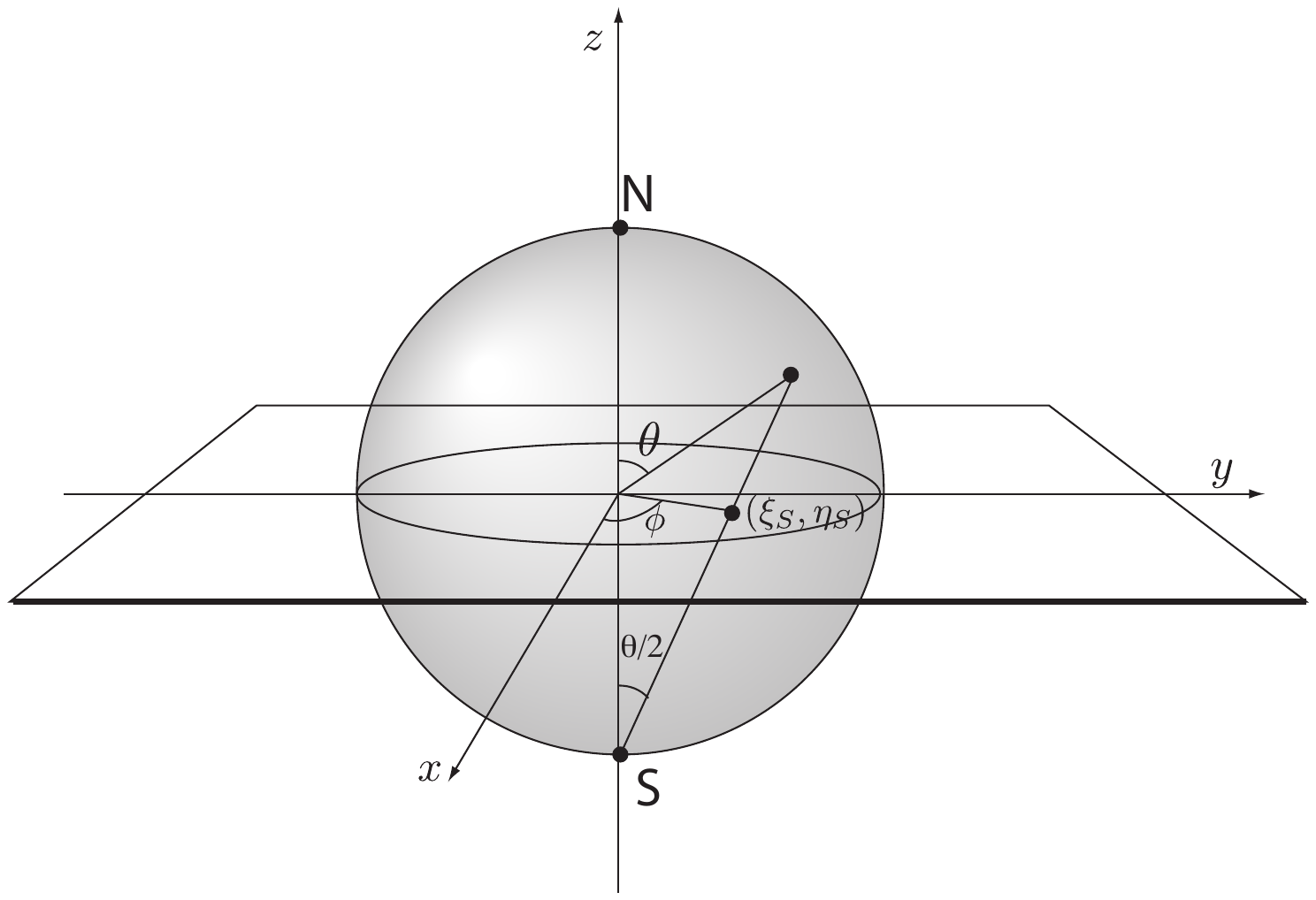}
\end{minipage}
 
\vskip-24pt 
\caption{In the stereographic projection of the Riemann sphere onto the complex plane from the south pole, complex numbers lying outside (resp., inside) the unit circle are projected from points in the lower (resp., upper) hemisphere. \index{Riemann sphere}}
\label{fig:RiemannS}
 
\end{figure}
%
%
%
%
%

\begin{example}\rm 
If we start with two identical circles in the $xz$--plane, of radius $r$ and
centered at $x=\pm 2r$, then rotate them round the $z$ axis in
$\mathbb{R}^3$, we get a torus, written $T^2$. The torus $T^2$ is a manifold.
\end{example}

\begin{exercise}
If we begin with a figure eight in the $xz$--plane, along the $x$ axis and
centered at the origin, and spin it round the $z$ axis in $\mathbb{R}^3$,
we get a ``pinched surface'' that looks like a sphere that has been
``pinched'' so that the north and south poles touch.  Is this a manifold?
Prove it.
\end{exercise}

\begin{answer}
The origin has a neighbourhood diffeomorphic to a double cone. This is
not diffeomorphic to $\mathbb{R}^2$. A proof of this is that, if the
origin of the cone is removed, two components remain; while if the origin
of $\mathbb{R}^2$ is removed, only one component remains.
\end{answer}

\begin{remark}\rm 
The sphere $S^2\subset \mathbb{R}^3$ will appear in several examples as a reduced space in which
motion takes place after applying a symmetry. Reduction by symmetry is
associated with a classical topic in  celestial mechanics known as normal
form theory. Reduction may be ``singular,'' in which case it leads to
``pointed'' spaces that are smooth manifolds except at one or more points.
For example different resonances of coupled oscillators correspond to the
following reduced spaces: 1:1 resonance -- sphere; 1:2 resonance --
pinched sphere with one cone point;  1:3 resonance -- pinched sphere with
one cusp point; 2:3 resonance -- pinched sphere with one cone point and 
one cusp point.
\end{remark}

\subsection{Motion: tangent vectors and flows}

Envisioning our later considerations of dynamical systems, we shall consider
motion along curves $c(t)$ parametrized by time $t$ on a smooth manifold $M$.
Suppose these curves are trajectories of a flow $\phi_t$ of a vector field. We
anticipate this means 
$\phi_t\left(c(0)\right) = c(t)$ and 
$\phi_t \circ \phi_s = \phi_{t+s}$ (flow property). The flow will be
tangent to $M$ along the curve. To deal with such flows, we will need the
concept of \emph{tangent vectors}. \index{manifold!tangent vectors}

Recall from \ref{tanbun-def} that 
the tangent bundle of $M$ is \[TM=\bigcup\limits_{x\in M} T_xM.\]
We will now add a bit more to that definition. The tangent bundle is an
example of a more general structure than a manifold.

\begin{definition}[Bundle]
A \emph{bundle} consists of a manifold $B$, another manifold $M$ called
the ``base space'' and a projection between them $\Pi{:\ } B\to M$. Locally,
in small enough regions of $x$ the inverse images of the projection $\Pi$
exist. These are called the fibres of the bundle. Thus, subsets of
the bundle $B$ locally have the structure of a Cartesian product. An
example is $(B,M,\Pi)$ consisting of
$(\mathbb{R}^2,\mathbb{R}^1,\Pi{:\ } \mathbb{R}^2\to\mathbb{R}^1)$. In this
case, $\Pi{:\ } (x,y)\in\mathbb{R}^2\to x\in\mathbb{R}^1$. Likewise, the
tangent bundle consists of $M,TM$ and a map $\tau_M{:\ } TM\to M$.
\end{definition}

Let $x = \left(x^1,\dots,x^n\right)$ be local coordinates on $M$, and let
$v = \left(v^1,\dots,v^n\right)$ be components of a tangent vector.
\[T_xM = \left\{ v \in \mathbb{R}^n \ \big| \ \frac{\partial f_i}{\partial
x} \cdot v = 0, i=1,\dots, m \right\}\] 
for 
\[M = \left\{ x \in
\mathbb{R}^n \ \big| \ f_i(x) = 0, i=1,\dots, m \right\}\]
These $2n$ numbers $(x,v)$ give local coordinates on $TM$, where
$\dim TM = 2 \dim M$. The \emph{tangent bundle projection} \index{manifold!tangent bundle projection}
is a map $\tau_M {:\ }  TM \to M$ which takes a tangent vector $v$ to a point
$x\in M$ where the tangent vector $v$ is attached (that is, $v\in T_x
M$). The inverse of this projection $\tau_M^{-1}(x)$ is called the \emph{fibre}  \index{manifold!fibre}
over $x$ in the tangent bundle.

\subsection{Summary of vector fields, integral curves and flows} \index{manifold!! vector fields, integral curves !flows}

\begin{definition}\label{VecField-defn}
A \emph{vector field} on a manifold $M$ is a map $X{:\ } M\to TM$ that assigns
a vector $X(x)$ at each point
$x\in M$. This implies that $\tau_M \circ X = \Id$.
\end{definition}

\begin{definition}
An \emph{integral curve} of $X$ with initial conditions $x_0$ at $t=0$ is a
differentiable map $c{:\ } ]a,b[ \to M$, where $]a,b[$ is an open interval
containing $0$, such that $c(0)=0$ and
$c'(t) = X\left(c(t)\right)$ for all $t\in ]a,b[$.
\end{definition}

\begin{remark}\rm 
A standard result from the theory of ordinary differential equations
states that $X$ being Lipschitz implies its integral curves are unique
and $C^1$ (see Coddington and Levinson~\cite{CoLe1984}). The integral curves $c(t)$ are differentiable for
smooth $X$.
\end{remark}

\begin{definition}
The \emph{flow} of $X$ is the collection of maps $\phi_t{:\ } M\to M$, where
$t\to\phi_t(x)$ is the integral curve  of $X$ with initial condition $x$.
\end{definition}

\begin{remark}\rm $\quad$

\begin{enumerate}
\item Existence and uniqueness results for solutions of $c'(t)=X(c(t))$
guarantee that flow $\phi$ of $X$  is smooth in $(x,t)$, for smooth $X$.
\item 
Uniqueness implies the flow property 
\[\phi_{t+s}=\phi_t \circ
\phi_s,
\eqno{(\mathrm{FP})}
\]
for initial condition $\phi_0=\Id$.
\item The flow property (FP) generalizes to the nonlinear case the
familiar linear situation where $M$ is a vector space, $X(x)=Ax$ is a
linear vector field for a bounded linear operator $A$, and
$\phi_t(x)=e^{At} x$.
\end{enumerate}
\end{remark}

\subsection{Differentials of functions and the cotangent bundle}

We are now ready to define differentials of smooth functions and the
cotangent bundle.

Let $f{:\ } M\to\R$ be a smooth function. We differentiate
$f$ at $x\in M$ to obtain $T_xf{:\ } T_xM\to T_{f(x)} \R$.
As is standard, we identify $T_{f(x)}\R$ with $\R$ itself, thereby
obtaining a linear map $df(x){:\ } T_xM\to \R$. The result $df(x)$ is an
element of the cotangent space $T_x^*M$, the dual space of the tangent space
$T_xM$. The natural pairing between elements of the tangent space and the
cotangent space is denoted as $\langle\cdot, \cdot\rangle{:\ } T_x^*M\times
T_xM\mapsto\mathbb{R}$.

In coordinates, the linear map $df(x){:\ } T_xM\to \R$ may be written as the
directional derivative,
\[
\langle df(x), v\rangle
=
df(x)\cdot v = \frac{\partial f}{\partial x^i}
\cdot v^i
,
\] 
for all $v\in T_xM$.
(Reminder: the summation convention is intended over repeated indices.) Hence,
elements $df(x)\in T_x^*M$ are dual to vectors $v\in T_xM$ with respect to the
pairing $\langle\cdot, \cdot\rangle$.

\begin{definition}
The symbol $df$ denotes the \emph{differential} of the function $f$.
\end{definition}

\begin{definition}
The dual space of the tangent bundle $TM$ is the \emph{cotangent bundle}
$T^*M$. That is, \index{manifold!cotangent bundle}
\[
(T_xM)^*=T_x^*M
\quad\hbox{and}\quad
T^*M=\bigcup_xT_x^*M
.
\]
\end{definition}
Thus, replacing $v\in T_xM$ with $df\in T_x^*M$, for all $x\in M$ and for
all smooth functions $f{:\ } M\to \R$, yields the \emph{cotangent bundle}
$T^*M$.

\subsubsection{Differential bases} When the basis of vector fields is Cartesion and denoted
as $\frac{\partial}{\partial x^i}$ for $i=1,\dots,n$, its dual basis may be
denoted as $dx^i$. In this notation, the differential of a function
at a point $x\in M$ is expressed as
\[
df(x)=\frac{\partial f}{\partial x^i} dx^i
\,.
\]

The corresponding pairing $\left<\cdot,\cdot\right>$ of bases is
written in Cartesion notation as
\[\left< dx^j, \frac{\partial}{\partial x^i}\right> = \delta_i^j \,.\] 
Here $\delta_i^j$ is the Kronecker delta, which equals unity for $i=j$ and
vanishes otherwise. That is, defining $T^*M$ requires a pairing
$\left<\cdot,\cdot\right>{:\ } T^*M\times TM\to \R$.

Although different pairings can exist, e.g., for curvilinear coordinates, Riemannian
manifolds, etc.,  for simplicity of notation in this text we will usually apply the Cartesian pairing, as above.

\newpage
\vspace{4mm}\centerline{\textcolor{shadecolor}{\rule[0mm]{6.75in}{-2mm}}\vspace{-4mm}}
\section{Derivatives of differentiable maps -- the tangent lift}

\secttoc

\textbf{What is this lecture about?} This lecture introduces the tangent and cotangent lifts
of differentiable maps between manifolds and discusses some of their properties obtained 
from the calculus chain rule.

\subsection{Derivatives of differentiable maps between manifolds}
We next define derivatives of differentiable maps between manifolds
(tangent lifts).

We expect that a smooth map $f{:\ } U\to V$ from a chart $U\subset M$ to a chart
$V\subset N$,  will lift to a map between the tangent bundles $TM$ and
$TN$ so as to make sense from the viewpoint of ordinary calculus,
\begin{align*}
U\times \R^m\subset TM 
&\longrightarrow V\times \R^n\subset TN \\
\big(q^1,\dots, q^m; X^1,\dots, X^m\big) 
&\longmapsto \big(Q^1,\dots, Q^n; Y^1,\dots, Y^n\big)
\,.\end{align*}
Namely, the relations between the vector field components should be
obtained from the differential of the map $f{:\ } U\to V$. Perhaps not
unexpectedly, these vector field components will be related by
\[
Y^i\frac{\partial }{\partial Q^i}
=
X^j\frac{\partial }{\partial q^j}
,\quad\hbox{so}\quad
Y^i = \frac{\partial Q^i}{\partial q^j} X^j
,
\] 
in which the quantity called the \emph{tangent lift} \index{manifold!tangent lift}
of the function $f$, denoted \index{tangent lift}
\[
Tf=\frac{\partial Q}{\partial q} 
\,,\]
arises from the chain rule and is equal to the Jacobian for the
transformation 
\[
Tf{:\ } TM\mapsto TN
\,.\]

The dual of the tangent lift is the cotangent lift, discussed further in \index{cotangent lift}
\ref{cot-lift}. Basically, the \emph{cotangent lift} of the function $f$, 
\index{manifold!cotangent lift}
\[
T^*f=\frac{\partial q}{\partial Q}
\]
arises from
\[
 \beta_i dQ^i = \alpha_j dq^j
,\quad\hbox{so}\quad
\beta_i=\alpha_j\frac{\partial q^j}{\partial Q^i}
\]
and $T^*f{:\ } T^*N\mapsto T^*M$. Note the directions of these maps:
\begin{align*}
Tf&{:\ } q,X\in TM \mapsto Q,Y\in TN \\
f&{:\ } q\in M \mapsto Q\in N \\
T^*f&{:\ } Q,\beta\in T^*N \mapsto  q,\alpha\in T^*M
&&\text{($T^*f$ map goes the other way)}\,.
\end{align*}

\subsection{Summary remarks about derivatives on manifolds}

\begin{definition}[Differentiable map]
A map $f{:\ } M\to N$ from manifold $M$ to manifold $N$ 
is said to be \emph{differentiable} (resp. $C^k$) \index{differentiable map}
if it is represented in local coordinates on $M$ and $N$ 
by differentiable (resp. $C^k$) functions. 
\end{definition}

\begin{definition}[Derivative of a differentiable map]
The \emph{derivative} of a differentiable map
\[f{:\ } M\to N\]
at a point $x\in M$ is defined to be the linear map
\[T_xf{:\ } T_xM\to T_xN\]
constructed, as follows. For $v\in T_xM$, choose a curve $c(t)$ that maps an open
interval $t\in(-\epsilon,\epsilon)$ around the point $t=0$ to the
manifold $M$:
$$c{:\ } (-\epsilon,\epsilon)\to M$$
with $c(0)=x$ and velocity vector $c'(0):= \frac{dc}{dt}\big|_{t=0}=v$.

Then $T_xf\cdot v$ is the velocity vector at $t=0$ of the curve $f\circ
c{:\ } \mathbb{R}\to N$. That is,
\[
T_xf\cdot v 
= \frac{d}{dt}f(c(t))\Big|_{t=0}
= \frac{\partial f}{\partial c}\frac{d}{dt}c(t)\Big|_{t=0}
\]
\end{definition}!

\begin{definition}\label{tang-lift-defn}
The union $Tf=\bigcup_xT_xf$ of the derivatives $T_xf{:\ } T_xM\to T_xN$
over points $x\in M$ is called the \emph{tangent lift} of the map
$f{:\ } M\to N$. \index{differentiable map! tangent lift} \index{tangent lift}
\end{definition}

\begin{remark}\rm 
The chain-rule definition of the derivative $T_xf$ of a differentiable
map at a point $x$ depends on the function $f$ and the vector $v$.  Other
degrees of differentiability are possible. For example, if
$M$ and
$N$ are manifolds and
$f{:\ } M\to N$ is of class $C^{k+1}$, then the tangent lift (Jacobian)
$T_xf{:\ } T_xM\to T_xN$ is $C^k$.
\end{remark}

\begin{exercise}
Let $\phi_t {:\ }  S^2 \to S^2$ rotate points on $S^2$ about a fixed axis
through an angle $\psi(t)$. Show that $\phi_t$ is the flow of a certain
vector field on $S^2$.
\end{exercise}

\begin{exercise}
Let $f {:\ }  S^2 \to\mathbb{R}$ be defined by $f(x, y, z) = z$. Compute $df$
using spherical coordinates $(\theta,\phi)$.
\end{exercise}

\begin{exercise}
Compute the tangent lifts for the two stereographic projections of
$S^2\to\mathbb{R}^2$ in \ref{2StereoProj-eg}. That is, assuming
$(x,y,z)$ depend smoothly on
$t$, determine:
\begin{enumerate}
\item
How $(\dot{\xi}_N,\dot{\eta}_N)$ depend on $(\dot{x},\dot{y},\dot{z})$.
Likewise for $(\dot{\xi}_S,\dot{\eta}_S)$.
\item
How $(\dot{\xi}_N,\dot{\eta}_N)$ depend on $(\dot{\xi}_S,\dot{\eta}_S)$.
\end{enumerate}
Hint: Recall $(1+z)(1-z)=1-z^2=x^2+y^2$ and
use $x\dot{x}+y\dot{y}+z\dot{z}=0$ when
$(\dot{x},\dot{y},\dot{z})$ is tangent to $S^2$ at $(x,y,z)$.
\end{exercise}

\newpage
\vspace{4mm}\centerline{\textcolor{shadecolor}{\rule[0mm]{6.75in}{-2mm}}\vspace{-4mm}}
\section{Lifted actions  and the Jacobi--Lie bracket on vector fields}

\secttoc

\textbf{What is this lecture about?} This lecture discusses the properties of the 
tangent and cotangent lifted actions of differentiable maps
on manifolds and their relations to the Jacobi--Lie bracket.

\subsection{Lifted actions}
\begin{definition} Let  $\Phi{:\ } G\times M\to M$ be a left action, and write
$\Phi_g(x)=\Phi(g,x)$ for $x\in M$.  The \emph{tangent lift action} \index{tangent lift action} 
of $G$ on the tangent bundle $TM$ is defined by $gv=T_x\Phi_g(v)$ for every $v\in T_xM$.
\end{definition}

\begin{remark}\rm  
In standard calculus notation, the expression for \emph{tangent lift} may be
written as
\[
T_x\Phi\cdot v 
= \frac{d}{dt}\Phi(c(t))\Big|_{t=0}
=
\frac{\partial \Phi}{\partial c}c'(t)\Big|_{t=0} 
=: 
D\Phi(x)\cdot v, \]
with $c(0)=x$ and $c'(0)=v$. 
Thus, $T_x\Phi=D\Phi(x)$ is nothing but the Jacobian on the map $\Phi$.
\end{remark} 

\begin{definition}
If $X$ is a vector field on $M$ and $\phi$ is a differentiable
map from $M$ to itself, then the \emph{push-forward} 
of $X$ by $\phi$ is the vector field $\phi_*X$ defined by
$\left(\phi_*X\right)\left(\phi(x)\right) =
T_x\phi\left(X(x)\right)$. Consequently, the following diagram commutes:
\index{differentiable map!push-forward}

\hspace{3cm}
\begin{picture}(150,100)(-60,0)
\put(20,75){\vector(1,0){65} }
\put(-10,72){$TM$}
\put(95,72){$TM$}
\put(40,85){$T\phi$}
\put(100,10){\vector(0,1){55} }
\put(110,30){$\phi_*X$}
\put(20,0){\vector(1,0){65} }
\put(45,10){$\phi$}
\put(0,10){\vector(0,1){55} }
\put(-5,-3){$M$}
\put(95,-3){$M$}
\put(-20,30){$X$}
\end{picture}
\vspace{5mm}

\noindent If $\phi$ is a diffeomorphism then the \emph{pull-back} $\phi^*X$ is also defined:  
$\left(\phi^*X\right)\left(x\right) =
T_{\phi(x)}\phi^{-1}\left(X\left(\phi(x)\right)\right)$.
Hence, one sees that push-forward by the map $\phi$ is the pull-back by the inverse map $\phi^{-1}$.
\index{differentiable map!pull-back}
\end{definition}

\begin{definition} \label{defgX}
Let  $\Phi{:\ } G\times M\to M$ be a left action, and write $\Phi_g(m)=\Phi(g,m)$. 
Then $G$ has a left action on $X\in\mathfrak{X}(M)$ (the set of vector fields on $M$)
by the push-forward: $gX = \left(\Phi_g\right)_* X$.
\end{definition}

\begin{definition} \label{definvt}
Let $G$ act on $M$ on the left.
A vector field $X$ on $M$ is \emph{invariant} with respect to this action
(we often say ``$G$--invariant'' if the action is understood) if $gX=X$ for all $g\in G$;
equivalently (using all of the above definitions!) $g\left(X(x)\right) = X(gx)$ for
all $g\in G$ and all $x\in X$.
\end{definition}


\begin{definition}
Consider the left action of $G$ on itself by left multiplication, $\Phi_g(h)=L_g(h) = gh$,
for $g,h\in G$.
A vector field on $G$ that is invariant with respect to this action is called 
\emph{left-invariant}. From definition \ref{definvt}, we see that $X$ is left-invariant
if and only if $g\left(X(h)\right) = X(gh)$, which in less compact notation means
$T_hL_g X(h) = X(gh)$.
The set of all such vector fields is written $\mathfrak{X}^L(G)$.
\end{definition}

\begin{proposition} Given a $\xi\in T_eG$, define $X_{\xi}^L(g)=g\xi$
(recall: $g\xi\equiv T_eL_g \xi$).
Then $X_{\xi}^L$ is the unique left-invariant vector field such that
$X_{\xi}^L(e)=\xi$.
\end{proposition}

\begin{proof}
To show that $\smash{X_{\xi}^L}$ is left-invariant, we need to show that
$g\big(\smash{X_{\xi}^L}(h)\big) = \smash{X_{\xi}^L}(gh)$ for every $g,h\in G$. This follows from the
definition of $\smash{X_{\xi}^L}$ and the associativity property of group actions:
\[
g\big(X_{\xi}^L(h)\big) = g (h \xi) = (gh) \xi = X_{\xi}^L(gh)\,.
\]
We repeat the last line in less compact notation:
\[
T_hL_g\big(X_{\xi}^L(h)\big) = T_hL_g (h \xi) = T_eL_{gh} \xi = X_{\xi}^L(gh)
\]
For uniqueness, suppose $X$ is left-invariant and $X(e)=\xi$. Then for any $g\in G$, we have
$X(g)=g(X(e))=g\xi=\smash{X_{\xi}^L}(g)$.
\end{proof}

\begin{remark}\rm 
Note that the map $\xi\mapsto \smash{X_{\xi}^L}$ is an vector space
isomorphism from $T_eG$ to $\mathfrak{X}^L(G)$.
\end{remark}

All of the above definitions have analogues for right actions. \index{right action}
The definitions of \emph{right-invariant}, $\mathfrak{X}^R(G)$ and
$\smash{X_{\xi}^R}$
use the right action of $G$ on itself defined by $\Phi(g,h)=R_g(h)=hg$.

\begin{exercise} There is a left action of $G$ on itself defined by
$\Phi_g (h)=hg^{-1}$.
\end{exercise}

We will use the map $\xi\mapsto \smash{X_{\xi}^L}$ to relate the Lie
bracket on $\mathfrak{g}$, defined as $[\xi,\eta] = \ad_{\xi} \eta$,
with the Jacobi--Lie bracket on vector fields.

\subsection{Jacobi--Lie bracket on vector fields}
\begin{definition} The \emph{Jacobi--Lie bracket} \index{Jacobi--Lie bracket} on $\mathfrak{X}(M)$
is defined in local coordinates by
\[ [X,Y]_{\text{J--L}} \equiv (DX)\cdot Y- (DY)\cdot X \]
which, in finite dimensions, is equivalent to
\[ [X,Y]_{\text{J--L}} \equiv -(X\cdot \nabla)Y+ (Y\cdot \nabla) X \equiv -[X,Y] \]
\end{definition}

\begin{theorem}[Properties of the Jacobi--Lie bracket] \label{JL}
\quad \index{Jacobi--Lie bracket!properties} 

\begin{enumerate}
\item
The Jacobi--Lie bracket satisfies
\[[X,Y]_{\text{J--L}}=\mathcal{L}_XY\equiv \frac{d}{dt}\Big|_{t=0} \Phi_t^* Y,\]
where $\Phi$ is the flow of $X$.
(This relation is coordinate-free, and can be used as an alternative definition.)
\item This bracket makes $\mathfrak{X}^L(M)$ a Lie algebra with 
$[X,Y]_{\text{J--L}}=-[X,Y]$, where $[X,Y]$ is the Lie algebra bracket on
$\mathfrak{X}(M)$.
\item $\phi_*[X,Y] = [\phi_*X,\phi_*Y]$ for any differentiable $\phi{:\ } M\to M$.
\end{enumerate}
\end{theorem}


\begin{theorem}\label{XLsub}
$\mathfrak{X}^L(G)$ is a subalgebra of $\mathfrak{X}(G)$.
\end{theorem}
\begin{proof} Let $X,Y\in \mathfrak{X}^L(G)$.
Using the last item of the previous theorem, and then the $G$ invariance of $X$ and $Y$,
gives the push-forward relations
\[
\left(L_g\right)_*[X,Y]_{\text{J--L}} 
= [\left(L_g \right)_*X,\left(L_g\right)_*Y]_{\text{J--L}} 
\]
for all $g\in G$.
Hence $[X,Y]_{\text{J--L}}\in \mathfrak{X}^L(G)$.
This is the second property in Theorem \ref{JL}.
\end{proof} 

\begin{theorem} \label{Lie--JL}
Set $\big[\smash{X_{\xi}^L},\smash{X_{\eta}^L}\big]_{\text{J--L}}(e)=[\xi,\eta]$ for every
$\xi,\eta\in \mathfrak{g}$, where the bracket on the right is the Jacobi--Lie bracket.
(One says: the Lie bracket on $\mathfrak{g}$ is the pull-back of the Jacobi--Lie bracket by
the map $\xi\mapsto \smash{X_{\xi}^L}$.)
\end{theorem}
\begin{proof} The proof of Theorem \ref{Lie--JL} for matrix Lie algebras is relatively easy: we have already seen that 
$\mathrm{ad}_AB = AB-BA$. On the other hand, since $X_A^L(C)=CA$ for all $C$, and this is 
linear in $C$, we have $DX_B^L(I) \cdot A=AB$, so
\begin{align*}
[A,B]=\big[X_A^L,X_B^L\big]_{\text{J--L}}(I) 
&= DX_B^L(I)\cdot X_A^L(I) - DX_A^L(I)\cdot X_B^L(I) \\
&=DX_B^L(I)\cdot A - DX_A^L(I)\cdot B = AB-BA
\,.\end{align*}
This is the third property of the Jacobi--Lie bracket listed in \ref{JL}.
For the general proof, see Marsden and Ratiu~\cite[Proposition~9.14]{MaRa1994}.
\end{proof} 

\begin{remark}\rm 
Theorem \ref{Lie--JL}, together with Item 2 in Theorem \ref{JL}, proves that the
Jacobi--Lie bracket makes $\mathfrak{g}$ into a Lie algebra.
\end{remark}

\begin{remark}\rm 
By Theorem \ref{XLsub}, the vector field
$\big[\smash{X_{\xi}^L},\smash{X_{\eta}^L}\big]$ is left-invariant.
Since $\big[\smash{X_{\xi}^L},\smash{X_{\eta}^L}\big]_{\text{J--L}}(e)=[\xi,\eta]$, it follows that
\[ \big[X_{\xi}^L,X_{\eta}^L\big] = X_{[\xi,\eta]}^L.  \]  
\end{remark}

\begin{definition} Let $\Phi{:\ } G\times M\to M$ be a left action, and
let $\xi\in \mathfrak{g}$.
Let $g(t)$ be a path in $G$ such that $g(0)=e$ and $g'(0)=\xi$. Then the
\emph{infinitesimal generator} \index{infinitesimal generator} of the action 
in the $\xi$ direction is the vector field $\xi_M$ on $M$ 
defined by 
\[
\xi_M(x) = \frac{d}{dt}\Big|_{t=0} \Phi_{g(t)}(x)
\,.\]
\end{definition}

\begin{remark}\rm 
Note: this definition does not depend on the choice of $g(t)$. For
example, the choice in Marsden and Ratiu~\cite{MaRa1994}  is $\exp(t\xi)$,
where $\exp$ denotes the exponentiation on Lie groups (not defined here).
\end{remark}

\begin{exercise} Consider the action of $SO(3)$ on the unit sphere $S^2$
around the origin, and let $\xi=(0,0,1)\hat{~}$. Sketch the vector field
$\xi_M$. (Hint: the vectors all point ``Eastward.'')
\end{exercise}

\begin{theorem}
For any left action of $G$, the Jacobi--Lie bracket of infinitesimal generators is related to the Lie bracket 
on $\mathfrak{g}$ as follows (note the minus sign):
\[
[\xi_M,\eta_M] = -[\xi,\eta]_M
\,.\]
\end{theorem}

For a proof, see Marsden and Ratiu~\cite[Proposition~9.3.6]{MaRa1994}.

\begin{exercise}
Express the statements and formulas of this lecture for the case of
$SO(3)$ action on its Lie algebra $\mathfrak{so}(3)$. (Hint: look at the previous
lecture.) Wherever possible, translate these formulas to $\mathbb{R}^3$ by
using the $\widehat{~}$ map: $\mathfrak{so}(3)\to\mathbb{R}^3$.

Write the Lie algebra for $\mathfrak{so}(3)$ using the Jacobi--Lie bracket in terms of
linear vector fields on $\mathbb{R}^3$. What are the characteristic curves
of these linear vector fields?
\end{exercise}

\begin{definition} Let $X$ and $Y$ be two vector fields on the same manifold $M$.
The \emph{Lie derivative} \index{Lie derivative} of $Y$ with respect to $X$ is 
$\mathcal{L}_XY\equiv \frac{d}{dt} \Phi_t^* Y\big|_{t=0}$,
where $\Phi$ is the flow of $X$.
\end{definition}

\begin{remark}\rm 
The Lie derivative $\mathcal{L}_XY$ is 
``the derivative of $Y$ in the direction given by $X$.''
Its definition is coordinate-independent. By contrast, $DY\cdot X$ (also
written as $X[Y]$) is also ``the derivative of $Y$ in the $X$ direction'',
but the value of $DY\cdot X$ depends on the coordinate system, and in
particular does not usually equal $\mathcal{L}_XY$ in the chosen
coordinate system.
\end{remark}

\begin{theorem}\label{JLbrkt-def}
$\mathcal{L}_XY = [X,Y]$, where the bracket on the right is the
\emph{Jacobi--Lie bracket}.
\end{theorem}

\begin{proof}
In the following calculation, we assume that $M$ is finite-dimensional, and we work in local coordinates. 
Thus we may consider everything as matrices, which allows us to 
use the product rule and
the identities $\big(M^{-1}\big)' = -M^{-1}M'M^{-1}$ and 
$\frac{d}{dt}\big(D\Phi_t(x)\big) = D\big(\frac{d}{dt}\Phi_t\big)(x)$.
\begin{align*}
\mathcal{L}_XY(x) &= \frac{d}{dt}\Phi_t^* Y(x)\Big|_{t=0}  \\
&=\frac{d}{dt}\big(D\Phi_t(x)\big)^{-1} 
Y\big(\Phi_t(x)\big) \Big|_{t=0} \\
&=\bigg[\bigg(\frac{d}{dt}\left(D\Phi_t(x)\right)^{-1}\bigg)
Y\big(\Phi_t(x)\big) 
+\big(D\Phi_t(x)\big)^{-1}\frac{d}{dt} Y\big(\Phi_t(x)\big) \bigg]_{t=0} \\
&=\bigg[-\big(D\Phi_t(x)\big)^{-1}
\bigg(\frac{d}{dt}D\Phi_t(x) \bigg)\big(D\Phi_t(x)\big)^{-1}
Y\big(\Phi_t(x)\big) \\[-1ex]
&\hspace{185pt}+ \big(D\Phi_t(x)\big)^{-1}\frac{d}{dt} Y\big(\Phi_t(x)\big)
\bigg]_{t=0} \\
&=\bigg[-\bigg(\frac{d}{dt}D\Phi_t(x)
\bigg)Y(x)  +\frac{d}{dt} Y\big(\Phi_t(x)\big)
\bigg]_{t=0} \\
&=-D\bigg(\frac{d}{dt}\Phi_t(x)\bigg|_{t=0}\bigg)
Y(x)  +DY(x)\bigg(\frac{d}{dt} \Phi_t(x)\bigg|_{t=0} \bigg)\\
&=-DX(x)\cdot Y(x) + DY(x)\cdot X(x)\\
&=[X,Y]_{\text{J--L}}(x)
\end{align*}
Therefore $\mathcal{L}_XY = [X,Y]_{\text{J--L}}$.
\end{proof}


\newpage
\vspace{4mm}\centerline{\textcolor{shadecolor}{\rule[0mm]{6.75in}{-2mm}}\vspace{-4mm}}
\section{Lie group action on its tangent bundle}
\noindent
\label{sec:11}

\secttoc

\textbf{What is this lecture about?} This lecture discusses the left-invariant and right-invariant
subalgebras of the Lie algebra of vector fields and explains the sign differences in their associated Jacobi--Lie brackets. 

\subsection{Definitions of actions}
\begin{definition}
A Lie group $G$ acts on its tangent bundle $TG$ by tangent lifts.%
\footnote{Recall that section \ref{tang-lift-defn} deals with tangent lifts of a
differentiable manifold.} Given $X\in T_hG$ we can consider the action
of $G$ on $X$ by either left or right translations, denoted as 
$T_hL_gX$ or $T_hR_gX$, respectively. These expressions may be
abbreviated as 
\[
T_hL_gX=L^*_gX=gX
\qquad\hbox{and}\qquad
T_hR_gX=R^*_gX=X g
.
\]
Left action of a Lie group $G$ on its tangent bundle $TG$ is
illustrated in the figure below.

\hspace{3cm}
\begin{picture}(150,100)(-60,0)
\put(20,75){\vector(1,0){65} }
\put(-10,72){$TG$}
\put(95,72){$TG$}
\put(40,85){$TL_g$}
\put(100,10){\vector(0,1){55} }
\put(110,30){$gX$}
\put(20,0){\vector(1,0){65} }
\put(45,10){$L_g$}
\put(0,10){\vector(0,1){55} }
\put(-5,-3){$G$}
\put(95,-3){$G$}
\put(-20,30){$X$}
\end{picture}
\vspace{5mm}

For matrix Lie groups, this action is just multiplication on the left or
right, respectively. 
\end{definition}

\subsection{Left- and right-invariant vector fields} \index{left- and right-invariant vector fields}
A vector field $X$ on $G$ is called left-invariant, if for every $g\in G$
one has $L^*_gX=X$, that is, if 
\[ (T_hL_g)X(h)=X(gh) \]
for every $h\in G$. 
The commutative diagram for a left-invariant vector field is illustrated
in the figure below.

\hspace{3cm}
\begin{picture}(150,100)(-60,0)
\put(20,75){\vector(1,0){65} }
\put(-10,72){$TG$}
\put(95,72){$TG$}
\put(40,85){$TL_g$}
\put(100,10){\vector(0,1){55} }
\put(110,30){$X$}
\put(20,0){\vector(1,0){65} }
\put(45,10){$L_g$}
\put(0,10){\vector(0,1){55} }
\put(-5,-3){$G$}
\put(95,-3){$G$}
\put(-20,30){$X$}
\end{picture}
\vspace{5mm}

\begin{proposition}
The set $\mathfrak{X}_L(G)$ of left-invariant vector fields on the
Lie group $G$  is a \emph{subalgebra} of $\mathfrak{X}(G)$ the set of all
vector fields on $G$.
\end{proposition}
\begin{proof}
If $X,Y\in\mathfrak{X}_L(G)$ and $g\in
G$, then 
\[ L_g^*[X,Y] = \big[L_g^*X,L_g^*Y\big] = [X,Y]. \]
Consequently, the Lie bracket $[X,Y]\in \mathfrak{X}_L(G)$. Therefore,
$\mathfrak{X}_L(G)$ is a subalgebra of $\mathfrak{X}(G)$, the set of all
vector fields on $G$. 
\end{proof}

\begin{proposition}
The linear maps
$\mathfrak{X}_L(G)$ and $T_eG$ are isomorphic as vector spaces.
\end{proposition}
\paragraph*{Demonstration of proposition.}
For each $\xi\in T_eG$, define a vector field $X_{\xi}$ on $G$ by letting 
$
X_{\xi}(g)=T_eL_g(\xi)
.
$
Then 
\begin{eqnarray*}
X_{\xi}(gh)&=&T_eL_{gh}(\xi)=T_e(L_g\circ L_h)(\xi) \\
&=& T_hL_g(T_eL_h(\xi))=T_hL_g(X_{\xi}(h))
,
\end{eqnarray*}
which shows that $X_{\xi}$ is left invariant. 
(This proposition is stated by Marsden and Ratiu~\cite[Chapter~9]{MaRa1994}, who refer to
Abraham and Marsden~\cite{AbMa1978} for the full proof.)

\subsection{Jacobi--Lie bracket of vector fields}
\begin{definition}[\emph{Jacobi--Lie bracket of vector fields}]
\label{Jac-Lie-bracket-defn}
Let $g(t)$ and $h(s)$ be
curves in $G$ with $g(0)=e$, $h(0)=e$ and define vector fields at the
identity of $G$ by the tangent vectors  $g'(0)=\xi$, $h'(0)=\eta$. Compute
the linearization of the Adjoint action of $G$ on $T_eG$ as 
\[
[\xi,\eta]:=
\frac{d}{dt}\frac{d}{ds}g(t)h(s)g(t)^{-1}\Big|_{s=0,t=0}
=
\frac{d}{dt}g(t)\eta g(t)^{-1}\Big|_{t=0}
=
\xi\eta-\eta\xi
.
\]
This is the \emph{Jacobi--Lie bracket} of the vector fields $\xi$ and
$\eta$.
\end{definition}

\begin{definition}
The \emph{Lie bracket} in $T_eG$ is defined by \index{Lie bracket}
\[
[\xi,\eta]:=[X_{\xi},X_{\eta}](e)
,
\]
for $\xi, \eta\in T_eG$ and for $[X_{\xi},X_{\eta}]$ the Jacobi--Lie bracket
of vector fields. This makes $T_eG$ into a Lie algebra. Note that 
\[
[X_{\xi},X_{\eta}]=X_{[\xi,\eta]}
,
\]
for all $\xi, \eta\in T_eG$.
\end{definition}

\begin{definition}
The vector space $T_eG$ with this Lie algebra structure is called the
\emph{Lie algebra of $G$} and is denoted by $\mathfrak{g}$. 
\end{definition}

If we let $\xi_L(g) = T_eL_g\xi$, then the Jacobi--Lie bracket of two such
left-invariant vector fields in fact gives the Lie algebra bracket: 
\[
[\xi_L,\eta_L](g) = [\xi,\eta]_L(g)
\,.\]

\begin{remark}\rm 
For the right-invariant case, the right hand side obtains a minus sign, namely,
\[[\xi_R,\eta_R](g) = -[\xi,\eta]_R(g).\]
The relative minus sign arises because of the difference in action
$(xh^{-1})g^{-1}=x(gh)^{-1}$ on the right versus $(gh)x = g(hx)$ on the
left.
\end{remark}

\subsection{Infinitesimal generators}
In mechanics, group actions 
often appear as symmetry transformations which arise through their
infinitesimal generators, defined as follows.
\begin{definition} Suppose $\Phi {:\ }  G \times M \to M$ is an action. For
$\xi\in \mathfrak{g}$, $\Phi^\xi(t,x) {:\ }  \mathbb{R}\times M \to M$ defined
by $\Phi^\xi(x) = \Phi(\exp{t\xi},x)=\Phi_{\exp{t\xi}}(x)$ is an
$\mathbb{R}$--action on $M$. In other words, $\Phi_{\exp{t\xi}}\to M$ is
a flow on $M$. The vector field on
$M$ defined by%
\footnote{Recall \ref{VecField-defn} of vector fields.}
\[
\xi_M(x) 
= 
\frac{d}{dt}\Big|_{t=0}
\Phi_{\exp{t\xi}}(x)
\]
is called the \emph{infinitesimal generator} of the action $\Phi {:\ }  G \times M \to M$
corresponding \index{infinitesimal generator} to $\xi$.
\end{definition}
The Jacobi--Lie bracket of infinitesimal generators is related to the Lie
algebra bracket as follows:
\[
[\xi_M,\eta_M]=-[\xi,\eta]_M
\,.\]
See, for example, Marsden and Ratiu~\cite[Chapter~9]{MaRa1994} for the
proof. 

%
%
%
\newpage
\vspace{4mm}\centerline{\textcolor{shadecolor}{\rule[0mm]{6.75in}{-2mm}}\vspace{-4mm}}
\section{Hamilton's principle on manifolds}
\label{Ham princ manifolds}

\secttoc

\textbf{What is this lecture about?} This lecture surveys the properties of Hamilton's principle on manifolds 
and discusses the implications of their Lie symmetries in Noether's theorem. 

\subsection{Hamilton's Principle of Stationary Action}
\begin{theorem}[Hamilton's Principle of Stationary Action] \label{teol331}
Let the smooth function $L{:\ } TQ\to\mathbb{R}$ be a Lagrangian on $TQ$. A
$C^2$ curve
$c{:\ }  [a , b]\rightarrow Q$ joining $q_a=c(a)$ to $q_b=c(b)$ satisfies the
 Euler--Lagrange equations if and only if
\[\delta\int_a^b L(c(t),{\dot c}(t)) 
+ \scp{p}{{\dot c}- \frac{dc}{dt}}
dt =0.\]
\end{theorem}

\begin{proof}
The meaning of the variational derivative in the statement is the
following. Consider a family of $C^2 $ curves $c(t,s)$ for
$|{s}| < \varepsilon$ satisfying $c_0(t) = c(t)$, $c(a,s) =
q_a $, and $c(b,s) = q_b $ for all ${s} \in (- \varepsilon,
\varepsilon)$. Then
\[
\delta\int_a^b L(c(t),{\dot c}(t)) dt : =
\frac{d}{d{s}}\Big|_{{s}= 0 }\int_a^b L(c(t,s),\dot{c}(t,s)) dt.
\]
Differentiating under the integral sign, working in local coordinates
(covering the curve $c(t) $ by a finite number of coordinate charts),
integrating by parts, denoting the variation as
\[
v(t): =\frac{d}{d{s}}\Big|_{{s}=0 } c(t,s),
\]
taking into account that $v(a) = v(b) = 0 $,
and applying the constraint ${\dot c}= \frac{dc}{dt}$ 
so that ${\dot v}= \frac{dv}{dt}$ yields
\[
\int_a^b\left(\frac{\partial L}{ \partial q^i} v^i + \frac{\partial L}{
\partial \dot{q}^i}\dot{v}^i  \right) dt
= \int_a^b\left(\frac{\partial L}{ \partial q^i} - \frac{d}{dt}
\frac{\partial L}{ \partial \dot{q}^i} \right) v^i dt
+
\scp{\frac{\partial L}{ \partial \dot{q}^i}}{v^i }\Big|_0^T
\,.\]
This vanishes for any $C^1 $ function $v(t) $ if and only if the
Euler--Lagrange equations hold and the endpoint terms vanish.
\end{proof}

\begin{remark}\rm  
The integral appearing in this theorem
\[
\mathcal{S}(c(\cdot)): = 
\int_a^b L(c(t),{\dot c}(t)) 
+ \scp{p}{{\dot c}- \frac{dc}{dt}}
dt 
\]
is called the \emph{action integral\/}. It is defined on $C^2$ curves
$c{:\ }  [a, b] \rightarrow Q $ with fixed endpoints, $c(a ) = q_a $ and
$c(b) = q_b $.
\end{remark} 

\begin{remark}\rm [Variational derivatives of functionals vs Lie
derivatives of functions]
The variational derivative of a functional 
$S[u]$ is defined as the linearization
\[
\lim_{\epsilon\to0}\frac{S[u+\epsilon v]-S[u]}{\epsilon}
=
\frac{d}{d\epsilon}\Big|_{\epsilon=0}S[u+\epsilon v]
=
\Big\langle \frac{\delta S}{\delta v},v\Big\rangle
.\quad
\]
Compare this to the expression for the Lie derivative of a function.
If $f$ is a real valued function on a manifold $M$ and $X$ is a vector
field on $M$, the Lie derivative of $f$ along $X$ is defined as the
directional derivative  
\[
\mathcal{L}_Xf=X(f):=\mathbf{d}f\cdot X
.
\]
If $M$ is finite-dimensional, this is
\[
\mathcal{L}_Xf=X[f]:=\mathbf{d}f\cdot X
=
\frac{\partial f}{\partial x^i}X^i
=
\lim_{\epsilon\to0}\frac{f(x+\epsilon X)-f(x)}{\epsilon}
.
\]
The similarity is suggestive: Namely, the Lie derivative of a function and
the variational derivative of a functional are both defined as
linearizations of smooth maps in certain directions.

\end{remark}

The next theorem emphasizes the role of Lagrangian
one-forms and two-forms in the variational principle. The following is a
direct corollary of the previous theorem.

\begin{theorem}\label{EL-manifold}
Given a $C^k$ Lagrangian $L{:\ } TQ\rightarrow \mathbb{R}$ for $k\geq 2$,
there exists a unique $C^{k-2}$ map ${\mathcal EL}(L){:\ } \ddot{Q}\rightarrow
T^*Q$, where
\[\ddot{Q}:=\bigg\{\frac{d^2q}{dt^2}\Big|_{t=0}\in T(TQ) ~\bigg|~
q(t) \text{ is a }\, C^2 \text{ curve in } Q \bigg\}\]
is a  submanifold of $T(TQ)$, and
a unique $C^{k-1}$ one-form $\Theta_L\in\Lambda^1(TQ)$, such that
for all $C^2$ variations $q(t,s)$ (defined on a fixed
$t$--interval) of $q(t,0)=q_0(t):=q(t)$, we have
\begin{align}
\delta \mathcal{S} :=
\frac{d}{ds}\Big|_{s=0}\mathcal{S}[c(\cdot ,s)] &=
\mathbf{D}\mathcal{S}[q(\cdot)]\cdot \delta q (\cdot) 
\\&\hspace{-15mm}= \int_a^b
\scp{{\mathcal EL}(L)\left(q, \dot{q}, \ddot{q}\right)}{\delta q}\, dt
+ \hspace{-1.3em} 
\underbrace{\scp{\Theta_L\big(q,\dot{q}\big)}{\delta
q}\Big|_a^b\ }_{\hbox{cf. \emph{Noether quantity}}}
\,.\nonumber\end{align}
In this equation, $\Theta_L\big(q,\dot{q}\big)$ is the \emph{fibre derivative} and $\delta q$ could for example be a \emph{Lie derivative}
\[
\Theta_L\big(q,\dot{q}\big) = \frac{\p L}{\p \dot{q}}
\quad\hbox{and}\quad
\delta q:
=\frac{d}{d s}\Big|_{s = 0} q(t,s).  \]
\end{theorem}


\subsection{Symmetries and Conservation Laws -- Noether's theorem}
In Theorem \ref{EL-manifold}, 
\begin{align}\label{Action-var}
\delta \mathcal{S}
&:=
\frac{d}{ds}\bigg|_{s=0}\mathcal{S}[c(\cdot ,s)]
=
\mathbf{D}\mathcal{S}[q(\cdot)]\cdot \delta q (\cdot) 
\\&= \int_a^b
{\mathcal EL}(L)\big(q, \dot{q}, \ddot{q}\big)\cdot\delta q\, dt
+\hspace{-1.3em}\underbrace{\Theta_L\left(q,\dot{q}\right)\cdot\delta
q\big|_a^b\ }_{\hbox{\emph{Noether quantity}}}
\nonumber\end{align}
where
\[\delta q(t)=\frac{d}{ds}\bigg|_{s = 0} q(t,s),\]
the map $\mathcal{EL}{:\ }  \ddot{Q}\rightarrow T^*Q$ is called the
\emph{Euler--Lagrange operator} \index{Euler--Lagrange operator} 
and its expression  in local coordinates is
\[
\mathcal{EL}(q, \dot{q}, \ddot{q})_{\,i} =
\frac{\partial L}{ \partial q^i} - \frac{d}{dt}
\frac{\partial L}{ \partial \dot{q}^i}
\,.
\]
One understands that the formal time derivative is taken in the
second summand and everything is expressed as a function of 
$(q, \dot{q}, \ddot{q})$.

\begin{theorem}[Symmetries and Conservation Laws Noether \cite{noether1918invariante}] 
If the action variation in equation \eqref{Action-var} vanishes
$\delta\mathcal{S}=0$ because of a symmetry transformation which does
\emph{not} preserve the end points and the Euler--Lagrange equations hold,
then the term marked \emph{cf. Noether's Theorem} must also vanish. However,
vanishing of this term now is interpreted as a constant of motion. Namely,
the term,
\[ A(v,w):= \langle \mathbb{F} L(v), w\rangle,
\quad\hbox{or, in coordinates}\quad A(q,\dot{q},\delta{q})
= \frac{\partial  L}{\partial \dot{q}^i}\,\delta{q}^i, \] 
is constant for solutions of the Euler--Lagrange equations. 

This result
first appeared in Noether \cite{noether1918invariante}. In fact, the result in
there is more general than this. In particular, in the PDE
(Partial Differential Equation) setting one must also include the
transformation of the volume element in the action principle. See,
for example, Olver~\cite{Ol2000} for good discussions of the history, framework and
applications of Noether's theorem.
\end{theorem}

\begin{exercise}
Show that conservation of energy results from Noether's Theorem if,
in Hamilton's principle, the variations are chosen as
\[\delta q(t)=\frac{d}{d s}\bigg|_{s = 0} q(t,s), \]
corresponding to symmetry of the Lagrangian under reparametrizations of
time along the given curve $q(t)\to q(\tau(t,s))$.
\end{exercise}

\subsection{The canonical Lagrangian one-form and two-form.}The one-form
$\Theta_L$, whose existence and uniqueness is guaranteed by \ref{EL-manifold}, 
appears as the boundary term of the derivative of the
action integral, when the endpoints of the curves on the configuration
manifold are free. In finite dimensions, its local expression is
\[ \Theta_L\big(q,\dot{q}\big) :=
\frac{\partial L}{\partial \dot{q}^i}\mathbf{d}q^i
\qquad\big(=p_i\big(q,\dot{q}\big)\,\mathbf{d}q^i\big).  \]
The corresponding closed two-form $\Omega_L=\mathbf{d}\Theta_L$ obtained
by taking its exterior derivative may be expressed as
\[ \Omega_L:=-\mathbf{d}\Theta_L=\frac{\partial^2 L}{\partial\dot{q}^i\partial
q^j}\mathbf{d}q^i\wedge \mathbf{d}q^j + \frac{\partial^2
L}{\partial\dot{q}^i\partial\dot{q}^j}\mathbf{d}q^i\wedge
\mathbf{d}\dot{q}^j
\qquad\big(=\mathbf{d}p_i\big(q,\dot{q}\big)\wedge\mathbf{d}q^i\big). \]
These coefficients may be written as the $2n \times 2n$
skew-symmetric matrix
\begin{equation}
\Omega_L=\left(\begin{matrix}
\mathcal{A} & \frac{\partial^2 L}{\partial\dot{q}^i\partial\dot{q}^j} \\
- \frac{\partial^2 L}{\partial\dot{q}^i\partial \dot{q}^j} &  0
\end{matrix}\right)
,
\label{none}
\end{equation}
where $\mathcal{A}$ is the skew-symmetric
$n{\times}n$ matrix 
$\smash{\big(\frac{\partial^2 L}{\partial\dot{q}^i\partial  q^j}\big)
-\big(\frac{\partial^2 L}{\partial\dot{q}^i\partial  q^j}\big)^T}$. 

Non-degeneracy of
$\Omega_L$ is equivalent to the invertibility of the matrix
$\smash{\big(\frac{\partial^2 L}{\partial\dot{q}^i\partial\dot{q}^j}\big)}$.

\begin{definition}
The \emph{Legendre transformation\/} $\mathbb{F}
L{:\ } TQ\rightarrow T^*Q$  is the smooth map near the
identity defined by \index{Legendre transformation\/} 
\begin{equation*}
\langle \mathbb{F}L(v_q),  w_q \rangle :=
\frac{d}{ds}\bigg|_{s=0}L(v_q+s w_q).
\end{equation*}
\end{definition}
In the finite dimensional case, the local expression of $\mathbb{F}L $
is
\[
\mathbb{F} L (q^i,\dot q^i)=\bigg(q^i,\frac{\partial L}
{\partial \dot{q}^i}\bigg)=(q^i,p_i(q,\dot{q})).
\]
If the skew-symmetric matrix \eqref{none} is invertible, the Lagrangian
$L$ is said to be \emph{regular\/}. In this case, by the implicit function 
theorem, $\mathbb{F}L$ is locally invertible. If $\mathbb{F}L$ is a
diffeomorphism,  $L$ is called \emph{hyperregular}.

\begin{definition}
Given a Lagrangian $L$, the \emph{action} of $L$ is  the map $A {:\ }  TQ
\rightarrow \mathbb{R}$ given by
\begin{equation}
\label{action of Lagrangian}
A(v):= \langle \mathbb{F} L(v),  v\rangle,
\quad\hbox{or, in coordinates}\quad 
A(q,\dot{q})=\frac{\partial  L}{\partial \dot{q}^i}\dot{q}^i,
\end{equation}
and the  \emph{energy\/} of $L$ is
\begin{equation}
\label{energy of Lagrangian}
E(v):= A(v)-L(v)
,\quad\hbox{or, in coordinates}\quad 
E(q,\dot{q})
=\frac{\partial  L}{\partial \dot{q}^i}\dot{q}^i
- L(q,\dot{q})
.
\end{equation}
\end{definition}

\subsection{Lagrangian vector fields and conservation laws}

\begin{definition}
A vector field $Z$ on $TQ$ is called a \emph{Lagrangian vector field\/}
\index{Lagrangian vector field}
if
\[
\Omega_L(v)(Z(v),w)= \langle \mathbf{d}E(v), w \rangle,
\]
for all $v\in T_qQ$, $w\in T_v(TQ)$.
\end{definition}
\begin{proposition}
The \emph{energy is conserved} along the flow of a Lagrangian vector field
$Z$.
\end{proposition}

\begin{proof} Let $v(t)\in TQ$ be an integral curve of $Z$. 
Skew-symmetry of $\Omega_L $ implies
\begin{align*}
\frac{d}{dt}E(v(t))=\langle \mathbf{d}E(v(t)),  \dot v(t) \rangle
&=\langle \mathbf{d}E(v(t)), Z(v(t))\rangle 
\\&=\Omega_L(v(t)) \left(Z(v(t)),Z(v(t))\right) = 0.
\end{align*}
Thus, the energy $E(v(t))$ is constant in time, $t$.
\end{proof}

\subsection{Equivalent dynamics for hyperregular Lagrangians and
Hamiltonians}

Recall that a Lagrangian $L$ is said to be \emph{hyperregular} if
its Legendre transformation $\mathbb{F}L{:\ } TQ\rightarrow T^*Q$  is
a diffeomorphism. \index{hyperregular Lagrangian}

The equivalence between the Lagrangian and Hamiltonian formulations for
hyperregular Lagrangians and Hamiltonians is summarized below, following
Marsden and Ratiu~\cite{MaRa1994}.

\begin{enumerate}
\item[(a)]
Let $L$ be a  hyperregular Lagrangian on $TQ$ and $H=E\circ (\mathbb{F}
L)^{-1}$, where $E$ is the energy of $L$ and
$(\mathbb{F}L)^{-1}{:\ } T^*Q\rightarrow TQ$ is the inverse of the Legendre
transformation. Then the Lagrangian vector field 
$Z$ on $TQ$ and the Hamiltonian vector field
$X_H$ on $T^*Q$ are related by the identity
$$(\mathbb{F} L)^*X_H=Z .$$
Furthermore, if $c(t)$ is an
integral curve of $Z$ and $d(t)$ an integral curve of $X_H$ with
$\mathbb{F} L(c(0))=d(0)$, then $\mathbb{F} L(c(t))=d(t)$  and their
integral curves coincide on the manifold $Q$. That is,
$\tau_Q(c(t))=\pi_Q(d(t))=\gamma(t)$, where $\tau_Q{:\ }  TQ \rightarrow Q $
and $\pi_Q{:\ }  T ^\ast Q \rightarrow Q $ are the canonical bundle projections.

In particular,  the pull-back of the inverse 
Legendre transformation $\mathbb{F} L^{-1}$ induces a
one-form
$\Theta$ and a closed two-form $\Omega $ on $T^*Q$ by
\begin{eqnarray*}
\Theta=(\mathbb{F} L^{-1})^*\Theta_L
,\qquad
\Omega=- \mathbf{d} \Theta=(\mathbb{F}L^{-1})^*\Omega_L
.
\end{eqnarray*}
In coordinates, these are the canonical presymplectic and
symplectic forms, respectively,
\begin{eqnarray*}
\Theta=p_i\,\mathbf{d}q^i
,\qquad
\Omega=- \mathbf{d} \Theta=\mathbf{d}p_i\wedge\mathbf{d}q^i
.
\end{eqnarray*}

\item[(b)]
A  Hamiltonian $H{:\ } T^*Q \rightarrow \mathbb{R}$ is said to be
\emph{hyperregular\/} if the smooth map $\mathbb{F} H{:\ } T^*Q\rightarrow TQ$,
defined by
\[
\langle  \mathbb{F}H(\alpha_q),  \beta_q\rangle
:= \frac{d}{ds}\bigg|_{s=0}
H(\alpha_q+s\beta_q),\qquad
\alpha_q, \beta_q \in T^*_qQ,\]
is a diffeomorphism.
Define the \emph{action\/} of $H$ by $G := \langle \Theta, 
X_H\rangle$. If $H$ is a hyperregular Hamiltonian then the energies of
$L$ and $H$ and the actions of $L$ and $H$ are related by
 $$E=H\circ (\mathbb{F} H)^{-1},\qquad \quad A=G\circ
 (\mathbb{F} H)^{-1}.$$
Also, the Lagrangian $L = A - E$ is hyperregular and  
$\mathbb{F} L = \mathbb{F} H^{-1}$.

\item[(c)]
These constructions define a bijective
correspondence between hyperregular Lagrangians and Hamiltonians.

\end{enumerate}

\begin{remark}\rm 
For thorough discussions of many additional results arising from the
Hamilton's principle for hyperregular Lagrangians see, for example,
\cite[Chapters~7 and~8]{MaRa1994} and \cite{HoScSt2009}.
\end{remark}

\newpage

\vspace{4mm}\centerline{\textcolor{shadecolor}{\rule[0mm]{6.75in}{-2mm}}\vspace{-4mm}}
\section{Euler--Lagrange equations on  manifolds}

\secttoc

\textbf{What is this lecture about?} This lecture is about the Euler--Lagrange 
approach to geodesic motion and its ramifications, such as the covariant derivative. 

\begin{definition}[Cotangent lift]\label{cot-lift}
Given two manifolds $Q$ and $S$ related by a diffeomorphism
$f{:\ } Q\mapsto S$, the \emph{cotangent lift} \index{cotangent lift}
$T^*f{:\ }  T^*S\mapsto T^*Q$
of $f$ is defined by 
\begin{equation}
\langle T^*f(\alpha),v\rangle = \langle \alpha,Tf(v)\rangle
\end{equation}
where 
\[ \alpha\in T^*_sS ,\quad v\in T_qQ ,\quad\hbox{and}\quad s=f(q).  \]
As explained by Marsden and Ratiu~\cite[Chapter~6]{MaRa1994}, cotangent lifts preserve the
\emph{action} of the Lagrangian $L$, which we write as
\begin{equation}\label{action-pair}
\langle \mathbf{p}, \mathbf{\dot{q}}\rangle =
\langle \alpha, \mathbf{\dot{s}}\rangle,
\end{equation}
where $\mathbf{p}=T^*f(\alpha)$ is the cotangent lift of $\alpha$ under
the diffeomorphism $f$ and $\mathbf{\dot{s}}=Tf(\mathbf{\dot{q}})$ is the
tangent lift of $\mathbf{\dot{q}}$ under the function $f$, which is
written in Euclidean coordinate components as $q^i\to s^i=f^i(\mathbf{q})$.
Preservation of the action in \eqref{action-pair} yields the
coordinate relations,
\begin{align*}
\text{(Tangent lift in coordinates)}\quad
\dot{s}^j&=\frac{\partial f^j}{\partial q^i}\dot{q}^i
\qquad\Longrightarrow
\\
\hspace{-5mm}
p_i&=\alpha_k\frac{\partial f^k}{\partial q^i}
\hspace{3mm}\text{(Cotangent lift in coordinates)}
\end{align*}
Thus, in coordinates, the cotangent lift is the inverse transpose of the
tangent lift.
\end{definition}

\begin{remark}\rm 
The cotangent lift of a function preserves the induced action one-form,
\begin{eqnarray*}\label{action-pair1}
\langle \mathbf{p}, \mathbf{d}\mathbf{q}\rangle
=
\langle \alpha, \mathbf{d}\mathbf{s}\rangle
,
\end{eqnarray*}
so it is a source of (pre-)symplectic transformations.
\end{remark}

\subsection{The classic Euler--Lagrange example: geodesic flow}
An important example of a Lagrangian vector field is the geodesic spray
of a Riemannian metric. A \emph{Riemannian
manifold} is a smooth manifold $Q$ endowed with a symmetric
nondegenerate covariant tensor $g$, which is positive definite. Thus, on
each tangent space $T_q Q$ there is a nondegenerate definite inner product
defined by pairing with $g(q)$. \index{Riemannian manifold!geodesic spray}

If $(Q, g)$ is a Riemannian manifold, there is a natural
Lagrangian on it given by the \emph{ kinetic energy\/} $K$ of the
metric $g$, namely,
$$K(v) : = \tfrac12 g(q)(v_q,v_q),
$$
for $q \in Q $ and $v_q \in T_q Q $. In
finite dimensions, in a local chart,
$$K(q, \dot{q})= \tfrac12 g_{ij}(q) \dot q^i \dot q^j. $$
The fibre derivative in this case is $\mathbb{F} K
(v_q) = g(q) (v_q, \cdot)$, for $v_q \in T_qQ$. In coordinates, this is 
\[
\mathbb{F} K (q,\dot q):=\left(q^i,\frac{\partial K}
{\partial \dot{q}^i}\right):=(q^i,g_{ij}(q) \dot q^j)=:(q^i,p_i).
\]
The Euler--Lagrange equations become the \emph{geodesic
equations\/} for  the Riemannian metric $g$, given (for finite dimensional $Q $ in
a local chart) by
\[\ddot{q}^i+\Gamma_{jk}^i\dot q^j\dot q^k=0,\quad  i
= 1 ,\ldots n,\]
where the three-index quantities
\[ \Gamma_{jk}^h=\tfrac12g^{hl}\bigg(\frac{\partial g_{jl}}{\partial
q^k}+ \frac{\partial g_{kl}}{\partial q^j}-\frac{\partial
g_{jk}}{\partial q^l}\bigg)
,\quad\hbox{with}\quad g_{ih}g^{hl}=\delta_i^l, \]
are the \emph{Christoffel symbols\/} \index{Christoffel symbols} of the Levi-Civita connection on $(Q,g)$.

\begin{exercise}
Explicitly compute the geodesic equation as an
Euler--Lagrange equation for the kinetic energy Lagrangian
$K(q, \dot{q})= \tfrac12 g_{ij}(q) \dot q^i \dot q^j$.
\end{exercise}

\begin{exercise}
For the kinetic energy Lagrangian $K(q, \dot{q})=
\frac{1}{2} g_{ij}(q) \dot q^i\dot q^j$ with $i,j=1,2,\dots,N$:
\begin{itemize}
\item
Compute the momentum $p_i$ canonical to $q^i$ for geodesic motion. 
\item
Perform the Legendre transformation to obtain the Hamiltonian for geodesic
motion.
\item
Write out the geodesic equations in terms of $q^i$ and its canonical
momentum $p_i$.
\item
Check directly that Hamilton's equations are satisfied.
\end{itemize}
\end{exercise}

\begin{exercise}
Consider the Lagrangian
\[
L_\epsilon(\mathbf{q},\mathbf{\dot{q}})
= 
\tfrac12 \|\mathbf{\dot{q}}\|^2
-
\tfrac{1}{2\epsilon}
(1 - \|\mathbf{q}\|^2)^2
\]
for a particle in $\mathbb{R}^3$. Let $\gamma_\epsilon(t)$ be the curve in
$\mathbb{R}^3$ obtained by solving the Euler--Lagrange equations for
$L_\epsilon$ with the initial conditions
$\mathbf{q}_0=\gamma_\epsilon(0), 
\mathbf{\dot{q}}_0 = \dot{\gamma}_\epsilon(0)$. Show that 
\[\lim_{\epsilon\to0}\gamma_\epsilon(t)\]
is a great circle on the two-sphere $S^2$, provided that $\mathbf{q}_0$ has
unit length and the initial conditions satisfy
$\mathbf{q}_0\cdot\mathbf{\dot{q}}_0 = 0$.
\end{exercise}

\begin{remark}\rm 
The Lagrangian vector field associated to $K(q, \dot{q})$ is called the
\emph{geodesic spray\/}. \index{geodesic spray}Since the Legendre transformation is a
diffeomorphism (in finite dimensions or in infinite dimensions if  the
metric is assumed to be strong), the geodesic  spray is always a second
order equation.
\end{remark}

\subsection{Covariant derivative}
The variational approach to geodesics recovers the classical
formulation using covariant derivatives, as follows. 
Let $\mathfrak{X}(Q)$ denote the set of vector fields on the manifold $Q$.
The \emph{covariant derivative} \index{covariant derivative}
\[
\nabla {:\ } \mathfrak{X}(Q)\times \mathfrak{X}(Q)\rightarrow
\mathfrak{X}(Q)\qquad
(X , Y) \mapsto \nabla_X(Y)
,
\]
 of the Levi-Civita connection on $(Q,g)$ is
given in local charts by
\[
\nabla_X(Y)=\Gamma_{ij}^kX^iY^j\frac{\partial}{\partial q^k}+
X^i\frac{\partial Y^k}{\partial q^i}\frac{\partial }{\partial q^k}
.
\]
If $c(t)$ is a curve on $Q$ and $Y\in \mathfrak{X}(Q)$, the covariant
derivative of $Y$ along $c(t)$ is defined by
$$\frac{DY}{Dt} := \nabla_{\dot c} Y,$$
or locally,
\[
\left(\frac{DY}{Dt}\right)^{\!k} =\Gamma_{ij}^k (c(t)) \dot c^i(t)
Y^j(c(t)) +
\frac{d}{dt} Y^k(c(t)).
\]
A vector field is said to be \emph{parallel transported\/} along
$c(t)$ if  \[\frac{DY}{Dt} =0.\] Thus $\dot c(t)$ is parallel
transported along $c(t)$ if and only if
$$\ddot{c}\,^i + \Gamma^i_{jk} \dot c^j\, \dot c^k = 0.
$$
In classical differential geometry a \emph{geodesic\/} is defined to be a 
curve $c(t)$ in $Q$ whose tangent vector $\dot{c}(t)$ is parallel
transported along $c(t)$. As the expression above shows, geodesics are
integral curves of the Lagrangian vector field defined by the kinetic
energy of $g$.

\begin{remark}\rm 
A classic problem is to determine the metric tensors $g_{ij}(q)$ for
which these geodesic equations admit enough additional
conservation laws to be integrable.
\end{remark}

\begin{exercise}
Consider a geodesic flow on a Riemannian manifold $M$ with metric $g_{ij}(x)$ for $x\in M$, 
so that $i,j=1,2,\dots,n$ for $\dim M = n$.
Show that if in Hamilton's principle the Lagrangian
\[
L(x,\dot{x}) = \frac12 \dot{x}^i g_{ij}(x) \dot{x}^j 
\]
admits an isometry group $G$, then the Poisson brackets among the conservation laws of the 
corresponding geodesic flow are isomorphic to the Lie algebra $\mathfrak{g}$ of the isometry group $G$. 
For this calculation, it will be helpful to recall the diamond $(\diamond)$ operation defined in equation 
\eqref{Noether-Lie-term} as
\[
\big\langle  p\diamond q \,,\, \xi \big\rangle_{\mathfrak{g}} 
:= 
\big\langle  p \,,\,-\, \mc{L}_\xi q\big\rangle_{TQ} 
\,,\]
where $\delta q = - \mc{L}_\xi q$ with $\xi\in \mathfrak{g}$ is the infinitesimal symmetry transformation 
of $q$ (i.e., minus the Lie derivative of $q$ in the direction $\xi$).
\end{exercise}

\eject

\begin{answer}
\begin{itemize}
\item
Hamilton's principle with Lagrangian $L(q,\dot{q})$ yields the Euler-Lagrange equations, plus an endpoint term obtained from integration by parts in time defined by the (nondegenerate) pairing 
\[
 \Big\langle  \frac{\p L}{\p \dot{q}} \,,\, \delta q\Big\rangle_{TQ} \Big|_a^b 
 =:  \big\langle  p \,,\, \delta q\big\rangle_{TQ} \Big|_a^b
\,.\]
\item
The \emph{Noether quantity} is defined to be  $\big\langle  p \,,\, \delta q\big\rangle_{TQ}$ for $\delta q =  -\,\mc{L}_\xi q$. The corresponding \emph{cotangent lift momentum map} is obtained in the form
\begin{align*}
\big\langle  p \,,\, \delta q\big\rangle_{TQ} &= \big\langle  p \,,\, -\,\mc{L}_\xi q\big\rangle_{TQ} 
=: \big\langle p\diamond q \,,\, \xi \big\rangle_{\mathfrak{g}}  
\\&=: \langle J(q,p),\xi \rangle_{\mathfrak{g}} = J_\xi(q,p)
\,,
\end{align*}
where $J_\xi(q,p)$ is the Noether Hamiltonian and one defines the diamond $(\diamond)$ operation as
\begin{align}
\big\langle  p\diamond q \,,\, \xi \big\rangle_{\mathfrak{g}} 
:= 
\big\langle  p \,,\,-\, \mc{L}_\xi q\big\rangle_{TQ} 
\,.
\label{diamond-def1}
\end{align}
Thus, the momentum map $J(q,p)$ of the isometry group $G$ of the geodesic Lagrangian $L_{geo}(q,\dot{q})$
is the quantity $J(q,p)= p\diamond q$. This form holds in general for geodesic problems.
\item
The canonical Poisson brackets $\{J_\xi (q,p),J_\eta (q,p)\}_{can}$ among the Noether Hamiltonians for the symmetries of the Lagrangian $L_{geo}(q,\dot{q})$ may be shown to be isomorphic to the Lie algebra $\mathfrak{g}$ of the isometry group $G$ via the  following calculation which uses the product rule for the canonical Poisson bracket,
\begin{align*}
\{J_\xi (q,p),J_\eta (q,p)\}_{can} &= \{J_\xi (q,p),\langle  p\diamond q \,,\, \eta \rangle_{\mathfrak{g} }\}_{can}
\\
&\hspace{-10mm}=\langle  \{J_\xi , p\}_{can} \diamond q +  p \diamond \{J_\xi , q \}_{can}
\,,\, \eta \rangle_{\mathfrak{g} }
\\
&\hspace{-10mm}=\langle  (-\mc{L}_\xi^Tp)\diamond q +  p \diamond (\mc{L}_\xi q) \,,\, \eta \rangle_{\mathfrak{g} }
\\
&\hspace{-10mm}=\langle  \mc{L}_\xi ( p \diamond q )\,,\, \eta \rangle_{\mathfrak{g} }
= \langle  \mathrm{ad}^*_\xi  ( p \diamond q )\,,\, \eta \rangle_{\mathfrak{g} }
\\
&\hspace{-10mm} = \langle    p \diamond q \,,\, \mathrm{ad}_\xi \eta \rangle_{\mathfrak{g} }
=  \langle    p  \,,\, -\,\mc{L}_{[\xi , \eta]}q \rangle_{TQ } 
\\
&\hspace{-10mm}= J_{[\eta,\xi]}(q,p)
\quad\hbox{(Anti-homomorphism)}
\,.\end{align*}
\end{itemize}

\end{answer}

\begin{definition}
A \emph{simple mechanical system\/}  \cite{Sm1970a,Sm1970b} \index{simple mechanical system}
is given by a Lagrangian of
the form $L(v_q) = K(v_q) - V(q) $, for $v_q \in T_qQ $. The smooth
function $V{:\ }  Q \rightarrow \mathbb{R}$ is called the \emph{potential
energy\/}. The total energy of this system is given by $E = K + V $
and the Euler--Lagrange equations (which are always second order for a
hyperregular Lagrangian) are
\[
\ddot{q}\,^i+\Gamma_{jk}^i\dot q^j\dot q^k +  g ^{il}
\frac{\partial V}{ \partial q ^l}=0,\quad  i  = 1 ,\ldots n,
\]
where $g^{ij}$ are the entries of the inverse matrix of the Riemannian metric $(g_{ij})$.
\end{definition}

\begin{exercise}[Gauge invariance] \index{gauge invariance} \label{gauge-ex} 
Show that the Euler--Lagrange equations are unchanged
under 
\begin{equation}
L(\mathbf{q}(t),\mathbf{\dot{q}}(t))
\rightarrow
L^\prime = L + \frac{d}{dt}\gamma(\mathbf{q}(t),\mathbf{\dot{q}}(t))
,
\end{equation}
for any function $\gamma{:\ }  \mathbb{R}^{6N} =
\{(\mathbf{q},\mathbf{\dot{q}}) \mid \mathbf{q}, \mathbf{\dot{q}} \in
\mathbb{R}^{3N} \} \rightarrow \mathbb{R}$.
\end{exercise}

\begin{exercise}[Generalized coordinate theorem]
\label{covar-ex} Show that the Euler--Lagrange equations are
\emph{unchanged in form} under any smooth invertible mapping
$f{:\ } \{\mathbf{q}\mapsto\mathbf{s}\}$. That is, with
\begin{equation}
L(\mathbf{q}(t),\mathbf{\dot{q}}(t))
=
\breve{L}(\mathbf{s}(t),\mathbf{\dot{s}}(t))
,
\end{equation}
show that
\begin{equation}
\frac{d}{dt}
\left(\frac{\partial L}{\partial
\mathbf{\dot{q}}}\right) 
-\frac{\partial L}{\partial \mathbf{q}}
=0
\quad\Longleftrightarrow\quad
\frac{d}{dt}
\left(\frac{\partial \breve{L}}{\partial
\mathbf{\dot{s}}}\right) 
-\frac{\partial \tilde{L}}{\partial \mathbf{s}}
=0
.
\end{equation}
\end{exercise}

\begin{exercise}
How do the Euler--Lagrange equations transform under
$\mathbf{q}(t)=\mathbf{r}(t)+\mathbf{s}(t)$?
\end{exercise}

\begin{exercise}[Other example Lagrangians]
Write the Euler--Lagrange equations, then apply the Legendre
transformation to determine the Hamiltonian and Hamilton's canonical
equations for the following Lagrangians. Determine which of them are
hyperregular.
\begin{itemize}
\item
$L(q, \dot{q})= \Big( g_{ij}(q) \dot q^i \dot q^j\Big)^{1/2}$ (Is it
possible to assume that $L(q,\dot{q})=1$? Why?)
\item
$L(\mathbf{q}, \mathbf{\dot{q}})
= -\big(1-\mathbf{\dot{q}}\cdot\mathbf{\dot{q}}\big)^{1/2}$
\item
$L(\mathbf{q}, \mathbf{\dot{q}})
= \frac{m}{2}\mathbf{\dot{q}}\cdot\mathbf{\dot{q}}
+
\frac{e}{c} \mathbf{\dot{q}}\cdot\mathbf{A}(\mathbf{q})$, for constants
$m$, $c$ and prescribed function $\mathbf{A}(\mathbf{q})$. How do the
Euler--Lagrange equations for this Lagrangian differ from free motion in a
moving frame with velocity $\frac{e}{mc}\mathbf{A}(\mathbf{q})$?
\item
Calculate the action and the energy for each of these Lagrangians..
\end{itemize}
\end{exercise}

\subsubsection{Example: Free special relativistic particle motion}

To illustrate the approach consider the Lagrangian $L(\mathbf{q}, \mathbf{\dot{q}})
= -\big(1-\mathbf{\dot{q}}\cdot\mathbf{\dot{q}}\big)^{1/2}$. 

\begin{enumerate} 
\item {\bf Action}
\[
S = \int_0^T L(q, \dot{q}) \,dt = -  \int_0^T \big(1-\mathbf{\dot{q}}\cdot\mathbf{\dot{q}}\big)^{1/2}\,dt
\]

\item {\bf Fibre derivative}
\[
\mathbf{p}
=
\frac{\partial L}{\partial \mathbf{\dot{q}}}
=
\frac{\mathbf{\dot{q}} }{ \sqrt{1-\mathbf{\dot{q}}\cdot\mathbf{\dot{q}} } } =: \gamma\, \mathbf{\dot{q}}
\quad\Longrightarrow \quad
\mathbf{\dot{q}} = \pm\,\frac{\mathbf{p} }{ \sqrt{1+\mathbf{p}\cdot\mathbf{p} }}
\]
so this Lagrangian is hyperregular, after making a choice of sign convention, that $\mathbf{p}\cdot \mathbf{\dot{q}}>0$, for example; so that $\gamma=\sqrt{1+\mathbf{p}\cdot\mathbf{p} }= 1/\sqrt{1-\mathbf{\dot{q}}\cdot\mathbf{\dot{q}}}$.
\item  {\bf Euler-Lagrange equations}\\
\[
\frac{d (\gamma\, \mathbf{\dot{q}}) }{dt} = 0
\]
\item  {\bf Hamiltonian and canonical equations}\\
The Hamiltonian for this system is
\[
H = \mathbf{p}\cdot \mathbf{\dot{q}} - L = \sqrt{1+|\mathbf{p}|^2 } = \gamma
\]
and its canonical equations are
\[
\frac{d\mathbf{q}}{dt} =  \frac{\partial H}{\partial \mathbf{p}}
= \frac{\mathbf{p} }{ \sqrt{1+|\mathbf{p}|^2 }}
\,,\qquad
\frac{d\mathbf{p}}{dt} = -\, \frac{\partial H}{\partial \mathbf{q}}= 0
\]

These equations represent uniform (force-free) motion in $\mathbb{R}^3$ of a relativistic particle with rest mass $m_0=1$, in units with $c=1$. 

In these units, the relation $H=\gamma$ is written as 
\[
H=\gamma\, m_0c^2=mc^2=E
\,.
\]
\end{enumerate}

\subsubsection{Example: charged particle in a magnetic field}
Consider a particle of charge $e$ and mass
$m$ moving in a magnetic field $\mathbf{B}$, where $\mathbf{B} =
\nabla \times\mathbf{A}$ is a given magnetic field on $\mathbb{R}^3$.
The Lagrangian for the motion is given by the ``minimal coupling''
prescription (jay-dot-ay)
\[
L(\mathbf{q}, \dot{\mathbf{q}})
=
\frac{m}{2}\|\dot{\mathbf{q}}\|^2 
+ \frac{e}{c} \mathbf{A}(\mathbf{q})\cdot\mathbf{\dot{q}}
,
\]
in which the constant $c$ is the speed of light. The derivatives of this
Lagrangian are
\[
\frac{\partial L}{\partial \mathbf{\dot{q}}}
=
m\mathbf{\dot{q}}+\frac{e}{c}\mathbf{A}
=:
\mathbf{p}
\quad\hbox{and}\quad
\frac{\partial L}{\partial \mathbf{q}}
=
\frac{e}{c}\nabla\mathbf{A}^T\cdot\mathbf{\dot{q}}
\]
Hence, the Euler--Lagrange equations
for this system are 
\begin{eqnarray*}
m\,\mathbf{\ddot{q}} =
 \frac{e}{c}
(\nabla\mathbf{A}^T\cdot\mathbf{\dot{q}}
-\nabla\mathbf{A}\cdot\mathbf{\dot{q}})
=
\frac{e}{c}\,\mathbf{\dot{q}}\times\mathbf{B}
\end{eqnarray*}
(Newton's equations for the Lorentz force).
The Lagrangian $L$ is hyperregular, because
\[ \mathbf{p}= \mathbb{F}L (\mathbf{q}, \mathbf{\dot{q}}) =
m\mathbf{\dot{q}}+\frac{e}{c}\mathbf{A}(\mathbf{q}) \]
has the  inverse
\[ \mathbf{\dot{q}} = \mathbb{F}H (\mathbf{q}, \mathbf{p}) =
\frac{1}{m}\left(\mathbf{p}- \frac{e}{c} \mathbf{A}(\mathbf{q})\right).  \]
The corresponding Hamiltonian is given by the invertible change of
variables,
\begin{equation}\label{Mag-Ham}
H(\mathbf{q},\mathbf{p}) 
= \mathbf{p} \cdot \dot{\mathbf{q}} - L(\mathbf{q}, \dot{\mathbf{q}})
= \frac{1}{2m}\left\| \mathbf{p} -
\frac{e}{c} \mathbf{A} \right\|^2 
.
\end{equation}
The Hamiltonian $H$ is hyperregular since
\[
\dot{\mathbf{q}} = \mathbb{F}H (\mathbf{q}, \mathbf{p}) =
\frac{1}{m}\left(\mathbf{p}- \frac{e}{c} \mathbf{A}\right)
\quad\hbox{has the inverse}\quad
\mathbf{p}= \mathbb{F}L (\mathbf{q}, \mathbf{\dot{q}}) = m
\dot{\mathbf{q}} + \frac{e}{c}\mathbf{A}.
\]
The canonical equations for this Hamiltonian recover Newton's equations
for the Lorentz force law. 

\newpage

\begin{shaded}
\subsubsection{Kaluza--Klein construction for charged particle in magnetic field}
Although the minimal-coupling Lagrangian is
not expressed as the kinetic energy of a metric, Newton's equations for
the Lorentz force law may still be obtained as geodesic equations. This is
accomplished by suspending them in a higher dimensional space via the
\emph{Kaluza--Klein construction\/}, which proceeds as follows.
\end{shaded}

Let $Q_{KK}$ be the manifold $\mathbb{R}^3\times S ^1$ with variables
$(\mathbf{q},\theta)$. On $Q_{KK} $ introduce the one-form $A +
\mathbf{d}\theta$ (which defines a connection one-form on the
trivial circle bundle $\mathbb{R}^3\times S^1 \rightarrow \mathbb{R}^3$)
and introduce the \emph{Kaluza--Klein Lagrangian} \index{Kaluza--Klein Lagrangian}
$L_{KK}{:\ } TQ_{KK}\simeq T\mathbb{R}^3\times TS ^1\mapsto \mathbb{R}$ as
\begin{equation}
\begin{aligned}
L_{KK}(\mathbf{q}, \theta, \dot{\mathbf{q}},  \dot \theta) &=
\tfrac{1}{2} m \|\dot{\mathbf{q}}\|^2 + \tfrac{1}{2}\big\|\big\langle A
+ \mathbf{d}\theta, (\mathbf{q}, \dot{\mathbf{q}}, \theta, \dot \theta)
\big\rangle \big\|^2 \nonumber \\
&= \tfrac{1}{2} m \|\dot{\mathbf{q}}\|^2 + \tfrac{1}{2}\big(\mathbf{A}
\cdot \dot{\mathbf{q}} + \dot \theta
\big)^2.
\end{aligned}
\end{equation}
The Lagrangian $L_{KK} $ is positive definite in
$(\dot{\mathbf{q}},\dot \theta) $; so it may be regarded as the kinetic
energy of a metric, the \emph{Kaluza--Klein metric\/} on $TQ_{KK} $.  
(This construction fits the idea of $U(1)$ gauge symmetry for
electromagnetic fields in $\mathbb{R}^3$.  It can be generalized to a
principal bundle with compact structure group endowed with a connection.
The Kaluza--Klein Lagrangian in this generalization leads to Wong's
equations for a color-charged particle moving in a classical Yang--Mills
field.) The Legendre transformation for
$L_{KK}$ gives the momenta
\begin{equation}
\label{KK Legendre1}
\mathbf{p} = m \dot{\mathbf{q}} + (\mathbf{A}\cdot \dot{\mathbf{q}} +
\dot \theta) \mathbf{A} \qquad \text{and} \qquad \pi =
\mathbf{A}\cdot \dot{\mathbf{q}} + \dot \theta.
\end{equation}
Since $L_{KK}$ does not depend on $\theta$, the Euler--Lagrange equation
\[
\frac{d}{dt}\frac{\partial L_{KK}}{\partial \dot\theta}  =
\frac{\partial L_{KK}}{\partial \theta} = 0
,
\]
shows that $\pi = \partial L_{KK}/\partial\dot \theta $ is conserved.
The \emph{charge\/} is now defined by $e: = c\pi$. The Hamiltonian
$H_{KK}$ associated to $L_{KK}$ by the Legendre transformation
\eqref{KK Legendre1} is
\begin{equation}
\begin{aligned}
H_{KK}(\mathbf{q}, \theta, \mathbf{p}, \pi) &= \mathbf{p}\cdot
\dot{\mathbf{q}} + \pi \dot \theta - L_{KK}(\mathbf{q}, \dot
{\mathbf{q}}, \theta, \dot \theta)  \\
&\hspace{-30mm} = \mathbf{p}\cdot \tfrac{1}{m}\left(\mathbf{p} - \pi\mathbf{A} \right) +
\pi(\pi- \mathbf{A}\cdot \dot{\mathbf{q}})
- \tfrac{1}{2} m \|\dot{\mathbf{q}}\|^2 - \tfrac{1}{2}\pi^2
\\
&\hspace{-30mm} =  \mathbf{p}\cdot \tfrac{1}{m}\left(\mathbf{p} - \pi\mathbf{A} \right)
+ \tfrac{1}{2}\pi^2
- \pi \mathbf{A} \cdot  \tfrac{1}{m}\left(\mathbf{p} -
\pi\mathbf{A} \right) - \tfrac{1}{2m}
\| \mathbf{p} - \pi \mathbf{A}\|^2 \\
&\hspace{-30mm} =  \tfrac{1}{2m} \| \mathbf{p} - \pi \mathbf{A}\|^2 + \tfrac{1}{2}\pi^2.
\end{aligned}
\end{equation}
On the constant level set $\pi = e/c$, the Kaluza--Klein Hamiltonian
$H_{KK}$ is a function of only the variables $(\mathbf{q}, \mathbf{p})$
and is equal to the Hamiltonian \eqref{Mag-Ham} for charged particle
motion under the Lorentz force up to an additive constant. This
example provides an easy but fundamental illustration of the geometry of
(Lagrangian) reduction by symmetry. The canonical equations for the
Kaluza--Klein Hamiltonian $H_{KK}$ now reproduce Newton's equations for the
Lorentz force law.

\begin{remark}\rm [Spherical pendulum]
Spherical pendulum dynamics is equivalent to a particle rolling on the interior 
of a spherical surface under gravity. Write down the Lagrangian and the equations
of motion for a spherical pendulum with $S^2$ as its configuration space.
Show explicitly that the Lagrangian is hyperregular. Use the Legendre
transformation to convert the equations to Hamiltonian form. Find the
conservation law corresponding to angular momentum about the axis of
gravity by ``bare hands'' methods.
\end{remark}

\begin{exercise}(Euler--Lagrange equations for differentially rotating frames.)\label{rot-ex}\\
The Lagrangian for a free particle of unit mass relative to a moving frame
is obtained by setting 
\[
L(\mathbf{\dot{q}},\mathbf{q},t)
=
\tfrac12\|\mathbf{\dot{q}}\|^2 + \mathbf{\dot{q}}\cdot\mathbf{R}(\mathbf{q},t)
\]
for a function $\mathbf{R}(\mathbf{q},t)$ which prescribes the space and
time dependence of the moving frame velocity. For example, a frame
rotating with time-dependent frequency $\Omega(t)$ about the vertical axis
$\mathbf{\hat{z}}$ is obtained by choosing
$\mathbf{R}(\mathbf{q},t)=\mathbf{q}\times\Omega(t)\mathbf{\hat{z}}$.
Calculate $\Theta_L\left(q,\dot{q}\right)$,
the Euler--Lagrange operator ${\mathcal EL}(L)\left(q,\dot{q},
\ddot{q}\right)$, the Hamiltonian and its corresponding canonical
equations.
\end{exercise}

\subsubsection{The free particle in $\mathbb{H}^2$: Part \#1}

\begin{exercise-white} 

The free particle in $\mathbb{H}^2$: Part \#1
\label{H2-halfplane}$\,$

%

In Appendix I of VI Arnold's book \cite{arnol2013mathematical}, we read.

\begin{quote}
EXAMPLE. We consider the upper half-plane $y>0$ of the plane of complex numbers $z=x+iy$ with the metric
\[
ds^2=\frac{dx^2+dy^2}{y^2}
\,.
\]
It is easy to compute that the geodesics of this two-dimensional Riemannian manifold are circles 
and straight lines perpendicular to the $x$-axis. Linear fractional transformations with real coefficients
\begin{equation}
z \to \frac{az+b}{cz+d}
\label{Mobius-xform}
\end{equation}
are isometric transformations of our manifold ($\mathbb{H}^2$), which is called the \emph{Lobachevsky plane}.
\index{Lobachevsky plane}
These isometric transformations of $\mathbb{H}^2$ have deep significance in physics. They correspond to the most general Lorentz transformation of space-time.
\end{quote}

{\begin{shaded}

Consider a free particle of mass $m$ moving on the Lobachevsky half-plane $\mathbb{H}^2$. Its Lagrangian is the kinetic energy corresponding to the Lobachevsky metric
Namely, 
\begin{equation}
L=\frac{m}{2}\left (\frac{\dot x^2 +\dot y^2}{y^2}\right ).
\label{Lag-H2}
\end{equation}

 \end{shaded} }

\begin{enumerate} [(A)]
\begin{shaded}
\item 

(1) Write the fibre derivatives of the Lagrangian (\ref{Lag-H2}) and 

(2) compute its Euler-Lagrange equations. 

These equations represent geodesic motion on $\mathbb{H}^2$. 

(3) Evaluate the Christoffel symbols. 
 \end{shaded} 
 
\begin{answer}$\,$\smallskip

Fibre derivatives:
\[
\frac{\partial L}{\partial \dot x} = \frac{m \dot x}{y^2} =: p_x
\quad\hbox{and}\quad
\frac{\partial L}{\partial \dot y} = \frac{m \dot y}{y^2} =: p_y
\]
Euler-Lagrange equations $\frac{d}{dt}\frac{\partial L}{\partial \dot x}=\frac{\partial L}{\partial x}$ and $\frac{d}{dt}\frac{\partial L}{\partial \dot y}=\frac{\partial L}{\partial y}$ yield, respectively:
\begin{equation}
\frac{d}{dt}\left(\frac{ \dot x}{y^2}\right) = 0
\quad\hbox{and}\quad
\frac{d}{dt}\left(\frac{ \dot y}{y^2}\right) = 
-\,\frac{\dot x^2 +\dot y^2}{y^3}
\label{Lag-H2-eqns}
\end{equation}
Expanding these equations yield the Christoffel symbols for the geodesic motion,
\[
\ddot x - \frac{2}{y} \dot x \dot y =0
\,,\quad
\ddot y + \frac{1}{y} \dot x^2 -\,\frac{1}{y} \dot y^2 =0
\,.
\]
Hence,
\[
\Gamma^1_{12} = -\,\frac{2}{y} = \Gamma^1_{21}
\,,\quad 
\Gamma^2_{11} = \frac{1}{y}
\,,\quad 
\Gamma^2_{22} = -\,\frac{1}{y}
\,.
\]
\end{answer}

\item
{\begin{shaded}

\fbox{Hint:} 
The Lagrangian in (\ref{Lag-H2}) is invariant under the group of linear fractional transformations with real coefficients.
These have an $SL(2,\mathbb{R})$ matrix representation
\begin{equation}
\begin{bmatrix}
a & b \\
c & d
\end{bmatrix}
\begin{bmatrix}
z \\ 1 
\end{bmatrix}
= \frac{az+b}{cz+d}
\label{SL2R-rep}
\end{equation}

(1) Show that the quantities
\begin{equation}
u = \frac{\dot x}{y}
\quad\hbox{and}\quad
v = \frac{\dot y}{y}
\label{invar-var}
\end{equation}
are invariant under a subgroup of these symmetry transformations. 

(2) Specify this subgroup in terms of the representation (\ref{SL2R-rep}). 

\end{shaded}
}

\begin{answer}$\,$\smallskip
 
The quantities (\ref{invar-var}) are invariant under a subgroup 
of translations and scalings.  
\begin{align*}
\begin{split}
T_\tau: (x,y) \mapsto (x+\tau, y) \qquad 
&\mbox{Flow of $X_T=\partial_x$,}
\\ &\mbox{$(\delta x,\delta y)=(1,0), \quad [X_T,X_S]=X_T$.} 
\\
S_\sigma: (x,y) \mapsto (e^\sigma x, e^\sigma y) \qquad 
&\mbox{Flow of $X_S =x\partial_x+ y\partial_y$}, 
\\ &\mbox{$(\delta x,\delta y)=(x,y)$.}
\end{split}
\end{align*}
These transformations are translations $T$ along the $x$ axis and scalings $S$ centered at $(x,y)=(0,0)$. They are represented by elements of (\ref{SL2R-rep}) as
\[
T = 
\begin{bmatrix}
1  & b \\
0 & 1
\end{bmatrix}
\quad\hbox{and}\quad
S = 
\begin{bmatrix}
a  & 0 \\
0 & 1
\end{bmatrix}
\]
That is, the transformations $T$ and $S$ are isometries of the metric $ds^2=(dx^2+dy^2)/y^2$ on $\mathbb{H}^2$ with $T: \, a=1=d,c=0, b\ne0$ and $S: a\ne0, b=0=c,\, d=1$.

\end{answer}$\,$

{\begin{shaded}
\item
(1) Use the invariant quantities $(u,v)$ in (\ref{invar-var}) as new variables in Hamilton's principle. 

\fbox{Hint:} the transformed Lagrangian is 
\[
\ell(u,v) = \frac{m}{2} (u^2+v^2)
\,.
\]

(2) Find the corresponding conserved Noether quantities. 
\end{shaded} }

\begin{answer}$\,$\smallskip

(1) The translations $T$ along the $x$ axis and scalings $S$ centered at $(x,y)=(0,0)$ leave invariant the quantities
\[
u = \frac{\dot x}{y}
\quad\hbox{and}\quad
v = \frac{\dot y}{y}
\,,\]
in terms of which the Lagrangian $L$ in (\ref{Lag-H2}) reduces to 
\[
\ell(u,v) = \frac{m}{2} (u^2+v^2)
\,.
\]
The reduced Hamilton's principle in the variables $u$ and $v$ yields,
\begin{align*}
0 = \delta S &= \delta \int^b_a \ell(u,v) \,dt
=  \int^b_a m(u\delta u +v\delta v) \,dt
\\&=
m  \int^b_a \frac{u}{y} (\delta\dot x - u\delta y)
+ \frac{v}{y} (\delta\dot y - v\delta y)\,dt
\\&=
-\,m  \int^b_a \left(\frac{d}{dt}\frac{u}{y}\right) \delta x  
+ \left(\frac{d}{dt}\frac{v}{y} + \frac{u^2+v^2}{y} \right) \delta y\,dt
+m \left[ \frac{u}{y} \,\delta x + \frac{v}{y} \,\delta y  \right]^b_a
\end{align*}
Thus, Hamilton's principle recovers equations (\ref{Lag-H2-eqns}) in the variables $u$ and $v$.\\

(2) Applying Noether's theorem to the endpoint term in these variables yields conservation of 
\[
C_T=\frac{u}{y}, \hbox{ for } (\delta x,\delta y)=(1,0)
 \hbox{ translations,} 
 \]
 and
 \[
C_S=\frac{ux + vy}{y} \hbox{ for } (\delta x,\delta y)=(x,y)
 \hbox{ scaling.}
 \]


\end{answer}

{\begin{shaded}
\item

Transform the Euler-Lagrange equations from $x$ and $y$ to the variables $u$ and $v$ that are invariant under the symmetries of the Lagrangian. \smallskip

Then:

(1) Show that the resulting system conserves the kinetic energy expressed in these variables. 

(2) Discuss its integral curves and critical points in the $uv$ plane. 

(3) Show that the $u$ and $v$ equations can be integrated explicitly in terms of $\mbox{sech}$ and $\mbox{tanh}$.

 \fbox{Hint:} In the $uv$ variables,  the Euler-Lagrange equations for the Lagrangian (\ref{Lag-H2}) are expressed as
\[
\frac{d}{dt}\frac{u}{y} = 0
\quad\hbox{and}\quad
\frac{d}{dt}\frac{v}{y} + \frac{u^2+v^2}{y} = 0
\,.
\]
Expanding these equations using $u = \dot x/y$ and $v = \dot y/y$ yields
\begin{equation}
\label{uv-eqns}
\dot u = uv\,, \qquad \dot v =-\,u^2
\end{equation}

\end{shaded} }

\begin{answer}$\,$\smallskip

(1) Equations (\ref{uv-eqns}) imply conservation of the kinetic energy
\[
\ell(u,v) = \frac{m}{2} (u^2+v^2) = E
\]
(2) The integral curves of the system of equations (\ref{uv-eqns}) in the $uv$ plane are either critical points along the axis $u=0$, or they are heteroclinic connections between these points that are semi-circles around the origin on level sets of the energy $E$. 
\smallskip

The critical points at $u=\dot x/y=0$ are relative equilibria of the system corresponding to vertical motion on the $xy$ plane. Those corresponding to ``upward motion" ($\dot y>0$) are unstable and the ones corresponding to ``downward motion"
($\dot y<0$) are stable.\smallskip

(3) The trial solutions $u=\mbox{tanh}$ and $v=\mbox{sech}$ quickly converge to the exact solutions of the $uv$ system.

\end{answer}

{\begin{shaded}
\item
(1) Legendre transform the Lagrangian (\ref{Lag-H2}) to the Hamiltonian side, obtain the canonical equations and 

(2) derive Poisson brackets for the variables $\{u,v\}$.  
\fbox{Hint:} $\{yp_x,yp_y\}=yp_x$.
\end{shaded} }

\begin{answer}$\,$\smallskip

The equations of motion on the Hamiltonian formulation are defined by introducing the momenta:
\begin{equation*}
p_x=\frac{\partial L}{\partial \dot x} =\frac{m\dot x}{y^2},
 \qquad p_y=\frac{\partial L}{\partial \dot y} =\frac{m\dot y}{y^2}, 
\end{equation*}
and the Hamiltonian
\begin{equation*}
H=\frac{y^2}{2m}  \left ({p_x^2 +p_y^2} \right ).
\end{equation*}
One gets
\begin{equation}
\label{E:Ham-Eqns}
\begin{split}
&\dot x =\frac{y^2p_x}{m} \qquad \dot p_x=0, \\
&\dot y =\frac{y^2p_y}{m}  \qquad \dot p_y =\frac{-y}{m}\left (p_x^2 +p_y^2 \right ).
\end{split}
\end{equation}
By defining
\begin{equation*}
u=yp_x/m, \qquad v=yp_y/m,
\end{equation*}
the Hamiltonian can be written as
\begin{equation*}
H=h(u,v)=\frac{1}{2}\left ( u^2 +v^2 \right ),
\end{equation*}
and the equations of motion  \eqref{E:Ham-Eqns} become, using $\{yp_x,yp_y\}=yp_x$, 
\begin{equation}
\label{E:Lie-Poisson}
\dot u = uv \qquad \dot v =-\,u^2.
\end{equation}

These equations are Hamiltonian with respect to the Lie-Poisson bracket
\begin{equation*}
\{ u , v\} = u,
\end{equation*}
and the reduced Hamiltonian $h(u,v)$ in terms of the invariant variables. Namely,
\[
\dot u = \{ u , h\} = uv
\quad
\dot v = \{ v , h\} = -\,u^2
\,.
\]

\end{answer}

\end{enumerate}

\end{exercise-white}


\vspace{1cm}

\subsubsection{The free particle in $\mathbb{H}^2$: Part \#2}

\begin{exercise-white}[The free particle in $\mathbb{H}^2$: Part \#2]$\,$

%

{\begin{shaded}

Consider the following pair of differential equations for $(u,v)\in \mathbb{R}^2$, 
\begin{align}
\dot{u} &=  uv
\,,\qquad
\dot{v} =   -\,  u^2
\,.
\label{2D-Toda}
\end{align}
These equations have discrete symmetries under combined reflection and time reversal, $(u,t)\to (-u,-t)$ and $(v,t)\to (-v,-t)$. (This is called $PT$ symmetry in the $(u,v)$ plane.)

\end{shaded} }

\begin{enumerate}[(A)]

{\begin{shaded}
\item
Find  $2\times2$ real matrices $L$ and $B$ for which the system (\ref{2D-Toda}) may be written as a commutator, namely, as
\[
\frac{dL}{dt}=[L,B]
\,.
\]
\fbox{Hint:} a basis for $2\times2$ real matrices is given by
\[
\sigma_1 = 
\begin{bmatrix}
0 & 1 \\
1 & 0 
\end{bmatrix}
\,,\quad
\sigma_2 =
\begin{bmatrix}
0 & 1 \\
-1 & 0 
\end{bmatrix}
\,,\quad
\sigma_3 = 
\begin{bmatrix}
1 & 0 \\
0 & -1 
\end{bmatrix}
\,.
\]

Explain what the commutator relation means and determine a constant of the motion from it.

\end{shaded}
}

\begin{answer}$\,$\medskip

We introduce two linear $2\times2$ matrices, one symmetric $(L^T=L)$ and the other skew-symmetric $(B^T=-B)$, as required for the commutator $[L,B]$ to be symmetric:
\begin{align*}
L
&=
\begin{bmatrix}
-v & u \\
u & v
\end{bmatrix}
=
u
\begin{bmatrix}
0 & 1 \\
1 & 0 
\end{bmatrix}
-
v
\begin{bmatrix}
1 & 0 \\
0 & -1 
\end{bmatrix}
=
u\sigma_1 -v\sigma_3
,
\\
B
&= \frac12
\begin{bmatrix}
0 & u \\
-u & 0 
\end{bmatrix}
= 
\frac{u}{2}
\begin{bmatrix}
0 & 1 \\
-1 & 0 
\end{bmatrix}
=
\frac{u}{2}\,\sigma_2
.
\end{align*}
Both matrices must be linear homogeneous, so that the commutator 
$[L,B]$ and time derivative $\frac{dL}{dt}$ can match powers using (\ref{2D-Toda}). 
The $\mathfrak{sl}(2,\mathbb{R})$ $\sigma$-matrices satisfy  
\[
[\sigma_1,\,\sigma_2] = 2\sigma_3
,\quad
[\sigma_2,\,\sigma_3] = 2\sigma_1
,\quad\hbox{and}\quad
[\sigma_3,\,\sigma_1] = -\,2\sigma_2
.\]
Thus, we find the commutator relation,
\begin{align*}
\frac{dL}{dt}
=
u^2 \sigma_3  +uv\,\sigma_1
=[L,\,B]
= 
\left[u\sigma_1 - v\sigma_3
,\,\frac{u}{2}\sigma_2\right]
.
\end{align*}

{\bf What the commutator relation means: isospectrality.} The commutator relation implies that 
the flow generates a similarity transformation of the $2\times2$ symmetric matrix $L(0)$. 
The traceless matrix $L(t)$ has one independent eigenvalue and the system (\ref{2D-Toda}) has only one conserved quantity. The conserved quantity is the determinant, $\det L(t)=\det L(0)$. However, this conservation law introduces no constraint on $u$.

\end{answer} 

{ \begin{shaded} 
\item
Write the system (\ref{2D-Toda}) as a double matrix commutator, $\frac{dL}{dt}=[L,[L,N]\,]$. In particular, find $N$ explicitly and explains what this means for the solutions.
\fbox{Hint:} compute $\frac{d}{dt}\,{\rm tr}\,(LN)$.

\end{shaded} }

\begin{answer} 
Substituting 
$
N := 
\begin{bmatrix}
a  & 0 \\
0  & b
\end{bmatrix}
$
into
$
\frac{dL}{dt}=[L,[L,N]\,]
$
yields $b-a=$, so for example we may set
\begin{align*}
N := 
\begin{bmatrix}
1  & 0 \\
0  & 2
\end{bmatrix}
.
\end{align*}
{\bf What this means for the solutions: Dissipative flow.}
The evolution by the double bracket relation $\frac{dL}{dt}=[L,[L,N]\,]$ is a \emph{dissipative flow} that decreases the quantity ${\rm tr}\,(LN)$ according to 
\[
\frac{d}{dt}\,{\rm tr}\,(LN)
= - \, {\rm tr}\,([L,\,N]^T[L,\,N])
\,,\]
until $L$ becomes diagonal and hence $[L,\,N]\to0$, because $N$ is diagonal.  Thus, the dynamics (\ref{2D-Toda}) becomes asymptotically steady as $L$  tends to a diagonal matrix. This means  the system (\ref{2D-Toda}) must asymptotically approach a stable equilibria that is consistent with its initial conditions and conservation laws.
For the present case, substituting the explicit forms of $L$ and $N$ yields 
\[
\frac{d}{dt}\,{\rm tr}\,(LN) = \frac12\,\dot{v}
= -\,\frac12\,u^2 = - \, {\rm tr}\,([L,\,N]^T[L,\,N])
= - \, \|[L,\,N]\|^2
\,,
\]
which holds by (\ref{2D-Toda}) and thus checks the previous calculation. In the present case, it will turn out that $\lim_{t\to\infty}u(t)=0$, which will verify $[L,\,N]\to0$, as the off-diagonal parts of $L$ will vanish asymptotically. 
\end{answer}

{ \begin{shaded} 
\item
Find explicit solutions and discuss their motion and asymptotic behaviour: 

(1) in time; and 

(2) in the $(u,v)$ phase plane. \textit{Hint:} keep the \textit{tanh} function in mind. 

\end{shaded} }

\begin{answer} $\,$

Keeping the $\tanh$ function in mind and recalling that
\[
\frac{d\tanh(ct)}{dt}=c\,{\rm sech}^2(ct)
\qquad
\frac{d\,{\rm sech}(ct)}{dt}=-\,c\,{\rm sech}(ct)\tanh(ct)
,\]
we find, for $u(0)=c$ and $v(0)=0$, 
\[
v(t) = - c\tanh(ct)
\quad\hbox{and}\quad
u(t) = c\,{\rm sech}(ct)
,\]
and of course we check, $2h=u^2+v^2=c^2(\tanh^2+\,{\rm sech}^2)=c^2$.

{\bf Motion and asymptotic behaviour.}
\begin{enumerate}
\item
{\bf In time:} We have $\lim_{t\to\infty}(u(t),v(t))=(0,-c)$. Consequently, the quantity $u(t)$ falls exponentially with time, from $u(0)$ toward the line of fixed points at $u=0$, while $u(t)$ goes to a constant equal to $-u(0)$.

\item
{\bf In the $(u,v)$ phase plane:} 
Since $h$ is conserved, the motion is along a family of semi-circles, each parameterised by its radius $c=\sqrt{2h}$, as
\[
u^2+v^2=c^2
\quad\hbox{for}\quad
u>0 
\quad\hbox{and}\quad
u<0 \,,
\]
lying in the upper and lower $(u,v)$ half planes. These semi-circular motions are mirror images, reflected across the vertical line of fixed points at $u=0$ in the $(u,v)$ plane. The equations of motion are $PT$-symmetric, so the fixed points along $u=0$ in the $(u,v)$ plane are stable for  $v<0$, and unstable for $v>0$. 

Thus, the two families of semi-circular motion both connect the line of fixed points at $u=0$ to itself. One family of semi-circles lies in the upper half $(u,v)$ plane, and the other lies symmetrically placed to complete the circles in the lower half $(u,v)$ plane. The flows along each reflection-symmetric pair of semi-circles pass in the same (negative) $v$ direction, from $v=c$ to $v=-c$.
\end{enumerate}
\end{answer} 

{ \begin{shaded} 
\item
Explain why the solution behaviour found in the previous part  is consistent with the behaviour predicted by the double bracket relation.
 \end{shaded} }
 
\begin{answer} $\,$

This analysis is consistent with the conclusion from the double-bracket relation $\frac{dL}{dt}=[L,[L,N]\,]$ that the dynamics of $L$-matrix 
\begin{align*}
L
=
\begin{bmatrix}
-v & u \\
u & v
\end{bmatrix}
\end{align*}
asymptotically becomes steady. In fact, since $\lim_{t\to\infty}u=0$ and $\lim_{t\to\infty}v(t)=-c$, the $L$-matrix asymptotically diagonalises and hence $[L,\,N]\to0$, because $N$ is diagonal.
\end{answer} 

\end{enumerate}
\end{exercise-white}

The next exercise illustrates the differences between Hamiltonian and dissipative evolution, written in matrix commutator form. 

{ \begin{shaded} 
\begin{exercise} {Commutator form of the 3D Volterra system.}$\,$\rm

Consider the dynamical system in $(x_1,x_2,x_3)\in\mathbb{R}^3$,
\begin{align}
\begin{bmatrix}
\dot{x}_1 \\ \dot{x}_2 \\ \dot{x}_3
\end{bmatrix}
=
\begin{bmatrix}
x_1x_2  \\  x_2x_3 - x_1x_2 \\ - x_2x_3
\end{bmatrix}
=
x_2
\begin{bmatrix}
x_1  \\ x_3 - x_1 \\   - x_3
\end{bmatrix}
\label{LV3}
\end{align}
This is a 3D version of the Volterra [1931] model of competition among species, which for more species is given by
\[
\dot{x}_n = x_n(x_{n+1}-x_{n-1})
,\quad
n=1,2,\dots,N, 
\quad\hbox{with}\quad
x_0=0 = x_{N+1}
.\]
\end{exercise} 
\end{shaded} }

\color{black}

\begin{enumerate}

{ \begin{shaded} 
\item
Find two conservation laws for the system (\ref{LV3}). 

\end{shaded} }
 
\begin{answer}$\,$

The flow of the vector field $(\dot{x}_1,\dot{x}_2,\dot{x}_3)\in T\mathbb{R}^3$ preserves the sum $H =  x_1 + x_2 + x_3$ and the product $C=x_1x_3$.

\end{answer}

{ \begin{shaded} 
\item
Verify that this system may be written in commutator form as
\[
\frac{dL}{dt}=[L,B]
\]
for the $4\times4$ matrices 
\begin{align*}
L &:= 
\begin{bmatrix}
x_1 & 0 & \sqrt{x_1x_2} & 0\\
0   & x_1+x_2 & 0 & \sqrt{x_2x_3} \\
\sqrt{x_1x_2} & 0 & x_2+x_3 & 0 \\
0 & \sqrt{x_2x_3} & 0 & x_3
\end{bmatrix}
\\ \\
B &:= \frac12
\begin{bmatrix}
0 & 0 & -\sqrt{x_1x_2} & 0\\
0  & 0 & 0 & -\sqrt{x_2x_3} \\
\sqrt{x_1x_2} & 0 & 0 & 0 \\
0 & \sqrt{x_2x_3} & 0 & 0
\end{bmatrix}
\end{align*}

\item
Explain how the two conservation laws for the sum $H =  x_1 + x_2 + x_3$ and the product $C=x_1x_3$ found earlier are related to the matrices $L$ and $B$. Explain what this means for the system in \eqref{LV3}.
\end{shaded} }

\begin{answer} $\,$

\qquad${\rm tr} L=2H=2(x_1+x_2+x_3)$ and $\det L = C^2=(x_1x_3)^2$. 

Preservation of the trace and determinant of the $4\times 4$ matrix $L$ means that two of its four eigenvalues are preserved. 

\end{answer}

{\begin{shaded} 
\item
Give the geometrical interpretation of the formula $\frac{dL}{dt}=[L,B]$ with $4\times4$ matrices $L$ and $B$.
\end{shaded} }

\begin{answer} 
The commutator form  has little effect on the spectrum of the $4\times 4$ symmetric matrix $L(t)$, because the 
system has only 2 conserved quantities. 
\end{answer}

{\begin{shaded} 
\item
Write the system (\ref{LV3}) as a double matrix commutator.
In particular, find the diagonal matrix $N$ for which $B=[L,N]$.
\end{shaded} }

\begin{answer} 
\[
\frac{dL}{dt} = [L,B]  =[L,[L,N]\,]\,.
\] 
\begin{align*}
N := 
\begin{bmatrix}
1  & 0 & 0 & 0\\
0  & 1 & 0 & 0 \\
0  & 0 & \frac12& 0 \\
0 & 0 & 0 & \frac12
\end{bmatrix}
\end{align*}
\end{answer}

{\begin{shaded} 
\item
Give the geometrical interpretation of the formula $\frac{dL}{dt}=[L,[L,N]\,]$.
\end{shaded} }

\end{enumerate}

\begin{answer} 
The evolution of the matrix $L(t)$ preserves its trace and determinant as it tends toward an equilibrium $x_2\to 0$ and matrix $L$ diagonalises to commute with diagonal $N$. The eigenvalues in the equilibrium diagonal form of $L$ are doubly degenerate. 
\end{answer}

\newpage
\vspace{4mm}\centerline{\textcolor{shadecolor}{\rule[0mm]{6.75in}{-2mm}}\vspace{-4mm}}
\section{Momentum maps}
\noindent
\label{sec:16}

\secttoc

\textbf{What is this lecture about?} This lecture describes the definitions and properties of 
cotangent-lift momentum maps and symplectic momentum maps. 

\subsection{The main idea}

Symmetries are often associated with conserved quantities.  For example, the flow
of any $SO(3)$--invariant Hamiltonian vector field on $T^*\mathbb{R}^3$
conserves angular momentum,
$\mathbf{q}\times \mathbf{p}$. More generally, given a Hamiltonian $H$ on
a phase space $P$, and a group action of $G$  on $P$ that conserves $H$,
there is often an associated ``momentum map'' $J{:\ } P\to \mathfrak{g}^*$ that
is conserved by the flow of the Hamiltonian vector field.


Note: all group actions in this section will be left actions until
otherwise specified.

Let $G$ be a Lie group, $\mathfrak{g}$ its Lie algebra, and
let $\mathfrak{g}^*$ be its dual. Suppose that $G$ acts symplectically on a
symplectic manifold $P$ with symplectic form denoted by $\Omega$. Denote
the infinitesimal generator associated with the Lie algebra element
$\xi$ by $\xi_P$ and let the Hamiltonian vector field associated
to a function $f {:\ }  P\to\mathbb{R}$ be denoted $X_f$, so that
$df=\iota_{X_f}\Omega$;  or, in equivalent expanded notation $df=X_f\contract\Omega$.%

\subsection{Definition, history and overview}
A \emph{momentum map}  $J {:\ }  P\to\mathfrak{g}^*$ is defined by the
condition relating the infinitesimal generator $\xi_P$ of a symmetry to
the vector field of its corresponding conservation law,
$\langle J,\xi \rangle$,
\[
\xi_P = X_{\langle J,\xi\,\rangle}
\]
for all $\xi\in \mathfrak{g}$. Here $\langle J,\xi\,\rangle {:\ } 
P\to\mathbb{R}$ is defined by the natural pointwise pairing.

A momentum map is said to be \emph{equivariant} 
\index{momentum map!equivariant}  when it is equivariant
with respect to the given action on
$P$ and the coadjoint action on $\mathfrak{g}^*$. That is,
\[
J(g \cdot p) = \Ad^*_{g^{-1}} J(p)
\]
for every $g\in G$, $p \in P$, where $g \cdot p$ denotes the action of $g$
on the point $p$ and where Ad denotes the adjoint action.

According to Weinstein~\cite{We1983a}, Lie~\cite{Lie1890} already knew that 
\index{momentum map!What Lie knew in 1890}
\begin{enumerate}
\item
An action of a Lie group $G$ with Lie algebra
$\mathfrak{g}$ on a symplectic manifold $P$ should be accompanied by such
an equivariant momentum map $J {:\ }  P\to\mathfrak{g}^*$ and 
\item
The orbits of this action are themselves symplectic manifolds. 
\end{enumerate}
The links with mechanics were developed in work by Lagrange, Poisson,
Jacobi and, later, by Noether. In particular, Noether showed that a momentum
map for the action of a group $G$ that is a symmetry of the Hamiltonian
for a given system is a \emph{conservation law} for that system.

In modern form, the momentum map and its equivariance were rediscovered
by Kostant~\cite{Ko1966} and Souriau~\cite{So1970} in the general
symplectic case, and by Smale~\cite{Sm1970a,Sm1970b} for the case of
the lifted action from a manifold $Q$ to its cotangent bundle $P =
T^*Q$. In this case, the equivariant momentum map is given explicitly by
\[
\langle J(\alpha_q),\xi\,\rangle  = \langle \alpha_q,\xi_Q(q)\rangle
,
\]
where $\alpha_q \in T^*Q$, $\xi\in\mathfrak{g}$, and where the angular
brackets denote the natural pairing on the appropriate spaces. See
Marsden and Ratiu~\cite{MaRa1994} and Ortega and Ratiu~\cite{OrRa2004} for additional history and
description of the momentum map and its properties. See the textbooks
\cite{holm2011geometricI,holm2011geometricII,HoScSt2009} and 
survey \cite{holm2011applications}  for further details.

\subsection{Hamiltonian systems on Poisson manifolds}

\begin{definition} A \emph{Poisson bracket} on a manifold $P$ is a 
skew-symmetric bilinear operation on \index{Poisson manifold}
$$\mathcal{F}(P) := \mathcal{C}^\infty\left(P,\R\right)$$
satisfying the Jacobi identity and the Leibniz identity,
\[
\{ FG,H\} = F\{G,H\} + \{F,H\} G
\]
The pair $\left(P,\{\cdot, \cdot \}\right)$ is called a \emph{Poisson manifold}.
\end{definition}

\begin{remark}\rm 
The Leibniz identity is sometimes not included in the definition. 
Note that bilinearity, skew-symmetry and the Jacobi identity are the axioms of a Lie algebra.
In what follows, a Poisson bracket is a binary operation that makes
$\mathcal{F}(P)$ into a Lie algebra and also satisfies the Leibniz
identity.
\end{remark}

\begin{exercise}
Show that the \emph{classical Poisson bracket}, defined in
cotangent-lifted coordinates 
\[\big(q^1,\dots,q^N,p_1,\dots,p_N\big)\]
on an $2N$--dimensional cotangent bundle $T^*Q$ by
\[
\left\{F,G\right\} = \sum_{i=1}^{N} \left( \frac{\partial F}{\partial q^i} \frac{\partial G}{\partial p_i} 
- \frac{\partial F}{\partial p_i} 
\frac{\partial G}{\partial q^i}  \right)
,
\]
satisfies the axioms of a Poisson bracket.
Show also that the definition of this bracket is independent of the choice of local coordinates 
$\left(q^1,\dots,q^N\right)$.
\end{exercise}

\begin{definition}
A \emph{Poisson map} between two Poisson manifolds is a map \index{Poisson manifold!Poisson map}
$$\varphi{:\ }  \big(P_1,\{\cdot, \cdot\}_1\big) \to \big(P_2,\{\cdot,
\cdot\}_2\big)$$
that preserves the brackets, meaning
\[
\{F\circ \varphi, G\circ \varphi\}_1 = \{ F,G\}_2\circ \varphi , \quad \quad \textrm{for all } 
F,G\in \mathcal{F}\left(P_2\right).
\]
\end{definition}

\begin{definition}
An action $\Phi$ of $G$ on a Poisson manifold \index{Poisson manifold!group action} $\left(  P,\left\{  ,\right\}  \right)
$ is \emph{canonical} if $\Phi_{g}$ is a Poisson map for every $g$, that is,
\[
\left\{  F\circ\Phi_{g},K\circ\Phi_{g}\right\}  =\left\{  F,K\right\}
\circ\Phi_{g}%
\]
for every $F,K\in \mathcal{F}(P)$.
\end{definition}

\begin{definition}
Let $\left(P,\{\cdot, \cdot\}\right)$ be a Poisson manifold, \index{Poisson manifold!Hamiltonian vector field} and let
$H{:\ } P\to \R$ be differentiable. 
The \emph{Hamiltonian vector field} for $H$ is the vector field $X_H$ defined by
\[
X_H(F) = \{F,H\}, \quad \quad \textrm{for any } F\in \mathcal{F}(P)
\]
\end{definition}

\begin{remark}\rm 
$X_H$ is well-defined because of the Leibniz identity and the correspondence between
vector fields and derivations (see Lee~\cite{Le2003}). 
\end{remark}

\begin{remark}\rm 
$X_H(F)= \pounds_{X_H}F = \dot F$, the Lie derivative of $F$ along the flow of $X_H$.
The equations \[\dot F = \{F,H\},\] called ``Hamilton's equations'', have
already appeared in theorem \ref{thm:lhequiv}, and are an equivalent definition of $X_H$.
\end{remark}

\begin{exercise} \label{canHam}
Show that Hamilton's equations for the classical Poisson bracket are the canonical 
Hamilton's equations,
\[
\dot q ^i = \frac{\partial H}{\partial p_i}\,, \qquad
\dot p_i = -\frac{\partial H}{\partial q^i}\,.
\]
\end{exercise}

\subsection{Infinitesimal invariance under Hamiltonian vector fields}

Let $G$ act smoothly on $P$,
and let $\xi \in \mathfrak{g}$. Recall (from \ref{sec:11}) that the infinitesimal
generator $\xi_{P}$ is the vector field on $P$ defined by 
\[ \xi_{P}(x)  =\frac{d}{dt}  {g(t)} x \bigg|_{t=0}, \]
for some path $g(t)$ in $G$ such that $g(0)=e$ and $g'(0)=\xi$.

\begin{remark}\rm  \label{mxinfgen}
For matrix groups, we can take $g(t) = \exp \left(t \xi\right)$.
This works in general for the exponential map of an arbitrary Lie group.
For matrix groups,
\[\xi_{P}\left(  \mathbf{x}\right)  
=\frac{d}{dt}  \exp(t \xi)  \mathbf{x} \bigg|_{t=0} 
= \xi \mathbf{x} 
\qquad\text{(matrix multiplication)}.
\] 
\end{remark}

%




\begin{definition} \label{infinv}
If $H{:\ } P\to \R$ is $G$--invariant, meaning that $H(gx)=H(x)$ for all $g\in G$ and $x\in P$,
then $\pounds_{\xi_P}H=0$ for all $\xi \in\mathfrak{g}$.
This property is called \emph{infinitesimal invariance}. \index{infinitesimal invariance}
\end{definition}

\begin{example}\rm [The momentum map for the rotation group]
Consider the cotangent bundle of ordinary Euclidean space $\mathbb{R}^3$.
This is the Poisson (symplectic) manifold with coordinates 
$(\mathbf{q},\mathbf{p})\in T^*\mathbb{R}^3\simeq\mathbb{R}^6$, equipped with the canonical
Poisson bracket. An element $g$ of the rotation group $SO(3)$
acts on $T^*\mathbb{R}^3$ according to
\[
g(\mathbf{q},\mathbf{p})=(g\mathbf{q},g\mathbf{p})
\] 
Set $g(t)=\exp(tA)$, so that $\frac{d}{dt}\big|_{t=0}g(t)=A$ and the
corresponding Hamiltonian vector field is
\[
X_A
=(\mathbf{\dot{q}},\mathbf{\dot{p}})
=(A\mathbf{q},A\mathbf{p})
\]
where $A\in \mathfrak{so}(3)$ is a skew-symmetric matrix. The corresponding
Hamiltonian equations read 
\[
\mathbf{\dot{q}}
=
A\mathbf{q}
=
\frac{\partial J_A}{\partial \mathbf{p}}
,\quad
\mathbf{\dot{p}}
=
A\mathbf{p}
=
-\frac{\partial J_A}{\partial \mathbf{q}}
.
\]
Hence,
\[
J_A \left(  \mathbf{q},\mathbf{p}\right)  
=-A\mathbf{p}\cdot \mathbf{q} = a_i\epsilon_{ijk}p_kq_j
= \mathbf{a}\cdot \mathbf{q}\times \mathbf{p}
.\]
for a vector $\mathbf{a}\in\mathbb{R}^3$ with components $a_i$, $i=1,2,3$.
So the momentum map for the rotation group is the
angular momentum $J=\mathbf{q}\times\mathbf{p}$.
\end{example}

\begin{example}\rm  Consider angular momentum $\mathbf{J}=\mathbf{q}\times \mathbf{p}$,
defined on $P=T^*\R^3$. 
%
For every $\xi\in \R^3$, define
\[
\mathbf{J}_{\xi}\left(  \mathbf{q},\mathbf{p}\right)  
:=\bs{\xi}\cdot\left(  \mathbf{q}\times \mathbf{p}\right) 
=\mathbf{p}\cdot\left( \bs{\xi} \times \mathbf{q}\right) 
\]
Using \ref{canHam} and \ref{mxinfgen},
$$X_{J_{\xi}} \left(\mathbf{q}, \mathbf{p}\right)
= \left( \frac{\partial J_{\xi}}{\partial \mathbf{p}}, 
-\frac{\partial J_{\xi}}{\partial \mathbf{q}}\right)
= \big(\bs{\xi} \times \mathbf{q}, \bs{\xi} \times \mathbf{p}\big)
=\widehat{\xi}_P\left( \mathbf{q}, \mathbf{p}\right),$$
where the last line is the infinitesimal generator corresponding to $\widehat{\xi}\in \mathfrak{so}(3)$.
Now suppose $H{:\ } P\to \R$ is $SO(3)$--invariant.
From \ref{infinv}, we have $\pounds_{\hat{\xi}} H = 0$.
It follows that 
\[
\pounds_{X_H} J_{\xi}
=\left\{  J_{\xi},H\right\}  
=-\left\{  H,J_{\xi}\right\}  
=-\pounds_{X_{J_{\xi}}} H
=-\pounds_{\xi_P} H
=0
\,.
\]
Since this holds for all $\xi$, we have shown that $J$ is conserved by the Hamiltonian flow. 
This is the Hamiltonian version of Noether's theorem. \index{Noether's theorem!Hamiltonian version}
\end{example}

\subsection{Defining momentum maps}

In order to generalise this example, we recast it using the 
hat map $\widehat{~}{:\ } \R^3\to \mathfrak{so}(3)$
and the associated map $\tilde{~}{:\ } \big(\R^3\big)^*\to \mathfrak{so}(3)^*$,
and the standard identification $\big(\R^3\big)^*\cong \R^3$ via the Euclidean dot product.
We consider $J$ as a function from $P$ to $\mathfrak{so}(3)^*$ given by
$J\big(  \mathbf{q},\mathbf{p}\big)  =
(\mathbf{q}\times\mathbf{p})\tilde{~}$.
For any $\xi = \widehat{\mathbf{v}}$, we define
$J_{\xi} (\mathbf{q},\mathbf{p})  
= \big<(\mathbf{q}\times \mathbf{p})\tilde{~}, \widehat{\mathbf{v}}
\big>
= (\mathbf{q}\times \mathbf{p}) \cdot {\mathbf{v}}$.
As before, we find that 
$X_{J_{\xi}}=\xi_P$ for every $\xi$, and $J$ is conserved by the Hamiltonian flow.
We take the first property, $X_{J_{\xi}}=\xi_P$, as the general definition of a momentum map.
The conservation of $J$ follows by the same Poisson bracket calculation as in the example;
the result is Noether's Theorem.


\begin{definition}
A \emph{momentum map} \index{momentum map!Poisson manifold} \index{Poisson manifold!momentum map}
for a canonical action of Lie group $G$ on Poisson manifold $P$ is a map
$J{:\ } P\rightarrow \mathfrak{g}^{\ast}$ such that,  for every $\xi \in \mathfrak{g}$, the map
$\smash{J_{\xi}}{:\ } P\to \R$
defined by $\smash{J_{\xi}}(p) = \left<J(p), \xi\right> $ satisfies
\[
X_{J_{\xi}}=\xi_{P}
\,.\]
\end{definition}

 
\begin{theorem}[Noether's Theorem]
Let $G$ act canonically on $(P,\{ \cdot, \cdot\})$ with momentum map
$J$. If $H$ is $G$--invariant, then $J$ is conserved by the flow of
$X_{H}$.
\end{theorem}

\begin{proof}
For every $\xi\in\mathfrak{g}$,
\[
\pounds_{X_H} J_{\xi}
=\left\{  J_{\xi},H\right\}  
=-\left\{  H,J_{\xi}\right\}  
=-\pounds_{X_{J_{\xi}}} H
=-\pounds_{\xi_P} H
=0
\,. 
\]
\end{proof}

\begin{exercise}
Show that momentum maps are unique up to a choice of a constant element of
\textrm{g}$^{\ast}$ on every connected component of $M$.
\end{exercise}




\begin{exercise}
Show that the $S^1$ action on the torus $T^2:=S^1\times S^1$ given by 
$\alpha \left(\theta,\phi\right) = \left(\alpha + \theta, \phi\right)$ is canonical
with respect to  the classical bracket (with $\theta,\phi$ in place of $q,p$), but
it does not have a momentum map.
\end{exercise}

\begin{exercise} \label{Petzval mommap}
Show that the Petzval invariant for Fermat's principle in axisymmetric,
translation-invariant media is a momentum map, $T^*\mathbb{R}^2\mapsto
sp(2,\mathbb{R})^*$ taking $(\mathbf{q},\mathbf{p})\mapsto(X,Y,Z)$. What is its
corresponding symmetry? What is its Hamiltonian vector field? 
\end{exercise}

\begin{theorem}[also due to Noether]\label{Cotlift-mommap}
Let $G$ act on $Q$, and
by cotangent lifts on $T^{\ast}Q$. 
Then $J{:\ } T^{\ast
}Q\rightarrow\mathfrak{g}^{\ast}$ defined by, for every $\xi\in\mathfrak{g}$,%
\[
J_{\xi}\left(  \alpha_{q}\right)  =\left\langle \alpha_{q},\xi_{Q}\left(
q\right)  \right\rangle ,\text{ for every }\alpha_{q}\in T_{q}^{\ast}Q,
\]
is a momentum map (the ``standard one'') 
for the $G$ action with respect to the classical Poisson bracket.
\end{theorem}


\begin{proof}
We need to show that $X_{J_{\xi}} = \xi_{T^*Q}$,
for every $\xi\in\mathfrak{g}$. From the definition of Hamiltonian vector fields, 
this is equivalent to showing that $\xi_{T^*Q}[F] = \{F,J_{\xi}\}$ for every $F\in \mathcal{F}(T^*Q)$.
We verify this for finite-dimensional $Q$ by using
cotangent-lifted local coordinates.
\begin{align*}
\frac{\partial J_{\xi}}{\partial p}(q,p) & = \xi_Q(q) \\
\frac{\partial J_{\xi}}{\partial q^i} (q,p)&= 
\left< p, \frac{\partial}{\partial q^i}\left(\xi_Q(q)\right)\right> \\
&\hspace{-8mm}=\left< p, \frac{\partial}{\partial q^i}\left( \left.\frac{\partial}{\partial t} \Phi_{\left(\exp (t\xi)\right)}( q) \right|_{t=0}\right) \right> 
=\left< p, \frac{\partial}{\partial t}\left( \left.\frac{\partial}{\partial q^i} 
\Phi_{\left(\exp (t\xi)\right)} (q) \right)\right|_{t=0} \right> \\
&\hspace{-8mm}=\left.\frac{\partial}{\partial t} \left< p, 
T\Phi_{\left(\exp (t\xi)\right)} \frac{\partial}{\partial q^i}(q)  \right> \right|_{t=0}
=\left.\frac{\partial}{\partial t} \left< T^*\Phi_{\left(\exp (t\xi)\right)} p, 
\frac{\partial}{\partial q^i}(q)  \right> \right|_{t=0} \\
&\hspace{-8mm}= \left< -\,\xi_{T^*Q}(q,p),\frac{\partial}{\partial q^i}(q)\right> \\
\frac{\partial J_{\xi}}{\partial q} (q,p) &= -\,\xi_{T^*Q}(q,p)
\end{align*}
So for every $F\in \mathcal{F}(T^*Q)$,
\begin{align*}
\xi_{T^*Q}[F] &= \frac{\partial}{\partial t} F\left(\exp (t\xi) q,
\exp(t\xi) p\right) \bigg|_{t=0} \\
&= \frac{\partial F}{\partial q}  \xi_Q(q) + 
\frac{\partial F}{\partial p}  \xi_{T^*Q}(q,p)
\\&= \frac{\partial F}{\partial q}  \frac{\partial J_{\xi}}{\partial p} -
\frac{\partial F}{\partial p}  \frac{\partial J_{\xi}}{\partial q}
= \{F,J_{\xi}\}
\,,\end{align*}
which completes the proof.
\end{proof}

\begin{example}\rm 
Let $G\subset M_{n}\left(  \mathbb{R}\right)  $ be a
matrix group, with cotangent-lifted action on
$(q,p)\in T^{\ast}\mathbb{R}^{n}$.  For every $\mathrm{g}\subset
M_{n}\left(\mathbb{R}\right)$, $q\mapsto g q$. The cotangent-lifted
action is $(q,p)\mapsto(g q,g^{-T}p)$. Thus, writing
$g=\exp(t\xi)$, the linearization of this group action yields the
vector field
\[
X_{\xi}=(\xi q,-\xi^T p)
\]
The corresponding Hamiltonian equations read
\[ \xi q=\frac{\partial J_{\xi}}{\partial p},\qquad
-\,\xi^T p=-\,\frac{\partial J_{\xi}}{\partial q} \]
This yields the momentum map $J(q,p)$ given by
\[ J_{\xi}\left(q,p\right)  
= \langle J(q,p),\xi\rangle
=p^{T}\xi_{Q}\left(  q\right)  =p^{T}\xi q. \]
In coordinates, $p^{T}\xi q=p_i\xi^i_jq^j$, so $J(q,p)=q^ip_j$.
\end{example}
\begin{exercise}
Calculate the momentum map of the cotangent lifted
action of the group of translations of $\mathbb{R}^{3}$.
\end{exercise}

\begin{answer}
The element $\mathbf{x}\in\mathbb{R}^{3}$ acts on
$\mathbf{q}\in\mathbb{R}^{3}$ by addition of vectors,
\[
\mathbf{x}\cdot(\mathbf{q})
=\mathbf{q}+\mathbf{x}
.
\]
The infinitesimal generator is 
$\lim_{\mathbf{x}\to0}\frac{d}{d\mathbf{x}}
(\mathbf{q}+\mathbf{x})=\Id $. Thus, $\xi_{\mathbf{q}}=\Id$ and
\[
\langle J_k,\xi\rangle
=\langle (\mathbf{q},\mathbf{p}),\xi_{\mathbf{q}}\rangle
=\langle \mathbf{p},\Id\rangle
=p_i\delta^i_k=p_k
\]
This is also Hamiltonian with $J_{\xi}=\mathbf{p}$, so that
$\{\mathbf{p},J_{\xi}\}=0$ and $\{\mathbf{q},J_{\xi}\}=\Id$.
\end{answer}

\begin{example}\rm \label{ex-ad-star}
Let $G$ act on itself by left multiplication, and by
cotangent lifts on $T^{\ast}G$. We first note that the infinitesimal action on
$G$ is
\[
\xi_{G}\left(  g\right)  =\frac{d}{dt} \exp(t\xi)
g\bigg|_{t=0}=TR_{g}\xi.
\]
Let $J_L$ be the  momentum map for this action.
Let $\alpha_{g}\in T_{g}^{\ast}G$. For every $\xi\in\mathfrak{g}$, we have
\[
\left\langle J_L\left(  \alpha_{g}\right)  ,\xi\right\rangle =\left\langle
\alpha_{g},\xi_{G}\left(  g\right)  \right\rangle =\left\langle \alpha
_{g},TR_{g}\xi\right\rangle =\left\langle TR_{g}^{\ast}\alpha_{g}%
,\xi\right\rangle
\]r
so $J_L\left(  \alpha_{g}\right)  =TR_{g}^{\ast}\alpha_{g}$. Alternatively,
writing $\alpha_{g}=T^{\ast}L_{g^{-1}}\mu$ for some $\mu\in\mathfrak{g}^{\ast
}$ we have
\[
J_L\left(  T^{\ast}L_{g^{-1}}\mu\right)  =TR_{g}^{\ast}T^{\ast}L_{g^{-1}}%
\mu=Ad_{g^{-1}}^{\ast}\mu.
\]
\end{example}

\begin{exercise}
Show that the momentum map for the right multiplication action $R_g (h)=hg$ is 
$J_R\left(  \alpha_{g}\right)  =TL_{g}^{\ast}\alpha_{g}$.
\end{exercise}

For matrix groups, the tangent lift of the left (or right) multiplication action is
again matrix multiplication. Indeed, to compute $TR_G(A)$ for any $A\in T_QSO(3)$, let
$B(t)$ be a path in $SO(3)$ such that $B(0)=Q$ and $B'(0)=A$. Then
\[
TR_{G}(A) =  \left.\frac{d}{dt} B(t)G\right|_{t=0} = AG.
\]
Similarly, $TL_G(A)=GA$.
To compute the cotangent lift similarly, we need to be able to consider elements of
$T^*G$ as matrices.  This can be done using any nondegenerate bilinear form on each
tangent space $T_QG$. 
We will use the Frobenius pairing, defined by 
\[
\langle\langle A,B\rangle\rangle
:= -\tfrac{1}{2}\tr \big(A^TB\big)
=-\tfrac{1}{2}\tr \big(AB^T\big).
\]
(The equivalence of the two formulas follows from the properties
$\tr(CD)=\tr(DC)$ and
$\tr(C^T)=\tr(C)$).

\begin{exercise}
Check that this pairing, restricted to $\mathfrak{so}(3)$, corresponds to the Euclidean inner
product  via the hat map.
\end{exercise}

\begin{example}\rm  \label{mommap-ex}
Consider the previous example for a \emph{matrix} group $G$. 
For any $Q\in G$, the pairing given above allows use to consider any
element $P\in T^*_QG$  as a matrix. The natural pairing of $T^*_QG$ with
$T_QG$ now has the formula,
\[
<P,A> = -\tfrac{1}{2}\tr\big(P^TA\big), \quad \textrm{for all } A \in T_QG.
\]
We compute the cotangent-lifts of the left and right multiplication actions:
\begin{align*}
\big<T^*L_Q(P),A\big> &= \big<P,TL_Q(A)\big> = \big<P,QA\big> \\ 
&= -\tfrac{1}{2} \tr\big(P^TQA\big)
= -\tfrac{1}{2} \tr\big(\big(Q^TP\big)^TA\big)
=\big<Q^TP,A\big> \\
\big<T^*R_Q(P),A\big> &= \big<P,TR_Q(A)\big> = <P,AQ> \\ 
&= -\tfrac{1}{2} \tr\big(P(AQ)^T\big)
= -\tfrac{1}{2} \tr\big(PQ^TA^T\big)
=\big<PQ^T,A\big>
\end{align*}
In summary,
\begin{align*}
T^*L_Q(P) = Q^TP \quad \textrm{and} \quad
T^*R_Q(P) = PQ^T
\end{align*}
We thus compute the momentum maps as
\begin{align*}
J_L(Q,P) &= T^*R_QP =PQ^T \\
J_R(Q,P) &= T^*L_{Q}P =Q^TP
\end{align*}
In the special case of $G=SO(3)$, these matrices $PQ^T$ and $Q^TP$ are skew-symmetric,
since they are elements of $\mathfrak{so}(3)$. Therefore,
\begin{align*}
J_L(Q,P) &= T^*R_QP =\tfrac{1}{2}\big(PQ^T - QP^T\big) \\
J_R(Q,P) &= T^*L_{Q}P = \tfrac{1}{2}\big(Q^TP - P^TQ\big)
\end{align*}
\end{example}

\begin{exercise}
Show that the cotangent lifted action on $SO(n)$ can be expressed 
as matrix multiplication:
\[
Q\cdot P = Q^TP
\,.\]
\end{exercise}

\subsection{Equivariance}

\begin{definition}[\textbf{Equivariance of momentum maps}] 
\index{momentum map!Poisson manifold!equivariant}
A momentum map is said to be \emph{equivariant} when it is equivariant
with respect to the given action on
$P$ and the coadjoint action on $\mathfrak{g}^*$. That is,
\[
J(g \cdot p) = \Ad_{g^{-1}}^* J(p)
\]
for every $g\in G$, $p \in P$, where $g \cdot p$ denotes the action of $g$
on the point $p$ and where Ad denotes the adjoint action.
\end{definition}

\begin{proposition}
All cotangent-lifted actions are $Ad^{\ast}$--equivariant.
\end{proposition}

\begin{proposition}
Every $Ad^{\ast}$--equivariant momentum map $J{:\ } P\to \mathfrak{g}^*$ is a Poisson map,
with respect to the `$+$' Lie--Poisson bracket on $\mathfrak{g}^*$.
\end{proposition}

\begin{exercise}
Prove the previous two Propositions.
\end{exercise}

\begin{exercise}
Show that the momentum map derived  from the cotangent 
Theorem \ref{Cotlift-mommap} is equivariant.
\end{exercise}

\begin{example}[\textbf{Momentum map for symplectic representations}]\rm 
\label{symp-mommap}

Let $(V, \Omega) $ be a symplectic vector space and let $G $ be a Lie
group acting linearly and symplectically on $V $. This action admits an
equivariant momentum map $\mathbf{J}{:\ }  V \rightarrow \mathfrak{g}$ given
by
\[
J^ \xi(v) =
\langle \mathbf{J}(v), \xi \rangle = \tfrac{1}{2}\Omega(\xi\cdot v , v ),
\]
where $\xi\cdot v $ denotes the Lie algebra representation of the
element $\xi \in  \mathfrak{g}$ acting on the vector $v \in V $. To verify
this, note that the infinitesimal generator $\xi_V(v) = \xi \cdot v $,
by the definition of the Lie algebra representation induced by the given
Lie group representation, and that $\Omega( \xi \cdot u, v ) = -\Omega(
u, \xi\cdot v ) $ for all $u, v \in V $. Therefore
\[
\mathbf{d}J^ \xi (u) (v) = \tfrac{1}{2}\Omega(\xi\cdot u , v ) +
\tfrac{1}{2}\Omega(\xi\cdot v, u ) = \Omega(\xi\cdot u , v ).
\]
Equivariance of $\mathbf{J}$ follows from the obvious relation $g ^{-1}
\cdot \xi \cdot g \cdot v = (\operatorname{Ad}_{g ^{-1}} \xi) \cdot v $
for any $g \in G $, $ \xi\in \mathfrak{g}$, and $v \in V $.
\end{example}

\begin{example}\rm [\textbf{Cayley--Klein parameters and the Hopf fibration}]
\label{CK-param} \index{Hopf fibration!Cayley--Klein parameters}

Consider the natural action of $SU(2)$ on $\mathbb{C}^2 $. Since this
action is by isometries of the Hermitian metric, it is automatically
symplectic and therefore has a momentum map $\mathbf{J}{:\ }  \mathbb{C} ^2
\rightarrow \mathfrak{su}(2)^\ast$ given in Example \ref{symp-mommap},
that is, 
\[
\langle \mathbf{J}(z, w), \xi \rangle = \tfrac{1}{2}
\Omega(\xi\cdot (z,w), (z, w)),
\]
where $z, w \in \mathbb{C}$ and $\xi \in \mathfrak{su}(2)$. Now the
symplectic form on $\mathbb{C}^2$ is given by minus the imaginary part of
the Hermitian inner product. That is, $\mathbb{C}^{n}$ has Hermitian
inner product given by 
\[
\mathbf{z} \cdot \mathbf{w}: = \sum_{j=1}^n z_j \overline{w}_j, 
\ \hbox{where}\ \
\mathbf{z} = (z_1, \dots, z_n), 
 \ \ \hbox{and} \ \ 
\mathbf{w}
=
(w_1, \dots, w_n) \in \mathbb{C}^n .
\] 
The symplectic form is thus
given by $\Omega(\mathbf{z}, \mathbf{w}) : = -\operatorname{Im}({\mathbf z}
\cdot \mathbf{w})$ and it is identical to the one given before on
$\mathbb{R}^{2n}$ by identifying $\mathbf{z} = \mathbf{u} + i
\mathbf{v} \in \mathbb{C}^n$ with $(\mathbf{u}, \mathbf{v}) \in
\mathbb{R}^{2n}$ and $\mathbf{w} = \mathbf{u}' + i \mathbf{v}' \in
\mathbb{C}^n$ with $(\mathbf{u}', \mathbf{v}') \in \mathbb{R}^{2n}$.

The Lie algebra $\mathfrak{su}(2)$ of $SU(2)$
consists of $2 \times 2 $ skew Hermitian matrices of trace zero. This Lie
algebra is isomorphic to $\mathfrak{so}(3) $ and therefore to
$(\mathbb{R}^3, \times )$ by the isomorphism given by the \textbf{tilde map},
\[
\mathbf{x} = (x^1, x^2,x^3) \in \mathbb{R}^3 
\quad\mapsto\quad
\widetilde{\mathbf{x}} : =
\tfrac{1}{2}
\left[
\begin{array}{cc}
-ix^3&-ix^1 - x^2\\
-ix^1 + x^2&ix^3
\end{array}
\right] \in \mathfrak{su}(2).
\]
Thus we have $[\widetilde{\mathbf{x}}, \widetilde{\mathbf{y}}] =
(\mathbf{x}\times \mathbf{y})\widetilde{\phantom{y}}$ for any
$\mathbf{x}, \mathbf{y} \in \mathbb{R}^3$. Other useful relations are
$\operatorname{det}(2\widetilde{\mathbf{x}}) = \| \mathbf{x}\|^2 $ and
$\operatorname{trace}(\widetilde{\mathbf{x}}\widetilde{\mathbf{y}}) =
-\frac{1}{2} \mathbf{x} \cdot \mathbf{y}$. 

Identify $\mathfrak{su}(2)^\ast$ with $\mathbb{R}^3$ by the \textbf{check map} $\check{(\,\cdot\,)}$:
$ \mu \in \mathfrak{su}(2)^\ast \mapsto \check{\mu} \in \mathbb{R}^3$
defined by
\[
\check{\mu} \cdot \mathbf{x}: = -2\langle \mu, \widetilde{\mathbf{x}}
\rangle
\]
for any $\mathbf{x} \in \mathbb{R}^3$.
With these notations, the
momentum map $\check{\mathbf{J}}{:\ }  \mathbb{C}^2 \rightarrow \mathbb{R}^3$
can be explicitly computed in coordinates: for any $\mathbf{x}\in
\mathbb{R}^3$ we have
\begin{align*}
\check{\mathbf{J}}(z, w) \cdot \mathbf{x} &=
-2\langle \mathbf{J}(z, w), \widetilde{\mathbf{x}} \rangle \\
&= \tfrac{1}{2}\operatorname{Im}\left(
\left[
\begin{array}{cc}
-ix^3&-ix^1 - x^2\\
-ix^1 + x^2&ix^3
\end{array}
\right]
\left[
\begin{array}{c}
z\\
w
\end{array}
\right]
\cdot
\left[
\begin{array}{c}
z\\
w
\end{array}
\right]
\right)\\
&= -\tfrac{1}{2}(2 \operatorname{Re}(w \overline{z}),
2\operatorname{Im}(w\overline{z}), |z|^2 - |w|^2) \cdot \mathbf{x}.
\end{align*}
Therefore
\[
\check{\mathbf{J}}(z, w) = -\tfrac{1}{2}(2w \overline{z}, |z|^2 - |w|^2) \in
\mathbb{R}^3.
\]
\begin{itemize}
\item
Thus, $\check{\mathbf{J}}$ is a Poisson map from
$\mathbb{C}^2 $, endowed with the  canonical symplectic structure, to
$\mathbb{R}^3 $, endowed with the $+$ Lie--Poisson structure.
\item
Therefore, $- \check{\mathbf{J}}{:\ }  \mathbb{C}^2 \rightarrow \mathbb{R}^3 $
is a canonical map, if $\mathbb{R}^3$  has the $-$ Lie--Poisson bracket
relative to which the free rigid body equations  are Hamiltonian.
\item
Pulling back the Hamiltonian $H(\boldsymbol{\Pi}) = \boldsymbol{\Pi}
\cdot {\mathbb I}^{-1} \boldsymbol{\Pi}/2$ to $\mathbb{C}^2$ gives a
Hamiltonian function (called collective) on $\mathbb{C}^2$. 
\item
The classical Hamilton equations for this function are therefore
projected by $-\check{\mathbf{J}}$ to the rigid body equations $\dot
{\boldsymbol{\Pi}} = \boldsymbol{\Pi} \times {\mathbb I}^{-1}
\boldsymbol{\Pi}$. 
\item
In this context, the variables $(z, w) $ are
called the \textbf{Cayley--Klein parameters\/}. 
\end{itemize}
\end{example} 

\begin{exercise}\rm
Show  that $-\check{\mathbf{J}}|_{S^3}{:\ }  S^3 \rightarrow S^2$
is the \emph{Hopf fibration\/}. In other words, {the momentum
map of the $SU(2)$--action on $\mathbb{C}^2$, the Cayley--Klein
parameters and the family of
Hopf fibrations on concentric three-spheres in $\mathbb{C}^2$ are all the
same map}.
\end{exercise}

\begin{exercise}\textbf{Optical traveling wave pulses} \,  \rm
The equation for the evolution of the complex amplitude of a polarized
optical traveling wave pulse in a material medium is given as
\[
\dot{z}_i 
=
\frac{1}{\sqrt{-1}}
\frac{\partial H}{\partial z_i^*}
\]
with Hamiltonian $H{:\ } \mathbb{C}^2\to\mathbb{R}$ defined by
\[
H
=
z_i^*\chi^{(1)}_{ij}z_j
+
3z_i^*z_j^*\chi^{(3)}_{ijkl}z_kz_l
\]
and the constant complex tensor coefficients $\chi^{(1)}_{ij}$ and
$\chi^{(1)}_{ijkl}$ have the proper Hermitian and permutation symmetries
for $H$ to be real. Define the Stokes vectors
by the isomorphism,
\[
\mathbf{u} = (u^1, u^2,u^3) \in \mathbb{R}^3 \mapsto
\widetilde{\mathbf{u}} : =
\frac{1}{2}
\left[
\begin{matrix}
-iu^3      &  -iu^1 - u^2 \\
-iu^1 + u^2&   iu^3
\end{matrix}
\right] \in \mathfrak{su}(2).
\]
\begin{enumerate}
\item
Prove that this isomorphism is an equivariant momentum map.
\item
Deduce the equations of motion for the Stokes vectors of this optical
traveling wave and write it as a Lie--Poisson Hamiltonian system. 
\item
Determine how this system is related to the equations for an $SO(3)$ rigid
body.
\end{enumerate}
\end{exercise}

\begin{exercise}

The formula determining the momentum map for the cotangent-lifted action
of a Lie group $G$ on a smooth manifold $Q$ may be expressed in terms of
the pairing $\langle\cdot ,\cdot \rangle{:\ }  
\mathfrak{g}^*\times\mathfrak{g}\mapsto\mathbb{R}$
as 
\[ \langle J, \xi\rangle = \langle p, \pounds_{\xi} q\rangle, \]
where $(q,p)\in T_q^*Q$ and $\pounds_{\xi} q$ is the infinitesimal generator of the action of
the Lie algebra element $\xi$ on the coordinate $q$. 

Define appropriate pairings and determine the momentum maps explicitly for the following
actions:
\begin{enumerate}
\item[(a)] 
$\pounds_{\xi} q=\xi\times q$ for $\mathbb{R}^3\times\mathbb{R}^3\mapsto\mathbb{R}^3$
\item[(b)] 
$\pounds_{\xi} q=\ad_{\xi} q$ for $\ad$--action
$\ad{:\ } \mathfrak{g}\times\mathfrak{g}\mapsto\mathfrak{g}$ in a Lie algebra
$\mathfrak{g}$
\item[(c)] 
$AqA^{-1}$ for $A\in GL(3,R)$ acting on $q\in GL(3,R)$ by matrix conjugation
\item[(d)] 
$Aq$ for left action of $A\in SO(3)$ on $q\in SO(3)$
\item[(e)] 
$AqA^T$ for $A\in GL(3,R)$ acting on $q\in \Sym(3)$, that is $q=q^T$.
\end{enumerate} 

\end{exercise}

\begin{answer}\quad
\begin{enumerate}
\item[(a)] 
$p \cdot \xi\times q = q \times p \cdot \xi \Rightarrow J = q \times p$. (The pairing
is scalar product of vectors.) 
\item[(b)] 
$\langle p, \ad_{\xi} q \rangle 
=
-\langle \ad^*_q p, \xi \rangle
\Rightarrow J = \ad^*_q p$ for the pairing
$\langle\cdot ,\cdot \rangle{:\ }  
\mathfrak{g}^*\times\mathfrak{g}\mapsto\mathbb{R}$
\item[(c)] 
Compute $T_e(AqA^{-1})=\xi q - q\xi = [\xi,q]$ for $\xi=A^\prime(0)\in gl(3,R)$ acting on
$q\in GL(3,R)$ by matrix Lie bracket $[\cdot, \cdot]$. For the matrix pairing 
$\langle A, B \rangle = \trace (A^TB)$, we have 
$$\trace \big(p^T[\xi,q]\big)=
  \trace\big(\smash{\big(pq^T-q^Tp\big)}^T\xi\big)\Rightarrow J = pq^T-q^Tp.$$
\item[(d)] 
Compute $T_e(Aq)=\xi q$ for $\xi=A^\prime(0)\in \mathfrak{so}(3)$ acting on
$q\in SO(3)$ by left matrix multiplication. For the matrix pairing 
$\langle A, B \rangle = \trace (A^TB)$, we have 
$$\trace \big(p^T\xi q\big)=\trace \big(\smash{\big(pq^T\big)}^T\xi\big)\Rightarrow J =
  \tfrac{1}{2}\big(pq^T-q^Tp\big),$$
where we have used antisymmetry of the matrix $\xi\in \mathfrak{so}(3)$.
\item[(e)] 
Compute $T_e\big(AqA^T\big)=\xi q + q \xi^T$ for $\xi=A^\prime(0)\in gl(3,R)$ acting on $q\in
\Sym(3)$. For the matrix pairing 
$\langle A, B \rangle = \trace (A^TB)$, we have 
$$\trace \big(p^T\big(\xi q + q \xi^T\big)\big)=\trace
\big(q\big(p^T+p\big)\xi\big)=\trace\big(\smash{\big(2qp\big)}^T\xi\big)\Rightarrow
J = 2qp,$$
where we have used symmetry of the matrix $\xi q + q
\xi^T$ to choose $p=p^T$. (The momentum canonical to the symmetric matrix $q=q^T$ should be
symmetric to have the correct number of components!)
\end{enumerate} 

\end{answer}

\begin{exercise}[$GL(n,\mathbb{R})$--invariant motions]\label{GLn-ex}
Begin with the Lagrangian
\[ L=\tfrac{1}{2}\tr \big(\dot{S}S^{-1}\dot{S}S^{-1}\big)
+ \tfrac{1}{2}\,\mathbf{\dot{q}}^T S^{-1}\mathbf{\dot{q}} \]
where $S$ is an $n\times n$ symmetric matrix and
$\mathbf{q}\in\mathbb{R}^n$ is an
$n$--component column vector.
\begin{enumerate}
\item Legendre transform to construct the corresponding Hamiltonian
and canonical equations.
\item Show that the system is invariant under the group action 
\[ \mathbf{q}\to A\mathbf{q} \quad\text{and}\quad S\to ASA^T \]
for any constant invertible $n\times n$ matrix, $A$.
\item Compute the infinitesimal generator for this group action and
construct its corresponding momentum map. Is this momentum map
equivariant?
\item Verify directly that this momentum map is a conserved $n\times n$
matrix quantity by using the equations of motion.
\item Is this system completely integrable for any value of $n>2$?
\end{enumerate}
\end{exercise}

\begin{answer}
The exercise at the end of section \ref{sec-BrokenSym} obtains and solves this same system 
for the dynamics of the mean and variance of a multivariate Gaussian process.
\index{broken symmetry!semidirect product}
\end{answer}
\end{comment}

\newpage
\vspace{4mm}\centerline{\textcolor{shadecolor}{\rule[0mm]{6.75in}{-2mm}}\vspace{-4mm}}
\section{Hamiltonian vector fields \& differential forms}
\label{sec-DiffForms}

\secttoc

\textbf{What is this lecture about?} This lecture discusses the coordinate-free approach to geometric mechanics
in the Lagrangian and Hamiltonian formulations. 

\subsection{Hamilton's principle for Euler-Lagrange \& Hamilton equations}
The Euler--Lagrange equations for the Lagrangian $L(q, \dot{q}): TQ\to \mathbb{R}$ follow from \textbf{Hamilton's principle}, as 
\begin{align}
\begin{split}
0 = \delta S &= \delta \int_0^T L(q, v_q) + \scp{p}{\frac{dq}{dt} - v_q } \, dt
\\&=
\int_0^T \scp{\frac{\p L}{\p q} - \frac{dp}{dt}}{\delta q} + \scp{\frac{\p L}{\p v_q } - p}{\delta v_q} 
+ \scp{\delta p}{\frac{dq}{dt} - v_q } + \underbrace{\scp{p}{\delta q}\Big|_0^T}_{\hbox{\emph{Noether term}}}
\,,\end{split} 
\label{sec6: HP-Lagrange}
\end{align}
Upon assuming that $\delta q$ vanishes at the endpoints in time, 
collecting terms in \eqref{sec6: HP-Lagrange} yields the \emph{Euler-Lagrange equations}, \eqref{ELeqns}.

To pass to the Hamiltonian side in canonical position-momentum $(q,p)$ coordinates,
we Legendre-transform to write the Lagrangian in \eqref{sec6: HPP} in  phase-space form as
\begin{align}
\begin{split}
0 = \delta S &= \delta \int_0^T  \scp{p}{v_q } - H(p,q) + \scp{p}{\frac{dq}{dt} - v_q } \, dt
\\&=
 \delta \int_0^T  \scp{p}{\frac{dq}{dt}}  - H(p,q) \, dt
\\&=
\int_0^T \scp{\delta p} {\frac{dq}{dt} - \frac{\p H}{\p p}} - \scp{\frac{dp}{dt} + \frac{\p H}{\p q}}{\delta q}  
+ \underbrace{\scp{p}{\delta q}\Big|_0^T}_{\hbox{\emph{Noether term}}}
\,,\end{split} 
\label{sec6: HP-Hamilton}
\end{align}
Variations of the Legendre-transformed phase-space Lagrangian $\de q$ and $\de p$ produce 
\emph{Hamilton's canonical equations} and the \textit{same} Noether term as before. Hamilton's equations lead to a vector field $X_H$ on phase space and we shall see that $X_H$ generates symplectic transformations, i.e., transformations that preserve the area in phase space. The Noether term leads to Noether's theorem on the Hamiltonian side.

\subsection{The differential}
The \emph{differential}, or \emph{exterior derivative} \index{differential!exterior derivative}
of a function $F$
on phase space with coordinates $(q,p)\in T^*Q$ is written
\begin{eqnarray*}
dF = F_q dq + F_p dp
,
\end{eqnarray*}
in which subscripts denote partial derivatives. For the Hamiltonian itself, the
exterior derivative yields
\begin{eqnarray*}
dH = H_q dq + H_p dp
= -\,\frac{dp}{dt} dq + \frac{dq}{dt} dp
\,,
\end{eqnarray*}
upon inserting the canonical equations 
\[
\frac{dq}{dt} = H_p \quad\hbox{and}\quad \frac{dp}{dt} = - H_q\,.
\]
The canonical equations also provide the definition of a Hamiltonian vector field (HVF)
\[
\frac{d}{dt} = \frac{dq}{dt}\p_q + \frac{dp}{dt} \p_p = H_p \p_q - H_q\p_p  =: \{\,\cdot\,,\, H\} : = X_H\,.
\]
This means that the action of the HVF $X_H$ on a phase space function $F$ 
yields its time derivative 
\[
\frac{d}{dt}F(q,p) = X_HF(q,p) = \{F\,,\, H\}
\,.
\]
The HVF $X_H$ may also act as a time derivative on differential forms defined on phase space. 
For example, $X_H$ acts on the time-dependent
one-form
$p\,dq(t)$ along solutions of Hamilton's equations as
\begin{align}
\begin{split}
X_H\big(p\,dq\big)
=
\frac{d}{dt}\big(p\,dq\big)
&=
\dot{p}\,dq
+
p\,d\dot{q}
\\
&=
\dot{p}\,dq-\dot{q}\,dp+d(p\dot{q})
\\
&=
-H_q dq
-H_p dp
 +d(p\dot{q})
\\
&=
d(-H+p\dot{q}) =: dL(q,\dot{q})
\,,\label{XH-pre-symp}
\end{split}
\end{align}
upon substituting Hamilton's canonical equations and applying the Legendre transformation.

\begin{exercise}
Show that the HVF $X_H$ commutes with its differential, or exterior derivative. Thus,
\begin{eqnarray*}
d(X_H F)  = X_H ( dF )
\,.
\end{eqnarray*}
\end{exercise}

The exterior derivative of the one-form $p dq$ yields the canonical, or
symplectic two-form%
\footnote{
The properties of differential forms are summarised in section \ref{ext-calculus-review} and 
developed for PDEs in \ref{diff-form-review}.
}
\begin{eqnarray}
d(p dq) = dp\wedge dq  = -\, dq\wedge dp
\,.\label{omega}
\end{eqnarray}
Here we have used the chain rule for the exterior derivative and its property that $d^2=0$. 
\begin{exercise}
Show that the property $d^2=0$ for the differential amounts to equality of cross derivatives for continuous functions.
\end{exercise}

Equation \eqref{omega} introduces the wedge product
$\wedge$, which combines two one-forms (the line elements $dq$ and $dp$) into a
two-form (the oriented surface element $dp\wedge dq = -dq\wedge dp$).  As a
result, the two-form
$\omega=dq\wedge dp$ representing area in phase space is \emph{conserved} along the
Hamiltonian flows generated by the Hamiltonian vector fields,
\begin{eqnarray*}
X_H\big(dq\wedge dp\big)
=
\frac{d}{dt}\big(dq\wedge dp\big)
=
0\,.
\end{eqnarray*}
\begin{proof}
The calculation in \eqref{XH-pre-symp} implies for $dp\wedge dq=-dq\wedge dp$ that
\[d\big(X_H\big(p\,dq\big)\big) = \frac{d}{dt}(dp\wedge dq)
= X_H(dp\wedge dq) = d^2L(q,\dot{q}) = 0\,.\] 
\end{proof} \vspace{-7mm}
Consequently, one finds the following.

\begin{theorem}[Poincar\'e's theorem]
Hamiltonian flows preserve the area in phase space.
\end{theorem}

\begin{definition}[Symplectic two-form]
The phase space area $\omega=dq\wedge dp$ is called the symplectic 2-form.
\end{definition}

\begin{definition}[Symplectic flows]
Flows that preserve area in phase space are said to be \emph{symplectic}.
\end{definition}

\begin{remark}\rm [Poincar\'e's theorem]
Hamiltonian flows are symplectic.
\end{remark}

\subsection{Short review of exterior calculus,
symplectic forms and\\ Poin\-car\'e's theorem in higher dimensions}
\label{ext-calculus-review}

Exterior calculus on symplectic manifolds is the geometric language of
Hamiltonian mechanics. As an introduction and motivation for more detailed study, we
begin with an overview.

In differential geometry, the operation of \emph{contraction} \index{contraction!equivalent notations} (denoted equivalently as
either $\iota_{X}\alpha$ or $X\contract\alpha$) introduces a pairing between  vector fields $X\in\mathfrak{X}(M)$ 
and differential forms $\alpha\in \Lambda(M)$.\footnote{The two notations $\iota$ for insertion and $\contract$ for contraction can be used interchangeably.
In this text we will choose the notation that seems most convenient for enhancing the legibility of any given presentation. 
In particular, we will tend to favour $\contract$ in the presence of subscripts and superscripts for general bases of vector fields and $k$-forms.}

Contraction is also called \emph{substitution}, or \emph{insertion} of a vector field into a
differential form. For example, there are the dual relations, 
\begin{eqnarray*}
\partial_{q}\contract dq=1
=\partial_{p}\contract dp
,\quad\hbox{and}\quad
\partial_{q}\contract dp=0
=\partial_{p}\contract dq
\end{eqnarray*}
A Hamiltonian vector field
\begin{eqnarray*}
X_H = \dot{q}\frac{\partial}{\partial q} +\dot{p}\frac{\partial}{\partial p}
= H_p\partial_q-H_q\partial_p =\{\cdot,H\}
\end{eqnarray*}
satisfies 
\begin{eqnarray*}
X_H\contract dq=H_p \quad\hbox{and}\quad X_H\contract dp=-H_q.
\end{eqnarray*}
\begin{definition}
The rule for the contraction operation, $\iota$ or $\contract$, also known as substitution of a vector field into
a differential form is to sum the substitutions of $X_H$ over the
permutations of the factors in the differential form that bring the
corresponding dual basis element into its leftmost position. 
\end{definition}
For example,
substitution of the Hamiltonian vector field $X_H$ into the symplectic
form  $\omega = dq\wedge dp$ yields
\begin{eqnarray*}
X_H\contract \omega = X_H\contract(dq\wedge dp) =
  (X_H\contract dq)\,dp - (X_H\contract dp)\,dq
\end{eqnarray*}
In this example, $X_H\contract dq=H_p$ and $X_H\contract dp=-H_q$, so
\begin{eqnarray*}
X_H\contract \omega = H_pdp + H_qdq = dH
\end{eqnarray*}
which follows because $\partial_q\contract dq=1=\partial_p\contract dp$
and $\partial_q\contract dp=0=\partial_p\contract dq$.

This calculation proves
\begin{theorem}[Hamiltonian vector field]
The Hamiltonian vector field $X_H=\{\cdot , H\}$ satisfies
\begin{eqnarray}
X_H\contract\omega&=&dH
\quad\hbox{with}\quad
\omega
=
dq\wedge dp
\label{HVF-def}
\end{eqnarray}
\end{theorem}
In fact, relation \eqref{HVF-def} may be taken as the \emph{definition} 
of a Hamiltonian vector field.

As a consequence of formula \eqref{HVF-def}, the flow of $X_H$ 
preserves the closed exact 2-form $\omega$ for any Hamiltonian
$H$. This preservation may be verified by a formal calculation using
\eqref{HVF-def}. Along
$(dq/dt,dp/dt)=(\dot{q},\dot{p})=(H_p,-H_q)$, we have
\begin{align*}
\frac{d\omega}{dt} = d\dot{q}\wedge dp + dq\wedge d\dot{p}
&= dH_p\wedge dp - dq\wedge dH_q 
\\&= d(H_p\,dp +H_q\, dq) = d(X_H\contract\omega) =d(dH) = 0
\,.\end{align*}
The first step here uses the chain rule for differential forms and the third and last
steps use the property of the exterior derivative $d$ that
$d^2=0$ for continuous forms. The latter is due to equality of cross
derivatives $H_{pq}=H_{qp}$ and antisymmetry of the wedge product:
$dq\wedge dp = - dp\wedge dq$.

Consequently, the
relation $d(X_H\contract\omega)=d^2H=0$ for Hamiltonian vector fields
shows the following.

\begin{theorem}[Poincar\'e's theorem for one degree of freedom]\label{Pthm-1dof}
The flow of a Hamiltonian vector field is \emph{symplectic}, which means it
preserves the phase-space area, or two-form, $\omega=dq\wedge dp$. 
\end{theorem}

\begin{definition}[Cartan's formula for the Lie derivative]
The operation of \emph{Lie derivative} of a differential form $\omega$ by a vector
field $X_H$ is defined by
\begin{eqnarray}\label{LieDerivGeomDef}
\pounds_{X_H}\omega
:=
d(X_H\contract\omega)
+X_H\contract d\omega
\end{eqnarray}
\end{definition}

\begin{corollary}
Because $d\omega=0$, the symplectic property $d\omega/dt=d(X_H\contract\omega)=0$
in Poincar\'e's theorem \ref{Pthm-1dof} may be rewritten using Lie derivative
notation as
\begin{eqnarray}\label{symp-thm}
\frac{d\omega}{dt}
=
\pounds_{X_H}\omega
:=
d(X_H\contract\omega)
+X_H\contract d\omega
=0.
\end{eqnarray}
\end{corollary}

\begin{remark}\rm \quad

\begin{itemize}
\item
Relation \eqref{symp-thm} associates Hamiltonian dynamics with the symplectic
flow in phase space of the Hamiltonian vector field 
\begin{eqnarray}
X_H = \{\cdot, H\}
= \frac{\partial H}{\partial p}\frac{\partial }{\partial q}
-
\frac{\partial H}{\partial q}\frac{\partial }{\partial p}
\,,
\label{HamVecCharEqns}
\end{eqnarray}
 which is divergenceless with respect to the symplectic form $\omega=dq\wedge dp$.
 That is, $d(X_H\contract\omega)=0$.
\item
The Lie derivative operation defined in  \eqref{symp-thm}
is equivalent to the time derivative along the characteristic paths
(flow) of the first order linear partial differential operator $X_H$, which are
obtained from its characteristic equations in \eqref{HamVecCharEqns}. This is
the \emph{dynamical meaning} of the Lie derivative $\pounds_{X_H}$ in
\eqref{LieDerivGeomDef} for which invariance $\pounds_{X_H}\omega=0$ gives the
geometric definition of  symplectic flows in phase space.
\end{itemize}
\end{remark}

\begin{theorem}[Poincar\'e's theorem for $N$ degrees of freedom]
For a system of $N$ particles, or $N$ degrees of freedom,
the flow of a Hamiltonian vector field preserves each subvolume in the
phase space $T^*\mathbb{R}^N$. That is, let 
$\omega_n\equiv dq_n\wedge dp_n$ be the symplectic form expressed
in terms of the position and momentum of the $n$th particle. Then
\begin{eqnarray*}
\frac{d\omega_M}{dt} = 0 ,\quad\hbox{for}\quad
\omega_M = \Pi_{n=1}^M\omega_n ,\quad\forall M\le N .
\end{eqnarray*}
\end{theorem}

The proof of the preservation of these \emph{Poincar\'e invariants}
$\omega_M$ with $M=1,2,\dots,N$ follows the same pattern as the
verification above for a single degree of freedom. Basically, this is because each factor 
$\omega_n = dq_n\wedge dp_n$ in
the wedge product of symplectic forms is preserved by its
corresponding Hamiltonian flow in the sum
\begin{eqnarray*}
X_H &=& \sum_{n=1}^M\Big( \dot{q_n}\frac{\partial}{\partial q_n}
+\dot{p_n}\frac{\partial}{\partial p_n} \Big) 
\\&=&
\sum_{n=1}^M\big( H_{p_n}\partial_{q_n}-H_{q_n}\partial_{p_n} \big) =
\sum_{n=1}^MX_{H_n} =\{\cdot , H\}
\end{eqnarray*}
That is, $\pounds_{X_{H_n}}\omega_M$ vanishes for each term in the sum
\[\pounds_{X_H}\omega_M=\sum_{n=1}^M\pounds_{X_{H_n}}\omega_M\]
since $\partial_{q_m}\contract dq_n=\delta_{mn}
=\partial_{p_m}\contract dp_n$
and $\partial_{q_m}\contract dp_n=0
=\partial_{p_m}\contract dq_n$.


\subsection{Quick summary of coordinate-free formulas for differential forms}

\begin{theorem}
The \emph{pull-back} $\phi_t^*$ of a smooth flow $\phi_t$ generated by the action of a smooth vector field $X$ 
on a smooth manifold $M$ commutes with the exterior derivative $d$, wedge product $\wedge$ and contraction $\contract$. \medskip
\end{theorem}

In other words, for $k$-forms $\alpha,\,\beta\in\Lambda^k(M)$, for each point $m\in M$, the pull-back $\phi_t^*$ satisfies
\begin{eqnarray*}
d(\phi_t^*\alpha)
&=&
\phi_t^*d\alpha
\,,\\
\phi_t^*(\alpha\wedge\beta)
&=&
\phi_t^*\alpha\wedge\phi_t^*\beta
\,,
\\
\phi_t^*(X\contract \alpha)
&=&
\phi_t^*X\contract \phi_t^*\alpha
\,.
\end{eqnarray*}
Thus, the exterior derivative $d$, wedge product $\wedge$ and contraction $\contract$ are said to be 
\textit{natural} under pullback by a smooth flow generated by a smooth vector field. 

In addition, the \emph{Lie derivative} $\pounds_X\alpha$ of a $k$-form $\alpha\in\Lambda^k(M)$ by the vector field $X$ tangent to the flow $\phi_t$ on $M$ is defined either dynamically or geometrically (by Cartan's formula) as
\begin{eqnarray}
\pounds_X\alpha
&=&
 \frac{d}{dt}\bigg|_{t=0}(\phi_t^*\alpha)
=
X \contract d\alpha
+d( X \contract \alpha)
,
\label{Dyn+Geom-def-LieDeriv}
\end{eqnarray}
in which the last equality is \emph{Cartan's geometric formula} for the Lie derivative 
\index{Lie derivative!Cartan's geometric formula}
introduced in \eqref{LieDerivGeomDef}. 

\begin{definition}\hskip-2pt
\emph{(The Lie chain rule)}$\,$\\ 	\index{Lie chain rule}
The tangent to the pull-back $\phi_t^*\alpha$ of a differential $k$-form $\alpha\in \Lambda^k$ is the pull-back 
of the Lie derivative of $\alpha$ with respect to the vector field (e.g., $X=\dot{\phi}_t\phi_t^{-1}$ for right invariance) 
that generates the flow, $\phi_t$:
\begin{eqnarray*}
 \frac{d}{dt}(\phi_t^*\alpha)
 =
 \phi_t^*\big(\pounds_X\alpha\big)
 \,.
\end{eqnarray*}

Likewise, for the push-forward, which is the pull-back by the inverse, we have
\begin{eqnarray*}
 \frac{d}{dt}((\phi_t^{-1})^*\alpha)
 =
-\, (\phi_t^{-1})^*\big(\pounds_X\alpha\big)
 \,.
\end{eqnarray*}
\vskip-28pt
\end{definition}
%
\begin{definition}\hskip-2pt
\emph{(Advected quantity)}$\,$\\ 	\index{advected quantity} \index{Lie chain rule}
An advected quantity is invariant along a flow trajectory. Hence, advected quantities satisfy
\[
 \alpha_0(x_0) = \alpha_t(x_t) = (\phi_t^*\alpha_t)(x_0)
 \,,\ \hbox{or equivalently,} \
\alpha_t (x_t) = (\alpha_0 \circ \phi_t^{-1})(x_t)
= ((\phi_t)_*\alpha_0) (x_t)
\,.\]
The dynamics of an advected quantity is given by the Lie chain rule for the push-forward as
\begin{eqnarray}
 \frac{d}{dt}\alpha_t(x_t)
 =
 \frac{d}{dt}(\phi_t)_*\alpha_0
 =
- \pounds_X\alpha_t
 \,.
 \label{LieChainRule-push}
\end{eqnarray}
The Lie chain rule implies the same advection dynamics in terms of the pull-back, 
\[
0 =  \frac{d}{dt}\alpha_0(x_0)  =  \frac{d}{dt} (\phi_t^*\alpha_t)(x_0)
= \phi_t^* (\partial_t + \pounds_X)\alpha_t(x_0)
= (\partial_t + \pounds_X)\alpha_t(x_t)
\,.\]
\vskip-28pt
\end{definition}
%

\begin{remark}\rm \hskip-2pt
These formulas will enable us later to write a coordinate-free formulation of ideal fluid mechanics with advected quantities, by introducing the advection dynamics as a constraint on Hamilton's principle for fluids, \cite{HoMaRa1998a}.
\vskip-28pt
\end{remark}

\subsection{Quick summary of coordinate-free Hamiltonian mechanics}\label{CoordFree-Ham}\large
\index{Hamiltonian mechanics!coordinate-free}
	A coordinate-free definition of the Poisson bracket can be formulated in terms of the operations of the \emph{differential} (or exterior derivative),  \emph{insertion} or \emph{contraction}, and \emph{Lie derivative}, denoted respectively as
\begin{align*}
	&{\rm d}: \mathfrak{X}(T^*M)\times \Lambda^k  \to \Lambda^{k+1}
	\\&
	\contract (\hbox{or } \iota): \mathfrak{X}(T^*M)\times \Lambda^k  \to \Lambda^{k-1}
	\\&
	\pounds_{X} = d(\iota_X)+ \iota_{X}d:\ \mathfrak{X}(T^*M)\times \Lambda^k  \to \Lambda^{k}
	\,,\end{align*}
	where $k$ is an even number for the cotangent bundle $T^*M$ of a manifold $M$.
For example, in two dimensions, insertion of the vector field $X=X^j\partial_j=X^1\partial_1+X^2\partial_2$ into the two-form $\alpha = \alpha_{jk}dx^j\,\wedge \,dx^k$
with $\alpha_{21}=-\,\alpha_{12}$ yields
\[
X\contract \alpha
=
 \iota_{X}  \alpha 
 = X^j\alpha _{j{i_2}}dx^{i_2}
= X^1\alpha_{12}dx^2+X^2\alpha_{21}dx^1
= \alpha_{12}(X^1dx^2-X^2dx^1)
\,.
\]

Consequently, the Poisson bracket can be defined by the insertion of Hamiltonian vector fields 
\[
X_{F} := \{\,\cdot\,,\,F \,\}  \quad\hbox{and}\quad X_{H} := \{\,\cdot\,,\,H \,\}
\]
into the closed symplectic 2-form $\omega\in \Lambda^2(T^*M)$ with $d\omega=0$, as 
\[
\{F, H\} 
= \iota_{X_H}\big(\iota_{X_F}\omega(\,\cdot\,,\,\cdot\,)\big)
= \omega(X_F,X_H)
= X_H\contract \big(X_F\contract \omega(\,\cdot\,,\,\cdot\,)\big)
\,.\] This formula is related to the original phase space coordinates by, 
	\[
	\omega = \sum_{i=1}^2dq_i\wedge dp_i = - \sum_{i=1}^2 dp_i\wedge dq_i
	\]
	and
	\[ 
	X_H =\{\,\cdot\,,H\}
	=\sum_{i=1}^2 \frac{\partial H}{\partial p_i}\frac{\partial }{\partial q_i}-\frac{\partial H}{\partial q_i}\frac{\partial }{\partial p_i}
	\,.\]
	Notice that the calculation of $\omega(X_F,X_H)$ in these coordinates yields 
\[
\{F, H\} = \iota_{X_H}\big(\iota_{X_F}\omega(\,\cdot\,,\,\cdot\,)\big)
= \iota_{X_H}\omega(\,X_F\,,\,\cdot\,)
= \iota_{X_H}dF
= \omega(X_F,X_H)
\,.\] 	
Consequently, the dynamics along the integral curves of $X_H$ are determined by \[dH =\omega(X_H\,,\,\cdot\,) = \iota_{X_H} \omega\,,\] in which the vector field $X_H$ is inserted ($\iota$) into the symplectic 2-form $\omega$ to create the exact 1-form $dH$. In the original coordinates, this is
\[
dH  = \iota_{X_H}(dq\wedge dp)
= \iota_{(\frac{\partial H}{\partial p}\frac{\partial }{\partial q}-\frac{\partial H}{\partial q}\frac{\partial }{\partial p})}(dq\wedge dp)
= \frac{\partial H}{\partial q}dq + \frac{\partial H}{\partial p}dp
\,.\] 
One can also check that $\omega(X_F,X_H) = \{F, H\}$ directly, as
\begin{align*}
\frac{dF}{dt} &= \omega(X_F,X_H) 
\\&= \iota_{X_H} (\iota_{X_F} \omega) = \iota_{X_H} dF
\\&=\iota_{(\frac{\partial H}{\partial p}\frac{\partial }{\partial q}
-\frac{\partial H}{\partial q}\frac{\partial }{\partial p})}\Big(\frac{\partial F}{\partial q}dq + \frac{\partial F}{\partial p}dp\Big) 
\\&= \frac{\partial H}{\partial p}\frac{\partial F}{\partial q} - \frac{\partial H}{\partial q}\frac{\partial F}{\partial p}
\\&=\{F, H\} \,.
\end{align*}

	\begin{definition}[Cartan's geometric definition of the Lie derivative]
	The coordinate-free expression 
	\[
	\pounds_{X_F}\omega=d(\iota_{X_F}\omega)+ \iota_{X_F}d\omega
	=d(X_F\contract \omega)+ X_F\contract d\omega
	\]
	is Cartan's geometric definition of the \emph{Lie derivative} of the symplectic 2-form $\omega$ with respect to the Hamiltonian vector field $X_F=\{\,\cdot\,,\,F\,\}$. 
	\end{definition}
	
	\begin{lemma}
	Since the symplectic form $\omega$ is closed ($d\omega=0$) and $\iota_{X_F} \omega=dF$ for a Hamiltonian vector field $X_F$, we have 
	\[
	\pounds_{X_F}\omega=d(\iota_{X_F}\omega)=d^2F=0
	\,.\]
	This means the symplectic form $\omega$ is locally invariant under the Lie algebra actions of Hamiltonian vector fields.  
	\end{lemma}	
	
	\begin{definition}[Symplectic flow]
	The finite transformation $\phi_\epsilon$ generated by the left-invariant Hamiltonian vector field $X_F=\phi_\epsilon^{-1}\phi'_\epsilon|_{\epsilon=0}$ is called a \emph{symplectic flow}. 
	\end{definition}

	\begin{theorem}[Symplectic flows preserve the symplectic 2-form $\omega$]
	A smooth symplectic flow 
	\[ {\phi_\epsilon^{\sss X_F}}:=\textrm{exp}(\epsilon X_F) \]
	generated by a (time-independent) Hamiltonian vector field $X_F$ given by 
	\[X_F=\frac{d}{d\epsilon}\phi_\epsilon^{\sss X_F}\big|_{\epsilon=0}=\{\,\cdot\,,\,F\,\}
	\,,\] 
	with $\iota_{X_F} \omega=dF$, preserves the symplectic 2-form $\omega$ under \emph{pull-back} by the flow ${\phi_{\scaleto{\epsilon}{4pt}}^{\sss X_F}}=\textrm{exp}(\epsilon X_F) $, defined as
	\[
	{{\phi_{\scaleto{\epsilon}{4pt}}^{\sss X_F}}}^*\omega(q,p) 
	:= \omega(\phi_\epsilon^{\sss X_F} q,\phi_\epsilon^{\sss X_F} p)
	\,.\] 
	\end{theorem}

	\begin{proof} 
	\[
	\frac{d}{d \epsilon}({{\phi_{\scaleto{\epsilon}{4pt}}^{\sss X_F}}}^*\omega)
	={{\phi_{\scaleto{\epsilon}{4pt}}^{\sss X_F}}}^*(\pounds_{X_F}\omega)
	= {{\phi_{\scaleto{\epsilon}{4pt}}^{\sss X_F}}}^*( d(\iota_{X_F}\omega)+ \iota_{X_F}d\omega )
	={{\phi_{\scaleto{\epsilon}{4pt}}^{\sss X_F}}}^* d(dF)=0
	\,,\]
	since $d\omega=0$ and $d(\iota_{X_F}\omega)=d^2F=0$. 
	\end{proof} 
	
	\emph{Remark.} In the context of pull-back by smooth flows here, the proof uses the dynamic definition of the Lie derivative,
	\[
	\pounds_{X}\omega=\frac{d}{d \epsilon}({\phi_{\scaleto{\epsilon}{4pt}}}^*\omega)\Big|_{\epsilon=0}
	\quad\hbox{with}\quad
	X=\phi_\epsilon^{-1}\phi'_\epsilon|_{\epsilon=0}\,,
	\]
	in the first step. In the second step, the proof uses the equivalence of the dynamic and Cartan definitions of the Lie derivative with respect to vector fields.\bigskip
	
	\begin{exercise} Demonstrate the equivalence of the dynamic and Cartan definitions of the Lie derivative $\pounds_X$ by calculating their actions on scalar functions. How is this result related to the familiar directional derivative of a scalar function?
	\end{exercise}

\subsection{Right-invariant symmetry-reduced Hamilton principle }
The right-invariant form of the Euler--Poincar\'e equation \eqref{EP-eqn-intro} follows from a constrained Hamilton--Pontryagin variational principle. Namely,
\index{Hamilton--Pontryagin!variational principle}\index{variational principle!Hamilton--Pontryagin}
\begin{align}
\begin{split}
0 = \delta S &= \delta \int_0^T \ell(\zeta) + \scp{\mu}{\dot{g}g^{-1} - \zeta} dt
\\&=  \int_0^T \scp{ \frac{\p \ell}{\p \zeta} - \mu}{\delta \zeta} 
+ \scp{\mu}{\frac{d \varkappa}{dt} - \text{ad}_\zeta\varkappa} 
+ \scp{\delta \mu}{\dot{g}g^{-1} - \zeta} dt
\\&=  \int_0^T \scp{ \frac{\p \ell}{\p \zeta} - \mu}{\delta \zeta} 
- \scp{\frac{d \mu}{dt} + \text{ad}^*_\zeta \mu}{\varkappa} 
+ \scp{\delta \mu}{\dot{g}g^{-1} - \zeta} dt
+ \underbrace{\scp{\mu}{\varkappa}\big|_0^T}_{\scp{\mu}{\varkappa} =: J_R^\varkappa}
\,.
\end{split}
\label{sec6: HPP}
\end{align}
where $\zeta= \dot{g}g^{-1} $ and the quantity $\varkappa := \delta{g}g^{-1} $ vanishes 
at the endpoints in time and we have used the right-invariant version of the relation from \eqref{EP-Var-Id},
\begin{align}
\delta (\dot{g}g^{-1}) = \frac{d \varkappa}{dt} - \text{ad}_{\dot{g}g^{-1}}\varkappa
\,.
\label{sec6: EP-Var-Id}
\end{align}
Hence we recover the right-invariant version of Euler--Poincar\'e equation in \eqref{EP-eqn-intro},
\begin{align}
\frac{d}{dt}\frac{\p \ell}{\p \zeta}  + \text{ad}^*_\zeta \frac{\p \ell}{\p \zeta}  = 0
\quad\hbox{with}\quad
\mu =  \frac{\p \ell}{\p \zeta} 
\,.\label{sec6: EP-Var-Id}
\end{align}

The Legendre-transformed Lagrangian for the Euler--Poincar\'e equations is given by 
\begin{align}
\begin{split}
0 = \delta S &= \delta \int_0^T \ell(\zeta) + \scp{\mu}{\dot{g}g^{-1} - \zeta} dt
   = \delta \int_0^T \scp{\mu}{\zeta} - h(\mu) + \scp{\mu}{\dot{g}g^{-1} - \zeta} dt
\\&= \delta \int_0^T \scp{\mu}{\dot{g}g^{-1} } - h(\mu) \,dt
\\&=  \int_0^T  \scp{\mu}{\frac{d \varkappa}{dt} - \text{ad}_\zeta\varkappa} 
+ \scp{\delta \mu}{\dot{g}g^{-1} - \frac{\partial h}{\partial \mu}} dt
\\&=  \int_0^T - \scp{\frac{d \mu}{dt} + \ad^*_\zeta \mu }{\varkappa} 
+ \scp{\partial \mu}{\dot{g}g^{-1} - \frac{\partial h}{\partial \mu}} dt + \scp{\mu}{\varkappa}\big|_0^T
\,,
\end{split}
\label{sec6: HPP}
\end{align}
where the quantities $\zeta= \dot{g} g^{-1}$ and $\varkappa := \delta{g}g^{-1} $ both vanish 
at the endpoints in time and we have used the right-invariant version of the relation from \eqref{EP-Var-Id},
\begin{align}
\delta (g^{-1}\dot{g}) = \frac{d \varkappa}{dt} - \text{ad}_{\zeta}\varkappa
\,.
\label{sec6: EP-Var-Id}
\end{align}
Thus, we recover the right-invariant version of the Lie-Poisson Hamiltonian form 
of the Euler--Poincar\'e equation in \eqref{EP-eqn-intro},
\begin{align}
\frac{d \mu}{dt} + \text{ad}^*_{ \frac{\partial h}{\partial \mu} } \mu  = 0
\quad\hbox{where we have used}\quad
\zeta = \frac{\partial h}{\partial \mu} \in \mathfrak{g}\,\simeq\ TM/G
\,.\label{sec6: EP-Var-Id}
\end{align}
\begin{exercise}
State the right-invariant Noether's theorem for symmetries and conservation laws on both the Lagrangian and Hamiltonian sides. Prove your statements. 

Hint: $\scp{\mu}{\varkappa} =: J_R^\varkappa$ is the Hamiltonian for the right-invariant momentum map, $\mu :=  {\p \ell}/{\p \zeta} = {\p \ell}/{\p (\dot{g}g^{-1})}$ which maps $T^*M\to T^*M/G \simeq \mathfrak{g}^*$.
\end{exercise}

\begin{exercise}
Prove that equation \eqref{sec6: EP-Var-Id} preserves $\Ad^*_{g_t} \mu \,$ with $\zeta = \dot{g} g^{-1}$
by showing that 
\[
\frac{d}{dt}\big(\Ad^*_{g_t} \mu\big) = Ad^*_{g_t}\Big(\frac{d \mu}{dt} + \text{ad}^*_{ \zeta } \mu\Big) = 0\,.
\]
\end{exercise}

\begin{exercise}
What is the Poisson bracket for the dynamics of $\mu\in \mathfrak{g}^*\,\simeq\ T^*M/G$ in \eqref{sec6: EP-Var-Id}?
\end{exercise}
\begin{answer}
Write $\frac{d}{dt}f(\mu)=\scp{\frac{\p f}{\p \mu}}{\frac{d\mu}{dt}}$ in Poisson bracket form, as
\begin{align}
\begin{split}
\frac{d}{dt}f(\mu) &= \scp{\frac{\p f}{\p \mu}}{\frac{d\mu}{dt}}
= - \scp{\frac{\p f}{\p \mu}}{ \text{ad}^*_{ \frac{\partial h}{\partial \mu} } \mu}
\\&= - \scp{ \text{ad}_{ \frac{\partial h}{\partial \mu} }\frac{\p f}{\p \mu}}{ \mu}
 = \scp{ \Big[ \frac{\partial h}{\partial \mu} \,,\, \frac{\p f}{\p \mu}\Big] }{ \mu}
\\&= - \scp{ \mu}{ \Big[ \frac{\partial f}{\partial \mu} \,,\, \frac{\p h}{\p \mu}\Big] }
=: \big\{ f\,,\, h\big\}_{LPB}(\mu)
\,.\end{split}
\label{sec6: LPB}
\end{align}
\end{answer}

\begin{exercise}
Prove that the LPB  (Lie Poisson Bracket) is Poisson, 
i.e., prove that $\{ f\,,\, h\}_{LPB}$ satisfies the properties of a Poisson bracket. 
\end{exercise}

\begin{answer}
$\{ f\,,\, h\}_{LPB}$ being a linear functional of the Lie algebra $\mathfrak{g}$ makes it easy to verify the Poisson properties.
Upon defining $J^{\xi_k}(\mu)= \scp{\mu}{\xi_k}$, $k=1,2,3$, we have
\[
\big\{ J^{\xi_2}\,,\, J^{\xi_3}\big\} = -\,\scp{\mu}{\big[ \xi_2 \,,\,\xi_3\big] }
\]
and
\[
\big\{ J^{\xi_1}\,,\,\big\{ J^{\xi_2}\,,\, J^{\xi_3}\big\}\big\} = \scp{\mu}{\big[ \xi_1\,,\,\big[ \xi_2 \,,\,\xi_3\big] \big]}
.\]
Consequently, vanishing of the sum of cyclic permutations of $k=1,2,3$ in the vector field commutators 
$[ \xi_1\,,\,[ \xi_2 \,,\,\xi_3] ]$ implies that the sum of cyclic permutations of the Lie Poisson brackets of their 
Hamiltonians $\{ J^{\xi_1}\,,\,\{ J^{\xi_2}\,,\, J^{\xi_3}\}\}$ also vanishes, because the Hamiltonians are linear functionals of 
$\mu\in\mathfrak{g}^*$.
\end{answer}

\newpage
\vspace{4mm}\centerline{\textcolor{shadecolor}{\rule[0mm]{6.75in}{-2mm}}\vspace{-4mm}}
\section{More about vector fields \& differential forms}
\label{diff-form-review}

\secttoc

\textbf{What is this lecture about?} This lecture demonstrates how to perform 
in coordinates the various coordinate-free calculations for differential forms 
discussed in the previous lecture.

\begin{remark}\rm 
For a survey of the basic definitions, properties, and operations on
differential forms, as well as useful of tables of relations between
differential calculus and vector calculus, see, {for example},
Bloch \cite[Chapter~2]{Bl2004}.
\end{remark}

\subsection{Vector fields and $1$--forms}

Let $M$ be a manifold. In what follows, all maps may be assumed to be $C^\infty$, although that's
not always necessary.

A \emph{vector field} on $M$ is a map $X{:\ } M\to TM$ such that $X(x)\in T_xM$ for every $x\in M$.
The set of all smooth vector fields on $M$ is written $\mathfrak{X}(M)$.
(``Smooth'' means differentiable or $C^r$ for some $r\le \infty$, depending on context.)

A \emph{(differential) $1$--form} on $M$ is a map $\theta{:\ } M\to T^*M$ such that $\theta(x)\in T_x^*M$ for every $x\in M$.

More generally, if $\pi{:\ } E\to M$ is a bundle, then a \emph{section} of
the bundle is a map $\varphi{:\ } M\to E$
such that $\pi\circ \varphi(x)=x$ for all $x\in M$. So a vector field is a section of the tangent bundle,
while a $1$--form is section of the cotangent bundle.

Vector fields can added and also multiplied by scalar functions $k{:\ } M\to \R$, as follows:
$\left(X_1 + X_2\right) (x) = X_1(x) + X_2(x), \ 
\left(kX\right)(x) = k(x) X(x)$.

Differential forms can added and also multiplied by scalar functions $k{:\ } M\to \R$, as follows:
$\left(\alpha + \beta\right) (x) = \alpha(x) + \beta(x), \
\left(k\theta\right)(x) = k(x) \theta(x)$.

We have already defined the push-forward and pull-back of a vector field.
The \emph{pull-back} of a $1$--form $\theta$ on $N$ by a map $\varphi{:\ } M\to N$ is
the $1$--form $\varphi^* \theta$ on $M$ defined by 
\[
\left(\varphi^* \theta \right)(x) \cdot v = \theta\left(\varphi(x) \right) \cdot T\varphi(v)
\]
The \emph{push-forward} of a $1$--form $\alpha$ on $M$
by a diffeomorphism $\psi{:\ } M\to N$ is the pull-back of $\alpha$ by $\psi^{-1}$.

A vector field can be \emph{contracted} with a differential form, using the pairing between
tangent and cotangent vectors: $\left(X\contract \theta\right)(x) = \theta(x)\cdot X(x)$.
Note that $X\contract \theta$ is a map from $M$ to $\R$.
Many books write $i_X\theta$ in place of $X\contract \theta$, and the contraction
operation is often called \emph{interior product}.

The \emph{differential} of $f{:\ } M\to\R$ is a $1$--form $df$ on $M$ defined by
\[ df(x)\cdot v = \frac{d}{dt} f(c(t)\Big|_{t=0} \]
for any $x\in M$, any $v\in T_xM$ and any path $c(t)$ in $M$ such that
$c(0)=0$ and $c'(0)=v$. The left hand side, $df(x)\cdot v$, means the pairing
between cotangent and tangent vectors, which could also be written $df(x)(v)$ or 
$\left<df(x),v\right>$.

Note:
\[
X\contract df = \pounds_X f = X[f]
\]

\begin{remark}\rm  $df$ is very similar to $Tf$,
but $Tf$ is defined for all differentiable $f{:\ } M\to N$, whereas $df$ is only defined when $N=\R$
(in this course, anyway).
In this case, $Tf$ is a map from $TM$ to $T\R$, and
$Tf(v) = df(x)\cdot v \in T_{f(x)}\R$ for every $v\in T_xM$
(we have identified $T_f(x)\R$ with $\R$.) 
\end{remark}

\subsection*{In coordinates\ldots}
Let $M$ be $n$--dimensional, and 
let $x^1,\dots,x^n$ be differentiable local coordinates for $M$. This
means that there's an open subset $U$ of $M$ and an open subset $V$ of
$\R^n$ such that the map $\varphi{:\ } U\to V$ defined by $\varphi(x) = \left(x^1(x),\dots,x^n(x)\right)$ is a diffeomorphism. In particular, 
each $x^i$ is a map from $M$ to $\R$, 
so the differential $dx^i$ is defined.
There is also a vector field $\smash{\frac{\partial}{\partial x^i}}$ for every $i$,
which is defined by 
$\smash{\frac{\partial}{\partial x^i}(x)} = \smash{\frac{d}{dt}
\varphi^{-1} \big(\varphi(x) + t \mathbf{e}_i\big)\big|_{t=0}}$,
where $\mathbf{e}_i$ is the $i^{\textrm{th}}$ standard basis vector.

\begin{exercise} Verify that 
\[ \frac{\partial}{\partial x^i} \contract dx^j \equiv \delta^i_j \]
(where $\equiv$ means the left hand side is a constant function with value
$\delta^i_j$).
\end{exercise}

\begin{remark}\rm  Of course, given a coordinate system
$\varphi=\big(x^1,\dots,x^n\big)$, it is usual to write $x=
\big(x^1,\dots,x^n\big)$, which means $x$ is identified with
$\big(x^1(x),\dots,x^n(x)\big)=\varphi(x)$.
\end{remark}

For every $x\in M$, the vectors $\frac{\partial}{\partial x^i}(x)$ form
a basis for $T_xM$, so every $v\in T^xM$ can be uniquely expressed as
$v=v^i \frac{\partial}{\partial x^i}(x)$.  This expression defines the
\emph{tangent-lifted coordinates} $x^1,\dots,x^n,v^1,\dots v^n$ on $TM$
(they are local coordinates, defined on $TU\subset TM$).

For every $x\in M$, the covectors $dx^i(x)$ form a basis for $T_x^*M$,
so every $\alpha \in T^xM$ can be uniquely expressed as $\alpha=\alpha_i dx^i(x)$.
This expression defines the \emph{cotangent-lifted coordinates} $x^1,\dots,x^n,\alpha_1,\dots,\alpha_n$
on $T^*M$ (they are local coordinates, defined on $T^*U\subset T^*M$).

Note that the basis $\big(\frac{\partial}{\partial x^i}\big)$ is
dual to the basis $\big(dx^1,\dots,dx^n\big)$, by the previous exercise.
It follows that
\[
\big(\alpha_i dx^i\big) \cdot \bigg(v^i \frac{\partial}{\partial x^i}\bigg)
= \alpha_i v^i 
\]
(we have used the summation convention).

In mechanics, the configuration space is often called $Q$, and the lifted coordinates are written:
$q^1,\dots, q^n,\dot{q}^1,\dots,\dot{q}^n$ (on $TQ$) and 
$q^1,\dots, q^n,p_1,\dots,p_n$ (on $T^*Q$).

\subsubsection{Why the distinction between subscripts and superscripts?} This distinction keeps track
of how tensor quantities vary when coordinates are changed (see next exercise).
One benefit is that using the summation convention gives
coordinate-independent answers.

\begin{exercise}
Consider two sets of local coordinates $q^i$ and $s^i$ on $Q$, related by
$\big(s^1,\dots, s^n\big) = \psi \big(q^1,\dots, q^n\big)$.
Verify that the corresponding tangent lifted coordinates $\dot{q}^i$ and $\dot{s}^i$
are related by
\[
\dot{s}^i =\frac{\partial \psi^i}{\partial q^j} \dot{q}^j.
\]
Note that the last equation can be written as $\mathbf{\dot s} = D\psi(q) \mathbf{\dot{q}}$,
where $\mathbf{\dot s}$ is the column vector $(\dot s^1,\dots \dot s^n)$,
and similarly for $\mathbf{\dot q}$.

Perform the calculation corresponding to change of variables on the cotangent bundle side. See Definition
\ref{cot-lift}.
\end{exercise}

\subsection{The next level: $TTQ$, $T^*T^*Q$, et cetera}

Since $TQ$ is a manifold, we can consider vector fields on it,
which are sections of $T(TQ)$. In coordinates, every vector field on
$TTQ$ has the form $X=a^i \smash{\frac{\partial}{\partial q^i}} + b^i
\smash{\frac{\partial}{\partial \dot q^i}}$, where the $a^i$ and $b^i$
are functions of $q$ and $\dot q$.  Note that the same symbol $q^i$ has
two interpretations: as a coordinate on $TQ$ and as a coordinate on $Q$,
so $\smash{\frac{\partial}{\partial q^i}}$ can mean a vector field $TQ$
(as above) or on $Q$.

The tangent lift of the bundle projection $\tau{:\ } TQ\to Q$ is a map
$T\tau{:\ } TTQ\to TQ$. If $X$ is written in coordinates as above, then
$T\tau\circ X= a^i \frac{\partial}{\partial q^i}$. A vector field
$X$ on $TTQ$ is \emph{second order} if $T\tau \circ X (v) = v$; or
in coordinates, $a^i = \dot q^i$. The name comes from the process of
reducing of second order equations to first order ones by introducing
new variables $\dot q^i = \frac{d q^i}{dt}$.

One may also consider $T^*TQ$, $TT^*Q$ and $T^*T^*Q$. However, the
sub\-script/su\-per\-script distinction is problematic there. 

\subsubsection{$1$--forms}

The $1$--forms on $T^*Q$ are sections of $T^*T^*Q$. Given cotangent-lifted
local coordinates
\[\big(q^1,\dots,q^n,p_1,\dots,p_n\big)\] on $T^*Q$, the general
$1$--form on $T^*Q$ has the form $a_i dq^i + b_i dp_i$, where $a_i$
and $b_i$ are functions of $(q,p)$ and we sum repeated indices over their range.  
The \emph{canonical $1$--form}
on $T^*Q$ is
\[\theta = p_i dq^i,\] 
also written
in the short form $\scp{p}{dq}$. 
Pairing $\theta(q,p)$ with an arbitrary tangent vector  
$v= a^j \smash{\frac{\partial}{\partial q^j}} 
+ b^j \smash{\frac{\partial}{\partial p^j}}\in T_{(q,p)}T^*Q$ gives
\[\big<\theta(q,p),v\big> =
\bigg<p_i dq^i, a^j \frac{\partial}{\partial q^j} 
+ b^j \frac{\partial}{\partial p^j}\bigg>
=p_i a^j\delta_j^i
= \bigg<p_i dq^i, a^j\frac{\partial}{\partial q^j} \bigg>
,\]

\subsubsection{$2$--forms}

Recall that a $1$--form on $M$, evaluated at a point $x \in M$, is a
linear map from $T_xM$ to $\R$.

A $2$--form on $M$, evaluated at a point $x \in M$, is a skew-symmetric
bilinear form on $T_xM$; and the bilinear form has to vary smoothly as $x$
changes.  (Bilinear forms may be skew-symmetric, symmetric
or neither; \emph{differential} forms are assumed to be skew-symmetric.)

The \emph{pull-back} of a $2$--form $\omega$ on $N$ by a map
$\varphi{:\ } M\to N$ is the $2$--form $\varphi^* \omega$ on $M$ defined by 
\[
\left(\varphi^* \omega \right)(x) \left( v,w\right) =
\theta\left(\varphi(x) \right) 
\left( T\varphi(v), T\varphi(w)\right)
\]
The \emph{push-forward} of a $2$--form $\omega$ on $M$
by a diffeomorphism $\psi{:\ } M\to N$ is the pull-back of $\omega$ by
$\psi^{-1}$.

A vector field $X$ can be \emph{contracted} with a $2$--form $\omega$ to
get a $1$--form 
$X\contract \omega$ defined by
\[
\left(X\contract \omega\right) (x) (v) = \omega(x)\left(X(x),v\right)
\]
for any $v\in T_xM$.
A shorthand for this is $\left(X\contract \omega\right) (v)=
\omega(X,v)$, or just
$X\contract \omega = \omega(X,\cdot)$.

The \emph{tensor product} of two $1$--forms $\alpha$ and $\beta$ is the
$2$--form $\alpha \otimes \beta$ defined by
\[
\left(\alpha \otimes \beta\right)(v,w) = \alpha(v) \beta(w)
\]
for all $v,w\in T^*_xM$.

The \emph{wedge product} of two $1$--forms $\alpha$ and $\beta$ is the skew-symmetric $2$--form 
$\alpha \wedge \beta$ defined by
\[
\left(\alpha \wedge \beta\right)(v,w) = \alpha(v) \beta(w) - \alpha(w) \beta(v)  \ .
\]

{\subsubsection{Exterior derivative}

The differential $df$ of a real-valued function is also called the exterior derivative of $f$.
In this context, real-valued functions can be called $0$--forms.
The exterior derivative is a linear operation from 
$0$--forms to $1$--forms that satisfies the Leibniz identity, a.k.a. the product rule,
\[
d(fg) = f\ dg + g \ df
\]
The exterior derivative of a $1$--form is an alternating $2$--form, defined as follows:
\[
d\big(a_i dx^i\big) = \frac{\partial a_i}{\partial x^j} dx^j \wedge dx^i .
\]
Exterior derivative is a linear operation from $1$--forms to $2$ forms.
The following identity is easily checked:
\[
d(df) = 0
\] 
for all scalar functions $f$.

\subsubsection{$n$--forms}

See Marsden and Ratiu~\cite{MaRa1994}, Lee~\cite{Le2003}, or Abraham
and Marsden~\cite{AbMa1978}.  Unless otherwise specified, $n$--forms are
assumed to be alternating.  Wedge products and contractions generalise.

It is a fact that all $n$--forms are linear combinations of wedge products of $1$--forms.
Thus we can define exterior derivative recursively by the properties
\begin{align*}
d(\alpha \wedge \beta) = d\alpha \wedge \beta + (-1)^k \alpha \wedge d\beta,
\end{align*}
for all $k$--forms $\alpha$ and all forms $\beta$, and 
\[ d\circ d = 0 \]
In local coordinates, if $\alpha = \alpha_{i_1\cdots i_k} dx^{i_1}
\wedge \cdots \wedge dx^{i_k}$ (sum over all $i_1< \cdots < i_k$), then
\[ d\alpha = \frac{\partial \alpha_{i_1\cdots i_k}}{\partial x^j}  dx^j
\wedge dx^{i_1} \wedge \cdots \wedge dx^{i^k} \]
The \emph{Lie derivative} of an $n$--form $\theta$ in the direction of
the vector field $X$ is defined as
\begin{align} \pounds_X \theta = \frac{d}{dt} \varphi_t^* \theta \bigg|_{t=0}, 
\label{Def-LieDeriv}
\end{align}
where $\varphi_t$ is the flow of $X=\dot{\varphi_t}\varphi_t^{-1}$ for right-invariant vector fields. 
Formula \eqref{Def-LieDeriv} follows from the \emph{Lie chain rule}, \index{Lie chain rule}
\begin{align} 
\frac{d}{dt} \varphi_t^* \theta = \varphi_t^*\Big(\pounds_{\dot{\varphi_t}\varphi_t^{-1}} \theta \Big)
\label{Def-LieChain Rule}
\end{align}


\begin{remark}\rm 
Pull-back commutes with the operations $d,\contract,\wedge$ and Lie derivative.
Consequently, the Lie derivative satisfies the product rule for any of these pullback properties.
\end{remark}

\subsubsection{Cartan's magic formula}
\[
\pounds _X \alpha = d\left(X\contract \alpha\right) + X\contract d\alpha
\] 
This looks even more magic when written using the notation $i_X\alpha = X\contract \alpha$:
\[
\pounds_X = d i_X + i_X d
\]
An $n$--form $\alpha$ is \emph{closed} if $d\alpha=0$, and \emph{exact} 
if $\alpha = d\beta$ for some $\beta$. All exact forms are closed (since
$d\circ d = 0$), but the converse is false. It is true that all closed
forms are \emph{locally} exact; this is the \emph{Poincar\'e Lemma}.

\newpage
\vspace{4mm}\centerline{\textcolor{shadecolor}{\rule[0mm]{6.75in}{-2mm}}\vspace{-4mm}}
\section{Euler--Poincar\'e (EP) reduction theorem}

\secttoc

\textbf{What is this lecture about?} This lecture states and proves the EP reduction theorem, 
which proceeds by applying reduction by Lie symmetry to Hamilton's principle. 

\begin{remark}\rm [Geodesic motion] As emphasized by Arnold~\cite{Ar1966}, in
many interesting cases, the Euler--Poincar\'e equations on the dual of
a Lie algebra $ \mathfrak{g}^* $ correspond to \textit{ geodesic motion}
on the corresponding group $G$. The relationship between the equations on
$\mathfrak{g}^*$ and on $G$ is the content of the basic
Euler--Poincar\'e theorem discussed later. Similarly, on the
Hamiltonian side, the preceding paragraphs described the relation between
the Hamiltonian equations on $ T ^\ast G $ and the Lie--Poisson equations
on $
\mathfrak{g}^\ast $.  The issue of geodesic motion is especially simple:
if either the Lagrangian on $\mathfrak{g}$ or the Hamiltonian on
$\mathfrak{g}^\ast$ is purely quadratic, then the corresponding motion on
the group is geodesic motion. \index{Euler--Poincar\'e equations!geodesic motion}
\end{remark}

\subsection{We were already speaking prose (EP)}
Many of our previous considerations may be recast immediately as Euler--Poincar\'{e} equations.
\begin{itemize}
\item Rigid bodies $\simeq$ \big(EP $SO(n)$\big), 
\item Affine invariant motions $\simeq$ \big(EP $G(A) = GL(n,\mathbb{R})\circledS\mathbb{R}^n$\big),
\item Heavy tops $\simeq$ \big(EP $SO(3)\times \mathbb{R}^3$\big), 
\item EP Diff
\end{itemize}

\subsection{Euler--Poincar\'{e} reduction}
\label{Elpor}

This lecture applies reduction by symmetry
to Hamilton's principle. For a $G$--invariant Lagrangian defined on $TG$,
this reduction takes Hamilton's principle from $TG$ to
$TG/G\simeq\mathfrak{g}$. Stationarity of the symmetry-reduced Hamilton's
principle yields the Euler--Poincar\'e equations on $\mathfrak{g}^*$. The corresponding
reduced Legendre transformation yields the Lie--Poisson Hamiltonian formulation of
these equations. \index{Euler--Poincar\'e equations!Lie-Poisson form} 
\index{Lie--Poisson Hamiltonian formulation!Euler--Poincar\'e equations}

\emph{Euler--Poincar\'{e} Reduction\/} starts with a {right}
(respectively, {left}) invariant Lagrangian $L{:\ } TG\rightarrow
\mathbb{R}$ on the tangent bundle of a Lie group $G $. This means that $L
(T_g R_h(v)) = L(v)$, respectively $L (T_g L_h(v)) = L(v)$, for all $g, h
\in G $ and all $v\in T_gG$. In shorter notation, right invariance of
the Lagrangian may be written as 
\begin{equation}
L(g(t) ,\dot g(t))=L(g(t)h ,\dot g(t)h)
,
\label{R-inv-Lag}
\end{equation}
for all $h\in G$.

\begin{theorem}[Euler--Poincar\'{e} Reduction]
\label{EP theorem}
 Let $G$ be a Lie group, $L {:\ }  TG \rightarrow \mathbb{R}$
a {right}-invariant Lagrangian, and $l: = L|_{\mathfrak{g}} {:\ }  \mathfrak{g}
\rightarrow \mathbb{R}$ be its restriction to $\mathfrak{g}$. For a
curve $g(t)\in G$, let 
\[
\xi(t) =  \dot g(t)\cdot g^{-1}(t) := T_{g(t)}
R_{g^{-1}(t)} \dot g(t) \in \mathfrak{g}
.
\]
Then the following four statements are equivalent:
\begin{enumerate}
\item[(i)] $g(t)$ satisfies the Euler--Lagrange
equations for Lagrangian $L$ defined on $G$.
\item[(ii)] The variational principle
$$\delta \int_a^b L(g(t) ,\dot g(t)) dt = 0$$
holds, for variations with fixed endpoints.
\item[(iii)] The ({right} invariant) \emph{Euler--Poincar\'{e}
equations\/} hold: \index{Euler--Poincar\'e equations!right invariance}
$$\frac{d}{dt} \frac{\delta l}{\delta \xi} =- \ad^*_{\xi}
  \frac{\delta l}{\delta \xi}.$$
\item[(iv)] The variational principle
$$\delta \int_a^b l (\xi(t)) dt = 0$$
holds on $\mathfrak{g}$, using variations of the form $\delta \xi =
\dot \eta - [\xi , \eta ]$, where $\eta(t)$ is an arbitrary path in
$\mathfrak{g}$ which vanishes  at the endpoints, that is, $\eta (a) = \eta
(b) = 0$.
\end{enumerate}
\end{theorem}

\begin{proof} The proof consists of three steps.

\textbf{Step I: Proof that (i) $\Longleftrightarrow$ (ii)}{\quad} 
This is Hamilton's principle: the Euler--Lagrange equations follow
from stationary action for variations $\delta g$ which vanish at the
endpoints. (See \ref{sec:13}.)

\textbf{Step II: Proof that (ii) $\Longleftrightarrow$ (iv)}{\quad} 
Proving equivalence of the variational principles (ii) on $TG$
and (iv) on $\mathfrak{g}$ for a right-invariant Lagrangian
requires calculation of the variations $\delta\xi$ of
$\xi = \dot{g}g^{-1}$ induced by $\delta g$. To simplify the exposition,
the calculation will be done first for matrix Lie groups, then generalized
to arbitrary Lie groups.

\textbf{Step IIA: Proof that (ii) $\Longleftrightarrow$ (iv) for a matrix
Lie group}{{\quad} d} For $\xi =
\dot{g}g^{-1}$, define $g_\epsilon(t)$ to be a family of curves in $G$
such that $g_0(t) = g(t)$ and  denote 
\[\delta g := \frac{dg_{\epsilon}(t)}{d\epsilon}\bigg|_{\epsilon
=0\normalsize.}\] 
The variation of $\xi$ is computed in terms of $\delta g$ as
\begin{equation}\label{ep1}
\delta \xi = \frac{d}{d\epsilon}\bigg|_{\epsilon = 0}
(\dot g_\epsilon g_\epsilon^{-1}) = 
\frac{d^2 g}{dt d\epsilon}\bigg|_{\epsilon = 0}g^{-1}
-\dot{g} g^{-1} (\delta g) g^{-1}.
\end{equation}
Set $\eta := g^{-1} \delta g$. That is, $\eta(t)$ is an arbitrary
curve in $\mathfrak{g}$ which vanishes at the endpoints. The time
derivative of $\eta$ is computed as
\begin{equation}\label{ep2}
\dot\eta = \frac{d \eta}{dt} = \frac{d}{d t}\bigg(\bigg(
\frac{d}{d\epsilon}\bigg|_{\epsilon = 0} g_\epsilon 
\bigg)g^{-1}\bigg) 
= \frac{d^2 g}{dt d\epsilon}\bigg|_{\epsilon = 0}g^{-1}
- (\delta g) g^{-1} \dot{g} g^{-1}.
\end{equation}
Taking the difference of \eqref{ep1} and \eqref{ep2} implies 
$$\delta \xi - \dot\eta = -\dot{g} g^{-1} (\delta g) g^{-1}
+ (\delta g) g^{-1} \dot{g} g^{-1}  = -\xi \eta + \eta \xi = -[\xi, \eta ].$$
That is, for matrix Lie algebras,
\[\delta \xi = \dot \eta - [\xi ,\eta ],\]
where $[\xi ,\eta ]$ is the matrix commutator.
Next, we notice that {right} invariance of $L$ allows one to change
variables in the Lagrangian by applying $g^{-1}(t)$ from the {right}, as
\[
L(g(t) ,\dot g(t))
=
L(e , \dot g(t)g^{-1}(t))
=: 
l (\xi(t))
.
\]
Combining this definition of the symmetry-reduced Lagrangian
$l{:\ } \mathfrak{g}\to\mathbb{R}$ together with the formula for variations 
$\delta \xi$ just deduced proves the equivalence of (ii) and
(iv) for matrix Lie groups.

\textbf{Step IIB: Proof that (ii) $\Longleftrightarrow$ (iv) for an
arbitrary Lie group}{\quad} 
The same proof extends to any Lie group $G$ by using the following lemma.

\begin{lemma}
\label{mixed partials lemma}
Let $g {:\ }  U \subset \mathbb{R}^2 \rightarrow G $ be a smooth map and
denote its partial derivatives by
\begin{equation}
\label{partial derivatives of g}
\xi(t, \varepsilon): = T_{g(t, \varepsilon)} 
R_{g(t,\varepsilon)^{-1}}
\frac{\partial g(t,\varepsilon)}{\partial t}, \qquad
\eta(t, \varepsilon): = T_{g(t, \varepsilon)} 
R_{g(t,\varepsilon)^{-1}}
\frac{\partial g(t,\varepsilon)}{\partial \varepsilon}.
\end{equation}
Then
\begin{equation}
\label{mixed partials}
\frac{\partial \xi}{\partial \varepsilon} - \frac{\partial
\eta}{\partial t} = -[ \xi, \eta],
\end{equation}
where $[ \xi, \eta]$ is the Lie algebra bracket on $\mathfrak{g}$. 
Conversely, if $U \subset \mathbb{R}^2 $ is simply connected and
$\xi, \eta{:\ } U \rightarrow \mathfrak{g}$ are smooth functions
satisfying \eqref{mixed partials}, then there exists a smooth
function $g {:\ }  U \rightarrow G $ such that  \eqref{partial
derivatives of g} holds.
\end{lemma}

\begin{proof}[Proof of \ref{mixed partials lemma}]
Write $\xi=\dot{g}g^{-1}$ and $\eta=g^\prime g^{-1}$ in natural notation
and express the partial derivatives $\dot{g}=\partial g/\partial t$ and
$g^\prime=\partial g/\partial \epsilon$ using the right translations as 
\[ \dot{g}=\xi\circ{g} \quad\hbox{and}\quad {g}^\prime=\eta\circ{g}. \]
By the chain rule, these definitions have mixed partial derivatives
\[ \dot{g}^\prime=\xi^\prime=\nabla\xi\cdot\eta
  \quad\hbox{and}\quad \dot{g}^\prime=\dot{\eta}=\nabla\eta\cdot\xi. \]
The difference of the mixed partial derivatives implies the desired
formula \eqref{mixed partials},
\[ \xi^\prime-\dot{\eta}=\nabla\xi\cdot\eta-\nabla\eta\cdot\xi
= -\,[ \xi, \eta] = -\,\ad_{\xi} \eta. \]
(Note the minus sign in the last two terms.)
\end{proof}

\textbf{Step III: Proof of equivalence (iii) $\Longleftrightarrow$ (iv)}
Let us show that the reduced variational principle produces the
Euler--Poincar\'e equations. \index{Euler--Poincar\'e equations!reduced variational principle}
\index{reduced variational principle!Euler--Poincar\'e equations}
We write the functional derivative of the
reduced action $S_{\mathrm{red}}=\smash{\int_a^b} l(\xi)\,dt$ with Lagrangian $l(\xi)$ in
terms of the natural pairing 
$\langle \cdot, \cdot\rangle$ between $\mathfrak{g}^\ast $
and
$\mathfrak{g}$ as
\begin{align}
\begin{split}
\delta \int_a^b l(\xi(t)) dt &=  \int_a^b \left\langle
\frac{\delta l}{\delta \xi}, \delta\xi\right\rangle dt
=  \int_a^b \left\langle \frac{\delta l}{\delta \xi} , \dot \eta -
\ad_{\xi} \eta\right\rangle dt\\
&=  \int_a^b \left\langle
\frac{\delta l}{\delta \xi}, \dot \eta \right\rangle dt -  \int_a^b
\left\langle \frac{\delta l}{\delta \xi} , \ad_{\xi}
\eta\right\rangle dt \\
& = -\int_a^b\, \left\langle 
 \frac{d}{dt} \frac{\delta l}{\delta \xi}
+ \ad^*_{\xi} \frac{\delta l}{\delta \xi},  \eta \right\rangle\,dt 
+ \left\langle \frac{\delta l}{\delta \xi}\,,\, \eta \right\rangle\bigg|_a^b
\\
& = -\int_a^b\, \left\langle 
 \frac{d}{dt} \frac{\delta l}{\delta \xi}
+ \ad^*_{\xi} \frac{\delta l}{\delta \xi},  \eta \right\rangle\,dt \,.
\end{split}
\label{proof-EPeqn}
\end{align}
The last equality follows from integration by parts and
vanishing of the variation $\eta(t)$  at the endpoints. Thus,
stationarity $\delta \smash{\int_a^b} l(\xi(t)) dt = 0$ for any $\eta(t) $ that
vanishes at  the endpoints is equivalent to
\begin{align}
\frac{d}{dt} \frac{\delta l}{\delta \xi} 
= -\ad^*_{\xi}\,\frac{\delta l}{\delta \xi},
\label{EPeqn-NothingAdvected}
\end{align}
which are the Euler--Poincar\'e equations arising from right-invariance of the \\Lagrangian in the Hamilton's principle for 
ideal fluid dynamics. \index{Euler--Poincar\'e equations!right invariance}
\end{proof}

\begin{exercise}\textit{Noether's theorem for Euler--Poincar\'e fluid equations}.\label{Ex Relabel Momap}\\
State and prove Noether's theorem arising from the right-invariance of the Lagrangian in equation \eqref{R-inv-Lag} under the relabelling transformation $g\to gh$ of the particle labels. 
\end{exercise}

\begin{answer}
An infinitesimal Lie symmetry transformation $\delta Q^A$ of the Lagrangian parcel labels $Q^A(\bx,t)$ in Euclidean space $\mathbb{R}^3$ with components $A=1,2,3$ obeying the Eulerian advection equation 
\[
\partial _t Q^A + \bu\cdot \nabla Q^A = 0\,,
\]
can be written as $\delta Q^A = v(Q^A)= \bv(\bx)\cdot \nabla Q^A$ for each component $A=1,2,3$ and a smooth divergence-free vector field $v$ with Euclidean components $v=\bv\cdot \nabla$. The corresponding momentum map is given by $L^2$ pairing 
\begin{align*}
\Big\langle P_A\,,\, v(Q^A) \Big\rangle_{L^2} 
&= \int_{\mathbb{R}^3} (P_A\nabla Q^A) \cdot \bv\,d^3x
= \int_{\mathbb{R}^3} v \contract (P_A\nabla Q^A\cdot d\bx) \otimes d^3x
\\&= \int_{\mathbb{R}^3} v \contract (P_A dQ^A) \otimes d^3x
=: \Big\langle P_A dQ^A\,,\, v \Big\rangle_{L^2}
\,.
\end{align*}
This is the momentum map for invariance of the Lagrangian under relabelling of the parcel labels $Q^A$ with $A=1,2,3$ in Eucldean coordinates.  
\end{answer}

\begin{exercise}\textit{The momentum map for an infinitesimal transformation of the velocity}.
What is the momentum map associated with an infinitesimal action of a diffeomorphism on the velocity vector field, $\xi$? 
Hint: look at the second-to-last line of equation \eqref{proof-EPeqn} in the proof of the Euler--Poincar\'e equation.
\end{exercise}

\begin{answer}
The endpoint term in the second-to-last line of equation \eqref{proof-EPeqn} in the proof of the Euler--Poincar\'e equation provides a hint because an infinitesimal transformation of the velocity may be written as $\eta:=-\pounds_v\xi$, so that 
\[
\left\langle \frac{\delta l}{\delta \xi}\,,\, \eta \right\rangle 
=
\left\langle \frac{\delta l}{\delta \xi}\,,\, -\pounds_v\xi \right\rangle 
=
\left\langle \frac{\delta l}{\delta \xi}\,,\, -\ad_v\xi \right\rangle 
=
\left\langle \frac{\delta l}{\delta \xi}\,,\, \ad_\xi v \right\rangle 
\]
when the variational vector field is defined as the infinitesimal Lie symmetry transformation for generated by the vector field $v$ for right-invariance of the fluid Lagrangian under relabelling of fluid parcels. With this formula in mind, we may define the diamond operation $(\diamond)$ as 
\begin{align*}
\left\langle \frac{\delta l}{\delta \xi}\,,\, -\pounds_v\xi  \right\rangle_{T^*G\times TG}
&=
\left\langle  \frac{\delta l}{\delta \xi}\,,\, \ad_\xi v  \right\rangle_{T^*G\times TG}
=
\left\langle \ad^*_\xi \frac{\delta l}{\delta \xi}\,,\, v \right\rangle_{\mathfrak{g}^* \times \,\mathfrak{g}}
\\&=
\left\langle \pounds_\xi \frac{\delta l}{\delta \xi}\,,\, v \right\rangle_{\mathfrak{g}^* \times \,\mathfrak{g}}
=:
\left\langle \frac{\delta l}{\delta \xi}\diamond \xi \,,\, v \right\rangle_{\mathfrak{g}^* \times \,\mathfrak{g}}
\,,\end{align*}
where the second pairing $\langle\,\cdot\,,\cdot\,\rangle_{\mathfrak{g}^* \times \,\mathfrak{g}}$ is between vector fields in $\mathfrak{g}$ and their $L^2$-dual quantities, the 1-form densities in $\mathfrak{g}^*$. 

\end{answer}

\begin{remark}\rm [{Left}-invariant ] \index{Euler--Poincar\'e equations!left invariance}
The same theorem holds for {left} invariant Lagrangians on $TG$, except
for a sign in the Euler--Poincar\'e equation, cf. \eqref{EPeqn-NothingAdvected},
$$
\frac{d}{dt} \frac{\delta l}{\delta \xi} = +\, \ad^*_{\xi}
\frac{\delta l}{\delta \xi}
,
$$
which arises because {left}-invariant variations satisfy 
$\delta \xi = \dot\eta + [\xi ,\ \eta ]$ (with the opposite sign).
\end{remark}

\begin{exercise}
Write out the corresponding proof of the Euler--Poincar\'e reduction
theorem for left-invariant Lagrangians defined on the tangent space $TG$
of a group $G$.
\end{exercise}

\subsubsection{Reconstruction} \index{reconstruction}
The procedure for reconstructing the solution
$v(t)\in T_{g(t)}G$ of the  Euler--Lagrange equations with initial
conditions $g(0) = g_0$ and $\dot{g}(0) = v_0 $ starting from the solution
of  the Euler--Poincar\'{e} equations is as follows. First, solve the
initial value problem for the {right}-invariant Euler--Poincar\'{e}
equations:
\begin{eqnarray*}
\frac{d}{dt} \frac{\delta l}{\delta \xi} 
= -\ad^*_{\xi} \frac{\delta l}{\delta \xi}
\quad\hbox{with}\quad
\xi(0) = \xi_0 : =  v_0 g_0^{-1}
\end{eqnarray*}
Then from the solution for $\xi(t)$ reconstruct the curve $g(t)$ on the
group by solving the ``linear differential equation with time-dependent
coefficients"
\[
\dot g(t) = \xi(t) g(t) 
\quad\hbox{with}\quad
g(0) = g_0.
\]
The Euler--Poincar\'{e} reduction theorem guarantees then that $v(t) =
\dot g(t) =  \xi(t) \cdot g(t) $ is a solution of the Euler--Lagrange
equations with initial condition $v_0 = \xi_0 g_0$.

\begin{remark}\rm 
Similar statements hold, with obvious changes for {left}-invariant
Lagrangian systems on $TG $.
\end{remark}

\subsection{Reduced Legendre transformation}
As in the equivalence relation between the Lagrangian and Hamiltonian
formulations discussed earlier, the relationship between
symmetry-reduced Euler--Poincar\'{e} and Lie--Poisson formulations is
determined by the Legendre transformation. 

\begin{definition}
The symmetric-reduced Legendre transformation 
$\mathbb{F}l {:\ } \mathfrak{g}\rightarrow\mathfrak{g}^*$ is
defined  by
$$
\mathbb{F} l (\xi) = \frac{\delta l}{\delta \xi} = \mu \,.
$$
\end{definition}

\subsubsection{Lie--Poisson Hamiltonian formulation} \index{Hamiltonian formulation!Lie--Poisson}
Let $h(\mu) := \langle \mu , \xi\rangle - l(\xi)$.
Assuming that $\mathbb{F} l$ is a diffeomorphism yields
$$\frac{\delta h}{\delta \mu} = \xi + \left\langle \mu, 
\frac{\delta
\xi}{\delta
\mu} \right\rangle -  \left\langle \frac{\delta l}{\delta \xi}, 
\frac{\delta \xi}{\delta
\mu} \right\rangle  = \xi .
$$
So the Euler--Poincar\'{e} equations for $l$ are equivalent to the
Lie--Poisson equations for $h$:
$$\frac{d}{dt} \left( \frac{\delta l}{\delta \xi}\right) = -{\mathrm ad}^*_{\xi}
\frac{\delta l}{\delta \xi} \Longleftrightarrow \dot \mu = -{\mathrm ad}^*_{\delta h/\delta \mu} \mu.
$$
The Lie--Poisson equations may be written in the Poisson bracket
form
\begin{equation} \label{lppoisson}
\dot{f} = \left\{ f, h \right\},
\end{equation}
where $f{:\ }  \mathfrak{g}^\ast \rightarrow \mathbb{R}$ is
an arbitrary smooth function and the bracket is the
(right) Lie--Poisson bracket given by
\begin{equation} 
\{f, h\}(\mu )  
=  \left\langle \mu , \left[ \frac{ \delta f}{\delta  \mu},
\frac{\delta  h}{\delta \mu } \right] \right\rangle 
=
-
\left\langle \mu , 
\operatorname{ad}_{\delta h/\delta \mu}
\frac{ \delta f}{\delta  \mu}\right\rangle
=
-
\left\langle \operatorname{ad}^*_{\delta h/\delta \mu}\mu , 
\frac{ \delta f}{\delta  \mu}\right\rangle
.
\end{equation} 
In the important case when $\ell$ is quadratic, the Lagrangian $L $
is  the quadratic form associated to a right invariant Riemannian
metric on $G$. In this case, the Euler--Lagrange equations for $L$
on $G$ describe geodesic motion relative to this metric and these
geodesics are then equivalently described by either the
Euler--Poincar\'e, or the Lie--Poisson equations.


\begin{exercise}
Compute the pure EP equations for geodesic motion on $SE(3)$. These equations turn out to be applicable to the motion of an ellipsoidal body through a fluid. 
\end{exercise}

\newpage

\section{EPDiff: \\ An Euler--Poincar\'e equation on the diffeomorphisms}

\secttoc

\textbf{What is this lecture about?} This lecture lays out the coordinate form (including tensor indices) 
arising in applying the Euler--Poincar\'e theorem for diffeomorphisms.

\subsection{The $n$--dimensional EPDiff equation}
Eulerian geodesic motion of a fluid in $n$ dimensions is
generated as an EP equation via Hamilton's principle, when the 
Lagrangian is given by the kinetic energy. The kinetic energy defines a
norm $\|\mathbf{u}\|^2$ for the Eulerian fluid velocity, taken as 
$\mathbf{u}(\mathbf{x},t){:\ } \mathbb{R}^n\times \mathbb{R}^1 \to \mathbb{R}^n$. The
choice of the kinetic energy as a positive functional of fluid velocity
$\mathbf{u}$ is a modeling step that depends upon the physics of the
problem being studied. We shall choose the Lagrangian,
\begin{equation}\label{Lag-ansatz-1}
\|\mathbf{u}\|^2 
= \int \mathbf{u}\cdot Q_{\mathrm{op}}\mathbf{u}\,d^nx
= \int \mathbf{u}\cdot\mathbf{m}\,d^nx
,\end{equation}
so that the positive-definite, symmetric, operator $Q_{\mathrm{op}}$ defines the
norm $\|\mathbf{u}\|$, for appropriate (homogeneous, say, or periodic)
boundary conditions.  The EPDiff equation is the Euler--Poincar\'e
equation for this Eulerian geodesic motion of a fluid. Namely, 
\begin{equation}\label{EP-eqn-abstract}
\frac{d}{dt}\frac{\delta \ell}{\delta \mathbf{u}}
+
\ad^*_{\mathbf{u}}
\frac{\delta \ell}{\delta \mathbf{u}}
=
0
,
\quad\hbox{with}\quad
\ell[\mathbf{u}]=\tfrac{1}{2}\|\mathbf{u}\|^2
.
\end{equation}
Here $\ad^*$ is the dual of the vector-field ad-operation (the
commutator) under the natural $L^2$ pairing
$\langle\cdot, \cdot\rangle$ induced by the variational derivative
$\delta\ell[\mathbf{u}]=\langle\delta\ell/\delta\mathbf{u}\,
, \delta\mathbf{u}\rangle$. This pairing provides the definition of
$\ad^*$,
\begin{equation}\label{ad*-eqn}
\langle\ad^*_{\mathbf{u}}\,\mathbf{m}, \mathbf{v}\rangle
=
-\langle\mathbf{m}, \ad_{\mathbf{u}}\mathbf{v}\rangle
,
\end{equation}
where $\mathbf{u}$ and $\mathbf{v}$ are vector fields, ${\mathrm ad}_{\mathbf{u}}\mathbf{v}
=[\mathbf{u},\mathbf{v}]$ is the commutator,
that is,
the \emph{Lie bracket\/} given in
components by (summing on repeated indices)
\begin{equation}\label{jlb-ad} [\mathbf{u}, \mathbf{v}]^i
= 
u^j\frac{\partial v^i}{\partial x^j} 
 - 
v^j\frac{\partial u^i}{\partial x^j}
,\quad\hbox{or}\quad
[\mathbf{u}, \mathbf{v}]
=
\mathbf{u}\cdot\nabla \mathbf{v}
-
\mathbf{v}\cdot\nabla \mathbf{u}
.
\end{equation}
The notation $
\operatorname{ad}_{\mathbf{u}} \mathbf{v} := [\mathbf{u}, \mathbf{v}]$
formally denotes the adjoint action of the \textit{ right\/} Lie algebra of
$\operatorname{Diff}(\mathcal{D})$ on itself,
and  $\mathbf{m}=\delta\ell/\delta\mathbf{u}$ is the fluid momentum, a
\emph{one-form density} whose co-vector components are also denoted as
$\mathbf{m}$. 

If $\mathbf{u} = u^j \partial/\partial x^j, 
\mathbf{m}  = m_i dx^i\otimes  dV$, then the preceding formula for $
\operatorname{ad}^\ast_{\mathbf{u}}(\mathbf{m}\otimes  dV)$ has the \emph{coordinate expression} in $\mathbb{R} ^n$,
\begin{eqnarray}\label{continuumcoadjoint-coord-form}
\big(\operatorname{ad}^\ast_{\mathbf{u}}\mathbf{m}\big)_i
dx^i\otimes  dV
&=&
\left (\frac{\partial}{\partial x^j}(u^jm_i) +
m_j \frac{\partial u^j}{\partial x^i}\right ) dx^i\otimes  dV
.
\end{eqnarray}
In this notation,  the abstract EPDiff equation \eqref{EP-eqn-abstract}
may be written explicitly in Euclidean coordinates as a partial
differential equation for a co-vector function
$$\mathbf{m}(\mathbf{x},t){:\ } R^{n}\times R^1 \to R^{n}.$$
Namely, 
\begin{equation}\label{ep-eqn-coord-form}
\frac{\partial }{\partial t}\mathbf{m} +
\underbrace{\mathbf{u}\cdot\nabla \mathbf{m}}_{\hbox{Convection}} +
\underbrace{\nabla \mathbf{u}^T\cdot\mathbf{m}}_{\hbox{Stretching}} + 
\underbrace{\mathbf{m}(\div\mathbf{u})}_{\hbox{Expansion}} =0,
\quad\text{with}\quad
\mathbf{m}= \frac{\delta \ell}{\delta \mathbf{u}}
=Q_{\mathrm{op}}\mathbf{u}
.
\end{equation}
To explain the terms in underbraces, we rewrite EPDiff as preservation of the one-form
density of momentum along the characteristic curves of the velocity. Namely, 
\begin{equation}\label{EPDiff-char-form}
\frac{d}{dt}\big(\mathbf{m}\cdot d\mathbf{x}\otimes dV\big)=0
\quad\hbox{along}\quad
\frac{d\mathbf{x}}{dt}=\mathbf{u}=G*\mathbf{m}.
\end{equation} 
This form of the EPDiff equation also emphasizes its nonlocality, since the velocity is
obtained from the momentum density by convolution against the Green's function $G$ of the
operator $Q_{\mathrm{op}}$. Thus, $\mathbf{u}=G*\mathbf{m}$ with
$Q_{\mathrm{op}}G=\delta(\mathbf{x})$, the
Dirac measure. We may check that this ``characteristic form'' of EPDiff
recovers its Eulerian form by computing directly,
\begin{align*}
\frac{d}{dt}\big(\mathbf{m}\cdot d\mathbf{x}\otimes dV\big)
&=
\frac{d\mathbf{m}}{dt}\cdot d\mathbf{x}\otimes dV
+
\mathbf{m}\cdot d\frac{d\mathbf{x}}{dt}\otimes dV
+
\mathbf{m}\cdot d\mathbf{x}\otimes \bigg(\frac{d}{dt}dV\bigg) \\
&\qquad\qquad\text{along }
\frac{d\mathbf{x}}{dt}=\mathbf{u}=G*\mathbf{m} \\
&\hspace{-10mm}=
\bigg(\frac{\partial }{\partial t}\mathbf{m} + 
\mathbf{u}\cdot\nabla \mathbf{m} + \nabla \mathbf{u}^T\cdot\mathbf{m} + 
\mathbf{m}(\div\mathbf{u}) \bigg)\cdot d\mathbf{x}\otimes dV =0.
\end{align*}

\begin{exercise}
Show that EPDiff may be written as 
\begin{equation}\label{EPDiff-Lie-form}
\Big(\frac{\partial}{\partial t}+\mathcal{L}_{\mathbf{u}}\Big)
\Big(\mathbf{m}\cdot d\mathbf{x}\otimes dV\Big)=0
,
\end{equation} 
where $\mathcal{L}_{\mathbf{u}}$ is the Lie derivative with respect to the vector field 
with components $\mathbf{u}=G*\mathbf{m}$. Hint: How does this Lie-derivative form of
EPDiff in \eqref{EPDiff-Lie-form} differ from its characteristic form
\eqref{EPDiff-char-form}?
\end{exercise}

EPDiff may also be written equivalently in terms of the operators div, grad and curl in
2D and 3D as
\begin{equation}\label{div-grad-curl-eqn}
\frac{\partial }{\partial t}\mathbf{m} 
- \mathbf{u}\times\curl\,\mathbf{m} 
+ \nabla (\mathbf{u}\cdot\mathbf{m}) 
+ \mathbf{m}(\div\,\mathbf{u})
=0.
\end{equation}
Thus, for example, its numerical solution would require an algorithm which has the
capability to deal with the distinctions and relationships among the operators div, grad
and curl.

\subsection{Deriving $n$--dimensional EPDiff equation as geodesic flow}
Let's derive the EPDiff equation \eqref{ep-eqn-coord-form} by following the proof of the
EP reduction theorem leading to the Euler--Poincar\'e equations for right invariance in
the form \eqref{EP-eqn-abstract}. Following this calculation for the present
case yields \index{Euler--Poincar\'e equations!right invariance} 
\begin{align*}
\delta \int_a^b l(\mathbf{u}) dt 
&=  \int_a^b \left\langle
\frac{\delta l}{\delta \mathbf{u}}, \delta\mathbf{u}\right\rangle dt
=  \int_a^b \left\langle 
  \frac{\delta l}{\delta \mathbf{u}} , \mathbf{\dot{v}} 
-
\ad_{\mathbf{u}} \mathbf{v}\right\rangle dt \\
&= \int_a^b \left\langle
\frac{\delta l}{\delta \mathbf{u}}, \mathbf{\dot{v}} \right\rangle dt -  
\int_a^b \left\langle \frac{\delta l}{\delta \mathbf{u}} , \ad_{\mathbf{u}}
\mathbf{v}\right\rangle dt 
\\&= -\int_a^b\, \left\langle 
  \frac{d}{dt} \frac{\delta l}{\delta \mathbf{u}} + 
\ad^*_{\mathbf{u}} \frac{\delta l}{\delta \mathbf{u}},  \mathbf{v} \right\rangle\,
dt,
\end{align*}
where $\langle\cdot, \cdot\rangle$ is the pairing between elements of
the Lie algebra and its dual. In our case, this is the $L^2$ pairing, for
example,
\[ \left\langle
\frac{\delta l}{\delta \mathbf{u}}, \delta\mathbf{u}\right\rangle =
\int \frac{\delta l}{\delta {u}^i}\,\delta {u}^i\,d^nx \]
This pairing allows us to compute the coordinate form of the EPDiff equation explicitly,
as
\begin{align*}
\int_a^b \left\langle
\frac{\delta l}{\delta \mathbf{u}}, \delta\mathbf{u}\right\rangle dt
&=
\int_a^b dt 
\int \frac{\delta l}{\delta {u}^i}
\bigg(
\frac{\partial v^i}{\partial t}
+
u^j\frac{\partial v^i}{\partial x^j} 
 - 
v^j\frac{\partial u^i}{\partial x^j}
\bigg)
d^nx\\
&= - \int_a^b dt \int \bigg\{
\frac{\partial }{\partial t}\frac{\delta l}{\delta {u}^i} +
\frac{\partial }{\partial x^j}\bigg( \frac{\delta l}{\delta {u}^i}u^j
\bigg) +
\frac{\delta l}{\delta {u}^j}\frac{\partial u^j}{\partial x^i} \bigg\} v^i
\,d^nx
\end{align*}
Substituting $\mathbf{m}=\delta l/\delta \mathbf{u}$ now recovers the coordinate forms for
the coadjoint action of vector fields in \eqref{continuumcoadjoint-coord-form} and the
EPDiff equation itself in \eqref{ep-eqn-coord-form}. When
$\ell[\mathbf{u}]=\frac{1}{2}\|\mathbf{u}\|^2$, EPDiff describes geodesic motion on the
diffeomorphisms with respect to the norm $\|\mathbf{u}\|$.

\begin{lemma} 
In Step IIB of the proof of the Euler--Poincar\'e reduction theorem
(that (ii) $\Longleftrightarrow$ (iv) for an arbitrary Lie group)
a certain formula for the variations for time-dependent vector fields
was employed. That formula was employed again in the calculation above as
\begin{equation}\label{mixed-partials}
\delta \mathbf{u}
=
\mathbf{\dot{v}} 
-
\ad_{\mathbf{u}} \mathbf{v}
.
\end{equation}
This formula may be rederived as follows in the present context.
We write $\mathbf{u}=\dot{g}g^{-1}$ and $\mathbf{v}=g^\prime g^{-1}$ in
natural notation and express the partial derivatives $\dot{g}=\partial
g/\partial t$ and $g^\prime=\partial g/\partial \epsilon$ using the right translations as 
\[
\dot{g}=\mathbf{u}\circ{g}
\quad\hbox{and}\quad
{g}^\prime=\mathbf{v}\circ{g}
.
\]
To compute the mixed partials, consider the chain rule for 
say $\mathbf{u}(g(t,\epsilon)\mathbf{x}_0)$ and set 
$\mathbf{x}(t,\epsilon)=g(t,\epsilon)\cdot\mathbf{x}_0$. 
Then,
\[
\mathbf{u}^\prime
=
\frac{\partial \mathbf{u}}{\partial \mathbf{x}}
\cdot
\frac{\partial \mathbf{x}}{\partial \epsilon} 
=
\frac{\partial \mathbf{u}}{\partial \mathbf{x}}
\cdot
g^\prime(t,\epsilon)\mathbf{x}_0
=
\frac{\partial \mathbf{u}}{\partial \mathbf{x}}
\cdot
g^\prime g^{-1}\mathbf{x}
=
\frac{\partial \mathbf{u}}{\partial \mathbf{x}}
\cdot\mathbf{v}(\mathbf{x})
.
\]
The chain rule for $\mathbf{\dot{v}}$  gives a similar formula
with $\mathbf{u}$ and $\mathbf{v}$ exchanged. Thus, the chain rule gives
two expressions for the mixed partial derivative
$\dot{g}^\prime$ as 
\[
\dot{g}^\prime=\mathbf{u}^\prime=\nabla\mathbf{u}\cdot\mathbf{v}
\quad\hbox{and}\quad
\dot{g}^\prime=\mathbf{\dot{v}}=\nabla\mathbf{v}\cdot\mathbf{u}
.
\]
The difference of the mixed partial derivatives then implies the desired
formula \eqref{mixed-partials}, since
\[
\mathbf{u}^\prime-\mathbf{\dot{v}}
=\nabla\mathbf{u}\cdot\mathbf{v}
-\nabla\mathbf{v}\cdot\mathbf{u}
= -[ \mathbf{u}, \mathbf{v}]
= -\ad_{\mathbf{u}}\mathbf{v}
.
\]
\end{lemma}

\newpage
\vspace{4mm}\centerline{\textcolor{shadecolor}{\rule[0mm]{6.75in}{-2mm}}\vspace{-4mm}}
\section{EPDiff solution behavior in 1D}

\secttoc

\textbf{What is this lecture about?} This lecture explores the construction of solutions for 
geodesic motion on the manifold of smooth invertible maps (diffeomorphisms).


In this lecture, we shall discuss the solutions of EPDiff for pressureless
compressible geodesic motion in one spatial dimension.  This is the \emph{EPDiff equation} in 1D,
\footnote{
A one-form density in 1D takes the form
$m(dx)^2$ and the EP equation is given by
\[
\frac{d}{dt}\big(m\,(dx)^2\big) =
\frac{dm}{dt}(dx)^2 + 2m\,(du)(dx) =0 \quad\hbox{with}\quad
\frac{d}{dt}dx=du=u_xdx \quad\hbox{and}\quad u=G*m, \]
where $G*m$ denotes convolution with a function $G$ on the real line.}
\begin{align}\label{1dEPDiff}
\partial_tm + \ad^*_um&=0,\quad\hbox{or, equivalently,} \\
\partial_tm+um_x+2u_xm&=0,\quad\hbox{with } m= Q_{\mathrm{op}}u.
\label{1dEPDiff-pde}
\end{align}
\begin{itemize}
\item
The EPDiff equation describes geodesic motion on the diffeomorphism group
with respect to a family of metrics for the fluid velocity $u(t,x)$, with
notation,
\begin{eqnarray}\label{mom-def-1D}
m
&=& \frac{\delta\ell}{\delta u} = Q_{\mathrm{op}}u
\quad\hbox{for a kinetic-energy Lagrangian}\quad
\\&&
\ell(u) 
= \tfrac{1}{2}\int u\,Q_{\mathrm{op}}u\,dx
= \tfrac{1}{2}\|u\|^2
.
\end{eqnarray}
\item
In one dimension, $Q_{\mathrm{op}}$ in equation \eqref{mom-def-1D} is a positive,
symmetric operator that defines the kinetic energy metric for the
velocity.  
\item
The EPDiff equation \eqref{1dEPDiff-pde} is written in terms of
the variable $m=\delta\ell/\delta u$. It is appropriate to call this variational
derivative $m$, because it is the momentum density associated with the
fluid velocity $u$. 
\item
Physically, the first nonlinear term in the EPDiff equation
\eqref{1dEPDiff-pde} is fluid  transport. 
\item
The coefficient 2 arises in the
second nonlinear term, because, in one dimension, two of the  summands in
$\ad^*_u\,m=um_x+2u_xm$ are the \textit{ same}, cf. equation
\eqref{continuumcoadjoint-coord-form}.  
\item
The momentum is expressed in terms of the velocity by
$m=\delta\ell/\delta{u}=Q_{\mathrm{op}}u$.
Equivalently, for solutions that vanish at spatial infinity, one may think of the velocity
as being obtained from the convolution,
\begin{equation}\label{filter-reln-1D}
u(x)=G*m(x)=\int G(x-y)m(y)\,dy
,
\end{equation}
where $G$ is the Green's function for the operator $Q_{\mathrm{op}}$ on the real
line. 
\item
The operator $Q_{\mathrm{op}}$ and its Green's function $G$ are chosen to be
even under reflection, $G(-x)=G(x)$, so that $u$ and $m$ have the same parity. Moreover, the
EPDiff equation \eqref{1dEPDiff-pde} conserves the total momentum $M=\int  m(y)\,dy$,
for any even Green's function.
\end{itemize}
\begin{exercise}
Show that equation \eqref{1dEPDiff-pde} conserves $M=\int  m(y)\,dy$ for any even Green's
function $G(-x)=G(x)$, for either periodic, or homogeneous boundary conditions.
\end{exercise}
\begin{itemize}
\item
The traveling wave solutions of 1D EPDiff when the Green's function $G$ is chosen to be
even under reflection are the ``pulsons,'' 
\[
u(x,t)=c\,G(x-ct)
.
\] 
\end{itemize}
\begin{exercise}
Prove this statement, that the traveling wave solutions of 1D EPDiff are pulsons when the
Green's function is even. What role is played in the solution by the Green's function being
even? Hint: Evaluate the derivative of an even function at $x=0$.
\end{exercise}
\begin{itemize}
\item
See Fringer and Holm~\cite{FrHo2001} and references therein for further discussions and
numerical simulations of the pulson solutions of the 1D EPDiff equation.
\end{itemize}
\vspace{4mm}\centerline{\textcolor{shadecolor}{\rule[0mm]{6.75in}{-2mm}}\vspace{-4mm}}

\subsection{Pulsons} 
The EPDiff equation \eqref{1dEPDiff-pde} on the real line has the
remarkable property that its solutions \emph{collectivise}%
\footnote{See Guillemin and Sternberg~\cite{GuSt1984} for discussions of the concept of collective variables
for Hamiltonian theories. We will discuss the collectivization for the EPDiff
equation later from the viewpoint of momentum maps.}
into the finite dimensional solutions of the ``$N$--pulson'' form that was discovered
for a special form of $G$ in Camassa and Holm~\cite{CaHo1993}, then was
extended for \textit{ any} even $G$ in Fringer and Holm~\cite{FrHo2001},
\begin{equation}\label{u-pulson}
u(x,t) = \sum_{i=1}^N p_i(t)\,G(x-q_i(t))
.
\end{equation}
Since $G(x)$ is the Green's function for the operator $Q_{\mathrm{op}}$, the
corresponding solution for the momentum $m=Q_{\mathrm{op}}u$ is given by a sum of
delta functions,
\begin{equation}\label{m-delta1}
m(x,t) = \sum_{i=1}^N p_i(t)\,\delta(x-q_i(t))
.
\end{equation}
Thus, the time-dependent ``collective coordinates'' $q_i(t)$ and $p_i(t)$
are the positions and velocities of the $N$ pulses in this solution. These
parameters satisfy the finite dimensional geodesic motion equations
obtained as canonical Hamiltonian equations 
\begin{eqnarray} \label{q-eqn} 
\dot{q}_i & = & \frac{\partial H_N}{\partial p_i}
= \sum_{j=1}^{N} p_j\, G(q_i-q_j), \\
\dot{p}_i & = & -\frac{\partial H_N}{\partial q_i}
= -p_i\sum_{j=1}^{N} p_j\, G\,'(q_i-q_j),
\label{p-eqn}
\end{eqnarray} 
in which the Hamiltonian is given by the quadratic form,
\begin{equation}\label{Ham-pulson}
H_N = \tfrac{1}{2} \sum_{i,j=1}^Np_i\,p_j\,
G(q_i-q_j)
.
\end{equation}

\begin{remark}\rm In a certain sense, equations \eqref{q-eqn}--\eqref{p-eqn}
comprise the analog for the peakon momentum relation \eqref{m-delta1} of
the ``symmetric generalized rigid body equations'' in \eqref{HamCan-eqns}.
\end{remark}

Thus, the canonical equations for the Hamiltonian $H_N$ describe the
nonlinear collective interactions of the $N$--pulson solutions of the
EPDiff equation \eqref{1dEPDiff-pde} as finite-dimensional geodesic
motion of a particle on an
$N$--dimensional surface whose co-metric is 
\begin{equation}\label{metric-pulson}
G^{ij}(q)=G(q_i-q_j)
.
\end{equation}
Fringer and Holm~\cite{FrHo2001} showed numerically that the $N$--pulson
solutions describe the emergent patterns in the solution of the
initial value problem for EPDiff equation \eqref{1dEPDiff-pde} with 
spatially confined initial conditions. 

\begin{exercise}
Equations \eqref{q-eqn}--\eqref{p-eqn} describe geodesic motion. 
\begin{enumerate}
\item
Write the Lagrangian and Euler--Lagrange equations for this motion. 
\item
Solve equations \eqref{q-eqn}--\eqref{p-eqn} for $N=2$ when
$\lim_{|x|\to\infty}G(x)=0$. 
\begin{enumerate}
\item
Why should the solution be described as exchange of momentum in elastic collisions?
\item
Consider both head-on and overtaking collisions.  
\item
Consider the antisymmetric case, when the total momentum
vanishes.
\end{enumerate}
\end{enumerate}
\end{exercise}

\subsubsection{Integrability} Calogero and Francoise~\cite{Ca1995,CaFr1996} found that for any finite number 
$N$  the Hamiltonian equations for $H_{N}$ in \eqref{Ham-pulson} are 
completely integrable in the Liouville  sense%
\footnote{A Hamiltonian system is integrable in the Liouville  sense, if
the number of compatible independent constants of motion in involution is the same 
as the number of its degrees of freedom.}
for
\begin{align*}
G &\equiv G_{1}(x)=\lambda + \mu \cos(\nu x) +
  \mu_{1} \sin(\nu |x|) \\[-1ex]
\text{and}\qquad
G &\equiv G_{2}(x)=\alpha + \beta |x| + \gamma x^{2},
\end{align*}
with $\lambda$, $\mu$, $\mu_{1}$, 
$\nu$, and $\alpha$, $\beta$, $\gamma$ 
being arbitrary constants, such that $\lambda$ and $\mu$ 
are real and $\mu_{1}$  and $\nu$ both real or both imaginary.%
\footnote{
This choice of the constants keeps $H_{N}$ real in \eqref{Ham-pulson}.} 
Particular cases of $G_{1}$ and $G_{2}$  are the peakons
$G_1(x)=e^{-|x|/\alpha}$ of  Camassa and Holm~\cite{CaHo1993} and  the compactons 
$G_2(x)=\max(1-|x|,0)$ of the Hunter--Saxton equation,
(see Hunter and Zheng~\cite{HuZh1994}).  The latter is the EPDiff equation \eqref{1dEPDiff-pde},
with $\ell(u)=\frac{1}{2}\int u_x^2\,dx$ and thus
$m=-u_{xx}$.

\subsubsection{Lie--Poisson Hamiltonian form of EPDiff} 
In terms of $m$, the conserved energy Hamiltonian for the EPDiff equation
\eqref{1dEPDiff-pde} is obtained by Legendre transforming the kinetic energy
Lagrangian, as
\[
h = \Big\langle \frac{\delta\ell}{\delta u},u\Big\rangle- \ell(u)
.\]
Thus, the Hamiltonian depends on $m$, as
\[
h(m)=\tfrac{1}{2}\int m(x)G(x-y)m(y)\,dxdy
,\]
which also reveals the geodesic nature of the EPDiff equation
\eqref{1dEPDiff-pde} and the role of $G(x)$ in the kinetic energy metric on
the Hamiltonian side. 

The corresponding \emph{Lie--Poisson bracket} for EPDiff as a Hamiltonian evolution
equation is given by,
\[
\partial_tm = \big\{m,h\big\} 
= - \ad^*_{\delta h/\delta m}m
= -(\partial{m}+m\partial)\frac{\delta h}{\delta m}
\quad\hbox{and}\quad
\frac{\delta h}{\delta m}=u
,
\]
which recovers the starting equation and indicates some of its connections with fluid
equations on the Hamiltonian side. For any two smooth functionals $f,h$ of $m$ in the space
for which the solutions of EPDiff exist, this Lie--Poisson bracket may be expressed as
\[
\big\{f,h\big\} 
= -\int \frac{\delta f}{\delta m}
(\partial{m}+m\partial)\frac{\delta h}{\delta m} dx
= -\int m \bigg[\frac{\delta f}{\delta m}
, \frac{\delta h}{\delta m}\bigg] dx
\]
where $[\cdot, \cdot]$ denotes the Lie algebra bracket of vector fields. That is,
\[
\bigg[\frac{\delta f}{\delta m}
, \frac{\delta h}{\delta m}\bigg]
=
\frac{\delta f}{\delta m}\partial\frac{\delta h}{\delta m}
-
\frac{\delta h}{\delta m}\partial\frac{\delta f}{\delta m}
.
\]

\begin{exercise}
What is the Casimir for this Lie--Poisson bracket? What does it mean from the viewpoint of
coadjoint orbits? 
\end{exercise}

\subsection{Peakons} \index{Camassa-Holm (CH) equation!Peakons, Lax pair}
The case $G(x)=e^{-|x|/\alpha}$ with a constant length scale $\alpha$ is the
Green's function for which the operator in the kinetic
energy Lagrangian \eqref{mom-def-1D} is
$ Q_{\mathrm{op}}=1-\alpha^2\partial_x^2$. For this (Helmholtz)
operator $ Q_{\mathrm{op}}$, the Lagrangian and corresponding kinetic energy
norm are given by, 
\[ \ell[u]
=\tfrac{1}{2}\|u\|^2
= \tfrac{1}{2}\int u Q_{\mathrm{op}}u\,dx
=\tfrac{1}{2}\int u^2+\alpha^2u_x^2\,dx,\quad\text{for}\quad
\lim_{|x|\to\infty}u=0 .\] 
This Lagrangian is the $H^1$ norm of the velocity in one
dimension. In this case, the EPDiff equation \eqref{1dEPDiff-pde} is also
the zero-dispersion limit of the completely integrable CH 
equation for unidirectional shallow water waves first derived in 
Camassa and Holm~\cite{CaHo1993},
\begin{equation} \label{CH-eqn}
m_t + um_x + 2 mu_x = -c_0u_x+\gamma{u}_{xxx},\qquad m=u-\alpha^2u_{xx}.
\end{equation}
This equation describes shallow water dynamics as completely integrable
soliton motion at quadratic order in the asymptotic expansion for
unidirectional shallow water waves on a free surface under gravity.
See Dullin, Gottwald and Holm~\cite{DuGoHo2001,DuGoHo2003,DuGoHo2004}
for more details and explanations of this asymptotic expansion for
unidirectional shallow water waves to quadratic order.

Because of the relation $m=u-\alpha^2u_{xx}$, equation \eqref{CH-eqn} is 
nonlocal. In other words, it is an integral-partial differential equation. 
In fact, after writing equation \eqref{CH-eqn} in the equivalent form,
\begin{equation} \label{CH-eqn-nonloc}
(1-\alpha^2\partial^2)(u_t+uu_x) =
-\partial\big(u^2+\tfrac12 \alpha^2 u_x^2\big)
-c_0u_x+\gamma{u}_{xxx},
\end{equation}
one sees the interplay between local and nonlocal linear dispersion in its phase
velocity relation,
\begin{equation} \label{disp-reln}
\smash{\frac{\omega}{k} =
\frac{c_0 - {\gamma}k^2}{1 + \alpha^2 k^2},}
\end{equation}
for waves with frequency $\omega$ and wave number $k$ linearized
around ${u}=0$. For $\gamma/c_0<0$, short waves and long
waves travel in the same direction. Long waves travel faster than short ones
(as required in shallow water) provided $\gamma/c_0 > -  \alpha^2$.
Then the phase velocity lies in the interval $\omega/k\in(-\gamma/\alpha^{\,2}, c_0]$.

The famous Korteweg--de\,Vries (KdV) soliton equation, 
\begin{equation} \label{KdV-eqn}
u_t+3uu_x
=
-c_0u_x+\gamma{u}_{xxx}
,
\end{equation}
emerges at \emph{linear} order in the asymptotic expansion for shallow
water waves, in which one takes $\alpha^2\to0$ in \eqref{CH-eqn-nonloc}
and \eqref{disp-reln}. In KdV, the  parameters $c_0$ and $\gamma$ are
seen as deformations of the \emph{Riemann equation},
\[
u_t+3uu_x=0
.
\]
The parameters $c_0$ and $\gamma$ represent linear wave dispersion, which modifies and
eventually balances the tendency for nonlinear waves to steepen and break. The parameter
$\alpha$, which introduces nonlocality, also regularizes this nonlinear tendency, even in
the absence of $c_0$ and $\gamma$.

\subsection{Integrability of the CH equation} 
\subsubsection{The bi-Hamiltonian property of CH.}
The CH equation \eqref{CH-eqn} may be written in bi-Hamiltonian form as
\begin{equation}\label{eq2}
 m_{t}=-(\partial-\partial^{3})\frac{\delta H_{2}[m]}{\delta m}=-(2\kappa \partial +\partial m+m\partial)\frac{\delta H_{1}[m]}{\delta m}\,,
\end{equation}

where the two Hamiltonians are given by
\begin{equation}\label{eq2a} H_{1}[m]
=
\tfrac{1}{2}\int m u\, dx
\qquad\hbox{and}\qquad
H_{2}[m]
=
\tfrac{1}{2}\int(u^{3}+uu_{x}^{2}+2\kappa u^{2})dx\,. 
\end{equation}

The integration is over the real line, for functions that decay sufficiently rapidly as $|x|\to  \infty $. (The integration is over  one period, for periodic functions.)

\subsubsection{The multi-Hamiltonian structure of CH.}
By Magri's theorem \cite{magri1978simple}, the bi-Hamiltonian property of CH in \eqref{eq2} implies an infinite sequence of conservation laws for it, known as a \textit{ multi-Hamiltonian structure} $H_n[m]$, $n=0,\pm1, \pm2,\ldots$,
such that 
\begin{equation}\label{eq2aa}
 (\partial-\partial^{3})\frac{\delta H_{n}[m]}{\delta m}
 =(2\kappa \partial +\partial m+m\partial)\frac{\delta H_{n-1}[m]}{\delta m}\,.
\end{equation}

\subsubsection{The Lax pair for CH.}
Its bi-Hamiltonian property also implies that the CH equation \eqref{CH-eqn} admits a \textit{Lax pair representation}, given by
\begin{align}\label{eq3} 
\Psi_{xx}&=\Big(\frac{1}{4}+\lambda
(m+\kappa)\Big)\Psi
 ,\\\label{eq4}
\Psi_{t}&=\Big(\frac{1}{2\lambda}-u\Big)\Psi_{x}+\frac{u_{x}}{2}\Psi+\gamma\Psi
,\end{align}

where $\kappa,\gamma$ are arbitrary real constants and the eigenvalue $\lambda$ is independent of time. The compatibility of the Lax pair ($\Psi_{xxt}=\Psi_{txx}$) for constant $\lambda$ means that the eigenvalue equation in \eqref{eq2aa} is \emph{isospectral}, i.e., its spectrum is invariant under the flow of the CH equation. The spectrum of \eqref{eq2aa} turns out to represent the speeds of the soliton solutions for at late times when they are sufficiently separated. Remarkably, when $\kappa=0$ the spectrum of the eigenvalue equation in \eqref{eq2aa} is purely \textit{discrete}, which means the solution in this case comprises only peakons. 

Many papers have been written to explore various features of the original CH equation.
See \cite{lundmark2022view} for a recent brief chronological discussion of the exploration of singular peakon solutions.

\newpage
\vspace{4mm}\centerline{\textcolor{shadecolor}{\rule[0mm]{6.75in}{-2mm}}\vspace{-4mm}}
\section{Diffeons -- singular momentum solutions of the EPDiff equation for
geodesic motion in higher dimensions} \label{sec-strings}
\index{diffeons! EPDiff equation}

\secttoc

\textbf{What is this lecture about?} This lecture extends the 1D singular geodesic solutions 
of EPDiff to higher dimensions.

As an example of the EP theory  in higher dimensions, we shall 
generalize the one-dimensional pulson solutions of the previous section to
$n$--dimensions. The corresponding singular momentum solutions of the EPDiff
equation in higher dimensions are called ``diffeons.''

\subsection{$n$--dimensional EPDiff equation}
Eulerian geodesic motion of a fluid in $n$--dimensions is
generated as an EP equation via Hamilton's principle, when the 
Lagrangian is given by the kinetic energy. The kinetic energy defines a norm
$\|\mathbf{u}\|^2$ for the Eulerian fluid velocity,
$\mathbf{u}(\mathbf{x},t){:\ } R^n\times R^1 \to R^n$. As mentioned earlier, the
choice of the kinetic energy as a positive functional of fluid velocity
$\mathbf{u}$ is a modeling step that depends upon the physics of the
problem being studied. Following our earlier procedure, as in equations
\eqref{Lag-ansatz-1} and \eqref{EP-eqn-abstract}, we shall choose
the Lagrangian,
\begin{equation}\label{Lag-ansatz-2}
\|\mathbf{u}\|^2 
= \int \mathbf{u}\cdot Q_{\mathrm{op}}\mathbf{u}\,d^nx
= \int \mathbf{u}\cdot\mathbf{m}\,d^nx
,\end{equation}
so that the positive-definite, symmetric, operator $Q_{\mathrm{op}}$ defines the
norm $\|\mathbf{u}\|$, for appropriate boundary conditions and the EPDiff
equation for Eulerian geodesic motion of a fluid emerges,
\begin{equation}\label{EP-eqn}
\frac{d}{dt}\frac{\delta \ell}{\delta \mathbf{u}}
+
\ad^*_{\mathbf{u}}
\frac{\delta \ell}{\delta \mathbf{u}}
=
0
,
\quad\hbox{with}\quad
\ell[\mathbf{u}]=\tfrac{1}{2}\|\mathbf{u}\|^2
.
\end{equation}

\subsubsection{Legendre transforming to the Hamiltonian side} 
The corresponding Legendre transform yields the following 
invertible relations between momentum and velocity,
\begin{equation}\label{Legendre-dual-rel}
\mathbf{m} =  Q_{\mathrm{op}}\mathbf{u}
\quad\hbox{and}\quad\
\mathbf{u} = G*\mathbf{m}
,
\end{equation}
where $G$ is the \emph{Green's function} for the operator $ Q_{\mathrm{op}}$, assuming
appropriate boundary conditions (on $\mathbf{u}$) that allow inversion of
the operator $ Q_{\mathrm{op}}$ to determine $\mathbf{u}$ from $\mathbf{m}$.

The corresponding \emph{Hamiltonian} is,
\begin{equation}\label{Ham-ansatz}
h[\mathbf{m}]
= \langle\mathbf{m}, \mathbf{u}\rangle
- \tfrac{1}{2}\|\mathbf{u}\|^2
= \tfrac{1}{2}\int \mathbf{m}\cdot G*\mathbf{m}\ d^nx
\equiv \tfrac{1}{2}\|\mathbf{m}\|^2,
\end{equation}
which also defines a norm $\|\mathbf{m}\|$ via a convolution kernel $G$
that is symmetric and positive, when the Lagrangian $\ell[\mathbf{u}]$ is a
norm. As expected, the norm $\|\mathbf{m}\|$ given by the
Hamiltonian $h[\mathbf{m}]$ specifies the velocity
$\mathbf{u}$ in terms of its Legendre-dual momentum $\mathbf{m}$ by the variational 
operation,
\begin{equation}\label{Ham-u-def}
\mathbf{u} 
= \frac{\delta h}{\delta \mathbf{m}}
= G*\mathbf{m}
\equiv \int G(\mathbf{x}-\mathbf{y})\,\mathbf{m}(\mathbf{y})\,d^ny
.
\end{equation}
We shall choose the kernel $G(\mathbf{x}-\mathbf{y})$ to be
translation-invariant (so Noether's theorem implies that total momentum
$\mathbf{M}=\int \mathbf{m}\,d^nx$ is conserved) and symmetric under
spatial reflections (so that $\mathbf{u}$ and $\mathbf{m}$ have the same
parity). 
 
After the Legendre transformation \eqref{Ham-ansatz}, the EPDiff equation
\eqref{EP-eqn} appears in its equivalent \emph{Lie--Poisson Hamiltonian
form} which is reminiscent of the EPSO(3) equation for rigid body motion
since $\delta h/\delta \mathbf{m}$ is a vector field,
\begin{equation}\label{LP-eqn}
\frac{\partial}{\partial t}\mathbf{m}
=
\{\mathbf{m},h\}
=
-
\ad^*_{{\delta h}/{\delta \mathbf{m}}}\mathbf{m}
=
-
\ad^*_{G*\mathbf{m}}\mathbf{m}
\,.
\end{equation}
Here the operation $\{\cdot,\cdot \}$ denotes the Lie--Poisson
bracket dual to the (right) action of vector fields amongst themselves by
vector-field commutation
\[
\{f, h\,\}
=
-
\left\langle
\mathbf{m}, \left[
\frac{\delta f}{\delta \mathbf{m}}
, 
\frac{\delta h}{\delta \mathbf{m}}
\right]
\right\rangle
\]
For more details and additional background
concerning the relation of classical EP theory to Lie--Poisson
Hamiltonian equations, see Holm, Marsden and Ratiu~\cite{MaRa1994,HoMaRa1998a}.

In a moment we will also consider the momentum maps for EPDiff.

\subsection{Diffeons: $n$--dimensional analogs of pulsons for the EPDiff 
equation} 

The momentum for the one-dimensional pulson solutions \eqref{m-delta1} on
the real line is  supported at points via the Dirac delta measures in its
solution ansatz,
\begin{equation}\label{pulson-ansatz}
m(x,t)=\sum_{i=1}^Np_i(t)\,\delta\big(x-q_i(t)\big)
,\quad
m\in{R^1}
.
\end{equation} 
We shall develop $n$--dimensional  analogs of these one-dimensional pulson
solutions for the Euler--Poincar\'e equation \eqref{div-grad-curl-eqn} by
generalizing this solution ansatz to allow measure-valued $n$--dimensional
vector solutions $\mathbf{m}\in{R^n}$ for which the Euler--Poincar\'e
momentum is supported  on co-dimension--$k$ \textit{ subspaces}
$R^{n-k}$ with integer $k\in[1,n]$. For example, one may consider the 
two-dimensional vector momentum $\mathbf{m}\in{R^2}$ in the plane that is
supported on one-dimensional curves (momentum fronts). Likewise, in three
dimensions, one could consider two-dimensional momentum surfaces (sheets),
one-dimensional momentum filaments, etc. The corresponding vector momentum
ansatz that we shall use is the following, cf. the pulson solutions
\eqref{pulson-ansatz},
\begin{equation}\label{m-ansatz}
\mathbf{m}(\mathbf{x},t) =
\sum_{i=1}^N\int\mathbf{P}_i(s,t)\,
\delta\big(\,\mathbf{x}-\mathbf{Q}\,_i(s,t)\,\big)ds,\quad
\mathbf{m}\in{R^n}.
\end{equation} 
Here, $\mathbf{P}_i,\mathbf{Q}_i\in{R^n}$ for $i=1,2,\dots,N$.
For example, when $n-k=1$, so that $s\in R^1$ is one-dimensional, the delta
function in solution \eqref{m-ansatz} supports an evolving family of
vector-valued curves, called \emph{momentum filaments}. (For
simplicity of notation, we suppress the implied subscript
$i$ in the arclength $s$ for each $\mathbf{P}_i$ and $\mathbf{Q}_i$.) The
Legendre-dual relations \eqref{Legendre-dual-rel} imply that the velocity
corresponding to the momentum filament ansatz \eqref{m-ansatz} is,
\begin{equation} \label{u-ansatz}
\mathbf{u}(\mathbf{x},t)
= G*\mathbf{m}
=
\sum_{j=1}^N\int\mathbf{P}_j(s^{\prime},t)\,
G\big(\,\mathbf{x}-\mathbf{Q}\,_j(s^{\prime},t)\,\big)ds^{\prime}
.
\end{equation} 
Just as for the 1D case of the pulsons, we shall show that substitution of
the $n$D solution ansatz \eqref{m-ansatz} and \eqref{u-ansatz} into the
EPDiff equation \eqref{ep-eqn-coord-form} produces canonical geodesic
Hamiltonian equations for the $n$--dimensional vector parameters
$\mathbf{Q}_i(s,t)$ and $\mathbf{P}_i(s,t)$, $i=1,2,\dots,N$.


\subsubsection{Canonical Hamiltonian dynamics of diffeon filaments in ${R^n}$}

For definiteness in what follows, we shall consider the
example of momentum filaments $\mathbf{m}\in{R^n}$ supported on
one-dimensional space curves in ${R^n}$, so $s\in{R^1}$ is the arclength
parameter of one of these curves. This solution ansatz is reminiscent of the
Biot--Savart Law for vortex filaments, although the flow is not
incompressible. The dynamics of momentum surfaces, for $s\in{R^k}$ with
$k<n$, follow a similar analysis.

Substituting the momentum filament ansatz \eqref{m-ansatz} for
$s\in{R^1}$ and its corresponding velocity \eqref{u-ansatz} into the
Euler--Poincar\'e equation \eqref{ep-eqn-coord-form}, then integrating against a
smooth test function $\phi(\mathbf{x})$ implies the following
canonical equations (denoting explicit summation on $i,j\in1,2,\dots N$),
\begin{align} 
\frac{\partial }{\partial t}\mathbf{{Q}}_i (s,t)
&=\sum_{j=1}^{N} \int\mathbf{P}_j(s^{\prime},t)\,
G(\mathbf{Q}_i(s,t)-\mathbf{Q}_j(s^{\prime},t))\,\big)ds^{\prime}
= \frac{\delta H_N}{\delta \mathbf{P}_i},\label{IntDiffEqn-Q}\\
\label{IntDiffEqn-P}
\frac{\partial}{\partial t}\mathbf{{P}}_i (s,t)
&= -\sum_{j=1}^{N} \int 
\big(\mathbf{P}_i(s,t)\!\cdot\!\mathbf{P}_j(s^{\prime},t)\big)
\frac{\partial }{\partial \mathbf{Q}_i(s,t)}
G\big(\mathbf{Q}_i(s,t)-\mathbf{Q}_j(s^{\prime},t)\big)\,ds^{\prime} \\
&= -\,\frac{\delta H_N}{\delta \mathbf{Q}_i}
,\qquad\text{(sum on $j$, no sum on $i$)}.\nonumber
\end{align}
The dot product $\mathbf{P}_i\cdot\mathbf{P}_j$ denotes the inner, or
scalar, product of the two vectors $\mathbf{P}_i$ and $\mathbf{P}_j$ in
$R^n$. Thus, the solution ansatz \eqref{m-ansatz} yields a closed set of
\emph{integro-partial-differential equations (IPDEs)} given by 
\eqref{IntDiffEqn-Q} and \eqref{IntDiffEqn-P} for the vector parameters
$\mathbf{Q}_i(s,t)$ and $\mathbf{P}_i(s,t)$ with $i=1,2\dots N$. These
equations are generated canonically by the following Hamiltonian function
$H_N{:\ } (R^n\times R^n)^{\otimes N}\to R$,
\begin{equation} \label{H_N-def}
H_N = \tfrac{1}{2}\!\int\!\!\!\!\int\!\!\sum_{i, j=1}^{N} 
\big(\mathbf{P}_i(s,t)\cdot\mathbf{P}_j(s^{\prime},t)\big) 
\,G\big(\mathbf{Q}_i(s,t)-\mathbf{Q}_{\,j}(s^{\prime},t)\big)
\,ds\,ds^{\prime}
.
\end{equation}
This Hamiltonian arises by substituting the momentum ansatz
\eqref{m-ansatz} into the Hamiltonian \eqref{Ham-ansatz} obtained from the
Legendre transformation of the Lagrangian corresponding to the kinetic
energy norm of the fluid velocity. Thus, the evolutionary IPDE system
\eqref{IntDiffEqn-Q} and \eqref{IntDiffEqn-P} represents canonically
Hamiltonian geodesic motion on the space of curves in $R^n$ with respect to
the co-metric given on these curves in \eqref{H_N-def}. The Hamiltonian
$H_N=\frac{1}{2}\|\mathbf{P}\|^2$ in \eqref{H_N-def} defines the norm
$\|\mathbf{P}\|$ in terms of this co-metric that combines
convolution using the Green's function $G$ and sum over filaments with the
scalar product of momentum vectors in $R^n$. 

\begin{remark}\rm  Note the Lagrangian property of the $s$ coordinate, since
\[
\frac{\partial }{\partial t}\mathbf{{Q}}_i (s,t)
=
\mathbf{u}(\mathbf{Q}_i (s,t),t) 
.
\]
\end{remark}

\subsection{Singular solution momentum map $\mathbf{J}_{\mathrm Sing}$ for diffeons}
\label{sing-mom-map-sec}


The diffeon momentum filament ansatz \eqref{m-ansatz} reduces,
and \emph{collectivizes} the solution of the geodesic EP PDE
\eqref{ep-eqn-coord-form}
in $n+1$ dimensions into the system \eqref{IntDiffEqn-Q} and
\eqref{IntDiffEqn-P} of $2N$ canonical evolutionary IPDEs. One can
summarize the mechanism by which this process occurs, by saying that the
map that implements the canonical $(\mathbf{Q},\mathbf{P})$ variables in
terms of singular solutions is a (cotangent bundle) momentum  map. Such
momentum maps are Poisson maps; so the canonical Hamiltonian
nature of the dynamical equations for
$(\mathbf{Q}, \mathbf{P})$ fits into a general theory which also
provides a framework for suggesting other avenues of
investigation.

\begin{theorem}\label{mom-map}
The momentum ansatz \textup{\eqref{m-ansatz}} for
measure-valued solutions of the \textup{EPDiff} equation
\textup{\eqref{ep-eqn-coord-form}}, defines an equivariant
momentum map
\[
\mathbf{J}_{\mathrm Sing}{:\ }  T ^{\ast} \operatorname{Emb}(S, \mathbb{R}^n)
\rightarrow
\mathfrak{X}^{\ast}(\mathbb{R}^n)
\]
that is called the \emph{singular solution momentum map} 
\index{momentum map! singular solution} 
in Holm and Marsden~\cite{HoMa2004}.
\end{theorem}

We shall explain the notation used in the theorem's statement in the
course of its proof. Right away, however, we note that the sense
of ``defines'' is that the momentum solution ansatz
\eqref{m-ansatz} expressing $\mathbf{m}$ (a vector function of
spatial position $\mathbf{x}$) in terms of
$\mathbf{Q}, \mathbf{P}$ (which are functions of $s$)
can be regarded as a map from the space of $(\mathbf{Q}(s),
\mathbf{P} (s))$ to the space of $\mathbf{m}$'s. This will turn
out to be the Lagrange-to-Euler map for the fluid description of
the singular solutions. The proof follows Holm and Marsden~\cite{HoMa2004}.


\begin{proof}For simplicity and without loss of
generality, let us take $N = 1$ and so suppress the index $a$.
That is, we shall take the case of an isolated singular
solution. As the proof will show, this is not a real
restriction. 

To set the notation, fix a
$k$--dimensional manifold
$S$ with a given volume element and whose points are denoted $s
\in S$. Let $\operatorname{Emb}(S, \mathbb{R}^n)$ denote the
set of smooth embeddings $\mathbf{Q}{:\ }  S \rightarrow \mathbb{R}^n$.
(If the EPDiff equations are taken on a manifold $M$, replace
$\mathbb{R}^n $ with $M$.) Under appropriate technical
conditions, which we shall just treat formally here,
$\operatorname{Emb}(S, \mathbb{R}^n)$ is a smooth manifold. (See,
for example, Ebin and Marsden~\cite{EbMa1970}, and Marsden and
Hughes~\cite{MaHu1983} for a discussion and references.)

The tangent space $T _{\mathbf{Q}} \operatorname{Emb}(S,
\mathbb{R}^n)$ to $\operatorname{Emb}(S, \mathbb{R}^n)$ at the
point $\mathbf{Q} \in \operatorname{Emb}(S, \mathbb{R}^n)$ is
given by the space of \emph{material velocity fields}, namely the
linear space of maps
$\mathbf{V}{:\ }  S \rightarrow \mathbb{R}^n$ that are vector fields
over the map $\mathbf{Q}$.  The dual space to this space will be
identified with the space of one-form densities over
$\mathbf{Q}$, which we shall regard as maps $\mathbf{P}{:\ }  S
\rightarrow \left(\mathbb{R}^n\right) ^{\ast}$. In summary, the
cotangent bundle $T ^{\ast} \operatorname{Emb}(S, \mathbb{R}^n)$
is identified with the space of pairs of maps $\left( \mathbf{Q},
\mathbf{P} \right)$.

This identification gives us the domain space for the singular solution momentum
map. Now we consider the action of the symmetry group.  Consider
the group $\mathfrak{G} = \operatorname{Diff}$ of diffeomorphisms
of the space $\mathfrak{S}$ in which the EPDiff equations are
operating, concretely in our case $\mathbb{R}^n$. Let it act on
$\mathfrak{S}$ by composition on the \textit{ left}. Namely for $\eta
\in \operatorname{Diff} (\mathbb{R}^n)$, we let 
\begin{equation} \label{action}
\eta \cdot \mathbf{Q} = \eta \circ \mathbf{Q}.
\end{equation} 
Now lift this action to the cotangent bundle $T ^{\ast} \operatorname{Emb}(S, \mathbb{R}^n)$
in the standard way (see, for instance, Marsden and Ratiu~\cite{MaRa1994} for this
construction). This lifted action is a symplectic (and hence Poisson)
action and has an equivariant momentum map. \textit{ We claim that
this momentum map is precisely given by the ansatz
\textup{\eqref{m-ansatz}}.}

To see this, one only needs to recall and then apply the general formula
for the momentum map associated with an action of a general Lie group
$\mathfrak{G}$ on a configuration manifold $Q$ and cotangent lifted to
$T^{\ast}Q$.

First let us recall the general formula. Namely, the momentum map  is
the map $\mathbf{J}{:\ }  T^{\ast}Q \rightarrow
\mathfrak{g}^\ast$  ($\mathfrak{g}^\ast$ denotes the dual
of the Lie algebra $\mathfrak{g}$ of $\mathfrak{G}$) defined by
\begin{equation} \label{momentummap}
\mathbf{J} (\alpha _q) \cdot \xi = \left\langle \alpha _q, \xi_Q (q)
\right\rangle,
\end{equation}
where $\alpha_q \in T ^{\ast} _q Q $ and $\xi \in \mathfrak{g}$, 
where $\xi _Q $ is the infinitesimal generator of the action of
$\mathfrak{G}$ on $Q$ associated to the Lie algebra element $\xi$,
and where
$\left\langle \alpha _q, \xi_Q (q)
\right\rangle$ is the natural pairing of an element of $T ^{\ast}_q
Q $ with an element of $T _q Q $.  

Now we apply this formula to the special case in which 
the group $\mathfrak{G}$ is the diffeomorphism group
$\operatorname{Diff} (\mathbb{R}^n)$, the manifold $Q$ is
$\operatorname{Emb}(S, \mathbb{R}^n)$ and where the action of
the group on
$\operatorname{Emb}(S, \mathbb{R}^n)$ is given by
\eqref{action}. The sense in which the Lie algebra of
$\mathfrak{G} = \operatorname{Diff}$ is the space
$\mathfrak{g} = \mathfrak{X}$ of vector fields is
well-understood. Hence, its dual $\mathfrak{g}^* = \mathfrak{X}^*$ is naturally regarded as
the space of one-form densities. The momentum map is thus a map
$\mathbf{J}{:\ }  T ^{\ast}
\operatorname{Emb}(S, \mathbb{R}^n)
\rightarrow \mathfrak{X}^{\ast}$.

With $\mathbf{J}$ given by \eqref{momentummap}, we only need to
work out this formula. First, we shall work out the infinitesimal
generators. Let $X \in \mathfrak{X}$ be a Lie algebra element. By
differentiating the action \eqref{action} with respect to $\eta$
in the direction of $X$ at the identity element we find that the
infinitesimal generator is given by
\[
X _{\operatorname{Emb}(S, \mathbb{R}^n)} (\mathbf{Q}) 
= X \circ \mathbf{Q}.
\]
Thus, taking $\alpha _q$ to be the cotangent vector
$(\mathbf{Q}, \mathbf{P})$, equation
\eqref{momentummap} gives
\begin{align*}
\left\langle \mathbf{J} (\mathbf{Q}, \mathbf{P} ),
X \right\rangle & 
= 
\left\langle (\mathbf{Q}, \mathbf{P}),
X \circ \mathbf{Q} \right\rangle \\
& = \int_{S} P _i(s) X ^i (\mathbf{Q}(s)) d^k s .
\end{align*}
On the other hand, note that the right hand side of
\eqref{m-ansatz} (again with the index $a$ suppressed,
and with $t$ suppressed as well), when paired with the
Lie algebra element $X$ is
\begin{align*}
\left\langle \int _S \mathbf{P}(s)\,
\delta \left( \mathbf{x}-\mathbf{Q}(s) \right) d^k s,
X \right\rangle & 
= 
\int _{\mathbb{R}^n} 
\int _S  \left( P _i (s)\,
\delta \left( \mathbf{x}-\mathbf{Q}(s) \right) d ^k s \right)
X ^i (\mathbf{x}) d ^n x \\
& = \int _S P _i (s) X ^i (\mathbf{Q} (s) d ^k s.
\end{align*}
This shows the expression given by
\eqref{m-ansatz} is equal to $\mathbf{J}$ and 
so the result is proved.
\end{proof}

The proof has shown the following basic fact.

\begin{corollary} \label{Poisson_mom-map}
The singular solution momentum map defined by the
singular solution ansatz \textup{\eqref{m-ansatz}}, namely,
\[
\mathbf{J}_{\mathrm Sing}{:\ }  
T ^{\ast} \operatorname{Emb}(S, \mathbb{R}^n)
\rightarrow
\mathfrak{X}(\mathbb{R}^n)^{\ast}
\]
is a Poisson map from the canonical Poisson structure on $T
^{\ast} \operatorname{Emb}(S, \mathbb{R}^n)$ to the Lie--Poisson
structure on $\mathfrak{X}(\mathbb{R}^n)^{\ast}$.
\end{corollary}

This is perhaps the most basic property of the singular solution
momentum map. Some of its more sophisticated properties are
outlined by Holm and Marsden~\cite{HoMa2004}.

\subsection*{Pulling back the equations.}
Since the solution ansatz \eqref{m-ansatz} has been shown
in the preceding Corollary to be a Poisson map, the pull back of
the Hamiltonian from $\mathfrak{X}^{\ast}$ to $T ^{\ast}
\operatorname{Emb}(S, \mathbb{R}^n)$ gives equations of motion on
the latter space that project to the equations on $\mathfrak{X}
^{\ast}$. 
\begin{quotation}
Thus, the basic fact that the momentum map $\mathbf{J}_{\mathrm Sing}$ 
is Poisson explains why the functions $\mathbf{Q}^a(s,t)$ and
$\mathbf{P}^a(s,t)$  satisfy canonical Hamiltonian equations.
\end{quotation}
Note that the coordinate $s\in{\mathbb{R}}^{k}$ that labels these
functions is a ``Lagrangian coordinate'' in the sense that it does
not evolve in time but rather labels the solution. 

In terms of the pairing
\begin{equation}
\langle\cdot, \cdot\rangle{:\ } 
\mathfrak{g}^*\times\mathfrak{g}\to{\mathbb{R}}
,
\end{equation}
between the Lie algebra $\mathfrak{g}$ (vector fields in $\mathbb{R}^n$)
and its dual $\mathfrak{g}^*$  (one-form densities in $\mathbb{R}^n$), the
following relation holds for measure-valued solutions
under the momentum map \eqref{m-ansatz},
\begin{equation}
\label{momentum-map-relation}
\begin{aligned}
\langle\mathbf{m}, \mathbf{u}\rangle
&= \int \mathbf{m}\cdot \mathbf{u}\,d^n\mathbf{x}
,\quad\hbox{$L^2$ pairing for }
\mathbf{m},\mathbf{u}\in{\mathbb{R}^n}, \\
&= {\int}{\int}\sum_{a, b=1}^{N}
\big(\mathbf{P}^a(s,t)\cdot\mathbf{P}^b(s^{\prime},t)\big)
G\big(\mathbf{Q}^a(s,t)-\mathbf{Q}^{\,b}(s^{\prime},t)\big) ds\,ds^{\prime} \\
&= {\int}\sum_{a=1}^{N}
\mathbf{P}^a(s,t)\cdot\frac{\partial\mathbf{Q}^a(s,t)}{\partial t} ds \\
&\equiv \langle\langle\mathbf{P}, \mathbf{\dot{Q}}\rangle\rangle,
\end{aligned}
\end{equation}
which is the natural pairing between the points $(\mathbf{Q},
\mathbf{P})
\in  T^{\ast} \operatorname{Emb}(S, \mathbb{R}^n)$ and $(\mathbf{Q},
\dot{\mathbf{Q}}) \in T \operatorname{Emb}(S, \mathbb{R}^n)$.  The momentum map 
relation \eqref{momentum-map-relation}  
corresponds to preservation of the action of the Lagrangian $\ell[\mathbf{u}]$ under
cotangent lift of $\operatorname{Diff} (\mathbb{R}^n)$.

When the Hamiltonian  $H[\mathbf{m}]$ defined on the dual
of the Lie algebra $\mathfrak{g}^*$ is pulled back to $ T^{\ast},
\operatorname{Emb}(S, \mathbb{R}^n)$ one recovers
\begin{equation}
H[\mathbf{m}] \equiv
\tfrac{1}{2}\langle\mathbf{m}, G*\mathbf{m}\rangle =
\tfrac{1}{2}\langle\!\langle\mathbf{P}, G*\mathbf{P}\rangle\!\rangle
\equiv H_N[\mathbf{P},\mathbf{Q}].
\label{geodesic-ham}
\end{equation}
In summary, in concert with the Poisson nature of the singular
solution momentum map, we see that the singular solutions in
terms of $\mathbf{Q}$ and $\mathbf{P}$ satisfy Hamiltonian
equations and also define an invariant solution set for the
EPDiff equations. In fact:
\begin{quotation}
This invariant solution set is a
special coadjoint orbit for the diffeomorphism group, as we
shall discuss in the next section.
\end{quotation}

\newpage
\vspace{4mm}\centerline{\textcolor{shadecolor}{\rule[0mm]{6.75in}{-2mm}}\vspace{-4mm}}
\section{The geometry of the momentum map} \label{geom_mommap}

\secttoc

\textbf{What is this lecture about?} This lecture
explores the geometry of the singular
solution momentum map discussed earlier in
a little more detail. The treatment is formal, in the sense that 
there are a number of
technical issues in the infinite dimensional case that will be left
open. We will mention a few of these as we proceed. 

\subsection{Coadjoint orbits} We claim  that \textit{ the
image of the singular solution momentum map is a coadjoint orbit
in $\mathfrak{X} ^{\ast}$.}  This means that (modulo some issues of
connectedness and smoothness, which we do not consider here) the
solution ansatz given by  \eqref{m-ansatz} defines a
coadjoint orbit in the space of all one-form densities, regarded as
the dual of the Lie algebra of the diffeomorphism group. 
These coadjoint orbits should be thought of as singular
orbits---that is, due to their special nature, they are not
generic. 

Recognizing them as coadjoint orbits is one way of
gaining further insight into why the singular solutions form
dynamically invariant sets---it is a general fact that coadjoint
orbits in $\mathfrak{g}^\ast$ are \textit{ symplectic submanifolds} of
the Lie--Poisson manifold $\mathfrak{g}^\ast$ (in our case
$\mathfrak{X}(\mathbb{R}^n)^{\ast}$) and, correspondingly, are dynamically invariant
for any Hamiltonian system on $\mathfrak{g}^\ast$.

The idea of the proof of our claim is simply this: whenever
one has an equivariant momentum map $\mathbf{J}{:\ }  P \rightarrow
\mathfrak{g}^\ast$ for the action of a group $G$ on a symplectic
or Poisson manifold $P$, and that action is transitive, then the \index{Poisson manifold!transitive group action}
image of $\mathbf{J}$ is an orbit (or at least a piece of an
orbit). This general result, due to Kostant, is stated more
precisely by Marsden and Ratiu~\cite[Theorem~14.4.5]{MaRa1994}. Roughly speaking, the
reason that transitivity holds in our case is because one can ``move
the images of the manifolds $S$ around at will with arbitrary velocity fields'' using
diffeomorphisms of $\mathbb{R}^n$. 

\subsection{The momentum map $\mathbf{J}_S$ and the Kelvin circulation theorem}
The momentum map $\mathbf{J}_{\mathrm Sing}$  involves
$\operatorname{Diff}(\mathbb{R}^n)$, the left action of the
diffeomorphism group on the space of embeddings
$\operatorname{Emb}(S,\mathbb{R}^n)$ by smooth maps of the target
space $\mathbb{R}^n$, namely,
\begin{equation}\label{leftDiff}
\operatorname{Diff}(\mathbb{R}^n){:\ } 
\mathbf{Q}\cdot\eta=\eta\circ\mathbf{Q},
\end{equation}
where, recall, $\mathbf{Q}{:\ } S\to \mathbb{R}^n$.
As above, the cotangent bundle $T ^{\ast}
\operatorname{Emb}(S,\mathbb{R}^n)$ is identified with the space of
pairs of maps $( \mathbf{Q}, \mathbf{P})$, with
$\mathbf{Q}{:\ } S\to \mathbb{R}^n$ and $\mathbf{P}{:\ } S\to T^*\mathbb{R}^n$.  

However, there is another momentum map $\mathbf{J}_S$  associated
with the \textit{ right action} of the diffeomorphism group of $S$ on
the embeddings
$\operatorname{Emb}(S,\mathbb{R}^n)$ by smooth maps of the
``Lagrangian labels'' $S$ (fluid particle relabeling by $\eta{:\ } S\to
S$). This action is given by
\begin{equation}\label{rightDiff}
\operatorname{Diff}(S){:\ }  
\mathbf{Q}\cdot\eta=\mathbf{Q}\circ\eta
.
\end{equation}
The infinitesimal generator of this right action is 
\begin{equation}\label{infgen-right}
X_{\operatorname{Emb}(S,\mathbb{R}^n)}(\mathbf{Q})
=
\frac{d}{dt}\bigg|_{t=0}\mathbf{Q}\circ\eta_t
=
T\mathbf{Q} \circ X.
\end{equation}
where $X \in \mathfrak{X}$ is tangent to the curve $\eta _t$ at $t
 = 0$. Thus, again taking $N = 1$ (so we suppress the index $a$) and
also letting $\alpha _q$ in the momentum map formula
\eqref{momentummap} be the cotangent vector
$(\mathbf{Q}, \mathbf{P})$, one computes $\mathbf{J}_S$:
\begin{align*}
\left\langle \mathbf{J}_S (\mathbf{Q}, \mathbf{P} ),
X \right\rangle & =  
\left\langle (\mathbf{Q}, \mathbf{P}), T\mathbf{Q}\cdot X \right\rangle \\
& =  \int_S P _i(s) \frac{\partial Q^i(s)}{\partial s^m} X ^m (s) \,d^ks \\
& =  \int_S X\Big( \mathbf{P}(s) \cdot d\mathbf{Q}(s)\Big) \,d^ks \\
& =  \left( \int_S \mathbf{P}(s) \cdot d\mathbf{Q}(s) \otimes\,d^ks ,
X (s) \right) \\
& = \langle \mathbf{P} \cdot d\mathbf{Q} , X \rangle .
\end{align*}
Consequently, the momentum map formula \eqref{momentummap} yields, cf. Exercise \ref{Ex Relabel Momap},
\begin{equation} \label{momentummap-Js}
\mathbf{J}_S (\mathbf{Q}, \mathbf{P} ) 
= 
\mathbf{P}\cdot d\mathbf{Q}
,
\end{equation}
with the indicated pairing of the one-form density $\mathbf{P}
\cdot d\mathbf{Q}$ with the vector
field $X$. 

We have set things up so that the following is true.
\begin{proposition} The momentum map $\mathbf{J}_S$ is preserved
by evolution equations \eqref{IntDiffEqn-Q}--\eqref{IntDiffEqn-P} for
$\mathbf{Q}$ and
$\mathbf{P}$.
\end{proposition}

\begin{proof} It is enough to notice that the Hamiltonian $H _N$
in equation \eqref{H_N-def}  is invariant under the cotangent lift
of the action of $\operatorname{Diff} (S)$; it merely amounts to the
invariance of the integral over $S$ under reparametrization;
that is, the change of variables formula; keep in mind that
$\mathbf{P}$ includes a density factor.
\end{proof}

\begin{remark}\rm $\quad$

\begin{itemize}
\item
This result is similar to the Kelvin--Noether theorem for circulation
$\Gamma$ of an ideal fluid, which may be written as 
\[\Gamma=\oint_{c(s)}D(s)^{-1}\mathbf{P}(s)\cdot d\mathbf{Q}(s)\]
for each Lagrangian circuit $c(s)$, where $D$ is the mass
density and $\mathbf{P}$ is again the canonical momentum density. 
This similarity should come as no surprise, because the
Kelvin--Noether theorem for ideal fluids arises from invariance of
Hamilton's principle under fluid parcel relabeling
by the \textit{ same} right action of the diffeomorphism group, as
in \eqref{rightDiff}.  

\item
Note that, being an equivariant momentum map, the map $\mathbf{J}_S$, as with
$\mathbf{J}_{\mathrm Sing}$,  is also a Poisson map. That is, substituting the canonical
Poisson bracket into relation \eqref{momentummap-Js}; that is, the relation
$\mathbf{M}(\mathbf{x}) = \sum_iP_i(\mathbf{x})\nabla Q^i(\mathbf{x})$ yields the
Lie--Poisson bracket on the space of $\mathbf{M}$'s. We use the different notations
$\mathbf{m}$ and $\mathbf{M}$ because these quantities are analogous to the body and
spatial angular momentum for rigid body mechanics. In fact, the quantity $\mathbf{m}$ given
by the solution Ansatz; specifically,
$\mathbf{m} = \mathbf{J}_{\mathrm Sing}(\mathbf{Q}, \mathbf{P})$ gives the singular solutions
of the EPDiff equations, while $\mathbf{M}(\mathbf{x}) = \mathbf{J}_S(\mathbf{Q},
\mathbf{P}) = \sum_iP_i(\mathbf{x})\nabla Q^i(\mathbf{x})$ is a conserved quantity.

\item
In the language of fluid mechanics, the expression of $\mathbf{m}$ in
terms of $(\mathbf{Q}, \mathbf{P})$ is an example of a   \emph{Clebsch
representation}, which expresses the solution of the EPDiff equations in
terms of canonical variables that evolve by standard canonical Hamilton
equations.  This has been known in the case of fluid mechanics for more
than 100 years.  For modern discussions of the Clebsch representation for
ideal fluids, see, for example, Holm and Kupershmidt~\cite{HoKu1983} and
Marsden and Weinstein~\cite{MaWe1983}. 

\item
One more remark is in order; namely the special case in which $S = M $ is
of course allowed. In this case, $ \mathbf{Q}$ corresponds to the map $\eta$ itself and
$\mathbf{P}$ just corresponds to its conjugate momentum. The quantity $\mathbf{m}$
corresponds to the spatial (dynamic) momentum density (that is, right translation of
$\mathbf{P}$ to the identity), while $\mathbf{M} $ corresponds to the conserved ``body''
momentum density (that is, left translation of $\mathbf{P}$ to the identity).
\end{itemize}
\end{remark}

\subsection{Brief summary}

$\operatorname{Emb}(S,\mathbb{R}^n)$ admits two group actions. These are: the group 
$\operatorname{Diff} (S)$ of diffeomorphisms of $S$, which acts by composition on
the \textit{ right}; and the group $\operatorname{Diff}(\mathbb{R}^n)$ which acts 
by composition on the \textit{ left}. The group
$\operatorname{Diff}(\mathbb{R}^n)$ acting from the left produces the singular solution
momentum map, $\mathbf{J}_{\mathrm Sing}$. The action of $\operatorname{Diff} (S)$ from the
right produces the conserved momentum map $\mathbf{J}_S{:\ }  T ^{\ast} 
\operatorname{Emb}(S, \mathbb{R}^n) \rightarrow
\mathfrak{X}(S)^{\ast}$. We now assemble both momentum maps into
one figure as follows \cite{HoMa2004}: \vspace{-3mm}

\begin{picture}(150,100)(-50,0)%
\put(100,75){$T^{\ast} \operatorname{Emb}(S,M)$} 

\put(78,50){$\mathbf{J}_{\mathrm Sing}$}        

\put(160,50){$\mathbf{J}_S$}   

\put(72,15){$\mathfrak{X} (M)^{\ast}$}       

\put(170,15){$\mathfrak{X}(S)^{\ast}$}       

\put(130,70){\vector(-1, -1){40}}  

\put(135,70){\vector(1,-1){40}}  

\end{picture}

For a full discussion of this dual pair for continuum dynamics, see \cite{gay2012dual}. 
We turn next to the Euler--Poincar\'e framework for continuum dynamics.

\newpage
\part{Euler--Poincar\'e framework of \\Continuum Partial Differential Equations}

\begin{figure}[H]
\begin{center}
\includegraphics[width=\textwidth]{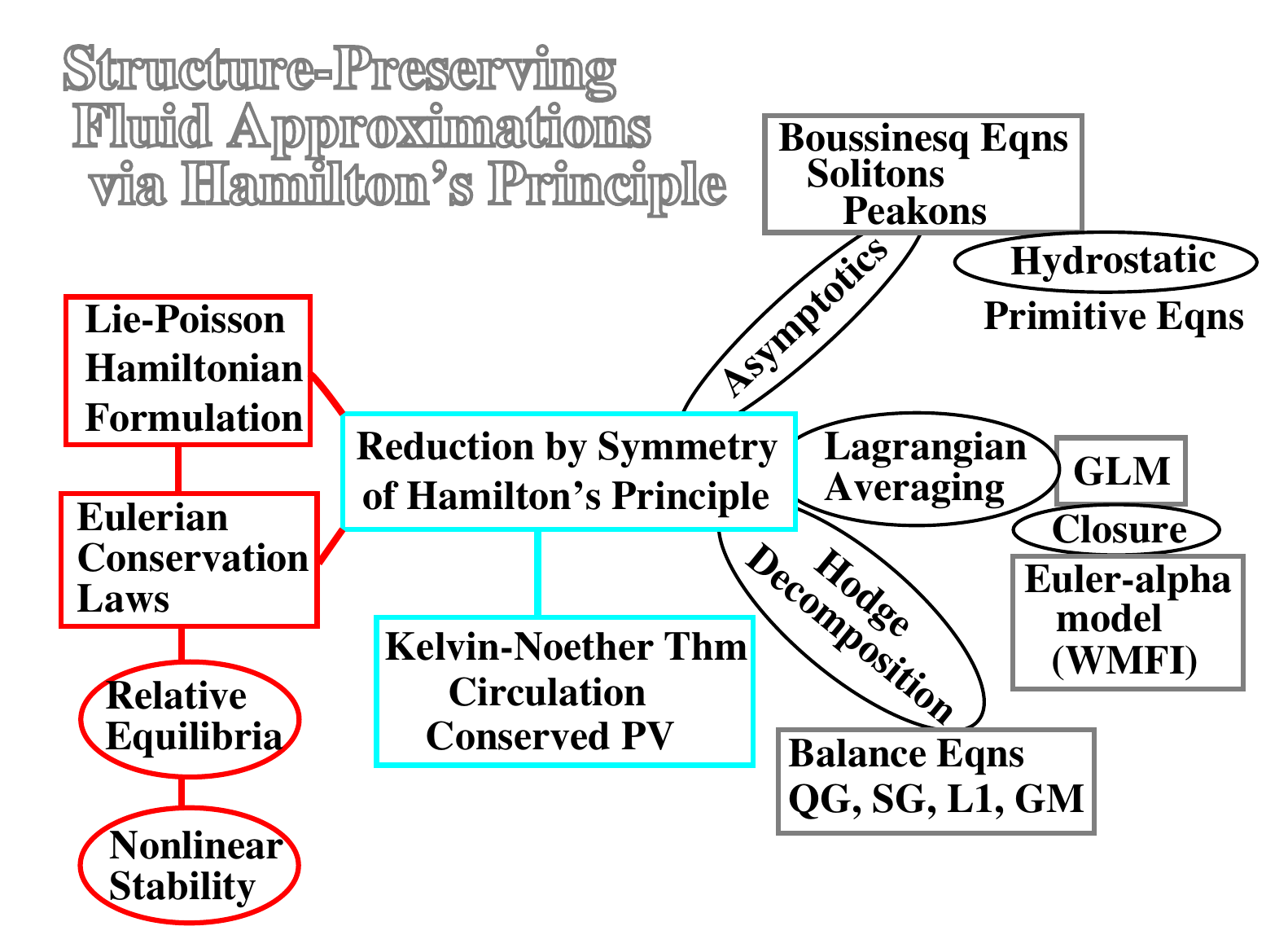}
\caption{Structure-preserving fluid approximations via Hamilton's principle}
\end{center}
\end{figure}\vspace{-5mm}
\vspace{4mm}\centerline{\textcolor{shadecolor}{\rule[0mm]{6.75in}{-2mm}}\vspace{-4mm}}

\newpage

\newpage
\vspace{4mm}\centerline{\textcolor{shadecolor}{\rule[0mm]{6.75in}{-2mm}}\vspace{-4mm}}
\section{Euler--Poincar\'e framework of fluid dynamics}
\label{sec-EPframe}

\secttoc

\textbf{What is this lecture about?} This lecture assembles the
Euler--Poincar\'e framework of the semidirect-product Lie--Poisson structure of 
ideal fluid dynamics, step by step from the set up
to the Kelvin-Noether theorem. \index{semidirect product!Kelvin theorem}

\subsection{Set up and basic assumptions}

Almost all fluid models of interest admit the following general
assumptions. These assumptions form the basis of the Euler--Poincar\'e
theorem for Continua that we shall state later in this section,
after introducing the notation necessary for dealing geometrically
with the reduction of Hamilton's Principle from the material (or
Lagrangian) picture of fluid dynamics, to the spatial (or Eulerian)
picture. This theorem was first stated and proved by Holm, Marsden and
Ratiu~\cite{HoMaRa1998a}, to which we refer for additional details,
as well as for abstract definitions and proofs.

\textbf{Basic assumptions underlying the Euler--Poincar\'e theorem for continua}
\begin{itemize}
\item There is a \textit{ right\/} representation of a Lie group $G$ on
the vector space $V$ and $G$ acts in the natural way on the \textit{
right\/} on $TG \times V^\ast$: $(U_g, a)h = (U_gh, ah)$.
\item The Lagrangian function $ L {:\ }  T G \times V ^\ast
\rightarrow \mathbb{R}$ is right $G$--invariant under the isotropy group of $a_0\in V ^\ast$.%
\footnote{For fluid dynamics, right $G$--invariance of the Lagrangian
function $L$ is traditionally called ``particle relabeling symmetry.''}
\item In particular, if $a_0 \in V^\ast$, define the
Lagrangian $L_{a_0} {:\ }  TG \rightarrow \mathbb{R}$ by
$L_{a_0}(U_g) = L(U_g, a_0)$. Then $L_{a_0}$ is right
invariant under the lift to $TG$ of the right action of
$G_{a_0}$ on $G$, where $G_{a_0}$ is the isotropy group of $a_0$.
\item  Right $G$--invariance of $L$ permits one to define the Lagrangian
on the Lie algebra $\mathfrak{g}$ of the group $G$. Namely,
$\ell{:\ }  {\mathfrak{g}} \times V^\ast \rightarrow \mathbb{R}$
is defined by, 
\[
\ell({u},a)
=
L\big(U_gg^{-1}(t), a_0g^{-1}(t)\big) 
= L(U_g, a_0)
,
\]
where
$
{u}=U_gg^{-1}(t)
$ and
$
a = a_0g^{-1}(t)
.
$
Conversely,  this relation defines for any
$\ell{:\ }  {\mathfrak{g}} \times V^\ast \rightarrow
\mathbb{R} $ a right $G$--invariant function
$ L {:\ }  T G \times V ^\ast
\rightarrow \mathbb{R} $.
\item For a curve $g(t) \in G, $ let
$u  (t) := \dot{g}(t) g(t)^{-1}$ and define the curve
$a(t)$ as the unique solution of the linear differential equation
with time dependent coefficients $\dot a(t) = -a(t)u (t)$, where the action
of an element of the Lie algebra $u\in\mathfrak{g}$ on an advected quantity
$a\in V^*$ is denoted by concatenation from the right.
The solution with initial condition $a(0)=a_0\in V^*$ can be
written as $a(t) = a_0g(t)^{-1}$.
\end{itemize}

\subsection{Notation for reduction of Hamilton's Principle by
symmetries}
\begin{itemize}
\item
Let $\mathfrak{g}(\mathcal{D})$ denote the space of
vector fields on $\mathcal{D}$ of some fixed differentiability class.
These vector fields are endowed with the \emph{ Lie bracket\/} given in
components by (summing on repeated indices)
\begin{equation}\label{jlb} [\mathbf{u}, \mathbf{v}]^i
= 
u^j\frac{\partial v^i}{\partial x^j} 
 - 
v^j\frac{\partial u^i}{\partial x^j}
.
\end{equation}
The notation $
\operatorname{ad}_{\mathbf{u}} \mathbf{v} := [\mathbf{u}, \mathbf{v}]$
formally denotes the adjoint action of the \textit{ right\/} Lie algebra of
$\operatorname{Diff}(\mathcal{D})$ on itself.

\item
Identify the Lie algebra of vector fields $\mathfrak{g}$ with its 
dual $\mathfrak{g}^\ast$ by using the $L^2$ pairing
\begin{equation}\label{l2p}
\left\langle \mathbf{ u},\mathbf{v}\right\rangle
=\int_{\mathcal{D}} \mathbf{ u}
\cdot
\mathbf{v}
\,  dV.
\end{equation}
%
\item
Let $\mathfrak{g}(\mathcal{D})^\ast$ denote the geometric dual
space of $\mathfrak{g}(\mathcal{D})$, that is,
$\mathfrak{g}(\mathcal{D})^\ast := \Lambda^1({\mathcal{D}}) \otimes
{\mathrm Den}({\mathcal{D}})$. This is the space of one--form densities on
$\mathcal{D}$. If $\mathbf{m}\, \otimes  dV \in
\Lambda^1({\mathcal{D}}) \otimes {\mathrm Den}({\mathcal{D}})$, then the
pairing of  $\mathbf{m} \otimes  dV$ with $\mathbf{u} \in
\mathfrak{g}(\mathcal{D})$ is given by the $L^2$ pairing,
\begin{equation}\label{continuumpairing}
\langle \mathbf{m} \otimes  dV, \mathbf{u} \rangle
= \int_{\mathcal{D}} \mathbf{m}\cdot \mathbf{u}\,  dV
\end{equation}
where $\mathbf{m}\cdot \mathbf{u}$ is the standard contraction of a
one--form $\mathbf{m}$ with a vector field $\mathbf{u}$. 
\item
For $\mathbf{u} \in \mathfrak{g}(\mathcal{D})$ and $\mathbf{m}\otimes 
dV \in \mathfrak{g}(\mathcal{D})^\ast$, the dual of the adjoint
representation is defined by
\begin{equation}\label{ad-star-def}
 \langle  \operatorname{ad}^\ast_{\mathbf{u}}(\mathbf{m}\otimes  dV),
\mathbf{v}
\rangle
= -\int_{\mathcal{D}}\mathbf{m} \cdot
\operatorname{ad}_{\mathbf{u}}\!\mathbf{v}\,dV 
= -\int_{\mathcal{D}}\mathbf{m}
\cdot [\mathbf{u}, \mathbf{v}]\,dV
\end{equation}
and its expression is
\begin{equation}\label{continuumcoadjoint}
 \operatorname{ad}^\ast_{\mathbf{u}}(\mathbf{m}\otimes  dV) 
= (\pounds_{\mathbf{u}}
\mathbf{m} + ( \div_{dV}\mathbf{u})\mathbf{m})\otimes  dV
= \pounds_{\mathbf{u}}(\mathbf{m}\otimes  dV),
\end{equation}
where  $\div_{dV}\mathbf{u}$ is the divergence of $\mathbf{u}$ relative
to the measure $dV$, that is, $\pounds_{\mathbf{u}}dV = (\div_{dV}\mathbf{u})dV$. Hence,
$\operatorname{ad}^\ast_{\mathbf{u}}$ coincides with the Lie-derivative
$\pounds_{\mathbf{u}}$ for one-form densities. 
\item
If $\mathbf{u} = u^j \partial/\partial x^j, 
\mathbf{m}  = m_i dx^i$, then the one--form factor in the preceding
formula for $ \operatorname{ad}^\ast_{\mathbf{u}}(\mathbf{m}\otimes  dV)$ has the
\emph{coordinate expression}
\begin{eqnarray}\label{continuumcoadjoint-coord}
\big(\operatorname{ad}^\ast_{\mathbf{u}}\mathbf{m}\big)_i dx^i
&=& \left ( u^j \frac{\partial
m_i}{\partial
x^j} + m_j \frac{\partial u^j}{\partial x^i} +
( \div_{dV}\mathbf{u})m_i \right )dx^i \\
&=&
\left (\frac{\partial}{\partial x^j}(u^jm_i) +
m_j \frac{\partial u^j}{\partial x^i}\right ) dx^i.
\end{eqnarray}
The last equality assumes that the divergence is taken
relative to the standard measure $dV = d^n\mathbf{x}$ in $\mathbb{R} ^n$.
(On a Riemannian manifold the metric divergence needs to be used.)
\end{itemize}

\subsection{Conventions and terminology in continuum mechanics}
Throughout the rest of the lecture notes, we shall follow
Holm, Marsden and Ratiu~\cite{HoMaRa1998a}  in using the conventions
and terminology for the standard quantities in continuum mechanics.

\begin{definition}
Elements of $\mathcal{D}$ representing the material particles of the system
are denoted by $X$; their coordinates $X^A, A=1,\ldots,n$ may thus be
regarded as the \emph{particle labels}. 
\begin{itemize}
\item
A \emph{configuration}, which we typically denote by
$\eta$, or $g$, is an element of $\operatorname{Diff}(\mathcal{D})$.
\item
A \emph{motion\/}, denoted as $\eta_t$ or alternatively as $g(t)$, is a time
dependent curve in $\operatorname{Diff}(\mathcal{D})$. 
\end{itemize}

\end{definition}


\begin{definition}
The \emph{Lagrangian}, or \emph{material velocity\/} $\mathbf{U}(X,t)$ of the
continuum along the motion
$\eta_t$ or $g(t)$ is defined by taking the time derivative of the motion
keeping the particle labels $X$ fixed:
\[
\mathbf{U}(X, t) := \frac{d\eta_t(X)}{dt}:=
\left.\frac{\partial}{\partial t}\right|_{X}\eta_t(X)
:= \dot{g}(t)\cdot X
.
\]
These are convenient shorthand notations for the
time derivative at fixed Lagrangian coordinate $X$.

Consistent with this definition of material velocity, the tangent space to
$\operatorname{Diff}(\mathcal{D})$ at $\eta \in
\operatorname{Diff}(\mathcal{D})$ is given by
\[
T_{\eta} \operatorname{Diff}(\mathcal{D})
= \{ \mathbf{U}_{\eta}{:\ }  {\mathcal{D}} \rightarrow T {\mathcal{D}}  \mid \mathbf{U}_{\eta}(X)
\in T_{\eta(X)}\mathcal{D} \}.
\]
Elements of $T_{\eta} \operatorname{Diff}(\mathcal{D})$ are usually
thought of as vector fields on $\mathcal{D}$ covering $\eta$. The
tangent lift of right translations on
$T\operatorname{Diff}(\mathcal{D})$ by $\varphi \in
\operatorname{Diff}(\mathcal{D})$
is given by
\[
\mathbf{U}_{\eta}\varphi := T_{\eta} R_\varphi (\mathbf{U}_{\eta})
= \mathbf{U}_{\eta} \circ \varphi.
\]
\end{definition}

\begin{definition}
During a motion $\eta_t$ or $g(t)$, the particle labeled by $X$ describes a
path in $\mathcal{D}$, whose points 
\[x(X, t):= \eta_t(X):=g(t)\cdot X,\]
are called the \emph{Eulerian} or \emph{spatial points\/} of this path,
which is also called the \emph{Lagrangian trajectory\/}, because a
Lagrangian fluid parcel follows this path in space. The derivative
$\mathbf{u}(x, t)$ of this path, evaluated at fixed Eulerian point $x$, is
called the \emph{Eulerian} or \emph{spatial velocity\/} of the system:
\[
\mathbf{u}(x, t)
:= \mathbf{u}(\eta_t(X), t)
:= \mathbf{U}(X, t)
:= \left.\frac{\partial}{\partial t}\right|_X\eta_t(X)
:= \dot{g}(t) \cdot X
:= \dot{g}(t)g^{-1}(t)\cdot x
.
\]
Thus the Eulerian velocity $\mathbf{u}$ is a time dependent vector
field on $\mathcal{D}$, denoted as $\mathbf{u}_t \in
\mathfrak{g}(\mathcal{D})$, where $\mathbf{u}_t(x) := \mathbf{u}(x, t)$.
We also have the fundamental relationships
\[
\mathbf{U}_t = \mathbf{u}_t \circ \eta_t
\quad\hbox{and}\quad
\mathbf{u}_t=\dot{g}(t)g^{-1}(t)
,
\]
where we denote $\mathbf{U}_t(X):= \mathbf{U}(X, t)$.
\end{definition}


\begin{definition}
The \emph{representation space} $V^\ast$ of
$\operatorname{Diff}(\mathcal{D})$ in continuum mechanics is often some
subspace of the tensor field densities  on $\mathcal{D}$, denoted as
$\mathfrak{T} (\mathcal{D})\otimes {\mathrm Den}({\mathcal{D}})$, and the
representation is given by pull back. It is thus a \textit{ right\/}
representation of
$\operatorname{Diff}(\mathcal{D})$ on
$\mathfrak{T}(\mathcal{D})\otimes {\mathrm Den}({\mathcal{D}})$. The right
action of the Lie algebra $\mathfrak{g}({\mathcal{D}})$ on $V^\ast$ is
denoted as \emph{concatenation from the right}. That is, we denote 
\[
a\mathbf{u} :=
\pounds_{\mathbf{u}} a
,
\]
which is the Lie derivative of the tensor field density
$a$ along the vector field $\mathbf{u}$.
\end{definition}

\begin{definition}
The \emph{Lagrangian of a continuum mechanical system} is a function 
\[
L{:\ } 
T\operatorname{Diff}(\mathcal{D}) \times V^\ast \rightarrow \mathbb{R} 
,
\]
which is right invariant relative to the tangent lift of right
translation of $\operatorname{Diff}(\mathcal{D})$ on itself and pull
back on the tensor field densities.
Invariance of the Lagrangian $L$ induces
a function $\ell{:\ }  \mathfrak{g}(\mathcal{D}) \times V^\ast
\rightarrow \mathbb{R} $ given by
\[
\ell(\mathbf{u}, a) 
= L(\mathbf{u}\circ \eta, \eta^\ast a)
= L(\mathbf{U}, a_0)
,
\]
where $  \mathbf{u} \in \mathfrak{g}({\mathcal{D}})$ and $  a \in V^\ast
\subset {\mathfrak{T}}({\mathcal{D}})\otimes {\mathrm Den}({\mathcal{D}})$,
and where $\eta^\ast a$ denotes the pull back of $a$ by the
diffeomorphism $\eta$ and $\mathbf{u}$ is the Eulerian velocity. 
That is, 
\begin{equation}\label{fund-relns}
\mathbf{U}=\mathbf{u}\circ \eta
\quad\hbox{and}\quad
a_0 = \eta^\ast a
.
\end{equation}
The evolution of $a$ is by right action, given by the equation
\begin{equation}\label{eqn-adv-qty}
\dot a = -{\pounds}_{\mathbf{u}}\, a = - a\mathbf{u}.
\end{equation}
The solution of this equation, for the
initial condition $a_0$, is 
\begin{equation}\label{soln-adv-qty}
a(t) = \eta_{t\ast} a_0=a_0g^{-1}(t)
,
\end{equation}
where the lower star denotes the push-forward operation and
$\eta_t$ is the flow of $\mathbf{u}=\dot{g}g^{-1}(t)$.
\end{definition}

\begin{definition}
\emph{Advected Eulerian quantities} are defined in continuum mechanics to
be those variables which are Lie transported by the flow of the Eulerian
velocity field. Using this standard terminology, 
equation \eqref{eqn-adv-qty}, or its solution \eqref{soln-adv-qty} states
that the tensor field density $a(t)$ (which may include mass density and
other Eulerian quantities) is advected.
\end{definition}

\begin{remark}\rm [Dual tensors]
As we mentioned, typically $V^\ast \subset
{\mathfrak{T}}({\mathcal{D}})\otimes {\mathrm Den}({\mathcal{D}})$ for
continuum mechanics.  On a general manifold, tensors of a given type have
natural duals. For example, symmetric covariant tensors are dual to
symmetric contravariant tensor densities, the pairing being given by the
integration of the natural contraction of these tensors. Likewise,
$k$--forms are naturally dual to
$(n-k)$--forms, the pairing being given by taking the integral of
their wedge product.
\end{remark}

\begin{definition}
The \emph{diamond operation} $\diamond$ between elements of $V$ and
$V^\ast$  produces an element of the dual Lie algebra
$\mathfrak{g}({\mathcal{D}})^\ast$ and is defined as 
\begin{equation}\label{continuumdiamond}
\langle b \diamond a, \mathbf{w}\rangle
= -\int_{\mathcal{D}} b \cdot \pounds_{\mathbf{w}}\,a\;,
\end{equation}
where $b\cdot \pounds_{\mathbf{w}}\,a$ denotes the contraction, as
described above, of elements of $V$ and elements of $V^\ast$ and
$\mathbf{w}\in\mathfrak{g}({\mathcal{D}})$. (These operations do \textit{ not}
depend on a Riemannian structure.)
\end{definition}

For a path $\eta_t \in \operatorname{Diff}(\mathcal{D})$, let 
$\mathbf{u}(x, t)$
be its Eulerian velocity and consider the curve $a(t)$ with initial
condition $a_0$ given by the equation
\begin{equation}
\p_t a + \pounds_{\mathbf{u}} a = 0.
\label{continuityequation}
\end{equation}
Let the Lagrangian $L_{a_0}(\mathbf{U}) := L(\mathbf{U}, a_0)$ be right-invariant
under $\operatorname{Diff}(\mathcal{D})$. We can now state the
Euler--Poincar\'{e} Theorem for Continua of Holm, Marsden and Ratiu~\cite{HoMaRa1998a}.

\subsection{Statement of the Euler--Poincar\'{e} Theorem for Continua}

\begin{theorem}[Euler--Poincar\'{e} Theorem for Continua]
\label{EPforcontinua}
Given a path $\eta_t$ in  $\operatorname{Diff}(\mathcal{D})$ with
Lagrangian velocity $\mathbf{U}$ and Eulerian velocity $\mathbf{u}$, the
following are equivalent:
\begin{enumerate}
\item [(i)] Hamilton's variational principle
\begin{equation} \label{continuumVP}
\delta \int_{t_1}^{t_2} L\left(X, \mathbf{U}_t (X),
a_0(X)\right)\,dt=0
\end{equation}
holds, for variations $\delta\eta_t$ vanishing at the endpoints.
\item [(ii)] $\eta_t$ satisfies the Euler--Lagrange
equations for $L_{a_0}$ on
$\operatorname{Diff}(\mathcal{D})$.
\item [(iii)] The constrained variational principle in
Eulerian coordinates
\begin{equation}\label{continuumconstrainedVP}
   \delta \int_{t_1}^{t_2} \ell(\mathbf{u},a)\ dt=0
\end{equation}
holds on $\mathfrak{g}(\mathcal{D}) \times V^\ast$, using
variations of the form
\begin{equation}\label{continuumvariations}
   \delta \mathbf{u} = \frac{\partial \mathbf{w}}{\partial t}
                   +[\mathbf{u},\mathbf{w}]
= \frac{\partial \mathbf{w}}{\partial t}
                   +{\mathrm\,ad\,}_{\mathbf{u}}\mathbf{w}
, \qquad
   \delta a = - \pounds_{\mathbf{w}}\,a,
\end{equation}
where $\mathbf{w}_t = \delta\eta_t \circ \eta_t^{-1}$ vanishes at
the endpoints.
\item [(iv)] The Euler--Poincar\'{e} equations for continua
\begin{equation}\label{continuumEP}
   \frac{\partial }{\partial t}\frac{\delta \ell}{\delta \mathbf{u}}
   = - \operatorname{ad}^{\ast}_{\mathbf{u}}\frac{\delta \ell}
        {\delta \mathbf{u}}
   +\frac{\delta \ell}{\delta a}\diamond a
   =-\pounds_{\mathbf{u}} \frac{\delta \ell}{\delta \mathbf{u}}
    +\frac{\delta \ell}{\delta a}\diamond a,
\end{equation}
hold, with auxiliary equations $(\partial_t + \pounds_{\mathbf{u}})a = 0$
for each advected quantity $a(t)$. The
$\diamond$ operation defined in \eqref{continuumdiamond} needs to be
determined on a case by case basis, depending on the nature of the tensor
$a(t)$. The variation $\mathbf{m}=\delta \ell/\delta \mathbf{u}$ is a
one--form density and we have used relation \eqref{continuumcoadjoint} in
the last step of equation \eqref{continuumEP}.
\end{enumerate}
\end{theorem}

We refer to Holm, Marsden and Ratiu~\cite{HoMaRa1998a} for the proof of this theorem in the abstract setting. We shall see some of the features of this result
in the concrete setting of continuum mechanics shortly.

\subsection{Discussion of the Euler--Poincar\'e equations}

The following string of equalities
shows \emph{directly} that (iii) is equivalent to (iv):
\begin{eqnarray}\label{continuumEPderivation}
0
&=&\delta \int_{t_1}^{t_2} l(\mathbf{u}, a) dt
=\int_{t_1}^{t_2}\left(\frac{\delta l}{\delta \mathbf{u}}\cdot
\delta\mathbf{u} +\frac{\delta l}{\delta a}\cdot \delta a\right)dt
\nonumber \\
&=&\int_{t_1}^{t_2} \left[\frac{\delta l}{\delta \mathbf{u}}
\cdot \left(\frac{\partial\mathbf{w}}
{\partial t}-\ad_{\mathbf{u}}\,\mathbf{w}\right) -\frac{\delta
l}{\delta a}\cdot \pounds_{\mathbf{w}}\, a \right]dt
\nonumber \\
&=&\int_{t_1}^{t_2} \mathbf{w}\cdot
\left[-\frac{\partial}{\partial t}
\frac{\delta l}{\delta\mathbf{u}} -\ad^*_{\mathbf{u}}\frac
{\delta l}{\delta\mathbf{u}} +\frac{\delta l}{\delta a} \diamond
a\right]dt.
\end{eqnarray}
The rest of the proof follows essentially the same track as the proof of the pure
Euler--Poincar\'e theorem, modulo slight changes to accommodate the advected quantities. 
\index{Euler--Poincar\'e equations!theorem}

In the absence of dissipation, most Eulerian fluid equations%
\footnote{Exceptions to this statement are certain multiphase fluids, and
complex fluids with active internal degrees of freedom such as liquid
crystals. These require a further extension, not discussed here.} 
can be written in the EP form in equation \eqref{continuumEP}, 
\begin{equation}\label{EPeqn}
\frac{\partial}{\partial t} \frac{\delta\ell}{\delta\mathbf{u}}
+
\ad_{\mathbf{u}}^*\frac{\delta\ell}{\delta\mathbf{u}}
=
\frac{\delta\ell}{\delta{a} }\diamond{a}
,\quad\hbox{with}\quad
\big(\partial_t + \pounds_{\mathbf{u}}\big)a = 0
.
\end{equation}
Equation \eqref{EPeqn} is \emph{Newton's Law}: The Eulerian time derivative
of the momentum density $\mathbf{m}=\delta\ell/\delta\mathbf{u}$ (a
one-form density dual to the velocity $\mathbf{u}$) is equal to the force
density $(\delta\ell/\delta a)\diamond a$, with the $\diamond$ operation
defined in \eqref{continuumdiamond}.
Thus, Newton's Law is written in the Eulerian fluid representation as%
\footnote{
In coordinates, a one-form density takes the form
$\mathbf{m}\cdot d\mathbf{x}\otimes{dV}$ and the EP equation
\eqref{continuumEP} is given  mnemonically by
\[ \frac{d}{dt}\Big|_{\mathrm{Lag}}\big(\mathbf{m}\cdot
d\mathbf{x}\otimes{dV}\big) = \underbrace{
\frac{d\mathbf{m}}{dt}\Big|_{\mathrm{Lag}}\hspace{-3mm}\cdot
d\mathbf{x}\otimes{dV}}_{\hbox{Advection}} +
\underbrace{\mathbf{m}\cdot d\mathbf{u}\otimes{dV}}_{\hbox{Stretching}} +
\underbrace{\mathbf{m}\cdot
d\mathbf{x}\otimes(\nabla\cdot\mathbf{u}){dV}}_{\hbox{Expansion}}
= \frac{\delta\ell}{\delta{a} }\diamond{a} \]
with
$\smash{\frac{d}{dt}\big|_{\mathrm{Lag}}}d\mathbf{x} :=
\big(\partial_t + \pounds_{\mathbf{u}}\big)d\mathbf{x}
=d\mathbf{u}=\mathbf{u}_{,j}dx^j$, upon using commutation of Lie derivative and exterior derivative.
Compare this formula with the definition of 
$ \operatorname{ad}^\ast_{\mathbf{u}}(\mathbf{m}\otimes  dV)$ in
equation \eqref{continuumcoadjoint-coord}.}
\begin{equation}\label{EPeqn-m}
\frac{d}{dt}\bigg|_{\mathrm{Lag}} \mathbf{m} :=
\big(\partial_t + \pounds_{\mathbf{u}}\big)\mathbf{m} =
\frac{\delta\ell}{\delta{a} }\diamond{a},\quad\hbox{with}\quad
\frac{d}{dt}\bigg|_{\mathrm{Lag}}a :=
\big(\partial_t + \pounds_{\mathbf{u}}\big)a = 0.
\end{equation}
\begin{itemize}
\item
The left side of the EP equation in \eqref{EPeqn-m} describes the fluid's
dynamics due to its kinetic energy. A fluid's kinetic energy typically
defines a norm for the Eulerian fluid velocity,
$KE=\smash{\frac12}\|\mathbf{u}\|^2$. The left side of the EP equation is the
\emph{geodesic} part of its evolution, with respect to this norm.
See Arnold and Khesin~\cite{ArKh1998} for discussions of this interpretation of ideal
incompressible flow and references to the literature. However, in a
gravitational field, for example, there will also be dynamics due to
potential energy.  And this dynamics will by governed by the right side of
the EP equation. 

\item
The right side of the EP equation in \eqref{EPeqn-m} modifies the geodesic
motion. Naturally, the right side of the EP equation is also a geometrical
quantity. The diamond operation $\diamond$ represents the dual of the Lie
algebra action of vectors fields on the tensor $a$.  Here
$\delta\ell/\delta{a}$ is the dual tensor, under the natural pairing
(usually, $L^2$ pairing) $\langle\cdot ,\cdot\,\rangle$ that is induced
by the variational derivative of the Lagrangian $\ell(\mathbf{u},a)$. The
diamond operation $\diamond$ is defined in terms of this pairing in
\eqref{continuumdiamond}. For the $L^2$ pairing, this is integration
by parts of (minus) the Lie derivative in \eqref{continuumdiamond}. 

\item
The quantity $a$ is typically a tensor (for example, a density, a scalar, or a
differential form) and we shall sum over the various types of tensors
$a$ that are involved in the fluid description.  The second equation in
\eqref{EPeqn-m} states that each tensor
$a$ is carried along by the Eulerian  fluid velocity $\mathbf{u}$. Thus,
$a$ is for fluid ``attribute,'' and its Eulerian evolution is given by 
minus its Lie derivative, $-\pounds_{\mathbf{u}}a$. That is, $a$ stands for
the set of fluid attributes that each Lagrangian fluid  parcel carries
around (advects), such as its buoyancy, which is determined by its
individual salt, or heat content, in ocean circulation.  

\item
Many examples of how equation \eqref{EPeqn-m}
arises in the dynamics of continuous media are given by Holm, Marsden and
Ratiu~\cite{HoMaRa1998a}.
The EP form of the Eulerian fluid description in \eqref{EPeqn-m} is
analogous to the classical dynamics of rigid bodies (and tops, under
gravity) in body coordinates. Rigid bodies and tops are also governed by
Euler--Poincar\'e equations, as Poincar\'e showed in a
\index{Euler--Poincar\'e equations!rigid body}
two-page paper with no references, over a century ago
\cite{poincare1901forme}. For modern discussions of the EP theory, see, for example, 
Marsden and Ratiu~\cite{MaRa1994}, or Holm, Marsden and Ratiu~\cite{HoMaRa1998a}. 
\end{itemize}

\begin{exercise}
For what types of tensors $a_0$ can one recast the EP equations for continua
\eqref{continuumEP} as geodesic motion, by using a version of the Kaluza--Klein
construction?
\end{exercise}
\vspace{4mm}\centerline{\textcolor{shadecolor}{\rule[0mm]{6.75in}{-2mm}}\vspace{-4mm}}

\subsection{Corollary: the EP theorem implies the Kelvin--Noether theorem}
\begin{corollary}[Kelvin--Noether Circulation Theorem]\label{KelThmforcontinua}
Assume $\mathbf{u}(x, t)$
satisfies the Euler--Poincar\'e equations for continua:
\[ \frac{\partial}{\partial t}\left(\frac{\delta\ell}{\delta\mathbf{u}}\right)
   = -\pounds_{\mathbf{u}} \left(\frac{\delta \ell}{\delta \mathbf{u}}\right)
    +\frac{\delta \ell}{\delta a}\diamond a \]
and the quantity $a$ satisfies the \emph{advection relation}
\begin{equation}\label{advect-def}
\frac{\partial a}{\partial t} +
\pounds_{\mathbf{u}} a = 0.
\end{equation}
Let $\eta_t$ be the flow of the Eulerian
velocity field $\mathbf{u}$, that is, $\mathbf{u}  =
(d\eta_t/dt)\circ \eta_t^{-1}$. Define the advected fluid loop 
$\gamma_t  := \eta_t\circ \gamma_0$ and the circulation map $I(t)$ by
\begin{equation}\label{circ-def}
I(t) = \oint_{\gamma_t }\frac{1}{D }\frac{\delta \ell}{\delta \mathbf{u}}
.\end{equation}
In the circulation map $I(t)$ the advected mass density $D_t$ satisfies
the push-forward relation $D_t=\eta_*D_0$. This implies the 
advection relation \eqref{advect-def} with $a=D$, namely, the continuity
equation,
\[
\partial_tD+\div D\mathbf{u}=0
.
\]
Then the map $I(t)$
satisfies the \emph{Kelvin circulation relation},
\begin{equation}\label{KN-theorem}
\frac{d}{dt}I(t) = \oint_{\gamma_t }
\frac{1}{D}\frac{\delta \ell}{\delta a}\diamond a\;.
\end{equation}
\end{corollary}
Both an abstract proof of the Kelvin--Noether Circulation Theorem and a
proof tailored for the case of continuum mechanical systems are given in
Holm, Marsden and Ratiu~\cite{HoMaRa1998a}. We provide a version of the
latter below.

\begin{proof} First we change variables in the expression
for $I(t)$:
\[ I(t) = \oint_{\gamma_t}\frac{1}{D_t}
\frac{\delta l}{\delta \mathbf{u}}
=\oint_{\gamma_0} \eta_t^\ast\left[\frac{1}{D_t}\frac{\delta l}
{\delta \mathbf{u}}\right] = \oint_{\gamma_0}
\frac{ 1 }{ D _0 } \eta_t^\ast\left[\frac{\delta
l} {\delta \mathbf{u}}\right]. \]
Next, we use the Lie derivative formula, namely
\[
\frac{d}{dt}\left(\eta_t^*\alpha_t\right) =
\eta_t^*\left(\frac{\partial}{\partial t}\alpha_t
+ \pounds_{\mathbf{u}} \alpha_t \right)\;,
\]
applied to a one--form density $\alpha_t$.
This formula gives
\begin{align*}
      \frac{d}{dt} I(t)
  = \frac{d}{dt}  \oint_{\gamma_0}
\frac{ 1 }{ D _0 } \eta_t^\ast\left[\frac{\delta
l} {\delta \mathbf{u}}\right]
  = \oint_{\gamma_0} \frac{1}{D _0} \frac{d}{dt}
\left( \eta_t^\ast\left[
\frac{\delta l} {\delta \mathbf{u}}\right]\right) \\
  = \oint_{\gamma_0} \frac{1}{D _0} \eta_t^*\left[
\frac{\partial}{\partial t}
\left(\frac{\delta l} {\delta \mathbf{u}}\right) +
\pounds_{\mathbf{u}}
\left(\frac{\delta l} {\delta \mathbf{u}} \right)\right].
\end{align*}
By the Euler--Poincar\'e equations \eqref{continuumEP}, this becomes
\[
      \frac{d}{dt} I(t)
  =   \oint_{\gamma_0} \frac{1}{D _0} \eta_t^*\left[
\frac{\delta  l}{\delta  a}
\diamond a \right] = \oint_{\gamma_t} \frac{1}{D _t} \left[
\frac{\delta  l}{\delta  a}
\diamond a \right],
\]
again by the change of variables formula.
\end{proof}

\begin{corollary}
Since the last expression holds for every loop $\gamma_t$, we may
write it as
\begin{equation}
\left(\frac{\partial}{\partial t} + \pounds_{\mathbf{u}} \right)
\frac{1}{D} \frac{\delta l}{\delta \mathbf{u}}
= \frac{1}{D} \frac{\delta  l}{\delta  a} \diamond a.
\label{KThm}
\end{equation}
\end{corollary}

\begin{remark}\rm 
The Kelvin--Noether theorem is called so here because its derivation relies
on the invariance of the Lagrangian $L$ under the particle relabeling
symmetry, and  Noether's theorem is associated with this symmetry. However,
the result \eqref{KN-theorem} is the \emph{Kelvin circulation theorem}: the
circulation integral $I(t)$ around any fluid loop ($\gamma_t$, moving with
the velocity of the fluid parcels $\mathbf{u}$) is invariant under the
fluid motion.  These two statements are equivalent.  We note that \emph{two
velocities} appear in the integrand $I(t)$: the fluid velocity $\mathbf{u}$
and $D^{-1}\delta\ell/\delta\mathbf{u}$. The latter velocity is the
momentum density $\mathbf{m}=\delta\ell/\delta\mathbf{u}$ divided by the
mass density $D$.  These two velocities are the basic ingredients for
performing  modeling and analysis in any ideal fluid problem. One
simply needs to put these ingredients together in the Euler--Poincar\'e
theorem and its corollary, the Kelvin--Noether theorem.
\end{remark}

\subsection{Active advection defined via composition of maps} 

In the Euler--Poincar\'e theorem, Eulerian fluid dynamics in a certain domain ${\cal D}$ is derived by applying for Hamilton's principle for a Lagrangian defined on 
$T(G \circledS V)/G =\mathfrak{g} \circledS V$ which leads to the Euler--Poincar\'e equations on $T^*(G \circledS V)/G = \mathfrak{g}^* \circledS V$.
One may ask, though, what happens when a degree of freedom being carried along in the frame of motion of an Euler--Poincar\'e flow in the Euler--Poincar\'e theorem 
is not just passively advected? Suppose instead that the advected degree of freedom has its own dynamical volition, as in the case of waves propagating through a flowing fluid?

\begin{exercise}
Formulate and analyse hybrid equations of motion on $\mathfrak{g}^* \circledS V_1 \times T^*V_2$ 

\begin{itemize}
\item
Use the Hamilton--Pontryagin principle to derive the equations of motion for an Euler--Poincar\'e system 
coupled to an additional \emph{canonical} degree of freedom whose symplectic dynamics takes place in
the frame of motion of the Euler--Poincar\'e system. 

The Hamilton--Pontryagin principle for such a system
is given by
\begin{align}
\begin{split}
0=\delta S = \delta \int_0^T \ell(u,a ) &+ \scp{m}{\dot{g}g^{-1}-u} + \scp{b}{a_0g^{-1} - a }
\\&+ \scp{p}{\p_t q + {\cal L}_u q} - h(q,p)\,dt
\,,\end{split}
\label{wave-action}
\end{align}
for $(a,b)\in T^*V_1$ and $(q,p)\in T^*V_2$
\item
Find the momentum $m\in \Lambda^1\otimes Den$ and $p \in T^*V_2$.
\item
Legendre transform to obtain the Hamiltonian in the variables 
$(m,q,p)$.
\item
Write the equations of motion for this system in Lie--Poisson bracket form.
\end{itemize}

\end{exercise}

\begin{answer}
\begin{itemize}
\item
The variations of the Lagrangian in Hamilton's principle \eqref{wave-action} 
\begin{align}
\begin{split}
\delta u:& \quad \frac{\delta \ell}{\delta u} - m   - p\diamond q = 0
\,,\\
\delta m:& \quad \dot{g}g^{-1} - u  = 0
\,,\\
\delta a:& \quad \frac{\delta \ell}{\delta a} - b  = 0
\,,\\
\delta b:& \quad a_0g^{-1} - a  = 0, \quad\Longrightarrow\quad  \p_t a + {\cal L}_{\dot{g}g^{-1}} a = 0
\,,\\
\delta  p:& \quad  \p_t q + {\cal L}_u q - \frac{\delta h}{\delta p} = 0
\,,\\
\delta  q:& \quad  -\p_t p + {\cal L}_u^T p - \frac{\delta h}{\delta q} = 0
\,,\\
\delta  g:& \quad  -(\p_t  + {\cal L}_{\dot{g}g^{-1}}) m + b\diamond a = 0
\,.
\end{split}
\label{wave-vars}
\end{align}
\textbf{Kelvin--Noether theorem.}
Because the mass density ($D$) is always among the variables $a$ advected by the flow, the last equation 
in \eqref{wave-vars} implies Kelvin's theorem:
\begin{align}
\frac{d}{dt} \oint_{C_t} D^{-1} m = \oint_{C_t} (\p_t  + {\cal L}_{\dot{g}g^{-1}}) (D^{-1} m) = \oint_{C_t} D^{-1}(b\diamond a)
\,,
\label{wave-Kelvin}
\end{align}
for a material loop pushed forward by the flow, as $C_t=g_*C_0$.

\item
The Legendre transform produces the system's Hamiltonian and its variations, as
\begin{align}
\begin{split}
h(m,a,q,p) &= \scp{m}{u} + \scp{p}{\p_tq} - \ell(u,a)
\\&\quad 
- \scp{p}{\p_t q + {\cal L}_u q} + h(q,p)
\\
&= \scp{m}{u} + \scp{p}{- {\cal L}_u q } - \ell(u,a) + h(q,p)
\\
&= \scp{m + p\diamond q }{u }  - \ell(u,a) + h(q,p)
\\
h(M,a,q,p) &=: \scp{M }{u }  - \ell(u,a) + h(q,p)
\\
\delta h(M,a,q,p)  &= \scp{\delta M}{u} + \scp{ M - \frac{\delta \ell}{\delta u} }{\delta u} 
- \scp{\frac{\delta \ell}{\delta a}}{\delta a}
\\&\quad + \scp{\delta  p}{- {\cal L}_u q + \frac{\delta h}{\delta p} } + \scp{- {\cal L}_u^T p + \frac{\delta h}{\delta q}}{\delta q} 
\,,\end{split}
\label{wave-LegTrans}
\end{align}
where the notation $M:= {\delta \ell}/{\delta u}  = m + p\diamond q$ for the total momentum has been introduced.

\item
The equations of motion for this hybrid system take the following block-diagonal bracket form, 
\begin{align}
\frac{d}{dt} \begin{pmatrix} M \\  a \\ p \\ q \end{pmatrix}  
= -  \begin{bmatrix} 
\p M + M \p & \Box \diamond a & 0 & 0  
\\ 
{\cal L}_{\Box}a & 0 & 0 & 0 
\\
0 & 0 & 0 & 1
\\
0 & 0 & -1 & 0
\end{bmatrix} 
\begin{pmatrix} 
\p h/\p M  = u \\ \p h/ \p a =  -\p \ell/ \p a   \\ \p h/ \p p  \\  \p h/ \p q
\end{pmatrix}
.\label{waveLP-brkt}
\end{align}
which comprises the sum of a Lie--Poisson bracket for the fluid and a symplectic bracket for the 
actively advected degree of freedom, coupled through their total momentum density, $M : = m + p\diamond q$.
\end{itemize}
\end{answer}

\begin{remark}\rm 
This exercise combines the dynamics of mass flow and dynamical advection, e.g., for wave propagation in a moving medium. 
This is accomplished by coupling the Hamilton--Pontryagin constrained Lagrangian (for the flow of the medium) 
with a phase-space Lagrangian (for the propagation of the waves in the frame of motion of the moving medium).
This coupling provides a means of describing how the reaction to wave propagation in the moving frame affects the motion
of the medium. This approach also applies to complex fluids such as liquid crystals (where order-parameter dynamics
takes place in a moving fluid) \cite{holm2002euler} and to superfluids (where the superfluid velocity is measured relative to the normal 
fluid motion) \cite{HoKu1982,holm1987superfluid}. These applications may all be recognised as examples of the 
composition of maps approach \cite{holm2023lagrangian}. \index{composition of maps}
\end{remark}


\newpage
\vspace{4mm}\centerline{\textcolor{shadecolor}{\rule[0mm]{6.75in}{-2mm}}\vspace{-4mm}}
\section{Euler--Poincar\'{e} theory of Geophysical Fluid Dynamics}
\label{sec-GFD-applic}

\secttoc

\textbf{What is this lecture about?} This lecture provides the geometric mechanics
framework for Geophysical Fluid Dynamics (GFD).

\subsection{Variational formulae in three dimensions} 
We compute explicit formulae for the variations $\delta a$ in the cases
that the set of tensors $a$ is drawn from a set of scalar fields and
densities on ${\mathbb{R}}^3$. We shall denote this symbolically by
writing
\begin{equation}
a\in\{b,D\,d^3x\}.
\label{Eul-ad-qts}
\end{equation}
We have seen that invariance of the
set $a$ in the Lagrangian picture under the dynamics of
$\mathbf{u}$ implies in the Eulerian picture that
\[
     \left( \frac{\partial}{\partial t}
         + \pounds_{\mathbf{u}} \right) \,a=0 ,
\]
where $\pounds_{\mathbf{u}}$ denotes Lie derivative with respect to the
velocity vector field $\mathbf{u}$. Hence, for a fluid dynamical
Eulerian action $\mathfrak{S}=\int\,dt\ \ell(\mathbf{u};b,D)$, the advected
variables
$b$ and $D$ satisfy the following Lie-derivative relations,
\begin{align}
\left(\frac{\partial}{\partial t}+ \pounds_{\mathbf{u}}\right) b=0
\,,\quad
&\text{or}\quad \frac{\partial b}{\partial t} = -\ \mathbf{u}\cdot\nabla\,b,
\label{eqn-b} \\
\left(\frac{\partial}{\partial t}+ \pounds_{\mathbf{u}}\right)D\,d^3x =0
\,, \quad
&\text{or}\quad \frac{\partial D}{\partial t}
= -\ \nabla\cdot(D\mathbf{u}).
\label{eqn-D}
\end{align}
In fluid dynamical applications, the advected Eulerian variables $b$
and $D\,d^3x$ represent the buoyancy $b$ (or specific entropy, for the
compressible case) and volume element (or mass density) $D\,d^3x$,
respectively. According to section \ref{EPforcontinua}, equation
\eqref{continuumconstrainedVP}, the variations of the tensor
functions $a$ at fixed $\mathbf{x}$ and $t$ are also given by Lie
derivatives, namely
$\delta a = -\pounds_{\mathbf{w}}\,a$, or
\begin{eqnarray}
\delta b
&=& -\pounds_{\mathbf{w}}\ b = -\mathbf{w}\cdot\nabla\,b,
\nonumber \\
\delta D\ d^3x&=&-\pounds_{\mathbf{w}}\,(D\,d^3x)
= -\nabla\cdot(D\mathbf{w})\ d^3x
.
\end{eqnarray}
Hence, Hamilton's principle \eqref{continuumconstrainedVP} with this
dependence yields
\begin{eqnarray}
0 &=&\delta \int dt\ \ell(\mathbf{u}; b,D)
\nonumber \\
&=&\int dt\ \bigg[\frac{\delta \ell}{\delta
\mathbf{u}}\cdot \delta\mathbf{u}
+\frac{\delta \ell}{\delta b}\ \delta b
+\frac{\delta \ell}{\delta D}\ \delta D
\bigg]
\nonumber \\
&=&\int dt\ \bigg[\frac{\delta \ell}{\delta \mathbf{u}}
\cdot \Big(\frac{\partial \mathbf{w}}{\partial t}
-\ad_{\mathbf{u}}\,\mathbf{w}\Big)
-\frac{\delta \ell}{\delta b}\ \mathbf{w}\cdot\nabla\,b
-\frac{\delta \ell}{\delta D}\ \Big(\nabla
\cdot(D\mathbf{w})\Big)\bigg]
\nonumber \\
&=&\int dt\ \mathbf{w}\cdot
\bigg[-\frac{\partial }{\partial t}
\frac{\delta \ell}{\delta \mathbf{u}}
-\ad^*_{\mathbf{u}}\ \frac{\delta \ell}{\delta \mathbf{u}}
-\frac{\delta \ell}{\delta b}\ \nabla\,b
+D\ \nabla\frac{\delta \ell}{\delta D}\bigg]
\nonumber \\
&=&-\int dt\ \mathbf{w}\cdot
\bigg[\Big(\frac{\partial }{\partial t}
 +\pounds_{\mathbf{u}} \Big)\frac{\delta \ell}{\delta \mathbf{u}}
+\frac{\delta \ell}{\delta b}\ \nabla\,b
-D\ \nabla\frac{\delta \ell}{\delta D}\bigg],
\label{eq-EP-Eul}
\end{eqnarray}
where we have consistently dropped boundary terms arising from
integrations by parts, by invoking natural boundary conditions.
Specifically, we may impose $\mathbf{\hat{n}}\cdot\mathbf{w}=0$ on the
boundary, where $\mathbf{\hat{n}}$ is the boundary's outward unit normal
vector and $\mathbf{w} = \delta\eta_t \circ \eta_t^{-1}$ vanishes at
the endpoints.

\subsection{Euler--Poincar\'e framework for geophysical fluid dynamics (GFD)}
The Euler--Poincar\'e equations for continua \eqref{continuumEP} may
now be summarized in vector form for advected Eulerian variables $a$ in the
set \eqref{Eul-ad-qts}. We adopt the notational convention of the
circulation map $I$ in equations \eqref{circ-def} and \eqref{KN-theorem}
that a one form density can be made into a one form (no longer a density)
by dividing it by the mass density $D$ and we use the Lie-derivative
relation for the continuity equation
$({\partial}/{\partial t}+\pounds_{\mathbf{u}})Dd^3x = 0$. Then, the
Euclidean components of the Euler--Poincar\'e equations for continua
in equation \eqref{eq-EP-Eul} are expressed in Kelvin theorem form
\eqref{KThfm} with a slight abuse of notation as
\begin{equation}
\Big(\frac{\partial }{\partial t} + \pounds_{\mathbf{u}}\Big)
\Big(\frac{1}{D}\frac{\delta \ell}{\delta
\mathbf{u}}\cdot d\mathbf{x}\Big)
\,+\,\frac{1}{D}\frac{\delta \ell}{\delta b}\nabla b \cdot d\mathbf{x}
\,-\nabla\Big(\frac{\delta \ell}{\delta D}\Big)\cdot d\mathbf{x}
 = 0,
\label{EP-Kthm}
\end{equation}
in which the variational derivatives of the Lagrangian $\ell$ are to be
computed according to the usual physical conventions, that is, as
Fr\'echet derivatives. Formula \eqref{EP-Kthm} is the
Kelvin--Noether form of the equation of motion for ideal continua.
Hence, we have the explicit Kelvin theorem expression, cf. equations
\eqref{circ-def} and \eqref{KN-theorem},
\begin{equation} \label{KN-theorem-bD}
\frac{d}{dt}\oint_{\gamma_t(\mathbf{u})} \frac{1}{D}\frac{\delta \ell}{\delta \mathbf{u}}\cdot d\mathbf{x}
= -\oint_{\gamma_t(\mathbf{u})}
\frac{1}{D}\frac{\delta \ell}{\delta b}\nabla b \cdot d\mathbf{x}\;,
\end{equation}
where the curve $\gamma_t(\mathbf{u})$ moves with the fluid velocity
$\mathbf{u}$. Then, by Stokes' theorem, the Euler equations generate
circulation of
$\mathbf{v}:=(D^{-1}\delta{l}/\delta\mathbf{u})$
whenever the gradients~$\nabla b$ and
$\nabla(D^{-1}\delta{l}/\delta{b})$ are not collinear. The
corresponding \emph{conservation of potential vorticity} $q$ on fluid
parcels is given by
\begin{equation} \label{pv-cons-EP}
\frac{\partial{q}}{\partial{t}}+\mathbf{u}\cdot\nabla{q} =
0,
\quad \hbox{where}\quad
{q}=\frac{1}{D}\nabla{b}\cdot\curl
\left(\frac{1}{D}\frac{\delta \ell}{\delta \mathbf{u}}\right).
\end{equation}
This is also called \emph{PV convection}.
Equations \eqref{EP-Kthm}--\eqref{pv-cons-EP} embody most of the panoply
of equations for GFD.  The vector form of equation \eqref{EP-Kthm} is,
\begin{eqnarray}\label{vec-EP-eqn}
\underbrace{\Big({\partial\over \partial t} + \mathbf{u}\cdot\nabla\Big)
\Big({1\over D}{\delta l\over \delta \mathbf{u}}\Big)
+{1\over D}{\delta l\over \delta u^j}\nabla u^j}
_{\hbox{Geodesic Nonlinearity: Kinetic energy}}
= \underbrace{\nabla {\delta l\over \delta D}
-{1\over D} {\delta l\over \delta b}\nabla b}_{\hbox{Potential energy}}
\end{eqnarray}
In geophysical applications, the Eulerian variable $D$ represents
the frozen-in volume element and $b$ is the buoyancy. In this case,
\emph{Kelvin's theorem} is
\[
{dI\over dt}=\int\!\!\!\int_{S(t)}
\nabla \left({1\over D} {\delta l\over \delta b}\right)
\times\nabla b \cdot d \mathbf{S} 
,
\]
with circulation integral
\[
I=\oint_{\gamma(t)} {1\over D}{\delta l\over \delta \mathbf{u}}\cdot
d\mathbf{x}
.
\]
\subsection{Euler's equations for a rotating stratified ideal
incompressible fluid}
\label{sec-Euler}

\subsubsection{The Euler--Boussinesq Lagrangian}
In the Eulerian velocity representation, we
consider Hamilton's principle for fluid motion in a three dimensional
domain with action functional ${S}=\int l\,dt$ and
Lagrangian
$l(\mathbf{u},b,D)$ given by
\begin{equation}
l(\mathbf{u},b,D) = \int \rho_0 D (1+b) \big(\tfrac12 |\mathbf{u}|^2
+ \mathbf{u}\cdot\mathbf{R}(\mathbf{x}) - gz\big) - p(D-1)\,d^{3}x,
\label{lag-v}
\end{equation}
where $\rho_{tot}=\rho_0 D (1+b)$ is the total mass density, $\rho_0$
is a dimensional constant and $\mathbf{R}$ is a given function of
$\mathbf{x}$. This variations at fixed $\mathbf{x}$ and $t$ of this
Lagrangian are the following, 
\begin{align}
\frac{1}{D}\frac{{\delta} l}{{\delta} \mathbf{u}}
  &= \rho_0(1{+}b)(\mathbf{u}{+} \mathbf{R}), &
\frac{{\delta} l}{{\delta} b}
  &= \rho_0 D\big(\tfrac{1}{2}|\mathbf{u}|^2
  {+} \mathbf{u}\cdot\mathbf{R} {-} gz\big), \nonumber \\
\frac{{\delta} l}{{\delta} D}
  &= \rho_0(1{+}b)\big(\tfrac{1}{2}|\mathbf{u}|^2
  {+} \mathbf{u}\cdot\mathbf{R} {-} gz\big) {-} p, &
\frac{{\delta} l}{{\delta} p} &= -(D{-}1).
\label{vds-1}
\end{align}
Hence, from the Euclidean component formula \eqref{vec-EP-eqn} for
Hamilton principles of this type and the fundamental vector identity, 
\begin{equation}
(\mathbf{b}\cdot\nabla)\mathbf{a} + a_j\nabla
b^j =-\ \mathbf{b}\times(\nabla\times \mathbf{a})
+ \nabla(\mathbf{b}\cdot\mathbf{a}),
\label{fvid}
\end{equation}
we find the motion equation for an Euler fluid in three dimensions,
\begin{equation}
\frac{d\mathbf{u}}{dt} - \mathbf{u} \times \curl \mathbf{R}
+g\mathbf{\hat{z}}
+ \frac{1}{\rho_0(1+b)}\nabla p = 0,
\label{Eul-mot}
\end{equation}
where $\curl\mathbf{R}=2\boldsymbol{\Omega}(\mathbf{x})$ is the
Coriolis parameter (that is, twice the local angular rotation
frequency). In writing this equation, we have used advection of
buoyancy,
$$
\frac{\partial{b}}{\partial{t}}+\mathbf{u}\cdot\nabla{b} = 0,
$$
from equation \eqref{eqn-b}. The pressure $p$ is determined by requiring
preservation of the constraint $D=1$, for which the continuity equation
\eqref{eqn-D} implies $\div \mathbf{u}=0$. The Euler motion equation
\eqref{Eul-mot} is Newton's Law for the acceleration of a fluid due to
three forces: Coriolis, gravity and pressure gradient. The dynamic
balances among these three forces produce the many circulatory flows of
geophysical fluid dynamics. The \emph{conservation of potential vorticity}
$q$ on fluid parcels for these Euler GFD flows is given by
\begin{equation} \label{pv-cons-EulGFD}
\frac{\partial{q}}{\partial{t}}+\mathbf{u}\cdot\nabla{q} =
0,
\quad \hbox{where, on using }D=1,\quad
{q}=\nabla{b}\cdot\curl
\big(\mathbf{u}+ \mathbf{R}\big).
\end{equation}

\begin{exercise}(Semidirect-product Lie--Poisson bracket for
compressible ideal fluids)$\quad$
\begin{enumerate}
\item
Compute the Legendre transform for the Lagrangian,
\[
l(\mathbf{u},s,D){:\ }  
\mathfrak{X}\times\Lambda^0\times\Lambda^3\mapsto\mathbb{R}\]
whose advected variables satisfy the auxiliary equations,
\[
\frac{\partial b}{\partial t}
= -\ \mathbf{u}\cdot\nabla\,s
,\qquad
\frac{\partial D}{\partial t}
= -\ \nabla\cdot(D\mathbf{u}).
\]
In which $s$ represents entropy per unit mass and $D$ represents mass density,
as thermodynamic variables.
\item
Compute the Hamiltonian, assuming the Legendre transform is a linear
invertible operator on the velocity $\mathbf{u}$. For definiteness
in computing the Hamiltonian, assume the Lagrangian is given by
\begin{equation}
l(\mathbf{u},b,D) = {\int}
D \big(\tfrac{1}{2} |\mathbf{u}|^2
+ \mathbf{u}\cdot\mathbf{R}(\mathbf{x})
 - e(D,s)\big)
\,d^{\,3}x,
\end{equation}
with prescribed function $\mathbf{R}(\mathbf{x})$ and specific internal
energy $e(D,b)$ satisfying the First Law of Thermodynamics,
\[
de=\frac{p}{D^2}dD + T ds
,
\]
where $p$ is pressure, $T$ temperature.
\item
Find the semidirect-product Lie--Poisson bracket for the Hamiltonian
formulation of these equations. 
\item
Does this Lie--Poisson bracket have Casimirs? If so, what are the
corresponding symmetries and momentum maps?
\end{enumerate}
\end{exercise}
\begin{answer}
The solution here is essentially the same as for the Euler--Boussinesq equations treated above. 
\end{answer}\index{semidirect product!momentum map}\index{momentum map!semidirect product}

\subsection{Hamilton-Poincar\'e reduction implies Lie-Poisson fluid equations}


In the Euler--Poincar\'e framework one starts with a Lagrangian defined on
the tangent bundle of a Lie group $G$
\[
L{:\ } TG\rightarrow\mathbb{R}
\]
and the dynamics is given by Euler--Lagrange equations arising from the variational
principle
\[
\delta\int_{t_0}^{t_1} L(g,\dot{g}) dt =0
\]
The Lagrangian $L$ is taken left/right invariant and because of this property
one can \emph{reduce} the problem obtaining a new system which is defined
on the Lie algebra $\mathfrak{g}$ of $G$, obtaining a new set of equations,
the Euler--Poincar\'e equations, arising from a reduced variational principle 
\index{variational principle!reduced}
\[
\delta\int_{t_0}^{t_1} l(\xi) dt =0
\]
where $l(\xi)$ is the reduced Lagrangian and $\xi\in\mathfrak{g}$.

\begin{exercise}
Is there a similar procedure for Hamiltonian systems? 
More precisely: given a Hamiltonian function
\[
H{:\ } T^*G\rightarrow\mathbb{R}
\]
defined on the cotangent bundle $T^*G$,
may one perform a similar procedure of reduction and derive the equations of motion on the dual of the Lie algebra $\mathfrak{g}^*$, provided the Hamiltonian
is again left/right invariant.
\end{exercise}

\begin{answer}
Hamilton--Poincar\'e reduction gives a positive answer to this question, in
the context of variational principles as it is done in the Euler--Poincar\'e
framework: we are going to explain how this procedure is performed.

More in general, we will also consider advected quantities belonging to a
vector space $V$ on which $G$ acts, so that the Hamiltonian is written in
this case as
\[
H{:\ } T^*G\times V^*\rightarrow \mathbb{R}
\]
(see Holm, Marsden and Ratiu~\cite{HoMaRa1998a,HoMaRa1998b}).
The space $V$ is regarded here exactly the same as in the Euler--Poincar\'e theory.\\
The equations of motion, that is, Hamilton's equations, may be derived from the
following variational principle
\[
\delta\int_{t_0}^{t_1} \{\langle p(t),\dot{g}(t) \rangle - H_{a_0} (g(t),p(t))\}\, dt =0
\,.\]
As we know from ordinary classical mechanics, ($\dot{g}(t)$ should
be considered as the tangent vector to the curve $g(t)$, so that $\dot{g}(t)\in
T_{\!g(t)} G$).
\end{answer}

\begin{exercise}
What happens if $H_{a_0}$ is left/right invariant?
\end{exercise}
\begin{answer}

It turns out that in this case the whole function
\[
F(g,\dot{g},p)=\langle p,\dot{g} \rangle - H_{a_0} (g,p)
\]
is also invariant. The proof is straightforward once the action is,
specified \footnote{ From here until the end of this lecture 
we will consider only \textit{left invariance}. Examples of right invariant Hamiltonians 
will be considered in later lectures.}
\[
h\,(g,\dot{g},p)=(hg, T_g L_h\, \dot{g}, T^*_{hg} L_{h^{-1}}\, p)
\]
where $T_g L_h {:\ }  T_g G \rightarrow T_{hg}G$ is the tangent of the left translation map $L_h\,g=hg\in{G}$ at the point $g$ and 
$T^*_{hg}L_{h^{-1}}{:\ } T^*_g G \rightarrow T^*_{hg}G$ is the dual of the
map $T_{hg}L_{h^{-1}} {:\ }  T_{hg}G\rightarrow T_g G$.

We now check that
\begin{align*}
\langle h\, p,  h\, \dot{g} \rangle
&=\langle T^*_{hg} L_{h^{-1}}\, p,  T_g L_h\, \dot{g} \rangle
\\
&=\langle p, T_{hg} L_{h^{-1}}\circ T_g L_h\, \dot{g} \rangle
\\
&=\langle p, T_g (L_{h^{-1}}\circ L_h)\, \dot{g} \rangle
=\langle  p, \dot{g} \rangle
\end{align*}
where the chain rule for the tangent map has been used. The same
result holds for the right action.\\
Due to this invariance property, one can write the variational principle
as
\[
\delta\int_{t_0}^{t_1} \{\langle \mu,\xi \rangle - h (\mu,a)\}\, dt =0
\]
with
\[
\mu(t)=g^{-1}(t)\,p(t)\in\mathfrak{g}^*,
\qquad
\xi(t)=g^{-1}(t)\,\dot{g}(t)\in\mathfrak{g},
\qquad
a(t)=g^{-1}(t)\,a_0\in V^*
\]
In particular $a(t)$ is the solution of
\[
\p_t{a}(t)=-\,\xi(t)\, a_0.
\]
where a Lie algebra action of $\mathfrak{g}$ on $V^*$ is implicitly defined.
In order to find the equations of motion one calculates the variations
\[
\delta\int_{t_0}^{t_1} \{\langle \mu,\xi \rangle - h (\mu,a)\}\, dt 
=
\int_{t_0}^{t_1} \left\{\langle \delta\mu,\xi \rangle + 
\langle \mu,\delta\xi \rangle  - 
\left\langle \delta\mu,\frac{\delta h}{\delta \mu}\right\rangle -
\left\langle \delta a,\frac{\delta h}{\delta a}\right\rangle \right\}\, dt
\]
As in the Euler--Poincar\'e theorem, we use the following expressions for the
variations,
\begin{align*}
\delta\xi=\p_t\eta + [\xi,\eta]\,,
\qquad
\delta a=-\eta a
\,,
\end{align*}
and using the definition of the diamond operator we find
\begin{align*}
&\int_{t_0}^{t_1} \left\{\langle \delta\mu,\xi \rangle + 
\langle \mu,\delta\xi \rangle  - 
\left\langle \delta\mu,\frac{\delta h}{\delta \mu}\right\rangle -
\left\langle \delta a,\frac{\delta h}{\delta a}\right\rangle \right\}\, dt\\
&=
\int_{t_0}^{t_1} \left\{\left\langle 
\delta\mu, \xi - \frac{\delta h}{\delta \mu} \right\rangle + 
\langle \mu, \dot\eta + \text{ad}_{\xi} \eta \rangle  + 
\left\langle \eta a,\frac{\delta h}{\delta a}\right\rangle \right\}\, dt\\
&=
\int_{t_0}^{t_1} \left\{\left\langle 
\delta\mu, \xi - \frac{\delta h}{\delta \mu} \right\rangle + 
\langle -\dot\mu + \text{ad}^*_{\xi} \mu, \eta \rangle  - 
\left\langle \frac{\delta h}{\delta a}\diamond a,  \eta\right\rangle \right\}\, dt.
\end{align*}
Consequently, the $\delta \mu$ variation yields
\[
\xi = \frac{\delta h}{\delta \mu}
\,,\]
and the equations of motion are given by
\[
\dot\mu = \text{ad}^*_{\xi} \mu - \frac{\delta h}{\delta a}\diamond a
\,,
\]
together with\footnote{Recall the concatenation notation for the Lie algebra action,
in which $\frac{\delta h}{\delta \mu}\, a\simeq \mathcal{L}_{\delta h/\delta \mu}a$.}
\[
\dot{a}=-\,\frac{\delta h}{\delta \mu}\, a
\,.
\]
The equations of motion written on the dual Lie algebra $\mathfrak{g}$ are
called \emph{Lie--Poisson} equations. 
\end{answer}

This exercise has proven the following theorem.

\begin{theorem}[Hamilton--Poincar\'e reduction theorem]
With the preceding notation, the following statements are equivalent:
\begin{enumerate}
\item
With $a_0$ held fixed, the variational principle
\[
\delta\int_{t_0}^{t_1} \{\langle p(t),\dot{g}(t) \rangle - H_{a_0} (g(t),p(t))\}\, dt =0
\]
holds, for variations $\delta{g(t)}$ of $g(t)$ vanishing at the endpoints.
\item
$(g(t),p(t))$ satisfies Hamilton's equations for $H_{a_0}$ on G.
\item
The constrained variational principle
\[
\delta\int_{t_0}^{t_1} \{\langle \mu(t),\xi(t) \rangle - h (\mu(t),a(t))\}\, dt =0
\]
holds for $\mathfrak{g}\times V^*$, using variations of $\xi$ and $a$ of the form
\begin{align*}
\delta\xi=\p_t\eta + [\xi,\eta],
\qquad
\delta a=-\eta a
\end{align*}
where $\eta(t)\in\mathfrak{g}$ vanishes at the endpoints
\item
The \emph{Lie--Poisson} equations hold on $\mathfrak{g}\times V^*$
\[
(\p_t\mu,\p_t{a}) = \left(\textnormal{ad}^*_{\xi} \mu - \frac{\delta h}{\delta a}\diamond a, -\frac{\delta h}{\delta \mu}\, a\right)
.\]
\end{enumerate}
\end{theorem}

\begin{remark}\rm 
One might have preferred to start with an invariant Hamiltonian defined on
\[
T^*(G\times V)=T^*G\times V\times V^*
\]
However, as mentioned by Holm, Marsden and
Ratiu~\cite{HoMaRa1998a,HoMaRa1998b}, such an approach turns out to be
equivalent to the treatment presented here.
\end{remark}

\begin{remark}\rm [Legendre transform]
Lie--Poisson equations may arise from the Euler--Poincar\'e setting by Legendre
transform
\[
\mu=\frac{\delta l}{\delta \xi}.
\]
If this is a diffeomorphism, then the Hamilton--Poincar\'e
theorem is equivalent to the Euler--Poincar\'e theorem.
\end{remark}
 
\begin{exercise}[Lie--Poisson structure] \index{Lie--Poisson structure!dual Lie algebra!$\mathfrak{g}^*\circledS V^*$}
Show that the space $\mathfrak{g}^*\times V^*$ is a Poisson manifold. \index{Poisson manifold!Lie-Poisson structure} 
\index{Poisson manifold!$\mathfrak{g}^*\times V^*$}
\end{exercise}
\begin{answer}\begin{align*}
\p_t{F}(\mu,a)  &=
\left\langle \p_t{\mu},\frac{\delta F}{\delta\mu}\right\rangle 
+\left\langle \p_t{a},\frac{\delta F}{\delta a}\right\rangle
\\
& =\left\langle 
\text{ad}_{\delta H/\delta\mu}^{\ast}\mu
-\frac{\delta H}{\delta a}\diamond a,
\frac{\delta F}{\delta\mu}
\right\rangle -
\left\langle
\frac{\delta H}{\delta\mu}a,\frac{\delta F}{\delta a}
\right\rangle
\\
& =\left\langle \mu,
\left[  \frac{\delta H}{\delta\mu},\frac{\delta F}{\delta\mu}\right]  
\right\rangle -
\left\langle 
\frac{\delta H}{\delta a}\diamond a,
\frac{\delta F}{\delta\mu}
\right\rangle -
\left\langle
\frac{\delta H}{\delta\mu}a,\frac{\delta F}{\delta a}
\right\rangle
\\
& =-
\left\langle 
\mu,\left[  \frac{\delta F}{\delta\mu},\frac{\delta H}{\delta\mu}\right] \right\rangle -
\left\langle 
\frac{\delta H}{\delta a}\diamond a,
\right\rangle -
\left\langle
\frac{\delta H}{\delta\mu}a,\frac{\delta F}{\delta a}
\right\rangle
\\
& =-\left\langle 
\mu,\left[  \frac{\delta F}{\delta\mu},\frac{\delta H}{\delta\mu}\right]
\right\rangle -
\left\langle a,\frac{\delta F}{\delta\mu}\frac{\delta H}{\delta a}-
\frac{\delta H}{\delta\mu}\frac{\delta F}{\delta a}\right\rangle
\end{align*}
One may directly verify that the bracket formula
\[
\{F,H\}(\mu,a)=
-\left\langle 
\mu,\left[  \frac{\delta F}{\delta\mu},\frac{\delta H}{\delta\mu}\right]
\right\rangle -
\left\langle a,\frac{\delta F}{\delta\mu}\frac{\delta H}{\delta a}-
\frac{\delta H}{\delta\mu}\frac{\delta F}{\delta a}\right\rangle
\]
satisfies the definition of a Poisson structure. 
This Lie--Poisson structure has been found during lectures for the simpler case without advected quantities.
In particular, this calculation verifies that \emph{any dual Lie algebra $\mathfrak{g}^*$
is a Poisson manifold}. \index{Poisson manifold! dual Lie algebra}
\end{answer}

\begin{remark}\rm [Right invariance]
It can be shown that for a right invariant Hamiltonian one has
\begin{align*}
\{F,H\}(\mu,a)&=+
\left\langle 
\mu,\left[  \frac{\delta F}{\delta\mu},\frac{\delta H}{\delta\mu}\right]
\right\rangle +
\left\langle a,\frac{\delta F}{\delta\mu}\frac{\delta H}{\delta a}-
\frac{\delta H}{\delta\mu}\frac{\delta F}{\delta a}\right\rangle
\\
(\p_t\mu,\p_t{a})&=-
\left(\textnormal{ad}^*_{\xi} \mu - \frac{\delta h}{\delta a}\diamond a, 
-\frac{\delta h}{\delta \mu}\, a\right)
\end{align*}
with all signs changed for a left invariant Hamiltonian.
\end{remark}

\newpage
\vspace{4mm}\centerline{\textcolor{shadecolor}{\rule[0mm]{6.75in}{-2mm}}\vspace{-4mm}}
\section{Five more continuum applications}

\secttoc

\textbf{What is this lecture about?} This lecture applies deterministic geometric mechanics 
to derive five continuum dynamics equations of current research interest ranging from kinetic theory
to nonlinear waves.

\subsection{The Vlasov equation in kinetic theory}
In plasma physics a major topic is collisionless particle dynamics in $N$-dimensions, whose
primary equation, the Vlasov equation, will be derived heuristically here. In this context, a central role is held by the probability distribution on phase space $P:=f(\mathbf{q,p},t)\,d^N{q}\wedge\,d^N{p}$.
Because the total probability is conserved, one may write the continuity equation
just as one does for the conservation of total mass in the context of fluid dynamics
\[
\big(\p_t + \mathcal{L}_u \big) P = 
\Big(\p_t{f}+\nabla_{(\mathbf{q}\,,\,\mathbf{p})} \cdot (\mathbf{u}\, f)\Big)\,d^N{q}\wedge\,d^N{p}=0
\,,\] 
where $\nabla_{(\mathbf{q}\,,\,\mathbf{p})}$ is the divergence in $2N$-dimensional phase space and $\mathbf{u}$ is a Hamiltonian vector field on phase space, which is given by the
single particle motion,
\[
\mathbf{u}=(\dot{\mathbf{q}},\dot{\mathbf{p}})\in\mathfrak{X}\big(T^*\mathbb{R}^N\big)
\,.\]
If we now assume that the generic single particle undergoes Hamiltonian
motion, the Hamiltonian function $h(\mathbf{q,p})$ can be introduced directly by means of the single particle Hamilton's equations
\[
\mathbf{u} :=
(\dot{\mathbf{q}},\dot{\mathbf{p}})=
\left( \frac{\partial h}{\partial \mathbf{p}}, 
-\frac{\partial h}{\partial \mathbf{q}}\right)
\,.\]
This implies that the vector field $\mathbf{u}$ is divergence free, $\nabla_{(\mathbf{q}\,,\,\mathbf{p})} \cdot \mathbf{u}=0$, provided the Hessian of the particle Hamiltonian $h$ is symmetric. Therefore, the Vlasov equation written in terms of the distribution function $f(\mathbf{q,p},t)$ 
emerges as
\[
\p_t f +\mathbf{u}\cdot\nabla_{(\mathbf{q}\,,\,\mathbf{p})} f=0
\,.\]
By expanding the Hamiltonian $h$ as the total single particle energy
\[
h(\mathbf{q,p})=\frac1{2m}\mathbf{p}^2+V(\mathbf{q},\mathbf{p})
\,,\]
one obtains the more common form of the Vlasov equation,
\[
\frac{\partial f}{\partial t}+
\frac{\mathbf{p}}{m}\cdot\frac{\partial f}{\partial \mathbf{q}}-
\frac{\partial V}{\partial \mathbf{q}}\cdot\frac{\partial f}{\partial \mathbf{p}}=0
\,.\]
For more information about what can be done with this hybrid phase-space 
approach to geometric mechanics, see \cite{tronci2010hamiltonian}.

\begin{exercise}
Cast the evolution equations for differentiable functionals $F$ of the 
Vlasov probability density $f$ into Lie--Poisson form.
\end{exercise}
\begin{answer}
First write the Vlasov equation in
terms of a generic single particle Hamiltonian $h(\mathbf{q,p})$ as
\[
\p_t f+\{f,h\}=0
\,,\]
and recall the canonical Poisson bracket
\[
\{f,h\}=
\frac{\partial f}{\partial \mathbf{q}}\cdot\frac{\partial h}{\partial \mathbf{p}}-
\frac{\partial f}{\partial \mathbf{p}}\cdot\frac{\partial h}{\partial \mathbf{q}}
\,.\]
Recall also that the canonical Poisson bracket endows
the set $\mathcal{F}(T^*\mathbb{R}^N)$ of phase space functions 
with a Lie algebra structure. Namely, for right action of the symplectic 
diffeomorphisms on $\mathcal{F}(T^*\mathbb{R}^N)$,
\[
[k,h]=\{k,h\} = \ad_hk
\,.\]
At this point, in order to look for a Lie--Poisson equation, one calculates
the coadjoint operator such that
\[
\langle f,\{h,k\} \rangle
=
\langle f,-\text{ad}_h k \rangle=
\langle -\text{ad}^*_h f,k \rangle=
\langle \{h,f\},k \rangle
\,,\]
where the last equality is justified by the Leibniz property of the Poisson
bracket, with the pairing defined as the $L^2$ pairing in phase space,
\[
\langle f,g \rangle=\int f\,g\,d^N{q}\wedge\,d^N{p}
\,.\]
This argument shows that the Vlasov equation can be written in the Lie--Poisson form
\[
\p_t f+\text{ad}^*_h\,f=0
\,.\]

The corresponding Lie-Poisson bracket for functionals $F$ and $H$ of the Vlasov probability density  $f$ is given by
\[
\
\frac{dF}{dt}=\big\{F,H\big\}_{LP} = - \scp{f}{\bigg[\frac{\delta F}{\delta f},\frac{\delta H}{\delta f}\bigg]}
= - \scp{\ad^*_{{\delta H}/{\delta f}} f}{\frac{\delta F}{\delta f}}
.\]
To check this result, notice that the Vlasov equation can now be rewritten as 
\[
\p_t f = \Big\{f,H\Big\} = -\, \ad^*_{{\delta H}/{\delta f}}\,f
\,,\]
by setting ${\delta H}/{\delta f}=h(\mathbf{q,p})$ and writing the functional $F$ as 
\[
F(f) 
= \int f(\mathbf{q'}\,,\,\mathbf{p'})\delta ((\mathbf{q'}\,,\,\mathbf{p'})-(\mathbf{q}\,,\,\mathbf{p}))\,d^N{q'}\wedge\,d^N{p'}
\,,\]
when performing the variation with respect to $f$.
\end{answer}

\subsection{Ideal barotropic compressible fluids in 3D}

The reduced Lagrangian for ideal barotropic compressible fluids is written as
\[
l(\mathbf{u},D)
=\int \big(\tfrac{1}{2}|\mathbf{u}|^2  - e(D)\big)
\,  Dd^3x
\]
where $|\mathbf{u}|^2 := u \contract u^\flat $ with $u^\flat =\mathbf{u}\cdot d\bs{x}$ and the velocity vector field $u \in\mathfrak{X}(M\!\!\subset\!\mathbb{R}^3)$ written in components as $u=\mathbf{u}\cdot \nabla$ has no normal components $u\contract d^3x = \mathbf{u}\cdot \nh \,dS = 0$ on the boundary $\partial M$ and $Dd^3x$ is the advected density,
which satisfies the \emph{continuity equation}
\[
\partial_t{Dd^3x} + \mathcal{L}_{u} (Dd^3x) = 0\,.
\]
Moreover, the internal energy satisfies the barotropic First Law of Thermodynamics
\[
de=-p(D)d(D^{-1})=\frac{p(D)}{D^2}\,dD
\]
for the pressure function $p(D)$. 
The reduced Legendre transform on this Lie algebra $\mathfrak{X}(\mathbb{R}^3)$
is given by
\[
m := \frac{\delta \ell}{\delta u}  = \mathbf{m}\cdot d\bs{x} \otimes d^3x  
=\mathbf{u}\cdot d\bs{x} \otimes  D d^3x
\]
and the Hamiltonian is then written as
\[
h(\mathbf{m},D)=
\langle \mathbf{m},\mathbf{u}\rangle-l(\mathbf{u},D)
\]
that is
\[
h(\mathbf{m},D)= \int \frac{1}{2D}\,|\mathbf{m}|^2\, + De(D) \,d^3x
\]
The Lie--Poisson equations in this case arise from the general theory,
\begin{align}\begin{split}
\partial_t{m}
&=
-\text{ad}^*_{\delta h/\delta m}\, m
-
\frac{\delta h}{\delta D}\diamond D
\,,\\
\partial_t{D}d^3x&=-\mathcal{L}_{\delta h/\delta m}\, Dd^3x
\,.\label{LP-Euler-Eqns}
\end{split}\end{align}
Earlier we found that the coadjoint action is given by the Lie derivative. On the other hand we may calculate the expression of the diamond operation from
its definition
\[
\left\langle 
\frac{\delta h}{\delta D}, -\pounds_{\eta} D
\right\rangle
=
\left\langle 
\frac{\delta h}{\delta D} \diamond D , \eta 
\right\rangle
\]
to find
\[
\left\langle 
\frac{\delta h}{\delta D}, -\div  D \eta
\right\rangle
=
\left\langle 
D\nabla\frac{\delta h}{\delta D}  , \eta 
\right\rangle
\]
Therefore, we have
\[
\frac{\delta h}{\delta D}\diamond D =  D \nabla\frac{\delta h}{\delta D}
\,,\]
where
\[
\delta h/\delta D= -\frac{|\mathbf{m}|^2}{2D^2} + \Big(e + \frac{p}{D}\Big)
\,.\]
Substituting into the momentum equation and using the First Law to find $d(e+p/D)=(1/D)dp$ yields
\begin{align}
(\partial_t + \mathcal{L}_{u})(\mathbf{m}/D)\cdot d\bs{x} 
=  - \nabla \big(e + p/D  - |\mathbf{u}|^2/2 \big)\cdot d\bs{x}
\,.\label{barotropic-motion-eqn}
\end{align}
Upon expanding the Lie derivative for $\mathbf{m}/D=\mathbf{u}$ and using the continuity equation for the density, this quickly becomes
\[
\partial_t{\mathbf{u}}=-\mathbf{u}\!\cdot\!\nabla{\mathbf{u}}
\, -
\frac{1}{D}\nabla p
\,,\]
which is Euler's equation for a barotropic fluid.

\begin{remark}\rm [Kelvin-Noether theorem for barotropic fluid dynamics]
Conservation of the circulation integral for the barotropic fluid motion equation 
in \eqref{barotropic-motion-eqn} is derived from
\begin{align}
\frac{d}{dt}\oint_{c(u)} \bs{u} \cdot d\bs{x}
= \oint_{c(u)} (\partial_t + \mathcal{L}_{u}) (\bs{u}\cdot d\bs{x} )
= -\oint_{c(u)} \mrm{d}\big(e + p/D  - |\mathbf{u}|^2/2 \big)
= 0
\,,
\label{barotropic-Kelvin-thm}
\end{align}
where $c(u)$ denotes a material loop moving with the barotropic fluid flow and the fundamental law 
of calculus (vanishing of the loop integral of the differential of a function) has been used. 
\end{remark}

\subsection{Euler's equations for 3D ideal incompressible fluid motion} 
The barotropic equations for compressible flows yield Euler's equations for ideal incompressible fluid motion in three dimensions when the internal energy in the reduced Lagrangian for ideal compressible fluids is replaced by the constraint $D=1$, as
\begin{align}
l(\mathbf{u},D)
=\int \frac{D}{2}|\mathbf{u}|^2 
- p(D-1)\,  d^3x
\,,
\label{RedLag-Euler}
\end{align}
where again $\mathbf{u}\in\mathfrak{X}(M\!\!\subset\!\mathbb{R}^3)$ is tangential on the boundary $\partial M$ and the advected density $D$ satisfies the continuity equation,
\begin{align}
\partial_t{D}+\div D\mathbf{u} =0
.
\label{continuity}
\end{align}
The continuity equation \eqref{continuity} enforces the divergence-free condition $\div \mathbf{u}=0$ when evaluated on the constraint $D=1$. The pressure $p$ in the reduced Lagrangian in \eqref{RedLag-Euler} is now a Lagrange multiplier, which is determined by the condition that incompressibility be preserved by the dynamics. 

As for the barotropic case, the Hamiltonian is obtained from
\[
h(\mathbf{m},D)=
\langle \mathbf{m},\mathbf{u}\rangle-l(\mathbf{u},D)
\]
and found with $m := \frac{\delta \ell}{\delta u}  = \mathbf{m}\cdot d\bs{x} \otimes d^3x$ to be
\[
h(\mathbf{m},D)= \int \frac{1}{2D}\,|\mathbf{m}|^2\, + p(D-1) \,d^3x
\,.\]
The Lie--Poisson equations in this case arise from the general theory,
\begin{align}\begin{split}
\partial_t{m}
&=
-\text{ad}^*_{\delta h/\delta m}\, m
-
\frac{\delta h}{\delta D}\diamond D
\,,\\
\partial_t({D}d^3x) &=-\mathcal{L}_{\delta h/\delta m}\, (Dd^3x)
\,.\label{LP-Euler-Eqns}
\end{split}\end{align}

In Lie-Poisson matrix operator form, this is 
\begin{equation}
\p_t 
\begin{pmatrix}
m_i \\ D
\end{pmatrix}
= -
\begin{bmatrix}
\p_j m_i + m_j\p_i & \Box\diamond D \\
\mathcal{L}_{\Box}D & 0 
\end{bmatrix}
\begin{pmatrix}
{\delta h/\delta m_j = m^j/D= u^j} \\ \delta h/\delta D = p -\,\frac12 |\bs{m}|^2/D^2 = p -\,\frac12 |\bs{u}|^2  
\end{pmatrix}
.\label{LP-Euler-eqn}
\end{equation}
The corresponding Lie-Poisson bracket is defined on the dual%
\footnote{Dual spaces are understood in the sense of $L^2$ pairing in the plane
denoted by angle brackets $\scp{\,\cdot}{\cdot\,}$.
} 
of the semidirect-product Lie algebra $\mathfrak{X}_{vol}\circledS \Lambda^0$   
\index{Lie algebra!semidirect product!$\mathfrak{X}_{vol}\circledS V^*$}
\index{Lie algebra!semidirect product!fluid dynamics}
of volume-preserving vector fields $\mathfrak{X}_{vol}(\mathbb{R}^3)$ acting on functions
$\Lambda^0(\mathbb{R}^3)$.
Dual coordinates: the 1-form density $m\in \mathfrak{X}_{vol} \otimes \mrm{Den}(\mathbb{R}^3)$ is dual to $\mathfrak{X}_{vol}(\mathbb{R}^3)$; and density $D$ is dual to scalar functions in $\mbb{R}^3$.

The Lie-Poisson motion equation for Euler's fluid equation in \eqref{LP-Euler-Eqns} for an incompressible fluid is given by
\[
\partial_t{\mathbf{u}}+\mathbf{u}\!\cdot\!\nabla{\mathbf{u}} + u_j\nabla u^j
= - \,\nabla \big(p - \tfrac12 |\mathbf{u}|^2\big)
\,,\]
or, in Lie-derivative form,
\begin{align}
\big(\partial_t + \mathcal{L}_u\big) (\mathbf{u}\cdot d\bx)
=
\big(\partial_t \bs{u} - \bs{u}\times \op{curl}\bs{u} + \frac12\nabla |\bs{u}|^2\big)\cdot d\bs{x}
= - \,d\big(p - \tfrac12 |\mathbf{u}|^2\big)
\,.
\label{Lie-motion}
\end{align}
The differential of equation \eqref{Lie-motion} and the property $d^2=0$ combine now to yield an equation for the vorticity,
$\bs{\omega}:=\curl \mathbf{u}$,
\[
\big(\partial_t + \mathcal{L}_u\big) \dd(\mathbf{u}\cdot d\bx)
=
\big(\partial_t + \mathcal{L}_u\big) (\curl \mathbf{u}\cdot d\mathbf{S})
= 0
\,.\]

Expanding out the Lie derivative in the previous formula and using $\op{div}\bs{\omega}=0$
yields the classic vortex advection and stretching formula for the vorticity 
$\bs{\omega}:=\curl \mathbf{u}$ in three dimensions,
\[
\big(\partial_t + \mathcal{L}_u\big) (\curl \mathbf{u}\cdot d\mathbf{S})
=
\big(\p_t \bs{\omega} + \bu\cdot\nabla\bs{\omega} - \bs{\omega}\cdot\nabla \bu \big)\cdot d\mathbf{S}  = 0
\,.\]
Since both $\bs{u}$ and $\bs{\omega}$ are divergence free, this equation may also be written in vector form as 
\[
\p_t \bs{\omega} - \op{curl} \big(\bs{u}\times\bs{\omega}\big) = 0\,.
\]
\begin{exercise}
Show that the Euler fluid equation \eqref{Lie-motion} implies conservation of both the kinetic energy 
\[
H := \int_{\cal D} \frac12 |\bs{u}|^2 \,d^3x
\,,\]
as well as the following interesting quantity known as the Hopf invariant, or \textit{helicity},
\[
\Lambda := \int_{\cal D} \bs{u}\cdot d\bs{x}\wedge \mrm{d}(\bs{u}\cdot d\bs{x})
= \int_{\cal D} \bs{u}\cdot \bs{\omega} \,d^3x
\,.\]
The Hopf invariant (or helicity) is a topological quantity which characterises the knottedness of the lines of vorticity. 
It is also the \textit{only Casimir} of the Lie-Poisson bracket for Euler's 3D fluid equations in \eqref{LP-Euler-eqn}.
\end{exercise}

\begin{exercise}
Another potential Poisson bracket candidate for vorticity $\bs{\omega}:=\curl \mathbf{u}$ dynamics of Euler's fluid equations might be written for real functionals $\Lambda$, $F$, $H$ of vorticity $\bs{\omega}$ as
\begin{align*}
\frac{dF}{dt} = \{F,H\}
&:= \int_{\cal D} \curl\frac{\de \Lambda}{\de \bs{\omega}} \cdot 
\curl\frac{\de F}{\de \bs{\omega}} \times  \curl\frac{\de H}{\de \bs{\omega}}\,d^3x
\\& = - 
\int_{\cal D} 
\frac{\de F}{\de \bs{\omega}} \cdot \curl \left( \frac{\de \Lambda}{\de \bs{u}}
 \times \frac{\de H}{\de \bs{u}}\right)\,d^3x
\,,
\end{align*}
with Hamiltonian $H= -\,\frac12\int_{\cal D} \bs{\omega}\cdot \Delta^{-1}\bs{\omega} 
= \frac12\int_{\cal D} |\bs{u}|^2d^3x$. 

Would this formulation make sense as a Poisson bracket for Euler's fluid equations? 
What would the Casimirs be for such a Poisson bracket?
\end{exercise}

\subsection{Euler's 3D fluid equations for variables depending only on the horizontal coordinates $(x,y)$} 

When the fluid velocity depends on only two planar dimensions with $(x,y)$ coordinates, its
Hodge decomposition may be written as
\begin{align}
\bu(x,y,t)=\nabla^\perp\psi(x,y,t) + v(x,y,t)\zh = (-\,\psi_y,\psi_x,v),
\quad\hbox{with}\quad
\div \mathbf{u}=0
\,,
\label{Hodge-decomp-2D}
\end{align}
where the \textit{grad-perp} operator $\nabla^\perp$ is defined as 
\[
\nabla^\perp:=\zh\times\nabla
\,.\]
The corresponding 3D vorticity is given by 
\begin{align}
\bs{\omega}  = \Delta \psi \,\zh - \nabla^\perp v  
\,,\label{vort-dyn-eqn}
\end{align}
The divergence-free velocity $\bu$ is written terms of the operator $\nabla^\perp=\zh\times\nabla$ acting on the stream function $\psi$ and on vertical velocity component $v$. The vorticity equation the becomes 
\begin{align}\begin{split}
0 &= \big(\partial_t + \mathcal{L}_u\big) (\bs{\omega}\cdot d\mathbf{S})
\\&=
\dd \big(((\partial_t + \mathcal{L}_u)(v\,dz)\big) + (\partial_t + \mathcal{L}_u)(\Delta\psi \,dx\wedge dy)
\\&=
\dd \big( \big(\partial_t v + J(\psi,v)\big)\,dz\big) + \big(\partial_t \omega + J(\psi,\omega) \big) \,dx\wedge dy
\,,\label{vort-dyn-eqn}
\end{split}\end{align}
where one defines $\omega:=\Delta\psi$ and $\psi=\Delta^{-1}\omega$, as well as  the notation
\[
J(\psi,\omega)=[\psi,\omega]=\psi_x\omega_y - \psi_y\omega_x
\,.\]
This calculation has applied commutation of Lie derivative and differential $\mrm{d}$, $\mrm{d}^2z=0$, and 
\begin{align}
0 = \big(\partial_t + \mathcal{L}_u\big) (dx\wedge dy)
=
\op{div}\bu \,dx\wedge dy
= 0\,.
\label{symplectic-motion}
\end{align}
\begin{exercise}
Explain why equation \eqref{symplectic-motion} holds and how it enters the calculation of \eqref{vort-dyn-eqn}.
\end{exercise}
From the result \eqref{vort-dyn-eqn} we have obtained two scalar advection equations. Namely, 
\begin{align}
v_t + [\psi,v] = 0 
\quad\hbox{and}\quad
\partial_t \omega + [\psi,\omega] = 0
\,.\label{vort-dyn-eqn2}
\end{align}
For the kinetic energy Hamiltonian for incompressible fluid flow given by
\begin{align}
H(\omega) = \frac12 \int_{\mathbb{R}^2} |\mathbf{u}|^2 + v^2 \,dxdy 
= \frac12 \int_{\mathbb{R}^2} \omega\Delta^{-1}\omega + v^2\,dxdy
\,,\label{Euler-vort-dyn-Ham}
\end{align}
one may write equations \eqref{vort-dyn-eqn2} in Lie-Poisson matrix form as 
\begin{equation}
\p_t 
\begin{pmatrix}
\omega \\ v 
\end{pmatrix}
= -
\begin{bmatrix}
[\Box, \omega] & [\Box, v] \\
[\Box, v ] & 0 
\end{bmatrix}
\begin{pmatrix}
{\delta H/\delta\omega = \psi} \\ {\delta H/\delta v = v}
\end{pmatrix}
= -
\begin{pmatrix}
[\psi, \omega] \\ [\psi, v] 
\end{pmatrix}
.\label{LP-vort-dyn-eqn}
\end{equation}
\begin{exercise}
Derive equation \eqref{LP-vort-dyn-eqn} and write its corresponding Lie-Poisson bracket. 
\end{exercise}
Thus, the vertical component $\omega\zh$ of the 3D vorticity $\bs{\omega}$ 
is advected actively by the 2D divergence-free velocity and the vertical  
component $v\zh$ of the 3D velocity is advected passively. 
The corresponding Lie-Poisson bracket is defined on the dual 
of the semidirect-product Lie algebra $\mathfrak{X}_{vol}\circledS \Lambda^2$  
of volume preserving vector fields $\mathfrak{X}_{vol}(\mbb{R}^2)$ acting on 2-forms 
$\Lambda^2(\mbb{R}^2)$, modulo exact 2-forms \cite{marsden1983coadjoint}.
Dual coordinates: the 2-form $\omega$ is dual to 1-forms; and scalar function $v$ is dual to densities 
$\mrm{Den}(\mathbb{R}^2)$. (In $\mbb{R}^2$, densities are also 2-forms.)


\subsection{A modified quasi-geostrophic (QG)  in the $\beta$-plane approximation } 

The quasi-geostrophic (QG) approximation is a basic tool for the analysis of meso- and large-scale motion in geophysical and astrophysical fluid dynamics \cite{pedlosky2013geophysical}.  Physically, QG theory applies when the motion is nearly in geostrophic balance; that is, when pressure gradients nearly balance the Coriolis force. Mathematically, in the simplest case of a barotropic fluid in a domain $\mathscr{D}$ on the plane $\mathbb{R}^2$ with coordinates $\left(x_1, x_2\right)$, QG dynamics in the $\beta$-plane approximation is expressed by the following evolution equation for the streamfunction $\psi$ of the geostrophic fluid velocity $\mathbf{u}=\hat{\boldsymbol{z}} \times \nabla \psi$,
\begin{align}
\p_t (\Delta \psi-\mathscr{F} \psi) + [\psi, \Delta \psi]+\beta \frac{\partial \psi}{\partial x_1}=0,
\label{QG-eqns1}
\end{align}
where $\partial_t$ is the partial time derivative, $\Delta$ is the planar Laplacian.Also, the brackets 
\[
[a, b] \equiv \partial(a, b) / \partial\left(x_1, x_2\right) = J(a,b)
\]
represent the Jacobi bracket (Jacobian) for functions $a$ and $b$ on $\mathbb{R}^2$. In other notation, $\beta$ is the gradient of the Coriolis parameter, $f$, taken as $f=f_0+\beta x_2$ in the $\beta$-plane approximation, with constants $\beta$ and $f_0$. (Neglecting $\beta$ gives the $f$-plane approximation.) 
The symbol  $\mathscr{F}$ in the QG equation \eqref{QG-eqns1} denotes the rotational Froude number, 
\begin{equation}
\mathscr{F}:=\frac{Fr^2}{Ro^2} =\frac{f_0^2L^2}{U^2} \frac{U^2}{g b_0}={L^2}/{L_R^2}=O(1)
\quad\hbox{and}\quad
L_R^2={g b_0}/{f_0^2}\,,
\label{GFDregime}
\end{equation}
where $R_o=U/f_0L$ is the Rossby number, $Fr=U/\sqrt{gb_0}$ is the Froude number,  $g$ is gravitational acceleration, $b_0$ is mean depth,
and $L$ and $U$ are typical lengths and velocities, respectively. Thus, the QG approximation with \emph{rotational} Froude number of order $O(1)$ 
corresponds to slow flows in a rapidly rotating frame of motion, such as the flows which occur in the ocean and atmosphere. Consequently, fluid dynamics in the regime
$\mathscr{F}=O(1)$ is often called \emph{Geophysical Fluid Dynamics} (GFD).

The $\mathrm{QG}$ equation \eqref{QG-eqns1} may be derived from the basic equations of rotating shallow water flow by proper scaling and subsequent asymptotic expansion in the Rossby number $Ro\ll1$ and $\mathscr{F}=O(1)$ in the GFD regime where the rotational Froude number $\mathscr{F}$ is given by the square of the ratio of the characteristic scale of the motion $L$ to the deformation radius $L_R$ defined in \eqref{GFDregime}. See, e.g., Pedlosky \cite{pedlosky2013geophysical}, Allen and Holm \cite{allen1996extended}, and Stegner and Zeitlin \cite{stegner1996asymptotic}. 
In the the $f$-plane approximation, equation \eqref{QG-eqns1} may be written in terms of the potential vorticity, $q$, as
\begin{align}
\p_tq + \mathbf{u} \cdot \nabla q=0\,,
\quad q := \Delta \psi-\mathscr{F} \psi + f  =: \mu + f
\,.\label{QG-eqns2}
\end{align}
Upon augmenting the QG kinetic energy Hamiltonian to account for the vertical  component of 
divergence-free 3D velocity as done in \eqref{Euler-vort-dyn-Ham} for the Euler fluid equation,
one has
\begin{align}\begin{split}
H(\mu,v) &= \frac12 \int_{\mathbb{R}^2} |\nabla \psi |^2 + \mathscr{F} \psi^2  + v^2 \,dxdy 
\\&= \frac12 \int_{\mathbb{R}^2} \mu\big(\Delta - \mathscr{F}\big)^{-1}\mu + v^2\,dxdy
\,.\label{QG-Ham}
\end{split}\end{align}
One may then write equations \eqref{QG-eqns2} in Lie-Poisson form as, cf. equation \eqref{LP-vort-dyn-eqn},
\begin{equation}
\p_t 
\begin{pmatrix}
\mu \\ v 
\end{pmatrix}
= -
\begin{bmatrix}
[\Box, q ] & [\Box, v] \\
[\Box, v ] & 0 
\end{bmatrix}
\begin{pmatrix}
{\delta H/\delta\mu = \psi} \\ {\delta H/\delta v = v}
\end{pmatrix}
= -
\begin{pmatrix}
[\psi, q] \\ [\psi, v] 
\end{pmatrix}
.\label{LP-vort-dyn-eqn2}
\end{equation}
Thus, the quantity $\mu= q-f$ with potential vorticity $q$ is advected \textit{actively} 
by the 2D divergence-free horizontal velocity $\zh\times \nabla \psi$ and the vertical  
component $v$ of the 3D velocity is advected \textit{passively}
as a \textit{diagnostic tracer} for the vertical component of the QG flow.

The Lie-Poisson bracket corresponding to the Poisson matrix in \eqref{LP-vort-dyn-eqn2} is 
\[
\big\{F\,,\,H \big\}(\mu,q) = -\,\int_{\cal D} 
\begin{pmatrix}
{\delta F/\delta\mu } \\ {\delta F/\delta v }
\end{pmatrix}
\begin{bmatrix}
[\Box, q ] & [\Box, v] \\
[\Box, v ] & 0 
\end{bmatrix}
\begin{pmatrix}
{\delta H/\delta\mu } \\ {\delta H/\delta v }
\end{pmatrix}
\,dx_1dx_2
\,.\]
This Lie-Poisson bracket is defined on the dual of the semidirect-product Lie algebra $\mathfrak{X}_{vol}\circledS \Lambda^2$  
of volume preserving vector fields $\mathfrak{X}_{vol}(\mbb{R}^2)$ acting on 2-forms 
$(\mu,v)\in\Lambda^2(\mbb{R}^2)$, modulo exact 2-forms 
\cite{weinstein1983hamiltonian,marsden1983coadjoint}.
Dual coordinates are: 2-form $\mu$ is dual to scalar functions; and scalar function $v$ (vertical velocity)  
is dual to densities $\mrm{Den}(\mathbb{R}^3)$. (In $\mbb{R}^2$, densities are also 2-forms.) 
For related treatments of the geometrical/variational approach to the QG model, see, e.g., 
\cite{holm1985nonlinear,HoMaRa1998a,virasoro1981variational,zeitlin1994differential}.

\begin{exercise}
What are the Casimirs for the Poisson matrix in equation \eqref{LP-vort-dyn-eqn2}?
\index{semidirect product!Casimirs}
\end{exercise}

\begin{answer}
\begin{itemize}
\item
The Casimirs for the Poisson matrix in \eqref{LP-vort-dyn-eqn2} are easily seen to be
\[
C_{\Phi,\Psi}(\mu,v) = \int_{\cal D} \Phi(v) + q \Psi(v) \, d^2x
\,.
\]
\item
If initially the vertical velocity $v$ vanishes initially, though, it will remain so,
for the Hamiltonian in \eqref{QG-Ham}, 
\[
v(\bx,0)=0\to v(\bx,t)=0
\,.\]
In this case, the variable $v$ can be ignored and the Lie Poisson bracket becomes simply,
\begin{align}
\big\{F,H\big\} = - \int_{\cal D} q \Big[ \frac{\delta F}{\delta q}\,,\, \frac{\delta H}{\delta q} \Big]\, d^2x
\,.
\label{q-LPB}
\end{align}
The Casimirs have direct effect on the potential vorticity 
\[
q = \mu+f = \Delta \psi-\mathscr{F} \psi + f
\]
if vertical velocity  $v$ vanishes initially, 
because the Casimirs for the Lie--Poisson bracket in \eqref{q-LPB} are
\begin{align}
C_{\Phi}(q) = \int_{\cal D} \Phi(q) \, d^2x
\,,
\label{q-Casimir}
\end{align}
provided the stream function $\psi$ is constant on the boundary, 
assuming a simply connected domain of flow.
\end{itemize}
\end{answer}



\newpage
\vspace{4mm}\centerline{\textcolor{shadecolor}{\rule[0mm]{6.75in}{-2mm}}\vspace{-4mm}}
\section{Dispersive shallow water (DSW) equations in 1D and 2D}\label{BKeqn-sec}

\secttoc

\textbf{What is this lecture about?} 
In this lecture a dynamical shift in the transport velocity in equation \eqref{DSW-transport-vel} 
for the shallow water dynamics in two spatial dimensions is found to introduce additive second-derivative 
terms in one dimension which lead to a completely integrable Hamiltonian nonlinear wave equation.

\subsection{Hamilton--Pontryagin derivation of 2D DSW equations}
\index{Hamilton--Pontryagin principle!DSW equations} \index{dispersive shallow water equations}

We consider the constrained reduced Lagrangian for ideal \emph{dispersive} shallow water (DSW) equations in two spatial dimensions, 
written in \emph{Hamilton--Pontryagin} form as%
\index{variational principle!Hamilton--Pontryagin}
\begin{align}
\begin{split}
l(\mathbf{u},\eta)
=\int_M &\big(\tfrac{1}{2}\eta|\mathbf{u}|^2  - \tfrac12 {\rm g}\big(\eta^2 { + \alpha^2|\nabla\eta|^2 } \big)\, d^2x
\\&
+ \scp{m}{\dot{g}g^{-1} - u - \kappa \nabla^\sharp \log \eta } + \scp{\phi}{\eta_0g^{-1} - \eta}
\,,\end{split}
\label{HP-DSW-Lag}
\end{align}
where $\scp{\cdot}{\cdot}$ denotes $L^2$ pairing.
In the Lagrangian  \eqref{HP-DSW-Lag},  the gravitational acceleration constant, ${\rm g}$, multiplies $\eta^2/2$ in the usual potential energy. { An energy penalty term $\alpha^2|\nabla\eta|^2$ with constant positive $\alpha^2$ of dimension $[L]^2$ has been added in the gravitational potential energy to control wave steepness and thereby ensure that the resulting variational equations are well posed in the sense of Hadamard}.  An additional transport velocity vector field $\kappa \nabla^\sharp \log \eta $ with constant $\kappa$ of the dimension of diffusivity $[L]^2/[T]$ has also been included to produce dispersion. When $\alpha^2$ and $\kappa$ are absent, then one acquires the standard theory of shallow water dynamics. 

In geometric notation, one has $|\mathbf{u}|^2 := u \contract u^\flat $ with $u^\flat =\mathbf{u}\cdot d\bs{x}$ and the velocity vector field $u \in\mathfrak{X}(M\!\!\subset\!\mathbb{R}^2)$ (written in components as $u=\mathbf{u}\cdot \nabla$) has no normal components on the boundary $\partial M$. That is, $u\contract d^2x = \mathbf{u}\cdot \nh \,ds = 0$ on the boundary. The columnar volume of fluid $\eta d^2x$ with local depth $\eta$ moves with the flow and so satisfies the \emph{continuity equation}
\begin{align}
\big(\p_t + \pounds_{\dot{g} g^{-1}} \big)(\eta d^2x) = 0
= \Big(\partial_t \eta + {\rm div}(\eta (\dot{g} g^{-1})\big)\Big)d^2x
\,.
\label{DSW-eta}
\end{align}
In this lecture, one takes advantage of the feature discussed in Remark  \ref{HP-feature} that the Hamilton--Pontryagin approach accommodates alternative  types of transport velocity $v=\dot{g} g^{-1}$ and advected quantities $a_t = a_0g^{-1}$.

Taking variations in the Hamilton--Pontryagin action integral with constrained Lagrangian $l(\mathbf{u},\eta)$ in \eqref{HP-DSW-Lag} yields 
\begin{align}
\begin{split}
0 = \delta S &= \delta \int_M l(\mathbf{u},\eta) \,dt
\\&
= \int_M \scp{\eta u^\flat - m}{\delta u} + \scp{B(\mathbf{u},\eta) - \phi }{\delta \eta}
+ \scp{\delta\phi}{\eta_0g^{-1} - \eta}
\\&\qquad \scp{-\,\big(\p_t + \pounds_{\dot{g} g^{-1}} \big) m  + \eta\, d B}{\xi} 
+ \scp{\delta m}{\dot{g}g^{-1} - u - \kappa \nabla^\sharp \log \eta } 
\,dt
\,.
\end{split}
\label{HP-DSW-var}
\end{align}
Here one has the following notation,
\begin{align}
\xi = \delta g g^{-1} 
\,,\quad 
\delta(\dot{g} g^{-1}) = \big(\p_t - \ad_{\dot g g^{-1}} \big)\xi
\,,\quad 
\delta \eta = \delta(\eta_0 g^{-1}) = - \, \pounds_\xi \eta
\,,\label{var-xi}
\end{align}
in which the $\phi$-constraint relation has been used.  The other constraints yield
\begin{align}
m = \eta u^\flat = \mathbf{u}\cdot d\bx \otimes \eta d^2x = \mathbf{m}\cdot d\bx \otimes d^2x
\,,\quad
\dot{g} g^{-1} = u - \kappa \nabla^\sharp \log\eta
\,,\label{var-m}
\end{align}
and also $B(\mathbf{u},\eta) = \phi$, where the Bernoulli function ${\cal B}(\mathbf{u},\eta)$ is given by
\begin{align}
{\cal B}(\mathbf{u},\eta) := \frac12 |\mathbf{u}|^2 - {\rm g}{(1 - \alpha^2\Delta)}\eta - \frac{\kappa}{\eta}\, {\rm div}(\eta \mathbf{u})
\,.\label{B-deta-DSW}
\end{align}
The full set of DSW equations may be written as
\begin{align}
\begin{split}
 \big(\p_t + \pounds_{\dot{g} g^{-1}} \big) u^\flat &= d{\cal B}
\,, \\
  \big(\p_t + \pounds_{\dot{g} g^{-1}} \big) \eta d^2x &= 0 
\,.
\end{split}
\label{B--DSW-eqns}
\end{align}
When $\kappa$ and $\alpha$ both vanish,  equations \eqref{B--DSW-eqns} recover the familiar
equations for shallow water waves.

\subsection{Kelvin's theorem and potential vorticity (PV) for DSW equations}
The motion equation in \eqref{B--DSW-eqns} yields the Kelvin theorem 
\begin{align}
\frac{d}{dt}\oint_{c(\dot{g} g^{-1})} u^\flat 
=
\oint_{c(\dot{g} g^{-1})} \big(\p_t + \pounds_{\dot{g} g^{-1}} \big) u^\flat 
= \oint_{c(\dot{g} g^{-1})}  d{\cal B} = 0
\,,\label{DSW-Kelvin}
\end{align}
where $c(\dot{g} g^{-1})$ is a material loop moving with the transport velocity vector field 
\begin{align}
v := \dot{g} g^{-1} = u - \kappa \nabla^\sharp \log\eta
\,,\label{DSW-transport-vel}
\end{align}
as evaluated by the constraint relation in \eqref{var-m}.

The Kelvin theorem for the dispersive shallow water equations 
may also be written in fluid dynamics terms as
\begin{align}
\frac{d}{dt}\oint_{c(v)}\mathbf{u}\cdot d\bx 
= 
\oint_{c(v)} \nabla {\cal B} \cdot d\bx = 0
\,.\label{DSW-Kelvin-fluid}
\end{align}
The motion equation  in \eqref{B-deta-DSW} also implies
\begin{align}
\big(\p_t + \pounds_{\dot{g} g^{-1}} \big) u^\flat  = d{\cal B}
\,,\label{DSW-umot-eqn}
\end{align}
whose exterior derivative yields
\begin{align}
\big(\p_t + \pounds_{\dot{g} g^{-1}} \big) du^\flat  = d^2{\cal B} = 0
\,.\label{DSW-umot-eqn}
\end{align}
In fluid dynamics terms with vorticity 2-form defined by $du^\flat =: \omega d^2x$ along with the continuity 
equation for the depth in \eqref{DSW-eta}, the previous equation implies the following advection law for 
the scalar potential vorticity (PV) defined as the function 
\[
q:=\omega/\eta
= (u_{2,1} - u_{1,2})/\eta
\,,\]
\begin{align}
\Big(\p_t + \mathbf{v} \cdot \nabla\Big)q =
\Big(\p_t + (\mathbf{u} - \kappa \nabla^\sharp \log\eta)\cdot \nabla\Big)q = 0
\,,\quad\hbox{with PV},\quad
q:=\omega/\eta
\,.\label{DSW-omega-eqn-fluid}
\end{align}
In turn, the PV advection equation \eqref{DSW-omega-eqn-fluid} implies conservation by the DSW equations in \eqref{B-deta-DSW} of the quantity 
\begin{align}
C_\Phi := \int_M \eta \Phi(q)\,d^2x
\,,\label{DSW-omega-eqn-fluid}
\end{align}
for any differentiable function $\Phi$.

\subsection{Hamiltonian formulation of the DSW equations}

By the Legendre transformation of the DSW Lagrangian in equation \eqref{HP-DSW-Lag}, one finds the DSW Hamiltonian 
\begin{align}
\begin{split}
h (m,\eta) &:= \scp{m}{\dot{g} g^{-1}} - l (u,\eta)
\\&:= 
\int_M \frac{|\mathbf{m}|^2}{2\eta} - \kappa \mathbf{m} \cdot\nabla^\sharp\log(\eta)  
+ \frac{\rm g}{2} (\eta^2{+\alpha^2|\nabla\eta|^2}) \, d^2x
\,,
\end{split}
\label{DSW-Ham}
\end{align} 
whose variational derivatives are given by
\begin{align}
\begin{split}
\frac{\delta h}{\delta \mathbf{m}} &= \mathbf{m}/\eta - \kappa \nabla^\sharp \log(\eta) 
= \mathbf{u} - \kappa \nabla^\sharp \log(\eta) 
\,,\\
\frac{\delta h}{\delta \eta} &= -\,\frac{|\mathbf{m}|^2}{2\eta^2} + \frac{\kappa}{\eta}{\rm div}\mathbf{m} 
+ {\rm g}{(1 - \alpha^2\Delta)}\eta
= -\,{\cal B} 
\,.
\end{split}
\label{DSW-omega-eqn-fluid}
\end{align}
The Lie--Poisson equations in this case arise from the standard theory,
\begin{align}\begin{split}
\partial_t{m}
+\text{ad}^*_{\delta h/\delta m}\, m
&=
-\,
\frac{\delta h}{\delta \eta}\diamond \eta
= 
-\,
\eta \nabla \frac{\delta h}{\delta \eta}
\,,\\
\partial_t({\eta}d^2x) + \mathcal{L}_{\delta h/\delta m}\, (\eta d^2x) &= 0
\,.\label{LP-DSW-Eqns}
\end{split}\end{align}

In Lie-Poisson matrix operator form, this is 
\begin{equation}
\p_t 
\begin{pmatrix}
m_i \\ \eta
\end{pmatrix}
= -
\begin{bmatrix}
\p_j m_i + m_j\p_i & \Box\diamond \eta \\
\mathcal{L}_{\Box}\eta & 0 
\end{bmatrix}
\begin{pmatrix}
\delta h/\delta m_j = v^j :=  u^{\,j} - \kappa \p^j\!\log(\eta) 
\\  
\delta h/\delta \eta = - {\cal B}  
\end{pmatrix}
.\label{LP-DSW-brkt} 
\end{equation}
The variables in this Lie-Poisson operator are defined on the dual%
\footnote{Dual spaces are defined for $L^2$ pairing in the planar domain $M$
denoted by angle brackets $\scp{\,\cdot}{\cdot\,}$.
} 
of the semidirect-product Lie algebra $\mathfrak{X}\circledS \Lambda^0$  
of vector fields $\mathfrak{X}(M)$ acting on functions $\Lambda^0(M)$.
Dual coordinates: the 1-form density $m\in \mathfrak{X}^* \otimes \mrm{Den}(M)$ 
is dual to the vector fields $\mathfrak{X}(M)$; and depth $\eta\in \mrm{Den}(M)$ is 
dual to scalar functions defined in domain $M$. \index{semidirect product!Lie algebra}

In the usual fluid dynamics notation, the system \eqref{LP-DSW-Eqns} becomes 
\begin{align}\begin{split}
\partial_t \mathbf{u}
+ (\mathbf{v}\cdot\nabla) \mathbf{u}+ u_j \nabla v^j
&=
\nabla {\cal B}
\,,\\
\partial_t\eta + {\rm div} (\eta \mathbf{v})&= 0
\,.\label{LP-DSW-System}
\end{split}\end{align}
where the transport velocity $ \mathbf{v} =  \mathbf{u} - \kappa \nabla \log\eta$ is given 
in \eqref{DSW-transport-vel} and the Bernoulli function ${\cal B}(\mathbf{u},\eta)$ is defined in \eqref{B-deta-DSW}.

\begin{exercise}
Calculate the transformation of variables $(\mathbf{m},\eta)\to (\mathbf{u}=\mathbf{m}/\eta,\eta)$
for the 2D version of the Lie--Poisson structure in \eqref{LP-DSW-brkt} and determine the equations
of motion in the variables $(\mathbf{u},\eta)$.
\end{exercise}

\begin{answer}
In two dimensions, the transformation of the Lie-Poisson operator becomes
\[
\begin{bmatrix}
\delta_{ki}/\eta   &	- m_k /\eta^2		
\\							
0	&	1		
\end{bmatrix}
\begin{bmatrix}	
\p_j m_i + m_j\p_i  &	\eta\p_i	
\\
\p_j \eta	&	0
\end{bmatrix}
\begin{bmatrix}
\delta_{jl}/\eta	&	0		
\\							
- m_l / \eta^2   &	1		
\end{bmatrix}
=
\begin{bmatrix}
(u_{k,l} - u_{l,k} )/\eta	 &	\p_k		
\\							
\p _l	&	0		
\end{bmatrix}
\,.\]
The Hamiltonian \eqref{DSW-Ham} in the variables $(\mathbf{u},\eta)$ is written as
\begin{align}
\tilde{h} (\mathbf{u},\eta) := 
\int_M \eta \frac{|\mathbf{u}|^2}{2} - \kappa\mathbf{u} \cdot\nabla\eta 
+ \frac{\rm g}{2} (\eta^2{+\alpha^2|\nabla\eta|^2}) \, d^2x
\,,
\label{DSW-Ham-u}
\end{align} 
with variations
\begin{align}
\delta\tilde{h} (\mathbf{u},\eta) &:= 
\int_M \eta\mathbf{v}\cdot\delta \mathbf{u}
+ \tilde{\cal B} \,\delta\eta \, d^2x
\,,\label{DSW-Ham-u-var}
\end{align} 
with transport velocity $ \mathbf{v} :=  \mathbf{u} - \kappa \nabla \log\eta$ and Bernoulli function,
cf. \eqref{B-deta-DSW},
\begin{align}
\tilde{\cal B}
= \frac12 |\mathbf{u}|^2 + \kappa {\rm div}\mathbf{u} + g(1-\alpha^2\Delta)\eta 
\,.\label{DSW-B-tilde}
\end{align} 
The equations of motion in the variables $(\mathbf{u},\eta)$ in \eqref{LP-DSW-System} can be rewritten as 
\begin{align}
\begin{split}
\p_t 
\begin{pmatrix}
u_k \\ \eta
\end{pmatrix}
&= -
\begin{bmatrix}
(u_{k,l} - u_{l,k} )/\eta	 &	\p_k		
\\							
\p _l	&	0		
\end{bmatrix}
\begin{pmatrix}
\delta h/\delta u_l = \eta v^l
\\
\delta h/\delta \eta = \tilde{\cal B}
\end{pmatrix}
,\\
\p_t 
\begin{pmatrix}
\mathbf{u} \\ \eta
\end{pmatrix}
&=
-
\begin{pmatrix}
-\mathbf{v}\times {\rm curl} \mathbf{u} + \nabla\tilde{\cal B}\
\\
{\rm div}(\eta \mathbf{v})
\end{pmatrix}
.\end{split}
\label{LP-DSW-System2}
\end{align} 
When $\kappa$ and $\alpha$ both vanish,  equations \eqref{LP-DSW-System2} recover the familiar
equations for shallow water waves. 

\end{answer}

\subsection{The BKBK equation: DSW in one dimension}
In one dimension, the DSW system \eqref{LP-DSW-System} becomes 
\begin{align}\begin{split}
\partial_t u + \p_x \big(uv - {\cal B}) &= 0 
\\& = \partial_t u + \p_x \big( u^2/2 + {\rm g}{(1 - \alpha^2\p_x^2)}\eta + \kappa u_x\big) 
\\&= \partial_t u + uu_x + {\rm g}{(1 - \alpha^2\p_x^2)}\eta_x + \kappa u_{xx} 
\,,\\
\partial_t\eta + \p_x \big( \eta v\big) &=0
\\& = \partial_t\eta + \p_x \big( \eta (u - \kappa \p_x \log\,\eta)\big),
\label{1D-DSW-System} 
\end{split}\end{align} 
where we have used the one-dimensional version of ${\cal B}$ defined in \eqref{B-deta-DSW} and the 
relation between $u$ and $v$; namely,  $\kappa \p_x \log(\eta) = u - v$.

The one-dimensional DSW system in \eqref{1D-DSW-System}, can be rewritten as,
\begin{align}\begin{split}
\partial_t u + uu_x + {\rm g}{(1 - \alpha^2\p_x^2)}\eta_x + \kappa u_{xx} &= 0
\,,\\
\partial_t\eta + \p_x ( \eta u) - \kappa \eta_{xx} &=0
\,.\label{BK-System-alpha} 
\end{split}
\end{align} 
After transforming one-dimensional dynamical variables $(m,\eta)\to (u,\eta)$ for the Hamiltonian 
in \eqref{DSW-Ham} and transforming Lie-Poisson operator in \eqref{LP-DSW-brkt} to the new variables, 
the system in \eqref{BK-System}  takes the following Hamiltonian form,
\begin{equation}
\p_t 
\begin{pmatrix}
u \\ \eta
\end{pmatrix}
= -
\begin{bmatrix}
0 & \p_x \\
\p_x & 0 
\end{bmatrix}
\begin{pmatrix}
\delta h/\delta u  = \eta u - \kappa \eta_x 
\\  
\delta h/\delta \eta = u^2/2 + {\rm g}{(1 - \alpha^2\p_x^2)}\eta + \kappa u_x
\end{pmatrix}
.\label{BK-System-Ham2} 
\end{equation}
 For $\alpha^2=0$, the system \eqref{BK-System-alpha} recovers the BKBK system,
 \begin{align}\begin{split}
\partial_t u + uu_x + {\rm g}\eta_x + \kappa u_{xx} &= 0
\,,\\
\partial_t\eta + \p_x ( \eta u) - \kappa \eta_{xx} &=0
\,,\label{BKBK-System} 
\end{split}
\end{align} 
and the equation \eqref{BK-System-Ham2} recovers one of the Hamiltonian formulations of the BKBK system.%
\footnote{See \cite{cheviakov2024analytical} for a compendium of nonlinear shallow water waves in one spatial dimension,
including the BKBK equation system \eqref{BKBK-System}.}

\begin{remark}\rm
Expressions \eqref{BKBK-System} and the Hamiltonian form \eqref{BK-System-Ham2} {with $\alpha^2=0$} 
have recovered the celebrated Broer-Kaup-Boussinesq-Kupershmidt (BKBK)
 system which for $\kappa = -1/2$ was shown in \cite{kupershmidt1985mathematics} to be a completely integrable 
 nonlinear wave system with three Hamiltonian structures in terms of variables $(u,\eta)$. In fact, in \cite{kupershmidt1985mathematics} 
 Kupershmidt even called the system in \eqref{BKBK-System} ``the richest integrable system known to date.''
\end{remark}

\begin{exercise}
Prove that the transformation of variables $(m,\eta)\to (u=m/\eta,\eta)$
takes the 1D version of the Lie--Poisson structure in \eqref{LP-DSW-brkt} into the constant--coefficient
Poisson operator in equation \eqref{BK-System-Ham2}. 
\end{exercise}

\begin{answer}
The transformation of the Poisson operator $(m,\eta)\to (u=m/\eta,\eta)$ is done by multiplying 
the Lie--Poisson operator from the left by the Jacobian matrix and from the right by its transpose. 
Explicitly, one computes
\[
\begin{bmatrix}
1/\eta   &	- m /\eta^2		
\\							
0	&	1		
\end{bmatrix}
\begin{bmatrix}	
\p m + m\p  &	\eta\p	
\\
\p \eta	&	0
\end{bmatrix}
\begin{bmatrix}
1/\eta	&	0		
\\							
- m / \eta^2   &	1		
\end{bmatrix}
=
\begin{bmatrix}
0	 &	\p		
\\							
\p 	&	0		
\end{bmatrix}
\,.\]
\end{answer}

\begin{exercise}
Linearise the system \eqref{BK-System-alpha} around the equilibrium solution $(u,\eta)=(0,\eta_0)$ where $\eta_0$ is the constant mean depth. 
Calculate the dispersion relation $\omega(k)$ and the phase velocity $c_p(k):=\omega(k)/k$ for travelling waves depending on space and time 
in the vicinity of this equilibrium as $\exp(i(kx-\omega t))$ for wave number $k$ and frequency $\omega$. 

In this context, what can go wrong?
\end{exercise}

\begin{answer}
For travelling waves in the vicinity of the constant equilibrium solution 
$(u,\eta)=(0,\eta_0)$, equation \eqref{BK-System-Ham2} yields
\begin{equation}
\begin{pmatrix}
\omega/k & 0
\\
0 & \omega/k 
\end{pmatrix}
\begin{pmatrix}
\tilde{u} \\ \tilde{\eta}
\end{pmatrix}
=
\begin{pmatrix}
 ik\kappa  &  {\rm g}({1+\alpha^2k^2)}
\\
\eta_0  & - ik\kappa   
\end{pmatrix}
\begin{pmatrix}
\tilde{u} \\ \tilde{\eta}
\end{pmatrix}
,\label{BK-System-lin} 
\end{equation}
for which a solution exists, provided
\begin{equation}
\det
\begin{pmatrix}
k^{-1}\omega - ik\kappa  &  {\rm g}({1+\alpha^2k^2)}
\\
\eta_0  & k^{-1}\omega + ik\kappa   
\end{pmatrix}
=
0\,.
\end{equation}
Hence, the dispersion relation is 
\begin{equation}
\omega^2(k^2) = k^2({\rm g}\eta_0({1+\alpha^2k^2)} - \kappa^2 k^2)
\,,
\label{disp-alpha}
\end{equation}
and the travelling wave solution is linearly well-posed ($\omega^2$ remains positive) provided $g\eta_0{(1+\alpha^2)} > \kappa^2$.

\end{answer}

\begin{exercise}
In the context of the dispersion relation in equation \eqref{disp-alpha} in the previous exercise, 
what can go wrong, if $\alpha^2=0$? Is this really a problem?
\end{exercise}

\begin{answer}

If $\alpha^2=0$, then the dispersion relation in equation \eqref{disp-alpha} becomes
\begin{equation}
\omega^2(k^2) = k^2({\rm g}\eta_0 - \kappa^2 k^2)
\,,
\label{disp-noalpha}
\end{equation}
and the BKBK system \eqref{BKBK-System} is \emph{linearly ill-posed} for higher wave numbers, $ k^2>g\eta_0 / \kappa^2 $,
independently of the sign of the parameter $\kappa$. Regardless of this linear ill-posedness, though, the solutions of the 
BKBK system in \eqref{BKBK-System}  are completely integrable as a Hamiltonian system for $\kappa = -1/2$, as 
was shown in \cite{kupershmidt1985mathematics}.

\end{answer}
 
As mentioned above, for $\alpha^2=0$ the system \eqref{BK-System-alpha} recovers the celebrated BKBK system \eqref{BKBK-System}, rewritten here as
 \begin{align}\begin{split}
\partial_t u + uu_x  &= -\, {\rm g}\eta_x  -\, \kappa u_{xx} 
\,,\\
\partial_t\eta + u\eta_x &= -\, \eta u_x  + \kappa \eta_{xx} 
\,.\label{BK-System}  
\end{split}
\end{align} 
The linearised dispersion relation is $\omega^2(k^2)= k^2(- \,\kappa^2 k^2 + {\rm g}\eta_0)$, 
which appears to be problematic for the linear travelling waves in this system as indicated
in the previous exercise. However, the nonlinear system is in fact completely integrable for $\kappa=-1/2$
\cite{kupershmidt1985mathematics}. 
This type of subtle analytical issue may not be uncommon with Boussinesq long wave equations.


For a recent review of Boussinesq long waves and their ``good" and ``bad"  analytical issues of the Boussinesq equations, 
see, e.g., \cite{klein2021nonlinear}. For further discussions of the ``good" and ``bad"  aspects of the BKBK system, see \cite{klein2024kaup}.

\begin{remark}\rm
Controlling the wave steepness with the $\alpha^2$ term introduced in the Hamilton--Pontryagin variational principle
in \eqref{HP-DSW-Lag} regularises both the ``good" and ``bad" DSW equations in 1D. They could  have also both
been  regularised by surface tension (introduced by adding $-\tau \eta_{xxxx}$ for positive parameter $\tau$ to the 
right-hand side of the $\eta$-equation in \eqref{BK-System}). However, if this were done, then the system would have  
no longer  been Hamiltonian. 
\end{remark}

\begin{remark}\rm
The primary lesson of this lecture is that the dynamical shift in the transport velocity in equation \eqref{DSW-transport-vel} 
for the shallow water dynamics in two dimensions introduces additive second-derivative terms in the BKBK system \eqref{BK-System} 
which produce proportional `bad' Boussinesq wave dispersion in one dimension. For further discussion of the BKBK system, particularly from  
from the viewpoint of generalised two-cocycles, see \cite{kupershmidt1992variational,kupershmidt2006extended}.
For recent discussions of `good' versus `bad' Boussinesq equations from the viewpoint of analysis of integrable shallow water
equations in one dimension, see e.g., \cite{charlier2023good,cheviakov2024analytical}. 

\end{remark}

\newpage

\newpage
\vspace{4mm}\centerline{\textcolor{shadecolor}{\rule[0mm]{6.75in}{-2mm}}\vspace{-4mm}}
\section{Rotating Shallow Magnetised Water  (RSW-MHD)}

\secttoc

\textbf{What is this lecture about?} This lecture introduces magnetohydrodynamics (MHD) into
rotating shallow water (RSW) dynamics via the Euler-Poincar\'e variational principle and derives their Hamiltonian formulation. 

\subsection{Rotating Shallow Water MHD (RSW-MHD)}\index{Rotating Shallow Water MHD!RSW-MHD}
Following \cite{gilman2000magnetohydrodynamic,dellar2002hamiltonian,dellar2003common},
the Rotating Shallow Water MHD (RSW-MHD) motion equation in a thin domain with bathymetry $\hbar(\bs{x})$ is written as
\begin{align}
 \partial_t \bs{u}+\bs{u} \cdot \nabla \bs{u}
+ 2\bs{\Omega} \times \bs{u}
= - g \nabla (\eta - \hbar(\bs{x})) + \bs{B} \cdot \nabla \bs{B} 
\,.\label{RSW-MHD-mot}
\end{align}
The dynamical RSW-MHD variables in \eqref{RSW-MHD-mot} denote the horizontal velocity $\bs{u}$, horizontal magnetic field $\bs{B}$, 
and local depth $\eta - \hbar(\bs{x})$. The parameters for Coriolis force and gravitational acceleration are, respectively, $2\bs{\Omega}$ and $g$.

Two advection relations hold as auxiliary equations for the RSW-MHD motion \eqref{RSW-MHD-mot},
\begin{align}
\begin{split}
&  \partial_t \eta+\nabla \cdot(\eta \bs{u})=0 \,,\\
&  \partial_t \bs{B} + \bs{u} \cdot \nabla \bs{B} - \bs{B} \cdot \nabla \bs{u} = 0
\,.
\end{split}
\label{RSW-MHD-aux}
\end{align}
Together, these auxiliary equations imply preservation of
\begin{align}
 \nabla \cdot (\eta \bs{B})=0
\,,\label{RSW-MHD-div}
\end{align}
which can therefore be regarded as a non-dynamical constraint on the initial values.

\begin{exercise}
Prove the previous statement. Hint: see Remark \ref{eta-B-proof}.
\end{exercise}

To put ourselves into the geometric mechanics framework, we shall begin by showing that the RSW-MHD equations 
in  \eqref{RSW-MHD-mot} and \eqref{RSW-MHD-aux} follow as Euler--Poincar\'e equations for Hamilton's principle 
with the action integral
\begin{align}\begin{split}
S= & \int_0^T \ell(\bs{u}, \eta, \bs{B}) d t 
\\
= & \int_0^T \!\!\!\int_{C S} 
\left(\frac{1}{2}|\bs{u}|^2 +\frac{1}{\op{Ro}} \bs{u} \cdot \bs{R}(\bs{x})
-\frac{1}{2 } |\bs{B}|^2
-\frac{1}{2 \mathrm{Fr}^2} (\eta-2 \hbar(\bs{x}))\right) \eta\, d^2x d t 
\,,\label{RSW-MHD-action}
\end{split}\end{align}
where $CS\in \mbb{R}^2$ denotes the horizontal cross-section. The RSW-MHD equations in \eqref{RSW-MHD-mot} and \eqref{RSW-MHD-aux} will be derived by first evaluating the variational derivatives for the Lagrangian in the action integral \eqref{RSW-MHD-action} as%
\footnote{In Euclidean coordinates, a vector field $u = \bu\cdot\nabla$ has an associated $1$-form $u^\flat = \bu\cdot d\bx$ where $d\bx$ is the dual basis to $\nabla$ under contraction, $\nabla\contract d\bx = Id$.
With this notation, the terms in action integral \eqref{RSW-MHD-action} involving inner product may be written in coordinate-free notation as $|\bu|^2 = u \contract u^\flat,\ |\bs{B}|^2 = B \contract B^{\flat}, \hbox{ and } \bu\cdot\bs{R} = u \contract R$.}
\begin{align}\begin{split}
\frac{1}{\eta} \frac{\delta l}{\delta \bs{u}} 
& = \Big(\bs{u}+\frac{1}{\op{Ro}} \bs{R}(\bs{x})\Big)\cdot \mb{d}\bs{x}
=: \bs{V}(\bs{x}, t) \cdot \mb{d}\bs{x} =: V^\flat \\
\frac{\delta l}{\delta \eta} & = \Big(\frac{1}{2}|\bs{u}|^2+\frac{1}{\op{Ro}} \bs{u} \cdot \bs{R}(\bs{x}) - \frac12 |\bs{B}|^2 -\frac{1}{\mathrm{Fr}^2} \big(\eta-\hbar(\bs{x})\big) \Big)
=:\beta(\bs{x}, t) \\
\frac{\delta l}{\delta \bs{B}{}} & =  \bs{B}\cdot \mb{d}\bs{x}\otimes \eta d^2x
=:  B^\flat \otimes \eta d^2x
\,.\label{Var-action}
\end{split}\end{align}
The required Euler--Poincar\'e variations are then given by their transformation properties,
\begin{align}\begin{split}
\delta u &= \p_t\xi - \ad_u\xi 
\,,\quad
\delta \eta = -{\cal L}_\xi \eta = - \op{div}(\eta \bs{\xi})\, d^2x
\,,
\\
\delta B{} &=  - {\cal L}_\xi B{} 
= \big(- \bs{\xi}\cdot\nabla \bs{B} + \bs{B}\cdot\nabla \bs{\xi}\big)\cdot\nabla
\,.\label{Var-formulas}
\end{split}\end{align}
The corresponding auxiliary equations are
\begin{align}\begin{split}
\p_t \eta &= -{\cal L}_u \eta = - \op{div}(\eta \bs{u})\, d^2x
\,,\\
\p_t B{} &= -{\cal L}_u B{} = \ad_u B{} = -\,[u,B{}]
\,,\\
\p_t \bs{B}\cdot\nabla
&= \big(- \bs{u}\cdot\nabla \bs{B} + \bs{B}\cdot\nabla \bs{u}\big)\cdot\nabla
\,.\label{aux-eqns}
\end{split}\end{align}
Finally, the Euler--Poincar\'e equations follow by direct calculation as
\begin{align}\begin{split}
0 = \delta S 
&= 
\int_0^T \scp{\frac{\delta \ell}{\delta u}}{\p_t\xi - \ad_u\xi}
+ \scp{\frac{\delta \ell}{\delta \eta}}{- {\cal L}_\xi\eta}
+ \scp{\frac{\delta \ell}{\delta B{}}}{- \ad_\xi B{}}dt
\\&=
\int_0^T \scp{- (\p_t + \ad^*_u)\frac{\delta \ell}{\delta u}
+ \frac{\delta \ell}{\delta \eta}\diamond \eta 
+ \frac{\delta \ell}{\delta B}\diamond B}{\xi}
dt
\\&=
\int_0^T \scp{- (\p_t + \ad^*_u)\frac{\delta \ell}{\delta u}
+ \frac{\delta \ell}{\delta \eta}\diamond \eta 
+ \frac{\delta \ell}{\delta B{}}\diamond B{}}{\xi}
\\&=
\int_0^T \scp{- (\p_t + {\cal L}_u) (\eta d^2x \otimes V^\flat) 
+ \eta d^2x \otimes \mb{d}\beta 
- {\cal L}_{B{}} (\eta d^2x \otimes B^\flat)}{\xi}
\\&=
\int_0^T \scp{\eta d^2x \otimes \big(- (\p_t + {\cal L}_u)  V^\flat 
+ \mb{d}\beta - {\cal L}_{B{}}B^\flat) \big)
- {\cal L}_{B{}} (\eta d^2x) \otimes B^\flat)
}{\xi}
\\&=
\int_0^T -\Big\langle \eta d^2x \otimes \big(
\p_t \bs{u} +\bs{u} \cdot \nabla \bs{u}+ 2\bs{\Omega} \times \bs{u}
\\&\hspace{2cm}
+ g \nabla (\eta - \hbar(\bs{x}))
- \bs{B} \cdot \nabla \bs{B} \big)
- \op{div}(\eta \bs{B})d^2x\otimes B^\flat
\,,\,\xi \Big\rangle
dt 
\,.
\end{split}
\label{Var-formulas}
\end{align}
\begin{remark}\label{eta-B-proof}\rm 
By virtue of the auxiliary equations in \eqref{aux-eqns} in geometric form, 
\[
(\p_t + {\cal L}_u)(\eta\,d^2x)= 0 
\quad\hbox{and}\quad
(\p_t + {\cal L}_u)B{} = 0
\,,\]
one proves advection of the divergence $\op{div}(\eta \bs{B}) $,
\begin{align}
   (\p_t + {\cal L}_u)d\big(B{}\contract (\eta\,d^2x)\big) = (\p_t + {\cal L}_u)( \op{div}(\eta \bs{B})  d^2x) = 0
\,.\label{Advect-div}
\end{align}
Therefore, if  the quantity $\op{div}(\eta \bs{B})$ in \eqref{Var-formulas} vanishes initially, then it will remain so. 
Consequently, one may ignore the last term in \eqref{Var-formulas} by assuming that it has vanished initially and will remain so. 
This calculation concludes the Euler--Poincar\'e variational derivation of the motion equation \eqref{RSW-MHD-mot} for RSW-MHD.
\end{remark}
\begin{remark}\rm 
The Kelvin-Noether theorem for RSW-MHD is given by
\begin{align}
\begin{split}
\frac{d}{dt}\oint_{c(u)} V^\flat &= \oint_{c(u)} ( \p_t + {\cal L}_u)V^\flat 
= \oint_{c(u)} (\mb{d}\beta - {\cal L}_{B{}}B^\flat) 
= - \oint_{c(u)} {\cal L}_{B{}}B^\flat 
\\&= - \oint_{c(u)} B{}\contract {\rm d} B^\flat 
= - \oint_{c(u)} \bs{B} \times {\rm curl} \bs{B}\cdot d\bs{x}
= \oint_{c(u)}\bs{J} \times \bs{B} \cdot d\bs{x}
\,,\end{split}
\label{KNthm-RSW-MHD}
\end{align}
where $V^\flat$ is defined in equation \eqref{Var-action}.
Thus, the circulation of an RSW-MHD flow in terms of $V^\flat$ is not conserved. 
\end{remark}

\begin{exercise} The Thermal Rotating Shallow Water MHD (TRSW-MHD) model is an extension of the RSW-MHD model 
to include horizontally varying buoyancy and an inert lower layer. 
The TRSW-MHD equations modify the RSW-MHD equations to include  the variable (nonnegative) buoyancy $\gamma^2(\bx,t) = (\bar{\rho} - \rho(\bx,t))/\bar{\rho}$, where  $\rho$ is the (time and space dependent) mass density of the active upper layer and $\bar{\rho}$ is the uniform mass density of the inert lower layer whose boundary is represented by bathymetry $\hbar(\bs{x})$. 

Hamilton's principle for the TRSW-MHD equations is the same as for the RSW-MHD equations in \eqref{RSW-MHD-action} except that in both the Lagrangian in \eqref{RSW-MHD-action} and the Hamiltonian in \eqref{RSWMHD-Ham} the factor $1/\mathrm{Fr}^2$ is replaced by $\gamma^2(\bx,t)/\mathrm{Fr}^2$ where the scalar buoyancy function  $\gamma^2(\bx,t)$ is advected dynamically by the TRSW-MHD flow velocity. 

Calculate the Euler-Poincar\'e equations and Lie-Poisson Hamiltonian equations for the TRSW-MHD model.
\end{exercise}

\subsection{Hamiltonian structure of RSW-MHD equations}}\index{Rotating Shallow Water MHD!Hamiltonian structure}

Next, we derive the Hamiltonian and the Lie-Poisson structure of the RSW-MHD equations inherited from being Euler--Poincar\'e equations. 
The Lie-Poisson properties of these equations are interesting (and helpful) in studying a variety of astrophysical phenomena, 
including the dynamics gravity waves and Alfv\'en waves on the solar tachocline \cite{gilman2000magnetohydrodynamic,dellar2002hamiltonian}. 

The Legendre transformation of the Lagrangian in Hamilton's principle \eqref{RSW-MHD-action} yields the Hamiltonian 
in terms of $m=\bs{m}\cdot d\bs{x}\otimes d^2x$, $B{}=\bs{B}\cdot\nabla$, and $\eta d^2x$,
\begin{align}
h (m, \eta, B) = \int \Big( \frac{1}{2 \eta^2}  |\bs{m}-\eta\bs{R}/Ro|^2 + \frac{1}{2 } |\bs{B}|^2
+ \frac{1}{2 \mathrm{Fr}^2} (\eta-2 \hbar(\bs{x}))\Big) \eta\, d^2 x 
 \,.\label{RSWMHD-Ham}
\end{align}
Variations are obtained as
\begin{align}
\delta h = \int u\contract \delta m + \eta \delta B \contract B^\flat + \phi \,\delta \eta 
\, d^2 x 
 \,.\label{RSWMHD-HamVar}
\end{align}
with Bernoulli function $\phi$ defined for this case as 
\begin{align}
\phi := - \,\frac{|\bs{u}|^2}{2} + \frac12 |\bs{B}|^2 + \frac{1}{ \mathrm{Fr}^2}  (\eta - \hbar(\bs{x}))
 \,.\label{RSWMHD-phi}
\end{align}
The corresponding semidirect-product Lie--Poisson Hamiltonian equations are give by 
\index{semidirect product!Lie Poisson bracket}
\begin{align}
\p_t
\begin{pmatrix}
m \\ B{} \\ \eta
\end{pmatrix}
&=
-
\begin{bmatrix}
\ad^*_{\Box}m 	& -\ad^*_{B{}} {\Box}    & {\Box} \diamond \eta 
\\ - \ad_{\Box} B{}    & 	0		 & 	0
\\ {\cal L}_{\Box} \eta		& 	0		 & 	0
\end{bmatrix}
\begin{pmatrix}
u \\  \eta d^2x \otimes B^\flat \\ \phi
\end{pmatrix}
\\&=
-
\begin{pmatrix}
{\cal L}_{u} (\eta d^2x \otimes u^\flat)  +  \eta d^2x \otimes{\cal L}_{B{}}B^\flat   +  \eta d^2x \otimes {\rm d} \phi  
\\ - \ad_{u} B{}    
\\ {\cal L}_{u} \eta	
\end{pmatrix}
 ,\label{RSWMHD-LPeqns}
\end{align}
where we have used the auxiliary equations and $\op{div}(\eta \bs{B})d^2x\otimes B^\flat=0$ to write 
\begin{align}
\begin{split}
{\cal L}_{u} m &= {\cal L}_{u} V^\flat \otimes \eta d^2x  + V^\flat \otimes {\cal L}_{u} (\eta d^2x ) 
\\
{\cal L}_{B{}}\big( \eta d^2x \otimes B^\flat \big) & =  \op{div}(\eta \bs{B})d^2x\otimes B^\flat + {\cal L}_{B{}}B^\flat \otimes \eta d^2x 
\\& 
=   {\cal L}_{B{}}B^\flat \otimes \eta d^2x  
\\&= \big(\bs{B}\cdot \nabla \bs{B} + \tfrac12\nabla |\bs{B}|^2 \big)\cdot d\bx \otimes \eta d^2x
\,. \end{split}
 \,.\label{RSWMHD-auxi}
\end{align}

Hence, we obtain the RSW-MHD equations in \eqref{RSW-MHD-mot} and \eqref{RSW-MHD-aux} as
\begin{align}
\p_t
\begin{pmatrix}
V^\flat \\ B{} \\ \eta
\end{pmatrix}
&=
-
\begin{bmatrix}
{\cal L}_{u}  V^\flat  +  {\cal L}_{B{}}B^\flat   +   {\rm d} \phi  
\\ - \ad_{u} B{}    
\\ {\cal L}_{u} \eta	
\end{bmatrix}
 .\label{RSWMHD-LPeqns2}
\end{align}

This lecture has used the Euler--Poincar\'e version of Hamilton's principle to derive the equations of Rotating Shallow Water MHD (RSW-MHD) introduced in \cite{gilman2000magnetohydrodynamic}. The Lie-Poisson brackets and Casimirs for the symmetry-reduced Hamiltonian formulation of RSW-MHD have also been 
considered in \cite{dellar2002hamiltonian}. For a review of potential applications of RSW-MHD in Solar plasma physics, see \cite{miesch2005large}. 

\begin{exercise}
Because the initial condition ${\rm div}(\eta \bs{B}) =0$ is advected by the RSW-MHD fluid flow, 
one may define a stream function for $\eta \bs{B} = \zh \times \nabla\psi $ and determine its advection relation. 
Transform variables to eliminate $\bs{B}$ for $\eta^{-1}\zh\times \nabla \psi$ and recalculate the Hamiltonian structure for RSW-MHD
in terms of $(m,\eta.\psi)$. Having calculated the SW-MHD equations explain the physics of its Kelvin circulation theorem. 
\end{exercise}

\begin{answer}
The Hamiltonian in terms of $m=\bs{m}\cdot d\bs{x}\otimes d^2x$, $B{}=\bs{B}\cdot\nabla$, and $\eta d^2x$
in equation \eqref{RSWMHD-Ham} transforms into
\begin{align}
h (m, \eta, B) = \int \Big( \frac{1}{2 \eta^2}  |\bs{m}-\eta\bs{R}/Ro|^2 
+ \frac{1}{2\eta^2 } |\nabla\psi|^2
+ \frac{1}{2 \mathrm{Fr}^2} (\eta-2 \hbar(\bs{x}))\Big) \eta\, d^2 x 
 \,.\label{RSWMHD-Ham-psi}
\end{align}
The variations of this Hamiltonian are obtained as
\begin{align}
\delta h = \int \bs{u}\cdot \delta \bs{m} - J\,\delta \psi + \tilde{\phi} \,\delta \eta 
\, d^2 x 
 \,,\label{RSWMHD-HamVar-psi}
\end{align}
where we have substituted 
\begin{align}
 - {\rm div}(\eta^{-1}\nabla\psi) = {\rm div}(\zh\times \bs{B}) = -\,\zh\cdot {\rm curl}\bs{B} = -\,J
 \,.\label{RSWMHD-HamVar-J}
\end{align}
The modified Bernoulli function $\tilde\phi$ in \eqref{RSWMHD-HamVar-psi} is defined for this case as 
\begin{align}
\tilde\phi := - \,\frac{|\bs{u}|^2}{2} - \frac{1}{2} |\bs{B}|^2 + \frac{1}{ \mathrm{Fr}^2}  (\eta - \hbar(\bs{x}))
 \,.\label{RSWMHD-phi-psi}
\end{align}
The corresponding semidirect-product Lie--Poisson Hamiltonian equations are given by 
\index{semidirect product!Lie Poisson bracket}
\begin{align*}
\p_t
\begin{pmatrix}
m \\ \psi \\ \eta
\end{pmatrix}
&=
-
\begin{bmatrix}
\ad^*_{\Box}m 	&  {\Box} \diamond \psi   & {\Box} \diamond \eta 
\\ {\cal L}_{\Box} \psi    	& 	0		 & 	0
\\ {\cal L}_{\Box} \eta		& 	0		 & 	0
\end{bmatrix}
\begin{pmatrix}
u \\  - J \\ \phi
\end{pmatrix}
\\
\p_t
\begin{pmatrix}
m_i \\ \psi \\ \eta
\end{pmatrix}
&=
-
\begin{bmatrix}
\p_j m_i + m_j\p_i 	&  -\psi_{,i}  & \eta\p_i
\\ \psi_{,j}    	& 	0		 & 	0
\\ \p_j\eta		& 	0		 & 	0
\end{bmatrix}
\begin{pmatrix}
u^j \\  - J \\ \tilde{\phi}
\end{pmatrix}
\\
\p_t
\begin{pmatrix}
\bs{u}\\ \psi \\ \eta
\end{pmatrix}
&=
-
\begin{pmatrix}
- (\bs{u}\cdot \nabla)\bs{u} + 2\bs{\Omega} \times \bs{u}  + \eta^{-1}J\,\nabla\psi  +  \nabla \big(- \frac{1}{2} |\bs{B}|^2 + \tfrac{1}{ \mathrm{Fr}^2}  (\eta - \hbar(\bs{x}))\big)\
\\ (\bs{u}\cdot\nabla) \psi  
\\ {\rm div}(\eta 	 \bs{u} )
\end{pmatrix}
.
\end{align*}
The Kelvin circulation theorem for RSW-MHD takes the following form, cf. equation \eqref{KNthm-RSW-MHD},
\begin{align}
\frac{d}{dt}\oint_{c(u)}\bs{V} \cdot d\bs{x} = -\,\oint_{c(u)} \eta^{-1}J\,\nabla\psi \cdot d\bs{x}
= \oint_{c(u)} (J\,\zh \times\bs{B})\cdot d\bs{x}
\,.\label{RSWMHD-Kelvin-psi}
\end{align}
Thus, perhaps not surprisingly, the $J\,\zh\times\bs{B}=\bs{B}\cdot\nabla\bs{B} - \tfrac12 \nabla|\bs{B}|^2$ magnetic force generates circulation also in the SW-MHD equations.
In particular, this means that the RSW-MHD equations do not advect a potential vorticity.

\end{answer}

\newpage

\newpage
\vspace{4mm}\centerline{\textcolor{shadecolor}{\rule[0mm]{6.75in}{-2mm}}\vspace{-4mm}}
\section{Incompressible 2D MHD Alfv\'en wave turbulence}

\secttoc

\textbf{What is this lecture about?} This lecture introduces the system of planar incompressible Hall magnetohydrodynamic (HMHD) nonlinear Alfv\'en wave 
dynamics, then discusses its subsystems and derives their Hamiltonian formulations. 

\subsection{Magnetohydrodynamics (MHD) systems in planar coordinates $(x,y)$} 

Alfvén waves \cite{alfven1942existence} sustainable in Hall Magnetohydrodynamics (HMHD) are the most common electromagnetic phenomena in magnetised plasmas. 
Nonlinear Alfvén waves play important roles in dispersing energy in a variety of plasma regimes, ranging from laboratory plasmas to space plasmas \cite{biskamp2003magnetohydrodynamic}. 
In particular, the Hall effect in magnetised plasma turbulence plays an important role in regulating the transport of energy in space and astrophysical plasmas; 
as discussed, e.g., \cite{mininni2007energy}. 

Planar Alfvén wave turbulence in quasi-neutral pla\}smas has been modelled by differential equations of the following form \cite{hazeltine1983reduced,hazeltine1985electromagnetic,hazeltine1985shear,holm1985hamiltonian,shukla1984nonlinear},
\begin{align}\begin{split}
& \partial_t \omega +\left\{\omega,\phi\right\}+\left\{A,J\right\}=0, \\ 
& \partial_t A+\{A,\phi\}-\alpha\{A,\chi\}=0, \\
& \partial_t \chi+\{\chi,\phi\}+\left\{A,J\right\}=0\,.
\label{ATurb-modEqs}
\end{split}\end{align}
Where $\omega$ denotes 2D vorticity; $A$ is 2D magnetic `vector' potential; $\chi$ is the `stream function' for the divergence free magnetic field, $\bs{B}=\nabla^\perp A$; and $J=\Delta A$ is the current density $\curl \bs{B} = \Delta A\,\zh$.

The bracket operation $\{\,\cdot\,,\,\cdot\, \}$ in these equations is the canonical Poisson bracket for functions $a, b$ in the $(x,y)$ plane,
$$ 
\{a, b\} := a_x b_y-a_y b_x 
\,.$$
Incompressible ideal fluid flow in the plane preserves the area element; so there is no wonder that it would be Hamiltonian. Under the $L^2$ pairing in $\mathbb{R}^2$ for functions $(a,b,c)$ vanishing at infinity we have the permutation identity
\begin{align}
\int_{\mathbb{R}^2} c \,\{a, b\}\,dx\wedge dy =: \scp{c}{\{a, b\} }
= \scp{a}{\{b, c\} } = \scp{b}{\{c, a\} }
\,.\label{Permutations_abc}
\end{align}
This identity follows easily from integration by parts when one recalls the relation of the canonical Poisson bracket in the plane to the Jacobian $J(a,b)=\{a,b\}$ of an area preserving transformation 
\[
\{a, b\}\,\,dx\wedge dy = J(a,b)\,\,dx\wedge dy = da\wedge db = -\,db\wedge da
\,,\]
so that integration by parts and use of $d^2b=0$ yields
\begin{align*}
\int_{\mathbb{R}^2} c \,\{a, b\}\,dx\wedge dy 
&= \int_{\mathbb{R}^2} c\, da\wedge db
= \int_{\mathbb{R}^2} c\, d (a\, db)
\\&= - \int_{\mathbb{R}^2} dc\wedge(a\, db)
= \int_{\mathbb{R}^2} a \,db \wedge dc 
 = \int_{\mathbb{R}^2} a \,\{b, c\}\,dx\wedge dy 
\,.\end{align*}

Physically in 2D Alfv\'en wave turbulence, the field $\phi(x,y,t)$  in the system \eqref{ATurb-modEqs} represents the electrostatic potential 
which acts as a hydrodynamics stream function 
for Hall drift waves in the plane, whose divergence-free drift velocity $\bs{v}$ with vorticity $\omega= \zh\cdot\curl\bs{v}$ is given respectively by
$$
\bs{v} = \nabla^\perp \phi = (-\phi_y,\phi_x)^T
\quad\hbox{and}\quad
\omega = \Delta \phi
\,.$$

The field $A(x,y,t)$ in the system denotes the normalised magnetic flux potential, so that the magnetic field in the plane $\bs{B}(x,y,t)$ and the current density $J = \zh\cdot\curl\bs{B}$ are given by
$$
\bs{B} =  \nabla^\perp A = (-A_y,A_x)^T
\quad\hbox{and}\quad
J = \Delta A
\,.$$

Finally, the field $\chi(x,y,t)$ in the system \eqref{ATurb-modEqs} denotes the normalised deviation of the charged particle density from its constant equilibrium value.

The details of the physical approximations which lead to the equations in \eqref{ATurb-modEqs}  are discussed in 
references \cite{hazeltine1985electromagnetic,hazeltine1985shear}.

\subsection{Subsystems of the nonlinear planar Alfv\'en wave equations}\label{sec: ATsubsystems} \index{Alfv\'en wave turbulence!subsystems}
A simplification of the nonlinear Alfv\'en wave equations \eqref{ATurb-modEqs} applies to plasma physics 
in the low-beta limit (weak magnetic fields). Subsets of equations \eqref{ATurb-modEqs} include both ideal 
reduced magnetohydrodynamics (RMHD), see e.g. \cite{kadomtsev1973nonlinear,strauss1976nonlinear,strauss1977dynamics}, 
and the Hasegawa-Mima (HM) equation, see e.g. \cite{hasegawa1978pseudo,hasegawa1979nonlinear}.

\textbf{Low-beta RMHD.}
The low-beta RMHD model\footnote{Low beta RMHD means MHD reduced to 2D in which $\beta=|\bs{B}||^2/p\ll1$ where $\beta$ is the ratio of magnetic field intensity to hydrodynamic pressure.}
results upon neglecting the constant parameter $\alpha$ appearing in the second equation of system \eqref{ATurb-modEqs} to find
\begin{align}\begin{split}
& \partial_t \omega +\left\{\omega,\phi\right\}+\left\{A,J\right\}=0, \\ 
& \partial_t A+\{A,\phi\}
= 0\,.
\label{RMHD-modEqs}
\end{split}\end{align}
This limit decouples the field $\chi$ in \eqref{ATurb-modEqs} from the other fields, $\omega$ and $A$. The evolution of $\omega$ and $A$ then constitutes the RMHD system. The RMHD system bas been used to simulate nonlinear shear Alfvén dynamics in tokamaks, see e.g.  \cite{hazeltine1985shear}.

The RMHD system for $\omega$ and $A$ with $\alpha=0$ in the system \eqref{ATurb-modEqs} has the following associated Lie-Poisson bracket 
\begin{equation}
\p_t 
\begin{pmatrix}
\omega \\ A
\end{pmatrix}
= -
\begin{bmatrix}
\{\square, \omega\} & \{\square, A\} \\
\{\square, A\} & 0 
\end{bmatrix}
\begin{pmatrix}
{\delta H/\delta\omega}=-\phi \\ {\delta H/\delta A}=-J
\end{pmatrix}
.\label{LPB-RMHD}
\end{equation}
This Lie-Poisson operator has the same structure as for 2D Euler flow with Hodge 
decomposition in \eqref{LP-vort-dyn-eqn}.

The Hasegawa-Mima (HM) equation is recovered from system \eqref{ATurb-modEqs} by assuming the linear relation,
$$
\phi - \alpha \chi = 0
\,.$$
The HM equation arises physically upon linearizing the adiabatic Maxwell-Boltzman limit for the electrons, see e.g., \cite{hazeltine1983reduced,shukla1984nonlinear}. After assuming this linear relation, the system \eqref{ATurb-modEqs} implies that $\partial_t \chi + \left[A,J\right]=0$ and $\partial_t A=0$. Hence, the field $A$ decouples from the RMHD system in \eqref{RMHD-modEqs}. The difference between the first and third equation in \eqref{ATurb-modEqs} then becomes
\begin{align}
\partial_tq+\{\phi, q \}=0
\quad\hbox{with}\quad
q:=\omega - \chi
\quad\hbox{and}\quad
\chi = \phi/\alpha
\,,
\label{eqn: HM-PV}
\end{align}
which is the HM equation for the electrostatic field. The HM equation describes ideal drift wave turbulence in a low-beta plasma, see e.g. \cite{hasegawa1978pseudo,hasegawa1979nonlinear}. 
\begin{remark}\rm 
The HM equation also has a hydrodynamic interpretation for geostrophic fluid dynamics, for example in oceanic and atmospheric physics, 
where the quantity $q:=\omega - \alpha^{-1} \phi$ is known as \textit{potential vorticity} (PV), see e.g. \cite{stern1975minimal,larichev1976strongly,flierl1980dynamics}.
\end{remark}

\subsection{Hamiltonian structure of nonlinear planar Alfv\'en wave dynamics}
The nonlinear Alfv\'en wave equations \eqref{ATurb-modEqs}  comprises a Hamiltonian system with the quadratic Hamiltonian,
\begin{align} \begin{split}
H(\omega,A,\chi) 
&= \frac12 \int_{\mathbb{R}^2} |\bs{v}|^2 + |\bs{B}|^2 + \alpha \chi^2 \,dxdy 
\\&= \frac12 \int_{\mathbb{R}^2} -\,\omega\Delta^{-1}\omega + |\nabla^\perp A|^2 + \alpha \chi^2 \,dxdy 
\,.\label{ATurb-Ham}
\end{split}\end{align}
Rearranging the equations into Hamiltonian form yields
\begin{align}
\p_t\begin{pmatrix} \omega \\  \\ \chi \\  \\ A \end{pmatrix}
&= - 
	\begin{bmatrix} 
	\{\,{\Box}, \omega\} &  \{\,{\Box}, \chi\}  & \{\,{\Box}, A\}
	\\ \\
	\{\,{\Box}, \chi\} & \{\,{\Box}, \chi\} &  \{\,{\Box}, A\}
	\\ \\
	\{\,{\Box},A\} & \{\,{\Box},A\} & 0
	\end{bmatrix} 
  \begin{pmatrix}
    \frac{\delta H}{\delta \omega} = -\,  \Delta^{-1}\omega =  -\, \phi \\ \\
    \frac{\delta H}{\delta \chi} = \alpha\chi \\ \\
    \frac{\delta H}{\delta A} = -\, \Delta A =   -\,J
    \end{pmatrix}
\label{Poisson-NestedSDP-Part1}
    \\& = -
    \begin{pmatrix} 
    \{\omega,\phi \} + \{\alpha\chi,\chi \} + \{A,J\}
    \\ \\
   \{\chi , \phi-\alpha\chi\}  + \{A,J\}
    \\ \\
   \{A,\phi-\alpha\chi \}
    \end{pmatrix}
.\label{Poisson-NestedSDP-Part2}
\end{align}
\begin{remark}\rm 
The matrix operator in square brackets in \eqref{Poisson-NestedSDP-Part1} defines a Lie-Poisson bracket 
$\{F,H\}(\omega,\chi,A)$ on the dual of the following \emph{nested} semidirect product Lie algebra
\index{Lie algebra!nested semidirect product!$\mathfrak{g}_1\circledS (\mathfrak{g}_2 \oplus \,V) \oplus  (\mathfrak{g}_2 \circledS V)$}
\begin{align}
\mathfrak{g}_1\,\circledS\, (\mathfrak{g}_2\, \oplus \,V) 
\oplus  (\mathfrak{g}_2\, \circledS\, V) 
\,,\label{NestedSDP-LP1}
\end{align}
with dual coordinates $\omega\in \mathfrak{g}^*_1$, $\chi\in \mathfrak{g}^*_2$ and $A\in V^*$. 
Equation \eqref{NestedSDP-LP1} represents the nested direct sum of two semidirect product Lie algebras. 
\index{semidirect product!Lie algebra}
\end{remark}

\begin{remark}\rm 
Exercise \ref{Ex42} provides a finite-dimensional example of such a Lie-Poisson bracket defined on the 
dual of a nested semidirect product Lie algebra.\index{semidirect product!Lie Poisson bracket}
\end{remark}

Here, the semidirect product actions are defined as follows. Let an element of $\mathfrak{g}_1\,\circledS\, (\mathfrak{g}_2\, \oplus \,V)$ be written as  $\left(X_1 ; (X_2;a_2)\right)$. For another element $\left(\bar{X}_1 ; (\bar{X}_2;\bar{a}_2)\right)$ with $(a_2,\bar{a}_2)\in V$ the first semidirect product action in equation \eqref{NestedSDP-LP1} is defined by
\begin{align}\begin{split}
&\Big[\big(X_1 ; (X_2;a_2)\big),\big(\bar{X}_1 ; (\bar{X}_2;\bar{a}_2)\big)\Big]
\\&:=\Big(\big[X_1, \bar{X}_1\big] ; \Big(\big[X_1, \bar{X}_2\big] - \big[\bar{X}_1, X_2\big] + \big[X_2, \bar{X}_2\big] 
; X_1(\bar{X}_2\bar{a}_2) - \bar{X}_1(X_2a_2)
\Big)\Big)\,,
\label{NestedSDP-action}
\end{split}\end{align}
where the action of elements of $\mathfrak{g}_1$ and $\mathfrak{g}_2$ on elements of $V$ are written as concatenation.

With these definitions of the semidirect product actions in equation \eqref{NestedSDP-action}, dual coordinates of \eqref{NestedSDP-LP1} are identified as follows: $\omega\in \mathfrak{g}^*_1$ is dual to $\mathfrak{g}_1$; $\chi\in \mathfrak{g}^*_2$ is dual to $\mathfrak{g}_2$; and $A\in V^*$ is dual to $V$. For further discussion of Poisson brackets in continuum physics which are associated to the duals of semidirect product Lie algebras, see e.g. \cite{HoMaRa1998a}.

The conservation of energy $H(\omega,A,\chi) $ in \eqref{ATurb-Ham} under the dynamics of system \eqref{Poisson-NestedSDP-Part1} is now an immediate consequence of the Hamiltonian formulation and skew symmetry of the Lie-Poisson bracket in \eqref{Poisson-NestedSDP-Part1}.

\textbf{Casimir functionals for $\{F,H\}(\omega,\chi,A)$.} As for the well known conservation of integrals of the potential vorticity 
in \eqref{eqn: HM-PV}, the system \eqref{ATurb-modEqs} conserves Casimir functionals of the form
\begin{align}
C_{FGK}=\int_{\mathbb{R}^2} \big[F(A)+\chi G(A)+K(\omega-\chi)\big] \,dxdy\,,
\label{NestedSDP-Casimirs}
\end{align}
for arbitrary differentiable functions $F, G$, and $K$. For example, taking $F=0, G=0$, and $K(\xi)=\xi^2$ yields the conserved quantity
$$
\int_{\mathbb{R}^2} (\omega-\chi)^2\,dxdy\,,
$$
which is analogous to the enstrophy invariant in GFD for the potential vorticity $q$-equation \eqref{eqn: HM-PV}.

Furthermore, being Casimirs of the Poisson bracket \eqref{Poisson-NestedSDP-Part1}, the functionals $C_{FGK}$ in \eqref{NestedSDP-Casimirs} are conserved in the sense that
$$
\{C_{FGK}, K\}=0, \quad \forall K(\omega,  \chi,A)
\,.$$

That is, the functionals $C_{FGK}$ are conserved for any Hamiltonian $K(\omega,  \chi,A)$, not just for the Hamiltonian $H$ in \eqref{ATurb-Ham}.


\begin{exercise}

Show that the change of field variables in the Lie-Poisson matrix of \eqref{Poisson-NestedSDP-Part1}
from $(\omega,\chi,A)$ to $(\mu, \chi, A)$  with $\mu=\omega-\chi$ yields the following  Lie-Poisson operator,
\begin{align}
\p_t\begin{pmatrix}\mu \\  \\ \chi \\  \\ A \end{pmatrix}
&= -
	\begin{bmatrix} 
	\mathrm{ad}^*_{\Box} \mu &  0  & 0
	\\ \\
	0 & \mathrm{ad}^*_{\Box}\chi &  \mathrm{ad}^*_{\Box}  A
	\\ \\
	0 & \mathrm{ad}^*_{\Box}A & 0
	\end{bmatrix} 
  \begin{pmatrix}
    \frac{\delta H}{\delta \mu}  \\ \\
    \frac{\delta H}{\delta  \chi}  \\ \\
    \frac{\delta H}{\delta  A} 
    \end{pmatrix},
\label{Poisson-NestedSDP-right-diag}
\end{align}
which is dual to the direct sum of volume-preserving vector fields $\mathfrak{g}_1\in\mathfrak{X}(\mathbb{R}^2)$ 
and the semidirect product Lie algebra of volume-preserving vector fields $\mathfrak{g}_2\in\mathfrak{X}(\mathbb{R}^2)$ 
acting on the vector space of scalar densities, $V$. Namely,
\index{semidirect product!Lie Poisson bracket}
\begin{align}
\mathfrak{g}_1 \oplus  (\mathfrak{g}_2\, \circledS\, V) 
\,.\label{NestedSDP-LP2}
\end{align}
with dual coordinates $\mu\in \mathfrak{g}^*_1$, $\chi\in \mathfrak{g}^*_2$ and $A\in V^*$. 

\end{exercise}

\begin{answer}
To prove this statement, notice that the Jacobian matrix for the transformation from $(\omega,\chi,A)$ to $(\mu, \chi, A)$  with $\mu=\omega-\chi$ is given by
\begin{align*}
J = 
	\begin{bmatrix} 
	1 &  -1\  & 0
	\\ 
	0 & \ 1\ &  0
	\\ 
	0 & \ 0\ &  1
	\end{bmatrix}.
\end{align*}
Multiplying the Poisson matrix in \eqref{Poisson-NestedSDP-Part1} by Jacobian $J$ from the left
 and by its transpose $J^T$ from the right produces the transformed Poisson matrix in 
 \eqref{Poisson-NestedSDP-right-diag}. This linear change of variables preserves the eigenvalues 
 of the Poisson matrix. In particular, such linear transformations preserve the matrix null eigenvectors.
 Hence, this linear change of variables preserves Casimirs.  

Consequently, one may check that the variational derivatives of the following functions 
$C_1,C_2,C_3$ are Casimirs, 
\begin{align}
C_1 = F_1(\mu)\,,\quad C_2 = \chi F_2(A)\quad \hbox{and}\quad C_3 = F_3(A),
\label{Poisson-NestedSDP-Casimirs}
\end{align}
where $F_1$, $F_2$ and $F_3$ are arbitrary differentiable functions of their arguments. 
That is, their variational derivatives are null eigenvectors of  Lie-Poisson bracket in 
\eqref{Poisson-NestedSDP-right-diag} as well as the transformed Lie-Poisson bracket 
in \eqref{Poisson-NestedSDP-right-diag}. This transformation mirrors the discussion in
section \ref{sec: ATsubsystems} of the physical arguments for reducing the system of Alfv\'en turbulence 
equations to simpler subsystems. 

\end{answer}


\begin{exercise}

Show that exercise \ref{Ex42} also admits a finite-dimensional example of transforming a Lie-Poisson bracket defined on the 
dual of a nested semidirect product Lie algebra into another Lie-Poisson bracket defined on the 
dual of a direct sum as in \eqref{Poisson-NestedSDP-right-diag}.\index{semidirect product!Lie Poisson bracket!nested}

\end{exercise}

\newpage


\newpage
\vspace{4mm}\centerline{\textcolor{shadecolor}{\rule[0mm]{6.75in}{-2mm}}\vspace{-4mm}}
\section{L\"ust Hall Magnetohydrodynamics (LHMHD)}\index{L\"ust Hall Magnetohydrodynamics!LHMHD}

\secttoc

\textbf{What is this lecture about?} This lecture demonstrate the process of constructing 
step-by-step the series of nested Lie-Poisson brackets for ideal hydrodynamics equations 
leading successively from Euler's fluid equations, to MHD, to Hall MHD (HMHD) and then to L\"ust Hall MHD (LHMHD).
In this series of models, each step breaks the original symmetry of ideal hydrodynamics further to introduce additional
physical effects and additional semidirect-product Poisson structures. 
For simplicity, we study reduced solutions depending only on planar $(x,y)$ coordinates.
\index{semidirect product!Lie Poisson bracket}

\subsection{Incompressible L\"ust Hall MHD (LHMHD) in two dimensions}

The Hall effect in a quasi-neutral plasma is a classical problem in plasma physics \cite{biskamp2003magnetohydrodynamic,lighthill1960studies}. 
The Hall effect produces an additional advective drift of the magnetic field lines induced by the electron fluid motion. 
The Hall drift velocity is proportional to the current density, which generates the magnetic force, that in turn generates 
fluid vorticity leading to MHD turbulence. 
The Hall effect is particularly interesting in the physics of the solar wind and space weather because it can both create 
small-scale structures that seed high-wavenumber instabilities and also create energy and magnetic field transfer from 
smaller scales to larger scales, \cite{mininni2007energy}. 

The nonlinear interactions of the slow-fast, large-small, resolved-unresolved decomposition of the mean and the fluctuating or turbulent components of the physical processes in MHD plasmas require a structure-preserving approach. Geometric mechanics models Hall MHD as a two-fluid ion-electron plasma, \cite{holm1987hall,holm1987superfluid}. 

L\"ust Hall magnetohydrodynamics (LHMHD) is a further generalisation that accounts for both the impact of Hall drift velocity and the influence of electron inertial effects \cite{lust1959ausbreitung}. 


Upon introducing the following incompressible flow Ansatz for 3D velocity $\bu(x,y,t)$ and magnetic field $\mb{B}(x,y,t)$,
LHMHD solutions which depend only on planar $(x,y)$ coordinates take the following form
\begin{align}
\begin{split}
    \mb{B}(x,y,t) = \nabla {A}(x,y,t)\times \wh{z} + b(x,y,t)\wh{z}\,,\\
    \bu(x,y,t) = -\nabla \phi(x,y,t)\times \wh{z} + v(x,y,t)\wh{z}\,.
\end{split}
\label{RLHMHD-solns}
\end{align}
The full 3D LHMHD equations originally derived in \cite{lust1959ausbreitung} may be written for the solutions in \eqref{RLHMHD-solns} 
depending only on planar $(x,y)$ coordinates via the canonical Poisson bracket $\{\,\cdot\,,\,\cdot\,\}$ as 
\cite{holm2024deterministic}
\begin{align}
    \begin{split}
        &\p_t \omega + \pb{\phi}{\omega} + \pb{\triangle{A}}{{A}} - \gamma^2\pb{\triangle^2{A}}{{A}} - \gamma^2\pb{b}{\triangle b} = 0\,,\\
        &\p_t v + \pb{\phi}{v} + \pb{{A}}{b} - \gamma^2\pb{{A}}{\triangle b} = 0\,,\\
        &\p_t {A} + \pb{\phi}{{A}} - \frac{R}{a}\pb{{A}}{b} + \frac{\gamma^2 R}{a}\pb{{A}}{\triangle b} = 0\,,\\
        &\p_t b + \pb{\phi}{b} + \pb{{A}}{v} - \frac{R}{a}\pb{\triangle {A}}{{A}} + \frac{\gamma^2 R}{a}\pb{\triangle^2 {A}}{{A}} - \frac{\gamma^2 R}{a}\pb{\triangle b}{b} = 0
    \,,\end{split}
\label{LRMHD-Ham}
\end{align}
in which the Hall parameter $R/a$ and the inertial parameter $\gamma^2$ are prescribed constants.
In the reduced dynamical variables in \eqref{RLHMHD-solns}, the conserved energy Hamiltonian for ideal 3D LHMHD in the literature, e.g., \cite{holm2024deterministic}, 
may be expressed as
\begin{align}
    H = \frac{1}{2}\int_{\mcal{D}} - \phi \omega + v^2 + \left(-{A}\triangle{A} + b^2\right) + \gamma^2\left(-b\triangle b + (\triangle {A})^2\right)\,d^2x\,.
\label{2d-LHMHD-Ham}
\end{align}
The variational derivatives of this Hamiltonian are given by
\begin{align}
    \begin{split}
        \frac{\delta H}{\delta \omega} = -\phi
        \,,\quad 
        \frac{\delta H}{\delta v} = v
        \,,\quad 
        \frac{\delta H}{\delta b} = b - \gamma^2\triangle b
        \,, \quad 
        \frac{\delta H}{\delta {A}} = - \triangle{A} + \gamma^2\triangle^2 {A}\,.
    \end{split}
\end{align}
In terms of these Hamiltonian variations, the reduced LHMHD equations take the following Lie-Poisson form
\begin{align}
    \begin{split}
        &\p_t \omega + \pb{\phi}{\omega} - \pb{\frac{\delta H}{\delta {A}}}{{A}} - \pb{\frac{\delta H}{\delta b}}{b} = 0\,,\\
        &\p_t v + \pb{\phi}{v} - \pb{\frac{\delta H}{\delta b}}{{A}} = 0\,,\\
        &\p_t b + \pb{\phi}{b} + \pb{{A}}{v} + \frac{R}{a}\pb{\frac{\delta H}{\delta {A}}}{{A}} + \frac{R}{a}\pb{\frac{\delta H}{\delta b}}{b} = 0\,,\\
        &\p_t {A} + \pb{\phi}{{A}} + \frac{R}{a}\pb{\frac{\delta H}{\delta b}}{{A}} = 0\,.
    \end{split}
\end{align}
One may write these equations in terms of a Lie-Poisson operator as
\begin{align}
    \p_t 
    \begin{pmatrix}
        \omega \\  v \\ b \\ {A}
    \end{pmatrix}
    =
    \begin{bmatrix}
        \pb{\Box}{\omega} & \pb{\Box}{v}  & \pb{\Box}{b} & \pb{\Box}{{A}}\\
        \pb{\Box}{v} & 0 & \pb{\Box}{{A}} & 0\\
        \pb{\Box}{b} & \pb{\Box}{{A}} & - Ra^{-1}\pb{\Box}{b} & - Ra^{-1}\pb{\Box}{{A}}\\
        \pb{\Box}{{A}} & 0 & - Ra^{-1}\pb{\Box}{{A}} & 0
    \end{bmatrix}
    \begin{pmatrix}
        -\phi \\ \delta H/\delta v  \\ \delta H/\delta b \\ \delta H/ \delta {A}
    \end{pmatrix}
\label{2dLMHD-LPop}
\end{align}
The Lie-Poisson bracket implied by this Lie-Poisson matrix is dual of the following 
nested semidirect product of two semidirect product Lie algebras \index{Lie algebra!nested semidirect product}
\index{Lie algebra!nested semidirect product!$\mathfrak{s}_1 \circledS \mathfrak{s}_2\quad \mathfrak{s}_i = \mathfrak{g}_i\circledS V_i $}
\begin{align}
    \mathfrak{s}_1 \circledS \mathfrak{s}_2\,, 
    \quad \text{where} \quad 
    \mathfrak{s}_i = \mathfrak{g}_i\circledS V_i
    \quad i=1,2
    \,.
\label{2nestedSDP-LMHD}
\end{align}

The skew symmetry of the Lie-Poisson operator \eqref{2dLMHD-LPop} under $L^2$ pairing guarantees that the Hamiltonian $\mathrm{H}(\omega,{A},b,v) $ in \eqref{2d-LHMHD-Ham} will be preserved by the dynamics it generates through the Lie-Poisson bracket associated with the Lie-Poisson matrix operator in \eqref{2dLMHD-LPop}. 

\begin{remark}\rm \textbf{Casimir integrals for reduced L\"ust Hall MHD.} \index{L\"ust Hall Magnetohydrodynamics!Casimirs}
Casimir integrals are conserved by any Hamiltonian, because their variational derivatives are null eigenvectors of the corresponding Lie-Poisson operator. In particular, the Casimir integrals of the Lie-Poisson bracket arising from the Lie-Poisson matrix in \eqref{2dLMHD-LPop} comprise a set of constants of motion for the system of 2D reduced L\"ust Hall MHD equations appearing in \eqref{LRMHD-Ham},
\[
C_\Phi(\omega,v) = 
\int_{\cal D} \omega \Phi_1(v) + \Phi_2(v)\,d^2x
\quad\hbox{and}\quad
C_\Theta(b,{A}) = 
\int_{\cal D} b \Theta_1({A}) + \Theta_2({A})\,d^2x
\,,\]
where $\Phi_1(v),\Phi_2(v),\Theta_1({A}),\Theta_2({A})$
are differentiable functions of their respective arguments.
\end{remark}

\subsection{Known subsystems of the reduced LHMHD equations}\label{sec: LHMDsubsystems}
\begin{itemize}
\item
The reduced Hall MHD equations emerge in the limit $\gamma^2 \rightarrow 0$ in the reduced LHMHD Hamiltonian in \eqref{2d-LHMHD-Ham}. 
These equations may also be obtained by modifying the Poisson structure in \eqref{2dLMHD-LPop} into block diagonal semidirect product form as follows
\index{semidirect product!Lie Poisson bracket}
\begin{align}
    \p_t 
    \begin{pmatrix}
        \omega \\  v \\ b \\ {A}
    \end{pmatrix}
    =
    \begin{bmatrix}
        \pb{\Box}{\omega} & \pb{\Box}{v}  & 0 & 0 \\
        \pb{\Box}{v} & 0 & 0 & 0 \\
        0 & 0 & - Ra^{-1}\pb{\Box}{b} & - Ra^{-1}\pb{\Box}{{A}}\\
        0 & 0 & - Ra^{-1}\pb{\Box}{{A}} & 0
    \end{bmatrix}
    \begin{pmatrix}
        -\phi \\ \delta H/\delta v  \\ \delta H/\delta b \\ \delta H/ \delta {A}
    \end{pmatrix}.
\label{2dLMHD-LPop-Hall-MHD}
\end{align}

\item
The reduced inertial MHD equations emerge in the limit $R/a \rightarrow 0$ in the Poisson operator in \eqref{2dLMHD-LPop} as
\begin{align}
    \p_t 
    \begin{pmatrix}
        \omega \\  v \\ b \\ {A}
    \end{pmatrix}
    =
    \begin{bmatrix}
        \pb{\Box}{\omega} & \pb{\Box}{v}  & \pb{\Box}{b} & \pb{\Box}{{A}}\\
        \pb{\Box}{v} & 0 & \pb{\Box}{{A}} & 0\\
        \pb{\Box}{b} & \pb{\Box}{{A}} & 0 & 0 \\
        \pb{\Box}{{A}} & 0 & 0 & 0
    \end{bmatrix}
    \begin{pmatrix}
        -\phi \\ \delta H/\delta v  \\ \delta H/\delta b \\ \delta H/ \delta {A}
    \end{pmatrix}.
\label{2dLMHD-inertial}
\end{align}
\item
In the absence of the variable $v$, the Lie-Poisson matrix operator in \eqref{2dLMHD-LPop} reduces to the Lie-Poisson matrix operator in \eqref{Poisson-NestedSDP-Part1} derived in \cite{hazeltine1985shear,hazeltine1985electromagnetic} for modelling Alfv\'en waves in plasmas,
\begin{align}
    \p_t 
    \begin{pmatrix}
        \omega \\ b \\ {A}
    \end{pmatrix}
    =
    \begin{bmatrix}
        \pb{\Box}{\omega}   & \pb{\Box}{b} & \pb{\Box}{{A}}\\
        \pb{\Box}{b}  & - Ra^{-1}\pb{\Box}{b} & - Ra^{-1}\pb{\Box}{{A}}\\
        \pb{\Box}{{A}}  & - Ra^{-1}\pb{\Box}{{A}} & 0
    \end{bmatrix}
    \begin{pmatrix}
        -\phi   \\ \delta H/\delta b \\ \delta H/ \delta {A}
    \end{pmatrix}.
\label{2dLMHD-LPop-Alfven}
\end{align}

\item
The reduced MHD equations \cite{kadomtsev1973nonlinear,strauss1976nonlinear,strauss1977dynamics} 
arise in this notation when $b$ is absent in \eqref{2dLMHD-LPop-Alfven} and with a relative minus sign convention, 
\begin{equation}
\p_t 
\begin{pmatrix}
\omega \\ {A}
\end{pmatrix}
= - 
\begin{bmatrix}
\{\square, \omega\} & \{\square, {A}\} \\
\{\square, {A}\} & 0 
\end{bmatrix}
\begin{pmatrix}
{\delta H/\delta\omega}=-\phi \\ {\delta H/\delta {A}}=-J
\end{pmatrix}
.\label{LPB-RMHD-LHMHD}
\end{equation}

\item
Finally, when all magnetic effects are suppressed, one returns to the Poisson structure for the vorticity dynamics 
of the planar Euler fluid equations. 

\end{itemize}

\begin{exercise}
Show that when $v$ is absent in the reduced LHMHD Lie--Poisson operator in \eqref{2dLMHD-LPop}, then it reduces to the Alfv\'en Lie--Poisson operator in \eqref{Poisson-NestedSDP-Part1}.
\end{exercise}

{\begin{exercise}
Show that Exercise \ref{Ex3NestedSDP} in paragraph \ref{subsec-5.4} provides a finite-dimensional example of a 
Lie-Poisson operator dual to the nested semidirect product of three semidirect product Lie algebras, 
as opposed to equation \eqref {2nestedSDP-LMHD} for LHMHD which involves 
the nested semidirect product of two semidirect product Lie algebras.
\end{exercise}

\subsection{Summary: whorls within whorls, within whorls}\index{whorls}
``Big whorls have little whorls, ..., and so on to viscosity." 
The whorls in this rhyme verse by L. F. Richardson \cite{richardson1922weather} refer to coherent circulations governed by Euler's fluid equation [1765] for ideal volume-preserving fluid flow. 
Euler's equation implies Kelvin's theorem \cite{kelvin1869vortex}:  that time-dependent Euler fluid flows ${\phi_t}$ preserve functionals of the  fluid velocity 1-form $u^\flat:=\bs{u}\cdot d\bs{x}$ defined on a space of loops $c_t={\phi_t}_*c_0$ pushed forward by $\phi_t$,
\[
\frac{d}{dt} \oint_{{\phi_t}_*c_0} u^\flat 
= 
\oint_{{\phi_t}_*c_0} (\p_t + {\cal L}_{u})u^\flat = 0
\,.\]
Here, ${\phi_t}_*c_0$ with $\phi_0=Id$ is the push-forward of the initial material loop  $c_0$ by the smooth invertible flow map $\phi_t$ depending on time $t$ 
\footnote{Semidirect-product action $G\circledS V$ of smooth invertible maps $G$ on vector spaces $V$ comprises the configuration manifold of Euler flow. Eulerian fluid dynamics is defined on $T(G \circledS V)/G =\mathfrak{g} \circledS V$ and $T^*(G \circledS V)/G = \mathfrak{g}^* \circledS V$.}
The proof of Kelvin's theorem follows from the \emph{Lie chain rule}  for the push-forward, recalled here from equation \eqref{LieChainRule-push},
\[
\frac{d}{dt}{\phi_t}_*u^\flat 
= - \,{\phi_t}_*({\cal L}_{u} u^\flat)
\,.\]
The Lie chain rule defines the coordinate-free \emph{Lie derivative} ${\cal L}_{u}$ as (minus) the tangent to the push-forward along the vector field $u := \bs{u}\cdot \nabla$ which generates the flow $\phi_t$ 
\footnote{In general, one uses Lie-derivative and differential  notation to make one's calculations coordinate-free.
This notation is appropriate here, since solutions of Euler's equations for homogeneous fluids follow geodesic paths on the manifold of smooth volume preserving invertible maps (diffeomorphisms) with respect to the metric given by the fluid kinetic energy \cite{arnold2014differential}.}. 

Vector-space valued advected fluid quantities evolve by push-forward $a_t={\phi_t}_*a_0\in V$ and satisfy $(\p_t + {\cal L}_{u})a_t = 0$. Their introduction reduces the Lie group symmetry of the Lagrangian $\ell(u,a_t)$ in Hamilton's principle $\delta S = \delta \int_0^T\ell({u},a_t)dt$ for Euler flow to the isotropy subgroup of the initial condition $a_0$. The remaining symmetry leads to the Kelvin-Noether theorem \cite{HoMaRa1998a}
\[
\frac{d}{dt} \oint_{{\phi_t}_*c_0} u^\flat 
= 
\oint_{{\phi_t}_*c_0} (\p_t + {\cal L}_{u})u^\flat 
= 
\oint_{{\phi_t}_*c_0} \frac{\delta \ell}{\delta a_t}\diamond a_t
\]
in which the diamond operation $(\diamond)$ is defined in terms of two pairings:
$\scp{b\diamond a}{\xi}_{\mathfrak{g}^*\times\mathfrak{g}} 
:= \scp{b}{-{\cal L}_{\xi} a_t}_{V^*\times V}$ for $\xi\in \mathfrak{X}({\cal D})$
a vector field defined on the flow domain, ${\cal D}$. 

The Lie symmetry-reduced Legendre transformation defines the corresponding Hamiltonian 
\[
h(m,a):=\scp{m}{u}_{\mathfrak{g}^*\times\mathfrak{g}}- \ell({u},a)
\quad\hbox{\&}\quad
m:= { \de\ell }/{\de u}
\]
on the Poisson manifold $(m,a)\in P=\mathfrak{g}^*\circledS V$, where $\circledS$ denotes semidirect action of the Lie algebra of vector fields $\mathfrak{g}=\mathfrak{X}({\cal D})$ on the vector space of advected quantities $a\in V({\cal D})$ with Lie-Poisson bracket in $L^2$ pairing \index{Poisson manifold!semidirect product} \index{semidirect product!Poisson manifold}
\[
\{ f , h \}_{LP} = - \scp{\mu}{[\de f / \de (m,a),\de h / \de (m,a)]}_{L^2}
\,.\]
Its semidirect-product (SDP) dynamics is denoted as
    \begin{equation*}
    \label{eqn:LPB}
    \frac{df}{dt} =
    \{ f , h \}_{LP}
    = - 
    \begin{pmatrix}
    \de f / \de m \\  \de f / \de a
    \end{pmatrix}^{\!\!T}
    \begin{bmatrix}
    {\cal L}_{\Box}m & \Box \diamond a 
    \\
    {\cal L}_{\Box}a & 0
    \end{bmatrix}
    \begin{pmatrix}
    \de h / \de m \\  \de h / \de a
    \end{pmatrix}.
    \end{equation*}
This is the Lie-Poisson Hamiltonian formulation of ideal fluid dynamics on Poisson 
manifolds obtained via Lie-symmetry reduction of Hamilton's principle in \cite{HoMaRa1998a}.

\textbf{Nested semidirect-product action for MHD plasma dynamics.}
In plasma dynamics approximated by the magnetohydrodynamics (MHD) equations, the magnetic field breaks Euler symmetry and is advected by the fluid motion map $\phi_t$. Hence, the Lie-Poisson Hamiltonian formulation derived in \cite{HoMaRa1998a} applies to MHD \cite{holm1987hall}.

Allowing for finite electron Larmor radius breaks the symmetry further, as the electrons in Hall MHD are regarded as a second fluid whose charge density is carried by the B-field. This is the classical Hall effect. The Hall effect shifts momentum to account for the composition of flow maps as the fluid velocity carries the B-field lines and the B-field carries the electron charge density. Thus, Hall MHD produces whorls within whorls, within whorls \cite{holm2023lagrangian}. \index{whorls!composition of maps} \index{composition of maps!whorls}

However, at the level of Hall MHD the Poisson bracket is block~diagonal. When  Hall MHD is extended further to L\"ust Hall MHD to account for finite ion Larmor radius effects, the remaining off-diagonal parts of the Lie-Poisson bracket get involved, as the nested SDP bracket \eqref{2nestedSDP-LMHD}  becomes fully entwined by the additional physics of finite ion Larmor radius. 

In summary, this lecture has demonstrated the process of constructing step-by-step the series of nested Lie-Poisson brackets for ideal hydrodynamics equations leading successively from Euler's fluid equations, to MHD, to Hall MHD and then to L\"ust Hall MHD, after having reduced the full versions of the theory from 3D on $(x,y,z)\in\mathbb{R}^3$ to 2D on $(x,y)\in\mathbb{R}^2$ by using the Hodge decomposition in equation \eqref{Hodge-decomp-2D}. For a review of potential applications of LHMHD in plasma physics, see \cite{miesch2005large}. For derivations using Hamilton's principle of the Hamiltonian dynamics of a charged fluid, including electro- and magnetohydrodynamics, see \cite{holm1986hamiltonian,holm1987hamiltonian}. For recent developments to make the transport operator in Hall MHD stochastic, see \cite{holm2024deterministic}.

\appendix

\newpage

\section{Appendix: Geometric Mechanics -- Definitions and Topics}\label{course-outline-app}

\subsection{A sketch of the lecture topics}

\begin{itemize}
\item Spaces -- Smooth Manifolds: locally $\mathbb{R}^n$ spaces on which the rules of calculus apply
\item Motion -- Flows $\phi_t\circ\phi_s=\phi_{t+s}$ of Lie groups acting on
smooth manifolds 
\item Laws of Motion and discussion of solutions
\item Newton's Laws
     \begin{itemize}
     \item
     Newton: $d\mathbf{p}/dt=\mathbf{F}$, for momentum $\mathbf{p}$ and prescribed force $\mathbf{F}$
(on $\mathbb{R}^n$ historically)
     \item
     Lie group invariant variational principles
          \begin{itemize}
          \item
          Euler--Lagrange equations: optimal ``action'' (Hamilton
principle)
          \item
          Geodesic motion -- optimal dynamics with respect to kinetic energy metric
          \end{itemize}
     \end{itemize}

     \item
     Lagrangian  and Hamiltonian Formalism
          \begin{itemize}
          \item Newton's Law of motion
          \item Euler--Lagrange theorem
          \item Noether theorem (Lie symmetries imply conservation laws)
          \item Euler--Poincar\'e theorem
          \item Kelvin--Noether theorem
          \end{itemize}
     \item
     Applications and examples
          \begin{itemize}
          \item Geodesic motion on a Riemannian manifold\footnote{A Riemannian
manifold is a smooth manifold $Q$ endowed with a symmetric nondegenerate 
covariant tensor $g$, which is positive definite. }
          \item Rigid body -- geodesic motion on Lie group $SO(3)$
          \item Other geodesic motion, for example, Riemann ellipsoids on Lie group $GL(3,R)$
          \item Heavy top
          \end{itemize}
     \item
     Lagrangian mechanics on Lie groups and Euler--Poincar\'e (EP) equations
          \begin{itemize}
          \item EP$(G)$, EP equations for geodesics on a Lie group $G$
          \item EPDiff$(\mathbb{R})$ for geodesics on Diff$(\mathbb{R})$
          \item Pulsons, the singular solutions of EPDiff$(\mathbb{R})$)
with respect to any norm
          \item Peakons, the singular solitons for
EPDiff$(\mathbb{R},H^1)$, with respect to
                the $H^1$ norm
          \item EPDiff$(\mathbb{R}^n)$ and singular geodesics
          \item Diffeons and momentum maps for EPDiff$(\mathbb{R}^n)$
          \end{itemize}
     \item
     Euler--Poincar\'e (EP) equations for continua
          \begin{itemize}
          \item EP semidirect-product reduction theorem
          \item Kelvin--Noether circulation theorem
          \item EP equations with advected parameters for geophysical fluid dynamics
          \end{itemize}
     \end{itemize}

%

\subsection{Hamilton's principle of stationary action}
\begin{tabular}{l|l}
\textbf{Lagrangians on $T\mathbb{R}^{3N}$} & \textbf{$G$--invariant Lagrangians on $TG$} \\ 
\hline   
Euler--Lagrange equations & Euler--Poincar\'e equations \\
Noether's theorem & Kelvin--Noether theorem \\
Symmetry $\implies$ cons. laws & Conservation laws emerge \\
Legendre transformation & Legendre transformation \\ 
Hamilton's canonical equations & Lie--Poisson equations \\
Poisson brackets & Lie--Poisson brackets \\
Symplectic manifold & Poisson manifold \\
Symplectic momentum map & Cotangent-lift momentum map \\
Reduction by symmetry & Reduction to coadjoint orbits
\end{tabular}

\subsection{Motivation for the geometric approach}

We begin with a series of outline sketches to motivate the geometric approach taken in the course and explain more about its content.

\subsubsection{Why is the geometric approach interesting to know?}
\begin{itemize}
\item Defines problems on manifolds
\begin{itemize}
\item coordinate-free
\begin{itemize}
\item needn't re-do calculations when changing coordinates
\item more compact
\item unified framework for expressing ideas and using symmetry
\end{itemize}
\end{itemize}
\item ``First principles'' approach
\begin{itemize}
\item variational principles - systematic -- unified approach \\
for example, similarity between tops and fluid dynamics (semi-direct product), and MHD, and \dots
\item POWER \\
Geometric constructions can give useful answers without us having to find and work with
complicated explicit solutions. For example, classifying rigid body equilibria and fluid equilibria, for example,
as critical points of the sum over their constants of motion, and then taking second variations to  derive conditions 
for their stability. 
\end{itemize}
\end{itemize}

\subsection{Potential Masters/PhD Research Topics in Geometric Mechanics}
\subsubsection*{Rigid body}
     \begin{itemize}
     \item
     Euler--Lagrange and Euler--Poincar\'e equations
     \item
     Kelvin--Noether theorem 
     \item
     Lie--Poisson bracket, Casimirs and  coadjoint orbits
     \item
     Reconstruction and momentum maps
     \item
     The symmetric form of the rigid body
     equations ($\dot{Q}=Q\Omega$, $\dot{P}=P\Omega$)
     \item
     $\mathbb{R}^3$ bracket and intersecting level surfaces
\[
\mathbf{\dot{x}}=\nabla{C}\times\nabla{H}
=\nabla(\alpha C + \beta H)\times\nabla(\gamma C + \epsilon H)
,
\quad\hbox{for}\quad
\alpha\epsilon-\beta\gamma=1
\]
\end{itemize}

Examples:
\begin{enumerate}
\item[(1)] Rigid body motion: rotations and translations.
\item[(2)] Fermat's principle and ray optics,
\item[(3)] Multicomponent dynamics,
\item[(4)] Composition of maps,
\end{enumerate}

\begin{itemize}
     \item
     $SU(2)$ rigid body, Cayley--Klein parameters and Hopf fibration 
     \item
     Higher dimensional rigid bodies
     \begin{itemize}
     \item Manakov integrable top on $O(n)$ and its spectral problem
     \end{itemize}
     \end{itemize}
\subsubsection*{Heavy top -- symmetry breaking}
     \begin{itemize}
     \item
     Euler--Poincar\'e variational principle for the heavy top
     \item 
     Kaluza--Klein formulation of the heavy top 
     \end{itemize}
\subsubsection*{Geometric formulations of physical models}
     \begin{itemize}
     \item Kirchhoff elastica, underwater vehicles, liquid crystals, stratified
     flows, polarization dynamics of telecom optical pulses, ideal fluid flows, geophysical fluid dynamics
\end{itemize}

\subsubsection*{General theory}
     \begin{itemize}
     \item
     Euler--Poincar\'e semidirect-product reduction theorem
     \item 
     Semidirect-product Lie--Poisson formulation
     \end{itemize}
     
\subsubsection*{Shallow water waves. and ideal fluid flows}
     \begin{itemize}
     \item 
     KdV and CH equation -- solitons and peakons (geodesics on the Bott-Virasoro group)
     \item 
     EPDiff equation -- (geodesics on manifold of smooth invertible maps)
     \end{itemize}
\subsubsection*{Fluid dynamics}
     \begin{itemize}
     \item
     Euler--Poincar\'e variational principle for incompressible and compressible ideal fluids
     \item
     Preparation for Stochastic Geometric Mechanics (discussed elsewhere, see \cite{holm2024deterministic})
     \end{itemize}

\subsection{Course outline}
\begin{itemize}
\item{Geometric Structures in Classical Mechanics}
\begin{itemize}
\item Vocabulary of counterpoints between mathematics and physics
\item Smooth manifolds
\begin{itemize}
\item locally $\mathbb{R}^n$ spaces which admit calculus, tangent vectors, action principles
\end{itemize}
\item Lie groups: Groups of transformations that depend smoothly on a set of parameters. \\
Lie groups are also manifolds.
\begin{itemize}
\item uniqueness of Lie group product implies flow property $\phi_{t+s} = \phi_t\circ\phi_s$
\item Lie symmetries encode conservation laws into geometry
\end{itemize}
\item Variational principles with Lie symmetries
\begin{itemize}
\item Euler--Lagrange equations $\rightarrow$ Euler--Poincar\'e equations (compact)
\item Two formulations (Lagrangian \& Hamiltonian) which are mutually beneficial

\begin{tabular}{l|l}
Lagrangian side: & Hamiltonian side \\ 
\hline   
Hamilton's principle & Lie--Poisson brackets \\
Noether's theorem & Lie symmetries $\implies$ cons. laws \\
Lie symmetries $\implies$ cons. laws & Momentum maps \\
composed of momentum maps & Jacobi identity
\end{tabular}

\end{itemize}
\end{itemize}

\item Geometric Formulations of Physical Applications and Modelling
\begin{itemize}
\item Newtonian mechanics
\item rigid body
\item heavy tops -- integrable cases
\item geometric ray optics
\item fluids -- Kelvin-Noether theorem
\item waves $\begin{cases} 
\text{shallow water waves} \\
\text{solitons}
\end{cases}$
\item Dispersive shallow water waves
\item Magnetohydrodynamics (MHD)
\item Magnetohydrodynamic rotating shallow water
\item 2D Hall MHD and Extended Hall MHD
\end{itemize}
\end{itemize}

\newpage
%

\vspace{4mm}\centerline{\textcolor{shadecolor}{\rule[0mm]{6.75in}{-2mm}}\vspace{-4mm}}


\bibliographystyle{siam}
\bibliography{link}


\printindex

\end{document}